%% file: main.tex
\documentclass[11pt,a4paper]{article}

\usepackage[top=12mm,bottom=12mm,left=30mm,right=30mm,head=12mm,includeheadfoot]{geometry}
\usepackage{microtype}

\usepackage{tocloft}

\usepackage[nottoc,notlot,notlof]{tocbibind}

\usepackage{titlesec}
\titlespacing*{\section}{0pt}{1.8\baselineskip}{\baselineskip}

\usepackage[utf8]{inputenc}
\usepackage[bitstream-charter]{mathdesign}
\usepackage[english]{babel}
\usepackage[T1]{fontenc}

\usepackage{graphicx}
\graphicspath{{./images/}}
\usepackage{booktabs}
\usepackage{longtable}
\usepackage{multirow}
\usepackage[width=.90\textwidth]{caption}

\usepackage{amsthm,amsmath,amssymb,amsfonts,bbm}
\usepackage{mathtools}
\usepackage{physics}
\usepackage{braket}
\usepackage{slashed}
\usepackage{relsize,exscale}
\usepackage{stackrel}
\usepackage{bm}
\usepackage{cancel}
\usepackage{youngtab}

\usepackage{setspace}
\usepackage{enumitem}
\usepackage{multicol}
\usepackage{mdframed}
\usepackage{tikz}
\usepackage{xcolor}
\usepackage{cite}

\usepackage{hyperref}
\hypersetup{
    colorlinks=true,
    linkcolor={red!50!black},
    citecolor={blue!50!black},
    urlcolor={blue!80!black}
}
\usepackage{cite}
\usepackage{doi}

\usepackage[noabbrev,nameinlink]{cleveref}

\DeclareSymbolFont{usualmathcal}{OMS}{cmsy}{m}{n}
\DeclareSymbolFontAlphabet{\mathcal}{usualmathcal}

\title{\bfseries Wess--Zumino terms in 0+1 $\bm{SU(N)}$ superspin systems}

\author{
Juan S. Morales\thanks{Corresponding author: \href{mailto:juan.morales@itp3.uni-stuttgart.de}{juan.morales@itp3.uni-stuttgart.de}}\\
\small Institute for Theoretical Physics III, Universit\"at Stuttgart, Pfaffenwaldring 57,\\
\small 70569 Stuttgart, Germany
\and
Mikhail N. Kiselev\\
\small The Abdus Salam International Centre for Theoretical Physics,\\
\small Strada Costiera 11, I-34151 Trieste, Italy
}

\date{}

\begin{document}

\maketitle

\begin{abstract}
These notes present a self-contained introduction to Wess--Zumino (WZ) terms in quantum systems with $SU(N)$ symmetry, with an emphasis on the interplay between geometry, topology, and condensed-matter applications. We begin with the $SU(2)$ spin coherent-state path integral, where the Berry phase appears as a WZ term encoding the symplectic structure of the Bloch sphere. This example is then used to introduce the geometric origin of topological terms, their relation to integral cohomology classes, and the role of the Berry curvature as the first Chern class of the canonical $U(1)$ bundle. We next discuss physical realizations in which such geometric terms affect dynamics, including adiabatic Berry phases and geometric quantum noise in magnetic quantum dots. A substantial part of the notes is devoted to the condensed-matter motivation for higher $SU(N)$ symmetries, covering $SU(N)$ Heisenberg models, $SU(4)$ spin-orbital and spin-pseudospin systems, multipolar exchange interactions, and higher-spin multipolar orders. Finally, we develop the $0+1$-dimensional $SU(N)$ superspin coherent-state construction, identify the phase space with $\mathbb{CP}^{N-1}$, and derive explicit local WZ terms for $SU(3)$ and $SU(4)$. The appendices provide the algebraic dictionaries needed to connect the abstract superspin language with concrete physical embeddings, including multipolar generator bases and several useful $SU(4)$ parametrizations.
\end{abstract}

\tableofcontents
\vspace{1em}

\input{chapter1_introduction}
\input{chapter2_su2_top_terms}

\input{chapter3_Su2_berry_phases}

\input{chapter3_SuN_condensed_matter}

\input{chapter4_5_SuN_generalization}
\appendix
\addtocontents{toc}{\protect\addvspace{0.8em}}
\addcontentsline{toc}{part}{Appendices}

\input{appendix1_multipolar_generators}
\input{appendix2_SU4_embeddings}
\input{appendix3_spin_covers_su-so}

\addtocontents{toc}{\protect\addvspace{0.8em}}

\input{final_remarks}
\input{acknoledgements}

\bibliographystyle{bibstyle}
\bibliography{references1}

\end{document}

%% file: chapter1_introduction.tex
\section{Introduction}
As it is well known, magnetism, as a fundamental interaction in condensed matter, finds its microscopic origin in the quantum mechanical behavior of particles (atoms or electrons), dictated by their spin and angular momentum \cite{blundell}. In this sense, the $SU(2)$ symmetry group, responsible for the rotational invariance of these degrees of freedom, provides the natural framework to describe both their dynamics under external fields and the exchange interactions that drive collective magnetic phenomena. It is then this universality of $SU(2)$ symmetry that lies at the core of our understanding of magnetic order and the quantum interplay of spins across very different systems. In particular, in the $S=1/2$ fundamental representation of $SU(2)$, the Hilbert space $\mathcal{H}_{1/2}$ admits a geometric description in terms of the Bloch sphere $S^2$, which can be seen as the manifold of coherent states. As we will show later, this point of view emerges naturally once one moves from the Hamiltonian to the Lagrangian formulation, where the classical phase space is identified with $S^2$.
What is interesting, however, is that this geometric reformulation does not come alone: it brings with it a non-trivial ``topological'' term in the action. This is the $SU(2)$ Wess-Zumino (WZ) term, which corresponds to the path-integral representation of the Berry phase and encodes the geometric structure underlying spin dynamics. While here we focus on the simplest case of a $S=1/2$ $(0+1)$-dimensional system, the same structure appears much more broadly, from topological insulators \cite{monaco} to fractionalization and skyrmion physics \cite{Aitchison, wilczek1989geometric}.

Our goal with these lecture notes is twofold. First, we revisit the derivation of the $SU(2)$ WZ-term in Section \ref{spin}, and then use it, in the first part of Section \ref{models}, to explore its physical consequences. This perspective naturally leads us to the notion of geometric quantum noise, and to the realization that topological terms play a fundamentally different role in the dynamics: rather than contributing to the energy, they encode the geometric structure of the phase space and thereby govern the equations of motion themselves. Along the way, we emphasize why such terms are called ``topological'' and how they can be constructed in a broader setting.\\

With this in mind, in the second part of these notes, we turn to the question of how this geometric and topological framework extends beyond the $SU(2)$ setting and naturally emerges in condensed matter systems with enlarged internal symmetries. In particular, we discuss in Section~\ref{sec:su_n_condensed_matter} why higher $SU(N)$ groups appear in systems with orbital degeneracy, multipolar exchange interactions, or strong correlation effects, and how, in these contexts, the notion of a ``spin'' must be generalized to higher-dimensional internal spaces. This also requires some care in the choice of generators: the same algebra may be written in a standard Gell--Mann basis, in a multipolar basis adapted to a spin-$S$ interpretation, or in tensor-product bases adapted to spin-orbital degrees of freedom. The corresponding dictionaries are collected in Appendices~\ref{app:suN-multipoles} and~\ref{app:su4-embeddings-parametrizations}. Starting from this motivation, we then move to the path integral construction of $SU(N)$ systems in Section~\ref{sec:su_n_superspin}, where we point out the specific way in which the geometric picture changes: instead of the Bloch sphere $S^2$, the coherent-state manifold becomes the complex projective space $\mathbb{CP}^{N-1}$, reflecting the intrinsic redundancy of such group coherent states. There, we clarify that even though the general structure remains the same --- a topological Wess--Zumino term together with a dynamical contribution --- the WZ term is now associated with the canonical Berry connection over $\mathbb{CP}^{N-1}$ and with the corresponding first Chern class. \\

\noindent We make this structure explicit by deriving the local form of the Wess--Zumino terms for the simplest non-trivial cases, namely $SU(3)$ and $SU(4)$. These examples allow us to see concretely how the geometry evolves with $N$, through the appearance of multi-chart phase spaces and higher-dimensional complex projective manifolds, while preserving a quantized index associated with $\pi_2(\mathbb{CP}^{N-1})=\mathbb{Z}$. Finally, by combining the WZ term with the expectation value of the Hamiltonian, we obtain the full effective Lagrangian and derive the corresponding classical equations of motion, thereby extending the familiar Bloch dynamics of $SU(2)$ spins to the $SU(N)$ superspin framework.
The appendices complement this construction by providing the algebraic background needed to translate between different physical realizations of the same $SU(N)$ structure. Appendix~\ref{app:suN-multipoles} reviews the multipolar organization of $SU(N)$ generators, with the explicit $SU(3)$ and $SU(4)$ decompositions into dipolar, quadrupolar, and octupolar sectors. Appendix~\ref{app:su4-embeddings-parametrizations} focuses on $SU(4)$ and compares several useful viewpoints: the generalized Gell--Mann basis, the spin-orbital tensor-product basis, the $SO(4)$ singlet-triplet organization, and Euler-angle parametrizations adapted to density matrices or transport problems. These appendices are not needed for the formal derivation of the WZ term, but they provide the dictionary for connecting the abstract superspin language with concrete condensed-matter settings.\\

\noindent Finally, a comment on style. Throughout these notes, we have tried to keep the derivations as explicit as possible and to avoid skipping intermediate steps. This is not only for completeness, but also because many of the structures discussed here — particularly those involving geometric and topological terms — can be subtle on a first encounter. In fact, much of this exposition is written with the perspective of someone trying to fully understand these constructions from the ground up. For this reason, several calculations are presented in detail, with the hope that the reader can follow the logic step by step without having to fill in large gaps.

%% file: chapter2_su2_top_terms.tex
\section{SU(2)-Spin and Topological Terms}\label{spin}
In this section, we will review the construction of the classical action (from the coherent path integral) for an $SU(2)$ spin theory. Then, we clarify the topological nature of one of its terms by introducing two families of topological field theories, the so-called $\theta$-terms and Wess-Zumino (WZ) TQFTs. We conclude our discussion by presenting the equivalence between such a WZ-term and the one coming from the $U(1)$ Monopole problem.
\subsection{Spin Path Integral}
Let us start by recalling that the $SU(2)$ group is defined (in the fundamental representation) by
\begin{equation}
    SU(2) = \Set{g\in M_{2\times 2}(\mathbb{C}) \: | \:  g^\dagger g = \mathbb{1}_{2} \: \text{ and} \: \det(g)=1}\:.
\end{equation}
As is well known, $SU(2)$ is the Lie group — that is, it is simultaneously a smooth manifold and a group — associated with the Lie algebra $\mathfrak{su}(2)$ generated by the three Pauli matrices (a three-dimensional vector space equipped with a product given by the Lie bracket/commutator). Mathematically, $\mathfrak{su}(2)$ consists of traceless anti-Hermitian $2\times2$ matrices; throughout this text, however, we adopt the equivalent Hermitian-generator convention commonly used in physics, with $S^i=\sigma_i/2$. Indeed, one way to see that such Lie algebra has dimension three is by counting the number of free parameters to specify a group element. In other words, the conditions $g^\dagger g = \mathbb{1}_{2}$ and $\det(g)=1$ impose constraints that leave three independent real parameters. An arbitrary Hermitian generator may therefore be written as
\begin{equation}
    A=a_1S^x+a_2S^y+a_3S^z,
    \qquad
    a_i\in\mathbb{R},
\end{equation}
and every element of $SU(2)$ can be obtained through the exponential map,
\begin{equation}
    g=e^{i\Vec{a}\cdot\Vec{S}}=e^{iA}.
\end{equation}
Under this Lie algebra basis, the Lie brackets of the generators fulfill 
\begin{equation}
    [S^i, S^j]=i\epsilon_{ijk}S^k
\end{equation}
where $\epsilon_{ijk}$ is the totally antisymmetric Levi-Civita tensor. However, upon introducing the basis $\Set{S^+=S^x+iS^y,\ S^-=S^x-iS^y, \ S^z}$, we have now
\begin{equation}
    [S^+, S^-]=2S^z \ \text{ and } \ [S^z, S^\pm]= \pm S^\pm.
\end{equation}
This motivates the definition of the structure constants of a Lie algebra in a given basis. Being $\{T^a\}_{a=1}^{\dim\mathfrak g}$ the generators/basis of a Lie algebra $\mathfrak{g}$, the structure constants $f^{abc}$ are defined as the coefficients of 
\begin{equation}
    [T^a, T^b]= i f^{abc}T^c. 
\end{equation}
\parskip=0.3cm 

Having introduced its algebraic structure, let us now turn to the physical relevance of $SU(2)$, which stems from its relation to the group of proper rotations in three-dimensional space,
\begin{equation}
    SO(3)
    =
    \Set{
    R\in M_{3\times 3}(\mathbb{R})
    \: | \:
    R^{\mathsf T}R=\mathbb{1}_3
    \ \text{and}\
    \det R=1
    }.
\end{equation}
Classically, rotations are described by $SO(3)$. In quantum mechanics, however, physical states correspond to rays in Hilbert space, so the unitary operators representing a rotation are defined only up to an overall phase. Consequently, rotational symmetry need not be realized through ordinary unitary representations of $SO(3)$, but more generally through continuous projective unitary representations,
\begin{equation}
    U(R_1)U(R_2)
    =
    e^{i\omega(R_1,R_2)}U(R_1R_2),
\end{equation}
where $\omega(R_1,R_2)$ is defined modulo $2\pi$ and satisfies the cocycle condition required by associativity. The key point is that such projective unitary representations of $SO(3)$ lift to ordinary unitary representations of its simply connected double cover $SU(2)$. Thus, the appearance of $SU(2)$ in quantum spin is not merely a convenient reformulation of the classical rotation group: it is necessary in order to include all physically admissible realizations of rotational symmetry, in particular those corresponding to half-integer spin. At the infinitesimal level, this relation follows from the Lie-algebra isomorphism
\begin{equation}
    \mathfrak{su}(2)\simeq\mathfrak{so}(3),
\end{equation}
since both algebras are characterized by the commutation relations
$[J_i,J_j]=i\epsilon_{ijk}J_k$. The groups themselves, however, are not globally isomorphic. Their relation may be seen from the conjugation action of $SU(2)$ on traceless Hermitian $2\times2$ matrices, which can be identified with vectors $\Vec{x}\in\mathbb{R}^3$ through $X(\Vec{x})=\Vec{x}\cdot\Vec{\sigma}$. This action induces a proper rotation $R_g\in SO(3)$ according to
\begin{equation}
    gX(\Vec{x})g^\dagger=(-g)X(\Vec{x})(-g)^\dagger=X(R_g\Vec{x})\:.
\end{equation}
Thus, $g$ and $-g$ induce the same spatial rotation. The resulting homomorphism $g\mapsto R_g$ from $SU(2)$ onto $SO(3)$ has kernel $\Set{\mathbb{1}_2,-\mathbb{1}_2}\simeq\mathbb{Z}_2$, and therefore $SU(2)$ is the double cover of $SO(3)$ (every spatial rotation has two representatives):
\begin{equation}
    SO(3)\simeq SU(2)/\mathbb{Z}_2.
\end{equation}
Accordingly, the irreducible quantum-spin sectors are obtained by classifying the irreducible unitary representations of $SU(2)$. These are labeled by
\begin{equation}
    S=0,\frac{1}{2},1,\frac{3}{2},\ldots,
\end{equation}
and act on the $(2S+1)$-dimensional Hilbert spaces $\mathcal{H}_S\simeq\mathbb{C}^{2S+1}$.
In particular, in the spin-$S$ representation $U_S$, the nontrivial element of the kernel acts as $U_S(-\mathbb{1}_2)=(-1)^{2S}\mathbb{1}_{\mathcal H_S}$. Hence, for integer $S$, the elements $g$ and $-g$ act identically and the representation descends to an ordinary representation of $SU(2)/\mathbb Z_2\simeq SO(3)$. For half-integer $S$, instead, they differ by a minus sign, so the representation does not descend to $SO(3)$, but defines a projective representation of it. Equivalently, a $2\pi$ rotation acts as
\begin{equation}
    U_S(2\pi)=(-1)^{2S}\mathbb{1}_{\mathcal H_S}.
\end{equation}
Integer-spin states are therefore invariant under a $2\pi$ rotation, whereas half-integer-spin states acquire a minus sign. Hence, restricting the quantum theory to ordinary representations of $SO(3)$ would incorrectly exclude half-integer spin. The use of $SU(2)$ incorporates both integer- and half-integer-spin sectors within ordinary unitary representation theory. A systematic discussion of this double covering, including its formulation as $SU(2)\simeq\operatorname{Spin}(3)$ and other $SU-SO$ relations, is presented in Appendix~\ref{app:low_rank_su_so}.

\subsubsection{Representation Theory of SU(2) in the Language of SU(N) Irreps}

Once identified as the carrier spaces of the irreducible unitary representations of $SU(2)$, the spin Hilbert spaces $\mathcal{H}_S$ can be placed within the general language of representation theory. A representation of a group $G$ on a vector space $V$ is a group homomorphism
\begin{equation}
    \rho:G\longrightarrow GL(V),
\end{equation}
which associates each element of $G$ with an invertible linear transformation of $V$. Such a representation is irreducible if the only invariant subspaces $W\subseteq V$ satisfying
\begin{equation}
    \rho(g)W\subseteq W
    \qquad
    \text{for all }g\in G
\end{equation}
are
\begin{equation}
    W=\Set{0}
    \qquad\text{and}\qquad
    W=V.
\end{equation}
More generally, one may ask: What are the irreducible representations of $SU(2)$? Are the Hilbert spaces $\mathcal{H}_S$ the only irreducible representations of $SU(2)$? To answer these questions, one has to consider the general construction of irreducible representations of compact Lie groups. In this work, however, we will only need the following general idea:
\begin{center}
    "the irreducible representations $\mathcal{H}_{c_2, \dots, c_{N}}$ of $SU(N)$ groups may be characterized by the eigenvalues $c_2, \ldots, c_N$ of the so called \textbf{N-1 Casimir operators} $\hat{C}_2, \ldots, \hat{C}_N$, which are algebraically-independent,  constructed from the generators of $\mathfrak{su}(N)$ and have the property of commuting with all of them (and therefore with the whole algebra). In this notation, one may write an irrep schematically as $\mathcal{H}_{c_2, \ldots, c_N}$, where $\hat{C}_m|\psi\rangle=c_m|\psi\rangle$ for $m=2, \ldots, N$ and $\forall |\psi\rangle \in \mathcal{H}_{c_2, \ldots, c_N}$."
\end{center}
The reason the labels start at $m=2$ is that the generators of $\mathfrak{su}(N)$ are traceless, so there is no independent linear Casimir. More concretely, if $X=x_a T^a \in \mathfrak{su}(N)$, the invariant polynomials
\begin{equation}
    \operatorname{Tr}\left(X^2\right), \operatorname{Tr}\left(X^3\right), \ldots, \operatorname{Tr}\left(X^N\right)
\end{equation}
generate the independent Casimir operators. Equivalently, one can write them schematically as
\begin{equation}
    \hat{C}_m=d_{a_1 \cdots a_m}^{(m)} T^{a_1} \cdots T^{a_m} \qquad [m=2, \ldots, N]\:\:,
\end{equation}
where $d_{a_1 \cdots a_m}^{(m)}$ is a completely symmetric invariant tensor of rank $m$. The quadratic case corresponds to the invariant tensor $\delta_{a b}$, giving $\hat{C}_2=\sum_a T^a T^a$. For $SU(3)$, there is in addition a cubic Casimir built from the symmetric tensor $d_{a b c}$, while for $SU(4)$ one also has an independent quartic Casimir, and so on.

In this sense, for the case of $SU(2)$, we have only one Casimir operator: the total angular momentum $\hat S^2=(S^x)^2+(S^y)^2+(S^z)^2$. In fact,  the quantum number that physicists call ``spin'' is the integer or half-integer label $S$ of the representation, while $S(S+1)$ is the corresponding eigenvalue of the quadratic Casimir. Equivalently, we will see in Sec.\ref{Sec:su_n_representation} that irreducible representations of $SU(N)$ may also be labeled through highest weights or \emph{Dynkin labels}, for which one first has to introduce the Cartan subalgebra.

The \textbf{Cartan subalgebra} $\mathfrak{h}$ of a (semisimple) Lie algebra $\mathfrak{g}$ is defined as the maximal commuting subalgebra of $\mathfrak{g}$ (i.e. the largest set of mutually commuting elements in some representation). Thus, the \textbf{Cartan operators} are defined as the generators $\Set{H_i}$ of this subalgebra $\mathfrak{h}$. Note that these are different from Casimirs since they only commute between each other, and the Casimirs commute with the whole algebra.  In our case of $SU(N)$, one can show that there are $N-1$ of such operators, which implies that for the simplest example of $SU(2)$, the only Cartan operator is indeed the choice of one single generator $S^i$ (traditionally taken as $S^z$). 

\noindent Now, since the Cartan generators $\Set{H_1, \dots, H_{N-1}}$ represent all the operators that can be simultaneously diagonalized, we know (due to the QM spectrum postulate) that their corresponding eigenvalues $\Set{\Set{\lambda_{j_1}}, \dots, \Set{\lambda_{j_{N-1}}}}$ are sufficient to index the basis states in the given representation (i.e. the basis elements of $\mathcal{H}_{c_1, \dots, c_{N-1}}$). This becomes clear in the case of  $SU(2)$, where we know that the z-spin projection $s_z$ describes uniquely the basis states 
\begin{equation*}
    \mathcal{H}_S=\text{Span}(\left\{\ket{S,s_z}\right\}) \ \  \text{ with } \ \  S^z\ket{S,s_z}=s_z \ket{S,s_z}
\end{equation*}
and again $S$ labels the irrep - $S^2\ket{S,s_z}=S(S+1)\ket{S,s_z}$. In the present text, we will restrict ourselves to the fundamental representation of $SU(N)$, which we will define later for the case $N>2$. But have in mind that, for $N=2$, the fundamental irrep is the one with the lowest dimension of $\mathcal{H}_S$, namely $S=1/2$, with $\dim=2$.

Having established the basis of $SU(N)$ representation theory, one fundamental feature that we will exploit later on is the existence of a \textbf{maximum weight state} $\ket{\uparrow} \in \mathcal{H}_{c_2, \dots, c_{N}}$  defined as the element with the biggest possible eigenvalues of Cartan operators (or equivalently being annihilated by the raising operators - see Sec.\ref{sec:su_n_superspin}) in any arbitrary irrep $\mathcal{H}_{c_2, \dots, c_{N}}$ of $SU(N)$. For instance, in any irrep of $SU(2)$, we have that such $\ket{\uparrow}$ corresponds to the totally polarized state $\ket{S, S}$.

\noindent Note that all our discussion about irreps is also valid for reducible representations once we take into account \textbf{the irrep decomposition theorem}: For any finite-dimensional unitary representation $V$ of a compact group $G$, there is a unique decomposition 
\begin{equation}
    V=V_1^{\bigoplus a_1}\oplus\dots \oplus V_k^{\bigoplus a_k},
\end{equation} 
where the $V_i$ are distinct irreducible representations and $a_i$ account for their multiplicities. In this line, what physicists call the "addition of angular momentum/spin" is indeed the process of finding the unique irrep decomposition of the tensor product of irreps. For instance, $\mathcal{H}_{1/2} \otimes \mathcal{H}_{1/2} = \mathcal{H}_{1} \oplus \mathcal{H}_{0}$, or in other words, a system of two $1/2$-spins can be expressed as the triplet $\mathcal{H}_{1}$ and the singlet $\mathcal{H}_{0}$.

\subsubsection{Generalized Coherent States}
In addition to irreps, we need another ingredient to follow the recipe of the spin path integral. This time, we are required to generalize the notion of coherent state to the case of a quantum system described by the action of $SU(2)$ or another Lie group $G$ \cite{classical_spin_Zhang}. Usually, we tend to think that the concept of a coherent state is exclusively related to quantum optics. Most of the time, coherent states are defined as the most classical (i.e., minimum uncertainty $\Delta x \Delta p = \hbar / 2$) state of a quantum harmonic oscillator. Originally, their construction was done as the eigenstates of the annihilation operator, such that they are generated by applying a "displacement operator" of the Heisenberg-Weyl group to the vacuum state of the harmonic oscillator. Recall that the Heisenberg-Weyl group $H_3$ is the Lie group generated by the Lie algebra $\mathfrak{h}_3$, which has generators $\Set{\mathbb{1}, \hat{x}, \hat{p}}$ the identity, the position, and the momentum operators. Note, then, that we can redefine these generators to $\Set{\mathbb{1}, \hat{a}, \hat{a}^\dagger}$ with $\hat{a}=\hat{x}+i\hat{p}$ and $\hat{a}^\dagger=\hat{x}-i\hat{p}$ ($\hbar=1$). In this way, the coherent states can be obtained from
\begin{equation}
    \ket{\alpha}=e^{\alpha \hat{a}^\dagger - \bar{\alpha} \hat{a}} \ket{0}, 
\end{equation}
where $e^{\alpha \hat{a}^\dagger - \bar{\alpha} \hat{a}} \in H_3$ is the so-called displacement operator. Indeed, this name is motivated by the fact that once we consider $\alpha=q+ip$, then the state $\ket{\alpha}=\ket{q,p}$ corresponds to the groundstate of a simple harmonic oscillator centered at $q$. However, the most important feature of this set of coherent states is that they form an overcomplete basis of the Hilbert space of a point particle in one dimension (the quantum system described by $H_3$), and therefore one can decompose the identity as 
\begin{equation}
    \hat{\mathbb{1}}=\int \frac{d p d q}{2 \pi}|p, q\rangle\langle p, q|. 
\end{equation}
These constructions motivate a more general definition of coherent states beyond eigenstates of the annihilation operator. Instead, we can replace them with the definition of "the state obtained by applying a displacement operator of the Heisenberg-Weyl group on the vacuum state of the harmonic oscillator." 

This was the key that brought Perelomov and Gilmore to generalize coherent states to arbitrary Lie groups $G$ \cite{Perelomov1986, gilmore1990}. The only thing we need to do is to apply "certain elements" of the group to a "certain distinguished state" to obtain any coherent state. Moreover, one can show that the set of coherent states provides the correct framework to prove Dirac’s conjecture: commutators are replaced by Poisson brackets in the classical limit.

\subsubsection{SU(2) Coherent States and Haar Measure}
Now, we are ready to start constructing the spin path integral. As the first step, let us recall that, in any irrep, any element of $SU(2)$ can be written as 
\begin{equation}\label{eq:euler-rep}
    \tilde{g}(\phi, \theta^\prime, \psi)=e^{-i \phi \hat{S}_3} e^{-i \theta^\prime \hat{S}_2} e^{-i \psi \hat{S}_3} \quad \text{ where } \quad \phi, \psi \in(0,2 \pi), \:\theta^\prime \in(0, \pi). 
\end{equation}
Therefore, following the above definition of the generalized coherent states, we propose the form of the $SU(2)$ coherent states as
\begin{equation}\label{eq:proposal}
    \ket{\tilde{g}}=\tilde{g}(\phi, \theta^\prime, \psi) \ket{\uparrow},
\end{equation}
where $\ket{\uparrow}$ is the maximum weight state of the irrep and $\tilde{g}(\phi, \theta^\prime, \psi)$ is an element of $SU(2)$ expressed as before. At this point, several questions arise: (1) Why the state $\ket{\uparrow}$ is the distinguished state from which we construct the coherent states? (2) Do we need all the elements of $SU(2)$ to construct all of them? Are there any redundancies that don't give new ones? If that is the case, (3) Which are those redundancies? 

Answering them, we have:
\begin{itemize}
    \item In the first place, one could argue that the selection of $\ket{\uparrow}$ is just a matter of choice justified by its convenient existence in all irreps of $SU(N)$. In other words, the reason lies in the fact that it marks a standardized starting point for all $SU(N)$ groups and irreps. However, this picture ignores the preferential position that this state has. For the case of the Heisenberg-Weyl group, we took the groundstate $\ket{0}$ of the harmonic oscillator because we know that it is a coherent state (Commutator $\to$ Poisson Bracket in the classical limit) and belongs to the Hilbert space. Now, with $\ket{\uparrow}$, the logic is the same. We selected it because of its possibility of recovering the classical dynamics of observables (Poisson bracket with $H$) by projecting the respective commutators in such a state. This means that from the beginning, we already know that in all irreps and all SU(N), $\ket{\uparrow}$  is a coherent state (constructed trivially from $\ket{\uparrow}=e^0 e^0 \ket{\uparrow}$).
    \item With respect to having redundancies, we should ask ourselves first what could be considered a redundancy. On one side, there is a $U(1)$ redundancy coming from the definition of physical states, which are defined up to complex phases $\left[\ket{\psi}\right]=\left[e^{i\theta}\ket{\psi}\right]$. However, from another perspective, there is another source of redundancy, arising from transformations that leave the state $\ket{\uparrow}$ invariant. The question now must be why leaving this state invariant should be a redundancy. This will take us to the idea that, in the end, these coherent states will be constructed to reproduce the quantum dynamics (in periodic time) of (super-)spin. Therefore, the elements $\tilde{g}(\phi, \theta^\prime, \psi)$ should be thought as time-dependent  $\tilde{g}(\phi(\tau), \theta^\prime(\tau), \psi(\tau))$ such that their evolution describes a path in $SU(2)$. However, if continuous deformation of those paths in $SU(2)$ gives us the exact same physical path of the spin states, then we should consider those elements as redundancies. In conclusion, \textit{the elements $\tilde{g}(\phi, \theta^\prime, \psi)$ which leave the physical state $[\ket{\uparrow}]$ invariant are redundancies for the path integral}. 
\end{itemize}

Later on,  when we deal with $SU(N)$, we will see how redundancies of the second kind appear, but for the moment, in $SU(2)$, we just have complex phase redundancies. Certainly, this $U(1)$-redundancy is extremely easy to see once we consider the $SU(2)$ Euler-representation (Eq.\ref{eq:euler-rep}) and the group action (Eq.\ref{eq:proposal}):
\begin{equation}
    \ket{\tilde{g}}= e^{-i \phi \hat{S}_3} e^{-i \theta^{\prime} \hat{S}_2} e^{-i \psi \hat{S}_3} \ket{\uparrow} = \left( e^{-i \phi \hat{S}_3} e^{-i \theta^\prime \hat{S}_2} \ket{\uparrow} \right) e^{-i \psi S} 
\end{equation}
where we have used $e^{-i \psi \hat{S}_3} \ket{\uparrow}= e^{-i \psi \hat{S}_3} \ket{S,S}= e^{-i \psi S}\ket{\uparrow}$. In this line, the only group elements that generate relevant physics to the path integral are of the form 
\begin{equation}
    g(\phi, \theta^\prime) =  e^{-i \phi \hat{S}_3} e^{-i \theta^\prime \hat{S}_2},
\end{equation}
and the \textbf{SU(2) coherent states}
\begin{equation}\label{eq:spin_cohe_exp}
    \boxed{|g(\phi, \theta^\prime)\rangle \equiv e^{-i \phi \hat{S}_3} e^{-i \theta^\prime \hat{S}_2}\ket{\uparrow}.}
\end{equation}

In a parallel to the previous discussion, we know that any basis element of $\mathcal{H}_S$ can be reached through the action of $SU(2)$ on $\ket{\uparrow}$ once we evoke irreducibility (since, in irreps, the span of the orbit is equal to the space). Therefore, one could be able to find a decomposition of the identity in terms of the "redundant" coherent states $\ket{\tilde{g}(\phi, \theta, \psi)}$. However, since the projectors fulfill 
\begin{equation}
    \ket{\tilde{g}}\bra{\tilde{g}}=\ket{g}e^{-i\psi S} e^{i\psi S}\bra{g}=\ket{g}\bra{g},
\end{equation}
then the decomposition of the identity in terms of the redundant coherent states is equivalent to the decomposition of the identity in terms of the actual coherent states. Indeed, a simple proof by Schur's lemma \cite{altland} gives us that such a decomposition is given by  
\begin{equation}\label{eq:deco_ident}
    \boxed{\mathbb{1}= \frac{1}{\text{vol}(G)} \int d\mu_2 \ \ket{g(\phi, \theta^\prime)}\bra{g(\phi, \theta^\prime)},}
\end{equation}
where $\text{vol}(G)$ denotes the volume of $SU(2)$, and the measure $d\mu_2$ is given by
\begin{equation}\label{eq:S3_measure}
    d \mu_2=\frac{1}{2}\cos(\frac{\theta^\prime}{2}) \sin( \frac{\theta^\prime}{2}) \ d \theta^\prime d \varphi_1 d \varphi_2. 
\end{equation}
and is called \textbf{Haar measure}. The formal definition of the Haar measure of a continuous group $G$ is the unique (up to positive multiplicative factors) bi-invariant measure. In practice, this means that integrating a function $f(g)$ is equal to integrating $f(gh)$ or $f(hg)$ for any element $h$ of the group. For the moment, note that this measure $d\mu_2$ coincides with the one induced by the round metric (redefining $\theta^\prime = 2\theta$)
\begin{equation}\label{eq:round_metric_S3_in_R4}
    \left|d s_2\right|^2=d \theta^2+\cos ^2(\theta) d \varphi_1^2+\sin ^2(\theta) d \varphi_2^2 
\end{equation}
of $S^3$ embedded in $\mathbb{R}^4$. Nevertheless, this is not surprising, since the 3-sphere gives the Lie group isomorphism  
\begin{equation}
    SU(2) \simeq S^3.
\end{equation}

Summarizing what we have so far, we find the $SU(2)$ coherent states and a decomposition of the identity on them. However, one would like to know where these coherent states live. Is there any mathematical structure behind them? In fact, a simple computation gives us some insights. From a practical point of view, we could say that the (classical-analog) degrees of freedom of our theory are enclosed on the coherent-state-expectation values of the three spin operators $\langle g(\phi, \theta)|\mathbf{S}| g(\phi, \theta)\rangle$. This will be more clear in the next subsection once we compute the spin Lagrangian, but for the time being, what is important to notice is that, using the intermediate identity $e^{-i \phi \hat{S}_i} \hat{S}_j e^{i \phi \hat{S}_i}=\hat{S}_j \cos \phi+\epsilon_{i j k} \hat{S}_k \sin \phi$, the mentioned expectation value is
\begin{equation}\label{eq:exp_value}
    \langle g(\phi, \theta^\prime)|\mathbf{S}| g(\phi, \theta^\prime)\rangle=S\left(\begin{array}{c}
\sin \theta^\prime \cos \phi \\
\sin \theta^\prime \sin \phi \\
\cos \theta^\prime
\end{array}\right) \equiv S \mathbf{n},
\end{equation}
which corresponds exactly to a 3D - unit vector $\mathbf{n}$ times the spin number $S$. This result leads us to think that the true degrees of freedom of the spin classical analog live in the unit sphere $S^2$. However, $S^2$ is not only a beautiful geometrical shape, and that's all. It encodes a nontrivial mathematical structure (the topology). Here, we refer the reader to Sec.\ref{sec:top_actions} for the treatment of the topological features of $S^2$, but for now, keep in mind that working in $S^2$ is not like working in $\mathbb{R}^2$ where all the points can be reached.    

\noindent In this line, the previous calculation showed us that the coherent expectation values of spin live in the sphere $S^2$; nevertheless, this does not answer the question of the space in which the coherent states themselves live. To do so, let's see an alternative way to construct the $SU(2)$ coherent states. This procedure will indeed standardize the construction of the $SU(N)$ coherent states in Sec.\ref{sec:su_n_superspin}.

Like before, we start from the fact that any element of $SU(2)$ can be written, this time in the fundamental representation $S=1/2$, as 
\begin{equation}
    \tilde{g}\left(\theta, \varphi_1, \varphi_2\right)=\left(\begin{array}{cc}
e^{i \varphi_1} \cos \theta & -e^{-i \varphi_2} \sin \theta \\
e^{i \varphi_2} \sin \theta & e^{-i \varphi_1} \cos \theta
\end{array}\right)
\end{equation}
where now the "half-polar angle" $\theta \in (0,\pi/2)$, and the "azimuthal angles" $\varphi_1,\varphi_2 \in (0, 2\pi)$. In addition, having the third Pauli matrix $S^z$ already diagonalized, the maximum weight state in this irrep is just
\begin{equation}
    \ket{\uparrow}=\begin{pmatrix} 1 \\ 0 \end{pmatrix},
\end{equation}
which means that the \textbf{"redundant" coherent states} are 
\begin{equation}\label{eq:redundant_cohe_2}
\boxed{\ket{\mathbf{\tilde{n}_4}} \equiv \left|\tilde{g}(\theta, \varphi_1, \varphi_2)\right\rangle=\tilde{g}\left|\uparrow\right\rangle=\left(\begin{array}{l}
e^{i \varphi_1} \cos \theta \\
e^{i \varphi_2}\sin \theta
\end{array}\right).}
\end{equation}
The question now is how we can account for the $U(1)$ redundancy of the coherent states, not in the states themselves, but in the group elements that defined them. Put differently, we have to express the elements $\tilde{g}(\theta, \varphi_1, \varphi_2)$ as some matrix on which a $U(1)$ element is operating. This is exactly 
\begin{equation}
    \tilde{g}\left(\theta, \varphi_1, \varphi_2\right)=\left(\begin{array}{cc}
\cos \theta & -e^{-i\left(\varphi_2-\varphi_1\right)} \sin \theta \\
e^{i\left(\varphi_2-\varphi_1\right)} \sin \theta & \cos \theta
\end{array}\right)\left(\begin{array}{cc}
e^{i \varphi_1} & 0 \\
0 & e^{-i \varphi_1}
\end{array}\right), 
\end{equation}
where we can see the action of the $U(1)$ factor in the right matrix. Now, what we want to do is say that no matter which angle $\varphi_1$ we choose, the group elements from which we construct the states are the same. Formally, this is done by taking the quotient $SU(2) / U(1)$, but what is really interesting is that this space is indeed
\begin{equation}
    \boxed{SU(2) / U(1) \simeq S^2 }
\end{equation}
a topological space isomorphic to the 2-sphere $S^2$. Therefore, it is not only the coherent state expectation values that live in $S^2$; the same coherent state space is $S^2$. Or even better, the group elements from which these states are constructed are the ones that constitute $S^2$. Finally, since in the quotient, the specific value of the representative doesn't matter, we can calmly take $\varphi_1 =0$, meaning that any \emph{coherent group element} $g\in SU(2)/U(1)$ is effectively described by  
\begin{equation}
    g\left(\theta, \varphi_2\right)=\left(\begin{array}{cc}
\cos \theta & -e^{-i \varphi_2} \sin \theta \\
e^{i \varphi_2} \sin \theta &  \cos \theta
\end{array}\right), 
\end{equation}
and the $SU(2)$ coherent states (in the fundamental irrep)
\begin{equation}\label{eq:irrep_spin_cohere}
\boxed{\ket{\mathbf{n}_2}\equiv \left|g (\theta, \varphi_2)\right\rangle=\left(\begin{array}{l}
 \ \ \: \cos \theta \\
e^{i \varphi_2}\sin \theta
\end{array}\right)}\:.
\end{equation}
At this point, one may wonder what happens to the geometric structure of the group manifold under this quotient. The round metric on $S^3\simeq SU(2)$ (Eq.\ref{eq:round_metric_S3_in_R4}) induces a natural metric on the quotient space $SU(2)/U(1)$ by projecting out the directions along the $U(1)$ fiber. More concretely, by introducing the combinations $\phi=\varphi_2-\varphi_1$ and $\psi=\varphi_1$, one identifies $\psi$ as the fiber redundant coordinate, while $(\theta,\phi)$ parametrize the physical space of coherent states. Eliminating the fiber direction, the induced metric on the quotient becomes ($\theta'=2\theta$)
\begin{equation}
    ds^2_{\rm FS}=d\theta^2+\sin^2\theta\cos^2\theta\,d\phi^2
=\frac14\left(d\theta'^2+\sin^2\theta'\,d\phi^2\right).
\end{equation}
which is, up to the overall radius normalization, the standard metric on the sphere $S^2$. From a more general perspective, this metric coincides with the \textbf{Fubini--Study metric} on $\mathbb{CP}^1$, and its associated volume form reproduces the integration measure over coherent states. As we will see later, this metric also defines a canonical closed two-form (the Kähler form), which underlies the geometric origin of the Wess--Zumino term.
\subsubsection{Construction of (Super)-Spin Path Integral}\label{sec:construction_superspin_path_integral}
Once we have found the $SU(2)$ coherent states, which describe physical paths in the spin space, we can describe the quantum mechanical evolution of the spin system through the imaginary-time path integral \cite{altland, berry_phase_ferro}. Even though here we will be thinking in $SU(2)$, keep in mind that the following procedure is exactly the same for an $SU(N)$ quantum system. The only thing one needs to change is to replace the $SU(2)$-coherent states $\ket{g(\theta, \varphi)}$ by their corresponding $SU(N)$-counterparts $\ket{g(\scriptstyle{2N-2 \text{ angles}})}$, that we will derive in Sec.\ref{sec:su_n_superspin}.

Our starting point is to consider the physical situation of a particle of spin $S=1/2$ subject to the Hamiltonian   
\begin{equation}
    H=\mathbf{B} \cdot \hat{\mathbf{S}},
\end{equation}
and our final goal is to find the path integral representation of the quantum partition function
\begin{equation}\label{eq:partition}
    Z=\Tr(e^{-\beta H})=\sum_{v\in basis} \braket{v|e^{-\beta H}|v},
\end{equation}
where we have identified periodic imaginary time as $\tau=it \in [0,\beta)$. Recall that this partition function is nothing but the propagator $U(q_f,-i\beta, q_i, 0)$, but replacing the position degrees of freedom by final and initial spin states (we will see that these are indeed the coherent group elements). In summary, Eq.\ref{eq:partition} is, in fact, the spin propagator $U(g_f,\beta, g_i, 0)$ (Green function) for the imaginary time evolution of an initial state $\ket{g_i(\theta, \varphi)}$ to a final state $\ket{g_f(\theta, \varphi)}$.

\noindent As usual in the path integral formulation, one considers the imaginary time slicing 
\begin{equation}
     0=\tau_0<\tau_1<\dots < \tau_i <\dots <\tau_{N-1}<\tau_N=\beta
\end{equation}
such that the whole evolution is broken into $N$ partial evolutions with time step $\epsilon=\beta/N$. This is written as 
\begin{equation}
    Z=\Tr(e^{-N\epsilon H })= \sum_{v\in basis} \braket{v|\mathbb{1}_{N}e^{-\epsilon H} \mathbb{1}_{N-1}e^{-\epsilon H}\dots e^{-\epsilon H} \mathbb{1}_1e^{-\epsilon H}|v},
\end{equation}
where we have conveniently inserted identities during each time step evolution. Hence, after using the coherent state decomposition of the identity (Eq. \ref{eq:deco_ident}), we have
\begin{equation}
\begin{split}
    Z=\sum_{v\in bas} \frac{1}{\text{vol}^N} \int dg_N \braket{v|g_N}\int dg_{N-1} & \braket{g_N|e^{-\epsilon H}|g_{N-1}} ... \int dg_i \braket{g_{i+1}|e^{-\epsilon H}|g_i} \\ & \ \ \ \ \ \ \ \ ... \int dg_1 \braket{g_{2}|e^{-\epsilon H}|g_1} \braket{g_{1}|e^{-\epsilon H}|v}.
\end{split}
\end{equation}
To proceed, we exploit first the fact that $\epsilon$ is infinitely small, such that the matrix elements can be approximated (at first order) as
\begin{equation}
    \braket{g_{i+1}|e^{-\epsilon \mathbf{B} \cdot \hat{\mathbf{S}}}|g_i}\approx e^{\braket{ g_{i+1} | g_i}-\braket{ g_i | g_i} - \epsilon \braket{ g_{i+1}|\mathbf{B} \cdot \hat{\mathbf{S}}| g_i}, }, 
\end{equation}
and second, we use $\ket{v}=\int dg_0 \braket{g_0|v}\ket{g_0}$ so that the last and first factors can be related by
\begin{equation}
\begin{split}
    \braket{v|g_N} ... \braket{g_{1}|e^{-\epsilon H}|v}& = \int dg_0 \braket{g_0 | g_N} \braket{v|g_0}... \int dg_0\braket{g_{1}|e^{-\epsilon H}|g_0} \braket{g_0|v} \\ &= \int dg_0 \braket{g_0|g_N} ... \braket{g_{1}|e^{-\epsilon H}|g_0} \\ &= \int dg_0 \ \delta_{0,N} \  ... \braket{g_{1}|e^{-\epsilon H}|g_0}.
\end{split}
\end{equation}
Putting everything together, taking the continuum limit, and reorganizing, we then arrive at the expression:
\begin{equation}
    Z=\lim _{N \rightarrow \infty} \int_{g_N=g_0} \prod_{i=0}^N d g_i \ \exp \left[-\epsilon \sum_{i=0}^{N-1}\left(-\frac{\braket{ g_{i+1} | g_i}-\braket{ g_i | g_i}}{\epsilon}+\braket{ g_{i+1}|\mathbf{B} \cdot \hat{\mathbf{S}}| g_i}\right)\right]\:.
\end{equation}
The remaining step is to evaluate the quantities above in the continuum limit ($\epsilon \to 0, \ N \to \infty$ but keeping constant $\epsilon N= \beta$):
\begin{itemize}
    \item The derivative from
\begin{equation}
\begin{split}
    \lim_{\epsilon \to 0} \frac{\braket{g_{i+1}|g_i} - \braket{g_i|g_i}}{\epsilon} & =\lim_{\epsilon \to 0} \frac{\braket{\uparrow|g_{i+1}^{\dagger}g_i|\uparrow}-\braket{\uparrow|g_i^{\dagger}g_i|\uparrow}}{\epsilon} \\ & = \braket{\uparrow | \left( \lim_{\epsilon \to 0} \frac{g_{i+1}^{\dagger}-g_i^\dagger}{\epsilon} \right) g_i| \uparrow} \\ & = \braket{\uparrow | \left[\partial_\tau g^{\dagger}(\tau)\right] g(\tau) |\uparrow } = \braket{\partial_\tau g | g}
\end{split}
\end{equation}\
    \item The integral from 
        \begin{equation}
            \lim_{\epsilon \to 0 } \sum_{i=0}^{N} \epsilon \ \  \xmapsto{ \ \ \ \ \ \ \ } \ \ \int_{0}^{\beta} d\tau
        \end{equation}
    \item The matrix elements ignoring different time steps
    \begin{equation}
        \lim_{\epsilon \to 0 } \braket{g_{i+1} | \mathbf{B} \cdot \hat{\mathbf{S}} | g_i } \ \  \xmapsto{ \ \ \ \ \ \ \ } \ \ 
    \braket{g(\tau) | \mathbf{B} \cdot \hat{\mathbf{S}} | g(\tau)}
    \end{equation}
\end{itemize}
Therefore, the final expression for the \textbf{Spin Path Integral} is given by
\begin{equation}\label{eq:spin_path_integral_gener}
    \boxed{\mathcal{Z}=\int_{g(\beta)=g(0)} D g \ \exp \left[-\int_0^\beta d \tau\left(-\braket{\partial_\tau g | g}+\langle g|\mathbf{B} \cdot \hat{\mathbf{S}}| g\rangle\right)\right]}\ \ ,
\end{equation}
where we now regard the coherent group elements $g \in SU(2)/U(1)\simeq S^2$ as maps $g(\tau)$ from the periodic time to the 2-sphere. This is:
\begin{equation}
    g: S^1 \to S^2.
\end{equation}
Consequently, it is important to notice that what the periodic imaginary time path integral is doing is no more than the sum over continuous deformations of loops in the 2-sphere, weighted by the exponential of minus \textbf{the spin action}
\begin{equation}\label{eq:spin_action}
    \boxed{S[g(\tau)]=\int_0^\beta d \tau\left(-\braket{\partial_\tau g | g}+\langle g|\mathbf{B} \cdot \hat{\mathbf{S}}| g\rangle\right).}
\end{equation}
The first term received the name of \textbf{Wess-Zumino}, and later on, we will see that only depends on the topology of $S^2$ and its choice of coordinates. 
\subsubsection*{Bloch Equations from the Spin action}
Having Eq.\ref{eq:spin_action}, Eq.\ref{eq:exp_value} and Eq.\ref{eq:spin_cohe_exp}, it is easy to compute
\begin{enumerate}
    \item $\braket{g|\mathbf{B} \cdot \hat{\mathbf{S}}|g}=\mathbf{B}\cdot\braket{g|\hat{\mathbf{S}}|g}=\mathbf{B}\cdot S\mathbf{n}=S B n_z = SB \cos(\theta^\prime)$
    \item $\partial_\tau g= \partial_\tau(e^{-i\phi \hat{S}_3} e^{-i\theta^\prime \hat{S}_2}) = -i \hat{S}_3 (\partial_\tau \phi) e^{-i\phi \hat{S}_3} e^{-i\theta^\prime \hat{S}_2} -i (\partial_\tau \theta^\prime) e^{-i\phi \hat{S}_3} \hat{S}_2 e^{-i\theta^\prime \hat{S}_2} $
    \item $\braket{\partial_\tau g | g}=i\dot{\phi} \braket{\uparrow|e^{i\theta^\prime \hat{S}_2} e^{i\phi \hat{S}_3} \hat{S}_3 e^{-i\phi \hat{S}_3} e^{-i\theta^\prime \hat{S}_2}|\uparrow} + i\dot{\theta^{\prime}} \braket{\uparrow|e^{i\theta^\prime \hat{S}_2} \hat{S}_2 e^{i\phi \hat{S}_3} e^{-i\phi \hat{S}_3} e^{-i\theta^\prime \hat{S}_2}|\uparrow} $ \\ $\ \ \text{                             } \ \ \ \ =i\dot{\phi} \braket{\uparrow|(\hat{S}_3 \cos\theta^{\prime}-\hat{S}_1 \sin\theta^{\prime})|\uparrow} + i\dot{\theta^{\prime}} \braket{\uparrow|\hat{S}_2|\uparrow}= iS \dot{\phi}\cos\theta^\prime$,
\end{enumerate}
which gives us what we call \textbf{the local representation of the spin action}
\begin{equation}
\boxed{S[\phi, \theta^\prime]=SB\int_0^\beta d\tau \cos\theta^\prime + iS\int_0^\beta d\tau (-\cos\theta^\prime)\partial_\tau \phi} \ \ . 
\label{eq:action_singl_spin_imagi}
\end{equation}
Here, recall the convention through the whole document $\theta^\prime \in (0,\pi)$, and sometimes (more important in Sec.\ref{sec:su_n_superspin}) $\theta=\theta^\prime/2 \in (0,\pi/2)$. 

At this point, having already the action, one would like to derive the classical Euler-Lagrange equations of motion. Therefore, taking the variation of the previous expression and making it equal to zero, 
\begin{equation}
    \delta S[\phi, \theta^\prime] = S \int_0^\beta d\tau \left[ (-B \sin\theta^\prime + i \sin\theta^\prime \dot{\phi}) \delta\theta^\prime + i (-\cos\theta^\prime) \delta \dot{\phi}\right]=0,
\end{equation}
we find the equations
\begin{equation}
    \left\{ 
    \begin{aligned} 
   -B \sin\theta^\prime + i \sin\theta^\prime \dot{\phi} & = 0 \xrightarrow{ \quad } \: \boxed{\phi=-iB\tau}  \\
   \partial_\tau(-i\cos \theta^\prime) = i \sin \theta^\prime \dot{\theta^\prime} &= 0 \xrightarrow{\quad  } \:\boxed{\theta^\prime=\text{constant}}
\end{aligned} \right. .
\label{eq:Bloch_eqtns_imagi}
\end{equation}
These are precisely the equations describing precession around the $z$-axis in imaginary time. Hence, the important check is that the dynamics obtained from the action agrees with the usual semiclassical spin dynamics. To see this, recall that the Bloch equations follow from taking the coherent-state expectation value of the Heisenberg equation of motion ($\hbar=e=m_e=1$),
\begin{equation}
\braket{g|\dot{\hat{\mathbf{S}}}|g}=\braket{g|\left[H,\mathbf{S}\right]|g} \xrightarrow{ \  \ \ \  } \dot{\mathbf{n}}=-i \mathbf{B} \times \mathbf{n} \xrightarrow{ \  \ \ \  }  \left\{ 
    \begin{aligned} 
    \dot{\phi} &= -iB\\
  \dot{\theta^\prime}&= 0\:,
\end{aligned} \right.
\end{equation}
after using the parametrization of Eq.\ref{eq:exp_value}, we immediately recover the above local equations. An important observation is that the Wess-Zumino term is necessary to obtain the correct classical equations; if this term were not there, we would obtain a dynamics that does not correspond to the classical analog obtained through the coherent state expectation values.

\noindent In the next section, we will see why one of these equations (the first one) cannot be valid for all the points in the $S^2$ sphere. As a spoiler, the answer will lie in the fact that the azimuthal angle is not well-defined in the poles, or mathematically, the 2-sphere cannot be covered by only one chart. Physically, we should think of this situation as the impossibility of having precession if our initial configuration is exactly a polarized state (i.e., an eigenvector $\ket{\uparrow}$ or $\ket{\downarrow}$).

\subsection{Topological Actions: Geometry and Topology}\label{sec:top_actions}
In this section, we introduce the mathematical machinery needed to understand the topological (and geometric) nature of certain terms that sometimes appeared in the classical action of some quantum systems \cite{altland}. Recall that even though the systems themselves are quantum, through the coherent state path integral, we are able to obtain the corresponding classical analog in the form of an action and a Lagrangian. 

\noindent Unfortunately, the amount of background needed to fully follow every detail of this section goes beyond the scope of these notes. Therefore, it is strongly recommended to first go through \cite{nakahara, atiyah, proseminar} to study the basics of differential geometry, connections and curvatures on a fiber bundle, and characteristic classes. From this point on, the text becomes more mathematical, and notions of the three mentioned topics will be assumed.

\subsubsection{\texorpdfstring{$\bm\theta$-terms}{theta-terms}, Chern classes and winding numbers}
As mentioned before, in some cases, the classical action contains terms that depend only on the topology of the field configuration space. To clarify this point, it is useful to think of a total action 
\begin{equation}
    S[\phi(x)]= S_0[\phi(x)]+S_{top}[\phi(x)]
\end{equation}
which has two contributions: $S_0[\phi(x)]$, which depends on the local geometric and dynamical structure of the theory, and $S_{top}[\phi(x)]$, which depends only on the topology of the target space $T$ and topological properties of the map
\begin{equation}
    \phi: M^d \to T, 
\end{equation}
rather than on the local details of the specific configuration. In the effective-field-theory setting (where we can take the compactification of $\mathbb{R}^d$ to $S^d$ due to some finite energy requirement -- $\phi$ constant at infinity), we often take the space $M^d$, where the field is defined, to be the $d$-sphere $S^d$. On the opposite side, the target space $T$ is usually constructed from a quotient $G/H$, where $G$ is one of the compact classical groups $U(N)$, $O(N)$, $SU(N)$ or $Sp(N)$, and $H$ is some subgroup thereof. Physically, this is due to the fact that we often work with theories with continuous symmetries described by the action of a Lie group $G$, but these symmetries may or may not be spontaneously broken down into a smaller group $H$, which, for our purposes, is not needed to describe new physical situations.

In this way, and using QFT slang, we say that the base manifolds $M^d$ of our theories are spheres, while the target manifolds $T$ are topological cosets $G/H$ that account for some physical redundancies. The key point here is that maps of the form 
\begin{equation}
    \phi: S^d \to G/H, 
\end{equation}
are classified by the $\mathbf{d^{th}}$\textbf{- Homotopy Group} of $T=G/H$, denoted as $\pi_d(T)$. As a rule in physics, we are only interested in the specific cases when these homotopy groups are indeed the integers $\mathbb{Z}$. Thus, the set of all maps $\Set{\phi: S^d\to T}$ is divided into subsets $\Set{\phi}_W$, called \textbf{topological sectors}, each of them characterized by an integer number $W \in \mathbb{Z}$. This integer $W$ receives different names depending on the context; some of the common ones are: \textbf{winding number, Chern number} or \textbf{topological charge}.

The standard example where these winding numbers appear is a theory $\phi: S^d \to S^d$, where we have that $\pi_d(S^d)=\mathbb{Z}$ for all $d$-dimensions. In this particular case, maps of $S^d$ into itself are classified according to how often they wrap around the $d$-sphere. In simple words, the winding number $W$, here, represents how many times the map $\phi$ completely covers the target sphere $S^d$. Special attention must be drawn to the cases $g:S^1\to S^1 \simeq U(1)$ and $g:S^3\to S^3\simeq SU(2)$, since they underlie the appearance of $U(1)$ monopoles and $SU(2)$ instantons, respectively. 

Once we know that the fields of our theories are classified by $\mathbb{Z}$, the next step is to know how to compute the corresponding integer $W$ for a specific field configuration $\phi_W$. In this text, we adopt two different viewpoints to compute this integer number, but keep in mind that there are several ways to arrive at the same answer. The first one will motivate the name of winding number since we will be working with the space $T$ as a geometrical shape $S^d$ (a manifold). By contrast, the second formalism is more closely related to a generalization of the form of Chern number, which strongly uses the fact that in the target space, we have a redundancy in the form of a Lie group $G$ with its corresponding Lie algebra $\mathfrak{g}$. 
\begin{itemize}
    \item \underline{As Winding number -- Geometric shape}
    
    We start the construction of $W$ in this approach from the fact that $T=G/H$ is a (compact) $d$-dimensional smooth manifold, and therefore, it has a $d$-differential volume form $\omega \in \Omega^d(T)$ which can be normalized to 1. This is
    \begin{equation}
        \int_T \omega =1\:\:. 
    \end{equation}
    Now, since our field $\phi: S^d \to T$ is a map between smooth manifolds, we can consider the induced map (pullback) in the cotangent spaces $\phi^*:T^*_{\phi(p)}T \to T^*_pS^d$, which naturally extends to differential forms as
    \begin{equation}
        \phi^*: \Omega^{n}(T) \to \Omega^n(S^d).
    \end{equation}
    In this way, the $d$-volume form in $T$, $\omega\in\Omega^d(T)$, can be "transported" or pulled back to an integrable $d$-form in $S^d$ by taking $\phi^*\omega\in\Omega^d(S^d)$. Thus, if $T$ is also regarded as a $d$-sphere $S^d$, and $\phi_1$ is an orientation-preserving diffeomorphism (like a change of coordinates), then the integral is still normalized to 1 (since the volume of the sphere is invariant under a change of coordinates) 
    \begin{equation}
        1=\int_{M=S^d} \phi^*_1 \omega. 
    \end{equation}
    This means indeed that any orientation-preserving diffeomorphic map $\phi_1$ is classified by the integer number $W=1$ (orientation-reversing diffeomorphism gives $W=-1$). Note that the diffeomorphism character can also be described as the situation in which the map $\phi_1$ covers exactly once the target $S^d$. 
    
    Following this idea, to generate higher integers, we should be able to cover the target $d$-sphere multiple times. Here is where the locality enters the game. Suppose we divide the base $d$-sphere into $W$ local open sets $\Set{U_i}_{i=1,...,W}$ (say they correspond to charts). Then, by definition of $d$-manifold, these open sets are diffeomorphic to $\mathbb{R}^d$, which, at the same time, is diffeomorphic to the target $d$-sphere without a single point (i.e. $S^d \smallsetminus \{p\}$). It follows that we can construct a map $\phi_W$ such that the restriction on each open $U_i$ is a diffeomorphism $\phi|_{U_i}: U_i \to S^d\smallsetminus \{p\}$. Therefore, if we want to compute the integral of the pullback volume form, we must use local integration
    \begin{equation}\label{eq:locali_windi}
        \int_{M=S^d} \phi_W^* \omega= \int_{U_1} \phi|_{U_1}^*\omega + \int_{U_2} \phi|_{U_2}^*\omega + ... + \int_{U_W} \phi|_{U_W}^*\omega \ ,
    \end{equation}
    where now each integral term is 
    \begin{equation}
        \int_{U_1} \phi|_{U_1}^*\omega = \int_{S^d \smallsetminus \{p\}} \omega = \int_{S^d} \omega = 1,
    \end{equation}
    after using diffeomorphic restrictions and the zero measure of single points (removing a single point does not change the volume). Consequently, we have found that Eq.\ref{eq:locali_windi} is indeed the winding integer number
    \begin{equation}
        \int_{M=S^d} \phi_W^* \omega= 1+1+...+1=W. 
    \end{equation}
    To summarize, in this approach, we have that the integers $W$, which classify the maps $\phi: M=S^d \to T\simeq S^d$, count how many times the map winds around the target space, and are computed through 
    \begin{equation}\label{eq:winding_number}
        \boxed{W=\int_M \phi^* \omega} \ \ .
    \end{equation}
    
    \item \underline{As Chern number -- Redundant Lie group}
    
    Even though the previous picture gave us an idea of how to compute these integers, its starting assumptions were too restrictive. In particular, the assumption that the base manifold $M^d=S^d$ and the target manifold $T^n$ have exactly the same dimension $d=n$. The question now is how to proceed when we have different dimensions, but keeping $\pi_d(T)=\mathbb{Z}$. The answer will take us to modern results in differential geometry and the theory of topological invariants (especially the celebrated Atiyah-Singer Index theorem). But to keep the answer short, we have to go through the computation of Chern numbers as the integrals of some characteristic volume forms (Chern classes). 
    
    Indeed, this formalism will give us the opportunity to classify maps that have a base manifold different from a $d$-sphere $S^d$. For instance, as we will see in Sec.\ref{sec:su_n_superspin}, maps defined in a general closed Riemann surface $\Sigma\simeq S^2$, $ \tilde n: \Sigma \rightarrow \mathbb{C P}^{N-1}$ are classified by the winding integer $W$, reflecting the nontrivial topology $\pi_2(\mathbb{C P}^{N-1})=\mathbb{Z}$ for all values of $N$. The key message is that this integer can be computed geometrically using the language of fibre bundles. In particular, $\mathbb{C P}^{N-1}$ admits a canonical $U(1)$ principal bundle, and the Berry connection $\mathcal{A}$ arises as the pullback of a connection on this bundle via the map $\tilde n: \Sigma \rightarrow \mathbb{C P}^{N-1}$. The corresponding curvature defines a two-form whose integral yields the first Chern number. To clarify all these concepts, let us briefly recall the theory of characteristic classes in principal bundles:
    \begin{enumerate}
        \item A $G$ principal bundle $P(M,G)$ (being $G$ a Lie group) is a geometric/topological construction which locally is done by gluing pieces of the form $U_i\times G$, where $U_i$ are open sets of the "base" manifold $M$. Indeed, the gluing instructions are maps $g_{i j}: U_i \cap U_j \rightarrow G$, such that in intersections $U_i \cap U_j$, these maps induce an equivalence relation in $\{U_i\times G\}$, in the way that $u_1=(p,g_1) \sim u_2=(p,g_2)$ if the gluing map connects the "group component" $u_1=(p,g_1)=(p,g(p)g_2)$. Then, the principal bundle is the collection of the points from $\{U_i\times G\}$ with the gluing regions identified.  
        \item Since locally, the tangent space of $P(M,G)$ is $T_uP\simeq T_pM \oplus \mathfrak{g}$, then we can choose (on each point) a decomposition $T_uP=H_uP\oplus V_uP$ such that (1) $V_uP\simeq \mathfrak{g}$ and (2) the decomposition is smooth (if we move to a neighboring point $p^\prime$ then the tangent vectors change smoothly). This decomposition is done by local connection 1-forms $\mathcal{A}_i$ defined on each local set $U_i$. However, these are not ordinary 1-forms; they are 1-forms evaluated (or with coefficients) in the Lie algebra $\mathfrak{g}$. This way, the spaces $H_uP$ are the kernels of the maps $\mathcal{A}_i(p)$, while $V_uP$ correspond to their images $\operatorname{Im}\mathcal{A}_i(p)$. Finally, since this construction is done locally, we should have some compatibility conditions in intersections (to be well-defined globally). This is the famous gauge transformation $\mathcal{A}_j=g^{-1}\mathcal{A}_ig+g^{-1}dg$.
        \item Once we have the local connection form $\mathcal{A}_i$, we define the local curvature 2-form by $\mathcal{F}=d\mathcal{A} + \mathcal{A}\wedge\mathcal{A}$. This form is indeed globally defined, such that in all open sets $U_i$ (under all local trivializations), we have the same equivalent (basis change) expression for $\mathcal{F}$. In other words, the curvature transforms covariantly under gauge transformations $\mathcal{F}_j=g^{-1}\mathcal{F}_i\:g$ (or in the case of G abelian, it is gauge invariant, like the EM tensor). 
        \item Having the 2-form $\mathcal{F}$, we can construct 2n-forms by taking $\mathcal{F}^n=\mathcal{F}\wedge ... \wedge \mathcal{F}$. Note that here, we are using the convention $w\wedge w= \frac{1}{2}\: w^\alpha\wedge w^\beta \:[T_\alpha,T_\beta]$. In particular, we claim that these 2n-forms are still gauge covariant, such that when we take $\operatorname{tr(\mathcal{F}^n)}$, we have a gauge invariant polynomial in the Lie algebra coefficients of $\mathcal{A}$. 
        \item Thus, we have a clue about constructing invariant polynomials on $\mathcal{A}$. They may be written in terms of a sum of products of $\text{tr}(\mathcal{F}^n)$. Indeed, the Chern-Weyl theorem establishes that any invariant polynomial induces a unique cohomology class. So, any invariant polynomial maps to a closed-form in $H^*(M)$, where $M$ is the base space of our fibre bundle. So it defines a cohomology class on $M$ that is independent of the chosen connection. These induced cohomology classes are called \textbf{characteristic classes}.
    \end{enumerate}
    In this way, one of the different characteristic classes that exist is the so-called \textbf{Chern character}, which is defined as the $2j-$forms 
    \begin{equation}
    \operatorname{ch}_j(\mathcal{F}) \equiv \frac{1}{j!} \Tr\left(\frac{i \mathcal{F}}{2 \pi}\right)^j,
\end{equation}
    where the $j$-exponent in the trace should be understood as taking the trace of $\mathcal{F}^j$ (where the wedge product is assumed) and multiplying it by $(i/2\pi)^j$. The point is that, in general, the Chern characters are not always the most convenient quantities for extracting integer invariants directly (e.g. not having traceless groups). Fortunately, they are the building blocks of the characteristic classes that do, the so-called \textbf{Chern classes}. Through a suitable expansion of the total Chern class (think about it as a generator)
    \begin{equation}
        C(\mathcal{F}) \equiv \operatorname{det}\left(I+\frac{\mathrm{i} \mathcal{F}}{2 \pi}\right) =1+C_1(\mathcal{F})+C_2(\mathcal{F})+... + C_{k}(\mathcal{F})
    \end{equation}
    one can obtain that the $j^{th}$ Chern classes are (here $k$ is some integer determined by the number of eigenvectors of $\mathcal{F}$)
    \begin{equation}
        \begin{aligned}
C_0(\mathcal{F}) & =1 =ch_0(\mathcal{F}) \\
C_1(\mathcal{F}) & =\frac{i}{2 \pi} \operatorname{Tr}(\mathcal{F}) = ch_1(\mathcal{F}) \\
C_2(\mathcal{F}) & =\frac{1}{2}\left(\frac{i}{2 \pi}\right)^2[\operatorname{Tr}(\mathcal{F}) \wedge \operatorname{Tr}(\mathcal{F})-\operatorname{Tr}(\mathcal{F}^2)] = \frac{1}{2} \operatorname{ch}_1^2(\mathcal{F}) - \operatorname{ch}_2(\mathcal{F}) \\
\ldots & 
\end{aligned}
    \end{equation}
Therefore, given a principal bundle over a 2$n$-dimensional manifold $M$, \textbf{the Chern number} $c_n$ is defined as the integral of the $n$-th Chern form over $M$,
\begin{equation}\label{eq:chern_number}
    \boxed{c_n=\int_{M} C_n(\mathcal{F})} \ \ .
\end{equation}
In the most general case, the integrality of $c_n$ follows from the fact that Chern classes are integral characteristic classes, but to give an analytic interpretation of such a result, the Atiyah-Singer Index Theorem proves that the expression in Eq.\ref{eq:chern_number} is exactly the index of some elliptic operator acting on the sections of the corresponding associated vector bundle. In simpler words, it is an integer because it is the difference between two dimensions (index of an operator). 
\end{itemize}

Having introduced these two equivalent ways (Eq.\ref{eq:winding_number} and Eq.\ref{eq:chern_number}) to obtain the integers $W$, we would now like the topological action $S_{top}[\phi]$ to depend only on the topology of the field $\phi$ (i.e. the topological sector $W$ to which $\phi$ belongs). In this way, we should have that the action $S_{top}[\phi_W]$ is indeed a function $F(W)$ over the winding numbers $W$. However, by the requirement of this $F$ to be well-defined, one can show that this function should be linear in $W$. That is, $S_{top}=F(W)=aW+b$. Therefore, ignoring the constant $b$ (which can be reabsorbed in path integral measure) and choosing $a=i\theta$ by the physical motivation of having a purely imaginary result, we have that the action for a \textbf{topological} $\bm{\theta}$\textbf{-term} is defined as
\begin{equation}
    \boxed{S_{\text {top }}[\phi]=i \theta W} \ \ .
\end{equation}
Note that by choosing the constant as imaginary, we ensure that the parameter $\theta$ is defined $mod-2\pi$ since taking the exponential leaves us with a phase of the form 
\begin{equation}
    e^{-iW(\theta+2\pi)}=e^{-iW\theta}e^{-i2\pi W} =e^{-iW\theta}.
\end{equation}
In this way, we refer to the parameter $\theta$ as an angle, and eventually, it is the only parameter that must be specified in the theory (it's indeed the coupling constant of our topological term). 

Similarly, a very important feature that describes these $\theta$-terms is the fact that they do not contribute to the classical equations of motion in the usual variational sense. Since we have an action that is represented by the integral over the closed manifold in which our theory is constructed, we can associate a topological lagrangian $\mathcal{L}_{top}$ which contributes to the total one $\mathcal{L}=\mathcal{L}_0 + \mathcal{L}_{top}$. However, once we take $\delta S=0$ or construct the Euler-Lagrange equations, we will find that the contributions of this term cancel. The reason behind this fact is that the topological action $S_{top}[\phi_W]$ is invariant under local deformations of fields, or in other words, continuous variations of one field $\phi_W$ will give other fields in the same homotopic sector $W$ such that the $\theta$-term is exactly the same. In order to change the value of the topological action, one would need non-continuous deformations, which are not included in $\delta S$. 

\noindent The important relevance of such terms is therefore not encoded at the level of the classical equations of motion, but rather in the structure of the path integral itself. Since configurations belonging to different homotopy sectors $W$ cannot be continuously deformed into one another, the functional integral naturally decomposes into a sum over disconnected sectors, each weighted by a phase $e^{- S_{\text {top }}\left[\phi_W\right]}$:
\begin{equation}
    \mathcal{Z}=\sum_{W \in \mathbb{Z}} e^{-i\theta W} \int D \phi_W \:\: e^{-S_0[\phi]}
\end{equation}
In this way, the topological term assigns a relative phase between distinct sectors of the theory, even though it remains invisible to local variations of the fields. Thus, these terms encode the global geometry of configuration space and directly influence the quantum interference pattern between topologically inequivalent configurations.

As examples of the computation of the mentioned integers $W$, here we will take into consideration theories where we have maps $g: S^1\to S^1$, $\tilde n:S^2 \to S^2$, and $g:S^3\to S^3$. The first case will be important for the construction of $U(1)$-monopoles, the second for the existence of non-trivial skyrmions, and the latter for the structure of $SU(2)$-instantons. 

\begin{itemize}
    \item[$\maltese$] \underline{$U(1)-S^2$ Monopole Bundle}
    
   For this first example, we consider a $U(1)$ principal bundle over the $2$-dimensional sphere $S^2$. Physically, this situation can be understood by considering a gauge field $A_\mu$ defined in three-dimensional space $\mathbb{R}^3\smallsetminus\set{0}$ and restricting its field strength to a closed surface $S^2$ surrounding the origin. In this sense, the sphere plays the role of a probe on which we measure the magnetic flux.  Since the surface is closed, we can compute the magnetic flux as $\Phi =\int_{S^2} \mathbf{B} \cdot \mathrm{d} \mathbf{S}$,where $\mathrm{d}\mathbf{S}$ is the surface element of $S^2$, and $\mathbf{B}$ is the magnetic field in the ambient space. Due to the radial orientation of the sphere, only the radial component of $\mathbf{B}$ contributes to the flux. Therefore, a configuration with nonzero flux through $S^2$ is equivalent to having a magnetic charge (monopole) located inside the sphere. In this way, the nontrivial topology of the $U(1)$ bundle is physically realized as the presence of a magnetic monopole.

    Mathematically, we know that the Chern number of this $P(S^2,U(1))$ principal bundle is given by
    \begin{equation}
    W_{P(S^2,U(1))}=\frac{\mathrm{i}}{2 \pi} \int_{S^2} \mathcal{F}=\int_{S^2} \operatorname{C}_1(\mathcal{F}),
    \end{equation}
    where the curvature two form is related to the connection through  $\mathcal{F}=d\mathcal{A}$, and at the same time, the connection with the vector potential by $\mathcal{A}=iA_\mu dx^\mu$. Additionally, with a bit of work, one can show that the flux integrand $\mathbf{B} \cdot \mathrm{d} \mathbf{S}$ is proportional to the curvature 2-form $\mathcal{F}$. Specifically, we find
    \begin{equation}
        \mathcal{F}=\frac{1}{2}\mathcal{F}_{\mu\nu} \ dx^\mu \wedge dx^\nu = \frac{i}{2}F_{\mu\nu} \ dx^\mu \wedge dx^\nu,
    \end{equation}
    where $F_{\mu\nu}$ is the electromagnetic tensor defined by $E_i=F_{i0}$ and $B_i=\frac{1}{2}\epsilon^{ijk}F_{jk}$. In this line, if the Chern number and magnetic flux integrands are proportional $ \mathcal{F} \propto \mathbf{B} \cdot \mathrm{d} \mathbf{S}$, and moreover, the former one is equal to an integer after integration, then the integral of second one should also be proportional to that integer $\int \mathbf{B} \cdot \mathrm{d} \mathbf{S} \propto W$. This will mark the whole point of our story: \textit{monopole magnetic flux is quantized because of the existence of Chern number}.  

    However, one may wonder why a nonzero magnetic flux through a closed surface does not contradict the usual description of electromagnetism in terms of a vector potential. The point is that for a nontrivial $U(1)$ bundle over $S^2$, the gauge potential cannot be defined globally as a smooth one-form on the whole sphere. Instead, it must be described in different local charts, which is naturally accommodated by the fact that $S^2$ cannot be covered by a single chart, and therefore requires at least two overlapping patches. Thus, in order to obtain a magnetic field of the form $\mathbf{B}=(g/r^2)\:\hat{\mathbf{e}}_r$, one should divide the expression for the vector potential in the two hemispheres 
    \begin{equation}\label{eq:poten_mono}
        \begin{aligned}
\boldsymbol{A}^{\mathrm{N}}(\boldsymbol{r}) & =\frac{g(1-\cos \theta)}{r \sin \theta} \hat{\boldsymbol{e}}_\phi \xrightarrow{\ \ \ \ } \mathcal{A}_{\mathrm{N}}=\mathrm{i} g(1-\cos \theta) \mathrm{d} \phi \\
\boldsymbol{A}^{\mathrm{S}}(\boldsymbol{r}) & =-\frac{g(1+\cos \theta)}{r \sin \theta} \hat{\boldsymbol{e}}_\phi \xrightarrow{\ \ \ \ } \mathcal{A}_{\mathrm{S}}=-\mathrm{i} g(1+\cos \theta) \mathrm{d} \phi.
\end{aligned}
    \end{equation}
Where indeed we should notice that to pass from one hemisphere to the other (in the intersection which is $S^1$), we should perform a gauge transformation $t_{NS}: S^1 \to U(1)$  (the denominated transition functions of our fiber bundle) 
    \begin{equation}
    \mathcal{A}_{\mathrm{N}}=t_{\mathrm{NS}}^{-1} \mathcal{A}_{\mathrm{S}} t_{\mathrm{NS}}+t_{\mathrm{NS}}^{-1} \mathrm{~d} t_{\mathrm{NS}}\:, \qquad \text{with } \:\: t_{N S}=e^{2 i g  \phi}\:.
    \end{equation}
    However, using the fact that we are dealing with an abelian group, we can simplify to $\mathcal{A}_{\mathrm{N}}-\mathcal{A}_{\mathrm{S}}=t_{\mathrm{NS}}^{-1} \mathrm{d} t_{\mathrm{NS}}$, which in our case of Eq.\ref{eq:poten_mono}, is $\mathcal{A}_{\mathrm{N}}-\mathcal{A}_{\mathrm{S}}=2ig \mathrm{~d}\phi$. In this way, we can compute the magnetic flux just by watching the intersection $S^1$ of the two hemispheres $U_N, U_S$ (with the help of Stokes' theorem)
    \begin{equation}\label{eq:stoke_mono}
    \begin{aligned}
\Phi & =\int_{S^2} \mathbf{B} \cdot \mathrm{d} \mathbf{S}=\int_{U_{\mathrm{N}}} \mathrm{d} A_{\mathrm{N}}+\int_{U_{\mathrm{S}}} \mathrm{d} A_{\mathrm{S}} \\
& =\int_{S^1} A_{\mathrm{N}}-\int_{S^1} A_{\mathrm{S}}=2 g \int_0^{2 \pi} \mathrm{d} \phi=4 \pi g, 
\end{aligned}
\end{equation}
where the minus sign at LHS of the 2nd line comes from the opposite relative orientation of $S^1$ with respect to $U_S$. This procedure indeed will mark a very deep feature about Chern numbers and, in general, $\theta$-terms: "\textit{One can compute integers by taking the difference between local expressions defined in two suitable hemispheres}."

Indeed, if we had not fixed the form of $\mathcal{A}_N$ and  $\mathcal{A}_S$ in advance, and we just aimed to construct an integer, we could have proceeded more abstractly. Since each hemisphere is contractible, one may choose a trivialization in which, for instance, $\mathcal{A}_S=0$. In this case, all the nontrivial information is encoded in the overlap, where the connection on the northern patch becomes a pure gauge $\mathcal{A}_N = t_{\mathrm{NS}}^{-1} \mathrm{d} t_{\mathrm{NS}}$. Therefore, the entire topology of the configuration is captured by the transition function $t_{NS}: S^1 \to U(1)\simeq S^1$, recovering the topological description of our old friend $\pi_1(S^1)=\mathbb{Z}$. The advantage of this point of view is that it admits a natural generalization for higher groups. For maps $g: S^1 \to U(N)$, one can define 
    \begin{equation}
        W[g]=\frac{i}{2 \pi} \int_{S^1} d x \operatorname{tr}\left(g^{-1} \partial_x g\right) = \frac{i}{2 \pi} \int_{S^1} dx\:\: \partial_x\log (\operatorname{det} g) 
    \end{equation}
    where the trace has been included to account for multiple generators. In this sense, the maps $g$ play the role of transition functions that encode how local trivializations are glued together in $P(S^2,U(N))$ bundles, and the resulting integers are precisely the first Chern numbers of their respective bundles. Note that the quantization in integer units reflects the fact that $\pi_1(U(N))=\mathbb{Z}$ for all $N\geq1$ and corresponds to the winding of the function $\det g \in U(1)$. 

\item[$\maltese$] \underline{$SU(2)-S^4$ Instanton Bundle}
    
    In our next relevant example, we have an $SU(2)$ principal bundle $P(S^4, SU(2))$ over $S^4$. Since we have a four-dimensional base manifold, we need to consider the second Chern class, which is now equivalent to the second Chern character due to the traceless property of $\mathfrak{su}(2)$. Then, the Chern number for this theory is written as
    \begin{equation}
    W_{P(S^4,\:SU(2))}=\frac{1}{2} \int_{S^4} \operatorname{tr}\left(\frac{i \mathcal{F}}{2 \pi}\right)^2=\int_{S^4} \operatorname{C}_2(\mathcal{F}).
    \end{equation}
    Now, in the same spirit as our previous discussion (see $U(1)$-monopole), we look for a way to express the integer in terms of a one-dimensional lower boundary theory. In this way, we consider the two hemispheres of $S^4$, denoted again by $U_{N, S}$, such that their intersection is (homotopically equivalent to) $S^3$. Using Poincaré Lemma, we know that the top Chern class is locally exact $\operatorname{tr}(\mathcal{F}^2)=dK$ for some $K_{N,S}\in \Omega^{4-1}(U_{N,S})$. Indeed, this $K$  received the name of \textbf{Chern-Simons 3-form}, and is given by
    \begin{equation}
        K_{N,S}=\operatorname{tr}\left[\mathcal{A}\wedge d\mathcal{A}+\frac{2}{3} \mathcal{A}^3\right]=\operatorname{tr}\left[\mathcal{A}\wedge\left(\mathcal{F}-\mathcal{A}^2\right)+\frac{2}{3} \mathcal{A}^3\right],
    \end{equation}
   where the local dependence of the connection $\mathcal{A}_{N,S}$  should be implicitly understood. This means that by expressing
    \begin{equation}\label{eq:stoke_insta}
        \int_{S^4} \operatorname{tr}(\mathcal{F}^2) = \int_{U_{\mathrm{N}}} dK_N + \int_{U_{\mathrm{S}}} dK_S=\int_{S^3} K_N - \int_{S^3} K_S,
    \end{equation}
    and arbitrarily considering a local trivialization with $A_S=0$ ($A_N=g^{-1}dg$), we have that
    \begin{equation}\label{eq:insta_local}
    W_{P(S^4,SU(2))} \propto -\frac{1}{3} \int_{S^3} \operatorname{tr} \mathcal{A}_N^3,
    \end{equation}
    where we have used that $\mathcal{F}=0$ in $S^3$. Note that this is the same mechanism as in $P(S^2, U(1))$: we pass all the construction freedom to the north hemisphere connection $\mathcal{A}_N$, defined by the maps $g: S^3 \to S^3 \simeq SU(2)$. Therefore, we recover the homotopy group $\pi_3(S^3)=\mathbb{Z}$ and the traditional winding number interpretation. As before, the advantage of this framework is precisely that, for a broader class of groups, this construction can be naturally generalized. In particular, whenever the target group $G$ admits a nontrivial third homotopy group $\pi_3(G)=\mathbb{Z}$, one can associate an integer to maps $g:S^3 \to G$ through the expression
\begin{equation}
W[g]=\frac{1}{24 \pi^2} \int d^3 x \ \epsilon^{i j k} \operatorname{tr}\left(\left(g^{-1} \partial_i g\right)\left(g^{-1} \partial_j g\right)\left(g^{-1} \partial_k g\right)\right).
\end{equation}
Note that this expression is precisely the local form of Eq.\ref{eq:insta_local}, obtained by expanding the wedge products. This formula provides an integral representation of the winding number of the map $g$, and generalizes the $SU(2)$ case discussed above to higher-dimensional Lie groups such as $SU(N)$ with $N\geq 2$. In this sense, the topological invariant is again encoded in the transition functions of $P(S^4,G)$ bundles, independently of the specific details of the group, as long as $\pi_3(G)$ is nontrivial.

    \item[$\maltese$] \underline{$S^2-S^2$ Gauss-Bonnet Theorem}
    
    Unlike our previous examples, where we established the correspondences
\begin{align*} &\bullet \quad \pi_1(S^1) =\mathbb{Z}\:\:\xleftrightarrow{\quad}\: \:\:U(1)\text{ bundle over }S^2 \:\xleftrightarrow{\quad}\:\text{ topol. sectors } \left\{A_\mu^W:S^2\to U(1)\right\}\\ &\bullet \quad \pi_3(S^3)=\mathbb{Z} \:\:\xleftrightarrow{\quad}\: \:\:SU(2)\text{ bundle over }S^4 \:\xleftrightarrow{\quad}\:\text{ topol. sectors } \left\{A_\mu^W:S^4\to SU(2)\right\}\:, \end{align*}
for the case of $\pi_2(S^2)=\mathbb{Z}$ we cannot directly appeal to a Lie group or principal bundle construction, since $S^2$ is not a group. The natural question is then: \textit{from where does the topological integer arise in this case?}

In the monopole and instanton examples, the physical setting was a gauge theory with dynamical gauge fields. These gauge fields (connections) $A_\mu$ are classified, modulo gauge transformations, into different topological sectors. Mathematically, these sectors correspond to inequivalent principal bundles $P(M,G)$, characterized by their Chern numbers $c_n(P)$. In this way, the topology is encoded in the global structure of the gauge bundle over spacetime.

However, for a map $\tilde n:S^2\to S^2$, the situation is fundamentally different. The primary topological object is not a gauge field or a bundle, but the map $\tilde n$ itself. The associated integer arises from how this map wraps the domain sphere around the target sphere. In this case, it is more natural to adopt the winding-number (pullback) construction, postponing the discussion of characteristic classes, although both perspectives are ultimately related.

To implement this construction, we recall that by denoting a point on the target sphere $S^2$ as $\mathbf{y}=(y_1,y_2,y_3)$, the normalized area form on the sphere is given by
\begin{equation}
\omega_{S^2}=\frac{1}{4\pi}\:\mathbf{y}\cdot(d\mathbf{y}\wedge d\mathbf{y})\:\:.
\end{equation}
Then, given a map $\tilde n:M^2\to S^2$, the pullback $\tilde n^*(\omega_{S^2})$ defines a two-form on the domain $M^2$. In local coordinates $x=(x^1,x^2)$, this leads to the expression
\begin{equation} \label{eq:winding_number_S2}
W[\tilde n]\:\:=\int_{M^2} \tilde n^* (\omega_{S^2})\:\:=\:\:\frac{1}{4 \pi} \int d^2 x \:\:\tilde n(x) \cdot\left[\partial_1 \tilde n(x) \times \partial_2 \tilde n(x)\right]\:.
\end{equation}

At this point, a common source of confusion arises. The structure of this expression resembles that of the Gauss–Bonnet invariant of a two-dimensional manifold,
\begin{equation}
\chi(M^2)=\frac{1}{2 \pi} \int_{M^2} K d A=\frac{1}{2 \pi} \int_{M^2} \mathbf{n} \cdot\left(\partial_1 \mathbf{n} \times \partial_2 \mathbf{n}\right) \:d^2 x
\end{equation}
where $K$ is the Gaussian curvature of the manifold $M^2$, and $dA$ is the corresponding area element. At first sight, this expression appears identical to the one obtained in Eq.\ref{eq:winding_number_S2}. This similarity often leads to the misleading conclusion that the quantization of $W[\tilde n]$ is a direct consequence of the Gauss--Bonnet theorem. However, this is not the correct interpretation. The Gauss--Bonnet theorem is a statement about the \textit{intrinsic geometry} of the manifold $M^2$, which requires the introduction of a Riemannian metric. In this setting, the Gaussian curvature $K$ is constructed from the Levi--Civita connection of the tangent bundle $TM$, and the unit vector $\mathbf{n}$ is not an independent field, but the normal vector determined by the embedding of the surface. The resulting integral computes the Euler characteristic $\chi(M^2)$, a topological invariant of the manifold itself.

In contrast, in Eq.\ref{eq:winding_number_S2}, the field $\tilde n$ is an independent map $\tilde n : M^2 \to S^2$, and the integer $W[\tilde n]$ characterizes the topology of this map rather than that of the manifold. More precisely, it computes the degree of $\tilde n$, counting how many times (with orientation) the domain $M^2$ wraps around the target sphere $S^2$. While $\chi(M^2)$ is fixed once the manifold is given, the degree $W[\tilde n]\in\mathbb{Z}$ depends on the specific configuration of the field and labels different homotopy classes in $\pi_2(S^2)=\mathbb{Z}$.

Nevertheless, the similarity between both expressions is not accidental. In the Gauss--Bonnet framework, the relevant object is the curvature two-form $\Omega$ of the tangent bundle $TM$, whose associated characteristic class is the Euler class,
\begin{equation}
e(TM)\;=\;\frac{1}{2\pi}\,\Omega \in H^2(M^2)\quad \xrightarrow[]{\:\text{s.t. Gauss-Bonnet }\:} \quad \chi(M^2)=\int_{M^2} e(TM)\:.
\end{equation}

On the other hand, in Eq.\ref{eq:winding_number_S2}, the integrand arises from the pullback of a closed two-form defined on the target sphere,
\begin{equation}
\omega_{S^2}=\frac{\mathbf{y}\cdot(d\mathbf{y}\wedge d\mathbf{y})}{4\pi} \in H^2(S^2)\quad \xrightarrow[]{\:\text{s.t. map degree }\:} \quad W[\tilde n]=\int_{M^2} \tilde n^*(\omega_{S^2})\in\mathbb{Z}\:.
\end{equation}

Thus, both constructions can be understood within the same general framework:
\begin{equation}
\text{topol. integer} \;=\;\int_{M^2} \omega\ 
\quad \xleftrightarrow[]{\qquad } \quad \omega \in H^2(\cdot)
\end{equation}
where $\omega$ is a representative of a cohomology class. The key difference lies in its origin: in Gauss--Bonnet, $\omega=e(TM)$ is built from the intrinsic geometry of the domain, while in Eq.\ref{eq:winding_number_S2}, $\omega=\tilde n^*(\omega_{S^2})$ is obtained from the target space via the map $\tilde n$. In fact, for the sphere $S^2$, these two forms are closely related. The curvature form of the target sphere $\Omega_{S^2}$ associated with the Levi--Civita connection satisfies
\begin{equation}
\omega_{S^2} \;\propto\; \Omega_{S^2}\:,
\end{equation}
which explains why both constructions lead to the same local density. However, their interpretation remains fundamentally different:
\begin{equation}
\begin{aligned}
\text{Gauss--Bonnet:} \quad & \int_{M^2} e(TM) \;\;\Rightarrow\;\; \chi(M^2) \\
\text{Winding number:} \quad & \int_{M^2} \tilde n^*(\omega_{S^2}) \;\;\Rightarrow\;\; \deg(\tilde n)\:.
\end{aligned}
\end{equation}

This perspective also clarifies the generalization to higher dimensions. The Gauss--Bonnet theorem extends to
\begin{equation}
\chi(M^{2n}) = \int_{M^{2n}} e(TM) 
\;\sim\; \int_{M^{2n}} \operatorname{Pf}(\Omega)\:,
\end{equation}
where the Euler class is expressed in terms of the Pfaffian of the curvature. And similarly, general constructions of the form
\begin{equation}
W[\Phi] = \int_{M^d} \phi^*(\omega)\:,
\qquad \omega \in H^d(N)\:,
\end{equation}
where the choice of $\omega$ is dictated by the geometric structure of the target space. In physical applications, this leads to different characteristic forms:
\begin{equation}
\begin{aligned}
\text{complex geometry} \;&\Rightarrow\; \text{Kähler form } \omega \\
\text{gauge theory} \;&\Rightarrow\; \text{Chern classes } C_n(F) \\
\text{gravity} \;&\Rightarrow\; \text{Euler class } e(TM)\:.
\end{aligned}
\end{equation}
These different cases will be discussed in more detail in the following section.
\end{itemize}

\subsubsection*{The unifying framework: Integral classes}
So far, we have encountered two apparently different mechanisms through which integer topological invariants arise. On the one hand, in theories whose fields are maps $\phi: M^d \rightarrow N$, the relevant integer is obtained by integrating the pullback of a distinguished differential form on the target space. On the other hand, in gauge theories, integer invariants appear as Chern numbers of nontrivial bundles, whose topology is encoded in transition functions carrying nontrivial homotopy data. Even though these two pictures were introduced separately, they are not independent constructions. Rather, they should be understood as different realizations of the same general principle: topological terms arise from the pullback of geometrically distinguished closed forms whose cohomology classes are \emph{integral}.

\begin{itemize}
    \item[-] The first situation is the natural one in nonlinear sigma models, coherent-state path integrals, and effective theories of order-parameter fields. In these cases, the dynamical variable is a field configuration $\phi: M^d \rightarrow N, $ where $M^d$ is the spacetime manifold, worldvolume, or Euclidean base space, while $N$ is the target manifold parametrizing the allowed values of the field. Typical examples are maps into spheres $S^n$, projective spaces $\mathbb{CP}^{N-1}$, or more general coset manifolds $G/H$. In this setting, any differential form $\omega \in \Omega^d(N)$ on the target may be pulled back to a $d$-form $\phi^* \omega$ on the base, so that its integral over $M^d$ defines a functional of the field configuration. The basic question is then: under which conditions does this quantity depend only on the topology of $\phi$, and when does it become an integer?

    \item[-] By contrast, the second situation arises in a gauge theory framework. Here, the fundamental variable is no longer a map into a target manifold, but a connection $\mathcal{A}$ on a principal bundle over spacetime. The corresponding curvature two-form, $\mathcal{F}=d\mathcal{A}+\mathcal{A} \wedge\mathcal{A},$ allows one to construct gauge-invariant differential forms (e.g. Chern forms $C_n(\mathcal{F})$) such that their integrals over the base manifold define topological invariants of the bundle. From the physical point of view, these integers label different gauge-field sectors that contribute separately in the path integral, while from the mathematical point of view, they measure the global twisting of the bundle. In local coordinates one works with the connection $\mathcal{A}$, but globally the topology is carried by the gluing data, namely the transition functions relating local trivializations.
\end{itemize}

At first sight, these two constructions seem to belong to different worlds: one is formulated in terms of maps $\phi: M^d \rightarrow N$, while the other is expressed through gauge connections and bundle curvature. However, the difference is more apparent than real. In both cases, the topological integer is obtained by integrating a differential form that is not arbitrary, but geometrically selected by the structure of the problem. In the sigma-model picture, this is essentially the construction principle, whereas in the gauge-theory picture, it is a bit more subtle to observe such pullback framework. A useful way to make this connection explicit is to recall that a gauge field is not, strictly speaking, a globally defined object on spacetime. Rather, it is a \emph{local representative} of a more fundamental geometric structure, namely a connection defined on a principal bundle $P\left(M^d, G\right)$. To make this precise, consider that a principal bundle is locally trivial, so that over each open set $U \subset M^d$, one can choose a \emph{section}
$$
s_i: U_i \rightarrow P
$$
which assigns to every point in spacetime a point in the total space of the bundle. Physically, this choice corresponds to fixing a local gauge in the local open set $U_i$. Given such a section, one can pull back the global connection one-form $\omega_P$ defined on $P$ to obtain a local gauge field on spacetime,
$$
\mathcal{A}_i=s^*_i \omega_P
$$
Similarly, the curvature two-form $\Omega_P$ on the bundle, which is globally defined, gives rise to the familiar field strength
$$
\mathcal{F}_i=s^*_i \Omega_P\:.
$$
In this way, the gauge-theory quantities $\mathcal{A}$ and $\mathcal{F}$ that appear in local expressions are nothing but pullbacks of globally defined geometric objects along a section of the bundle.

From this perspective, the apparent difference with the sigma-model picture disappears. In both cases, the relevant quantities are obtained by pulling back differential forms defined on a geometric space: in the sigma-model case, this space is the target manifold $N$, while in gauge theory it is the total space of the principal bundle $P$. The role of the field $\phi: M^d \rightarrow N$ is now played locally by a choice of sections $s_i: M^d \rightarrow P$, and the characteristic forms constructed from the curvature are pulled back along this section to spacetime. So all the non-trivial topological properties of gauge quantities $\mathcal{A, F}$ are expressed in the topological freedom that we have to define local sections in a global space (the transition functions). Therefore, even in gauge theory, the topological invariant $\int_{M^d} C_n(\mathcal{F})$ should be understood as arising from the pullback of a geometrically defined form, rather than as a purely local construction in terms of the gauge field. The local expressions in terms of $\mathcal{A}$ and $\mathcal{F}$ simply provide a convenient representative of this underlying geometric structure.

This resolves the apparent difference we have between the winding number and Chern number pictures; but the fundamental questions remain: what are the required properties of the distinguished form $\omega\in \Omega^{d}(N)$ such that $\int_M \phi^* \omega$ only depends on the topology of $\phi$, and when does it become an integer? To understand why such expressions are topological, let us first focus on the role of closed differential forms. Suppose that $\omega \in \Omega^d(N)$ is closed, $d \omega=0$, and consider the functional
$$
I[\phi]=\int_{M^d} \phi^* \omega
$$
If $M^d$ is closed, then $I[\phi]$ is invariant under smooth deformations of the map $\phi$. In other words, changing the field continuously without crossing a singular configuration does not modify the value of the integral. This is the differential-form origin of homotopy invariance: \textbf{closed forms produce functionals that depend only on the homotopy class of the field configuration}, rather than on its local details. Nevertheless, closure alone is not enough for our purposes. A closed form indeed gives a homotopy-invariant quantity, but nothing forces its integral to be quantized. Multiplying a closed form by an arbitrary real number preserves closure, while generally destroying any possible integrality. Therefore, if we wish to obtain integer-valued topological invariants, we must refine the previous notion and identify a distinguished subclass of closed forms: those whose cohomology classes are \emph{integral}.

Concretely, a closed $d$-form $\omega$ on a manifold $N$ is said to be an \textbf{integral form} if all its periods over closed $d$-dimensional submanifolds are integers,
\begin{equation}
    \int_{\Sigma^d} \omega \:\:\:\in \mathbb{Z}
    \qquad
    (\forall\text{ closed } d \text{-dimensional } \Sigma^d \subset N)\:.
\end{equation}
More generally, one may allow oriented $d$-dimensional cycles (elements of the $d$-dimensional homology group of $N$), but for our purposes, it is sufficient to think of smooth closed submanifolds. This condition can be understood more systematically in terms of cohomology. The usual de Rham cohomology group $H^d(N,\mathbb{R})$ classifies closed differential forms up to exact ones, and therefore forms a continuous vector space: if $\omega$ is a representative, so is $\lambda\,\omega$ for any $\lambda\in\mathbb{R}$. In this sense, $H^d(N,\mathbb{R})$ contains no intrinsic notion of quantization. However, within this continuous space there exists a distinguished discrete subset, given by the integral cohomology group $H^d(N,\mathbb{Z})$. One may think of it as a lattice embedded inside $H^d(N,\mathbb{R})$, selecting those classes whose integrals over closed $d$-dimensional domains are quantized. Therefore, saying that a differential form $\omega$ is integral means that its de Rham cohomology class belongs to this lattice,
\begin{equation}
    [\omega] \in H^d(N,\mathbb{Z}) \subset H^d(N,\mathbb{R})\:.
\end{equation}

The practical importance of this definition is immediate: if $\omega$ is integral and $\phi: M^d \rightarrow N$ is any smooth map from a closed oriented manifold, then
\begin{equation}
    \boxed{
    \int_{M^d} \phi^* \omega = W \:\:\in \mathbb{Z}
    \quad \Rightarrow \quad
    S_{\text{top}} = i\theta W
    }\:\:.
\end{equation}
Thus, integral forms are precisely the closed forms that produce quantized topological $\theta$-terms after pullback. Note how if $M^d=S^d$, then the quantization depends only on the homotopy class of $\phi:S^d\to N$. So, whenever the relevant homotopy group is $\pi_d(N)= \mathbb{Z}$, the integral above measures exactly the corresponding winding or degree integer. Therefore, the relation between homotopy and differential forms is not accidental: the integral form translates the abstract classification by homotopy classes into an explicit local density that can be integrated.

The previous unification, however, still leaves an important question unanswered: where do such integral forms come from in practice? In physical applications, they do not appear arbitrarily. Rather, they arise from the geometric structures naturally associated with the problem. The relevant examples for our discussion are gauge bundles, tangent bundles in gravity, and the canonical line bundles that accompany complex and projective manifolds.
\begin{itemize}
    \item[-] As mentioned many times now, in gauge theories, the distinguished forms are the characteristic forms built from the curvature $\mathcal{F}$ of a connection. After the appropriate normalization, the Chern forms represent integral cohomology classes, and their integrals give the Chern numbers of the bundle. In this way, the quantization of topological charges in monopoles, instantons, and related gauge configurations is ultimately traced back to the integrality of the corresponding characteristic classes.
    \item[-] In gravity, the relevant bundle is no longer an internal gauge bundle, but the tangent bundle $TM$ of spacetime itself. The Levi--Civita connection defines a curvature two-form from which one constructs characteristic forms such as the Euler form or the Pontryagin forms. Their integrals yield topological invariants of the manifold, such as the Euler characteristic through the Gauss-Bonnet theorem. Thus, the same mechanism reappears: the geometry selects canonical closed forms whose cohomology classes are integral.
    \item[-] In complex geometry, and especially in the coherent-state manifolds relevant for quantum mechanics, the distinguished form is often a closed two-form associated with a canonical $U(1)$ bundle. For projective spaces such as $\mathbb{CP}^{N-1}$, this form is the Kähler form, which in the physical language becomes the Berry curvature. Its cohomology class is integral, and therefore its pullback by a field configuration leads to quantized topological terms. This is precisely the geometric origin of Berry-phase and Wess-Zumino contributions in coherent-state path integrals.
\end{itemize}

\subsubsection{Wess--Zumino Terms}
In the previous section, we saw that some integer topological invariants can be rewritten by passing from an integral over the whole space (typically a $d$-sphere $S^d$) to a difference of local integrals evaluated on overlapping patches. Explicitly, this is what happened in Eq.\ref{eq:stoke_insta} and Eq.\ref{eq:stoke_mono}, where the global integer was recovered from local data defined on the two hemispheres and their common boundary. Our task now is to formalize this mechanism as the general framework behind the definition of \textbf{Wess--Zumino (WZ) terms}.

\noindent The basic idea is the following. Suppose we have a field configuration $\phi:M^d\to T$ in $d$ spacetime dimensions, and suppose moreover that the target space $T$ admits a distinguished closed $d+1$-form $\omega\in\Omega^{d+1}(T)$ whose cohomology class is integral. Then, whenever the field $\phi$ can be extended to a $d+1$-dimensional manifold $U^{d+1}$ such that
\begin{equation}
    \partial U=M^d,
\end{equation}
one may define a WZ functional by integrating the pullback of $\omega$ over the filling manifold $U$. In this sense, a WZ term in $d$ dimensions is constructed from the same geometric object that, one dimension higher, would define a topological $\theta$-term. This is the precise sense in which WZ terms are descendants of higher-dimensional topological terms.

Schematically, this is exactly what happened in our previous examples when we considered $S^{d+1}$ and took as filling manifolds the north and south hemispheres $U_\pm$, whose common boundary is the equator,
\begin{equation}
    \partial U_\pm=M^d \simeq S^d.
\end{equation}
In that situation, the relevant $d+1$-form on the target is closed, and locally exact on each patch. Therefore, by Poincaré's lemma and Stokes' theorem, one may pass from an integral over the $d+1$-dimensional filling to an integral over its $d$-dimensional boundary. In the gauge-theory language this mechanism appears through a characteristic form such as a Chern form,
\begin{equation}\label{eq:wz_idea_correct}
    \int_U C_{d+1}= \int_{\partial U=M} K,
\end{equation}
where locally $C_{d+1}=dK$. The left-hand side is the \emph{extended representation} of the WZ term, while the right-hand side is its \emph{local representation}. 

To make the construction more transparent, let us now describe it in the winding-number picture. We begin with the original field $\phi:M^d\to T$ and introduce an extension
\begin{equation}\label{eq:extended_fields_correct}
\begin{split}
    \tilde{\phi}_U: U^{d+1} &\to T\\
    y &\mapsto \tilde{\phi}_U(y), \qquad \text{s.t. }\:\: \tilde{\phi}|_{\partial U=M}=\phi\:\:.
\end{split}
\end{equation}
In simple situations, one may think of this extension as a homotopy between the original field $\phi$ and a constant map. Writing the filling manifold $U^{d+1}\simeq [0,1]\times M^d$ schematically as an interpolation parameter $s\in[0,1]$ together with the boundary coordinates $x\in M^d$, one may represent the extension as $\tilde{\phi}(s,x)$ with
\begin{equation}
    \tilde{\phi}(1,x)=\phi(x),
    \qquad
    \tilde{\phi}(0,x)=\phi_0,
\end{equation}
where $\phi_0$ is a constant map to a chosen reference point of the target space. This picture is particularly intuitive when the target is a sphere, but the construction itself is more general. Then, the \textbf{extended representation of the WZ functional} is defined by
\begin{equation}\label{eq:global_repWZ_correct}
    \boxed{\Gamma_U[\phi]\equiv \int_U \tilde{\phi}^{*}_U\:\omega}\:,
\end{equation}
and since $\omega$ is closed, one may locally write $\omega=d\kappa_\alpha$ on a suitable patch $V_\alpha\subset T$ of the target space. Therefore, if the image of the extension field lies inside such a patch $\text{Im}(\tilde \phi_U) \subseteq V_\alpha$, Stokes' theorem gives the \textbf{WZ local representation}
\begin{equation}\label{eq:local_repWZ_correct}
    \boxed{\Gamma_\alpha[\phi]\equiv \int_M \phi^*\kappa_\alpha.}
\end{equation}
Thus, the two forms of the WZ functional are completely analogous to what we found before: the extended expression is written in one dimension higher $U^{d+1}\simeq [0,1]\times M^d$, while the local one is written directly on the physical spacetime manifold $M^d$.

In the special case in which $M^d=S^d$ and the filling is taken to be either hemisphere of $S^{d+1}$, one may define
\begin{equation}
    \Gamma_{U_\pm}[\phi]\equiv \pm\int_{U_\pm}\tilde{\phi}^*_{\pm}\:\omega,
\end{equation}
where the relative sign compensates for the opposite orientation of the two hemispheres, so that both choices induce the same orientation on the common boundary $S^d=\partial U_\pm$. In this way, the hemisphere construction should be understood as a particularly useful realization of the more general definition in Eq.\ref{eq:global_repWZ_correct}. However, in order to pass from this extended expression to a local one, one must introduce a second ingredient which is logically independent of the choice of hemisphere in the domain: a local cover of the target space $T$. Choosing an open cover $\{V_\alpha\}$ of $T$ together with local potentials $\kappa_\alpha$ with $\omega|_{V_\alpha}=d\kappa_\alpha$, we need to see whether or not it is possible to fulfill $\text{Im}(\tilde{\phi}_{\pm})\subseteq V_\alpha$.

\noindent The important point here is that the hemisphere choice $U_\pm\subset S^{d+1}$ and the patch choice $V_\alpha\subset T$ refer to different spaces, and therefore are independent. The first tells us how we extend the field in the domain, while the second tells us in which local chart of the target we describe the closed form $\omega$ by a potential $\kappa_\alpha$. Thus, even if we decide to use the north hemisphere $U_+$ as the filling manifold, the image of the corresponding extension $\tilde{\phi}_{+}$ may still lie in either one of several target patches, depending on the specific configuration. Therefore, if the image of the chosen extension satisfies $\mathrm{Im}(\tilde{\phi}_{\pm})\subset V_\alpha$, then Stokes' theorem gives
\begin{equation*}
    \Gamma_{U_\pm}[\phi]
    =\pm\int_{U_\pm}\tilde{\phi}_{\pm}^*\omega
    =\pm\int_{U_\pm}\tilde{\phi}_{\pm}^*(d\kappa_\alpha)
    =\int_{S^d}\phi^*\kappa_\alpha=\Gamma_\alpha[\phi],
\end{equation*}
where the last equality uses $\partial U_\pm=S^d$ with the induced boundary orientation.

Let us now illustrate this point in the simplest case when the target itself is the extension sphere, $T=S^{d+1}$ (so we consider globally extended maps $\tilde \phi: S^{d+1}\to S^{d+1}$). In that situation, the natural open cover is given by the two standard contractible patches
\begin{equation}
    V_N=S^{d+1}\smallsetminus\{p_S\},
    \qquad
    V_S=S^{d+1}\smallsetminus\{p_N\},
\end{equation}
with local forms $\kappa_N$ and $\kappa_S$ such that
\begin{equation}
    \omega=d\kappa_N \quad \text{in }V_N,
    \qquad
    \omega=d\kappa_S \quad \text{in }V_S.
\end{equation}
Now the conditions $\mathrm{Im}(\tilde{\phi}_{U_\pm})\subset V_N$ or $\mathrm{Im}(\tilde{\phi}_{U_\pm})\subset V_S$ have a very concrete meaning: they say that the image of the interpolating $(d+1)$-dimensional surface $\mathrm{Im}(\tilde\phi_\pm)$ in the target avoids one of the two poles. Physically, this means that the extension can be chosen so that, during the interpolation $s\in[0,1]$ from the reference configuration $\phi_0$ (constant function mapping to one point in $T$) to the boundary field $\phi(x)$, the image never crosses the excluded pole:
\begin{equation}
    p_{S,N} \notin \set{\tilde\phi_\pm(s,x)} \qquad (\forall s,x \in [0,1]\times M^d)
\end{equation}
In that case, a single local potential $\kappa_N$ or $\kappa_S$ is enough to describe the whole WZ functional. Notice, however, that the choice of filling hemisphere $U_\pm$ still does not determine whether one must use $\kappa_N$ or $\kappa_S$. For a given extension in the north hemisphere of the \emph{domain}, the image may avoid either the north or the south pole of the \emph{target}, depending on how the extension is chosen. Therefore, the target-space patch is not fixed by the hemisphere of the domain. What matters is only whether the image of the chosen extension fits inside one contractible patch of the target. In practice, one usually chooses whichever target patch makes the local description possible. If the image of $\tilde{\phi}_{U_+}$ lies inside both $V_N$ and $V_S$, then both local descriptions are valid and differ by an exact term on the overlap $V_N\cap V_S$ (which will produce an integer --Eq.\ref{eq:different_homotopy_extensions} but changing homotopy extensions by local potentials). If it lies only inside one of them, then that patch must be used. 

At this point, an important subtlety appears. The functional $\Gamma[\phi]$ is not, in general, strictly independent of the chosen extension $\tilde{\phi}$. Indeed, if $\tilde{\phi}_1$ and $\tilde{\phi}_2$ are two different extensions of the same boundary field $\phi$, then their difference is
\begin{equation}\label{eq:different_homotopy_extensions}
    \Gamma_1[\phi]-\Gamma_2[\phi]
    =
    \int_{U_1}\tilde{\phi}_1^*\omega
    -
    \int_{U_2}\tilde{\phi}_2^*\omega.
\end{equation}
Hence, by gluing $U_1$ and $U_2$ along their common boundary with opposite orientation, one obtains a closed $(d+1)$-dimensional manifold. Therefore, the difference above is equal to the integral of $\omega$ over a closed manifold, and since $[\omega]$ is integral, we conclude that
\begin{equation}\label{eq:wz_integer_ambiguity}
    \Gamma_1[\phi]-\Gamma_2[\phi]\in\mathbb{Z}.
\end{equation}
So the WZ functional is not a single-valued real number, but rather is defined only up to an integer ambiguity. This is precisely the reason why the quantum theory, and not the bare functional itself, is the truly well-defined object. Having defined the WZ functional $\Gamma[\phi]$, the corresponding \textbf{topological action for a WZ term} is taken to be
\begin{equation}\label{eq:def_wz_action}
    \boxed{S_{WZ}[\phi]=2\pi i\,k\,\Gamma[\phi],}
\end{equation}
where $k\in\mathbb{Z}$. With this normalization, the path-integral weight is independent of the choice of extension, since under
\begin{equation}
    \Gamma[\phi]\longrightarrow \Gamma[\phi]+n,
    \qquad n\in\mathbb{Z},
\end{equation}
one has
\begin{equation}
    e^{-S_{WZ}}
    \longrightarrow
    e^{-2\pi i k(\Gamma+n)}
    =
    e^{-2\pi i k\Gamma}\,e^{-2\pi i kn}
    =
    e^{-S_{WZ}}.
\end{equation}
In this line, the integer $k$ receives the name of the \textbf{level} of the Wess-Zumino theory.

\vspace{10pt}
At this point, the main features of WZ terms can be summarized as follows:
\begin{enumerate}
    \item A WZ term is constructed from a closed integral $(d+1)$-form $\omega$ on the target space and an extension of the physical field to one dimension higher. In this sense, it is the natural descendant of a higher-dimensional topological term.
    
    \item It admits two equivalent representations: the extended representation of Eq.\ref{eq:global_repWZ_correct}, which involves a filling manifold $U^{d+1}$ and an extended field $\tilde{\phi}_U$, and the local representation of Eq.\ref{eq:local_repWZ_correct}, which is written directly on the physical manifold $M^d$ in terms of a local potential $\kappa_\alpha$ such that $\omega=d\kappa_\alpha$.
    
    \item Unlike ordinary $\theta$-terms on closed manifolds, WZ terms generally do contribute to the equations of motion. Even though their definition is controlled by global topology, their variation is local and produces nontrivial dynamical terms. This is precisely what happens, for example, in the derivation of Bloch equations from spin coherent-state actions.
    
    \item The WZ functional is defined only modulo integers, and for this reason the coupling multiplying it must be quantized. With the convention of Eq.\ref{eq:def_wz_action}, this quantization is simply
    \begin{equation}
        k\in\mathbb{Z},
    \end{equation}
    and the resulting integer $k$ is called the level of the theory.
\end{enumerate}

\subsubsection{SU(2) Wess-Zumino}
Let's recap the result that we obtained in the path integral section. We claimed earlier that the second term in
\begin{equation}
    S[\phi, \theta^\prime]=SB\int_0^\beta d\tau \cos\theta^\prime + iS\int_0^\beta d\tau (-\cos\theta^\prime)\partial_\tau \phi, 
\end{equation}
corresponds to a WZ term. Now it is time to see why this is WZ and identify the underlying topology in the $SU(2)$ spin case. We then focus on studying the expression 
\begin{equation}\label{eq:WZ_SU2}
    S_{WZ}[\phi,\theta^\prime]=iS\int_0^\beta d\tau (-\cos\theta^\prime)\partial_\tau \phi.
\end{equation}
First, recall that the classical phase space of spin coherent states is $S^2$ and, consequently, cannot be described with just one chart. We cannot describe all coherent states with a single set of angles $\{(\phi,\theta^\prime)\}$; we will always miss a "pole". The question is then how this issue manifests itself in the topological action Eq.\ref{eq:WZ_SU2}. A beautiful way to clarify this point is, indeed, going back to classical mechanics and studying the structure of the emergent phase space $S^2$.

By looking at Eq.\ref{eq:WZ_SU2}, we can recognize the structure of a canonical term $\int d\tau \ p \dot{q}$ if we define $p=\cos\theta^\prime$ and $q=\phi$ (WZ  plays the role of a \emph{symplectic potential}). Therefore, our phase space is described by $(q,p)=(\phi,\cos\theta^\prime)$, which now can be seen as the local coordinates of $S^2$. In this way, when we have $\theta^\prime=0$ we are in the North Pole, while having $\theta^\prime=\pi$ means the South Pole. The problem comes when we consider the azimuthal angle $\phi$. Because of the non-existence of a diffeomorphism between $\mathbb{R}^2$ and $S^2$, we are forced to consider (at least) two charts of $S^2$. This means that there is no way to have well-defined coordinates $\cos\theta^\prime$ and $\phi$ for the two poles at the same time. Or, in even simpler words, in the poles, there is no sense of azimuthal angle since these points are totally defined by $\cos\theta^\prime=\pm 1$. Then, all the problem in Eq.\ref{eq:WZ_SU2} comes from the factor $\partial_\tau \phi = \dot{\phi}$, since in the poles this quantity is totally ill-defined. Recall that, this fact indeed has a very strong physical meaning in the sense that if we have completely polarized states ($\ket{\uparrow}$ or $\ket{\downarrow}$), the spin is fixed along the quantization axis and there is no precession, so the azimuthal angle $\phi$ ceases to be a meaningful dynamical variable, and the solution $\phi=-iB\tau$ is not valid. Consequently, we conclude that result Eq.\ref{eq:WZ_SU2} is only valid in the region $S^2\smallsetminus \{p_N,p_S\}$ (the sphere without north and south poles).

However, one can always add a total derivative to the action without changing its value. In this way, we may add a term of the form $\pm \partial_\tau \phi$, which leads to the equivalent expressions
\begin{equation}
    S_{WZ}^{N/S}[\phi,\theta^\prime]
    =
    - iS\int_0^\beta d\tau \,(\cos\theta^\prime \mp 1)\,\partial_\tau \phi.
\end{equation}
With this choice, the problematic factor $\partial_\tau \phi$ is effectively regularized at one of the poles: the expression with the minus sign is regular at $\theta^\prime=0$ (north pole), while the one with the plus sign is regular at $\theta^\prime=\pi$ (south pole). This structure should immediately remind us of the monopole potentials $\int_{S^1}\mathcal{A}_{N/S}$, whose explicit local forms are
\begin{equation}
    \begin{array}{lr}
\mathcal{A}_{\mathrm{N}}= iS(1-\cos \theta^\prime)\, \mathrm{d} \phi & \theta^\prime \neq \pi \\
\mathcal{A}_{\mathrm{S}}=- iS(1+\cos \theta^\prime)\, \mathrm{d} \phi & \theta^\prime \neq 0,
\end{array}
\end{equation}
which are related on the overlap by the gauge transformation
\begin{equation}
    \mathcal{A}_{\mathrm{S}}
    =
    \mathcal{A}_{\mathrm{N}} + e^{2iS\phi}\, d(e^{-2iS\phi})
    =
    \mathcal{A}_{\mathrm{N}} - 2iS\,d\phi.
\end{equation}
Therefore, in the spirit of Eq.\ref{eq:stoke_mono} and Eq.\ref{eq:wz_idea_correct}, we conclude that the \textbf{local representation of the SU(2) WZ term} is given by
\begin{equation}
    \boxed{
    S_{WZ}^{N/S}[\phi,\theta^\prime]
    =
    - iS\int_0^\beta d\tau \,(\cos\theta^\prime \mp 1)\,\partial_\tau \phi
    =
    \int_{S^1} \mathcal{A}_{N/S}
    }
    \label{eq:wz_local_rep_su2}
\end{equation}
where the two expressions correspond to the north and south patches of the target sphere. The difference between these two local descriptions captures the global topology of the configuration,
\begin{equation}
    S_{WZ}^{N}[\phi,\theta^\prime]
    -
    S_{WZ}^{S}[\phi,\theta^\prime]
    =
    \int_{S^1} (\mathcal{A}_N - \mathcal{A}_S)
    =
    4\pi i S
    =
    2\pi i W,
\end{equation}
with $W\in \mathbb{Z}$. Hence, we recognize the same mechanism encountered in the general theory of WZ terms: the functional is defined only up to integer ambiguities associated with different local descriptions. Requiring that the quantum weight $e^{-S_{WZ}}$ is independent of this ambiguity implies
\begin{equation}
    e^{-i\,4\pi S}=1 \quad \Rightarrow \quad 2S\in\mathbb{Z}.
\end{equation}
Thus, we recover the well-known half-integer quantization of spin, $S\in \tfrac{\mathbb{Z}}{2}$. In particular, for the fundamental representation one has $S=\tfrac{1}{2}$, corresponding to $W=1$. 

Having explored the local representation and its associated Chern-number interpretation of the WZ term, the remaining step is to describe its global formulation in terms of winding numbers and homotopy groups. To do so, we start from Eq.\ref{eq:action_singl_spin_imagi} and rewrite it in terms of the physical unit vector field
\begin{equation}
    \frac{\braket{g|\mathbf{S}|g}}{S}=\mathbf{n}:S^1 \to S^2,
\end{equation}
so that the action becomes
\begin{equation}
    S[\mathbf{n}]
    =
    S\int_0^\beta d\tau \,(\mathbf{B}\cdot \mathbf{n})
    +
    iS\int_0^\beta d\tau \,(-\cos\theta^\prime)\,\partial_\tau \phi.
\end{equation}
The question of why the WZ term cannot be written globally in terms of the one-dimensional field $\mathbf{n}$ is related to the fact that in one dimension it can only be expressed locally in terms of a potential (the Berry connection). A global and coordinate-independent definition requires passing to one higher dimension. Therefore, we consider homotopic extensions of the field (as in Eq.\ref{eq:extended_fields_correct})
\begin{equation}
    \tilde{\mathbf{n}}_\pm:[0,1]\times S^1 \to S^2,
    \qquad \text{with } \:\:\begin{cases}
        \tilde{\mathbf{n}}_\pm(1, \tau)=\mathbf{n}(\tau) \\
        \tilde{\mathbf{n}}_\pm(0, \tau)=\mathbf{n}^{}_0 \quad [\text{poles }p_N,p_S] \:.
    \end{cases}
\end{equation}
where $\mathbf{n}_0$ is a constant point choice (the poles) on $S^2$. Since the whole lower boundary $s=0$ is collapsed to a single point in the target, the cylinder is effectively turned into a punctured sphere (remove an infinitesimal 2-ball in the other pole) after quotienting that boundary circle to a point. In other words, one should think of
$$
\left([0,1] \times S^1\right) /\left(\{0\} \times S^1\right) \simeq S^2 \smallsetminus B_\epsilon(p_{N/S})\simeq U_{\pm}\:\:,
$$
as the two hemispheres $U_{\pm}$ which cover $S^2$. Thus, by gluing the extensions $\tilde{\mathbf{n}}_\pm$, we can recover the usual construction of global maps $\tilde{\mathbf{n}}:S^2\to S^2$, and therefore the homotopy group $\pi_2(S^2)=\mathbb{Z}$, with the corresponding winding number expression Eq.\ref{eq:winding_number_S2} (which, as discussed previously, should not be confused with the Gauss--Bonnet theorem, despite the similarity of local expressions).

\noindent Hence, the \textbf{extended representation of the SU(2) WZ term} is given by
\begin{equation}
    \boxed{S_{WZ}[\mathbf{\tilde n}]
    =
    i S \int_0^1 d s \int_0^\beta d \tau \ 
    \tilde{\mathbf{n}} \cdot\left(\partial_s \tilde{\mathbf{n}} \times \partial_\tau \tilde{\mathbf{n}}\right)}\:,
\end{equation}
which provides a global and geometrically well-defined expression for the WZ functional.

At this stage, the $SU(2)$ spin case can be reinterpreted in a way that will be particularly useful for the generalization to $SU(N)$. The key point is that the WZ term is not just a topological functional written on $S^2$, but the manifestation of an underlying \textbf{$U(1)$ bundle structure} over the coherent-state manifold. As already discussed, coherent states are defined only up to an overall phase,
\begin{equation}
    \ket{g}\sim e^{i\alpha}\ket{g},
\end{equation}
so the physical configuration space is not the Hilbert space itself, but the corresponding projective space whose elements are Hilbert rays $[e^{i\alpha}\ket{g}]$. For spin coherent states, this space is
\begin{equation}
    \mathbb{CP}^1 \simeq S^2.
\end{equation}
This means that the Berry phase appearing in the local representation of the WZ term should be understood as the holonomy of a $U(1)$ connection over $\mathbb{CP}^1$. In the present case, this bundle is precisely the one associated with the Hopf fibration
\begin{equation}
    S^1 \longrightarrow S^3 \longrightarrow S^2 \simeq \mathbb{CP}^1,
\end{equation}
where the fiber $S^1\simeq U(1)$ corresponds to the physically irrelevant phase of the states. In this language, the local one-forms $\mathcal{A}_{N/S}$ introduced in Eq.\ref{eq:wz_local_rep_su2} are local representatives of a Berry connection, while their curvature $\mathcal{F}=d\mathcal{A}$ is a globally defined two-form on the target space. This two-form is proportional to the standard area form on $S^2$, and at the same time it is the canonical Kähler form of $\mathbb{CP}^1$. Therefore, the WZ term may be viewed equivalently as
\begin{align*}
    & - \quad \text{the integral of a local connection along the physical path, or}\\
    & - \quad \text{the integral of its curvature over the extended surface.}
\end{align*}
This is exactly the same local/global mechanism discussed in the general theory of WZ terms.

 It is important to stress that, in the present $SU(2)$ case, several geometric structures collapse into the same object. Since $\mathbb{CP}^1$ is two-dimensional, the Berry curvature $\mathcal{F}$, the Kähler form $\omega_K$, and the normalized area form on the round sphere $\omega_{S^2}$ are all proportional. Moreover, because the tangent bundle of $S^2$ also has its Euler class in degree two $e(TS^2)\in H^{2}(S^2,\mathbb{Z})$, the same local density appears in the Gauss--Bonnet theorem. This is why, only in this special case, the winding-number density, the monopole curvature, the Kähler form, and the Euler-class density take the same local form.

\noindent However, this coincidence is special to $\mathbb{CP}^1\simeq S^2$. For higher groups $SU(N)$, we will find that coherent states live in general projective spaces $\mathbb{CP}^{N-1}$, and therefore, the role of the WZ term is still played by a closed integral \emph{two-form}, namely the Fubini--Study Kähler form, which is also the curvature of the canonical $U(1)$ bundle:
\begin{equation}
    (\text{Kähler})\quad \omega_K \in H^{2}(\mathbb{CP}^{N-1},\mathbb{Z}) \quad \xleftrightarrow[]{\quad \propto \quad } \quad (\text{Berry})\quad \mathcal{F} \in  H^{2}(\mathbb{CP}^{N-1},\mathbb{Z})
\end{equation}
By contrast, the Euler class of the tangent bundle $e(T\mathbb{CP}^{N-1})$ lives in degree $H^{2(N-1)}(\mathbb{CP}^{N-1},\mathbb{Z})$ and is therefore no longer the same object. Thus, the geometric structure that survives in the $SU(N)$ generalization is not the accidental coincidence with Gauss--Bonnet, but rather the existence of a canonical $U(1)$ bundle whose curvature defines the relevant integral Kähler form.

This is the reason why the $SU(2)$ WZ term should be regarded as the first member of a broader family of coherent-state topological terms: the sphere $S^2$ will be replaced by $\mathbb{CP}^{N-1}$, the monopole connection by the Berry connection of the canonical $U(1)$ bundle, and the area form by the Fubini--Study Kähler form. In this sense, the whole $SU(N)$ construction is already encoded in the geometric content of the simple spin-$\tfrac{1}{2}$ example.

%% file: chapter3_Su2_berry_phases.tex
\section{Some Physical Models} \label{models}
Having presented the $SU(2)$ WZ term together with its geometric and topological meaning, we now turn to the question of what this term actually does in concrete physical systems. The aim of this section is, first, to show how the WZ term reappears in problems where its presence has direct dynamical consequences. To this end, we begin with the simplest situation of fermionic particles interacting with a classical magnetic moment, and then move to the more realistic case of a magnetic quantum dot coupled to a metallic lead. These two examples will naturally take us through adiabatic evolution, Berry phases, dissipation, and fermionic functional integrals, providing a first view of how geometric terms shape physical dynamics rather than merely decorate the action.

\noindent At the same time, this section also serves as a transition toward the broader $SU(N)$ setting developed in the following chapters. After exploring the $SU(2)$ case and its physical consequences, we will discuss why higher $SU(N)$ symmetries become relevant in condensed matter systems with orbital, pseudospin, or multipolar degrees of freedom. In this way, the discussion will move from the familiar language of ordinary spin systems to $SU(N)$ Heisenberg models, $SU(4)$ spin-orbit and spin-pseudospin symmetries, and the multipolar exchange framework that motivates the superspin construction of Section\ref{sec:su_n_superspin}.

\subsection{Adiabatic SU(2) Berry Phase}
We start the discussion of WZ examples with the simplest case we can consider: fermionic $1/2$ particles with diagonalized energy spectrum $\xi_k=\epsilon_k-\mu$ interacting with a classical (but not fixed) magnetic moment $\bm{\mu}_{class}$. This means we are thinking in a Hamiltonian of the form 
\begin{equation}
    H=\sum_{k,\sigma} \xi_k \ a^\dagger_{k\sigma} a_{k\sigma} \: + \:  \tilde{\gamma} \ \bm{\mu}_{class}\cdot \mathbf{S} \: = \: \sum_{k,\sigma} \xi_k \ a^\dagger_{k\sigma} a_{k\sigma} \: + \:  \tilde{\gamma} \ \bm{\mu}_{class}\cdot \left[\frac{\hbar}{2}\sum_{k,\sigma_1, \sigma_2} a^\dagger_{k\sigma_1}\bm{\sigma}_{\sigma_1\sigma_2}\  a_{k\sigma_2}
    \right].
\end{equation}
with the classical magnetic moment $\bm{\mu}_{class}=|\bm{\mu}_{class}|\mathbf{n}=\mu_{class} \bm{n}$ having fixed magnitude but fluctuating direction $\mathbf{n}$. Therefore, the term $\tilde{\gamma} \ \bm{\mu}_{class}\cdot \mathbf{S}$, can be rewritten as $(\hbar=1)$
\begin{equation}
    \tilde{\gamma} \ \mathbf{\mu}_{class}\cdot \mathbf{S} = \big(\tilde{\gamma}\frac{1}{2}\mu_{class}\big) \sum_{k,\sigma_1, \sigma_2} a^\dagger_{k\sigma_1}\mathbf{n}\cdot\bm{\sigma}_{\sigma_1\sigma_2}\  a_{k\sigma_2} = \sum_k\gamma 
\begin{pmatrix}
a^{\dagger}_{k\uparrow} & a^{\dagger}_{k\downarrow} 
\end{pmatrix} \ \mathbf{n}\cdot \bm{\sigma}
\begin{pmatrix}
a_{k\uparrow} \\
a_{k\downarrow} 
\end{pmatrix},
\end{equation}
defining $\gamma$ such that it absorbs the remaining constants. In this way, through the coherent path integral \cite{altland}, we find that the real and imaginary time actions are 
\begin{equation}
\boxed{
\begin{aligned}
    iS[\overline\psi,\psi, \mathbf{n}]
    &= i \sum_k\int_{t_i}^{t_f} dt \ \overline{\psi}_k
    \left(\partial_t-\xi_k-\gamma \ \mathbf{n} \cdot \bm{\sigma}\right) \psi_k\\
    - S_{eucl}[\overline\psi,\psi, \mathbf{n}]
    &= - \sum_k\int_0^\beta d \tau \ \overline{\psi}_k
    \left(\partial_\tau+\xi_k+\gamma \ \mathbf{n} \cdot \bm{\sigma}\right) \psi_k 
\end{aligned}}
\label{eq:berry_action}
\end{equation}
where the times are connected by $t=-i\tau$, and the Grassmann variables $\overline{\psi}_k=(\overline{\psi}_{k\uparrow} \ \: \overline{\psi}_{k\downarrow})$ \ ; \ $\psi_k=(\psi_{k\uparrow} \ \: \psi_{k\downarrow})^{T}$.

\subsubsection{Making sense of the Partition Function on a Time-Evolution problem} \label{sec:evolu_parti}
Here, it is worth pausing for a moment and clarifying the physical meaning of the previous real and imaginary time actions, corresponding to the two path integrals that we have at our disposal. When one is interested in transition amplitudes between prescribed initial and final states under real-time evolution, the natural object is the real-time path integral
\begin{equation}
    \bra{\text{out}}U(t_f,t_i)\ket{\text{in}}=\frac{1}{\mathcal{N}}\int_{\text{fields}(t_{i,f})=\text{in,out}} D[\text{fields}] \: e^{ \: i  S[ \text{fields}]},
    \label{eq:real_pi}
\end{equation}
where $\mathcal{N}$ is a normalization constant which, for our present purposes, will not play any relevant role. On the other hand, when the system is assumed to be in thermal equilibrium, the central object is the partition function
\begin{equation}
    \label{eq:imagina-path-integral}
    Z=\text{tr}(e^{-\beta H})=\int D[\text{fields}] \ e^{-S_{eucl}^{0\to\beta}[\text{fields}]},
\end{equation}
which is formulated in periodic imaginary time. These two expressions describe different physical situations, and therefore, relating them is not automatic. The natural question, then, is under which circumstances one can pass from one formulation to the other, or equivalently, when it is legitimate to perform a Wick rotation and still compute the quantities of interest. But before addressing this point, let us specify what we want to compute in our problem. Our final goal is to derive the $SU(2)$ Berry phase in the path integral formulation, so we are naturally interested in the phase accumulated by an instantaneous eigenstate under adiabatic evolution. Therefore, we must think of a setup in which we have an initial state $\ket{\psi(t_i)}=\ket{n}$, where $\ket{n}$ is an instantaneous eigenstate of a Hamiltonian $H(\lambda)$ depending on a set of slowly varying parameters $\lambda=\{\lambda_j\}_{j\in I}\in\mathcal M$. Then, in the adiabatic regime, the system remains in the same instantaneous eigenstate up to a phase, so that
\begin{equation}
    \ket{n} \xmapsto{\quad \quad} \ket{\psi(t_f)} =  e^{i\phi_n(t_f,t_i)} \ket{n}
    \label{eq:evol_phase_state}
\end{equation}
where the time evolution phase $\phi_n$ includes dynamical $\theta_{_\text{dyn}}$ and geometrical/Berry $\gamma_{_\text{Berry}}$ parts
\begin{equation}
\begin{split}
    \phi_n(t_f,t_i)&=\theta_{_\text{dyn}}(t_f,t_i) + \gamma_{_\text{Berry}}(t_f,t_i)  \\ &= -\int_{t_i}^{t_f}  E_n\left(\scriptstyle{\lambda\left(t\right)}\right) d t \:\:+\:\: i \int_{t_i}^{t_f}\Big\langle n( {\scriptstyle \lambda\left(t\right) } ) \Big| \frac{d}{d t} \Big| \: n({\scriptstyle\lambda(t)})\Big\rangle d t\:.
\end{split}
\label{eq:evol_phase}
\end{equation}
As mentioned, the decomposition above is valid provided the Hamiltonian varies sufficiently slowly so that transitions between different instantaneous eigenstates are suppressed. In other words, for $m\neq n$, the specific condition for adiabaticity is written as 
\begin{equation}
    \frac{\left|\left\langle m(t)\right|\dot H(t)\left|n(t)\right\rangle\right|}{\left|E_m(t)-E_n(t)\right|^2}
    \ll 1.
    \label{eq:adiabatic_cond}
\end{equation}
A particularly important situation is that of \textbf{cyclic adiabatic evolution}, namely when the parameters follow a closed path in parameter space, $\lambda(t_i)=\lambda(t_f)$. In that case, the system returns to the same instantaneous eigenstate, again up to a phase. The total phase still contains both dynamical and geometric contributions, but now the geometric one becomes
\begin{equation}
    \gamma_{_\text{Berry}}
    =
    i\oint_C
    \big\langle n(\lambda)\big|
    \partial_{\lambda^j}
    \big|n(\lambda)\big\rangle
    d\lambda^j\:,
\end{equation}
where $C\subset\mathcal M$ denotes the closed path described by the parameters. In this light, what we would like to compute is the diagonal matrix element of the adiabatic evolution operator,
\begin{equation}
    \bra{n}U_{\text{adiab}}(t_f,t_i)\ket{n}=e^{i\phi_n(t_f,t_i)}\:,
\end{equation}
and, from it, isolate the geometric contribution. In what follows, we focus on the geometric part of the effective action, i.e., on the contribution that remains after the usual dynamical phase has been separated. Hence, using Eq.\ref{eq:real_pi}, the previous object can be written as a real-time functional integral. Since we are considering cyclic evolution, the fields satisfy periodic boundary conditions in time, and therefore
\begin{equation}
    \bra{n}U_{\text{adiab}}(t_f,t_i)\ket{n}
    =
    \oint_{\text{flds}(t_f)=\text{flds}(t_i)}
    D[\text{fields}] \:
    e^{i\int_{t_i}^{t_f}dt\,L(t)}\:\:\propto\:\:e^{i \gamma_{_\text{Berry}}}.
\end{equation}
So, in our present problem of a classical spin interacting with a bath of $S=1/2$ fermions, we would like to derive (starting with real-time action in Eq.\ref{eq:berry_action})
\begin{equation}
    \bra{n}U_{\text{adiab}}(t_f,t_i)\ket{n}
    =
    \oint D[\overline\psi_k,\psi_k]\:
    e^{iS[\overline\psi_k,\psi_k,\mathbf n]}
    \ \leadsto\
    e^{i\phi_n(t_f,t_i)},
\end{equation}
so that the Berry phase is encoded in the effective action obtained after integrating out the fermions. However, the immediate difficulty is, of course, that the real-time functional integral is highly oscillatory. One is therefore tempted to perform the Wick rotation $t=-i\tau$ and work instead with the Euclidean functional integral
\begin{equation}
    \oint D[\overline\psi_k,\psi_k]\:
    e^{-S_{eucl}[\overline\psi_k,\psi_k,\mathbf n]}.
\end{equation}
But here the subtlety becomes clear: 
\begin{center}
    \textit{the Berry phase is a quantity defined from real-time adiabatic evolution, whereas the partition function is an equilibrium object in imaginary time. So why should the latter know anything about the former?}
\end{center}
To answer this, we now recall the mechanism by which both formulations project onto the ground-state sector \cite{bertlmann, fradkin2021quantum}. A convenient way to see how this projection works is to consider the real-time transition amplitude in the presence of external sources $J=(\eta,\overline\eta)$ acting only in a finite interval $[t_1,t_2]\subset [t_i,t_f]$. Splitting the evolution operator as
\begin{equation}
    U(t_f,t_i)
    =
    U(t_f,t_2)\:\:U^{(J)}(t_2,t_1)\:\:\:U(t_1,t_i),
\end{equation}
and inserting complete sets of coherent states at the intermediate times $t_1,t_2$, we obtain
\begin{equation}
\begin{aligned}
\bra{\text{out}}U(t_f,t_i)\ket{\text{in}}
&=
\int \mathcal D[\overline\psi_{t_1},\psi_{t_1}, \overline\psi_{t_2},\psi_{t_2}]
\:\:\:
\bra{\text{out}}U(t_f,t_2)\ket{\overline\psi_{t_2},\psi_{t_2}}
\\
&\qquad\qquad\cdot\:
\bra{\overline\psi_{t_2},\psi_{t_2}}U^{(J)}(t_2,t_1)\ket{\overline\psi_{t_1},\psi_{t_1}}\:\:\:\:\:
\bra{\overline\psi_{t_1},\psi_{t_1}}U(t_1,t_i)\ket{\text{in}}\:\:.
\end{aligned}
\end{equation}
Then, expanding the two outer matrix elements in the energy eigenstates basis $\{\ket{\phi_n}\}_{n\in I}$ of the (free) Hamiltonian, gives
\begin{equation}
    \begin{aligned}
 \left\langle \text{out} \Big| U(t_f,t_2) \Big| \overline{\psi}_{t_2}, \psi_{t_2}\right\rangle & =\sum_n \left(\phi_n^{out}\right)^*  \left(\phi_n^{{}_{\overline{\psi}_{t_2}, \psi_{t_2}}}\right) e^{-i E_n(t_f-t_2) } \\
\left\langle \overline{\psi}_{t_1}, \psi_{t_1} \Big| U(t_1,t_i) \Big| \text{in} \right\rangle& =\sum_m   \left(\phi_m^{{}_{\overline{\psi}_{t_1}, \psi_{t_1}}}\right)^* \left(\phi_m^{in}\right) \ e^{-i E_m\left(t_1-t_i\right)},
\end{aligned}
\label{eq:analyt_cont_energ}
\end{equation}
Here is when one sees the problem immediately: in the limits $t_f\to+\infty$ and $t_i\to-\infty$, the factors $e^{-iE_n(t_f-t_2)}$ and $e^{-iE_m(t_1-t_i)}$ are purely oscillatory, so the large-time behavior does not by itself select any particular state. In fact, these contributions only oscillate faster and faster, and therefore, the limit is not well-defined as a convergent object. However, it is essential to recall that quantum field theory requires a precise prescription to define time-ordered amplitudes and ensure causality. This is implemented through the standard $i\epsilon$ prescription, or equivalently by giving the large real-time interval a small negative imaginary part.

\noindent In practice, for a large time interval $T$, one replaces $T\to T(1-i\eta)$, with $\eta>0$. Then a generic oscillatory factor becomes $e^{-iE_nT}\to e^{-iE_nT}e^{-\eta E_nT}$. After factoring out the ground-state contribution, this can be written as
\begin{equation}
    e^{-iE_nT}e^{-\eta E_nT}
    =
    e^{-iE_0T}e^{-\eta E_0T}
    \,
    e^{-i(E_n-E_0)T}
    e^{-\eta(E_n-E_0)T}.
\end{equation}
Thus, in the limit $T\to\infty$, all excited states with $E_n>E_0$ are exponentially suppressed relative to the ground state. This is the sense in which the $i\epsilon$ prescription, or equivalently a small imaginary-time component in the evolution, selects the vacuum sector. This infinitesimal deformation is not an arbitrary modification, but rather the prescription that fixes the analytic structure of the theory, determines the position of poles in propagators, and guarantees the proper causal behavior of Green's functions. The precise mathematical justification of such $i\epsilon$ prescriptions belongs to the theory of distributions \cite{bogol, finite_qed}, in particular to the analytic continuation of otherwise ill-defined products involving step functions and Green's functions.

Consequently, the main effect of this prescription at the level of Eq.~\ref{eq:analyt_cont_energ} is not merely to damp the oscillations, but to suppress excited-state contributions relative to the lowest-energy state. Assuming for simplicity that the ground state is non-degenerate, the only surviving contribution in this asymptotic-time limit is precisely the ground state $n=0$. This is the content of the Gell-Mann--Low construction: once the correct causal prescription is implemented, interacting amplitudes can be reduced to vacuum matrix elements. In other words, in this limit, one obtains schematically
\begin{equation}
    \label{eq:vacuum-to-vacuum}
    \bra{\text{out}}U(t_f,t_i)\ket{\text{in}}
    \:\xrightarrow[\:\:t_i\to-\infty\:\:] {\:\:t_f\to+\infty\:\:}\: \langle 0| U^{(J)}\left(t_2, t_1\right)|0\rangle
    \:\xrightarrow[\:\:t_1\to-\infty\:\:] {\:\:t_2\to+\infty\:\:}\:
    {}_J\langle0|0\rangle _J \:,
\end{equation}
where ${}_J\langle0|0\rangle _J$ is defined as the vacuum persistence amplitude for the case in which the source $J$ acts in the full time interval. Therefore, the asymptotic real-time evolution does not simply oscillate indefinitely, but instead projects onto the ground-state sector once the causal structure of the theory is properly taken into account.

After introducing ${}_J\langle0|0\rangle_J$, the natural next step is to explain why this vacuum object is the one we actually want. Here, it is convenient to slightly generalize our notation and let $J$ denote a generic set of external sources. In particular, besides the fermionic sources $J=(\eta,\overline\eta)$ introduced before, one may also consider sources coupled directly to operators of the theory. The point is that, once the source $J$ is coupled linearly to the fields or operators, the source-dependent vacuum amplitude becomes a generating object for correlation functions. Schematically, if the source enters the Hamiltonian or the action through a term of the form
\begin{equation}
    H^{(J)}(t)=H_0+J(t) \hat{\mathcal{O}}(t)\:,
\end{equation}
then differentiating the vacuum persistence amplitude with respect to $J$ brings down insertions of the operator $\hat{\mathcal{O}}$. In this way, one has
\begin{equation}
\frac{\delta ({}_J\langle0|0\rangle_J)}{\delta J(t)}
\quad\xrightarrow[J=0]{\quad\sim\quad}\quad
\langle0|T\hat{\mathcal O}(t)|0\rangle
\end{equation}
and, more generally,
\begin{equation}
\frac{\delta^n({}_J\langle0|0\rangle_J)}{\delta J(t_1)\cdots \delta J(t_n)}
\quad\xrightarrow[J=0]{\quad\sim\quad}\quad
\langle0|T\hat{\mathcal O}(t_1)\cdots \hat{\mathcal O}(t_n)|0\rangle\:.
\end{equation}
This is the reason why the amplitude ${}_J\langle0|0\rangle _J$ is so central: it is not merely a vacuum-to-vacuum overlap, but the object from which the relevant observables are generated. However, one should recall that the vacuum $|0\rangle$ appearing above is the ground state of the free Hamiltonian, whereas the physical ground state of a full interacting theory is a many-body state $|\Omega\rangle$. The role of the Gell-Mann--Low machinery is precisely to bridge these two descriptions. One does not need the full proof here; what matters for us is the logic behind it. Since the asymptotic time evolution projects onto the vacuum sector, the source-generated amplitudes computed with $|0\rangle$ can be normalized in such a way that they reproduce the expectation values in the interacting ground state $|\Omega\rangle$. In other words, the theorem states schematically that
\begin{equation}
\frac{{}_J\langle0|T\hat{\mathcal O}(t_1)\cdots \hat{\mathcal O}(t_n)|0\rangle_J}
{{}_J\langle0|0\rangle_J}
\Bigg|_{J\to0}
\:=\:\:
\langle\Omega|T\hat{\mathcal O}(t_1)\cdots \hat{\mathcal O}(t_n)|\Omega\rangle.
\end{equation}
Thus, after the appropriate normalization by the vacuum persistence amplitude, the free-vacuum functional machinery reproduces the correlation functions of the interacting ground state. This is the precise sense in which the asymptotic real-time evolution, together with the $i \epsilon$ prescription, isolates the same physical sector that one would like to study in the full interacting problem.

Now the reason why this discussion is useful for our purposes becomes transparent. The Berry phase is indeed defined through real-time adiabatic evolution, but the previous argument shows that, once one restricts to the asymptotic vacuum sector, the relevant information is already encoded in vacuum amplitudes and their source derivatives. At this stage, one may then ask whether the same vacuum projection can be implemented in a more convenient way. The answer is yes: this is precisely what imaginary-time evolution does. Indeed, under the Wick rotation $t=-i \tau$, the evolution operator becomes
\begin{equation}
e^{-iH(t_f-t_i)}
\quad\Longrightarrow\quad
e^{-H(\tau_f-\tau_i)}.
\end{equation}
and if we expand an arbitrary state in the exact energy eigenbasis, $\ket{\psi}=\sum_n c_n \ket{\phi_n}$, its Euclidean evolution is
\begin{equation}
\begin{aligned}
e^{-H\tau}\ket{\psi}
&=
\sum_n c_n e^{-E_n\tau}\ket{\phi_n}
=
e^{-E_0\tau}
\sum_n c_n e^{-(E_n-E_0)\tau}\ket{\phi_n}
\\
&\quad \xRightarrow[\tau\to+\infty]{}\quad
c_0\: e^{-E_0\tau}\ket{\phi_0},
\end{aligned}
\end{equation}
provided $c_0\neq 0$. Therefore, imaginary-time evolution automatically suppresses all excited states and projects onto the ground state. In this way, the Euclidean functional integral (Eq.\ref{eq:imagina-path-integral}) \emph{implements the same vacuum-selection mechanism} that in real time required the $i\epsilon$ prescription (Eq.\ref{eq:vacuum-to-vacuum}). 

\noindent This is the bridge that we needed. Even though the Berry phase is originally a real-time quantity, the Euclidean formalism accesses the same ground-state sector and therefore allows us to derive the corresponding effective action in a technically controlled way. Once the fermions are integrated out in imaginary time, \emph{the geometric contribution that survives in the effective action is precisely the Euclidean representative of the Berry/Wess--Zumino term}. In this sense, the partition function is not replacing the real-time definition of the Berry phase; rather, it is providing a more convenient route to extract the same ground-state geometric information.

\subsubsection{Derivation of the many-body WZ-term}
Having discussed why we can employ the partition function to compute the Berry phase, our starting point then now is
\begin{equation}
    Z=\int D\mathbf{n} \int D[\overline{\psi}_k,\psi_k] \ e^{ - \sum_k\int_0^\beta d \tau \ \overline{\psi}_k\left(\partial_\tau+\xi_k+\gamma \ \mathbf{n} \cdot \bm{\sigma}\right) \psi_k } = \int D\mathbf{n} \ \prod_k \text{det}(\partial_\tau+\xi_k+\gamma \ \mathbf{n} \cdot \bm{\sigma}),
\end{equation}
where in the last equality we used Gaussian integration. Note that we can express the determinant as $\text{det}A=e^{\text{ln (}\text{det}A)}=e^{\text{tr}(\text{ln}A)}$, such that 
\begin{equation}
    Z=\int D\mathbf{n} \prod_k e^{\text{tr} \:  \text{ln}(\partial_\tau+\xi_k+\gamma \ \mathbf{n} \cdot \bm{\sigma})} = \int D\mathbf{n} \ \ e^{- \sum_k -\text{tr} \:  \text{ln }(\partial_\tau+\xi_k+\gamma \ \mathbf{n} \cdot \bm{\sigma})}.
\end{equation}
Then, the effective action for the semiclassical magnetic moment unit vector $\mathbf{n}$ is defined as
\begin{equation}
    S_{\text{eff}}[\mathbf{n}]=-\sum_k  \text{tr} \:  \text{ln }(\partial_\tau+\xi_k+\gamma \ \mathbf{n} \cdot \bm{\sigma}),
    \label{eq:effe_act_berry}
\end{equation}
after the integration of fermionic fields. Now, since $\mathbf{n}$ is a unit vector in $S^2$, then any direction $\mathbf{n}$ can be written as a rotation matrix $\hat{R}\in SO(3)$ acting on the z-basis vector $e_3$. This is $\mathbf{n}=\hat{R}e_3$. However, any rotation $\hat{R}$ induces a $SU(2)$ conjugation in vector operators, so that the dot product $\mathbf{n} \cdot \bm{\sigma}$ of any direction $\mathbf{n}$ can be rewritten as 
\begin{equation}
    \mathbf{n} \cdot \bm{\sigma}=\left(\hat{R}e_3\right) \cdot \bm{\sigma}=e_3 \cdot U_R \bm{\sigma} U^{-1}_R= U_R \sigma_3 U^{-1}_R.
\label{eq:conjug_SU}
\end{equation}
Here, conjugation doesn't mean $f(z)=z^*$. Conjugation in group theory means the map $\Phi_g(u)=gug^{-1}$. 

For instance, if the standard polar representation $\mathbf{n}=(\sin \theta^\prime \cos \phi, \sin \theta^\prime \sin \phi, \cos \theta^\prime)^T$ is used, then the choice
\begin{equation}
    U_{\mathbf{n}}=e^{-i \phi \hat{S}_3} e^{-i \theta^\prime \hat{S}_2} e^{-i \psi \hat{S}_3}
    \label{eq:rot_frame2}
\end{equation}
of the $SU(2)$ operator is sufficient to fulfill Eq.\ref{eq:conjug_SU}. Note that this choice is not unique since we could also have chosen $U=-U_\mathbf{n}$ and obtain the same transformation. Indeed, this highlights that $SU(2)$ is the double cover of $SO(3)$, such that in any 3D rotation $\hat{R}$ induces two spin-rotations $U_{\pm}$ matrices.

Having said that, we can replace Eq.\ref{eq:conjug_SU} into Eq.\ref{eq:effe_act_berry} and pass from a $\mathbf{n}$-dependent action to an $U$-dependent action:
\begin{equation}
     S_k[U]=-\operatorname{tr} \ln \left(\partial_\tau+\xi_k+\gamma U \sigma_3 U^{-1}\right),
\end{equation}
where the subscript $\mathbf{n},R$ is not needed due to the not fixing of $\mathbf{n}$. Note, however, that even though we are denoting the action depending on a $SU(2)$ matrix, the path integral doesn't run over the whole $SU(2)$, only over the elements corresponding to the $\mathbf{n}\in S^2$ vectors (i.e., the coherent state manifold $SU(2)/U(1)$). 

\noindent To proceed, we utilize the conjugation invariance of the trace $\text{tr}(A)=\text{tr}(gAg^{-1})=\text{tr}(g^{-1}Ag)$, such that 
\vspace{-.2cm}
\begin{equation}
    S_k[U]=-\operatorname{tr} \ln \left[ U^{-1} \left(\partial_\tau +\xi_k+\gamma  U \sigma_3 U^{-1}\right)U\right]  = -\operatorname{tr} \ln \left(\partial_\tau+\xi_k+\gamma \sigma_3+U^{-1} \dot{U}\right),
    \label{eq:rot_frame3}
\end{equation} \vspace{0.3cm}
after using $(U^{-1}\partial_\tau U)(v)=U^{-1}[\dot{U}v+U(\partial_\tau v)]=(\partial_\tau + U^{-1}\dot{U})(v)$ in the last equality. At this point, we are ready to deal with the adiabatic assumptions. First, let's conveniently rewrite the $k-$term of the total effective action as
\begin{equation}
    S_k[U]=-\operatorname{tr \: ln} \left\{\left(\partial_\tau + \xi_k +\gamma \sigma_3\right)\left[\mathbb{1}+\left(\partial_\tau + \xi_k +\gamma \sigma_3\right)^{-1}(U^{-1}\dot{U})\right]\right\},
\end{equation}
such that we can define $\mathcal{D}_0:=\left(\partial_\tau + \xi_k +\gamma \sigma_3\right)$ and express it in the form (using linearity of the trace and properties of logarithm)
\begin{equation}
\begin{split}
    S_k[U]=&-\operatorname{tr \: ln} \left\{\mathcal{D}_0\left[\mathbb{1}+\mathcal{D}_0^{-1}(U^{-1}\dot{U})\right]\right\}=-\operatorname{tr \: ln}[\mathcal{D}_0] -\operatorname{tr \: ln}\left[\mathbb{1}+\mathcal{D}_0^{-1}(U^{-1}\dot{U})\right] \\
    =& \: \: \operatorname{tr \: ln}[\mathcal{D}_0^{-1}]- \operatorname{tr \: ln}\left[\mathbb{1}+\mathcal{D}_0^{-1}(U^{-1}\dot{U})\right].
\end{split}
\end{equation}
The first term, indeed, is no more than the free propagation of the fermionic metallic fields $\psi_k$, and therefore, it will not matter for expectation values of the semiclassical magnetic moment $\bm{\mu}$ (normalization will cancel the $e^{\sum_k \operatorname{tr ln (\partial_\tau +\xi_k+\gamma\sigma_3)^{-1}}}$ factor). In this line, we remain only with
\begin{equation}
    S_k[U]=-\operatorname{tr \: ln}\left[\mathbb{1}+\mathcal{D}_0^{-1}(U^{-1}\dot{U})\right].
    \label{eq:tr_ln_action}
\end{equation}
The next step is then trying to use the expansion 
\begin{equation}
    \operatorname{ln}(1+x)=x-\frac{x^2}{2}+\frac{x^3}{3}+ \:... \:\approx x \:\: \:\:\: \:\:\text{ for }\:|x|<<1,
\end{equation}
However, to use such an approximation, we have to argue that the absolute value of the diagonal elements of such an operator are much smaller than one: $|\mathcal{D}_0^{-1} (U^{-1}\dot{U})|<<1$. Here, therefore, it is important to recognize $\mathcal{D}_0^{-1}$ as the integral operator defined by the action of the Green matrix function $\hat{G}_0(\tau,\tau ')$ \textemdash{}
of the matrix differential operator $\left(\partial_\tau + \xi_k +\gamma \sigma_3\right)$ \textemdash{}  on the two-component spinor wavefunctions: 
\begin{equation}
    \begin{split}
        \mathcal{D}_0^{-1} : & \: \:\mathcal{H}\otimes \mathbb{C}^2 \longrightarrow \mathcal{H}\otimes \mathbb{C}^2 \\
        & \begin{pmatrix} \psi_+(\tau ) \\ \psi_-(\tau ) \end{pmatrix} \longmapsto \int d\tau ' \: \hat{G}_0(\tau,\tau ') \begin{pmatrix} \psi_+(\tau ') \\ \psi_-(\tau ') \end{pmatrix},
    \end{split}
\end{equation}
where the integration is understood component by component. In this way, we explicitly see that by taking the Fourier series of the matrix elements (in Matsubara frequencies)
\begin{equation}
    G_0^{ij}(\tau, \tau') = \frac{1}{\beta}\sum_{\omega_n}  e^{-i\omega_n (\tau - \tau')} G_0^{ij}(\omega_n), 
\end{equation}
we get the Green's equation
\begin{equation}
    \left(-i\omega_n + \xi_k +\gamma \sigma_3\right) G_0(\omega_n)=\mathbb{1}
\end{equation}
with the operator 
\begin{equation}
    -i\omega_n + \xi_k+ \gamma \sigma_3 =
\begin{bmatrix}
-i\omega_n + \xi_k + \gamma & 0 \\
0 & -i\omega_n + \xi_k - \gamma
\end{bmatrix}.
\end{equation}
This means that in Matsubara frequency space, the $k-$Green's function is written as
\begin{equation}
    G_0(\omega_n) =
\begin{bmatrix}
\frac{1}{-i\omega_n + \xi_k + \gamma} &0 \\
0 &\frac{1}{-i\omega_n + \xi_k - \gamma}
\end{bmatrix},
\end{equation}
and therefore, the matrix elements of $\mathcal{D}_0^{-1}$ are of the form 
\begin{equation}
\begin{split}
\langle\psi^{\dagger},\mathcal{D}_0^{-1}\psi\rangle \: &=\int_0^\beta d\tau \begin{pmatrix} \psi^\dagger_+(\tau ) \\ \psi_-^\dagger(\tau ) \end{pmatrix}\int_0^\beta d\tau ' \: \hat{G}_0(\tau,\tau ') \begin{pmatrix} \psi_+(\tau ') \\ \psi_-(\tau ') \end{pmatrix} \\
&=\sum_{\omega_n} \left\{ \frac{1}{\sqrt{\beta}}\int_0^\beta d\tau \begin{pmatrix} \psi^\dagger_+(\tau ) \\ \psi_-^\dagger(\tau ) \end{pmatrix} e^{-i\omega_n \tau} \:\:\: G_0(\omega_n) \:\:\:\frac{1}{\sqrt{\beta}}\int_0^\beta d\tau' \begin{pmatrix} \psi^+(\tau' ) \\ \psi_-(\tau' ) \end{pmatrix} e^{i\omega_n \tau'} \right\} \\
&=\sum_{\omega_n}\begin{pmatrix} \psi^\dagger_{+,n} \\ \psi_{-,n}^\dagger \end{pmatrix} G_0(\omega_n) \begin{pmatrix} \psi_{+,n} \\ \psi_{-,n}\end{pmatrix} \\
&=\sum_{\omega_n}\begin{pmatrix} \psi^\dagger_{+,n} \\ \psi_{-,n}^\dagger \end{pmatrix} \begin{bmatrix}
\frac{1}{-i\omega_n + \xi_k + \gamma} &0 \\
0 &\frac{1}{-i\omega_n + \xi_k - \gamma}
\end{bmatrix} \begin{pmatrix} \psi_{+,n} \\ \psi_{-,n}\end{pmatrix}.
\end{split}
\label{eq:elem_first}
\end{equation}

\begin{figure}[h!]
    \centering
    \includegraphics[width=0.8\linewidth]{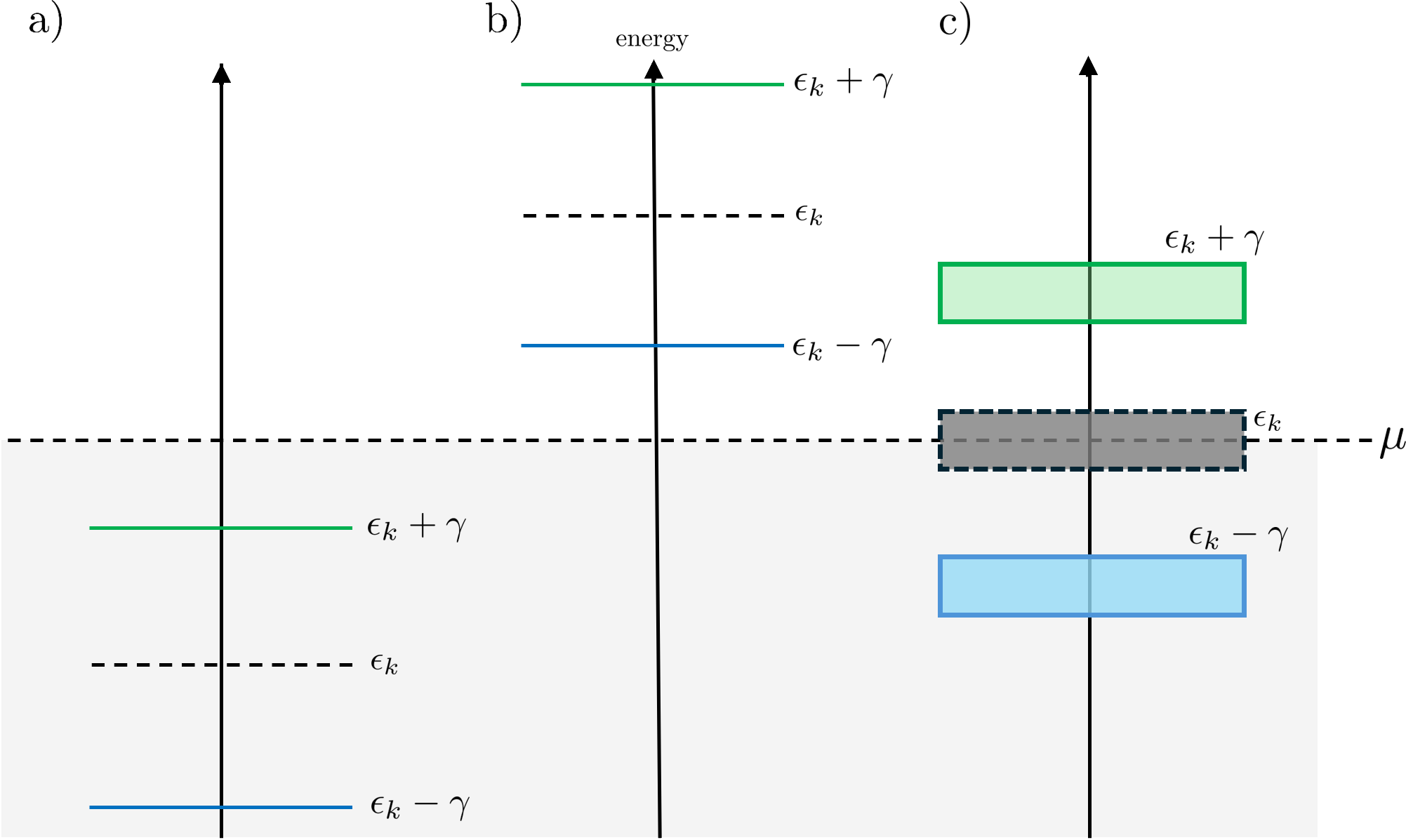}
    \caption{Possible energy configurations (assuming $\gamma>0)$: a) Both spin levels occupied. b) Both unoccupied. c) Upper level unoccupied but lower occupied.}
    \label{fig:energy_cases}
\end{figure}

In the same way, the matrix elements of the operator $U^{-1}\dot{U}=U^{-1}\partial_\tau U$ are obtained through the Fourier transformation as $\left(U^{-1}=\frac{1}{\sqrt{\mathrm{T}}}\sum_{\tilde{\omega}'}V_{\tilde{\omega}'} e^{i\tilde{\omega}'\tau} \ \ \ \text{and } \ \ \ U=\frac{1}{\sqrt{\mathrm{T}}}\sum_{\tilde{\omega}}U_{\tilde{\omega}} e^{i\tilde{\omega}\tau} \right) $
\begin{equation}
\begin{split}
    \langle \psi^{\dagger},(U^{-1}\dot{U})\psi\rangle &=\int_0^\beta d\tau \left[ \frac{1}{\sqrt{\beta}}\sum_{\omega_m}\psi_m^\dagger e^{i\omega_m \tau} \right]
    \frac{1}{\mathrm{T}}\sum_{\tilde{\omega}',\tilde{\omega}}(i\tilde{\omega})\hat{V}_{\tilde{\omega}'} \hat{U}_{\tilde{\omega}} \: e^{i(\tilde{\omega}'+\tilde{\omega})\tau}
    \left[
    \frac{1}{\sqrt{\beta}} \sum_{\omega_n}\psi_n e^{-i\omega_n \tau} \right] \\
    & =\sum_{\omega_m,\omega_n} \: \sum_{\tilde{\omega}',\tilde{\omega}} (i\tilde{\omega}) \: \psi_m^\dagger \hat{V}_{\tilde{\omega}'} \hat{U}_{\tilde{\omega}} \psi_n \: \frac{1}{\beta \mathrm{T}} \int_0^\beta d\tau \: e^{i(\tilde{\omega}'+\tilde{\omega}+\omega_m-\omega_n)\tau },
\end{split}
\end{equation}
where $\tilde{\omega}',\tilde{\omega}$ correspond to Fourier frequencies of the classical magnetic moment $\bm{\mu}_{class}=\mu \: \mathbf{n}(\tau)$, $\mathrm{T}$ to the period of such classical degree of freedom, and as usual $ \omega_{m,n}$ to fermionic Matsubara frequencies. Then, in order to proceed, we have to evoke the \textbf{adiabatic approximation} applied to the dynamics of the classical magnetic moment. First, we say that $\bm{\mu}_{class}(\tau)$ varies over much larger time scales than the fermionic fields $\psi,\overline{\psi}$, then causing that the only non-vanishing Fourier components $\hat{V}_{\tilde{\omega}'}, \hat{U}_{\tilde{\omega}}$ are the ones corresponding to frequencies $\tilde{\omega}',\tilde{\omega}$ much smaller than the Matsubara frequencies: $\tilde{\omega}',\tilde{\omega}\ll \omega_m,\omega_n$. This means that for the  time scales of the fermionic fields (the relevant scale in solving the fermionic path integral), the moment frequencies can be approximately taken as $\tilde{\omega}'+\tilde{\omega}=\epsilon \to 0$, and therefore, (in a first view)
\begin{equation}
\begin{split}
     \langle \psi^{\dagger},(U^{-1}\dot{U})\psi \rangle &=\sum_{\omega_m,\omega_n} \: \sum_{\tilde{\omega},\epsilon} (i\tilde{\omega}) \: \psi_m^\dagger \hat{V}_{\epsilon-\tilde{\omega}} \hat{U}_{\tilde{\omega}} \psi_n \: \frac{1}{\beta \mathrm{T}} \int_0^\beta d\tau \: e^{i(\epsilon+\omega_m-\omega_n)\tau }\\
     &=\sum_{\omega_m,\omega_n} \: \sum_{\tilde{\omega},\epsilon\to0}\frac{(i\tilde{\omega})}{\mathrm{T}} \: \psi_m^\dagger \hat{V}_{\epsilon-\tilde{\omega}} \hat{U}_{\tilde{\omega}} \psi_n \delta_{\omega_m,\omega_n-\epsilon}\\
     &\propto \sum_{n} \psi^\dagger_n \left[\sum_{\tilde{\omega}}(i\tilde\omega)\hat{V}_{-\tilde{\omega}} \hat{U}_{\tilde{\omega}}\right] \psi_n
\end{split}
\label{eq:elem_2nd}
\end{equation}
where the summation over $\epsilon\to0$ in the last line has been omitted for the time being. Thus, with the results of Eq.\ref{eq:elem_first} and Eq.\ref{eq:elem_2nd}, we see clearly that the matrix elements of $\mathcal{D}_0^{-1} (U^{-1}\dot{U})$ are of the order
\begin{equation}
\begin{split}
    \psi_n^\dagger\left[\mathcal{D}_0^{-1} (U^{-1}\dot{U})\right]\psi_n \:&\propto \: \psi_n^\dagger\left[ \frac{i\tilde{\omega}}{-i\omega_n+\xi_k\pm\gamma}\right]\psi_n\\&=\psi_n^\dagger\left[\frac{\tilde{\omega}}{\sqrt{\omega_n^2+(\xi_k\pm\gamma)^2}}e^{i\frac{\pi}{4}}e^{i\tan^{-1}(\frac{\omega_n}{\xi_k\pm\gamma})}\right]\psi_n.
\end{split}
\end{equation}
Now, since we are interested in the low temperature regime, where $T\ll |\xi_k\pm \gamma |$ and then $\omega_n\ll|\xi_k\pm \gamma |$, we explicitly see that 
\begin{equation}
    |\mathcal{D}_0^{-1} (U^{-1}\dot{U})| \underset{\sim}{\propto} \frac{\tilde{\omega}}{|\xi_k\pm\gamma|} \approx  \frac{\tilde{\omega}}{\gamma},
\end{equation}
after taking into account that only energies $\epsilon_k$ close to the metallic chemical potential $\mu$ ($\xi_k=\epsilon_k-\mu)$ can effectively participate in the interaction (later we will discuss this fact - see Fig.\ref{fig:energy_cases}). Now, by evoking again the adiabatic approximation where the frequency scales $\tilde{\omega}$ in which $\bm{\mu}_{class}$ varies are so slow that they do not induce transitions between spin states $\epsilon_k\pm\gamma$, we have that 
\begin{equation}
    \frac{\tilde{\omega}}{\gamma} \ll 1 \ \ \ \Longrightarrow{} \ \ \ \operatorname{ln}\left[\mathbb{1}+\mathcal{D}_0^{-1}(U^{-1}\dot{U})\right]=\ \mathcal{D}_0^{-1}(U^{-1}\dot{U})\ + \ \mathcal{O}(\tilde{\omega} / \gamma)^2
\end{equation}
such that, finally, Eq.\ref{eq:tr_ln_action} reduces to 
\begin{equation}
     S_k[U]\approx-\operatorname{tr} \left[\mathcal{D}_0^{-1}(U^{-1}\dot{U})\right].
\end{equation}
By all the above discussion, we have seen that both $\mathcal{D}_0^{-1}$ and $(U^{-1}\dot{U})$ are diagonal in the Matsubara space, meaning that 
\begin{equation}
\begin{split}
    S_k[U]&=-\sum_n \operatorname{tr}\left\{\begin{bmatrix}
\frac{1}{-i\omega_n + \xi_k + \gamma} &0 \\
0 &\frac{1}{-i\omega_n + \xi_k - \gamma}
\end{bmatrix} \sum_{\tilde{\omega},\epsilon\to0}\frac{(i\tilde\omega)}{\mathrm{T}}\hat{V}_{\epsilon-\tilde{\omega}} \hat{U}_{\tilde{\omega}}
\right\}\\
&=-\sum_{\tilde{\omega},\epsilon\to0}\frac{(i\tilde\omega)}{\mathrm{T}} \: \operatorname{tr}\left\{\begin{bmatrix}
\sum_n\frac{1}{-i\omega_n + \xi_k + \gamma} &0 \\
0 &\sum_n\frac{1}{-i\omega_n + \xi_k - \gamma}
\end{bmatrix}\begin{bmatrix} V_{\epsilon-\tilde{\omega}}^{00} &V_{\epsilon-\tilde{\omega}}^{01} \\
V_{\epsilon-\tilde{\omega}}^{10} & V_{\epsilon-\tilde{\omega}}^{11}
\end{bmatrix} \begin{bmatrix} U_{\tilde{\omega}}^{00} &U_{\tilde{\omega}}^{01} \\
U_{\tilde{\omega}}^{10} & U_{\tilde{\omega}}^{11}
\end{bmatrix}\right\}\\
&=-\sum_{\tilde{\omega},\epsilon\to0}\frac{(i\tilde\omega)}{\mathrm{T}} \: \operatorname{tr}\left\{\begin{bmatrix}
-\beta n_F(\epsilon_k+\gamma) &0 \\
0 & -\beta n_F(\epsilon_k-\gamma)
\end{bmatrix}\hat{V}_{\epsilon-\tilde{\omega}} \hat{U}_{\tilde{\omega}}\right\}\\
&=\operatorname{tr}\left\{\begin{bmatrix}
n_F(\epsilon_k+\gamma) &0 \\
0 & n_F(\epsilon_k-\gamma)
\end{bmatrix} \sum_{\tilde{\omega},\epsilon\to0}\frac{(i\tilde\omega)}{\mathrm{T}} \hat{V}_{\epsilon-\tilde{\omega}} \hat{U}_{\tilde{\omega}} \: \beta \right\}\\
&=\operatorname{tr}\left\{\begin{bmatrix}
n_F(\epsilon_k+\gamma) &0 \\
0 & n_F(\epsilon_k-\gamma)
\end{bmatrix} \sum_{\tilde{\omega},\epsilon\to0}\frac{(i\tilde\omega)}{\mathrm{T}} \hat{V}_{\epsilon-\tilde{\omega}} \hat{U}_{\tilde{\omega}} \: \int_0^\beta d\tau \: e^{i\epsilon \tau } \right\} \\ 
&= \int_0^\beta d\tau \:\: \operatorname{tr}\left\{\begin{bmatrix}
n_F(\epsilon_k+\gamma) &0 \\
0 & n_F(\epsilon_k-\gamma)
\end{bmatrix} (U^{-1}\dot{U}) \right\}.
\end{split}
\end{equation}
At this point, once the fermi distribution is regarded as the step function for the low $T$ regime, three cases are identified (see Fig.\ref{fig:energy_cases}):
\begin{enumerate}[label=\alph*)]
    \item Both levels are occupied ($\epsilon_k\pm\gamma<\mu$) causing
    \begin{equation}
        S_k[U]=\int_0^\beta d\tau \: \operatorname{tr}\left\{\mathbb{1}\cdot (U^{-1}\dot{U})\right\}=\int_0^\beta d\tau \: \operatorname{tr} (U^{-1}\dot{U}).
    \end{equation}
    \item Both levels are unoccupied ($\epsilon_k\pm\gamma>\mu$), and then
    \begin{equation}
        S_k[U]=\int_0^\beta d\tau \: \operatorname{tr}\left\{0\cdot (U^{-1}\dot{U})\right\}=0.
    \end{equation}
    \item Upper spin state unoccupied ($\epsilon_k+\gamma>\mu$) but lower spin state occupied ($\epsilon_k-\gamma<\mu$), thus the action
    \begin{equation}
        S_k[U]=\int_0^\beta d\tau \: \operatorname{tr}\left\{\begin{bmatrix}
0 &0 \\
0 & 1
\end{bmatrix}(U^{-1}\dot{U})\right\}=\int_0^\beta d\tau \: \bra{\downarrow}U^{-1}\dot{U}\ket{\downarrow}
    \end{equation}
    Note how this situation is more probable for energies close to the chemical potential, such that our assumption of $|\xi_k\pm \gamma|\approx \gamma$ lies exactly in this case. 
\end{enumerate}

For computing the matrix elements of $U^{-1}\dot{U}$, then once more, we make use of the Euler parametrization after taking the $U(1)$ quotient $[U]=e^{-i \phi \hat{S}_3} e^{-i \theta' \hat{S}_2}$:
\begin{equation*}
    \begin{split}
        \dot{U}&=-i\dot{\phi}\hat{S}_3 e^{-i \phi \hat{S}_3} e^{-i \theta' \hat{S}_2}-i\dot{\theta'}e^{-i \phi \hat{S}_3} \hat{S}_2 e^{-i \theta' \hat{S}_2}\, ,\\
        U^{-1}\dot{U}&=-i\left[\dot{\phi}e^{i \theta' \hat{S}_2}\hat{S}_3 e^{-i \theta' \hat{S}_2}+\dot{\theta'} \hat{S}_2\right].
    \end{split}
\end{equation*}
Using BCH formula $e^{i \theta' \hat{S}_2} \hat{S}_3 e^{-i \theta' \hat{S}_2} = \cos\theta' \hat{S}_3 - \sin\theta' \hat{S}_1$, we see the diagonal elements
\begin{equation*}
\begin{split}
    \bra{\uparrow}U^{-1}\dot{U}\ket{\uparrow}&=-i\:\bra{\uparrow}\left[\dot{\phi}\cos\theta' \hat{S}_3 - \sin\theta' \hat{S}_1+\dot{\theta'} \hat{S}_2\right]\ket{\uparrow}=-\frac{i}{2}\dot{\phi}\cos\theta'\\
    \bra{\downarrow}U^{-1}\dot{U}\ket{\downarrow}&=-i\:\bra{\downarrow}\left[\dot{\phi}\cos\theta' \hat{S}_3 - \sin\theta' \hat{S}_1+\dot{\theta'} \hat{S}_2\right]\ket{\downarrow}=\frac{i}{2}\dot{\phi}\cos\theta'.
    \end{split}
\end{equation*}
Therefore, summarizing the mentioned cases, we have
\begin{equation}
\boxed{
\begin{aligned}
    S_k^{(\epsilon_k\pm\gamma<\mu)}[\phi,\theta']
    &=S_k^{(\epsilon_k\pm\gamma>\mu)}[\phi,\theta']=0\\
    S_k^{(\epsilon_k-\gamma<\mu<\epsilon_k+\gamma)}[\phi,\theta']
    &=-\frac{i}{2}\int_0^\beta d\tau \: (1-\cos\theta')\partial_\tau \phi
\end{aligned}}\:,
\end{equation}
where in the last expression, we have introduced the usual total derivative term. This result is already highly non-trivial. Even though the adiabatic evolution suppresses real transitions between occupied and unoccupied levels, integrating out the fermions still leaves a non-vanishing contribution in the effective action. In other words, the fermions do not change their occupation, but they do accumulate a geometric phase as the classical magnetic moment slowly moves on the sphere. As a consequence, the partition function acquires the form
\begin{equation}
    Z=\int D\bm{n} \: e^{-\sum_k^{'}\left(-\frac{i}{2}\right)\int d\tau (1-\cos \theta ') \partial_\tau \phi}.
\end{equation}
where $\sum_k'$ denotes the sum over those $k$-states for which one of the two spin split levels is occupied while the other remains empty. Now, comparing with Eq.\ref{eq:wz_local_rep_su2}, we immediately recognize that for each such $k$ the effective action is precisely the local form of the Wess--Zumino term of a single spin-$1/2$, up to an overall sign that we will discuss in a moment. The question, however, is exactly the one raised in Sec.\ref{sec:evolu_parti}: why does an equilibrium imaginary-time functional integral reproduce the same geometric phase that one usually associates with adiabatic real-time evolution? The answer is now clear. In Sec.\ref{sec:evolu_parti} we saw that, once the $i\epsilon$ prescription is implemented, the asymptotic real-time evolution projects onto the ground-state sector, and that imaginary-time evolution implements precisely the same projection in a technically simpler way. Therefore, even though the Berry phase is originally defined through real-time adiabatic transport, the Euclidean functional integral still has access to the same geometric information, because both formulations isolate the same low-energy vacuum sector. From this point of view, the term above should not be interpreted as a thermal correction, but rather as the Euclidean realization of the geometric phase accumulated by the adiabatically evolving fermionic ground state. In this sense, our result admits two complementary readings:
\begin{enumerate}
    \item $[\textbf{Statistical/Euclidean Interpretation}]$
    After Wick rotation, integrating out the fast fer-mionic degrees of freedom produces an effective action for the slow classical field $\mathbf n(\tau)$. The surviving imaginary contribution is a topological term, namely the Euclidean representative of the Berry/Wess--Zumino phase associated with the adiabatic ground-state bundle. Hence, thermal virtual fluctuations around the many-body ground state (equilibrium state) cause the appearance of Berry phases. 
    \item $[\textbf{Dynamical/Real-time Interpretation}]$ 
    In the original dynamical problem, the slowly varying classical moment drags the fermionic spin sector adiabatically. Although no real transitions occur between the split levels, the occupied fermionic state follows the instantaneous eigenbasis and accumulates a geometric phase. Tracing out the fermions transfers this phase to the effective dynamics of the classical moment. Hence, starting with the many-body groundstate $\ket{GS}$ 
    \begin{equation}
    \ket{GS}=\Pi_{(\epsilon_k+\sigma_k\gamma)\leq \mu}\ket{k,\sigma_k,\bm{\mu}_0} 
    \end{equation}
    (filled Fermi sea and antiparallel configuration of metallic spin $\bm{S}$ and classical moment $\bm{\mu}_{class}$), we would like to compute the adiabatic survival amplitude of just the classical moment part. This means a matrix element $\bra{\bm\mu_{0}}U_{adiab}(+\infty,-\infty)\ket{\bm\mu_{0}}$ where we have traced out all the possible final metallic configurations
    \begin{equation}
    \bra{\bm\mu_{0}}U_{adiab}\ket{\bm\mu_{0}}=\sum  _{k',\sigma_{k'}}\bra{k',\sigma_{k'},\bm\mu_{0}}U_{adiab}\ket{GS}.
    \end{equation}
    Here is exactly when intuition makes sense. Even though a dynamical metallic transition cannot happen because intermediate variation of $\bm{\mu}_{class}$ is so slow, the fact that there are states unoccupied which are felt by the system introduces the mentioned complex phases. Now, in the adiabatic limit, the fast degrees of freedom are able to explore the entire phase space such that at least they feel its topology but not the value of the unoccupied energies. Under this idea, the fast degrees of freedom can transfer such topological information of the phase space to the slow-varying magnetic moment by means of the interaction. This also clarifies why only the intermediate case $\epsilon_k-\gamma<\mu<\epsilon_k+\gamma$ contributes. If both levels are occupied, the geometric phases of the two spin projections cancel each other; if both are empty, there is no contribution at all. Only when one level is occupied and the other one is empty does a net Berry phase remain in the effective action.
\end{enumerate}
We can now understand the sign of the result. For $\gamma>0$, the coupling $\gamma\,\mathbf n\cdot\boldsymbol{\sigma}$ favors the fermionic spin to align antiparallel to the classical magnetic moment in the occupied branch. Therefore, when the fermionic spin follows a closed path on $S^2$, the classical moment effectively feels the opposite orientation, which explains why the induced action is $$S_k[\mathbf n]=-S_{WZ}[\mathbf n].$$ If instead one had $\gamma<0$, the aligned branch would be occupied, the relevant projection would be $\bra{\uparrow}U^{-1}\dot U\ket{\uparrow}$, and the sign would be reversed:
$$S_k[\mathbf n]=+S_{WZ}[\mathbf n].$$
Thus, the overall sign of the induced WZ term simply reflects whether the coupling favors alignment or anti-alignment between the fermionic spin and the classical moment.

\subsection{Geometric Quantum Noise}
Once we have seen how Berry's phase emerges in the most simple (but non-trivial) situation in equilibrium many-body, we now move on and study the physical consequences of such WZ term's appearance. In Eq.\ref{eq:action_singl_spin_imagi},\ref{eq:Bloch_eqtns_imagi}, we have already derived the classical equations of motion (in imaginary time) of a magnetic moment $\bm{S}=S\mathbf{n}(\theta',\phi)$ in the presence of a magnetic field $\mathbf{B}\cdot\bm{S}$. Rotating back to real time and reintroducing fundamental constants, these equations then read as
\begin{equation}
    \frac{d \bm{S}}{d t}=- \left(\frac{ge}{2m_e}\right) \bm{S}\times \bm{B}=-\gamma \: \bm{S}\times \bm{B}\:.
\end{equation}
As always, this describes the usual precession of spins $1/2$ around a magnetic field $\bm{B}=B\hat{z}$. However, this idealized precession picture breaks down in real materials, where spins are not perfectly isolated. Interactions with the environment, such as spin-lattice coupling, impurities, and thermal fluctuations, lead to energy dissipation and decoherence. To account for these effects, one must go beyond the pure precession dynamics by including damping and stochastic terms. This motivates the introduction of the Landau-Lifshitz-Gilbert-Langevin framework, which captures both relaxation processes and the influence of disorder and noise.

\subsubsection{LLG-Langevin Equations}
To describe the realistic dynamics of spins in magnetic materials, where dissipation and fluctuations are unavoidable, we use the Landau-Lifshitz-Gilbert (LLG) equation extended by a Langevin term:
\begin{equation} 
\boxed{
\frac{d\bm{S}}{dt} = -\gamma \left[\bm{S} \times \bm{B}\right] \:+\: \frac{\alpha}{S} \left[ \bm{S} \times \frac{d\bm{S}}{dt} \right] \:+\: \bm{\eta}(t) \:} \: .
\end{equation}

\noindent In addition to the first standard Bloch term, we now have two extra contributions. The first of these new summands, proportional to a damping coefficient $\alpha>0$, is known as the Gilbert damping term and captures the tendency of a spin to gradually align with the external magnetic field $\bm{B}$. Hence, its function is modeling relaxation processes, ensuring that the spin eventually reaches equilibrium and energy is dispersed into the surroundings. On the other hand, the third term $\bm{\eta}(t)$ is known as the Langevin term, which accounts for stochastic torques resulting from thermal fluctuations or disorder effects in a real material. It is typically assumed to be sampled from a Gaussian white noise distribution with zero mean and correlation functions determined by the fluctuation-dissipation theorem. This is
\begin{equation}
    \left\{ 
    \begin{aligned} 
   & \left\langle\eta_i(t)\right\rangle=0 \ \ \ \ \ \ \ \ \ \ \ \ \ \ \ \ \ \ \ \ \ \ \ \ \ \ \ \ \ \text{ (Gaussian white noise)}  \\  &\left\langle\eta_i(t)\eta_j(t')\right\rangle=2D\delta_{ij} \delta (t-t') \\
   &\ D\propto \frac{\alpha k_BT}{\gamma S} \ \ \ \ \  \ \ \  \ \ \ \ \ \ \ \  \ \ \  (\text{Fluctuation-Dissipation theorem}).
\end{aligned} \right.
\end{equation}
\noindent Therefore, the introduction of the Langevin term entails numerous physical effects. First, and more importantly, we have the loss of deterministic behavior in spin dynamics: each realization of the stochastic torque leads to a different spin trajectory, thereby transforming the problem into one of a statistical nature. This randomness reflects the intrinsic uncertainty introduced by interactions with the environment and requires ensemble averaging to extract physically meaningful observables. In the absence of damping, the noise would induce perpetual wandering of the spin on the Bloch sphere, with no mechanism to restore equilibrium. However, when damping is present alongside the Langevin force, a balance is eventually achieved, while the noise constantly perturbs it, resulting in a dynamic steady state characterized by fluctuating contributions around their thermal expectation value. 

Now, while the LLG-Langevin equation is often introduced phenomenologically, a central question is how the stochastic torques arise from a microscopic quantum description of the environment. In this vein, the essence of the so-called \textit{Geometric Quantum Noise} lies in demonstrating that the spin $S U(2)$ geometric phase (i.e., the Wess-Zumino term) determines the structure of the random Langevin forces $\boldsymbol{\eta}(t)$. To this end, and drawing on the seminal work of \cite{kiselev}, we will focus on the case of a magnetic quantum dot with a large total spin $\boldsymbol{S}$ that is tunnel-coupled to a metallic lead. We will then show that, in the quantum regime---where the precession frequency exceeds the temperature---the stochastic spin torques are significantly shaped by the Berry phase accumulated during the precession. However, before addressing this point, we must first explore the physics of quantum dots and the field theoretical framework of systems out of equilibrium.  

\subsubsection{The Universal Hamiltonian of Quantum Dots}
Quantum dots (QDs) are artificial nanoscale structures that exhibit discrete energy levels, much like atoms. However, unlike natural atoms, the properties of quantum dots can be tuned by external parameters such as size, shape, and material composition, making them ideal candidates for studying mesoscopic physics, i.e., systems whose size lies between microscopic and macroscopic scales.

\noindent Intuitively,  QDs can be thought of as nanoscopic boxes that trap electrons by restricting their motion in all three spatial dimensions. In semiconductor devices, this is typically achieved by electrostatic gates that locally deplete a two-dimensional electron gas, or by nanofabrication techniques such as etching or heterostructure growth that define a small electronic island \cite{michler2009single}. These techniques create a spatially confined region in which the motion of electrons is restricted, effectively producing a potential well whose shape and depth can be tuned experimentally. As a result, the electronic states in quantum dots become quantized, yielding a discrete energy spectrum. In particular, when the characteristic size of the dot is comparable to or smaller than the electron’s de Broglie wavelength, quantum effects dominate, leading to rich phenomena that exceed classical expectations. However, modeling quantum dots in this regime, is particularly challenging due to several factors:
\begin{itemize}
    \item[-] \textbf{[}\underline{Disorder and Chaotic dynamics}\textbf{]}: the disorder arising from structural imperfections and impurities leads to irregular confining potentials. This results in chaotic single-particle electron dynamics, making exact modeling intractable. 
    \item[-] \textbf{[}\underline{Coulomb Interactions}\textbf{]}: as the QD-size decreases, the Coulomb repulsion between electrons becomes more significant, leading to complex many-body effects.  A paradigmatic example is the \textit{Coulomb Blockade}, where the addition of an extra electron to the dot requires a charging energy of $E_C=\frac{e^2}{2C}$, with $C$ the capacitance of the dot. When $E_C \gg k_B T$, electrons from the metallic leads can only tunnel into the dot if they possess sufficient energy to overcome this charging cost, strongly suppressing transport through the device.
    
    \item[-] \textbf{[}\underline{Spin Exchange Interactions}\textbf{]}: electron–electron interactions can also generate exchange processes between spins. In particular, spin exchange processes between localized electrons in the dot and conduction electrons in the leads can produce \textit{Kondo physics}; meanwhile, exchange interactions among the dot electrons themselves can favor collective spin configurations such as ferromagnetic or antiferromagnetic states.
    
    \item[-] \textbf{[}\underline{Strong Correlations}\textbf{]}: in the regime of very strong Coulomb repulsion and low electron densities, QD electrons can arrange themselves into ordered, crystalline-like patterns. This is known as a Wigner molecule.
\end{itemize}
All in all, the complicated form of interactions among QDs, coupled with the inevitable presence of disorder, leads to a highly complex and irregular energy spectrum. This complexity indeed goes beyond simple and predictable patterns. The energy levels are not evenly spaced; they exhibit a highly intricate distribution, so that even small changes in the dot's geometry or the distribution of impurities can drastically alter the energy spectrum. Consequently, traditional methods that rely on solving the Schrödinger equation for a well-defined potential become computationally intractable. Instead, a statistical approach must be assumed in order to make general predictions about the system’s behavior. 

This is where \textit{Random Matrix Theory} (RMT) comes into the story. Starting with the full Hamiltonian for a quantum dot:
\begin{equation}
    H=\sum_{i, j, \sigma} h_{i j} c_{i \sigma}^{\dagger} c_{j \sigma}+\frac{1}{2} \sum_{\alpha \beta \gamma \delta } V_{\alpha \beta \gamma \delta} \: c_{\alpha }^{\dagger} c_{\beta }^{\dagger} c_{\delta } c_{\gamma } \ \ ,
    \label{eq:detail_QD_hamil}
\end{equation}
in the RMT approach, we would like to consider $h_{ij}$ (the single-particle matrix element between localized states $i$ and $j$) and $V_{\alpha \beta \gamma \delta}$ (the interaction matrix elements with confining-spin indices $\alpha$) as random variables chosen from a specific ensemble \cite{Beenakker1997RandomMatrixTheory, Aleiner_2002}. Upon ensemble averaging, most terms are suppressed, and only those invariant under basis rotations remain. The resulting effective description captures the statistical properties of the dot without requiring detailed knowledge of its microscopic disorder. This is precisely the idea behind what is known as the \textbf{Universal Hamiltonian for Quantum Dots}: specific details of the dot’s geometry or impurity configuration do not matter, and only the symmetries of the system do.  Under this principle, the microscopical Hamiltonian in Eq.\ref{eq:detail_QD_hamil} can be rewritten in a simplified effective way, capturing the essential physics of electrons in a QD:
\begin{equation}
\boxed{
H_{\text{univ.QD}}=H_0+H_C+H_J+H_\lambda
}\ \ , \ \ \text{ where}
\end{equation}
\begin{itemize}
    \item The noninteracting part reads as ($\epsilon_\alpha\equiv$ energy of spin-degenerate single particle levels)
    \begin{equation}
        H_0=\sum_{\alpha, \sigma} \epsilon_\alpha a_{\alpha, \sigma}^{\dagger} a_{\alpha, \sigma}.
    \end{equation}
    \item The charging interaction term, accounting for the Coulomb blockade, is
    \begin{equation}
        H_C=E_C\left(\hat{N}-N_0\right)^2 \ \ \ ; \ \ \ \hat{N}=\sum_{\alpha, \sigma} a_{\alpha, \sigma}^{\dagger} a_{\alpha, \sigma}.
    \end{equation}
     Here, $E_C$ denotes the QD charging energy, $N_0$ the background/offset charge controlled by external gates, and $\hat{N}$ the total number of electrons operator.
    \item The term
        \begin{equation}
           H_J=-J \hat{\mathbf{S}}^2 \ \ \ ; \ \ \ \mathbf{S} \equiv \frac{1}2{} \sum_{\alpha, \sigma_1, \sigma_2} a_{\alpha, \sigma_1}^{\dagger} \boldsymbol{\sigma}_{\sigma_1, \sigma_2} a_{\alpha, \sigma_2}
        \end{equation}
    represents the ferromagnetic $(J>0)$ exchange interaction within the dot, being $\hat{\mathbf{S}}$ the operator of the total QD spin.

    \item Lastly, the interaction in the Cooper channel is described by
    \begin{equation}
        H_\lambda=\lambda T^{\dagger} T, \quad T=\sum_\alpha a_{\alpha, \uparrow} a_{\alpha, \downarrow}\ ,
    \end{equation}
    where $\lambda$ characterizes the effective interaction strength, and $T$ accounts for possible pairing correlations between electrons occupying the same orbital with opposite spins, i.e., singlet (conventional) superconducting correlations within the dot.
\end{itemize}

Now, there are cases in which one doesn't need to include all four terms to obtain a correct, effective description. For the dots fabricated in a 2D electron gas, the interaction in the Cooper channel is typically repulsive and, therefore, renormalizes to zero \cite{Aleiner_2002} \cite{saha2012quantum}. On the other hand, for the case of 3D quantum dots realized as small metallic grains, the interaction in the Cooper channel can be attractive, giving rise to a competition between superconductivity and ferromagnetism. However, if that is the case, we assume here the presence of a weak magnetic field $\bm{B}$ that breaks time-reversal symmetry and causes the suppression of the Cooper channel. Recall that time-reversal symmetry is crucial for the formation of Cooper pairs: a pair consists of time-reversed partners $(\mathbf{k}, \uparrow)$ and $(-\mathbf{k}, \downarrow)$, so that when the Zeeman splitting is present, the pairing turns out to be energetically unfavorable.

\noindent Similarly, once the number of particles in a QD is fixed, the charging Hamiltonian becomes a fixed charging energy $E_{\text{charging}}$ which only shifts the entire energy spectrum. Hence, under the suppressing conditions of Cooper and Charging interactions, the Hamiltonian that describes an isolated QD reads as
\begin{equation}
    \boxed{H_{\mathrm{dot}}=\sum_{\alpha, \sigma} \epsilon_\alpha a_{\alpha, \sigma}^{\dagger} a_{\alpha, \sigma}-J \boldsymbol{S}^2+\boldsymbol{B}\cdot \boldsymbol{S}\:,\:}
    \label{eq:hamil_dot}
\end{equation}
where again $\boldsymbol{S}$ is the operator of the total spin on the quantum dot, $\boldsymbol{B}$ is the external magnetic field, and $J>0$ is the corresponding ferromagnetic exchange constant. That said, even though electrons in a QD are confined to a finite region of space, they are not localized within it. Instead, their wavefunctions extend across the entire dot, so the electrons remain delocalized and behave as itinerant particles. As a consequence, exchange interactions can act collectively on these states, and the dot may undergo a magnetic phase transition in which ferromagnetism emerges due to the $-J \boldsymbol{S}^2$ exchange interaction. This tendency is quantified by the \textbf{Stoner criterion}, which states that a system develops a net large magnetic moment when the condition
\begin{equation}
    U \cdot \nu\left(\varepsilon_F\right) \gtrsim 1
    \label{eq:stoner_instabil}
\end{equation}
is met or approached. Here, $U$ stands for the strength of the exchange interaction (on-site energy), and $\nu\left(\varepsilon_F\right)$ for the density of states at the Fermi level. In what follows, we will assume that we have a QD close to the Stoner instability, i.e. the product $U \cdot \nu\left(\varepsilon_F\right)$ is close to 1 (but still < 1 ). This means the system is not yet ferromagnetic, but it is characterized by a large total spin which is very susceptible to spin fluctuations: small perturbations (like thermal fluctuations, disorder, or quantum fluctuations) can lead to significant spin polarization or even fluctuating local moments.

Thus, the physical scenario we will be studying consists of a quantum dot tunnel-coupled to a non-magnetic metallic lead, where the QD hosts a large collective spin due to the enhanced exchange interaction, while the coupling to the lead introduces dissipation and spin relaxation. Consequently, the system is described by the Hamiltonian
\begin{equation}
    H=H_{\mathrm{dot}}+H_{\text {lead }}+H_{\mathrm{tun}}
\end{equation}
where $H_{\text{dot}}$ captures the internal magnetic dynamics of the dot (Eq.\ref{eq:hamil_dot}), $H_{\text{lead}}$ describes the free electrons in the metallic reservoir, and $H_{\text{tun}}$ accounts for electron tunneling between the two subsystems:
\begin{equation}
\begin{split}
    H_{\text{lead}}&=\sum_{\gamma, \sigma} \epsilon_\gamma c_{\gamma, \sigma}^{\dagger} c_{\gamma, \sigma} \\
    H_{\text{tun}}&=\sum_{\alpha, \gamma, \sigma} V_{\alpha, \gamma} a_{\alpha, \sigma}^{\dagger} c_{\gamma, \sigma}+\text { h.c. }
\end{split}
\end{equation}

\subsubsection{Out-of-Equilibrium Field Theory}
Having discussed the physics of quantum dots, the second ingredient we will need for deriving Geometric Quantum Noise is the field theoretical approach to non-equilibrium dynamics, a.k.a. Keldysh model. This formalism extends the standard techniques of quantum field theory to systems driven out of equilibrium, where traditional equilibrium methods such as Matsubara imaginary time or real-time adiabatic evolution are no longer applicable. At its core, the Keldysh approach provides a framework to compute expectation values and correlation functions in real time, making it particularly powerful for analyzing open quantum systems, quantum transport, and dissipation. By introducing a closed time contour and duplicating the degrees of freedom, we can study both the system's unitary dynamics and the influence of its environment within a unified theoretical structure. Here we will be following \cite{kamenev2011field}.

\begin{figure}[h!]
    \centering
    \includegraphics[width=0.85\linewidth]{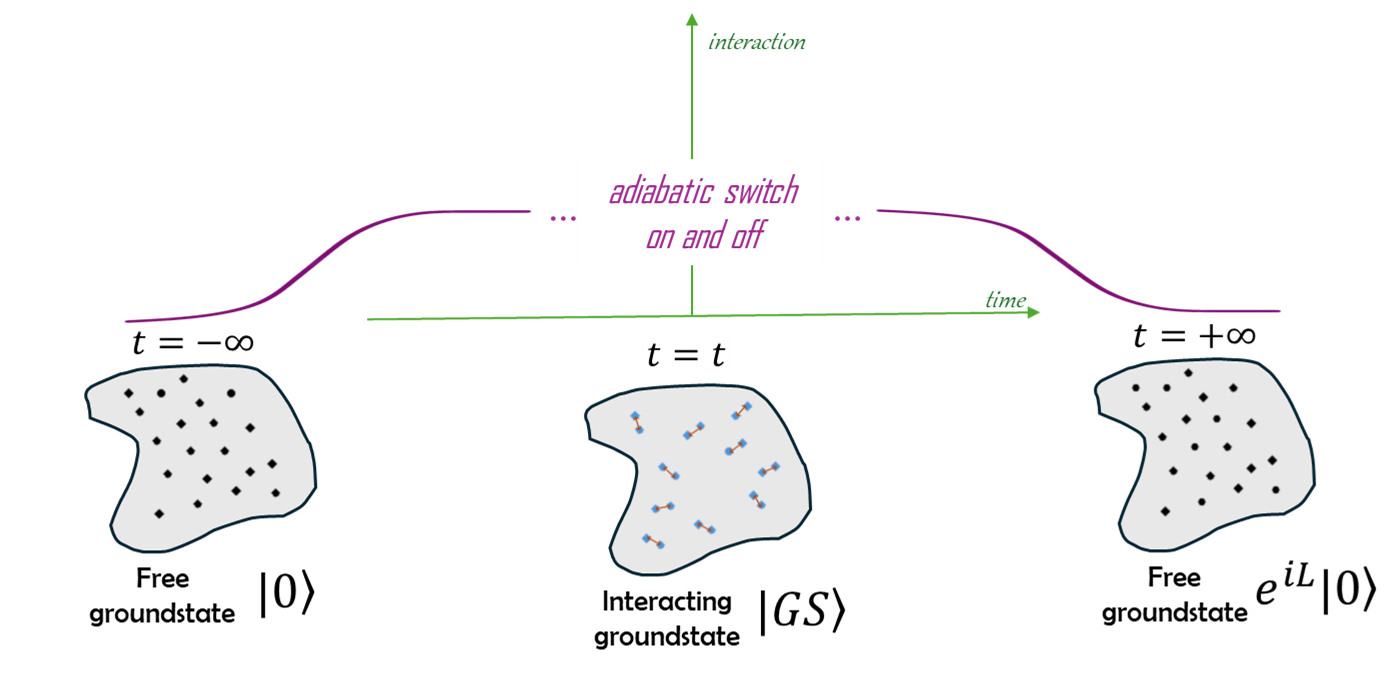}
    \caption{Adiabatic evolution of instantaneous groundstate.}
    \label{fig:adiabatic_evolu}
\end{figure}

As any other real-time QFT formalism, the final objective of the Keldysh model is to predict the dynamics of observables' expectation values $\langle\hat{O}\rangle(t)$. Therefore, having an initial state $\rho(-\infty)=\ket{\psi}\bra{\psi}$ in the $t=-\infty$ distant past, and taking into account that the density matrix evolves as ($\hbar=1$)
\begin{equation}
    \rho (t)=U_{t,-\infty}\ket{\psi}\bra{\psi}U_{t,-\infty}^\dagger=U_{t,-\infty}\rho(-\infty) U_{-\infty,t}\: \ \ \ \text{with} \ \ U_{t, t_0}=T e^{-i\int_{t_0}^t H\left(t^{\prime}\right) d t^{\prime}}
\end{equation}
the previous time-dependent quantum average values are written in the form
\begin{equation}
\langle\hat{O}\rangle(t)=\operatorname{Tr}\left[\hat{O}\:U_{t,-\infty}\rho(-\infty) U_{-\infty,t}\right].
\end{equation}
As always, even though physically, the trace of the density matrix should be $\operatorname{Tr}[\rho(t)]=1$, during functional expansions (as in path integral perturbation theory), the normalization can get distorted and the quantity $\expval{\hat{O}\rho(t)}$  might get contributions that need to be consistently canceled. So, it's safer to include $\operatorname{Tr}[\rho(t)]$ in the denominator until the very end of calculations, and then check whether it simplifies to 1. Then, 
\begin{equation}
\langle\hat{O}\rangle(t)=\frac{\operatorname{Tr[\hat{O}\rho (t)]}}{\operatorname{Tr}[\rho(t)]}=\frac{\operatorname{Tr}\left[\hat{O}\:U_{t,-\infty}\rho(-\infty) U_{-\infty,t}\right]}{\operatorname{Tr}\left[U_{t,-\infty}\rho(-\infty) U_{-\infty,t}\right]},
\end{equation}
which, after using the cyclic property of the trace, turns to (for simplicity $\rho:=\rho(-\infty)$)
\begin{equation}
\boxed{\langle\hat{O}\rangle(t)=\frac{\operatorname{Tr}[\overbrace{U_{-\infty,t}}^{\text {backward }} \hat{O} \overbrace{U_{t,-\infty}}^{\text {forward}} \ \rho \:]}{\operatorname{Tr}\left[U_{t,-\infty} \rho U_{-\infty, t}\right]}}
\label{eq:time_exp_value}
\end{equation}
Here is where the main difference with thermal Matsubara and adiabatic real-time QFTs comes into play. In the former formalism, it is assumed that the studied system has already thermalized, such that by definition, it is in equilibrium and can be described by statistical mechanics (i.e., the partition function). Similarly, in the latter, we consider that the interacting Hamiltonian is adiabatically turned on from $-\infty \mapsto t$, and then adiabatically turned off from $t\mapsto\infty$, with no possibility of having any time dependence that can take the system out of its instantaneous groundstate. This means, first, that the groundstate $ |\mathrm{GS}\rangle$ of an interacting many-body system is obtained by evolving the known and simple groundstate $|0\rangle$ of the non-interacting case
\begin{equation}
    \left\{ 
    \begin{aligned} 
   & |\mathrm{GS}\rangle=U_{t,-\infty}|0\rangle \\  
   &U_{+\infty,-\infty}|0\rangle=\mathrm{e}^{\mathrm{i} L}|0\rangle
\end{aligned} \right. \ \ \ \ , 
\label{eq:adiabatic}
\end{equation}
and second, the evolution of the free groundstate $|0\rangle$ upon adiabatic switching on and off brings the system
back into the state $|0\rangle$, up to a phase factor $\mathrm{e}^{\mathrm{i} L}$. See Fig.\ref{fig:adiabatic_evolu} for a pictorial description. 
This, indeed, is the assumption that eventually takes us to the celebrated Gell-Mann and Low theorem:
\begin{equation}
    \langle\mathrm{GS}| \hat{\mathcal{O}}|\mathrm{GS}\rangle=\frac{\langle 0| U_{+\infty, t} \hat{\mathcal{O}} U_{t,-\infty}|0\rangle}{\langle 0| U_{+\infty,-\infty}|0\rangle} \ .
\end{equation}
However, with regard to non-equilibrium dynamics, Eq.\ref{eq:adiabatic} is no longer a valid physical assumption and therefore the last theorem will not apply. Opposite to what we have on it, where only forward evolution matters, for the non-equilibrium framework, both forward and backward evolution operators will matter. To see this, let's note that we can insert the identity 
\begin{equation}
    \mathbb{1}=U_{\infty,t}^{\dagger} \:U_{\infty, t}=U_{t,\infty} \: U_{\infty, t}
\end{equation} 
next to one of the sides of $\hat{O}$ in Eq.\ref{eq:time_exp_value}. Hence, depending on where we are inserting the identity, let's define:
\begin{equation}
     \left\{ 
    \begin{aligned} 
   &\langle\hat{O}\rangle^{\text{forward}}(t)=
   \frac{\operatorname{Tr}[U_{-\infty,t} \: \mathbb{1} \: \hat{O} U_{t,-\infty} \ \rho \:]}{\operatorname{Tr}\left[U_{t,-\infty} \rho U_{-\infty, t}\right]}=\frac{\operatorname{Tr}[U_{-\infty,\infty} U_{\infty, t} \hat{O} U_{t,-\infty} \ \rho \:]}{\operatorname{Tr[\rho_{(-\infty)}]}} \\  
   \\
   &\langle\hat{O}\rangle^{\text{backward}}(t)=\frac{\operatorname{Tr}[U_{-\infty,t} \hat{O} \: \mathbb{1} \: U_{t,-\infty} \ \rho \:]}{\operatorname{Tr}\left[U_{t,-\infty} \rho U_{-\infty, t}\right]}=\frac{\operatorname{Tr}[U_{-\infty,t}\hat{O} U_{t,\infty} U_{\infty,-\infty} \ \rho \:]}{\operatorname{Tr[\rho_{(-\infty)}]}} \ \ .
\end{aligned} \right.
\end{equation}
\begin{figure}[h!]
    \centering
    \includegraphics[width=0.90\linewidth]{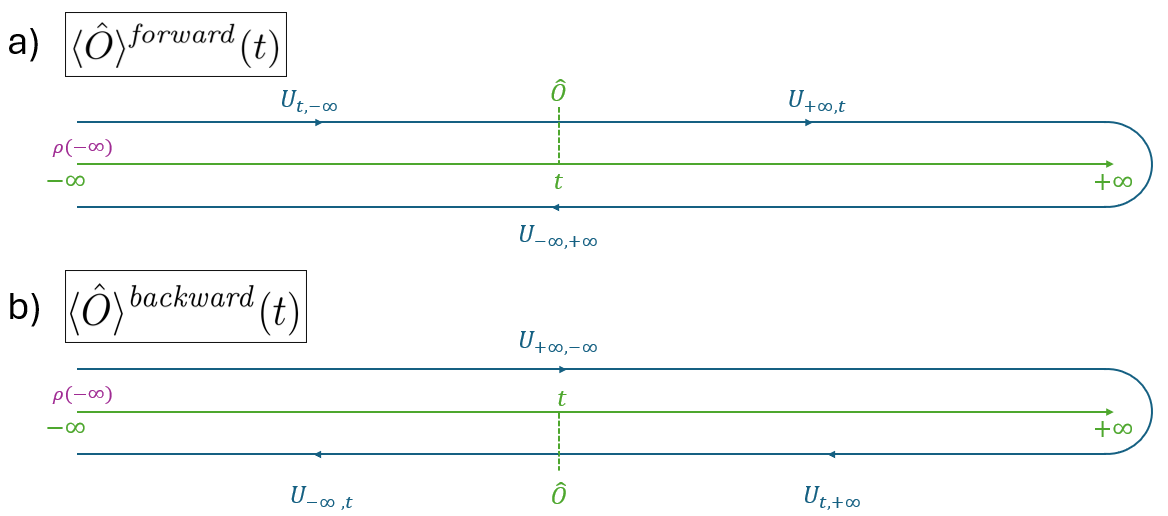}
    \caption{Forward and backward insertions of the an observable operator $\hat O$ on the closed real-time contour.}
    \label{fig:closed_contour}
\end{figure}

\noindent Basically, according to which branch of the evolution we insert in the observable operator, we will compute $\langle\hat{O}\rangle^{\text{forward}}(t)$ or $\langle\hat{O}\rangle^{\text{backward}}(t)$. Note how, until this moment, we have worked with the same Hamiltonian for the two different contours (or more specifically, the two different insertions in the big closed contour). Now, the idea is to invert the game in the way that we can have only one \textbf{ closed contour} $\bm{\mathcal{C}}$ \textbf{evolution operator} 
\begin{equation}
    U_\mathcal{C}=U_{-\infty,\infty} \: U_{\infty,-\infty} 
\end{equation}
but two different Hamiltonians $H^{\text{forward}}$ and $H^{\text{backward}}$ which generate or produce the expectation values $\langle\hat{O}\rangle^{\text{forward}}(t)$ and $\langle\hat{O}\rangle^{\text{backward}}(t)$, respectively. In this vein, we define the modified Hamiltonians as 
\begin{equation}
    \hat{H}_V^{\text{forward}/\text{backward}}(t)=\hat{H}(t) \pm \hat{O}\: V(t),
\end{equation}
where the quantity $V$ has the same source character as $J$ for the case of a scalar field. So the idea now is that on each contour branch, we have two different Hamiltonians producing different expectation values. Consequently, since each branch is half of the closed contour $\mathcal{C}$, we consider averaging the contributions in a new quantity defined as the Keldysh expectation value:
\begin{equation}
    \langle\hat{O}\rangle^{\text {Keld}}(t)=\frac{1}{2}\left(\langle\hat{O}\rangle_{(t)}^{\text {forward }}+\langle\hat{O}\rangle_{(t)}^{\text {backward}}\right)
\end{equation}
Under this framework, now the closed contour evolution operator $U_\mathcal{C}[V]$:
\begin{itemize}
    \vspace{-0.35cm}
    \setlength\itemsep{0.15em}
    \item[-] depends on the source $V$,
    \item[-] is composed of two subevolutions evolving with different Hamiltonians, 
    \item[-] and, different from the equilibrium case, $U_\mathcal{C}\neq 1$. 
\end{itemize}
Hence, it is with respect to this new closed contour evolution operator $U_\mathcal{C}[V]$ that we define the generating functional $Z[V]$ for the Keldysh model as 
\begin{equation}
    Z[V] \equiv \frac{\operatorname{Tr}\left[\hat{U}_\mathcal{C}(V) \: \rho_{(-\infty)}\right]}{\operatorname{Tr}[\rho_{(-\infty)}]}
\end{equation}
(Note that since we are not in the equilibrium case, it doesn't have the same meaning as the partition function). Therefore, with the above definition of the Keldysh expectation value, one can show that 
\begin{equation}
\boxed{ 
    \langle\hat{O}\rangle_{(t)}^{\text {Keld }}=\left.\frac{i}{2} \frac{\delta Z[V]}{\delta V(t)}\right|_{V=0}} \ \ .
\end{equation}

\noindent In general, one can say that the Keldysh technique is composed of two things:
\begin{figure}[h!]
    \centering
    \includegraphics[width=0.95\linewidth]{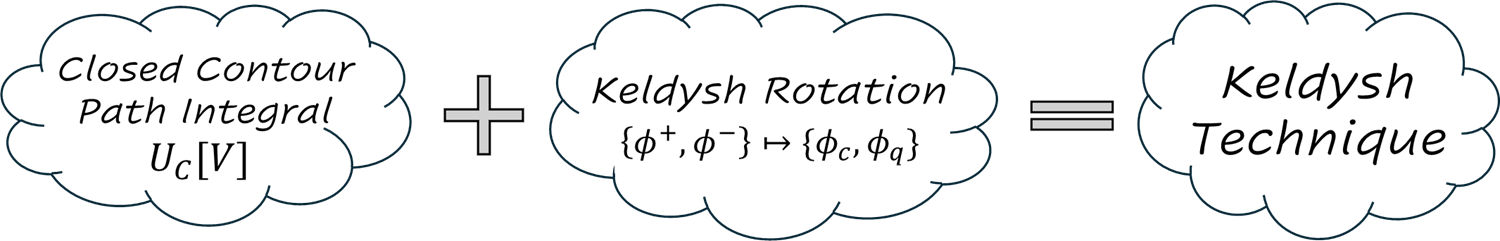}
    \label{fig:diagram_keldysh}
    \caption{Schematic structure of the Keldysh technique}
\end{figure}

\noindent Until now, we have completed half the way (the one that is common to fermions and bosons). In what follows, we will focus on how to apply the so-called Keldysh rotation to out-of-equilibrium bosonic degrees of freedom; nevertheless, keep in mind that there is also its fermionic counterpart. 

Similar to what one does in the Matsubara/real time path integrals, here we divide the time closed contour into $2N-2$ small intervals $\set{t_1\to -\infty , \dots, t_N,t_{N+1},\dots , t_{2N}\to -\infty}$ of time length $\Delta t$, with the particularity of having $t_N=t_{N+1}\to \infty $ (half of the closed evolution which goes to infinity). Then, by inserting instantaneous resolutions of the identity in coherent states, the "partition function" acquires the form (assuming bosonic complex fields)
\begin{equation}
    Z[V]=\frac{\operatorname{Tr}\left\{U_{\mathcal{C}}[V] \: \rho\right\}}{\operatorname{Tr}\{\rho\}}=\int \mathbf{D}[\bar{\phi}(t), \phi(t)] \:  \mathrm{e}^{\mathrm{i} S[\bar{\phi}, \phi, V]} \ .
\end{equation}
It is important to remember that each bosonic field $\phi(t),\bar{\phi}(t)$ has the two components $\phi^{u}$ and $\phi^{d}$ which reside on the forward and backward parts of the time contour, respectively. Then, the "non-interacting non-sourced" part of the action can be written as
\begin{equation}
    S[\bar{\phi}, \phi]=\int_{-\infty}^{+\infty} \mathrm{d} t\left[\bar{\phi}^{u}_{(t)}\left(\mathrm{i} \partial_t-\xi\right) \phi^{u}_{(t)}-\bar{\phi}^{d}_{(t)}\left(\mathrm{i} \partial_t-\xi\right) \phi^{d}_{(t)}\right],
\end{equation}
where the relative minus sign comes from the reversed direction of the time integration on the backward part of the contour. At this point, one would be tempted to think that the $\phi^{u}(t)$ and $\phi^{d}(t)$ fields are completely uncorrelated, but in the process of discretization, one finds that there are matrix elements connected them, and therefore, all the correlators
\begin{equation}
   \left\langle\phi_j^{u} \bar{\phi}_{j^{\prime}}^{d}\right\rangle \ \ \ , \ \ \   \left\langle\phi_j^{d} \bar{\phi}_{j^{\prime}}^{u}\right\rangle
   \ \ \ , \ \ \
   \left\langle\phi_j^{u} \bar{\phi}_{j^{\prime}}^{u}\right\rangle 
   \ \ \ , \ \ \
   \left\langle\phi_j^{d} \bar{\phi}_{j^{\prime}}^{d}\right\rangle
   \label{eq:original_correlators}
\end{equation}
are not necessarily zero. We now introduce a new pair of fields according to the so-called \textbf{Keldysh Rotation}:
\begin{equation}
\boxed{
\begin{aligned}
    \phi^{\mathrm{cl}}(t)=\frac{1}{\sqrt{2}}\left(\phi^{u}_{(t)}+\phi^{d}_{(t)}\right)  \quad \quad \quad \bar{\phi}^{\mathrm{cl}}(t)=\frac{1}{\sqrt{2}}\left(\bar{\phi}^{u}_{(t)}+\bar{\phi}^{d}_{(t)}\right) \\
    \phi^{\mathrm{q}}(t)=\frac{1}{\sqrt{2}}\left(\phi^{u}_{(t)}-\phi^{d}_{(t)}\right) \quad \quad \quad 
    \bar{\phi}^{\mathrm{q}}(t)=\frac{1}{\sqrt{2}}\left(\bar{\phi}^{u}_{(t)}-\bar{\phi}^{d}_{(t)}\right) 
\end{aligned}} \quad . 
\end{equation}
The superscripts “cl”
and “q” stand for the classical and the quantum components of the fields, respectively. In a very rough first look, their meaning should be evident, in the sense that the classical field $\phi^{\mathrm{cl}}(t)$ represents the average (a classical trajectory), and the quantum field $\phi^{\mathrm{q}}(t)$ encodes fluctuations and the system's response to external perturbations.  

In this line, we now have correlators of the form 
\begin{equation}
    \left\langle\phi^\alpha_{(t)} \bar{\phi}^\beta_{\left(t^{\prime}\right)}\right\rangle = \left(\begin{array}{cc}\left\langle\phi^{\mathrm{cl}}_{(t)} \bar{\phi}^\mathrm{cl}_{\left(t^{\prime}\right)}\right\rangle & \left\langle\phi^{\mathrm{cl}}_{(t)} \bar{\phi}^\mathrm{q}_{\left(t^{\prime}\right)}\right\rangle \\ \left\langle\phi^{\mathrm{q}}_{(t)} \bar{\phi}^\mathrm{cl}_{\left(t^{\prime}\right)}\right\rangle & 0\end{array}\right) \equiv \mathrm{i} G^{\alpha \beta}_{\left(t, t^{\prime}\right)}=\left(\begin{array}{cc}\mathrm{i} G^{\mathrm{K}}_{\left(t, t^{\prime}\right)} & \mathrm{i} G^{\mathrm{R}}_{\left(t, t^{\prime}\right)} \\ \mathrm{i} G^{\mathrm{A}}_{\left(t, t^{\prime}\right)} & 0\end{array}\right)
\end{equation}
 where the fact that the $(q, q)$ correlator is zero is a manifestation of the constraints followed by the original correlators in Eq.\ref{eq:original_correlators}. Here, superscripts R, A and K stand for the retarded, advanced and Keldysh components of the Green function, respectively. So, the idea with the Keldysh rotation is that, if one wishes to extend the treatment of a Matsubara problem/action of the form 
 \begin{equation}
     S_{\text{Mtsbra}}[\bar{\phi},\phi, ]= \int_0^\beta d \tau \:\bar{\phi}_{(\tau)} \: \hat{G}^{-1}(\tau) \: \phi_{(\tau)} = \sum_{\omega_n} \bar{\phi}_n  G^{-1}(i\omega_n) \phi_n \:  ,
 \end{equation}
to non-equilibrium or time-dependent conditions, one can double the number of degrees of freedom (correspondingly doubling the action) $\phi \rightarrow \vec{\phi}=\left(\phi^{\mathrm{cl}}, \phi^{\mathrm{q}}\right)^{\mathrm{T}}$, analytically continue in Fourier frequency space $G^{R/A}(\omega)=G(i\omega_n\to\omega\pm i0^+)$, and consider fields $\bar{\phi},\phi$ as functions of the real time $t$ or real frequency $\omega$. Consequently, the general form of the quadratic time translationally invariant Keldysh action is
\begin{equation}
    S\left[\phi^{\mathrm{cl}}, \phi^{\mathrm{q}}\right]=\iint_{-\infty}^{+\infty} \mathrm{d} t \mathrm{~d} t^{\prime}\left(\bar{\phi}^{\mathrm{cl}}, \bar{\phi}^{\mathrm{q}}\right)_t\left(\begin{array}{cc}0 & {\left[G^{-1}\right]^{\mathrm{A}}} \\ {\left[G^{-1}\right]^{\mathrm{R}}} & {\left[G^{-1}\right]^{\mathrm{K}}}\end{array}\right)_{t, t^{\prime}}\binom{\phi^{\mathrm{cl}}}{\phi^{\mathrm{q}}}_{t^{\prime}}
    \label{eq:keldysh_action}
\end{equation}
where the diagonal Keldysh component, $\left[G^{-1}\right]^{\mathrm{K}}$, of the quadratic form is found from the condition
\begin{equation}
    G^{\mathrm{K}} * \left[G^{\mathrm{A}}\right]^{-1}+G^{\mathrm{R}} * \left[G^{-1}\right]^{\mathrm{K}}=0,
\end{equation}
being $*$ the convolution operation. Hence, when one reformulates a Matsubara action in terms of classical and quantum fields (Keldysh rotation) in a non-equilibrium context, the action in Eq.\ref{eq:keldysh_action} acquires a specific matrix structure that reflects the system’s dynamics, causality, and statistical state: 
\begin{itemize}
    \item[\checkmark] \textbf{The cl-cl component vanishes:} for purely classical configurations ($\phi^{\mathrm{q}}=0$), the forward and backward time-contour contributions cancel. This implies
\begin{equation}
    S\left[\phi^{\mathrm{cl}}, 0\right]=0,
\end{equation}
    and reflects the fact that classical fields alone do not generate quantum dynamics.

    \item[\checkmark] \textbf{$\bm{\mathrm{cl}-\mathrm{q}}$ and $\bm{\mathrm{q}-\mathrm{cl}}$ components encode causality:} These off-diagonal terms are Hermitian conjugates and appear as retarded (lower triangular) and advanced (upper triangular) structures in the time domain. They ensure that responses occur only after perturbations (i.e., causal behavior), and their structure also prevents the cl-cl component from being renormalized, maintaining the exact identity above even when interactions are present.

     \item[\checkmark] \textbf{The $\bm{\mathrm{q}-\mathrm{q}}$ component controls fluctuations and statistics:} This term is anti-Hermitian and ensures the functional integral converges (the equivalent of the life-time of an excitation). In non-interacting systems, it is infinitesimally small in magnitude; however, when interactions or coupling to reservoirs are included, it becomes finite, encoding how the system is populated and how it exchanges energy or particles with its environment.
\end{itemize}

\noindent At this point, all the setup is the same as the Matsubara formalism after an analytical continuation. Then, to make the entire construction meaningful, one should introduce source fields in the form of external time-dependent potentials $V(t)$ which couple to observables $\hat{O}$ in interacting sourced Hamiltonians 
\begin{equation}
    H_V=V(t) \hat{O}.
    \label{eq:sourced_hamiltonian}
\end{equation}
For instance, an interaction that depends on the number of particles $H_V=V(t)\hat{n}=V(t) b^\dagger b$ will allow us to compute expectation values of number of photons $\left\langle n\right\rangle_{(t)}$ in a non-equilibrium context. As previously discussed, even though any classical external field $V(t)$ is the same on both branches of the contour, it is convenient to allow $V^{u}(t)$ and $V^{d}(t)$ to be distinct. Hence, we can consider the Keldysh rotation
\begin{equation}
    V^{\mathrm{cl}}(t)=\frac{1}{2}\left[V^{u}(t)+V^{d}(t)\right] ; \quad V^{\mathrm{q}}(t)=\frac{1}{2}\left[V^{u}(t)-V^{d}(t)\right],
\end{equation}
where $V^{u/d}(t)$ is the source potential on the forward (backward) branch of the contour. The key feature here is that if $V^{\mathrm{q}}=0$ the source potential is the same on the two branches, $V^{u}(t)=V^{d}(t)$, and thus the evolution operator brings the system exactly to its initial state, i.e. $\hat{U}_{\mathcal{C}}\left[V^{\mathrm{cl}}\right]=\hat{1}$. Therefore, one crucially needs the fictitious potential $V^{\mathrm{q}}(t)$ to generate observables out of equilibrium.

Having established the sourced Hamiltonian in Eq.\ref{eq:sourced_hamiltonian}, the sourced generating functional $Z[V]$ reads as  
\begin{equation}
    Z[V]=\int D[\bar{\phi}, \phi] \: \mathrm{e}^{i S[\bar{\phi}, \phi]+i S_V[\bar{\phi}, \phi]}
\end{equation}
where the equilibrium bare action $S[\bar{\phi}, \phi]$ is given by 
Eq.\ref{eq:keldysh_action} and the sourced $S_V[\bar{\phi},\phi]$ action is (for a source-density coupling)
\begin{equation}
\begin{split}
     S_V[\bar{\phi},\phi]&=-\int_\mathcal{C} dt \: V(t) \bar{\phi}(t) \phi(t)-\int_{-\infty}^{+\infty} \mathrm{d} t\left[V^{+} \bar{\phi}^{+} \phi^{+}-V^{-} \bar{\phi}^{-} \phi^{-}\right]\\
    &=-\int_{-\infty}^{+\infty} \mathrm{d} t\left[V^{\mathrm{cl}}\left(\bar{\phi}^{+} \phi^{+}-\bar{\phi}^{-} \phi^{-}\right)+V^{\mathrm{q}}\left(\bar{\phi}^{+} \phi^{+}+\bar{\phi}^{-} \phi^{-}\right)\right]\\
    &=-\int_{-\infty}^{+\infty} dt \:\left(\bar{\phi}^{\mathrm{cl}}, \bar{\phi}^{\mathrm{q}}\right)_t\left(\begin{array}{cc}V^\mathrm{q}(t) & V^{\mathrm{cl}}(t) \\ V^{\mathrm{cl}}(t) & V^{\mathrm{q}}(t)\end{array}\right)\binom{\phi^{\mathrm{cl}}}{\phi^{\mathrm{q}}}_{t} \: .
\end{split}
\end{equation}
With all this machinery, one is ready to study the non-equilibrium dynamics of observables by computing  connected/irreducible time-dependent correlators (cumulants) in the form of
\begin{equation}
    \left\langle (O-\expval{O})^n  \right\rangle(t)= \left(\frac{i}{2}\right)^n \: \left.\frac{\delta^n \ln Z[V^{\mathrm{cl}},V^{\mathrm{q}}]}{\delta\left[V^{\mathrm{q}}(t)\right]^n}\right|_{V=0}
\end{equation}
where, as usual, the average value is calculated from $\langle O(t)\rangle=(i / 2) \:\delta Z\left[V^{\mathrm{q}}\right] /\left.\delta V^{\mathrm{q}}(t)\right|_{V=0}$.
Here is where the power of the Keldysh technique becomes evident. Since any probability distribution is uniquely determined by its cumulants, then computing the cumulant expansion of observables gives us access to their probability densities. This is a non-trivial answer, since the Gibbs weight $e^{-\beta \hat{H}}$ is no longer true for a non-thermalized system, and therefore one may obtain non-Gibbs factors. Through this scope, we define the \textit{full counting statistics} of an operator $\hat{O}$ at time $t_0$ as the probability $\mathcal{P}(\lambda)$ which gives exactly 
\begin{equation}
\left\langle\hat{O}^k\left(t_0\right)\right\rangle=\int \mathrm{d}\lambda \: \lambda^k \: \mathcal{P}(\lambda) .
\end{equation}
and which has a $\eta$-parametrized generating function 
\begin{equation}
    Z[\eta]\equiv\int d\lambda \: e^{i\eta\lambda} \mathcal{P}(\lambda) = \sum_k \frac{(i\eta)^k}{k!} \left\langle\hat{O}^k(t_0)\right\rangle.
\end{equation} 
Indeed, the meaning of such a parameter $\eta$ is nothing but a particular
realization of the quantum source field $V^{\mathrm{q}}(t)$, tailored to generate the appropriate statistics. This is why $\eta$ is often called the counting field, and has the meaning of an auxiliary parameter introduced to keep track of the statistics over a given time interval of a particular observable, typically the number of particles transferred, energy exchanged, or charge current. 

This completes our introduction to the Keldysh formalism. Its flexibility makes it particularly well-suited for analyzing non-equilibrium phenomena, allowing direct access to real-time dynamics, fluctuation statistics, and the structure of quantum noise. Crucially, it remains robust even in the presence of interactions or time-dependent driving. With the formal groundwork now established, we proceed to apply it to the study of geometric quantum noise, where these tools reveal subtle and intrinsically quantum effects.

\subsubsection{Effective Action and Noisy Classical Equations}
Recall the system we want to study: a magnetic quantum dot tunnel-coupled to a non-magnetic metallic lead, where the dot hosts a large collective spin due to enhanced exchange interactions (i.e., close to a Stoner instability), and the coupling to the lead introduces dissipation and facilitates spin relaxation. Hence, we have a Hamiltonian of the form
\begin{equation}
    H=H_{\mathrm{dot}}+H_{\text {lead }}+H_{\mathrm{tun}}
\end{equation}
where
\begin{equation}
    H_{\mathrm{dot}}=\sum_{\alpha, \sigma} \epsilon_\alpha a_{\alpha, \sigma}^{\dagger} a_{\alpha, \sigma}-J \boldsymbol{S}^2+\boldsymbol{B}\cdot\boldsymbol{S} \:\: \Bigg| \:\:\: 
    H_{\text{lead}}=\sum_{\gamma, \sigma} \varepsilon_\gamma c_{\gamma, \sigma}^{\dagger} c_{\gamma, \sigma} \:\:\: \Bigg|\: 
    H_{\text{tun}}=\sum_{\alpha, \gamma, \sigma} V_{\alpha, \gamma} a_{\alpha, \sigma}^{\dagger} c_{\gamma, \sigma}+\text {h.c.}
\end{equation}
For this model, then, we have that the Keldysh generating functional $Z[\tilde{V}]$ is written as 
\begin{equation}
    Z[V]=\int D[\bar\Psi,\Psi]_{\text{dot}} \: D[\bar{\psi},\psi]_{\text{lead}} \: \: e^{iS[\bar\Psi,\Psi,\bar{\psi},\psi]+iS_{\tilde{V}}}
    \label{eq:contour_action_system}
\end{equation}
where the fictitious sourced $S_{\tilde{V}}$ action couples to non-equilibrium observables and the time translational invariant closed contour action (before Keldysh rotation) is
\begin{equation}
iS[\bar\Psi,\Psi,\bar{\psi},\psi]=iS_{\text{dot}}[\bar{\Psi},\Psi]+iS_{\text{lead}}[\bar{\psi},\psi] + iS_{\text{coupling}}[\bar\Psi,\Psi,\bar{\psi},\psi]
\end{equation}
with 
\begin{equation}
    \begin{split}
        S_{\text{dot}}[\bar{\Psi},\Psi]&=\oint_\mathcal{C} dt \Big[ \sum_{\alpha,\sigma} \bar{\Psi}_{\alpha\sigma}(i\partial_t-\epsilon_\alpha)\Psi_{\alpha\sigma}\:\:+ \: J\bm{S}^2 \: - \bm{B}\cdot \bm S    \Big] \\
        S_{\text{lead}}[\bar{\psi},\psi]&=\oint_\mathcal{C} dt \Big[ \sum_{\gamma,\sigma} \bar\psi_{\gamma\sigma}(i\partial_t-\varepsilon_\gamma)\psi_{\gamma\sigma}\Big]
        \\
        S_{\text{coupling}}[\bar\Psi,\Psi,\bar{\psi},\psi]&=\oint_\mathcal{C} dt \Big[- \sum_{\alpha,\gamma,\sigma} \left\{V_{\alpha\gamma}\bar\Psi_{\alpha\sigma}\psi_{\gamma\sigma}+V^{*}_{\alpha\gamma}\bar\psi_{\gamma\sigma}\Psi_{\alpha\sigma}\right\}\Big]
    \end{split}
\end{equation}
Now, since the term $\bm{S}^2$ stands for a quartic interaction in fermionic dot fields, we consider a \textbf{Hubbard-Stratonovich (mean-field) transformation} of the form
\begin{equation}
    (\text{ctn})\times  e^{i\frac{\lambda^2}{2}(\bar\psi\bm A\psi)^2}=\int D\bm\phi \: e^{- i\left(\frac{|\bm\phi|^2}{2} + \lambda \bm\phi \cdot \bar\psi\bm A\psi\right)},
\end{equation}
giving rise to
\begin{equation}
   e^{iJ \bm S^2}=e^{i\frac{J/2}{2}(\sum_\alpha\bar\Psi_\alpha\bm{\sigma} \Psi_\alpha)^2}= {  2\sqrt{\frac{J}{2}}}\int D\bm\phi \:  e^{- i\left(\frac{|\bm\phi|^2}{2} + \sqrt{\frac{J}{2}} \bm\phi \cdot \sum_\alpha\bar\Psi_\alpha\bm \sigma\Psi_\alpha\right)},
\end{equation}
and, with $\bm{\mathcal{M}}=2\sqrt{\frac{J}{2}}\bm\phi$, to
\begin{equation}
    \boxed{
    e^{iJ \bm S^2}=\int D\bm{\mathcal{M}} \:\:  e^{- i\left(\frac{|\bm{\mathcal{M}}|^2}{4J} +  \sum_\alpha\bar\Psi_\alpha \frac{\boldsymbol{\mathcal{M}}\cdot\bm{\sigma}}{2}\Psi_\alpha\right)}}
\end{equation}
 where $\bm{\mathcal{M}}$ plays the role of the bosonic magnetization field. Under this transformation, the dot action is now written as 
\begin{equation}
\begin{split}
    S_{\text{dot}}[\bar{\Psi},\Psi]&=\oint_\mathcal{C} dt \Big[ \sum_{\alpha} \bar{\Psi}_{\alpha}(i\partial_t-\epsilon_\alpha)\Psi_{\alpha}\:- \:\frac{1}{2} \sum_\alpha\bar\Psi_\alpha \boldsymbol{\mathcal{M}}\cdot\bm{\sigma}\Psi_\alpha \: - \bm{B}\cdot\frac{1}{2}\sum_\alpha\bar\Psi_\alpha \bm{\sigma}\Psi_\alpha    \Big]\\
    &=\oint_\mathcal{C} dt \Big[ \sum_{\alpha} \bar{\Psi}_{\alpha}\left(i\partial_t-\epsilon_\alpha-\frac{(\bm{\mathcal{M}}+\bm{B})\cdot\bm{\sigma}}{2}\right)\Psi_{\alpha}    \Big]\:,
\end{split}
\end{equation}
with the understanding that each fermionic field is a spinor of the form
\begin{equation}
    \bar\Psi_\alpha=\left(\bar{\Psi}_\alpha ^{\uparrow}\: \:,\:\bar{\Psi}_\alpha ^{\downarrow}\right) \quad ; \quad  
    \Psi_\alpha=\binom{\Psi_\alpha ^{\uparrow}}{\Psi_\alpha ^{\downarrow}}.
\end{equation}
In this line, by constructing the combined dot-lead vector representation $(\Psi_\alpha \:,\psi_\gamma )^T$, we write the non-sourced closed contour action (Eq.\ref{eq:contour_action_system}) as
\begin{equation}
    iS=i \oint_{\mathcal{C}} d t\left[\sum_{\alpha, \gamma}\left(\bar{\Psi}_\alpha, \bar{\psi}_\gamma\right)
    \left(\begin{array}{cc}i \partial_t-\epsilon_\alpha-\frac{(\bm{\mathcal{M}}+\bm{B})\cdot\bm{\sigma}}{2} & -V_{\alpha \gamma} \\ -V_{\alpha \gamma}^{*} & i \partial_t-\varepsilon_\gamma\end{array}\right)\binom{\Psi_\alpha}{\psi_\gamma}\right] - i\oint_{\mathcal{C}} d t\frac{|\bm{\mathcal{M}}|^2}{4J}.
\end{equation}
To make progress, we use the fact 
\begin{equation}
    \operatorname{tr} \ln \left(\begin{array}{ll}A & B \\ C & D\end{array}\right)=\operatorname{tr} \ln D+\operatorname{tr} \ln \left(A-B D^{-1} C\right)
    \label{eq:tr_ln_identity}
\end{equation}
\vspace{-0.35cm}
\begin{mdframed}[
  linewidth=0.5pt,
  roundcorner=5pt,
  innerleftmargin=10pt,
  innerrightmargin=10pt,
  innertopmargin=5pt,
  innerbottommargin=5pt,
  linecolor=black,
  tikzsetting={dash pattern=on 10pt off 2pt}
]
\textbf{\underline{(Proof)}} \:  We consider the matrix multiplication (assuming $D$ is invertible):
$$ \begin{pmatrix} A & B \\ C & D \end{pmatrix} \begin{pmatrix} \mathbb{1} & 0 \\ -D^{-1}C & \mathbb{1} \end{pmatrix} =  \begin{pmatrix} A - BD^{-1}C & B \\ 0 & D \end{pmatrix} $$
Then, by taking $\operatorname{ln}\:\operatorname{det}$, we have
$$ \operatorname{ln} \left[ \operatorname{det}\begin{pmatrix} A & B \\ C & D \end{pmatrix} \operatorname{det}\begin{pmatrix} \mathbb{1} & 0 \\ -D^{-1}C & \mathbb{1} \end{pmatrix} \right] = \ln \big[  \det(A - BD^{-1}C) \det D \big] . $$
and, after applying properties of the logarithm, we show ($\ln \det = \operatorname{tr \ln }$)
$$ \ln \det \left(\begin{array}{ll}A & B \\ C & D\end{array}\right)=\ln \det D+\ln \det \left(A-B D^{-1} C\right) $$
\end{mdframed}
to write the effective action for the bosonic field $\bm{\mathcal{M}}$, after integration over the dot-lead fermionic degrees of freedom, as
\begin{equation}
\begin{split}
    iS_{\text{eff}}&=\operatorname{tr} \ln \left[\left(\begin{array}{cc}G_{\text {dot }}^{-1} & -\hat{V} \\ -\hat{V}^{\dagger} & G_{\text {lead }}^{-1}\end{array}\right)\right]-i \oint_{\mathcal{C}} d t \frac{|\mathcal{M}|^2}{4 J}\\
    &=\operatorname{tr} \ln \left[G_{\text {lead }}^{-1}\right]+\operatorname{tr} \ln \left[G_{\text {dot }}^{-1}-V G_{\text {lead }}V^\dagger\right]-i \oint_{\mathcal{C}} d t \frac{|\mathcal{M}|^2}{4 J}.
\end{split}
\label{eq:wave_hand_1}
\end{equation}
Note that here we have defined 
\begin{equation*}
    G_{\text {dot }}^{-1}=i \partial_t-\epsilon_\alpha-\frac{(\bm{\mathcal{M}}+\bm{B})\cdot\bm{\sigma}}{2}\quad \quad ; \quad  G_{\text {lead }}^{-1}=i \partial_t-\varepsilon_\gamma \quad \: ; \quad \hat{V}=(V_{\alpha\gamma})_{\alpha\gamma}.
\end{equation*}
Furthermore, we will now redefine the magnetization $\bm{\mathcal{M}}$, to absorb the external $\bm B$ in $\bm M=\bm{\mathcal{M}}+\bm B$, and consider the self-energy $\Sigma$ term as $\Sigma=V G_{\text {lead }} V^\dagger$ . Consequently, the final form of the action reads as 
\begin{equation}
    \boxed{\: 
    iS_{\text{eff}}=\operatorname{tr} \ln \left[G_{\text {dot }}^{-1}-\Sigma\right]-i \oint_{\mathcal{C}} d t\left[\frac{M^2}{4 J}-\frac{\boldsymbol{B}\cdot \boldsymbol{M}}{2 J}\right]\:}\:,
    \label{eq:wave_hand2}
\end{equation}
where the $\operatorname{tr} \ln \left[G_{\text {lead }}^{-1}\right]$ and $B^2$ terms have been ignored just because they do not bring any new to the physics and are not relevant for our discussion. 

 At this point, the procedure will look very similar to what we did in deriving the Berry phase: consider an adiabatic evolution approximation and introduce a rotating frame via unitary operators $U$. Let's go with the latter first so we can obtain a general expression, which can then be the subject of approximations.
\begin{itemize}
    \item \textbf{\underline{Rotating Frame}}\\
    As the first step, let's parametrize the magnetization field $\bm{M}(t)$ in terms of its magnitude and direction  $\bm{M}(t)=M(t) \mathbf{n} (t)$. Then, in the same spirit of Eq.\ref{eq:conjug_SU},\ref{eq:rot_frame2},\ref{eq:rot_frame3}, we know that we can write any direction field $\mathbf{n}(t)$ as
    \begin{equation*}
        \begin{aligned}
            &\mathbf{n}(t)=(\sin \theta \cos \phi, \sin \theta \sin \phi, \cos \theta)^T{(t)}=\hat{R}(t)e_z \qquad \:\:\:\:\Big[\hat{R}\in SO(3)\Big]\:\:,\\
            \text{which implies}\:&\\
            &\mathbf{n} \cdot \bm{\sigma}=U_R\: \sigma_z \:U^{-1}_R \qquad \quad \:\Big[\text{with }\:\: U_R=e^{-i \phi \hat{S}_3} e^{-i \theta \hat{S}_2} e^{-i \psi \hat{S}_3}\in SU(2)\Big]\:\:.
        \end{aligned}
    \end{equation*}
    Then, by also rotating the spin-fermionic degrees of freedom and using the invariance of the trace, we have
    \begin{equation}
    \begin{aligned}
        \operatorname{tr} \ln \left[U^{-1}(G_{\text {dot }}^{-1}-\Sigma)U\right]&= \operatorname{tr} \ln \left[U^{-1}\left(i \partial_t-\epsilon_\alpha-\frac{M}{2}U\sigma_zU^{-1}-\Sigma\right)U\right]\\
        &=\operatorname{tr} \ln \left[U^{-1}(i \partial_t) U-\epsilon_\alpha-\frac{M}{2}\sigma_z-U^{-1}\Sigma U\right]\\
        \operatorname{tr} \ln \left[G_{\text {dot }}^{-1}-\Sigma\right]&=\operatorname{tr} \ln \left[i \partial_t+iU^{-1} \dot{U}-\epsilon_\alpha-\frac{M}{2}\sigma_z-U^{-1}(V G_{\text {lead }} V^\dagger) U\right]\:\:.
    \end{aligned}
    \end{equation}
    Now, let's define $Q=-iU^{-1} \dot{U}$, such that we can write compactly
    \begin{equation}
        \begin{aligned}
             \operatorname{tr} \ln \left[G_{\text {dot }}^{-1}-\Sigma\right]&=\operatorname{tr} \ln \left[i \partial_t-\epsilon_\alpha-\frac{M}{2}\sigma_z-(-iU^{-1} \dot{U})-U^{-1}\Sigma U\right]\\
             &=\operatorname{tr} \ln \left[ G_{\text {dot,z }}^{-1}-Q-U^{-1}\Sigma U\right]\:\:.
        \end{aligned}
    \end{equation}
    The main problem here is to compute the derivative $\dot{U}$ in a certain gauge $\psi$. Recall that we have already discussed that the factor $e^{-i\psi \hat{S}_3}\in U(1)$ is a gauge degree of freedom that depends on the selection of $\psi\in \mathbb{R}$. Moreover, when we have $U(t)$ parameterizing $\mathbf{n}(t)$, we can think of changing the gauge frame with time, which means $\psi(t)$. Hence, at different times we can have different gauges $\psi(t_1)$ and $\psi(t_2)$. Hence, for the following calculations its convenient to choose that the gauge field $\psi(t)$ is coupled to the angle $\phi(t)$, such that all the freedom is carried by a new vector field $\chi(t)$ as
    \begin{equation}
        \psi(t)=\chi(t)-\phi(t)\:\:.
    \end{equation}
    Hence, a long calculation ends up with 
    \begin{equation}
    \begin{aligned}
        Q=&-i\left(e^{-i \phi \sigma_z/2} e^{-i \theta \sigma_y/2} e^{-i (\chi-\phi) \sigma_z/2}\right)^{-1} \partial_t \left(e^{-i \phi \sigma_z/2} e^{-i \theta \sigma_y/2} e^{-i (\chi-\phi) \sigma_z/2}\right)\\
        =& \underbrace{\left[\dot{\phi}\left(1-\cos \theta\right)-\dot{\chi}\right] \frac{\sigma_z}{2}}_{\text{diagonal in z-basis}} \:\: -\underbrace{\:\frac{1}{2} e^{i \chi \sigma_z}\left[\dot{\theta} \sigma_y-\dot{\phi} \sin \theta \sigma_x\right] e^{i \phi \sigma_z}}_{\text{combine }\ket{\uparrow}, \ket{\downarrow}} \\
        =&: \hspace{1.5cm}Q_{||} \hspace{1.4cm}+ \hspace{1.9cm} Q_{\perp}\:\:\:,
    \end{aligned}
    \label{eq:computing_Q}
    \end{equation}
    where we have identified the two contributions $Q_{||}$ and $Q_{\perp}$ from the expressions above. In conclusion, we have
\[
\boxed{\:
\begin{aligned}
&\text{ Effective action (before Keldysh rotation - i.e. same as thermal PI)}\\
\:iS_{\text{eff}}&=\operatorname{tr} \ln \left[ G_{\text {dot,z }}^{-1}-Q_{||}-Q_{\perp}-U^{-1}\Sigma U\right]-i \oint_{\mathcal{C}} d t\left[\frac{M^2}{4 J}-\frac{\boldsymbol{B}\cdot \boldsymbol{M}}{2 J}\right]\: \\
 &\text{ where}\\
 &\qquad\qquad \bullet \quad G_{\text {dot,z }}^{-1}=i \partial_t-\epsilon_\alpha-\frac{M}{2}\sigma_z\\
 &\qquad\qquad \bullet \quad Q_{||}=\left[\dot{\phi}\left(1-\cos \theta\right)-\dot{\chi}\right] \frac{\sigma_z}{2}\\
 &\qquad\qquad \bullet \quad Q_{\perp}=-\frac{1}{2} e^{i \chi \sigma_z}\left[\dot{\theta} \sigma_y-\dot{\phi} \sin \theta \sigma_x\right] e^{i \phi \sigma_z}\\
 &\qquad\qquad \bullet \quad \Sigma_{\alpha \sigma, \alpha^{\prime} \sigma^{\prime}}\left(t, t^{\prime}\right)=\delta_{\sigma \sigma^{\prime}} \sum_\gamma V_{\alpha \gamma} G_{\text {lead}, \gamma}\left(t, t^{\prime}\right) V_{\alpha^{\prime} \gamma}^* \\
 &\text{Note that even though } \:\Sigma \text{ is diagonal in spin space, we keep } \:U^{-1}\Sigma U\:\:.
\end{aligned}
\:}
\]
    \item \textbf{\underline{Keldysh rotation}}\\
    The previous procedure was carried out formally as in the equilibrium path integral. However, in the closed-time-contour (Keldysh) formalism, every bosonic collective field
    $$
        M(t),\qquad \theta(t),\qquad \phi(t),\qquad \chi(t)
    $$
    must be doubled, since it lives independently on the forward and backward branches of the contour:
    $$
        M^u,M^d,\qquad \theta^u,\theta^d,\qquad \phi^u,\phi^d,\qquad \chi^u,\chi^d.
    $$
    Now, for any of these bosonic contour fields $X\in\{M,\theta,\phi,\chi\}$ we define their classical and quantum components as
    \begin{equation}
        X_c=\frac{X_u+X_d}{2},\qquad X_q=X_u-X_d.
    \end{equation}
    In this line, before the Keldysh rotation, any contour kernel $D$ is represented in the $(u,d)$ fermionic basis by the diagonal matrix
    \begin{equation}
    S=\int d t\left(\bar{\psi}_u, \bar{\psi}_d\right)\left(\begin{array}{cc}
D\left[X_u\right] & 0 \\
0 & -D\left[X_d\right]
\end{array}\right)\binom{\psi_u}{\psi_d} \quad \Rightarrow\quad 
        \hat D_{ud}=
        \begin{pmatrix}
            D_u & 0\\
            0 & -D_d
        \end{pmatrix}\:.
    \end{equation}
    Now, in order to rotate from $(u,d)$ fermionic basis to the $(c,q)$ fermionic basis, one needs to consider the standard Keldysh rotation matrix
    \begin{equation}
        L=\frac{1}{\sqrt 2}
        \begin{pmatrix}
            1&1\\
            1&-1
        \end{pmatrix},
        \qquad L^{-1}=L,
    \label{eq:def_L_rotation}
    \end{equation}
    such that in the basis $(\psi_c,\psi_q )^{T}$, the integral kernels (propagators) acquire the form
    \begin{equation}
        \tilde D=L\hat D_{ud}L^{-1}
        =D_c \tau_x+\frac{D_q}{2} \tau_0
        =
        \begin{pmatrix}
            D_q/2 & D_c \\
            D_c & D_q/2
        \end{pmatrix}_{K}.
    \end{equation}
     where $\set{\tau_{0,x,y,z}}$ are the Pauli matrices (+identity) in Keldysh space. To be completely clear, let's apply the previous rotation to each object entering the effective action.
    \begin{enumerate}
        \item[(i)] \underline{\textit{The dot inverse propagator }$G^{-1}_{\text{dot},z}$}\hfill\\ 
        \vspace{-0.1cm}
        
        Before the Keldysh rotation,
    \begin{equation}
        \hat G^{-1}_{\text{dot},z;ud}
        =
        \begin{pmatrix}
            i\partial_t-\epsilon_\alpha-\dfrac{M_u}{2}\sigma_z & 0\\[1mm]
            0 & -i\partial_t+\epsilon_\alpha+\dfrac{M_d}{2}\sigma_z
        \end{pmatrix}.
    \end{equation}
    but rotating to the $(c,q)$ basis gives
    \begin{equation*}
\begin{aligned}
\tilde G^{-1}_{\text{dot},z}
&=
L\hat (G^{-1}_{\text{dot},z;ud})L^{-1}
\\[10pt]
&=
\left(i \partial_t-\epsilon_\alpha-\frac{M_c}{2} \sigma_z\right) \tau_x
-\frac{M_q}{4} \sigma_z \tau_0 .
\end{aligned}
\end{equation*}

\begin{equation}
\boxed{
\tilde G^{-1}_{\text{dot},z}
=
\begin{pmatrix}
-\frac{M_q}{4} \sigma_z & i \partial_t-\epsilon_\alpha-\frac{M_c}{2} \sigma_z \\[6pt]
i \partial_t-\epsilon_\alpha-\frac{M_c}{2} \sigma_z & -\frac{M_q}{4} \sigma_z
\end{pmatrix}_K
}
\end{equation}
    \vspace{0.1cm}
    \item[(ii)] \underline{\textit{The Berry/gauge term } $Q=Q_{||}+Q_\perp$}\hfill\\ 
\vspace{-0.1cm}

Since $Q$ is a bosonic functional of the angles $(\theta,\phi,\chi)$, it is also doubled on the contour. However, because it enters the \emph{quadratic contour kernel} of the fermionic action, the lower branch appears with the usual minus sign coming from the reversed orientation of the backward contour. Therefore, the correct object in the $(u,d)$ basis is
\begin{equation}
    \hat Q_{ud}=
    \begin{pmatrix}
        Q_u & 0\\
        0 & -Q_d
    \end{pmatrix},
    \qquad
    Q_u\equiv Q[\theta_u,\phi_u,\chi_u],\quad
    Q_d\equiv Q[\theta_d,\phi_d,\chi_d].
\end{equation}
Introducing the classical and quantum components
\begin{equation}
    Q_c=\frac{Q_u+Q_d}{2},
    \qquad
    Q_q=Q_u-Q_d,
\end{equation}
one may rewrite the previous matrix as
\begin{equation}
    \hat Q_{ud}
    =
    \begin{pmatrix}
        Q_c+\dfrac{Q_q}{2} & 0\\[2mm]
        0 & -Q_c+\dfrac{Q_q}{2}
    \end{pmatrix}
    =
    Q_c\,\tau_z+\frac{Q_q}{2}\,\tau_0.
\end{equation}
And after Keldysh rotation, the matrix becomes
\begin{equation}
    \tilde Q
    =L\hat Q_{ud}L^{-1}
    =Q_c\tau_x+\frac{Q_q}{2}\tau_0
    =
    \begin{pmatrix}
        Q_q/2 & Q_c\\
        Q_c & Q_q/2
    \end{pmatrix}_{K}.
\end{equation}
Decomposing further into the longitudinal and transverse components, one has
\begin{equation} 
    \tilde Q=\tilde Q_{||}+\tilde Q_\perp, \qquad \text{with } \quad
    \begin{cases}
        \tilde Q_{||}
        =
        Q_{c,||}\tau_x+\dfrac{Q_{q,||}}{2}\tau_0,
        \\[2mm]
        \tilde Q_{\perp}
        =
        Q_{c,\perp}\tau_x+\dfrac{Q_{q,\perp}}{2}\tau_0
    \end{cases}\:\:,
\end{equation}
where
\begin{equation}
    Q_{c,||}=\frac{Q_{u,||}+Q_{d,||}}{2},
    \qquad
    Q_{q,||}=Q_{u,||}-Q_{d,||},
\end{equation}
\begin{equation}
    Q_{c,\perp}=\frac{Q_{u,\perp}+Q_{d,\perp}}{2},
    \qquad
    Q_{q,\perp}=Q_{u,\perp}-Q_{d,\perp}.
\end{equation}
Hence, explicitly,
\begin{equation}
    \boxed{\tilde Q=
    \begin{pmatrix}
        \dfrac{1}{2}(Q_{q,||}+Q_{q,\perp}) & Q_{c,||}+Q_{c,\perp}\\[2mm]
        Q_{c,||}+Q_{c,\perp} & \dfrac{1}{2}(Q_{q,||}+Q_{q,\perp})
    \end{pmatrix}_{K}}\:\:.
\end{equation}

    \item[(iii)] \underline{\textit{The rotation matrix } $U$}\hfill\\ 
        \vspace{-0.1cm}
        
    The spin rotation itself is also branch-dependent on the contour, but since it is a contour field (rather than a quadratic kernel in the action), it does \emph{not} acquire the extra minus sign on the backward branch. Therefore,
\begin{equation}
    \hat U_{ud}=
    \begin{pmatrix}
        U_u & 0\\
        0 & U_d
    \end{pmatrix},
    \qquad
    \hat U^{-1}_{ud}=
    \begin{pmatrix}
        U_u^{-1} & 0\\
        0 & U_d^{-1}
    \end{pmatrix}.
\end{equation}
Defining
\begin{equation}
    U_c=\frac{U_u+U_d}{2},\qquad U_q=U_u-U_d,
\end{equation}
one obtains
\begin{equation}
    \boxed{\tilde U=L\hat U_{ud}L^{-1}
    =U_c\tau_0+\frac{U_q}{2}\tau_x
    =
    \begin{pmatrix}
        U_c & U_q/2\\
        U_q/2 & U_c
    \end{pmatrix}_{K}}\:\:,
\end{equation}
and similarly,
\begin{equation}
    \boxed{\tilde{U}^{-1}
    =
    (U^{-1})_c\tau_0+\frac{(U^{-1})_q}{2}\tau_x
    =
    \begin{pmatrix}
        (U^{-1})_c & (U^{-1})_q/2\\
        (U^{-1})_q/2 & (U^{-1})_c
    \end{pmatrix}_{K}}\:\:.
\end{equation}
    
\item[(iv)] \underline{\textit{The lead self-energy} $\Sigma=V G_{\text {lead }} V^\dagger$}\hfill\\ 
        \vspace{-0.1cm}

Different from the objects considered above, the Keldysh structure of the self-energy is more subtle. Therefore, to correctly perform the Keldysh rotation, it is essential to carefully revisit the structure of Eq.\ref{eq:wave_hand_1},\ref{eq:wave_hand2}. Let's return to the object
\begin{equation}
\operatorname{tr} \ln \left[\left(\begin{array}{cc}
G_{\text {dot }}^{-1} & -\hat{V} \\
-\hat{V}^{\dagger} & G_{\text {lead }}^{-1}
\end{array}\right)\right].
\end{equation}
It is important to emphasize that this $2\times 2$ matrix structure lives \emph{only} in the dot/lead sector. Each entry is itself an operator acting on the full space
\begin{equation}
\mathcal{H}_{\mathrm{dot/lead}}
\otimes 
\mathcal{H}_{\mathrm{orbital}}
\otimes 
\mathcal{H}_{\mathrm{spin}}
\otimes 
\mathcal{H}_{\mathrm{time\ contour}}.
\end{equation}

In particular, the inverse lead Green operator $G^{-1}_{\text{lead}}$ is a contour-local differential operator. Writing it explicitly in the $(u,d)$ contour basis, one has
\begin{equation}
\hat{G}_{\text {lead},ud}^{-1}(t,t')
=
\begin{pmatrix}
(i\partial_t-\varepsilon)\,\delta(t-t') & 0 \\
0 & -(i\partial_t-\varepsilon)\,\delta(t-t')
\end{pmatrix}.
\end{equation}
The appearance of the delta function reflects the fact that $G^{-1}$ is \emph{local on the contour}, while the relative minus sign originates from the reversed orientation of the backward branch. Now, when applying the $\operatorname{tr}\ln$ identity (Eq.\ref{eq:tr_ln_identity}), what is implicitly performed is a factorization of determinants in the dot/lead sector:
\begin{equation}
\begin{aligned}
    \text{det}_{\text{dot}\oplus\text{lead}}(\mathbb{M})
\:\:\:&=\:\:\:
\text{det}_{\text{lead}}(G_{\text{lead}}^{-1})
\:\:\cdot\:\:
\text{det}_{\text{dot}}(G_{\text{dot}}^{-1}-VG_{\text{lead}}V^{\dagger})\\[2pt]
\Rightarrow\quad\ln\text{det}_{\text{dot}\oplus\text{lead}}(\mathbb{M})
\:\:\:&=\:\:\:
\ln\text{det}_{\text{lead}}(G_{\text{lead}}^{-1})
\:\:+\:\:
\ln \text{det}_{\text{dot}}(G_{\text{dot}}^{-1}-\Sigma)
\end{aligned}
\end{equation}
and therefore, in the exponential (i.e., action)
\begin{equation}
\operatorname{tr}_{\text{dot}\oplus\text{lead}}\ln(\mathbb{M})
=
\operatorname{tr}_{\text{lead}}\ln(G_{\text{lead}}^{-1})
+
\operatorname{tr}_{\text{dot}}\ln(G_{\text{dot}}^{-1}-\Sigma).
\end{equation}

At this point, the self-energy is defined as
\begin{equation}
\Sigma = \hat V\, G_{\text{lead}}\, \hat V^\dagger,
\end{equation}
which involves the \emph{inverse} of the operator $G_{\text{lead}}^{-1}$. This is the crucial difference with the previous quantities: while $G^{-1}$ is local and branch-diagonal, its inverse is not.

More precisely, given the contour differential operator
\begin{equation}
G_{\text{lead},\mathcal C}^{-1}\left(t, t_1\right)=\left(i \partial_t-\varepsilon\right) \:\:\delta_{\mathcal{C}}\left(t-t_1\right)
\end{equation}
its inverse is the contour Green function $G_{\text{lead},\mathcal C}(t,t')$, defined by
\begin{equation}
\begin{aligned}
    \int_{\mathcal C} dt_1\:
G_{\text{lead},\mathcal C}^{-1}(t,t_1)\,
G_{\text{lead},\mathcal C}(t_1,t')&=\delta_{\mathcal C}(t-t')\\
\int_{\mathcal{C}} d t_1\:(i \partial_t-\varepsilon) \delta\left(t-t_1\right) \:G_{\text{lead},\mathcal{C}}\left(t_1, t^{\prime}\right)&=\\
(i \partial_t-\varepsilon)\left[\int_{\mathcal{C}} d t_1 \delta\left(t-t_1\right) \:G_{\mathcal{C}}\left(t_1, t^{\prime}\right)\right]&=\\[8pt]
 \Rightarrow\:\:\left(i \partial_t-\varepsilon\right) \:\:G_{\text{lead},\mathcal{C}}\left(t, t^{\prime}\right)=\delta\left(t-t^{\prime}\right)  \\[6pt]  \xrightarrow[]{\text{solution}}\:\: G_{\text{lead},\mathcal{C}}\left(t, t^{\prime}\right)=-i\:\Big\langle T_{\mathcal{C}}\:[\psi(t) \psi^{\dagger}(t^{\prime})]\Big\rangle&
\end{aligned}
\end{equation}
Hence, the inverse of the quadratic lead integral kernel is the contour-ordered propagator of the lead fermionic fields. Importantly, even though the inverse operator $G^{-1}_{\mathcal C}$ is local in time and diagonal in the contour branch space, its inverse $G_{\mathcal C}(t,t')$ is a nonlocal object that encodes the propagation of fermions along the entire Keldysh contour. Specifically, since $t,t'$ can be in the (up,down) branches, and every point on the lower one is later on the contour than every point on the upper one, we can write the 
\begin{equation}
    G_{\text{lead};u d}^{(t, t')}=-i\left(\begin{array}{cc}
\Big\langle T [\psi(t) \psi^{\dagger}(t^{\prime})]\Big\rangle & -\Big\langle\psi^{\dagger}(t') \psi(t)\Big\rangle \\
\Big\langle\psi(t) \psi^{\dagger}(t')\Big\rangle & \Big\langle\tilde{T} [\psi(t) \psi^{\dagger}(t')]\Big\rangle
\end{array}\right)\equiv \left(\begin{array}{ll}
G^T & G^{<} \\
G^{>} & G^{\tilde{T}}
\end{array}\right)
\end{equation}
where $T$ is the usual time-ordered operator applied in the upper branch, and $\tilde{T}$ is the anti-time-ordered operator applied in the lower branch. We note then that the off-diagonal components are generically nonzero, since they originate from contour ordering between different branches and reflect the statistical occupation of the lead. In particular, for a non-interacting fermionic level $\varepsilon$, one finds explicitly
\begin{equation}
\begin{aligned}
G^{<}(t,t') &= i\, f(\varepsilon)\, e^{-i\varepsilon (t-t')}\\
G^{>}(t,t') &= -i\,[1-f(\varepsilon)]\, e^{-i\varepsilon (t-t')},
\end{aligned}
\end{equation}
where $f(\varepsilon)$ is the Fermi-Dirac distribution. Having determined the inverse of $G^{-1}_{\text{lead}}$, namely the contour Green function $G_{\text{lead}}$, we can now perform the Keldysh rotation. Applying the transformation to the $(u,d)$ matrix yields
\begin{equation}
\tilde{G}_{\text{lead}}
=
L (G_{\text{lead};u d}^{(t, t')}) L^{-1}
=
\begin{pmatrix}
G^K_{\text{lead}} & G^R_{\text{lead}} \\
G^A_{\text{lead}} & 0
\end{pmatrix},
\end{equation}
where the components are given by
\begin{equation}\left\{
\begin{aligned}
 iG^R_{\text{lead}} (t,t') &= i(G^T - G^{<})=\Big\langle T [\psi(t) \psi^{\dagger}(t^{\prime})]\Big\rangle+\, f(\varepsilon)\, e^{-i\varepsilon (t-t')}\\
iG^A_{\text{lead}} (t,t') &= i(G^T - G^{>})=\Big\langle T [\psi(t) \psi^{\dagger}(t^{\prime})]\Big\rangle-\,[1-f(\varepsilon)]\, e^{-i\varepsilon (t-t')}\\
iG^K_{\text{lead}} (t,t') &= i(G^{>} + G^{<})=\:\:[1-2f(\varepsilon)]\, e^{-i\varepsilon (t-t')}.
\end{aligned}\right.
\label{eq:green_function_keldysh_lead}
\end{equation}
In this basis, the Green function is naturally decomposed into response functions ($G^{R,A}$) and the fluctuation component ($G^K$), which will directly determine the dissipative and noisy contributions to the effective action.

Finally, since the tunneling amplitude $V$ does not carry any nontrivial structure in contour space, its representation in the $(u,d)$ basis is simply proportional to the identity in Keldysh space, $\hat V_{ud}=V\tau_0$, and remains unchanged under the Keldysh rotation. Consequently, all nontrivial Keldysh structure of the self-energy $\Sigma=V G_{\text{lead}} V^\dagger$ originates entirely from the Green function $G_{\text{lead}}$. This leaves us with the expression for the Keldysh rotated self-energy term as
\begin{equation}
    \boxed{\tilde{\Sigma}(t,t')=V \big[\tilde{G}_{\text{lead}}(t,t')\big] V^{\dagger}=V\begin{pmatrix}
G^K_{\text{lead}} & G^R_{\text{lead}} \\[3pt]
G^A_{\text{lead}} & 0
\end{pmatrix} V^{\dagger}\equiv \left(\begin{array}{cc}
\Sigma^K_{(t,t')} & \Sigma^R_{(t,t')} \\[3pt]
\Sigma^A_{(t,t')} & 0
\end{array}\right)}\:.
\end{equation}

\item[(iv)] \underline{\textit{Pure Bosonic term} $\oint_{\mathcal{C}} d t\left[M^2-2\boldsymbol{B}\cdot \boldsymbol{M}\right]$}\hfill\\ 
        \vspace{-0.1cm}

Since the contour integral splits as
\begin{equation}
    \oint_{\mathcal{C}} d t=\int d t_u-\int d t_d
\end{equation}
we have that
\begin{equation*}
    \begin{aligned}
        i \oint_{\mathcal{C}} d t\left[\frac{M^2}{4 J}-\frac{\mathbf{B} \cdot \mathbf{M}}{2 J}\right]&=i \int d t\left[\frac{M_u^2}{4 J}-\frac{\mathbf{B} \cdot \mathbf{M}_u}{2 J}\right]-i \int d t\left[\frac{M_d^2}{4 J}-\frac{\mathbf{B} \cdot \mathbf{M}_d}{2 J}\right]\\
        &=i \int d t\left[\frac{M_u^2-M_d^2}{4 J}-\frac{\mathbf{B} \cdot \mathbf{M}_u-\mathbf{B} \cdot \mathbf{M}_d}{2 J}\right]\:\:.
    \end{aligned}
\end{equation*}
Computing $M_u^2-M_d^2=\left(M_c+M_q / 2\right)^2-\left(M_c-M_q / 2\right)^2=2 M_c M_q$, we get the final result
\begin{equation}
    \boxed{i \oint_{\mathcal{C}} d t\left[\frac{M^2}{4 J}-\frac{\mathbf{B} \cdot \mathbf{M}}{2 J}\right]=i \int d t\left[\frac{M_c M_q}{2 J}-\frac{\mathbf{B} \cdot \mathbf{M}_q}{2 J}\right]}
\end{equation}  
\end{enumerate}
\vspace{5pt}
With all these new expressions, we summarize the generalization to the Keldysh space as
\begin{equation}
\boxed{\:
\begin{aligned}
&\textbf{ Effective action (after Keldysh rotation)}\\
\quad &\qquad i\mathcal{S}_{\text{eff}}
=
\operatorname{tr} \ln \left[
\tilde{G}_{\mathrm{dot}, z}^{-1}
-\tilde{Q}
-\tilde{U}^{-1} \tilde{\Sigma} \tilde{U}
\right]
+i \int d t \,\frac{\boldsymbol{B} \cdot\boldsymbol{M}_q}{2 J}
-i \int d t \,\frac{M_c M_q}{2 J}
\\[6pt]
&\text{ where}\\
&\qquad\qquad \bullet \quad 
\tilde G^{-1}_{\text{dot},z}
=
\begin{pmatrix}
-\dfrac{M_q}{4} \sigma_z 
&
i \partial_t-\epsilon_\alpha-\dfrac{M_c}{2} \sigma_z \\[6pt]
i \partial_t-\epsilon_\alpha-\dfrac{M_c}{2} \sigma_z 
&
-\dfrac{M_q}{4} \sigma_z
\end{pmatrix}_K
\\[12pt]
&\qquad\qquad \bullet \quad 
\tilde Q=
\begin{pmatrix}
\dfrac{1}{2}(Q_{q,||}+Q_{q,\perp}) 
&
Q_{c,||}+Q_{c,\perp}\\[2mm]
Q_{c,||}+Q_{c,\perp} 
&
\dfrac{1}{2}(Q_{q,||}+Q_{q,\perp})
\end{pmatrix}_{K}
\\[12pt]
&\qquad\qquad \bullet \quad 
\tilde U
=
\begin{pmatrix}
U_c & U_q/2\\
U_q/2 & U_c
\end{pmatrix}_{K},
\qquad
\tilde{U}^{-1}
=
\begin{pmatrix}
(U^{-1})_c & (U^{-1})_q/2\\
(U^{-1})_q/2 & (U^{-1})_c
\end{pmatrix}_{K}
\\[12pt]
&\qquad\qquad \bullet \quad 
\tilde{\Sigma}(t,t')
=
V
\begin{pmatrix}
G^K_{\text{lead}} & G^R_{\text{lead}} \\[3pt]
G^A_{\text{lead}} & 0
\end{pmatrix} V^{\dagger}
=
\begin{pmatrix}
\Sigma^K_{(t,t')} & \Sigma^R_{(t,t')} \\[3pt]
\Sigma^A_{(t,t')} & 0
\end{pmatrix}_K
\\[6pt]
&\text{All objects now live in Keldysh space and encode both response and fluctuations.}
\end{aligned}
\:}
\label{eq:final_keldysh_action}
\end{equation}

\vspace{5pt}

\item \textbf{\underline{Adiabatic Approximation}}

So far, we haven't done any approximations, and the action in Eq.\ref{eq:final_keldysh_action} completely governs the out-of-equilibrium dynamics of magnetization magnitude $M(t)$ and magnetization direction $\mathbf{n}(t)$ [through the time dependence of angles $\theta(t), \phi(t)$]. To discuss the adiabatic regime, the relevant slow/fast separation must be formulated at the level of the physical direction field $\mathbf n(t)$, rather than directly in terms of the angular variables $\theta(t)$ and $\phi(t)$, since the parametrization $\mathbf n(t)=(\sin\theta\cos\phi,\,\sin\theta\sin\phi,\,\cos\theta)$ is nonlinear. Accordingly, we decompose
$$
\mathbf n(t)=\mathbf n_{\text{slow}}(t)+\mathbf n_{\text{fast}}(t),
$$
where $\mathbf n_{\text{slow}}(t)$ contains the low-frequency components and $\mathbf n_{\text{fast}}(t)$ the high-frequency ones. More precisely, in frequency space, one may write schematically
$$
\mathbf n(t)=\int d\omega\,\mathbf n(\omega)e^{-i\omega t},
$$
and separate the modes according to whether $|\omega|$ is smaller or larger than the characteristic thermal and magnetic-field scales of our problem
\begin{equation}
    \mathbf n_{\text{slow}}(t) \:\:\xleftrightarrow[]{\quad} \:\:\omega \lesssim \:\max\left[\frac{k_B T}{\hbar}, \frac{g \mu_B B}{\hbar}\right] \quad \qquad \mathbf n_{\text{fast}}(t) \:\:\xleftrightarrow[]{\quad} \:\:\omega \gg \left[\frac{k_B T}{\hbar}, \frac{g \mu_B B}{\hbar}\right]\:.
\end{equation}
With this splitting, one can now show that by performing the integration over the fast degrees of freedom, and obtaining an effective action for the slow modes 
\begin{equation}
    \int \mathcal{D} \mathbf{n}=\int \mathcal{D} \mathbf{n}_{\text {slow }} \int \mathcal{D} \mathbf{n}_{\text {fast }} \:\:\xrightarrow[]{\text{ integrating }} \:\:\:\int \mathcal{D}\mathbf{n}_{\text {fast }} e^{iS} \propto e^{iS_{\text{eff}}}\:\:,
\end{equation}
we generate an effective potential $V_{\text{eff}}$ for $M$
\begin{equation}
    S_{\mathrm{eff}} \supset-\int d t V_{\mathrm{eff}}(M)\:.
\end{equation}
The crucial point here is that for quantum dots close to the Stoner instability (Eq. \ref{eq:stoner_instabil}), even on the paramagnetic side, interaction effects strongly enhance spin fluctuations in such a way that the magnetization amplitude is effectively stabilized around a large and well-defined average value. Within our formalism, this mechanism is captured by the fact that, after the integration over fast angular modes of $\mathbf{n}(\mathrm{t})$, the effective potential $V_{\text {eff }}(M)$ favors a finite magnetization amplitude, and as a result, $M(t)$ fluctuates around some mean value $M_0$. Therefore, the longitudinal fluctuations become massive (i.e., energetically costly), while the angular degrees of freedom remain soft.
In other words, at low energies and long times, the dynamics of our system is dominated by slow variations of the direction $\mathbf n(t)$, while fluctuations of the magnitude $M(t)$ are strongly suppressed. 

Hence, it is a very good approximation to neglect the residual dynamics of the longitudinal mode and replace
\begin{equation}
    M(t) \approx M_0\qquad \text{throughout the contour}\:.
\end{equation}
This means, in Keldysh space, we have 
\begin{equation}
    M_u(t)=M_d(t)=M_0 \quad \Rightarrow \quad M_c(t)=M_0,\qquad M_q(t)=0,
\end{equation}
so that the amplitude is frozen on both branches of the contour, and the pure bosonic term in the magnetization effective action vanishes
\begin{equation}
    i \int d t \:\:\:M_q\left(\frac{\mathbf{B} \cdot \tilde{\mathbf{n}}_q}{2 J}-\frac{M_c}{2 J}\right)=0\:\:.
\end{equation}

In conclusion, in the adiabatic regime where longitudinal fluctuations are suppressed, and the magnetization amplitude is globally fixed, the effective action reduces to
\begin{equation}
\begin{aligned}
&\textbf{Keldysh Effective action (frozen M- magnitude)}\\[4pt]
\quad& \qquad \qquad \quad i\mathcal{S}_{\text{eff}}
=
\operatorname{tr} \ln \left[
\tilde{G}_{\mathrm{dot}, z}^{-1}
-\tilde{Q}
-\tilde{U}^{-1} \tilde{\Sigma} \tilde{U}
\right]
\\[6pt]
&\text{ where}\\
&\quad \bullet \quad 
\tilde G^{-1}_{\text{dot},z}
=
\begin{pmatrix}
0 
&
i \partial_t-\epsilon_\alpha-\dfrac{M_0}{2} \sigma_z \\[6pt]
i \partial_t-\epsilon_\alpha-\dfrac{M_0}{2} \sigma_z 
&
0
\end{pmatrix}_K \qquad
\end{aligned}
\end{equation}
and all the other quantities/matrices remain the same.\\

However, this is still not the full content of the adiabatic approximation. The point is that, even after fixing the magnetization amplitude to $M_0$, the effective action still contains $\tilde{Q}$,$\tilde{\Sigma}$ contributions, and in general, the resulting functional determinant cannot be evaluated exactly. Therefore, the problem must now be treated perturbatively by expanding the $\operatorname{tr} \ln$ functional.

But it is extremely important to note that, although there are indeed two conceptually different expansions - namely, an expansion in $\tilde{Q}$, corresponding to an adiabatic (gradient) expansion, and an expansion in $\tilde{\Sigma}$, corresponding to weak tunneling - these two expansions cannot be naively performed independently. The reason is that the action is invariant under the choice of the gauge field $\chi(t)$, while the operator $\tilde{Q}$ explicitly depends on this choice. As a consequence, a naive truncation of the expansion simultaneously in $\tilde{Q}$ and $\tilde{\Sigma}$ generally violates this gauge invariance. In particular, one can show that the expansion in $\tilde{U}{ }^{-1} \tilde{\Sigma} \tilde{U}$ is gauge invariant only if all orders in $\tilde{Q}$ are taken into account.

Therefore, a consistent organization of the perturbative expansion requires keeping $\tilde{Q}$ formally resummed inside the propagator, and performing the expansion only in the tunneling self-energy. To this end, we rewrite the effective action as
$$
i \mathcal{S}_{\mathrm{eff}}=\operatorname{tr} \ln \left[\left(\tilde{G}_{\mathrm{dot}, z}^{-1}-\tilde{Q}\right)-\tilde{U}^{-1} \tilde{\Sigma} \tilde{U}\right]
$$
and define the $Q$-dressed propagator
\begin{equation}
\tilde{\mathcal{G}}_Q:=\left(\tilde{G}_{\mathrm{dot}, z}^{-1}-\tilde{Q}\right)^{-1} \quad \xrightarrow[]{\qquad} \quad i \mathcal{S}_{\mathrm{eff}}=\operatorname{tr} \ln \left[\tilde{\mathcal{G}}^{-1}_Q-\tilde{U}^{-1} \tilde{\Sigma} \tilde{U}\right]\:\:.
\end{equation}
Then, by using the "tr ln"-identity
\begin{equation}
    \operatorname{tr} \ln (A-X)=\operatorname{tr} \ln A-\operatorname{tr}\left(A^{-1} X\right)-\frac{1}{2} \operatorname{tr}\left(A^{-1} X A^{-1}X\right)+\cdots\:\:,
\end{equation}
the \textbf{expansion in the dot-lead tunneling is gauge-invariant}
\begin{equation}
    \boxed{i \mathcal{S}_{\mathrm{eff}}\:\:=\:\:\operatorname{tr} \ln \left(\tilde{\mathcal{G}}_Q^{-1}\right)-\operatorname{tr}\left(\tilde{\mathcal{G}}_Q \tilde{U}^{-1} \tilde{\Sigma} \tilde{U}\right)-\frac{1}{2} \operatorname{tr}\left(\tilde{\mathcal{G}}_Q \tilde{U}^{-1} \tilde{\Sigma} \tilde{U} \tilde{\mathcal{G}}_Q \tilde{U}^{-1} \tilde{\Sigma} \tilde{U}\right)+\cdots}
    \label{eq:tunneling_expansion}
\end{equation}
The problem is that, even though $\tilde{Q}$ is formally needed to all orders inside $\tilde{\mathcal{G}}_Q$, one would still like to evaluate the theory in the adiabatic regime, where the direction field $\mathbf{n}(t)$ varies slowly in time. To make this statement more explicit, let us formally expand the $Q$-dressed propagator as
\begin{equation}
\tilde{\mathcal{G}}_Q=\left(\tilde{G}_{\mathrm{dot}, z}^{-1}-\tilde{Q}\right)^{-1}=\:\:\tilde{G}_{\mathrm{dot}, z}+\tilde{G}_{\mathrm{dot}, z} \tilde{Q} \tilde{G}_{\mathrm{dot}, z}+\tilde{G}_{\mathrm{dot}, z} \tilde{Q} \tilde{G}_{\mathrm{dot}, z} \tilde{Q} \tilde{G}_{\mathrm{dot}, z}+\cdots
\end{equation}

Since the operator $\tilde{Q}$ is constructed from $Q=-i U^{-1} \partial_t U$ every insertion of $\tilde{Q}$ carries one time derivative of the slow collective coordinates parameterizing the direction $\mathbf{n}(t)$. Therefore, the previous series may be interpreted as a gradient expansion in time:
\begin{equation}
\begin{aligned}
    \tilde{G}_{\mathrm{dot}, z} \quad &\xleftrightarrow[]{\qquad} \quad (\partial_t\mathbf{n})^0\\
    \tilde{G}_{\mathrm{dot}, z} \tilde{Q} \tilde{G}_{\mathrm{dot}, z} \quad &\xleftrightarrow[]{\qquad} \quad (\partial_t\mathbf{n})^1\\[2pt]
    \tilde{G}_{\mathrm{dot}, z} \tilde{Q} \tilde{G}_{\mathrm{dot}, z} \tilde{Q} \tilde{G}_{\mathrm{dot}, z} \quad &\xleftrightarrow[]{\qquad} \quad (\partial_t\mathbf{n})^2 \:\:,
\end{aligned}
\end{equation}
and so on. From the viewpoint of adiabaticity alone, this would suggest that higher-order terms become progressively less important, and that one should keep only the first few terms of the expansion. 

However, this is precisely where the tension with gauge invariance appears. The operator $\tilde{Q}$ depends explicitly on the gauge field $\chi(t)$, and therefore any arbitrary truncation of the series above generally produces expressions that depend on the chosen gauge. In other words, although the full resummed object $\tilde{\mathcal{G}}_Q$ is gauge covariant and leads to a gauge-invariant effective action, a naive truncation of its expansion in powers of $\tilde{Q}$ need not preserve this property. Thus, the adiabatic intuition suggests keeping only the lowest orders in $\tilde{Q}$, while gauge invariance requires, at least formally, that all orders in $\tilde{Q}$ be retained.

The way to reconcile these two facts is to exploit the gauge freedom in $\chi(t)$ in such a way that the effect of $Q$ is minimized from the outset. To see how this works, recall the decomposition
$$
\begin{aligned}
\tilde{Q}=\tilde{Q}_{\|}+\tilde{Q}_{\perp}\:\:\: \qquad \text{with } \quad
    \begin{cases}
        \tilde Q_{||}
        =
        Q_{c,||}\tau_x+\dfrac{Q_{q,||}}{2}\tau_0,
        \\[2mm]
        \tilde Q_{\perp}
        =
        Q_{c,\perp}\tau_x+\dfrac{Q_{q,\perp}}{2}\tau_0
    \end{cases}\:\:,\\[6pt] \text{where } \quad \begin{cases}
    Q_{c,||}(t)=\displaystyle{\frac{1}{2}\sum_{s=u,d}}[\dot{\phi}_s(1-\cos \theta_s)-\dot{\chi}_s] \frac{\sigma_z}{2}\\[1pt] 
    Q_{q,||}(t)=\displaystyle{\sum_{s=u,d}}(-1)^s \:[\dot{\phi}_s(1-\cos \theta_s)-\dot{\chi}_s] \frac{\sigma_z}{2}\\[1pt] 
    Q_{c, \perp}(t)=-\displaystyle{\frac{1}{2}\sum_{s=u,d}}\frac{1}{2} e^{i \chi_s \sigma_z}\left[\dot{\theta}_s \sigma_y-\dot{\phi}_s \sin \theta_s \sigma_x\right] e^{i \phi_s \sigma_z}\\[1pt] 
    Q_{q, \perp}(t)=-\displaystyle{\sum_{s=u,d}}(-1)^s\frac{1}{2} e^{i \chi_s \sigma_z}\left[\dot{\theta}_s \sigma_y-\dot{\phi}_s \sin \theta_s \sigma_x\right] e^{i \phi_s \sigma_z}
\end{cases}\:\:.
\end{aligned}
$$
Here, $(-1)^{u}=+1$, while $(-1)^{d}=-1$. The physical distinction between these two pieces is important. The transverse term $Q_{\perp}$ (no matter which Keldysh component) is off-diagonal in the instantaneous $z$-basis and mixes the local spin-up and spin-down states. By contrast, the longitudinal term $Q_{||}$ is diagonal in spin space and acts only through $\sigma_z$. Accordingly, it does not induce spin flips, but rather enters as a branch-dependent phase field. Indeed, inside the inverse propagator it appears schematically in spin space as
$$
\tilde{G}_{\mathrm{dot}, z}^{-1}-\tilde{Q} \quad\sim\quad i \partial_t-\epsilon_\alpha-\frac{M_0}{2} \sigma_z-Q_{\|}-Q_{\perp},
$$
so that $Q_{||}$ effectively shifts the energies of spin eigenstates with opposite signs, and in this sense, it behaves like a time-dependent spin-dependent chemical potential. Now, since this piece is also gauge dependent through $\chi(t)$, and in the Keldysh formalism the classical component encodes the physical trajectory of the system (while the quantum component parametrizes fluctuations around it), one would ideally like to choose a gauge in which the classical contribution of this unphysical phase field vanishes. 

But, before discussing the gauge fixing, it is useful to note that, if one momentarily ignores the gauge-invariance issue and simply performs a gradient counting, then $Q_{\perp}$ contributes only at second order in the adiabatic expansion. The reason is that $\tilde{G}_{\text {dot }, z}$ is diagonal in spin space, whereas $Q_{\perp}$ is off-diagonal. Therefore, at linear order one has
$$
\operatorname{tr}_{\mathrm{spin}}\left(\tilde{G}_{\mathrm{dot}, z} \tilde{Q}_{\perp}\right)=0
$$
so the first non-vanishing contribution involving $Q_{\perp}$ arises only from terms of the form
$$
\operatorname{tr}\left(\tilde{G}_{\mathrm{dot}, z}  \tilde{Q}_{\perp} \tilde{G}_{\mathrm{dot}, z}  \tilde{Q}_{\perp}\right)
$$
which are of order $\left(\partial_t \mathbf{n}\right)^2$. Thus, in the strict adiabatic limit, the transverse part is automatically subleading. The potentially dangerous term is instead $Q_{\|}$, because being diagonal in spin space it contributes already at first order.

This observation clarifies the strategy: one should use the gauge freedom in $\chi(t)$ to suppress the leading longitudinal contribution as much as possible. Formally, from
$$
Q_{\|}=[\dot{\phi}(1-\cos \theta)-\dot{\chi}] \frac{\sigma_z}{2}
$$
one sees that the naive choice
$$
\dot{\chi}=\dot{\phi}(1-\cos \theta)
$$
would make $Q_{\|}=0$. However, in the Keldysh formalism this choice cannot be imposed independently on the forward and backward branches, because the gauge field must satisfy the boundary condition
$$
\chi_q(t= \pm \infty)=0 .
$$
Therefore, the appropriate procedure is not to set $Q_{||}$ identically to zero on each branch, but rather to choose $\chi$ so that the classical component $Q_{c, ||}$ vanishes while preserving the required boundary conditions on the contour.
A convenient gauge choice accomplishing this is
\begin{equation}
\dot{\chi}_c(t)=\dot{\phi}_c(t)[1-\cos\theta_c(t)],
\qquad
\chi_q(t)=\phi_q(t)[1-\cos\theta_c(t)]\:\:
\label{eq:gauge_choice_chi}
\end{equation}
under which one has
\begin{equation}
Q_{c,||}=0\:\:,\qquad \quad Q_{q,||} =\frac{1}{2}\sigma_z\:\sin\theta_c
\left[
\dot\phi_c\theta_q-\dot\theta_c\phi_q
\right]\:\:.
\label{eq:Qparallel_q_final}
\end{equation}
This is precisely the desired outcome. The classical longitudinal piece, which would otherwise act as a gauge-dependent spin-dependent chemical-potential shift, is eliminated, while the remaining quantum component is already of the appropriate Keldysh form: it is linear in the quantum fields $\phi_q,\theta_q$ and therefore compatible with the expansion around the physical saddle at small values of, again, $\phi_q,\theta_q$.

In this way, the apparent conflict between adiabaticity and gauge invariance is resolved. Gauge invariance tells us that $\tilde{Q}$ must be kept formally resummed inside $\tilde{\mathcal{G}}_Q$, whereas adiabaticity tells us that the physical effect of $\tilde{Q}$ should be small. The correct implementation of both requirements is therefore to choose a gauge in which the leading longitudinal contribution is minimized, and after this gauge fixing, the only first-order contribution comes from the quantum component $Q_{q,||}$, while $Q_{\perp}$ remains subleading and enters only at second order in the gradient expansion.

Consequently, in the leading adiabatic regime the dynamics is governed by the residual longitudinal quantum term in Eq.\ref{eq:Qparallel_q_final}, together with the tunneling expansion in $\tilde{\Sigma}$, whereas the transverse piece $Q_{\perp}$ may be neglected at lowest order. This means in equations having
\begin{equation}
\begin{aligned}
    \bullet\quad \:&\operatorname{tr} \ln \left(\tilde{\mathcal{G}}_Q^{-1}\right) = \operatorname{tr} \ln \left(\tilde{G}_{\mathrm{dot}, z}^{-1}-\tilde{Q}\right) \approx \:\:\operatorname{tr} \ln \left(\tilde{G}_{\mathrm{dot}, z}^{-1}\right)-\operatorname{tr}\left(\tilde{G}_{\mathrm{dot}, z}\tilde Q_{||}\right)\\[2pt]
    \bullet\quad \:& \tilde{\mathcal{G}}_Q=\left(\tilde{G}_{\mathrm{dot}, z}^{-1}-\tilde{Q}\right)^{-1}\approx\:\:\tilde{G}_{\mathrm{dot}, z}+\tilde{G}_{\mathrm{dot}, z} \tilde{Q} \tilde{G}_{\mathrm{dot}, z}\\[4pt]
    \bullet\quad \:& -\operatorname{tr}\left(\tilde{\mathcal{G}}_Q \tilde{U}^{-1} \tilde{\Sigma} \tilde{U}\right)\approx - \operatorname{tr}\left(\tilde{G}_{\mathrm{dot}, z} \tilde{U}^{-1} \tilde{\Sigma} \tilde{U}\right) - \operatorname{tr}\left(\tilde{G}_{\mathrm{dot}, z}\tilde{Q} \tilde{G}_{\mathrm{dot}, z} \tilde{U}^{-1} \tilde{\Sigma} \tilde{U}\right)\:\:,
\end{aligned}
\end{equation}
and hence, in the light of Eq.\ref{eq:tunneling_expansion}, we obtain at first low orders
\begin{equation}
\begin{aligned}
    i \mathcal{S}_{\mathrm{eff}}\:\:&=\:\:\operatorname{tr} \ln \left(\tilde{\mathcal{G}}_Q^{-1}\right)-\operatorname{tr}\left(\tilde{\mathcal{G}}_Q \tilde{U}^{-1} \tilde{\Sigma} \tilde{U}\right) + \cdots \\[6pt]
    &\approx \:\: \underbrace{\operatorname{tr} \ln \left(\tilde{G}_{\mathrm{dot}, z}^{-1}\right)}_{0^{\text{th}} \text{ order}}-\underbrace{\operatorname{tr}\left(\tilde{G}_{\mathrm{dot}, z}\tilde{Q}_{||}\right)}_{\propto \:(\partial_{\mathrm{t}}\mathbf{n})} - \underbrace{\operatorname{tr}\left(\tilde{G}_{\mathrm{dot}, z} \tilde{U}^{-1} \tilde{\Sigma} \tilde{U}\right)}_{\mathcal{O}(1^{\text{st}}) \text{ weak tunneling}} - \underbrace{\operatorname{tr}\left(\tilde{G}_{\mathrm{dot}, z}\tilde{Q} \tilde{G}_{\mathrm{dot}, z} \tilde{U}^{-1} \tilde{\Sigma} \tilde{U}\right)}_{(\partial_{\mathrm{t}}\mathbf{n})\times\:\mathcal{O}(1^{\text{st}}) \text{ weak tunneling}}\:.
\end{aligned}
\end{equation}
All in all, neglecting the field-independent zeroth-order contribution and retaining only the leading order in both the adiabatic (time-gradient) expansion and in the weak coupling to the leads, the effective action simplifies considerably. In this regime, only terms linear in $(\partial_{\mathrm{t}}\mathbf{n})$ and linear in $\Sigma$ are considered, while higher-order gradient contributions as well as mixed terms involving both gradients and tunneling are discarded. Furthermore, after the gauge fixing discussed above, the only surviving geometric contribution is the longitudinal quantum component $Q_{q,||}$, whereas the transverse part $Q_{\perp}$ and the classical longitudinal component $Q_{c,||}$ are suppressed. As a result, the low-energy magnetization dynamics is governed by the following
\begin{equation}
\boxed{\:
\begin{aligned}
&\qquad \qquad \textbf{Adiabatic Keldysh Effective action (leading order)}\\[4pt]
\quad & \qquad \quad \qquad \qquad  i\mathcal{S}_{\text{eff}}
=
-\operatorname{tr}\left(\tilde{G}_{\mathrm{dot}, z} \tilde{Q}_{\|}\right)-\operatorname{tr}\left(\tilde{G}_{\mathrm{dot}, z} \tilde{U}^{-1} \tilde{\Sigma} \tilde{U}\right)
\\[6pt]
&\text{ where}\\
&\qquad\qquad \bullet \quad 
\tilde G_{\text{dot},z}
=
\begin{pmatrix}
0
&
i \partial_t-\epsilon_\alpha-\dfrac{M_0}{2} \sigma_z \\[6pt]
i \partial_t-\epsilon_\alpha-\dfrac{M_0}{2} \sigma_z 
&
0
\end{pmatrix}^{-1}_K
\\[12pt]
&\qquad\qquad \bullet \quad 
\tilde{Q}_{||}=\frac{1}{2}
\begin{pmatrix}
Q_{q,||}
&
0\\[2mm]
0 
&
Q_{q,||}
\end{pmatrix}_{K}\:\:, \qquad Q_{q,||} =\frac{1}{2}\sigma_z\:\sin\theta_c
\left[
\dot\phi_c\theta_q-\dot\theta_c\phi_q
\right] \\[6pt]
&\quad \text{and } \:(\tilde{U}, \tilde{U}^{-1}, \tilde{\Sigma})\: \text{ as defined previously (Eq.\ref{eq:final_keldysh_action})}\:.
\end{aligned}
\:}
\label{eq:final_keldysh_adiabatic_action}
\end{equation}
Now, the two contributions appearing in the above effective action have a clear physical interpretation. The first term, arising from the geometric part of the action, gives rise to the Berry phase (Wess-Zumino-Novikov-Witten) term that governs the coherent spin dynamics. The second term, arising from the coupling to the leads, generates an AES-type effective action, encoding dissipation and fluctuations due to electron tunneling. In the following, we derive these two contributions explicitly.

\item \textbf{\underline{WZNW/Berry Phase Action}}

A natural way to proceed is to evaluate explicitly the first contribution in Eq.\ref{eq:final_keldysh_adiabatic_action}, namely
\begin{equation}
    iS_{\mathrm{WZNW}}
    \equiv
    -\operatorname{tr}\!\left(\tilde{G}_{\mathrm{dot}, z}\tilde{Q}_{||}\right)\:.
\end{equation}
To do so, we first perform the trace in Keldysh space. Using
\begin{equation}
    \tilde G_{\text{dot},z}
    =
    \begin{pmatrix}
        G^K_{\text{dot},z} & G^R_{\text{dot},z}\\
        G^A_{\text{dot},z} & 0
    \end{pmatrix}
    \quad \xrightarrow[\times\tilde Q_{||}]{\text{ mat. product }} \qquad 
    =
    \frac{1}{2}
    \begin{pmatrix}
        G^K_{\text{dot},z}Q_{q,||} & G^R_{\text{dot},z}Q_{q,||}\\
        G^A_{\text{dot},z}Q_{q,||} & 0
    \end{pmatrix}_K
\end{equation}
Hence, taking first the trace over Keldysh indices, only the upper-left entry contributes, and therefore
\begin{equation}
    iS_{\mathrm{WZNW}}
    =
    -\frac{1}{2}\int dt\;
    \operatorname{tr}_{\alpha,\sigma}\!\left[
        G^K_{\text{dot},z}(t,t)\,Q_{q,||}(t)
    \right],
    \label{eq:wznw_first_reduction}
\end{equation}
where $\operatorname{tr}_{\alpha,\sigma}$ denotes the remaining trace over orbital and spin indices.

Now, in the rotated $z$-basis, the dot inverse propagator is diagonal in spin space,
\begin{equation}
    i\partial_t-\epsilon_\alpha-\frac{M_0}{2}\sigma_z
    =
    \begin{pmatrix}
        i\partial_t-\epsilon_{\alpha,\uparrow} & 0\\
        0 & i\partial_t-\epsilon_{\alpha,\downarrow}
    \end{pmatrix} \:,\qquad \text{with } \:\:\begin{cases}
        \epsilon_{\alpha,\uparrow}=\epsilon_\alpha+\frac{M_0}{2}\\\epsilon_{\alpha,\downarrow}=\epsilon_\alpha-\frac{M_0}{2}
    \end{cases}\:\:.
\end{equation}
Therefore, in exactly the same way as in Eq.\ref{eq:green_function_keldysh_lead}, the Keldysh Green function of the dot is obtained by the replacement $\varepsilon_\gamma\to \epsilon_{\alpha,\sigma}$, giving
\begin{equation}
    iG^K_{\text{dot},z;\alpha\sigma}(t,t')
    =
    \big[1-2f(\epsilon_{\alpha,\sigma})\big]e^{-i\epsilon_{\alpha,\sigma}(t-t')} \quad \xrightarrow[]{\:\:t=t'\:\:}\quad 1-2f(\epsilon_{\alpha,\sigma})
\end{equation}
Substituting now $Q_{q,||}=\frac{1}{2}\sigma_z\,\sin\theta_c\left[\dot\phi_c\,\theta_q-\dot\theta_c\,\phi_q\right]$ into Eq.\ref{eq:wznw_first_reduction}, we obtain
\begin{equation}
    \begin{aligned}
        iS_{\mathrm{WZNW}}
        &=
        -\frac{1}{4}\int dt\;
        \sin\theta_c
        \left[
            \dot\phi_c\,\theta_q-\dot\theta_c\,\phi_q
        \right]
        \operatorname{tr}_{\alpha,\sigma}
        \left[
            G^K_{\text{dot},z}(t,t)\sigma_z
        \right]\\[4pt]
        &=
        -\frac{1}{4}\int dt\;
        \sin\theta_c
        \left[
            \dot\phi_c\,\theta_q-\dot\theta_c\,\phi_q
        \right]
        \sum_\alpha
        \left[
            G^K_{\alpha\uparrow}(t,t)-G^K_{\alpha\downarrow}(t,t)
        \right].
    \end{aligned}
    \label{eq:wznw_before_sum}
\end{equation}
Using the equal-time form of the Keldysh Green functions, we then obtain
\begin{equation}
    \sum_\alpha
    \left[
        G^K_{\alpha\uparrow}(t,t)-G^K_{\alpha\downarrow}(t,t)
    \right]
    =
    -2i
    \sum_\alpha
    \left[
        f\!\left(\epsilon_\alpha-\frac{M_0}{2}\right)
        -
        f\!\left(\epsilon_\alpha+\frac{M_0}{2}\right)
    \right].
\end{equation}
At low temperature, this sum counts precisely the number $N(M_0)$ of orbital levels lying in the Stoner energy window of width $M_0$ around the Fermi energy. Defining therefore
\begin{equation}
    2S:=N(M_0)=\sum_\alpha
    \left[
        f\!\left(\epsilon_\alpha-\frac{M_0}{2}\right)
        -
        f\!\left(\epsilon_\alpha+\frac{M_0}{2}\right)
    \right]
\end{equation}
we obtain
\begin{equation}
    \sum_\alpha
    \left[
        G^K_{\alpha\uparrow}(t,t)-G^K_{\alpha\downarrow}(t,t)
    \right]
    =
    -4iS\:\:,
\end{equation}
and finally, inserting this back into Eq.\ref{eq:wznw_before_sum}, the Berry phase term becomes
\begin{equation}
    \boxed{
    iS_{\mathrm{WZ}}
    =
    iS\int dt\;
    \sin\theta_c
    \left[
        \dot\phi_c\,\theta_q-\dot\theta_c\,\phi_q
    \right]}
\end{equation}
which is the Wess-Zumino action written in a different gauge, and governing the coherent spin dynamics of the dot. This suggests an important character of the WZ action: despite the system being intrinsically out of equilibrium and coupled to dissipative environments, the leading contribution to the adiabatic dynamics retains the purely geometric character of the Berry phase. In other words, the topological action is robust, and more importantly, the coherent part of the dot magnetization dynamics remains protected and can be described in terms of an effective semiclassical spin with quantized magnitude $S$, even in the presence of non-equilibrium effects. 

\item \textbf{\underline{AES-like Action}}

Once we compute the first term in Eq.\ref{eq:final_keldysh_adiabatic_action}, our next step is to finish the expression of $iS_{\text{eff}}$ and compute the second term. For the time being, let us simply denote it by
\begin{equation}
    iS_{\mathrm{AES}}
    \equiv-\operatorname{tr}\!\left(\tilde{G}_{\mathrm{dot},z}\tilde{U}^{-1}\tilde{\Sigma}\tilde{U}\right)\:;
    \label{eq:def_AES_like_term}
\end{equation}
the reason for this name is its relation to the standard $U(1)$ AES action, in the sense, that our result will represent a $SU(2)$ generalization of it. The exact connection and the introduction to the $U(1)$ AES result will be discussed only after the computation is in place. In other words, we first derive the object explicitly, and only afterwards come back to its physical interpretation.

To make the structure of this term (Eq.\ref{eq:def_AES_like_term}) transparent, it is convenient to write the operator trace in the time representation. Since $\tilde U(t)$ is local in time, whereas $\tilde G_{\mathrm{dot},z}(t_1,t_2)$ and $\tilde\Sigma(t_2,t_1)$ are nonlocal kernels, one may formally write
\begin{equation}
    iS_{\mathrm{AES}}
    =
    -\int dt_1dt_2\;
    \operatorname{tr}_{K,\alpha,\sigma}
    \!\left[
        \tilde G_{\mathrm{dot},z}(t_1,t_2)\,
        \tilde U^{-1}(t_2)\,
        \tilde\Sigma(t_2,t_1)\,
        \tilde U(t_1)
    \right].
    \label{eq:aes_time_representation_1}
\end{equation}
However, at this point one must proceed with some care. Although Eq.\ref{eq:aes_time_representation_1} is correct, the trace above is still taken in the \emph{fermionic} Keldysh space. By contrast, the AES term that we want to extract is a \emph{bosonic} effective action for the contour fields $(U_u,U_d)$, and therefore its final kernel must satisfy the usual causal structure of Keldysh actions. In particular, the normalization condition of the closed-time-contour formalism requires that the action vanish when the two contour fields coincide, $U_u=U_d$, or equivalently $U_q=0$. This property is not manifest if one tries to evaluate Eq.\ref{eq:aes_time_representation_1} directly in the rotated $(c,q)$ matrix notation.

For this reason, the correct strategy is to temporarily return to the original contour $(u,d)$ representation and carry out the computation there. Only after the action has been rewritten explicitly in terms of the forward and backward contour fields will we perform the Keldysh rotation at the \emph{bosonic} level. In this way, the causal structure of the AES kernel is recovered correctly and no spurious $cc$-component is generated.

Using
$$
    \oint_{\mathcal C} dt=\sum_{s=u,d}\eta_s\int dt,
    \qquad \eta_u=+1,\quad \eta_d=-1,
$$
the AES contribution may therefore be rewritten as
\begin{equation}
\begin{aligned}
    iS_{\mathrm{AES}}
    =
    -\sum_{s,s'=u,d}\eta_s\eta_{s'}
    \int dt_1dt_2\;
    \operatorname{tr}_{\alpha,\sigma}
    \!\left[
        G^{ss'}_{\mathrm{dot},z}(t_1,t_2)\,
        U_{s'}^{-1}(t_2)\,
        \Sigma^{s's}(t_2,t_1)\,
        U_s(t_1)
    \right].
\end{aligned}
\end{equation}
Now, a key observation is that the contour structure of the previous expression can again be reorganized into a bilinear form. Indeed, by expanding in the branch indices and grouping terms appropriately, one finds schematically
\begin{equation}
\begin{aligned}
\sum_{s,s'}\eta_s\eta_{s'}\,
G^{ss'} U_{s'}^{-1}\Sigma^{s's} U_s
=
\left(\:U_u^{-1}\:\:\:U_d^{-1}\:\right)
\left(\begin{array}{cc}
G^{T}\Sigma^{T} & -\,G^{>}\Sigma^{<}\\[4pt]
-\,G^{<}\Sigma^{>} & G^{\tilde T}\Sigma^{\tilde T}
\end{array}\right)
\binom{U_u}{U_d}\:,
\end{aligned}
\end{equation}
where we have used the standard contour notation $G^{uu}=G^{T}$, $G^{ud}=G^{<}$, $G^{du}=G^{>}$,  $G^{dd}=G^{\tilde T}$ (and similarly for $\Sigma$). This allows us to rewrite the AES action in the compact bilinear form
\begin{equation}
\begin{aligned}
iS_{\mathrm{AES}}
=
-\int dt_1dt_2\;
\operatorname{tr}_{\alpha,\sigma}\left[
\left(\:U_u^{-1}\:\:\:U_d^{-1}\:\right)_{t_2}
\begin{pmatrix}
\mathcal{A}_{uu} & \mathcal{A}_{ud}\\
\mathcal{A}_{du} & \mathcal{A}_{dd}
\end{pmatrix}_{t_1,t_2}
\binom{U_u}{U_d}_{t_1}
\right],\\[9pt]
\text{with }\quad
\begin{pmatrix}
\mathcal{A}_{uu} & \mathcal{A}_{ud}\\
\mathcal{A}_{du} & \mathcal{A}_{dd}
\end{pmatrix}
=
\begin{pmatrix}
G^{T}\Sigma^{T} & -\,G^{>}\Sigma^{<}\\
-\,G^{<}\Sigma^{>} & G^{\tilde T}\Sigma^{\tilde T}
\end{pmatrix}.\hspace{50pt}&
\end{aligned}
\end{equation}
At this point, it is convenient to express the time-ordered components in terms of the physical $<,>$ Green functions. Using the standard relations ($t_{12}:=t_1-t_2$)
\begin{equation}
    \begin{aligned}
        G^{T}=\Theta(t_{12})G^{>}+\Theta(t_{21})G^{<},
\qquad
G^{\tilde T}=\Theta(t_{21})G^{>}+\Theta(t_{12})G^{<}\:,\\
\Sigma^{T}=\Theta(t_{12})\Sigma^{>}+\Theta(t_{21})\Sigma^{<},
\qquad
\Sigma^{\tilde T}=\Theta(t_{21})\Sigma^{>}+\Theta(t_{12})\Sigma^{<}\:,
    \end{aligned}
\end{equation}
one finds that the diagonal entries can be reorganized entirely in terms of the two combinations
\begin{equation}
\alpha^{>}(t_1,t_2):=G^{>}(t_1,t_2)\Sigma^{<}(t_2,t_1),
\qquad
\alpha^{<}(t_1,t_2):=G^{<}(t_1,t_2)\Sigma^{>}(t_2,t_1).
\end{equation}
In terms of these objects, the contour kernel takes the form
\begin{equation}
\begin{aligned}
\begin{pmatrix}
\mathcal{A}_{uu} & \mathcal{A}_{ud}\\
\mathcal{A}_{du} & \mathcal{A}_{dd}
\end{pmatrix}
=
\begin{pmatrix}
\Theta(t_{12})\alpha^{>}+\Theta(t_{21})\alpha^{<} & -\,\alpha^{>}\\
-\,\alpha^{<} & \Theta(t_{21})\alpha^{>}+\Theta(t_{12})\alpha^{<}
\end{pmatrix}.
\end{aligned}
\end{equation}

This representation makes the structure of the AES kernel transparent: all entries are built from the two basic processes encoded in $\alpha^{>}$ and $\alpha^{<}$, while the time-ordering information is carried entirely by the step functions.

We are now ready to perform the Keldysh rotation in the $U$-sector. Using the modified definition of Eq.\,\eqref{eq:def_L_rotation}, we rotate the contour kernel as
\begin{equation}
\begin{aligned}
    \alpha(t_1,t_2)
&= L^{T}\mathcal{A} \:
L=\begin{pmatrix}
    1&1\\
    \frac{1}{2}&-\frac{1}{2}
\end{pmatrix}\begin{pmatrix}
\mathcal{A}_{uu} & \mathcal{A}_{ud}\\
\mathcal{A}_{du} & \mathcal{A}_{dd}
\end{pmatrix}\begin{pmatrix}
    1&\frac{1}{2}\\
    1 & -\frac{1}{2}
\end{pmatrix}\\[10pt]
&=\begin{pmatrix}
\mathcal{A}_{uu} + \mathcal{A}_{ud} + \mathcal{A}_{du} + \mathcal{A}_{dd} &
\frac{1}{2}\left(\mathcal{A}_{uu} + \mathcal{A}_{du} - \mathcal{A}_{ud} - \mathcal{A}_{dd}\right) \\[5pt]
\frac{1}{2}\left(\mathcal{A}_{uu} - \mathcal{A}_{du} + \mathcal{A}_{ud} - \mathcal{A}_{dd}\right) &
\frac{1}{4}\left(\mathcal{A}_{uu} - \mathcal{A}_{du} - \mathcal{A}_{ud} + \mathcal{A}_{dd}\right)
\end{pmatrix}\\[10pt]
&=\begin{pmatrix}
    0 & \Theta(t_{12}) (\alpha^> - \alpha^<) \\[5pt]
    -\Theta(t_{21}) (\alpha^> - \alpha^<) & \frac{1}{2}(\alpha^> + \alpha^<)
\end{pmatrix}=:\begin{pmatrix}
0 & \alpha_A\\
\alpha_R & \alpha_K
\end{pmatrix}_{t_1,t_2}
\end{aligned}
\end{equation}
With this result at hand, the structure of the effective action becomes completely transparent: inserting the rotated kernel into the bilinear form, we arrive at
\begin{equation}
\boxed{\:
\begin{aligned}
&\qquad\qquad \text{AES-like Action in Keldysh space}\\[2pt]
\quad & 
iS_{\mathrm{AES}}
=
-\int dt_1dt_2\;
\operatorname{tr}_{\alpha,\sigma}\left[
\begin{pmatrix}
U^{-1}_c &U^{-1}_q
\end{pmatrix}_{t_2}\:
\begin{pmatrix}
0 & \alpha_A\\
\alpha_R & \alpha_K
\end{pmatrix}_{t_1,t_2}\:
\begin{pmatrix}
    U_c\\U_q
\end{pmatrix}_{t_1}
\right]\\[3pt]
&\text{where}\\
&\qquad\quad \bullet \quad \alpha_A(t_1,t_2)=\Theta(t_{12})\big[G^{>}_{(t_1,t_2)}\Sigma^{<}_{(t_2,t_1)}-G^{<}_{(t_1,t_2)}\Sigma^{>}_{(t_2,t_1)}\big]\\[5pt]
&\qquad\quad \bullet \quad \alpha_R(t_1,t_2)=-\Theta(t_{21})\big[G^{>}_{(t_1,t_2)}\Sigma^{<}_{(t_2,t_1)}-G^{<}_{(t_1,t_2)}\Sigma^{>}_{(t_2,t_1)}\big]
\\[5pt]
&\qquad\quad \bullet \quad \alpha_K(t_1,t_2)=\tfrac{1}{2}\big[G^{>}_{(t_1,t_2)}\Sigma^{<}_{(t_2,t_1)}+G^{<}_{(t_1,t_2)}\Sigma^{>}_{(t_2,t_1)}\big]
\end{aligned}}
\label{eq:aes_cq_bilinear}
\end{equation}
The remaining step is to perform the trace over the orbital dot space, $\operatorname{tr}_\alpha(\cdots)$. Since the matrices $U^{-1}$ and $U$ act only in spin space, this trace factorizes, and the action reduces to
\begin{equation}
    \begin{aligned}
        iS_{\mathrm{AES}}=-\int dt_1dt_2\;
\operatorname{tr}_{\sigma}\left[
\begin{pmatrix}
U^{-1}_c &U^{-1}_q
\end{pmatrix}_{t_2}\:
\begin{pmatrix}
0 & \operatorname{tr}_{\alpha}\alpha_A\\
\operatorname{tr}_{\alpha}\alpha_R & \operatorname{tr}_{\alpha}\alpha_K
\end{pmatrix}_{t_1,t_2}\:
\begin{pmatrix}
    U_c\\U_q
\end{pmatrix}_{t_1}
\right]\\
    \end{aligned}
\end{equation}
which means that the only relevant quantities that we have to compute are
\begin{equation}
\operatorname{tr}_\alpha\left[G^{>}_{(t_1,t_2)}\Sigma^{<}_{(t_2,t_1)}\right] \:\:,\qquad \quad\operatorname{tr}_\alpha\left[G^{<}_{(t_1,t_2)}\Sigma^{>}_{(t_2,t_1)}\right]\:.
\end{equation}
Here, we must recall that since the dot Green function is diagonal in both spin and dot-orbital indices, and the lead self-energy does not affect the spin space 
\begin{equation}
    \begin{aligned}
(G^{\lessgtr}_{\text{dot,z}})_{\alpha\alpha',\sigma\sigma'}=\delta_{\sigma\sigma'}\delta_{\alpha\alpha'}(G^{\lessgtr}_{\text{dot,z}})_{\alpha,\sigma} 
\qquad \Sigma_{\alpha\alpha'}=\sum_\gamma V_{\alpha\gamma}G_{\text{lead},\gamma}V^*_{\alpha'\gamma} \otimes \mathbb{1}_{\text{spin}}\:\:,
    \end{aligned}
\end{equation}
the action is fully determined by the two fundamental matrices elements
\begin{align}
    & \bullet \quad \:\: \begin{aligned}[t]
        \operatorname{tr}_\alpha\left[G^{>}_{\sigma}{(t_1,t_2)}\Sigma^{<}{(t_2,t_1)}\right]&=\sum_\alpha\delta_{\alpha\alpha'}\:G^{>}_{\sigma,\alpha}(t_1,t_2)\sum_\gamma V_{\alpha\gamma}G_{\text{lead},\gamma}^{<}(t_2,t_1)V^*_{\alpha'\gamma}\\
        &=\sum_{\alpha,\gamma} |V_{\alpha,\gamma}|^2G^{>}_{\sigma,\alpha}(t_1,t_2) G_{\text{lead},\gamma}^{<}(t_2,t_1)
    \end{aligned}\\[10pt]
    & \bullet \quad \:\: 
        \operatorname{tr}_\alpha\left[G^{<}_{\sigma}{(t_1,t_2)}\Sigma^{>}{(t_2,t_1)}\right]=\sum_{\alpha,\gamma} |V_{\alpha,\gamma}|^2G^{<}_{\sigma,\alpha}(t_1,t_2) G_{\text{lead},\gamma}^{>}(t_2,t_1)\:\:.
\end{align}
To proceed, we now evaluate these two building blocks explicitly. Since the dot and the lead are both described as noninteracting fermionic reservoirs in the rotated $z$-basis, their lesser and greater Green functions depend only on the time difference, which we denote by $\tau=t_1-t_2$,
\begin{equation}
\begin{aligned}
\begin{aligned}
G_{\alpha \sigma}^{>}(\tau)
&= -i\left(1-f_{d}(\epsilon_{\alpha \sigma})\right)
e^{-i \epsilon_{\alpha \sigma}\tau} \\
G_{\alpha \sigma}^{<}(\tau)
&= i f_{d}(\epsilon_{\alpha \sigma})
e^{-i \epsilon_{\alpha \sigma}\tau}
\end{aligned}
\qquad
\begin{aligned}
G_{\text{lead}, \gamma}^{>}(\tau)
&= -i\left(1-f_{l}(\varepsilon_\gamma)\right)
e^{-i \varepsilon_\gamma (-\tau)} \\
G_{\text{lead}, \gamma}^{<}(\tau)
&= i f_{l}(\varepsilon_\gamma)
e^{-i \varepsilon_\gamma (-\tau)}
\end{aligned}
\end{aligned}\:\:.
\end{equation}
Substituting these expressions into the previous sums, one sees that each kernel depends only on the energy difference $\epsilon_{\alpha\sigma}-\varepsilon_\gamma$. Hence, by further adopting the standard structureless tunneling approximation, and taking the respective continuum limits
\begin{equation}
\left|V_{\alpha \gamma}\right|^2 \;\approx\; |V|^2,
\qquad
\sum_{\alpha} \;\rightarrow\; \int d\epsilon\, \rho_{\mathrm{dot}}^{\sigma}(\epsilon),
\qquad
\sum_{\gamma} \;\rightarrow\; \int d\varepsilon\, \rho_{\mathrm{lead}}(\varepsilon)\:\:,
\end{equation}
the orbital trace is converted into an energy integral, and the three Keldysh kernels 
\begin{equation}
\begin{aligned}
\alpha_{A,\sigma}(\tau)
&= \Theta(\tau)\, |V|^2
\int d\epsilon\, d\varepsilon \;
\rho_{\text{dot}}^{\sigma}(\epsilon)\,
\rho_{\text{lead}}(\varepsilon)\;
\left[ f_l(\varepsilon) - f_d(\epsilon) \right]\;
e^{-i(\epsilon-\varepsilon)\tau}
\\[4pt]
\alpha_{R,\sigma}(\tau)
&= -\Theta(-\tau)\, |V|^2
\int d\epsilon\, d\varepsilon \;
\rho_{\text{dot}}^{\sigma}(\epsilon)\,
\rho_{\text{lead}}(\varepsilon)\;
\left[ f_l(\varepsilon) - f_d(\epsilon) \right]\;
e^{-i(\epsilon-\varepsilon)\tau}
\\[4pt]
\alpha_{K,\sigma}(\tau)
&= \frac{|V|^2}{2}
\int d\epsilon\, d\varepsilon \;
\rho_{\text{dot}}^{\sigma}(\epsilon)\,
\rho_{\text{lead}}(\varepsilon)\;
\left[ f_l(\varepsilon) + f_d(\epsilon)
- 2 f_d(\epsilon) f_l(\varepsilon) \right]\;
e^{-i(\epsilon-\varepsilon)\tau}
\end{aligned}
\end{equation}
Finally, in the wide-band limit, the densities of states are treated as approximately constant within the relevant energy window, and then obtaining
\begin{equation}
\begin{aligned}
\alpha_{A,\sigma}(\tau)
&= \left(|V|^2 \rho_{\text{dot}}^{\sigma}\,
\rho_{\text{lead}}\right) \:\:\left[\Theta(\tau)
\int d\epsilon\, d\varepsilon \;
\left[ f_l(\varepsilon) - f_d(\epsilon) \right]\;
e^{-i(\epsilon-\varepsilon)\tau}\right]
\\[4pt]
\alpha_{R,\sigma}(\tau)
&= -\left(|V|^2 \rho_{\text{dot}}^{\sigma}\,
\rho_{\text{lead}}\right) \:\:\left[\Theta(-\tau)
\int d\epsilon\, d\varepsilon \;
\left[ f_l(\varepsilon) - f_d(\epsilon) \right]\;
e^{-i(\epsilon-\varepsilon)\tau}\right]
\\[4pt]
\alpha_{K,\sigma}(\tau)
&= \frac{1}{2} \left(|V|^2 \rho_{\text{dot}}^{\sigma}\,
\rho_{\text{lead}}\right)
\int d\epsilon\, d\varepsilon \;
\left[ f_l(\varepsilon) + f_d(\epsilon)
- 2 f_d(\epsilon) f_l(\varepsilon) \right]\;
e^{-i(\epsilon-\varepsilon)\tau}\:\:.
\end{aligned}
\end{equation}
Having reached this point, it is convenient to define $g_{\sigma}=|V|^2\rho_{\text{dot}}^{\sigma}\,
\rho_{\text{lead}}$ and change integration variables from the microscopic energies $(\epsilon,\varepsilon)$ to their average $E=(\epsilon+\varepsilon)/2$ and difference $\omega=\epsilon-\varepsilon$, writing
\begin{equation}
\begin{aligned}
\alpha_{A,\sigma}
&= g_\sigma \:\Theta(\tau)
\int d\omega \: dE \;
\left[ f_l\left(E-\frac{\omega}{2}\right) - f_d\left(E+\frac{\omega}{2}\right) \right]\;
e^{-i\omega\tau}
\\[4pt]
\alpha_{R,\sigma}
&= - g_\sigma \: \Theta(-\tau)
\int d\omega \: dE \;
\left[ f_l\left(E-\frac{\omega}{2}\right) - f_d\left(E+\frac{\omega}{2}\right) \right]\;
e^{-i\omega\tau}
\\[4pt]
\alpha_{K,\sigma}
&= \frac{g_\sigma}{2} 
\int d\omega \: dE \;
\left[ f_l\left(E-\frac{\omega}{2}\right) + f_d\left(E+\frac{\omega}{2}\right)
- 2 f_d\left(E+\frac{\omega}{2}\right) f_l\left(E-\frac{\omega}{2}\right) \right]
e^{-i\omega\tau}\:.
\end{aligned}
\end{equation}
Hence, we note that the kernels acquire the form of a Fourier transform in the transferred energy $\omega$, while the average energy $E$ only enters as a spectral weight. So far, the expression obtained above is completely general and allows for arbitrary distribution functions $f_d$ and $f_l$, thus describing a genuinely nonequilibrium situation. However, in order to make contact with the standard AES result, it is instructive to consider the equilibrium limit in which both the dot and the lead are characterized by the same Fermi distribution, $f_d=f_l\equiv f$. Physically, this means that the fermionic degrees of freedom act as an equilibrium bath, while the collective spin field $U\sim\mathbf n$ remains fully dynamical.

In this case, it is convenient to isolate the dependence on the transferred energy by introducing the equilibrium spectral weights, which can be evaluated explicitly using standard identities 
\begin{equation}
\begin{aligned}
\bullet \quad \mathcal J(\omega)
&:=\int dE\,
\left[
f\!\left(E-\frac{\omega}{2}\right)-f\!\left(E+\frac{\omega}{2}\right)
\right]=\omega\:\:,\\[10pt]
\bullet \quad\mathcal N(\omega)
&:=\frac{1}{2}\int dE\,
\left[
f\!\left(E-\frac{\omega}{2}\right)+f\!\left(E+\frac{\omega}{2}\right)
-2f\!\left(E+\frac{\omega}{2}\right)f\!\left(E-\frac{\omega}{2}\right)
\right]\\&\qquad\qquad\qquad\qquad\qquad\qquad\qquad\qquad\quad\:\:=\frac{1}{2}\,
\omega \coth\!\left(\frac{\omega}{2T}\right)\:\:.
\end{aligned}
\end{equation}
In this way, the three kernels then take the compact form
\begin{equation}
\begin{aligned}
\alpha_{A,\sigma}(\tau)
&=g_\sigma\Theta(\tau)\int d\omega\, \mathcal J(\omega)e^{-i\omega\tau},\\[4pt]
\alpha_{R,\sigma}(\tau)
&=-g_\sigma\Theta(-\tau)\int d\omega\, \mathcal J(\omega)e^{-i\omega\tau},\\[4pt]
\alpha_{K,\sigma}(\tau)
&=g_\sigma\int d\omega\, \mathcal N(\omega)e^{-i\omega\tau}\:,
\end{aligned}
\end{equation}
such that the action in Eq.\ref{eq:aes_cq_bilinear} can written as
\begin{equation}
\boxed{
\begin{aligned}
&\hspace{115pt} \textbf{AES-like Action}\\[2pt]
\quad & 
iS_{\mathrm{AES}}=
-\int dt_1dt_2\;
\operatorname{tr}_{\sigma}\left[
\begin{pmatrix}
U^{-1}_c &U^{-1}_q
\end{pmatrix}_{t_2}\:
\begin{pmatrix}
0 & \hat{g}\:\tilde{\alpha}_A\\
\:\hat{g}\:\tilde{\alpha}_R & \hat{g}\:\tilde{\alpha}_K
\end{pmatrix}_{t_1,t_2}\:
\begin{pmatrix}
    U_c\\U_q
\end{pmatrix}_{t_1}
\right]\\[3pt]
&\text{where}\\
&\bullet \quad \hat{g}=\begin{pmatrix}
    g_\uparrow & 0\\ 0 & g_\downarrow\end{pmatrix}\hspace{60pt}
\qquad \bullet \quad \tilde{\alpha}_R(\tau)=-\Theta(-\tau) \int d\omega\: \omega \:e^{-i\omega\tau} 
\\[5pt]
&\bullet \quad \tilde{\alpha}_A(\tau)=\Theta(\tau) \int d\omega\: \omega \:e^{-i\omega\tau}  \hspace{16pt}
 \bullet \quad \tilde{\alpha}_K(\tau)= \int d\omega \: \frac{\omega}{2}\,
 \coth\!\left(\frac{\omega}{2T}\right)e^{-i\omega\tau}\:.
\end{aligned}}
\end{equation}
If one only wants the $SU(2)$-invariant part, then decomposing
\begin{equation}
    \hat{g}=\begin{pmatrix}
    g_\uparrow & 0\\ 0 & g_\downarrow\end{pmatrix}=\left(\frac{g_\uparrow+g_\downarrow}{2}\right)\mathbb{1}\:\:+\:\:\left(\frac{g_\uparrow-g_\downarrow}{2}\right)\sigma_z\:\:,
\end{equation}
the second term vanishes under the trace, and the final result is expressed as 
\begin{equation}
    \boxed{iS_{\mathrm{AES}}=
-g\int dt_1dt_2\;
\operatorname{tr}_{\sigma}\left[
\begin{pmatrix}
U^{-1}_c &U^{-1}_q
\end{pmatrix}_{t_2}\:
\begin{pmatrix}
0 & \tilde{\alpha}_A\\
\tilde{\alpha}_R & \tilde{\alpha}_K
\end{pmatrix}_{t_1,t_2}\:
\begin{pmatrix}
    U_c\\U_q
\end{pmatrix}_{t_1}
\right]}
\end{equation}
with $g:=\frac{1}{2}|V|^2\,
\rho_{\text{lead}}\left(\rho_{\text{dot}}^{\uparrow}+\rho_{\text{dot}}^{\downarrow}\right)$. Once this final result is obtained, the structure of the effective action becomes particularly suggestive. By having the kernels $\tilde{\alpha}_{R / A / K}$  fully determined by the universal spectral functions $\omega$ and $\omega \operatorname{coth}(\omega / 2 T)$, and entering in the action in a bilinear form with the characteristic causal Keldysh structure, we identify an extremely precise hallmark of an electronic environment acting as a dissipative bath for a collective degree of freedom.

This observation naturally leads us to recall that an identical structure appears in the well-known Ambegaokar-Eckern-Schön (AES) theory. There, one considers a small metallic island coupled to external leads, where Coulomb interactions give rise to charging effects. After decoupling the interaction and performing a gauge transformation, the tunneling amplitudes acquire phase factors $e^{\pm i\phi(t)}$, with $\phi(t)$ conjugate to the charge on the island. Integrating out the fermions then yields an effective action for this phase variable that is nonlocal in time and whose kernel is expressed in terms of the same combinations of lesser and greater Green functions as above.

In particular, the resulting AES action takes a bilinear form in the contour fields $e^{ \pm i \phi}$, with retarded, advanced, and Keldysh components that, in equilibrium, reduce exactly to the functions proportional to $\omega$ and $\omega \operatorname{coth}(\omega / 2 T)$. Comparing with the result derived here, we see that the correspondence is immediate: the scalar phase factor $e^{i \phi(t)} \in U(1)$ is replaced by the matrix-valued rotation $U(t) \in SU(2)$, while the dissipative kernel retains the same universal structure. In this sense, the action obtained above can be understood as a direct non-Abelian generalization of the original AES theory. 
\end{itemize}
At this point, it is natural to collect the two contributions derived above into a single low-energy Keldysh action. Restricting to the $SU(2)$-invariant part of the AES kernel, the leading adiabatic effective theory for the collective spin dynamics of the dot takes the form
\begin{equation}
\boxed{
\begin{aligned}
iS_{\mathrm{eff}}
&=
iS_B+iS_{\mathrm{WZ}}+iS_{\mathrm{AES}}\\[4pt]
&=
-i S \gamma B \int d t \sin \theta_c \theta_q\:\:+\:\:iS\int dt\;
\sin\theta_c
\left[
\dot\phi_c\,\theta_q-\dot\theta_c\,\phi_q
\right]
\\[4pt]
&\qquad\qquad
-g\int dt_1dt_2\;
\operatorname{tr}_{\sigma}\left[
\begin{pmatrix}
U^{-1}_c &U^{-1}_q
\end{pmatrix}_{t_2}\:
\begin{pmatrix}
0 & \tilde{\alpha}_A\\
\tilde{\alpha}_R & \tilde{\alpha}_K
\end{pmatrix}_{t_1,t_2}\:
\begin{pmatrix}
    U_c\\U_q
\end{pmatrix}_{t_1}
\right],
\end{aligned}}
\label{eq:full_geometric_noise_action}
\end{equation}
where the first term corresponds to the restoring of the action related to the magnetic field (in z-direction), the second term to the geometric Wess-Zumino contribution, and the last term to the dissipative AES kernel generated by the tunneling electrons. In this form, the physical content of the theory becomes transparent: the coherent precessional dynamics is encoded in the Berry-phase term, while the electronic environment contributes both a retarded response, through $\tilde{\alpha}_{R/A}$, and fluctuations, through $\tilde{\alpha}_K$.

The next step is then to extract from Eq.\ref{eq:full_geometric_noise_action} the corresponding semiclassical equations of motion. In the Keldysh formalism, the semiclassical equations of motion follow from expanding the action in the quantum components of the fields and imposing stationarity with respect to them
\begin{equation}
    \frac{\delta (i S_{\text {eff}})}{\delta \phi_q(t)}=\frac{\delta (i S_{\text {eff}})}{\delta \theta_q(t)}=0\:\:.
\end{equation}
To make this structure explicit, it is convenient to rewrite the AES contribution by separating the terms linear and quadratic in the quantum fields. Using the bilinear form of Eq.\ref{eq:full_geometric_noise_action}, one can schematically decompose
\begin{equation}
\begin{aligned}
iS_{\mathrm{AES}}
&=
-g\int dt_1dt_2\;
\operatorname{tr}_{\sigma}\Big[
U_c^{-1}(t_2)\,\tilde{\alpha}_A(t_1-t_2)\,U_q(t_1)\\
&\hspace{70pt}
+\,U_q^{-1}(t_2)\,\tilde{\alpha}_R(t_1-t_2)\,U_c(t_1)\\
&\hspace{70pt}
+\,U_q^{-1}(t_2)\,\tilde{\alpha}_K(t_1-t_2)\,U_q(t_1)
\Big]\:,
\end{aligned}
\label{eq:aes_split_RK}
\end{equation}
where the $U,U^{-1}$-matrix fields must be evaluated in the gauge defined in Eq.\ref{eq:gauge_choice_chi}. Note how the first two terms are linear in the quantum component $U_q$ and encode the retarded/advanced response of the electronic bath, while the last term is quadratic in $U_q$ and originates entirely from the Keldysh kernel $\tilde{\alpha}_K$. Focusing first on the retarded/advanced contributions, one has ($\tau=t_1-t_2, t=t_2$)
\begin{equation*}
    \begin{aligned}
        \bullet \quad I_A:&=\int dt d\tau \:\tilde{\alpha}_A(\tau) \operatorname{tr}_\sigma\left[U_c^{-1}(t) U_q(t+\tau)\right]\\
        &=\int d t \int_0^{\infty} d \tau \left(\int d \omega \:\omega e^{-i \omega \tau} \right)\operatorname{tr}_\sigma\left[U_c^{-1}(t) U_q(t+\tau)\right]\\
        &=\int d t \int_0^{\infty} d \tau \:\:[2\pi i \delta'(\tau)] \operatorname{tr}_\sigma\left[U_c^{-1}(t) U_q(t+\tau)\right]\\
        &=-\pi i \int d t \:\partial_\tau \operatorname{tr}_\sigma\left[U_c^{-1}(t) U_q(t+\tau)\right]_{\tau=0}=-\pi i \int d t \operatorname{tr}_\sigma\left[U_c^{-1}(t) \partial_t U_q(t)\right]\\[8pt]
        \bullet \quad I_R:&=\int dt d\tau\: \tilde{\alpha}_R(\tau)\operatorname{tr}_\sigma\left[U_q^{-1}(t) U_c(t+\tau)\right]\\
        &=-2 \pi i \int d t \int_{-\infty}^0 d \tau \:\delta^{\prime}(\tau) \operatorname{tr}_\sigma\left[U_q^{-1}(t) U_c(t+\tau)\right] =\pi i \int d t \operatorname{tr}_\sigma\left[U_q^{-1}(t) \partial_t U_c(t)\right]
    \end{aligned}
\end{equation*}
and passing to $(u,d)$ variables $\big[\operatorname{tr}(U_s^{-1} \partial_t U_s)=0,\:\:s=u,d\big]$, one can see
\begin{equation}
\begin{aligned}
    I_A&=-\frac{\pi i}{2} \int d t\left[\operatorname{tr}\left(U_d^{-1} \partial_t U_u\right)-\operatorname{tr}\left(U_u^{-1} \partial_t U_d\right)\right]=I_R\:\:.
\end{aligned}
\end{equation}
Then, using the equivalent form to the expression for $(U,U^{-1})$ in Eq.\ref{eq:computing_Q}, 
\begin{equation}
\begin{aligned}
    &U_s=\left(\begin{array}{cc}
e^{-i \chi_s / 2} c_s & -e^{-i \phi_s+i \chi_s / 2} s_s \\
e^{i \phi_s-i \chi_s / 2} s_s & e^{i \chi_s / 2} c_s
\end{array}\right), \quad \quad U_s^{-1}=\left(\begin{array}{cc}
e^{i \chi_s / 2} c_s & e^{-i \phi_s+i \chi_s / 2} s_s \\
-e^{i \phi_s-i \chi_s / 2} s_s & e^{-i \chi_s / 2} c_s
\end{array}\right)\:\:\\
    &\hspace{60pt}\text{with shorthand }\quad c_s:=\cos \frac{\theta_s}{2} \quad s_s:=\sin \frac{\theta_s}{2} \quad s=u, d\:\:\:,
\end{aligned}
\end{equation}
together with our gauge fixing Eq.\ref{eq:gauge_choice_chi} and the condition of small quantum components
\begin{equation}
    \begin{gathered}
\sin \frac{\theta_q}{2} \simeq \frac{\theta_q}{2} \qquad \cos \frac{\theta_q}{2} \simeq 1 \\
\sin \left[\frac{\phi_q}{2}\left(1 \pm \cos \theta_c\right)\right] \simeq \frac{\phi_q}{2}\left(1 \pm \cos \theta_c\right) \qquad \cos \left[\frac{\phi_q}{2}\left(1 \pm \cos \theta_c\right)\right] \simeq 1\:\:,
\end{gathered}
\end{equation}
one can show that the retarded and advanced contributions simplify to
\begin{equation}
    \begin{aligned}
        I_A+I_R&=\frac{\pi i}{2} \int d t\left[\dot{\theta}_c \theta_q+\dot{\phi}_c \phi_q \sin ^2 \theta_c\right]\\[4pt]
        &\qquad \Rightarrow\quad iS_{\text{AES}}=-\frac{\pi i g}{2} \int d t\left[\dot{\theta}_c \theta_q+\dot{\phi}_c \phi_q \sin ^2 \theta_c\right] \:+\:iS_{\text{AES}}^{K}\:\:.
    \end{aligned}
\end{equation}
Therefore, by grouping the coefficients of $\theta_q$ and $\phi_q$, the full effective action (Eq.\ref{eq:full_geometric_noise_action}) becomes
\begin{equation}
    \boxed{\:\:\begin{aligned}
iS_{\text{eff}}\:=\:\:& i \int d t \:\:\theta_q\left[S \sin \theta_c\left(\dot{\phi}_c-\gamma B\right)-\frac{\pi g}{2} \dot{\theta}_c\right] \\
& \quad+i \int d t \:\:\phi_q\:\:\sin\theta_c\left[-S  \dot{\theta}_c-\frac{\pi g}{2}\dot{\phi}_c \sin  \theta_c\right]\:+\:iS_{\text{AES}}^K\:\:.\:\:
\end{aligned}}
\end{equation}
The remaining task is then to compute
$$
iS^K_{\mathrm{AES}} = -g\int dt_1dt_2\; \operatorname{tr}_{\sigma}\left[ U_q^{-1}(t_2)\,\tilde{\alpha}_K(t_1-t_2)\,U_q(t_1) \right],
$$
which is quadratic in quantum fields and therefore cannot contribute directly to the deterministic saddle-point equations in the same way as the retarded/advanced part did. Instead, this term is precisely the one that encodes fluctuations. In other words, while the $R/A$ sector produced the dissipative drift terms, the Keldysh sector is the origin of the stochastic forces.

\noindent To make this structure explicit, it is convenient to expand the matrix $U_q$ on the basis of Pauli matrices. To linear order in the quantum components, one may write
\begin{equation}
    \begin{aligned}
        &\hspace{30pt} U_q=A_0^q \:\sigma_0+i \sum_{j=x, y, z} A_j^q \: \sigma_j \quad \xrightarrow[]{\qquad } \:\:\: U_q^{-1}=A_0^q \sigma_0-i \sum_{j=x, y, z} A_j^q \sigma_j\\[8pt]
        &\text{with } \\&\qquad \quad A_0^q =
        -\tfrac{1}{2}\,\theta_q\,\sin(\tfrac{\theta_c}{2})\cos(\tfrac{\chi_c}{2})
        -\,\phi_q\,\cos(\tfrac{\theta_c}{2})\sin^2(\tfrac{\theta_c}{2})\sin(\tfrac{\chi_c}{2}) \\
        &\qquad \quad A_x^q =
        \tfrac{1}{2}\,\theta_q\,\cos(\tfrac{\theta_c}{2})\sin(\phi_c-\tfrac{\chi_c}{2})+\,\phi_q\,\sin(\tfrac{\theta_c}{2})\cos^2(\tfrac{\theta_c}{2})\cos(\phi_c-\tfrac{\chi_c}{2})\\
        &\qquad \quad A_y^q = -\tfrac{1}{2}\,\theta_q\,\cos(\tfrac{\theta_c}{2})\cos(\phi_c-\tfrac{\chi_c}{2})+\,\phi_q\,\sin(\tfrac{\theta_c}{2})\cos^2(\tfrac{\theta_c}{2})\sin(\phi_c-\tfrac{\chi_c}{2})\\
        &\qquad\quad A_z^q = \tfrac{1}{2}\,\theta_q\,\sin(\tfrac{\theta_c}{2})\sin(\tfrac{\chi_c}{2}) -\,\phi_q\,\cos(\tfrac{\theta_c}{2})\sin^2(\tfrac{\theta_c}{2})\cos(\tfrac{\chi_c}{2})
    \end{aligned}
\end{equation}
With this parametrization, the trace over spin space becomes immediate, since only the scalar products of the coefficients survive. One thus obtains
\begin{equation}
\begin{aligned}
\operatorname{tr}_\sigma\left[U_q^{-1}\left(t_2\right) U_q\left(t_1\right)\right]=2&\sum_{j=0, x, y, z} A_j^q\left(t_2\right) A_j^q\left(t_1\right)\\& \Rightarrow \:\:iS_{\text{AES}}^K=-2g\int dt_1dt_2\,\:\tilde{\alpha}_K(t_1-t_2)\: \bm{A}^{q}(t_2)\cdot\bm{A}^{q}(t_1)\:.
\end{aligned}\end{equation}
At this stage, the physical content of the Keldysh contribution becomes completely transparent. The action is nonlocal in time and quadratic in the quantum fields, exactly as expected for a noise kernel generated after integrating out an electronic bath. This is the standard situation in which one performs a Hubbard-Stratonovich decoupling: instead of working with a nonlocal quadratic form in $\mathbf A_q$, one introduces an auxiliary stochastic field that couples linearly to it. Concretely, we decouple the Keldysh part as
\begin{equation} 
e^{iS^K_{\mathrm{AES}}} = \int \mathcal{D}\bm{\xi}\; e^{ i\int dt\;\bm{\xi}(t)\cdot\mathbf{A}_q(t)}\:\: e^{-\:\frac{1}{8 g} \int d t_1 d t_2\:\tilde{\alpha}_K^{-1}(t_1-t_2) \bm\xi(t_1)\cdot\bm\xi(t_2) }
\end{equation} 
In this way, the nonlocal quadratic term is traded for a linear coupling of the quantum fields to an auxiliary vector field $\bm\xi(t)$. By construction, this field is Gaussian, has zero mean, and its correlations are fixed by the inverse kernel appearing in the decoupling. Equivalently,
 \begin{equation} \langle \xi_i(t)\xi_j(t') \rangle = 4g\,\delta_{ij}\,\tilde{\alpha}_K(t-t')\:,\qquad \:\:\left\langle\xi_j(t)\right\rangle=0\:\:.
 \label{eq:noise_correlator} 
 \end{equation} 
So this is precisely the stochastic sector of the theory: the same Keldysh kernel that in the effective action measures fluctuations now becomes the correlator of the random force. In other words, the noise is not inserted by hand; it is generated microscopically by the electronic environment.

After this decoupling, the full effective action can be written in the equivalent stochastic form
\begin{equation}
    \boxed{\:\:\begin{aligned} &\textbf{Stochastic (Adiabatic) Keldysh Magnetization Action}\\
iS_{\text{eff}}\:=\:\:& i \int d t \:\:\theta_q\left[S \sin \theta_c\left(\dot{\phi}_c-\gamma B\right)-\frac{\pi g}{2} \dot{\theta}_c\right] \\
& \quad+i \int d t \:\:\phi_q\:\:\sin\theta_c\left[-S  \dot{\theta}_c-\frac{\pi g}{2}\dot{\phi}_c \sin  \theta_c\right]\\[4pt]
& \qquad\:\:+i\int dt \sum_{j=0,x,y,z} \xi_j(t) A_j^q(t)\\[5pt] &\qquad\qquad -\:\frac{1}{8g} \int dt_1dt_2 \sum_{j=0,x,y,z}\xi_j(t_1)\: \tilde{\alpha}_K^{-1}(t_1-t_2)\:\xi_j(t_2)\:\:.\:\:
\end{aligned}}
\end{equation}
Now the route to the equations of motion is the same as before: one varies with respect to the quantum components $\theta_q$ and $\phi_q$. The only difference is that the Keldysh sector no longer appears as a quadratic nonlocal term, but as a linear coupling to the fluctuating field $\bm\xi$. As a consequence, the saddle-point equations become stochastic equations. Carrying out the variation, one finds
\begin{equation}
\begin{aligned}
    &\bullet \quad \frac{\delta\left(i S_{\mathrm{eff}}\right)}{\delta \theta_q(t)}=0\quad \xrightarrow[]{\quad} \quad \underbrace{S\dot{\phi}_c \sin \theta_c}_{\text{WZ}}-\underbrace{S\gamma B \sin \theta_c}_{\text{B-field}}-\underbrace{\frac{\pi g}{2} \dot{\theta}_c}_{\text{Adv/Ret}}=\underbrace{f_\phi}_{\text{K-noise}}\\[8pt]
    &\bullet \quad \frac{\delta\left(i S_{\mathrm{eff}}\right)}{\delta \phi_q(t)}=0\quad \xrightarrow[]{\quad} \quad -\underbrace{S \dot{\theta}_c \sin \theta_c}_{\text{WZ}}\:\:-\:\:\underbrace{\frac{\pi g}{2} \dot{\phi}_c \sin^2\theta_c}_{\text{Adv/Ret}}\:\:=\underbrace{f_\theta}_{\text{K-noise}}
\end{aligned}
\end{equation}
The meaning of these two equations is now very clear. The Wess-Zumino term provides the coherent precessional dynamics, the retarded/advanced AES sector generates the dissipative Gilbert-like contributions, and the Keldysh part appears as the fluctuating torque. In this sense, the nonequilibrium bath produces both friction and noise, exactly as one expects from a Langevin description, but here with a structure fixed by the geometry of the spin rotation matrix.

Finally, after dividing by the appropriate factors and rewriting the fluctuating terms explicitly in terms of the components of $\bm\xi$, one arrives at the semiclassical stochastic equations of motion:
\begin{equation}
    \boxed{\:\:\begin{aligned} &\qquad\quad\textbf{Adiabatic and Non-equilibrium Semiclassical EOM}\\[4pt]
\:\:(i)\quad &\boxed{\sin \theta_c\left(\dot{\phi}_c-\gamma B\right)-\frac{\pi g}{2S} \dot{\theta}_c  =\eta_\phi}\\
&\hspace{50pt}\text{with }\quad \eta_\phi
= -\tfrac{1}{2S}\cos\tfrac{\theta_c}{2}\big[\xi_x\sin(\phi_c-\tfrac{\chi_c}{2})-\xi_y\cos(\phi_c-\tfrac{\chi_c}{2})\big]
\\&\hspace{190pt}-\tfrac{1}{2S}\sin\tfrac{\theta_c}{2}\big[\xi_z\sin\tfrac{\chi_c}{2}-\xi_0\cos\tfrac{\chi_c}{2}\big]\:\:\:\\[10pt]
\:\:(ii)\quad &\boxed{\dot{\theta}_c+\frac{\pi g}{2S} \dot{\phi}_c\sin \theta_c   =\eta_\theta}\\
&\hspace{50pt}\text{with }\quad \eta_\theta = \tfrac{1}{2S}\cos\tfrac{\theta_c}{2}\big[\xi_x\cos(\phi_c-\tfrac{\chi_c}{2})+\xi_y\sin(\phi_c-\tfrac{\chi_c}{2})\big]\\&\hspace{190pt} - \tfrac{1}{2S}\sin\tfrac{\theta_c}{2}\big[\xi_z\cos\tfrac{\chi_c}{2}+\xi_0\sin\tfrac{\chi_c}{2}\big]\:\:.\:
\end{aligned}}
\end{equation}
These are precisely the adiabatic nonequilibrium analogues of the LLG-Langevin equations. The important point, however, is that the noise terms $\eta_\phi$ and $\eta_\theta$ are not structureless random forces. Their form is inherited from the geometry of the $SU(2)$ rotation, and therefore from the Berry-phase sector itself. This is the sense in which the noise is genuinely geometric: the stochastic torques generated by the electronic bath are filtered through the same rotational structure that governs the coherent spin dynamics.

%% file: chapter3_SuN_condensed_matter.tex
\subsection{SU(N) Condensed Matter Physics}\label{sec:su_n_condensed_matter}
Contrary to what many people might expect, $SU(N)$ symmetry groups with $N>2$ are not exclusive to high-energy physics. They also play a central role in condensed matter, where they provide a powerful language to describe systems with rich internal structure. For a long time, however, such high-symmetry models were regarded as idealized constructions—useful mainly as theoretical tools to simplify strongly interacting problems or to organize large-$N$ expansions. This perspective has gradually changed. Already in the early 2000s, it was realized that $SU(4)$ Heisenberg-type models can emerge as effective descriptions of materials with coupled spin and orbital (or pseudospin) degrees of freedom. In these systems, the internal Hilbert space is naturally enlarged, and the resulting symmetry is not imposed by hand, but appears as an approximate low-energy invariance. From this point of view, $SU(N)$ models are no longer just mathematical extensions of the familiar $SU(2)$ case, but rather controlled frameworks that capture genuinely new physical regimes where quantum fluctuations, entanglement, and competing orders are significantly enhanced. At the same time, it is worth mentioning that in the last decade $SU(N)$ symmetries have also become experimentally accessible in a much more direct way through ultracold alkaline-earth(-like) atomic gases, where the decoupling between nuclear and electronic degrees of freedom leads to nearly perfect realizations of $SU(N)$-symmetric interactions with $N$ as large as $10$ \cite{Gorshkov2010,Cazalilla2014, Zhang2014}. Although we will not cover these systems here, they provide a remarkable platform where many of the theoretical ideas discussed in this section can be tested in a highly controlled setting.

Under this framework, in this section, we will develop the role of $SU(N)$ symmetry from two complementary perspectives. First, we introduce $SU(N)$ Heisenberg models as a systematic generalization of quantum magnetism, emphasizing how large-$N$ limit provides access to quantum-disordered phases such as valence-bond solids and spin liquids. We then turn to the more physical setting of $SU(4)$ spin–orbital (or spin–pseudospin) systems, where the symmetry emerges from microscopic multi-orbital Hubbard models and leads naturally to Kugel–Khomskii-type exchange interactions. Finally, we show how these $SU(N)$ Hamiltonians can be reinterpreted in terms of higher-spin and multipolar exchange interactions, providing a direct bridge between abstract symmetry-based constructions and experimentally relevant degrees of freedom. 
Our goal throughout is not to present exhaustive derivations, but to build intuition for why higher $SU(N)$ symmetries are both natural and useful in condensed matter, and how they open the door to forms of quantum order that go beyond the conventional $SU(2)$ paradigm.

\subsubsection{SU(N) Heisenberg models}\label{sec:sun_heisenberg_models}
The study of quantum magnetism is full of fascinating phenomena such as frustration, exotic magnetic order, and fractional excitations. These effects become particularly rich in two-dimensional systems, where many of the standard tools—like exact methods in one dimension or simple approximations in high dimensions—tend to fail. Even numerical approaches are often limited by the rapid growth of the Hilbert space or by the infamous sign problem.

However, a natural way to gain insight into this complex landscape is to approach the problem from a different angle: instead of working strictly within the usual $SU(2)$ spin symmetry, one can generalize the internal symmetry of the system to $SU(N)$. In this context, the so-called large-$N$ approximation has become a widely used framework for studying strongly interacting quantum systems over the past few decades. Originally developed in high-energy physics, it was later adapted to condensed matter problems, where it has been applied to Kondo and Anderson impurity models, heavy-fermion systems, and even strongly correlated electrons in high-$T_c$ superconductors. In this sense, it is worth emphasizing that in many of these applications the underlying true physical symmetry remains the spin-$SU(2)$, so that $N$ is not literally large. Nevertheless, the $1/N$ expansion provides an easy method for obtaining simple mean field theories, which have been found to be surprisingly successful in capturing both ordered and disordered magnetic phases, offering valuable insights into the collective behavior of quantum spin systems even at modest values of $N$. That is how the large-$N$ approach should be understood: less as a quantitative limit and more as a powerful organizing principle for strongly correlated systems.

Generally speaking, an $SU(N)$ Heisenberg model is constructed by assigning to each lattice site an internal Hilbert space transforming under an $SU(N)$ symmetry (a representation space). Indeed, the motivation behind such constructions can be traced back to a central question in quantum magnetism: how can one destabilize conventional Néel order and access more exotic quantum phases? Inspired by the fact that tuning exchange constants in a model of interacting spins can destabilize the usual Néel antiferromagnetic order and potentially realize exotic quantum ground states, leading theoretical physicists in the late 1980s began thinking about models that mimic the physics of melting a Néel antiferromagnetic state. In the $SU(2)$ spin-$1/2$ Heisenberg model, in a square lattice for instance, 
\begin{equation}
\label{eq:SU(2)_Heisenberg_Hamiltonian}
    H=J \sum_{\langle i j\rangle} \vec{S}_i \cdot \vec{S}_j
\end{equation}
with close to one electron per lattice site (relevant to cuprates), it is well known that for nearest-neighbor exchange $J>0$, the ground state exhibits long-range Néel order at zero temperature. Now, introducing next-nearest-neighbor (N.N.) interactions
\begin{equation}
    H=J_1 \sum_{\langle i j\rangle} \vec{S}_i \cdot \vec{S}_j+J_2 \sum_{\langle\langle i j\rangle\rangle} \vec{S}_i \cdot \vec{S}_j\:,
\end{equation}
one can show that for special choices of $(J_2,J_1)$, the square lattice becomes frustrated: spins cannot simultaneously minimize all pairwise interactions, and Néel order is inhibited. Similarly, if instead of including the next-N.N. direct exchange interactions, one considers higher-order superexchange processes, as ring exchange,
\begin{equation}
    H=J \sum_{\langle i j\rangle} \vec{S}_i \cdot \vec{S}_j+K \sum_{\square}\left(P_{i j k l}+ \text{h.c.}\right)\:,
\end{equation}
where $P_{ijkl}$ cyclically permutes spins around a plaquette, one can show that large values of $K$ destabilize Néel order and drive the system toward highly entangled quantum states. These include the celebrated \textit{resonating valence bond} (RVB) states, which generalize the idea of singlet pairing beyond static dimers.

\noindent However, these mechanisms rely on fine-tuning interactions, and even then, quantum fluctuations are not always strong enough to fully suppress magnetic order—especially in unfrustrated two-dimensional lattices. This naturally leads to a more systematic question:
\begin{center}
\textit{Can we enhance quantum fluctuations in a controlled way to explore new ground states?}
\end{center}
A remarkably simple answer is:
\begin{center}
\textit{Replace $SU(2)$ spins by $SU(N)$ degrees of freedom and study the limit $N\to\infty$.}
\end{center}
At first sight, this may resemble the classical limit of $SU(2)$ obtained by taking $S\to\infty$, since both increase the number of states per site. However, the two limits are fundamentally different. In the $SU(2)$ case, increasing $S$ suppresses quantum fluctuations. Indeed, for the rescaled spin operators,
\begin{equation}
    \left[\frac{S^i}{S}, \frac{S^j}{S}\right]
    =\frac{i \epsilon^{i j k}}{S} \frac{S^k}{S},
\end{equation}
the commutator vanishes as $S\to\infty$, and the system becomes effectively classical. In contrast, the large-$N$ limit of $SU(N)$ introduces additional internal degrees of freedom—new ``flavors'' rather than larger spins. These extra channels allow for a proliferation of quantum correlations and entanglement, often enhancing rather than suppressing fluctuations. As a result, the system can access a much broader landscape of quantum-disordered phases, including spin liquids and valence-bond solids, which are difficult or impossible to realize within the conventional $SU(2)$ spin-$\tfrac{1}{2}$ framework.

\subsubsection*{Berry phases in 2D SU(2) antiferromagnets: hedgehogs and VBS structure}
However, there is a deeper perspective behind this idea which is often less emphasized. The motivation for extending $SU(2)$ to $SU(N)$ is not only to enhance quantum fluctuations, but rather to better understand the role of \emph{Berry phases} in interacting spin systems. As we will now see, once the problem is formulated in terms of a path integral, these geometric phases give rise to nontrivial topological structures that strongly constrain the low-energy physics. Once one moves from the operator formulation of Eq.\ref{eq:SU(2)_Heisenberg_Hamiltonian} to the spin-coherent state description, the partition function of the interacting spin system can be written as a path integral over unit vectors $\mathbf{\hat{n}}_j(\tau) \in S^2$ (and their homotopic extensions $\mathbf{\tilde{n}}_j$), where each vector represents the local orientation of a spin at site $j$. Therefore, the resulting action naturally splits into the corresponding dynamical and topological many-body parts
\begin{equation}
\label{eq:spin_interactive_action_solids}
    \mathrm{S}^{\text{2-dim}}_{\text{Heisnbrg}}=\underbrace{i S\int_0^\beta d\tau \int_0^1  du \:\sum_j \mathbf{\tilde{n}}_j \cdot\left(\partial_u \mathbf{\tilde{n}}_j \times \partial_\tau \mathbf{\tilde{n}}_j\right)}_{\text{topological}}\:\:\:-\:\:\:\underbrace{JS^2  \int_0^\beta d \tau \sum_{j} \sum_{\mu=\hat x,\hat y} \: \mathbf{\hat{n}}_j \cdot \mathbf{\hat{n}}_{j+e_\mu}}_{\text{dynamical}}\:\:.
\end{equation}
At this stage, the problem is still formulated on the lattice. However, to access the low-energy, long-wavelength physics, one can introduce a continuum description by assuming that the spin configuration varies smoothly in space. In an antiferromagnet, this requires some care: the ground state is staggered rather than uniform. Therefore, one separates the rapidly oscillating Néel structure from smooth fluctuations by writing
\begin{equation}
    \begin{gathered}
\hat{n}_j= (-1)^{j_x+j_y}\: \mathbf{n}\left(x_j\right) \sqrt{1-\left(\frac{a^2}{S}\right)^2 \mathbf{L}^2\left(x_j\right)}+\left(\frac{a^2}{S}\right) \mathbf{L}\left(x_j\right) \\
\text { with }\:\:\:|\mathbf{n}|^2(x)=1 \quad \text{ and } \quad  \mathbf{n}(x) \cdot \mathbf{L}(x)=0\:\:.
\end{gathered}
\end{equation}
Here, $\mathbf{n}(x, \tau)$ is the slowly varying Néel field, while the factor $(-1)^{j_x+j_y}$ restores the alternating order. The field $\mathbf{L}(x, \tau)$, on the other hand, describes small uniform canting fluctuations, and its prefactor $\frac{a^2}{S}$ makes explicit that it is parametrically small in the continuum limit (in fact carries the meaning of a smooth magnetization density per unit cell). The constraints ensure that $\mathbf{n}$ remains a unit vector and that $\mathbf{L}$ lies in the tangent plane, so that fluctuations only affect the orientation of the spin. In this way, making use of the gradient expansion, one can systematically expand the fields $\mathbf{n}\left(x_j+\hat{e}_\mu\right)$ and $\mathbf{L}\left(x_j+\hat{e}_\mu\right)$ around $x_j$, retaining only the leading contributions in powers of the lattice spacing $a$. This procedure introduces two small parameters: the spatial gradients (controlled by $a$ ) and the canting amplitude $\frac{a^2}{S} \mathbf{L}$, both of which are assumed to be small in the low-energy limit. At the same time, the constraint $|\mathbf{n}(x)|^2=1$ can be implemented through a Lagrange multiplier field $\lambda(x, \tau)$, while $\mathbf{L}$ appears only quadratically in the action. As a consequence, $\mathbf{L}$ acts as an auxiliary field and can be integrated out exactly. Performing this procedure, together with the continuum limit $\sum_j \rightarrow \int d^2 x / a^2$, one arrives at an effective action written entirely in terms of the Néel field $\mathbf{n}(x, \tau)$. 

\noindent The dynamical contribution then reduces to the well-known $\bm{O(3)}$ \textbf{nonlinear sigma model} (NLSM),
\begin{equation}
\boxed{
S_{\text{dyn}}[\mathbf{n}]
=\frac{1}{2g}\int_{0}^\beta d\tau \int d^2x \big[
(\partial_\tau \mathbf{n})^2 + c^2 (\nabla \mathbf{n})^2
\big],
\qquad \text{with} \quad |\mathbf{n}|^2=1}\:\:,
\end{equation}
where the coupling constant $g$ and velocity 
$c$ are functions of the microscopic parameters. So, at this level, the physics is entirely controlled by smooth fluctuations of the Néel field, and the resulting theory describes the stability of long-range antiferromagnetic order. However—and this is the crucial point—the topological term does not reduce entirely to a smooth continuum expression. While in the derivation of the above $O(3)$ NLSM, there is indeed a continuous term coming from the original topological action in Eq.\ref{eq:spin_interactive_action_solids}, the AFM gradient expansion of such a term produces an additional, intrinsically lattice-scale contribution:
\begin{equation}
\begin{aligned}
\label{eq:berry_phase_interactive_spins}
&\boxed{\mathrm{S}_{\text{top}}\:\:\xRightarrow[]{\quad}\:\: \mathrm{S}^\prime_{\text{top}}=
i4\pi S \:\sum_j (-1)^{j_x+j_y}\:W_0[\mathbf{\tilde n}(x_j)]} \\
\text{with }\:\:\:W_0[&\mathbf{\tilde n}(x_j)]:=\frac{1}{4\pi}\int_0^\beta d\tau \int_0^1 du\:\:\: 
\mathbf{\tilde n}(x_j)\cdot
\Big[\partial_u \mathbf{\tilde n}(x_j) \times \partial_\tau \mathbf{\tilde n}(x_j)\Big]\:.
\end{aligned}
\end{equation}
Here, $W_0[\mathbf{\tilde n}]$ is indeed the topological winding number associated with the mapping of the homotopically extended site configurations into the sphere ($\mathbf{\tilde n}(x_j): S^1\times [0,1]\to S^2$). 

Hence, this is exactly where the structure of the theory becomes particularly illuminating. On the one hand, the dynamical NLSM favors smooth configurations of the Néel field and therefore stabilizes antiferromagnetic order. On the other hand, the Berry phase contribution is purely topological and depends only on global properties of the field configurations. The natural question is then:
\begin{center}
\textit{Does this lattice topological term affect the low-energy physics, or is it irrelevant?}
\end{center}
\noindent The answer depends crucially on dimensionality:
\begin{itemize}
    \item In one spatial dimension (not the case in previous expressions), the sum over sites can be converted into a smooth integral, and the topological term reduces to the familiar $\theta$-term of the $O(3)$ NLSM,
\begin{equation}
S_{\text{top}}^{(1+1)}= i\theta W[\mathbf n], \qquad \theta=2\pi S,
\end{equation}
with $W\in\mathbb Z$ the winding number of the map $\mathbf n: S^2\to S^2$. This immediately leads to the celebrated distinction between integer and half-integer spins, where destructive interference of topological sectors gives rise to qualitatively different ground states.
\item In two spatial dimensions, however, the situation is more subtle. The staggered sum in $\mathrm S^\prime_{\text{top}}$ cannot be written as a simple continuum topological invariant. Instead, its contribution depends sensitively on the spacetime structure of the field configurations. In particular, one finds that for \emph{smooth} configurations of $\mathbf n(x,y,\tau)$, the topological term vanishes identically:
\begin{equation}
\mathrm S^\prime_{\text{top}}\big[\mathbf n(x,y,\tau)\big]=0 \qquad \text{(smooth configurations)}\:,
\end{equation}
which means that, at the level of smooth fluctuations, the system is entirely described by the $O(3)$ NLSM, and the Néel ordered state emerges as the natural ground state.
\end{itemize}

\noindent But this is not the end of the story for the 2D case. In particular, if the Berry phase term is to have any physical effect, it must originate from configurations that go beyond the smooth continuum description. The problem is that one may wonder under what physical conditions it is justified to go beyond smooth field configurations, since after all, the continuum description was introduced precisely under the assumption that the Néel field varies slowly in space and time. 

\noindent The first important observation is that the field $\mathbf{n}(x,\tau)$ is not a fundamental microscopic variable, but rather a coarse-grained description of the underlying lattice spins. Strictly speaking, the only degrees of freedom that are physically well-defined are the spin orientations at the lattice sites ${\mathbf{x}_j}$. The continuum field $\mathbf{n}(x,\tau)$ is therefore an interpolation, and as such, it is not required to remain well-defined everywhere in spacetime. In particular, there is no obstruction to configurations in which $\mathbf{n}(x,\tau)$ becomes ill-defined at isolated points located, for instance, at the centers of plaquettes in the dual lattice. A second, equally important point is that the path integral is formulated in \emph{imaginary time}. This means that we are not describing a single physical trajectory, but rather summing over all possible configurations of the field $\mathbf{n}(x,\tau)$ compatible with the boundary conditions. From this perspective, there is no reason to restrict the functional integral to strictly continuous configurations. On the contrary, configurations that are smooth almost everywhere but develop localized singularities are perfectly admissible, and—as we have just seen—are in fact the only ones capable of producing a nontrivial contribution to the topological term.

These considerations naturally lead us to enlarge the space of field configurations and allow for \emph{isolated singular events in spacetime}. In the present context, such singularities occur at points $(x_0,y_0,\tau_0)$ where the Néel field becomes ill-defined. Geometrically, these configurations correspond to the so-called \textbf{hedgehog singularities}, which can be interpreted as spacetime monopoles of the order parameter field. To understand their role, it is useful to recall that at each fixed time slice $\tau$, the field defines a map
$\mathbf n_\tau:\mathbb{T}^2 \rightarrow S^2$,
which is characterized by an integer topological invariant—the \emph{skyrmion number},
\begin{equation}
\mathrm{Skyr}[\mathbf n_\tau]
=\frac{1}{4\pi}\int d^2x \:\:
\mathbf n_\tau \cdot \big[\partial_x \mathbf n_\tau \times \partial_y \mathbf n_\tau\big]\:\:\:\in\mathbb Z\:\:.
\end{equation}
Hence, for smooth evolution in $\tau$, this integer cannot change, and in this line, a hedgehog event precisely corresponds to a spacetime point where this conservation law is violated. In other words, hedgehogs act as tunneling events between configurations with different skyrmion numbers:
\begin{equation}
\Delta \mathrm{Skyr} = \pm Q.
\end{equation}
Now, in order to compute the staggered sum in Eq.\ref{eq:berry_phase_interactive_spins}, it is convenient to introduce the solid-angle variable
$\mathrm A(x_j):=4\pi\,W_0[\tilde{\mathbf n}(x_j)]$,
so that the Berry phase term becomes
\begin{equation}
\label{eq:alternating_sum_of_enclosed_areas_better}
S'_{\mathrm{top}}
=
iS\sum_{j_x,j_y}(-1)^{j_x+j_y}\,\mathrm A(x_j)\:.
\end{equation}
Thus, the problem reduces to understanding how the enclosed solid angles $\mathrm A(x_j)$ are modified in the presence of hedgehog events. A useful point of view is to regard $e^{i\mathrm A(\cdot)}$ as a phase field taking values on a circle of length $4\pi$, since solid angles are defined only modulo $4\pi$. In this sense, a hedgehog event of charge $Q$ located at $(x_0, y_0, \tau_0)$ appears as a vortex-like singularity of $\mathrm A(\cdot)$, namely
\begin{equation}
\frac{1}{4\pi}\oint_{C\ni x_0}\nabla \mathrm A\cdot d\mathbf x
=
Q \quad \xRightarrow[]{\quad}
\quad
\Delta \mathrm A=\pm 4\pi Q,
\end{equation}
where $C$ is a small loop enclosing the singularity. Therefore, once a hedgehog is present, $\mathrm A(\cdot)$ is no longer a single-valued function, and one must introduce branch cuts across which the solid angle jumps by multiples of $4\pi$.

This is the crucial observation: the staggered sum in Eq.\ref{eq:alternating_sum_of_enclosed_areas_better} receives no contribution from smooth regions, but only from those bonds crossed by the branch cuts connecting skyrmions and antiskyrmions. Furthermore, since such singularities cannot sit on the original spin sites, but only in between them, it is natural to reorganize the sum over $j$-sites into a sum over plaquettes $P_n$ of the dual lattice. In this way, one finds
\begin{equation}
\sum_j(-1)^{j_x+j_y}\mathrm A(x_j)
=
\frac{1}{4}\sum_{P_n}\big(\text{contribution of plaquette }P_n\big),
\end{equation}
where only plaquettes pierced by a hedgehog event contribute. At this point, the only remaining information is the position of the hedgehog on the dual lattice. Because the dual lattice splits into four inequivalent plaquette classes, each singular plaquette $P_n$ carries a phase label $\eta_n=0,1,2,3$, or equivalently
\begin{equation}
\zeta_n:=e^{i\pi\eta_n/2}
\in
\{1,\;i,\;-1,\;-i\},
\qquad
\eta_n=
\begin{cases}
0,&(\mathrm{even},\mathrm{even})\\
1,&(\mathrm{even},\mathrm{odd})\\
2,&(\mathrm{odd},\mathrm{odd})\\
3,&(\mathrm{odd},\mathrm{even})
\end{cases}.
\end{equation}
Accordingly, the contribution of a plaquette containing a hedgehog of charge $Q_n$ can be written, modulo $4\pi$, as \cite{haldane1988}
\begin{equation}
\text{(contribution of }P_n)\:
=\:
4\pi(\eta_n+4k)\,Q_n ,
\qquad k\in\mathbb Z,
\end{equation}
so that the full staggered sum and its exponential weight become
\begin{equation}
\begin{aligned}
    \sum_{j}(-1)^{j_x+j_y}\mathrm A(x_j)
=
\frac{1}{4}\sum_n4\pi(\eta_n+4k)\,Q_n\\
\qquad
e^{-S'_{\mathrm{top}}}
=
e^{-iS\sum_n\pi(\eta_n+4k)Q_n}
=
\prod_n
\left(e^{i\frac{\pi}{2}\eta_n}\right)^{-2SQ_n}.
\end{aligned}
\end{equation}
In other words, each hedgehog event carries a local Berry phase factor determined only by which one of the four dual-sublattice plaquettes it occupies:
\begin{equation}
\boxed{
e^{-S'_{\mathrm{top}}}
=
\prod_{n\:\in\:\text{singular plaq.}}
\left(\zeta_n\right)^{2SQ_n},
\qquad \text{ with }\quad 
\zeta_n\in\{1,i,-1,-i\}
}\:\:.
\end{equation}
This is the essential result. The lattice Berry phase turns the gas of hedgehog tunneling events into an interference problem: different singular configurations do not simply contribute with the same sign, but acquire position-dependent phases $1,i,-1,-i$, making an interference pattern that may destabilize the AFM order and could lead to a new quantum disordered phase. In particular, as long as the hedgehog events are energetically costly and therefore strongly suppressed, the system remains deep in the Néel phase. In that regime, the dominant configurations in the path integral are smooth, and the long-wavelength physics is well captured by the $O(3)$ NLSM alone. However, once these singular tunneling events may proliferate, the lattice Berry phase interferences become decisive, and the antiferromagnetic order is destabilized, making the system enter a quantum-disordered regime.

This is the general setting in which \textbf{valence-bond-solid (VBS) order} naturally emerges. In fact, one of the central lessons of the modern \emph{deconfined quantum criticality} (DQCP) picture is that the fate of the disordered phase cannot be understood solely from the smooth N\'eel fluctuations: one must also take into account the quantum interference between proliferating hedgehog events. In the N\'eel phase they are irrelevant; in the disordered phase they become the relevant topological objects. In this line, the partition function must be understood as a sum over all allowed spacetime configurations of smooth fields, together with their possible hedgehog and anti-hedgehog events,
\begin{equation}
    Z=\sum_{\text{all configurations}} e^{-S'_{\mathrm{top}}}\,e^{-S_{\mathrm{dyn}}}\:,
\end{equation}
where a given singular configuration is specified by the number of singular events $N_{\text{hed}}$, their topological charges $Q_n$, and their positions on the dual lattice, encoded by the phase labels $\zeta_n$ --- the last point being essential, since the Berry phase factor depends precisely on which dual-sublattice plaquettes they occupy. In addition, because of the periodicity in imaginary time, the total topological charge must vanish, and therefore hedgehogs and anti-hedgehogs must appear in equal numbers. 

\noindent With all of these specifications, if one factors out the smooth dynamical contribution, the topological part of the partition function takes the schematic form
\begin{equation}
    Z \propto \sum_{\substack{
    \frac{N_{\mathrm{hed}}}{2}=\# \text{ hedgehogs}\\
    \frac{N_{\mathrm{hed}}}{2}=\# \text{ anti-hedgehogs}
    }}
    \:\sum_{\substack{\text{position phase}\\\zeta_n=1,i,-1,-i}}
    \:\:\prod_{\substack{\text{plaquettes } \{n\}\\ \text{with hedgehogs}}}
    \left(\zeta_n\right)^{2SQ_n},
\end{equation}
with $Q_n\in\mathbb{Z}$ the topological charge of each singular event (change of skyrmion number), and $\zeta_n\in\{1,i,-1,-i\}$ encoding the position of the hedgehog on the dual lattice. Now, deep in the quantum-disordered VBS phase, where these singularities proliferate, this Berry-phase factor is precisely the most relevant part for determining which hedgehog configurations survive after summation. Different configurations do not add coherently: depending on the values of $S,Q_n$, the phases $1,i,-1,-i$ can cancel each other for certain periodicities of the hedgehog distribution, while reinforcing others. Therefore, the proliferated phase is not arbitrary --- its symmetry and degeneracy are selected by the interference pattern imposed by the Berry phases.

To make this mechanism explicit, it is sufficient to consider the simplest nontrivial sector, namely configurations with one hedgehog–anti-hedgehog pair,
\begin{equation}
    \begin{cases}
        N_{\mathrm{hed}}=2\\
        Q_1=+Q\\
        Q_2=-Q
    \end{cases}
    \quad \Longrightarrow \quad
    e^{-S'_{\mathrm{top}}}
    =
    (\zeta_1)^{2S Q}\,(\zeta_2)^{-2S Q}\:.
\end{equation}
In this case, summing over all possible positions of the pair amounts to summing independently over the four dual-sublattice phases,
\begin{equation}
    \sum_{\zeta_1} (\zeta_1)^{\,2SQ}
    =
    (1)^{2SQ}+(-1)^{2SQ}+(i)^{2SQ}+(-i)^{2SQ}\:.
\end{equation}
A straightforward evaluation shows that this sum vanishes unless
\begin{equation}
    2SQ \equiv 0 \quad (\mathrm{mod}\;4)\:.
\end{equation}
Therefore, only those tunneling processes for which the total Berry phase is a multiple of $2\pi$ survive the sum over positions, while all others are suppressed by destructive interference. This simple condition indeed has immediate and profound consequences. Depending on the value of the spin $S$, it imposes strong constraints on the allowed values of the topological charge $Q$:
\begin{center}
\begin{tabular}{c|c|c}
\textbf{Spin type} & \textbf{Condition on $Q$} & \textbf{Allowed values}\\
\hline
Half-integer & $Q \equiv 0 \;(\mathrm{mod}\;4)$ & $\pm4,\pm8,\dots$\\
Odd integer & $Q \equiv 0 \;(\mathrm{mod}\;2)$ & $\pm2,\pm4,\dots$\\
Even integer & no restriction & any $Q\in\mathbb{Z}$
\end{tabular}
\end{center}
In other words, the Berry phase acts as a selection rule that suppresses low-charge hedgehog events in a spin-dependent way. The minimal allowed value of $Q$ --- that is, the smallest tunneling process that survives interference --- depends crucially on $S$. Moreover, since these hedgehog events correspond to tunneling processes between configurations with different skyrmion numbers, the minimal allowed value of $Q$ sets the natural periodicity with which the system can fluctuate between topological sectors. As a consequence, it determines the \emph{number of distinct configurations that remain degenerate under such tunneling processes}. Importantly, these degeneracies should be understood as degeneracies of \emph{low-lying states}, rather than strictly of the ground state. The reason is that the above analysis is formulated within the semiclassical path-integral description, where one captures the dominant tunneling processes between nearly degenerate configurations. The true ground state may be a symmetric superposition of these configurations, but the Berry phase structure ensures that only certain patterns --- with a characteristic multiplicity fixed by the minimal allowed $Q$ --- can appear at low energies.
It is precisely in this sense that the Berry phase selects the structure of the quantum-disordered phase: by suppressing certain tunneling processes and allowing others, it enforces a discrete set of competing configurations, which ultimately manifest as the characteristic degeneracies and symmetry-breaking patterns of valence-bond-solid states.

The lesson of this $SU(2)$ analysis is therefore clear: the nontrivial structure of the disordered phase is ultimately fixed by Berry-phase interference between singular topological events. The next question is how to generalize this physics to a framework where it can be studied more systematically.

\subsubsection*{Large-N generalization as a systematic treatment of Berry phase terms}
At this point, it is worth stepping back and reconnecting with the original motivation behind the $SU(N)$ generalization. The discussion above makes it clear that the low-energy physics of quantum antiferromagnets is not controlled solely by smooth fluctuations of the order parameter, but crucially by the structure of Berry phases associated with singular tunneling events. In the $SU(2)$ case, these effects already lead to highly nontrivial interference phenomena, ultimately selecting the structure of the quantum-disordered phase. However, within $SU(2)$, the treatment of these topological effects remains rather indirect. The Berry phase appears as a discrete lattice contribution, and its consequences must be inferred through semiclassical arguments such as the ones presented above. In this sense, one of the central motivations for extending the symmetry to $SU(N)$ is that it provides a framework in which both the smooth dynamics and the topological contributions can be treated on equal footing in a controlled approximation.

Indeed, $SU(N)$ spin systems admit a natural formulation in terms of parton constructions, where the spin degrees of freedom are fractionalized into auxiliary fermionic or bosonic fields subject to a local constraint. In this representation, the emergence of gauge structures becomes explicit, and the corresponding effective theories naturally take the form of NLSM or gauge theories supplemented by Berry phase terms. Importantly, in the large-$N$ limit, fluctuations around saddle-point configurations are systematically suppressed, allowing for a controlled analysis of both the dynamical and topological sectors. From this perspective, the Berry phase effects that, in the $SU(2)$ case, manifest as subtle interference patterns between hedgehog events, are reinterpreted in $SU(N)$ as topological terms associated with emergent gauge fields and fractionalized excitations. In particular, the proliferation or suppression of monopole-like events --- the direct analogs of hedgehogs --- can be studied systematically within the large-$N$ expansion, providing a powerful handle on the stability of different quantum phases.

In this way, the $SU(N)$ generalization should be understood not merely as a method to enhance quantum fluctuations, but as a framework in which the interplay between geometry, topology, and strong correlations can be explored in a controlled and transparent manner. The rich structure uncovered above in the $SU(2)$ case thus finds a natural and more tractable extension in the large-$N$ limit, where both conventional order and exotic quantum-disordered phases can be analyzed on equal footing.

\subsubsection*{Constructing SU(N) Hamiltonians: symmetry and representation-theoretic perspective}

Having motivated the $SU(N)$ generalization, it is time to ask how to generalize the ideas of $SU(2)$ antiferromagnets in the presence of Berry phase fluctuations to higher symmetries. Here, rather than introducing the $SU(N)$ structure directly, it is useful to first revisit the $SU(2)$ case from a more structural point of view. The essential ingredient of any $SU(2)$ Heisenberg model is not its explicit form in terms of spin components, but rather the way symmetry is implemented. On each lattice site, the local Hilbert space transforms as a representation of $SU(2)$ (for instance, spin-$S$), and the interaction between neighboring sites is constructed to be invariant under global $SU(2)$ rotations. Explicitly, the Hamiltonian 
\begin{equation}
    H = J \sum_{\langle i j\rangle}H_{ij}=J \sum_{\langle i j\rangle} \vec{S}_i \cdot \vec{S}_j \quad \xrightarrow[]{\:\:\text{has in general}\:\:} \quad H_{ij} = \sum_{a,b=1}^{3} c_{ab} \:S_i^a S_j^b\:,
\end{equation}
which is nothing but the contraction of generators of the $\mathfrak{su}(2)$ Lie algebra on neighboring sites. Here, the coefficients $c_{ab}$ account for the different expressions that the bond interaction term $H_{ij}$ admits depending on the chosen basis of $\mathfrak{su}(2)$:
\begin{equation}
    \begin{cases}
        (S^x,S^y,S^z) \quad \Longrightarrow \quad H_{ij}=S_i^xS_j^x + S_i^yS_j^y +S_i^zS_j^z\\
        (S^+,S^-,S^z) \quad \Longrightarrow \:\:\: H_{ij}= \frac{1}{2} (S_i^{+} S_j^{-}+S_i^{-} S_j^{+}+2 S_i^z S_j^z)\:.
    \end{cases}
\end{equation}
Now, beyond symmetry invariance, there is a second, equally important property. On a bipartite lattice, the antiferromagnetic interaction favors configurations in which neighboring spins combine into singlets. In representation-theoretic terms, this means that if the local Hilbert spaces at sites $i$ and $j$ are $\mathcal{H}_A$ and $\mathcal{H}_B$, respectively, then their tensor product contains the trivial representation (spin $S=0$ irrep),
\begin{equation}
    \mathcal{H}_{ij}
    =
    \mathcal{H}_A \otimes \mathcal{H}_B
    =
    \mathbf{1} \oplus \dots \:,
\end{equation}
and the interaction energetically favors the singlet sector. Note that in the conventional $SU(2)$ homogeneous antiferromagnet, where all lattice sites carry the same spin-$S$ representation,  this requirement is automatically satisfied since the addition of angular momentum guarantees 
\begin{equation}
    \mathcal{H}_S \otimes \mathcal{H}_S
    =
    \mathcal{H}_0 \oplus \mathcal{H}_1 \oplus \cdots \oplus \mathcal{H}_{2S},
\end{equation}
so that the singlet sector $\mathcal{H}_0\simeq\mathbf{1}$ is always present.

This observation provides the guiding principle for constructing $SU(N)$ generalizations. We seek a Hamiltonian satisfying two key requirements:
\begin{align*}
    \begin{cases}
        \:\:\bullet \quad \text{it is invariant under global } SU(N) \text{ transformations},\\
     \:\:\bullet \quad\text{it favors the formation of singlets between neighboring sites}.
    \end{cases}
\end{align*}
Then, a natural starting point is therefore to consider interactions of the form
\begin{equation}
\label{eq:ansatz_for_su(n)_hamilto}
    H_{ij}^{SU(N)}
    =
    \sum_{a=1}^{N^2-1}
    c_{ab} \,
    T_i^a \,
    \overline{T}_j^{\,b},
\end{equation}
where $\{T^a\}$ and $\{\overline{T}^b\}$ are generators of $\mathfrak{su}(N)$ acting on the local Hilbert representation spaces of sites $i$ and $j$, respectively. However, not all choices of representations lead to a nontrivial singlet sector. In contrast to the $SU(2)$ case, where any two identical spin representations can combine into a singlet, for $SU(N)$ the existence of a singlet in the tensor product is highly restrictive. In particular, the tensor product $\mathcal{H}_A \otimes \mathcal{H}_B$ contains the trivial representation only if $\mathcal{H}_B$ is the conjugate representation of $\mathcal{H}_A$.

\noindent This immediately leads to a natural bipartite construction: one assigns a $SU(N)$ representation $\mathcal{H}_A$ to sublattice $A$, and its conjugate representation $\overline{\mathcal{H}_A}$ to sublattice $B$. In this way, any bond Hilbert space decomposes as
\begin{equation}
    \mathcal{H}_{ij}
    =
    \mathcal{H}_A
    \otimes
    \overline{\mathcal{H}_A}
    =
    \mathbf{1} \oplus \dots \:,
\end{equation}
and the interaction can favor the projection onto the singlet sector, in direct analogy with the $SU(2)$ antiferromagnet. Indeed, this is the fundamental structural difference between $SU(2)$ and $SU(N)$ magnetism: while in $SU(2)$ the singlet appears automatically in the tensor product of identical representations, in $SU(N)$ its existence requires pairing representations with their conjugates. As we will now see, implementing this construction in a concrete and tractable way naturally leads to the parton formulation of $SU(N)$ spin systems.

The next question is how to realize the generators $T^a$ and $\overline{T}^a$ in Eq.\ref{eq:ansatz_for_su(n)_hamilto} explicitly, and in a way that it can be generalized for all $SU(N)$. A natural starting point then is given by the matrix units of $U(N)$,
\begin{equation}
    (E_{\alpha\beta})_{\mu\nu}=\delta_{\alpha\mu}\delta_{\beta\nu},
    \qquad \alpha,\beta=1,\dots,N,
\end{equation}
which form a complete basis of operators on the local Hilbert space for the case of the $SU(N)$ fundamental representation $\mathcal{H}_A=\mathbb{C}^N=\bm N$. In this language, a useful observation is that the non-diagonal matrices $E_{\alpha\beta}$ with $\alpha\neq\beta$ already behave as the raising and lowering generators of $\mathfrak{su}(N)$. The only subtlety arises in the diagonal sector. While $U(N)$ contains $N$ diagonal generators $E_{\alpha\alpha}$, the algebra $\mathfrak{su}(N)$ requires them to be traceless. This is achieved by removing the identity component and defining the operators 
\begin{equation}
    \hat S^\beta_{\alpha}
    :=
    E_{\alpha\beta}-\frac{\delta_{\alpha\beta}}{N}\,\mathbb{1}_N,
    \qquad
    \alpha,\beta=1,\dots,N,
\end{equation}
so that all generators are now traceless. This is exactly where the notation gets confusing. At first sight, this construction seems to produce $N^2$ generators: $N(N-1)$ off-diagonal ones and $N$ diagonal ones. However, everything makes sense once one recalls that the set of diagonal operators $\{\hat S^\alpha_{\alpha}\}$ is not an independent set, since they satisfy the constraint
\begin{equation}
    \sum_{\alpha=1}^N \hat S^\alpha_{\:\alpha}=0,
\end{equation}
and therefore only $N-1$ of them are linearly independent, as required for $\mathfrak{su}(N)$. To clarify this point, it is valuable to see how this structure becomes particularly transparent in the familiar $SU(2)$ case. There, the non-diagonal and diagonal generators take the respective form
\begin{equation}
\begin{aligned}
\begin{array}{cc}
\hat S^2_{1}=E_{12}= \begin{pmatrix} 0 & 1\\ 0 & 0 \end{pmatrix}=S^+
&\qquad \hat S^1_{1}=E_{11}-\tfrac{1}{2}\mathbb{1}_2=
\tfrac{1}{2}\begin{pmatrix}1 & 0\\0 & -1\end{pmatrix}=S^z
\\[10pt]\hat S^1_{2}=E_{21}=\begin{pmatrix} 0 & 0\\ 1 & 0\end{pmatrix}=S^-
&\qquad \hat S^2_{2}=E_{22}-\tfrac{1}{2}\mathbb{1}_2=\tfrac{1}{2}\begin{pmatrix}-1 & 0\\0 & 1\end{pmatrix}=-S^z\:\:,
\end{array}
\end{aligned}
\end{equation}
so that only one single independent diagonal operator remains: $S^z$. In fact, combining these contributions, one recovers the usual longitudinal interaction as
\begin{equation}
    2S_i^z S_j^z
    =
    \hat S^1_{1}(i)\hat S^1_{1}(j)
    +
    \hat S^2_{2}(i)\hat S^2_{2}(j),
\end{equation}
and therefore, the standard form of the $SU(2)$ Heisenberg Hamiltonian can be rewritten as
\begin{equation}
    H_{ij}= \frac{1}{2} (S_i^{+} S_j^{-}+S_i^{-} S_j^{+}+2 S_i^z S_j^z)=\frac{1}{2} \sum_{\alpha,\beta=1}^{2}
    \hat S^\beta_{\alpha}(i)\,
    \hat{S}^\alpha_{\beta}(j)\:\:,
\end{equation}
showing explicitly how the apparently redundant diagonal sector reorganizes into the standard $SU(2)$ structure.

Hence, motivated by this observation, it is natural to rewrite the interaction bond term in a unified index notation. However, in doing so, one must be careful to keep track of the representation space on which the generators act. In the $SU(N)$ case, the operators on the two sublattices belong to conjugate representations, and therefore they are not identical objects. To make this explicit, let us denote by $\rho_i$ and $\rho_j$ the representations acting on the local Hilbert spaces $\mathcal{H}_i$ and $\mathcal{H}_j=\overline{\mathcal{H}_i}$, respectively. Then, the bond interaction should be understood as an operator on the tensor product space $\mathcal{H}_i\otimes \overline{\mathcal{H}_i}$ of the form
\begin{equation}
    H_{ij}=\frac{1}{N}
    \sum_{\alpha,\beta=1}^{N}
    \rho_i\big(\hat S^\beta_{\alpha}\big)
    \otimes
    \rho_j\big(\hat S^\alpha_{\beta}\big)\:\:.
\end{equation}
In this expression, even though the same algebraic generators $\hat S^\beta_{\alpha}$ defined in the fundamental irrep $\mathbb{C}^N$ appear on both sites, they are represented differently: on sublattice of $i$-site (say $A$), they act in a representation $\mathcal{H}_A$, while on sublattice of $j$-site (say $B)$ they act in the conjugate one $\overline{\mathcal{H}}_A$. In practice, one often suppresses the explicit notation of the representation maps and simply writes
\begin{equation}
    H_{ij}=\frac{1}{N}
    \sum_{\alpha,\beta=1}^{N}
    S^\beta_{\alpha}(i)\,
    \bar{S}^\alpha_{\beta}(j),
\end{equation}
with the understanding that $S(i)$ and $\bar S(j)$ act in conjugate representations on the two sublattices. In this form, the Hamiltonian makes no explicit reference to a particular basis of generators and treats diagonal and off-diagonal components on equal footing, while still encoding the crucial bipartite structure. This is summarized as
\begin{equation} \label{eq:su_N_heisenberg}
\boxed{\:
    \begin{aligned}
    \textbf{SU(N) He}&\textbf{isenberg model}\\
    &H^{SU(N)}=\frac{J}{N}\sum_{\alpha,\beta=1}^{N}
    S^\beta_{\alpha}(i)\,
    \bar{S}^\alpha_{\beta}(j)\\[5pt]
    \text{ with }\:\:S^\beta_{\alpha}
    :&=
    E_{\alpha\beta}-\frac{\delta_{\alpha\beta}}{N}\,\mathbb{1}_N \:\:\text{ and } \:\: S^\beta_{\alpha}(i):=\rho_i(S^\beta_\alpha)\:.\:
    \end{aligned}
}
\end{equation}
It is worth pausing at this point to make explicit how the above expression indeed reproduces the expected structure of an $SU(N)$ invariant interaction. In particular, one can naturally separate the diagonal and non-diagonal contributions,
\begin{equation}
\begin{aligned}
    H_{ij} =\frac{1}{N}\Bigg[\sum_{\alpha\neq\beta}\rho_i\big(\hat S^\beta_{\alpha}\big)\otimes\rho_j\big(\hat S^\alpha_{\beta}\big)+\sum_{\alpha=1}^{N}\rho_i\big(\hat S^\alpha_{\alpha}\big) \otimes\rho_j\big(\hat S^\alpha_{\alpha}\big)\Bigg]\:.
\end{aligned}
\end{equation}
The first term corresponds to the non-diagonal generators, which play the role of raising and lowering operators, directly generalizing the $S^+S^-$ structure of $SU(2)$. The second term contains the diagonal sector. However, as discussed above, the set $\{\hat S^\alpha_{\alpha}\}$ is constrained and spans only an $(N-1)$-dimensional subspace. Therefore, one can always choose a basis $\{T^a\}_{a=1}^{N-1}$ of traceless diagonal generators such that
\begin{equation}
    \sum_{\alpha=1}^{N}
    \rho_i\big(\hat S^\alpha_{\alpha}\big)
    \rho_j\big(\hat S^\alpha_{\alpha}\big)
    \;\Longrightarrow\;
    \sum_{a=1}^{N-1}
    c_a\,
    \rho_i(T^a)\,
    \rho_j(T^a)\:,
\end{equation}
for some coefficients $c_a$ that depend on the chosen basis. In this way, the Hamiltonian can be understood as the contraction of all generators of $\mathfrak{su}(N)$,
\begin{equation}
    H_{ij}
    \;\propto\;
    \text{(raising/lowering terms)}
    +
    \sum_{a=1}^{N-1}
    c_a\,
    \rho_i(T^a)\,
    \rho_j(T^a)\:,
\end{equation}
making explicit its $SU(N)$ invariant structure. A final remark concerns the precise normalization of the diagonal generators. Suppose that instead of defining the traceless operators as
$$
    \hat S^\beta_{\alpha}=E_{\alpha\beta}-\frac{\delta_{\alpha\beta}}{N}\,\mathbb{1}_N \quad \xrightarrow[\:\text{ different subtraction }\:]{\:\text{one considers slightly }\:} \quad \hat S^\beta_{\alpha}
    =E_{\alpha\beta}-\frac{\delta_{\alpha\beta}}{2}\,\mathbb{1}_N\:.
$$
In that case, the generators are no longer strictly traceless, and the diagonal sector acquires an additional contribution proportional to the identity. As a consequence, the Hamiltonian is modified only by a constant shift,
\begin{equation}
    H_{ij}
    \;\longrightarrow\;
    H_{ij}
    +
    d\mathbb{1}\:\:,
\end{equation}
which has no effect on the dynamics and can be safely ignored. Therefore, different choices of normalization in the diagonal sector are physically equivalent, up to an irrelevant additive constant.

\subsubsection*{Fermionic and Bosonic parton representations}
With this understanding, the structure of the $SU(N)$ Heisenberg interaction is now completely clear: it is the natural contraction of generators acting in conjugate representations, with raising/lowering processes and diagonal interactions treated on equal footing. Indeed, the previous discussion makes clear that everything is ultimately controlled by the algebra generated by the operators $\hat S^\beta_{\alpha}$. Therefore, rather than introducing new degrees of freedom in an ad hoc way, a more natural strategy is to look for explicit realizations of these generators in terms of more elementary operators. A key observation is that the matrix units $E_{\alpha\beta}$ themselves admit a simple bilinear representation. Indeed, consider a set of operators $a_\alpha(i)$ carrying a flavor index $\alpha=1,\dots,N$, and define
\begin{equation}
    E_{\alpha\beta}(i)
    \;=\;
    a^\dagger_{\alpha}(i)\,a_{\beta}(i)\:.
\end{equation}
Independently of whether the operators $a_\alpha$ satisfy fermionic or bosonic commutation relations, these bilinears obey the closed algebra
\begin{equation}
    \big[ E_{\alpha\beta}(i), E_{\gamma\delta}(i)\big]
    =
    \delta_{\beta\gamma} E_{\alpha\delta}(i)
    -
    \delta_{\alpha\delta} E_{\gamma\beta}(i)\:,
\end{equation}
which is precisely the $\mathfrak{u}(N)$ Lie algebra. In this way, the generators of $\mathfrak{su}(N)$ can be realized as
\begin{equation}
    \hat S^\beta_{\alpha}(i)
    =
    a^\dagger_{\alpha}(i)\,a_{\beta}(i)
    -
    \frac{\delta_{\alpha\beta}}{N}\,
    \hat n(i)\:\:,
    \quad \text{with }\quad 
    \hat n(i)=\sum_{\gamma=1}^N a^\dagger_{\gamma}(i)a_{\gamma}(i)\:.
\end{equation}
At this level, the construction is purely algebraic and does not depend on the statistics of the operators $a_\alpha$. However, the choice between fermions and bosons becomes crucial once one specifies the physical Hilbert space through constraints on the local occupation number.

\begin{itemize}
    \item In the fermionic case ($\{a_\alpha\}\to\{c_\alpha\}$), one typically imposes a fixed occupation constraint
    \begin{equation}
    \hat n(i)=m\quad \Longrightarrow \quad \mathcal{F}^{\text{fer}}_i= \ket{0}\oplus\cdots\oplus\boxed{ \mathcal{H}^{(m)}_{i,\text{fer}}}\oplus\cdots\oplus \mathcal{H}^{(N)}_{i,\text{fer}} 
    \end{equation}
    so that the physical Hilbert space is obtained by selecting states with exactly $m$ fermions per site. This construction naturally generates totally antisymmetric irreducible representations of $SU(N)$, corresponding to Young tableaux of a single column of length $m$:
    \begin{equation}
\mathcal{H}^{\text{phys}}_i=\mathcal{H}^{(m)}_{i,\text{fer}}=\bigwedge\nolimits^m \left(\mathbb{C}^N_{\text{flavor}}\right)\hspace{8pt}\sim\:\:\vcenter{\hbox{\yng(1,1,1)}}\;\; \text{\small ($m$ boxes, single column)} 
\end{equation}

    \item In contrast, if the operators $\{a_\alpha\}$ are taken to be bosonic $\{b_\alpha\}$, the same bilinear construction leads to a different class of representations. Imposing again a fixed occupation constraint
    \begin{equation}
    \hat n(i)=n_c\quad \Longrightarrow \quad \mathcal{F}_i^{\text{bos}}= \ket{0}\oplus\cdots\oplus\boxed{ \mathcal{H}^{(n_c)}_{i,\text{bos}}}\oplus\cdots\oplus \mathcal{H}^{(N)}_{i,\text{bos}}
    \end{equation}
    the resulting Hilbert space is now fully symmetric under exchange of flavors, and therefore realizes totally symmetric irreducible representations of $SU(N)$ (Young tableaux with a single row of length $n_c$):
    \begin{equation}
\mathcal{H}^{\text{phys}}_i=\mathcal{H}^{(n_c)}_{i,\text{bos}} = \mathrm{Sym}^{n_c} \left(\mathbb{C}^N_{\text{flavor}}\right) \hspace{6pt} \sim\:\:\: \yng(4) \;\; \text{\small ($n_c$ boxes, single row)}
    \end{equation}
\end{itemize}

It is important to stress that the above construction describes the local Hilbert space on a single site only. In the antiferromagnetic bipartite problem, however, one must still implement the conjugate-representation structure discussed before. For the fermionic construction, this is achieved in a particularly simple way: if sites on sublattice $A$ carry the antisymmetric irrep
$$
\mathcal{H}_A^{\mathrm{phys}} =
\bigwedge\nolimits^{m}\!\left(\mathbb{C}^N_{\mathrm{flavor}}\right),
$$
then sites on sublattice $B$ must carry the conjugate one, which is realized by fixing the occupation to
\begin{equation}
    \hat n(j)=N-m \quad \:\:(j\in B) \qquad \xRightarrow[]{\quad} \quad \mathcal{H}_B^{\mathrm{phys}}=\bigwedge\nolimits^{N-m}\!\left(\mathbb{C}^N_{\mathrm{flavor}}\right)
    \cong \overline{\bigwedge\nolimits^{m}\!\left(\mathbb{C}^N_{\mathrm{flavor}}\right)} \:\:.
\end{equation}
In this way, each nearest-neighbor bond combines $m$ fermions on one site with $N-m$ fermions on the other, so that the total number of fermions across the bond is $N$, namely the number of flavors. This is precisely the condition that allows the bond Hilbert space
\begin{equation}
    \mathcal{H}_{ij} \:\:=\:\bigwedge\nolimits^{m}\!\left(\mathbb{C}^N_{\mathrm{flavor}}\right)
    \;\otimes\;
    \bigwedge\nolimits^{N-m}\!\left(\mathbb{C}^N_{\mathrm{flavor}}\right)\:\:\simeq\:\: \mathbf{1} \:\oplus\:\dots
\end{equation}
to contain an $SU(N)$ singlet sector. The singlet is therefore not the whole bond space, but a distinguished invariant channel inside it, and it is exactly this channel that the antiferromagnetic interaction tends to favor.

The situation is more subtle in the bosonic construction. While fixing the occupation number $n_c$ naturally generates totally symmetric representations,
$$
\mathcal{H}_{A}^{\mathrm{phys}} = \mathrm{Sym}^{n_c}\!\left(\mathbb{C}^N_{\mathrm{flavor}}\right),
$$
this procedure by itself does not produce the conjugate representation required on the opposite sublattice. In contrast to the fermionic case, where antisymmetry relates $\bigwedge^{m}$ and $\bigwedge^{N-m}$, symmetric representations do not admit such a simple identification. As a consequence, implementing the bipartite structure in the bosonic framework requires a different strategy. One must explicitly distinguish the degrees of freedom on the two sublattices. A convenient choice is to introduce two independent bosonic species: one transforming in the fundamental representation on sublattice $A$, and the other transforming in the conjugate (anti-fundamental) representation on sublattice $B$. Denoting these operators by $b_\alpha(i)$ on $i\in A$, and $\bar b_\alpha(j)$ on $j\in B$, one defines
\begin{equation}
    E_{\alpha\beta}(i)= b^\dagger_\alpha(i)\,b_\beta(i),
    \qquad
    \overline E_{\alpha\beta}(j)= -\,\bar b^\dagger_\beta(j)\,\bar b_\alpha(j),
\end{equation}
since the conjugate representation is implemented by reversing the contraction structure of indices as
\begin{equation}
    \begin{aligned}
        b_\alpha \text { transforms in the fundamental } (\bm N) \qquad & \qquad \bar{b}_\alpha \text { transforms in the conjugate }(\bm{\overline{N}}) \\
        b_\alpha \longrightarrow U_{\alpha \beta} \:b_\beta \hspace{80pt} & \hspace{45pt} \bar{b}_\alpha \longrightarrow U_{\alpha \beta}^* \bar{b}_\beta=U^\dagger_{\beta\alpha}\bar{b}_\beta\quad.
    \end{aligned}
\end{equation}
with the additional minus sign reflecting the standard relation $\overline T^a=-(T^a)^*$ for generators in the conjugate representation.
Imposing again fixed occupation constraints on each sublattice,
\begin{equation}
    \sum_{\alpha} b^\dagger_\alpha(i)b_\alpha(i)=n_c,
    \qquad
    \sum_{\alpha} \bar b^\dagger_\alpha(j)\bar b_\alpha(j)=n_c,
\end{equation}
the local Hilbert spaces are now given by symmetric representations on both sublattices, but transforming in conjugate ways under $SU(N)$. In this manner, the tensor product
\begin{equation}
    \mathcal{H}_{ij}
    \:\:=\:\:
    \mathrm{Sym}^{n_c}(\mathbb{C}^N)
    \otimes
    \overline{\mathrm{Sym}^{n_c}(\mathbb{C}^N)}\:\:\simeq \:\:\mathbf{1}\oplus\dots
\end{equation}
again contains a singlet sector, ensuring that the antiferromagnetic interaction can favor its formation.

\noindent So, in contrast with the fermionic case, where the conjugation is encoded in the occupation constraint itself, the bosonic construction makes the bipartite structure explicit at the level of the operators. This distinction will play an important role in the physical interpretation: fermionic partons naturally lead to large-$N$ descriptions of valence-bond states, while bosonic partons provide a convenient framework for describing magnetically ordered phases and their quantum critical points.

Nevertheless, the previous fermionic and bosonic constructions are still limited to very special families of irreducible representations: the fermionic realization naturally generates totally antisymmetric representations (single-column Young tableaux of height $m$), while the bosonic one produces totally symmetric representations (single-row tableaux of length $n_c$). In order to access a broader class of $SU(N)$ representations — in particular, general rectangular Young tableaux with $m$ rows and $n_c$ columns (and their conjugate of $N-m$ rows and $n_c$ columns)— one must enlarge the Hilbert space with auxiliary degrees of freedom. The standard way to do so is to introduce an additional internal label, which we generically denote by $A$:
\begin{equation}
    a_\alpha(i)\;\longrightarrow\; a_{\alpha A}(i)\:.
\end{equation}
Here, the flavor index $\alpha=1,\dots,N$ continues to transform under $SU(N)$, while $A$ labels an auxiliary internal space that will later be projected out by suitable local constraints. At the level of a single particle, the local Hilbert space is thus enlarged as $\mathcal{H}_i^{(1)} =\mathbb{C}^N_{\text{flavor}} \otimes \mathbb{C}^{d_A}_{\text{aux}}$, where the dimension $d_A$ of the auxiliary space depends on the chosen parton construction. For fermions one introduces $d_A=n_c$ colors, while for bosons one instead takes $d_A=m$. In both cases, the purpose of the auxiliary index is the same: it enlarges the Fock space so that, after imposing suitable local constraints, one can isolate a nontrivial irreducible representation of $SU(N)$ which is no longer restricted to be purely symmetric or purely antisymmetric in flavor space. The way this works, however, is quite different for fermions and bosons. Let us begin with the fermionic construction:
\begin{itemize}
    \item  \textbf{Fermionic construction with $\bm{n_c}$ colors}\\
We now replace the fermionic operators by colored ones,
\begin{equation}
    c_\alpha(i)\;\longrightarrow\; c_{\alpha a}(i),
    \qquad
    \alpha=1,\dots,N,\qquad a=1,\dots,n_c,
\end{equation}
so that on each site the one-particle Hilbert space becomes
\begin{equation}
    \mathcal{H}_{i,\mathrm{fer}}^{(1)}
    =
    \mathbb{C}^N_{\mathrm{flavor}}
    \otimes
    \mathbb{C}^{n_c}_{\mathrm{color}}\:.
\end{equation}
Accordingly, the full local fermionic Fock space is generated from
$\mathbb{C}^N_{\mathrm{flavor}}\otimes\mathbb{C}^{n_c}_{\mathrm{color}}$, and fixing the total number of fermions to be $m n_c$ selects the sector
\begin{equation}
    \mathcal{H}_{i,\mathrm{fer}}^{(m n_c)}
    =
    \bigwedge\nolimits^{m n_c}
    \Big(
    \mathbb{C}^N_{\mathrm{flavor}}
    \otimes
    \mathbb{C}^{n_c}_{\mathrm{color}}
    \Big)\:.
\end{equation}
At this point, however, the space is still much larger than the physical irrep that we want. The essential restriction is that the auxiliary color degree of freedom is unphysical, and therefore physical states must be invariant under local $SU(n_c)$ rotations in color space. Equivalently, one imposes the color-singlet constraint
\begin{equation}
\label{eq:fermionic_color_constraint}
    \sum_{\alpha=1}^N
    c_{\alpha a}^\dagger(i)c_{\alpha b}(i)
    =
    m\,\delta_{ab},
\end{equation}
which says that each color sector contains exactly $m$ fermions, and at the same time projects onto states transforming trivially under the local color group. In particular, the generators
\begin{equation}
    G_{ab}(i)
    =
    \sum_{\alpha=1}^N
    c_{\alpha a}^\dagger(i)c_{\alpha b}(i)
    -
    m\,\delta_{ab}
\end{equation}
annihilate physical states, so that the physical subspace is precisely the color-singlet sector.

\noindent The representation content of this constrained fermionic space is made transparent by the standard decomposition
\begin{equation}
\label{eq:fermionic_schur_decomposition}
    \bigwedge\nolimits^{m n_c}
    \big(
    \mathbb{C}^{N}_{\text{flavor}}\otimes \mathbb{C}^{n_c}_{\text{color}}
    \big)
    =
    \bigoplus_{\lambda}\:\:
    S_\lambda(\mathbb{C}^{N}_{\text{flavor}})\otimes S_{\lambda'}(\mathbb{C}^{n_c}_{\text{color}}),
\end{equation}
where the sum runs over Young diagrams $\lambda$ with $m n_c$ boxes, $S_\lambda(\cdot)$ denotes the Schur functor associated with $\lambda$, and $\lambda'$ is the transposed Young diagram. In this notation, $S_\lambda(\mathbb{C}^N_{\text{flavor}})$ is the irreducible representation of $SU(N)$ labeled by $\lambda$, while $S_{\lambda'}(\mathbb{C}^{n_c}_{\text{color}})$ is the corresponding representation of the auxiliary color group $SU(n_c)$.

\noindent The color-singlet constraint now has a very precise meaning: in the decomposition of Eq.\eqref{eq:fermionic_schur_decomposition}, one must keep only the component for which the color factor $S_{\lambda'}(\mathbb{C}^{n_c}_{\text{color}})$ is trivial. Then, the only way to obtain a color singlet from $m n_c$ boxes is that $\lambda'$ be a rectangular tableau with $n_c$ rows and $m$ columns, i.e. a tableau of shape $\lambda'=(m^{n_c})$. Equivalently, the corresponding flavor tableau $\lambda$ is the transposed rectangle with $m$ rows and $n_c$ columns:
\begin{equation}
    \lambda=(n_c^{\,m}),
    \qquad
    \lambda'=(m^{\,n_c}).
\end{equation}
Therefore, after projection onto the color-singlet sector, the physical fermionic Hilbert space on a site becomes
\begin{equation}
\label{eq:fermionic_rectangular_irrep}
    \mathcal{H}_{i,\mathrm{fer}}^{\mathrm{phys}}
    \simeq
    S_{(n_c^{m})}\left(\mathbb{C}^N_{\mathrm{flavor}}\right)\:,
\end{equation}
namely the rectangular irreducible representation of $SU(N)$ with $m$ rows and $n_c$ columns.

\noindent This construction is particularly illuminating in the familiar $SU(2)$ case. There, taking $N=2$ and $m=1$, the tableau $(n_c)$ is simply a single row of length $n_c$, so that
\begin{equation}
    \mathcal{H}_{i,\mathrm{fer}}^{\mathrm{phys}}
    \simeq
    \mathrm{Sym}^{n_c}\!\left(\mathbb{C}^2\right)\simeq \mathcal{H}_{S},
\end{equation}
which is precisely the spin-$S$ irreducible representation given $n_c=2S$. In other words, the colored fermionic construction recovers arbitrary spin-$S$ representations of $SU(2)$ only after projecting onto the color-singlet sector.

\item  \textbf{Conjugate representation on the bipartite lattice (fermions)}\\
Once the single-site construction is understood, the bipartite antiferromagnetic structure is implemented exactly as before, but now at the level of rectangular tableaux. On sublattice $A$ one takes the rectangular representation with $m$ rows and $n_c$ columns,
\begin{equation}
    \mathcal{H}_{A}^{\mathrm{phys}}
    \simeq
    S_{(n_c^{\,m})}\!\left(\mathbb{C}^N_\text{flavor}\right),
\end{equation}
while on sublattice $B$ one must place its conjugate representation, which is the rectangle with $N-m$ rows and the same $n_c$ columns:
\begin{equation}
    \mathcal{H}_{B}^{\mathrm{phys}}
    \simeq
    S_{(n_c^{\,N-m})}\!\left(\mathbb{C}^N\right)
    \cong
    \overline{
    S_{(n_c^{\,m})}\!\left(\mathbb{C}^N\right)
    }.
\end{equation}
Equivalently, in the fermionic parton language this is realized by imposing
\begin{equation}
\begin{aligned}
    \sum_\alpha c_{\alpha a}^\dagger(i)c_{\alpha b}(i)=m\,\delta_{ab}
    \qquad i\in A,\\
    \sum_\alpha c_{\alpha a}^\dagger(j)c_{\alpha b}(j)=(N-m)\,\delta_{ab}
    \qquad j\in B.
\end{aligned}
\end{equation}
Thus, each color carries $m$ fermions on sublattice $A$ and $N-m$ fermions on sublattice $B$, which is the colored version of the fundamental/conjugate pairing discussed earlier.

\item  \textbf{Bosonic construction with $\bm{m}$ colors}\\
The bosonic case proceeds in parallel, but with an important interchange in roles. One now introduces bosonic operators
\begin{equation}
    b_\alpha(i)\;\longrightarrow\; b_{\alpha a}(i)
    \quad \text{ }\begin{cases}
        \alpha=1,\dots,N\\ a=1,\dots,m
    \end{cases} \quad \xRightarrow[]{\:\:\text{so that }\:\:} \quad \mathcal{H}_{i,\mathrm{bos}}^{(1)}
    =
    \mathbb{C}^N_{\mathrm{flavor}}
    \otimes
    \mathbb{C}^{m}_{\mathrm{color}}\:.
\end{equation}
Fixing the total number of bosons to be $m n_c$, one selects the sector
\begin{equation}
    \mathcal{H}_{i,\mathrm{bos}}^{(m n_c)}
    =
    \mathrm{Sym}^{m n_c}
    \Big(
    \mathbb{C}^N_{\mathrm{flavor}}
    \otimes
    \mathbb{C}^{m}_{\mathrm{color}}
    \Big) =
    \bigoplus_\lambda
    S_\lambda(\mathbb{C}^N_{\mathrm{flavor}})\otimes S_\lambda(\mathbb{C}^{m}_{\mathrm{color}})\:,
\end{equation}
where the relevant decomposition contains the same Young diagram $\lambda$ appearing in both flavor and color spaces, in contrast to the fermionic case, where we have the transpose $\lambda'$. Then, to isolate the rectangular $m\times n_c$ tableau in flavor space, one imposes the bosonic constraint
\begin{equation}
\label{eq:bosonic_color_constraint}
    \sum_{\alpha=1}^N
    b_{\alpha a}^\dagger(i)b_{\alpha b}(i)
    =
    n_c\,\delta_{ab},
\end{equation}
which says that each color contains exactly $n_c$ bosons. This selects the component in which the color tableau is a rectangle with $m$ rows and $n_c$ columns, and therefore the flavor tableau is the same rectangle. Hence, the physical bosonic Hilbert space is again
\begin{equation}
\label{eq:bosonic_rectangular_irrep}
    \mathcal{H}_{i,\mathrm{bos}}^{\mathrm{phys}}
    \simeq
    S_{(n_c^{\,m})}\!\left(\mathbb{C}^N_{\mathrm{flavor}}\right).
\end{equation}
\item \textbf{Conjugate representation on the bipartite lattice (bosons)}\\
Just as in the fermionic case, the single-site construction is not yet enough: in the antiferromagnetic problem one must still place conjugate representations on the two sublattices. For bosons, however, the conjugation is not implemented by changing the occupation number, but rather by changing the transformation law of the elementary operators themselves. Thus, on sublattice $A$ one takes bosons $b_{\alpha a}(i)$ transforming in the fundamental representation of $SU(N)$, while on sublattice $B$ one introduces independent bosons $\bar b_{\alpha a}(j)$ transforming in the conjugate representation:
\begin{equation}
    b_{\alpha a}(i)\longrightarrow U_{\alpha\beta}\,b_{\beta a}(i),
    \qquad
    \bar b_{\alpha a}(j)\longrightarrow U^*_{\alpha\beta}\,\bar b_{\beta a}(j).
\end{equation}
In this way, the local one-particle Hilbert spaces are
\begin{equation}
    \mathcal{H}^{(1)}_{A,\mathrm{bos}}
    =
    \mathbb{C}^N_{\mathrm{flavor}}\otimes\mathbb{C}^m_{\mathrm{color}},
    \qquad
    \mathcal{H}^{(1)}_{B,\mathrm{bos}}
    =
    \overline{\mathbb{C}^N_{\mathrm{flavor}}}\otimes\mathbb{C}^m_{\mathrm{color}}\:,
\end{equation}
so that the distinction between the two sublattices is already built at the microscopic level. One then imposes the same bosonic color constraint on both sublattices,
\begin{equation}
    \sum_{\alpha=1}^N b^\dagger_{\alpha a}(i)b_{\alpha b}(i)=n_c\,\delta_{ab},
    \qquad
    \sum_{\alpha=1}^N \bar b^\dagger_{\alpha a}(j)\bar b_{\alpha b}(j)=n_c\,\delta_{ab},
\end{equation}
which fixes $n_c$ bosons per color and projects onto the rectangular tableau with $m$ rows and $n_c$ columns. As a consequence, the physical Hilbert spaces on the two sublattices become
\begin{equation}
    \mathcal{H}^{\mathrm{phys}}_{A,\mathrm{bos}}
    \simeq
    S_{(n_c^{\,m})}\!\left(\mathbb{C}^N_{\mathrm{flavor}}\right),
    \qquad
    \mathcal{H}^{\mathrm{phys}}_{B,\mathrm{bos}}
    \simeq
    S_{(n_c^{\,m})}\!\left(\overline{\mathbb{C}^N_{\mathrm{flavor}}}\right)
    \cong
    \overline{
    S_{(n_c^{\,m})}\!\left(\mathbb{C}^N_{\mathrm{flavor}}\right)
    }.
\end{equation}
Therefore, in the bosonic construction the same rectangular Young tableau appears on both sublattices, but with opposite transformation properties under $SU(N)$. This is the bosonic analog of the fermionic relation
$
S_{(n_c^{\,N-m})}(\mathbb{C}^N)\cong \overline{S_{(n_c^{\,m})}(\mathbb{C}^N)}
$,
except that now the conjugation is carried explicitly by the operators rather than by changing the occupation number.
\end{itemize}
\noindent So, although the fermionic and bosonic constructions start from very different statistics and impose different local constraints, they can both be used to generate the same rectangular irreducible representation of $SU(N)$. The difference is in how that representation is carved out of the enlarged auxiliary space: in the fermionic case through antisymmetrization and projection to a color singlet, and in the bosonic case through symmetrization together with fixed boson number per color. This distinction is precisely what later makes the two formulations naturally suited to different physical regimes.

What remains is to rewrite the $U(N)$ matrix units, and therefore the $SU(N)$ generators, in the enlarged color space. The rule is simple: once the auxiliary index is introduced, the bilinears must be summed over color, since the physical generators act only on flavor and are color singlets. Thus, the elementary operators $E_{\alpha\beta}$ are promoted as
\begin{equation}
    E_{\alpha\beta}(i)
    \;\longrightarrow\;
    \sum_{a} a^\dagger_{\alpha a}(i)\,a_{\beta a}(i)\:,
\end{equation}
so that the corresponding traceless generators are obtained, as before, by subtracting the appropriate multiple of the identity fixed by the local constraint.

\noindent For the fermionic construction, this gives
\begin{equation}
\label{eq:fermionic_generators_colored}
\begin{aligned}
    S^\beta_{\alpha}(i)
    &=
    \sum_{a=1}^{n_c}
    c^\dagger_{\alpha a}(i)c_{\beta a}(i)
    -
    \delta_{\alpha\beta}\frac{m\,n_c}{N},
    \qquad i\in A,\\[4pt]
    \bar S^\beta_{\alpha}(j)
    &=
    \sum_{a=1}^{n_c}
    c^\dagger_{\alpha a}(j)c_{\beta a}(j)
    -
    \delta_{\alpha\beta}\frac{(N-m)\,n_c}{N},
    \qquad j\in B,
\end{aligned}
\end{equation}
together with the color constraints
\begin{equation}
\label{eq:fermionic_constraints_AB}
\sum_{\alpha=1}^N c^\dagger_{\alpha a}(i)c_{\alpha b}(i)=m\,\delta_{ab},
\qquad
\sum_{\alpha=1}^N c^\dagger_{\alpha a}(j)c_{\alpha b}(j)=(N-m)\,\delta_{ab}.
\end{equation}
In other words, the generators are obtained by summing the flavor bilinears over all colors, while the local constraints ensure that the resulting representation is precisely the rectangular tableau with $m$ rows and $n_c$ columns on sublattice $A$, and its conjugate with $N-m$ rows and $n_c$ columns on sublattice $B$.

\noindent In the bosonic construction the logic is analogous, but now the conjugation is implemented directly at the operator level. Summing again over color, one obtains
\begin{equation}
\label{eq:bosonic_generators_colored}
\begin{aligned}
    S^\beta_{\alpha}(i)
    &=
    \sum_{a=1}^{m}
    b^\dagger_{\alpha a}(i)b_{\beta a}(i)
    -
    \delta_{\alpha\beta}\frac{m\,n_c}{N},
    \qquad i\in A,\\[4pt]
    \bar S^\beta_{\alpha}(j)
    &=
    -\sum_{a=1}^{m}
    \bar b^\dagger_{\beta a}(j)\bar b_{\alpha a}(j)
    +
    \delta_{\alpha\beta}\frac{m\,n_c}{N},
    \qquad j\in B,
\end{aligned}
\end{equation}
subject to
\begin{equation}
\label{eq:bosonic_constraints_AB}
\sum_{\alpha=1}^N b^\dagger_{\alpha a}(i)b_{\alpha b}(i)=n_c\,\delta_{ab},
\qquad
\sum_{\alpha=1}^N \bar b^\dagger_{\alpha a}(j)\bar b_{\alpha b}(j)=n_c\,\delta_{ab}.
\end{equation}
Here the minus sign in $\bar S^\beta_{\alpha}$ is the characteristic signature of the conjugate representation. The subtraction terms again make the generators traceless, exactly as in the abstract definition of $\mathfrak{su}(N)$.

\noindent It is worth noting, however, that in the bosonic literature one often drops these diagonal subtraction terms and works instead with the non-traceless bilinears alone. The reason is simple: once the boson-number constraints in Eq.\eqref{eq:bosonic_constraints_AB} are imposed, the trace part is just a fixed constant on the physical Hilbert space. Therefore, replacing
\begin{equation}
    S^\beta_{\alpha}(i)
    \;\to\;
    \sum_{a=1}^{m} b^\dagger_{\alpha a}(i)b_{\beta a}(i),
    \qquad
    \bar S^\beta_{\alpha}(j)
    \;\to\;
    -\sum_{a=1}^{m}\bar b^\dagger_{\beta a}(j)\bar b_{\alpha a}(j)
\end{equation}
changes the Heisenberg Hamiltonian only by an additive constant, which has no physical effect. This is precisely the convention often adopted in Schwinger-boson treatments.

\noindent Collecting everything, the two standard large-$N$ parton realizations of the $SU(N)$ antiferromagnet can be summarized as
\begin{equation}
\boxed{
\begin{aligned}
&\quad\textbf{SU(N) Heisenberg model in Fermionic partons} \\[2pt]
&\hspace{118pt}H^{\mathrm{fer}}_{SU(N)}
=
\frac{J}{N}\sum_{\langle ij\rangle}\sum_{\alpha,\beta=1}^{N}
S^\beta_{\alpha}(i)\,\bar S^\alpha_{\beta}(j),\\[3pt]
&\text{ with traceless } SU(N) \text{ generators}\\[3pt]
&\:\:\:S^\beta_{\alpha}(i)
=
\sum_{a=1}^{n_c}c^\dagger_{\alpha a}(i)c_{\beta a}(i)
-
\delta_{\alpha\beta}\frac{m\,n_c}{N},
\qquad
\bar S^\beta_{\alpha}(j)
=
\sum_{a=1}^{n_c}c^\dagger_{\alpha a}(j)c_{\beta a}(j)
-
\delta_{\alpha\beta}\frac{(N-m)\,n_c}{N},\\[3pt]
&\text{ and local constraints}\\[3pt]
&\hspace{70pt} \sum_{\alpha} c^\dagger_{\alpha a}(i)c_{\alpha b}(i)=m\,\delta_{ab},
\qquad
\sum_{\alpha} c^\dagger_{\alpha a}(j)c_{\alpha b}(j)=(N-m)\,\delta_{ab}\:.
\end{aligned}}
\end{equation}

\begin{equation}
\boxed{
\begin{aligned}
&\quad\textbf{SU(N) Heisenberg model in Bosonic partons} \\[2pt]
&\hspace{118pt}H^{\mathrm{bos}}_{SU(N)}
=
\frac{J}{N}\sum_{\langle ij\rangle}\sum_{\alpha,\beta=1}^{N}
S^\beta_{\alpha}(i)\,\bar S^\alpha_{\beta}(j),\\[3pt]
&\text{ with traceless } SU(N) \text{ generators}\\[3pt]
&\qquad S^\beta_{\alpha}(i)
=
\sum_{a=1}^{m}b^\dagger_{\alpha a}(i)b_{\beta a}(i)
-
\delta_{\alpha\beta}\frac{m\,n_c}{N},
\qquad
\bar S^\beta_{\alpha}(j)
=
-\sum_{a=1}^{m}\bar b^\dagger_{\beta a}(j)\bar b_{\alpha a}(j)
+
\delta_{\alpha\beta}\frac{m\,n_c}{N},\:\:\\[3pt]
&\text{ and local constraints}\\[3pt]
&\hspace{70pt} \sum_{\alpha} b^\dagger_{\alpha a}(i)b_{\alpha b}(i)=n_c\,\delta_{ab},
\qquad
\sum_{\alpha} \bar b^\dagger_{\alpha a}(j)\bar b_{\alpha b}(j)=n_c\,\delta_{ab}.
\end{aligned}}
\end{equation}

\subsubsection*{SU(N) Magnets: Large-N, VBS order and Deconfined Quantum Criticality}

At this point, having constructed the family of $SU(N)$ Heisenberg Hamiltonians and their parton realizations, the natural question is whether the geometric mechanism discussed in the $SU(2)$ case survives in this more general setting. In particular, does the antiferromagnet still carry a Berry-phase structure, or was that feature special to the $O(3)$ coherent-state description? The answer is that the same pattern reappears in a more general geometric form.

\noindent To see how this emerges, it is useful to recall the basic logic of coherent-state constructions. One starts from a fixed highest-weight state $\ket{\Psi_0}$ and generates the full set of coherent states by acting with unitary transformations \cite{read1989some}. In this way, the local degrees of freedom are parametrized by the orbit of this reference state under $SU(N)$. For the rectangular $m\times n_c$ representation considered in the semiclassical antiferromagnet, this may be written schematically as
\begin{equation}
   |q\rangle = e^{\left(q_\mu^{\lambda}\hat S_{\lambda}^{\mu}-q_\mu^{\lambda *}\hat S_{\mu}^{\lambda}\right)}|\Psi_0\rangle.
\end{equation}
Here the indices run as $\lambda=1,\dots,m$ and $\mu=m+1,\dots,N$, so that the complex parameters $q_\mu^\lambda$ form an $m\times (N-m)$ matrix. These variables provide a local parametrization of the coherent-state manifold around the reference configuration. In Sec.\,\ref{sec:su_n_superspin}, we will return to this construction in detail for the simplest case of a single $SU(N)$ degree of freedom in the fundamental representation. In that situation, the coherent-state manifold is $\mathbb{CP}^{N-1}$. By contrast, the present semiclassical antiferromagnetic problem involves the more general rectangular $m\times n_c$ representation, and therefore the natural target space is no longer projective space but the Grassmannian. Thus, the discussion here should be viewed as the general framework, while the later section will provide the explicit and more elementary realization of the same logic in the special case $m=1$.

\noindent Hence, in the semiclassical treatment of Read and Sachdev, instead of working directly with the states, it is convenient to describe their expectation values in terms of a matrix field. Concretely, evaluating the generators on a coherent state yields
\begin{equation}
\langle q|\hat S^\beta_{\alpha}|q\rangle = \frac{n_c}{2}\, Q^\beta_{\alpha},
\end{equation}
which defines a Hermitian matrix $Q$ of the form
\begin{equation}
Q=U\Lambda U^\dagger,
\qquad
\Lambda=
\begin{pmatrix}
\mathbb 1_m & 0\\
0 & -\mathbb 1_{N-m}
\end{pmatrix},
\qquad 
U = \exp
\begin{pmatrix}
0 & q \\
- q^\dagger & 0
\end{pmatrix},
\qquad
Q^2=\mathbb 1,
\end{equation}
where $U\in U(N)$ encodes the local orientation of the coherent state, while $\Lambda$ represents the reference configuration.

Now, since different choices of $U$ related by right multiplication with $U(m)\times U(N-m)$ leave $Q$ invariant, the physical configurations are naturally identified with the coset space
\begin{equation}
Q(x,\tau)\in \frac{U(N)}{U(m)\times U(N-m)}
\equiv \mathrm{Gr}(m,\mathbb C^N).
\end{equation}
Therefore, the sphere $S^2$ of the $SU(2)$ N\'eel vector is replaced by a Grassmannian target space. However, this replacement is not merely geometric: as discussed in Sec.\,\ref{sec:top_actions}, such manifolds admit canonical closed two-forms whose cohomology classes are integral. This is ultimately rooted in the fact that their second homotopy group is nontrivial,
\begin{equation}
\pi_2\!\left(\mathrm{Gr}(m,\mathbb C^N)\right)=\mathbb Z\qquad \Big[\forall N, \quad 1 \le m \le N-1\Big],
\end{equation}
so that maps from space-time into the target manifold can be classified by an integer topological charge. In this sense, the Grassmannian plays exactly the same role as $S^2\simeq\mathbb{CP}^1$, but at a higher level of complexity.

The important point is that the structure of the path integral remains the same: the action again separates into a dynamical part and a topological one. More precisely, for each site the coherent-state path integral takes the form
\begin{equation}
S^{(1)}_i[Q]
=
\underbrace{
\int_0^\beta d\tau\int_0^1 du\;
\frac{n_c}{4}\,
\mathrm{Tr}\!\left(
\tilde Q\,\partial_u \tilde Q\,\partial_\tau \tilde Q
\right)}_{\text{Berry/Wess--Zumino term}}
\;-\;
\underbrace{\int_0^\beta d\tau\,H(Q(\tau))}_{\text{dynamical term}},
\end{equation}
where $\tilde Q(\tau,u)$ is a homotopic extension such that $\tilde Q(\tau,1)=Q(\tau)$ and $\tilde Q(\tau,0)=\Lambda$. At this stage, one should recognize that this term is not an ad hoc construction, but precisely the Wess--Zumino functional associated with a closed integral form on the target space. In the language developed previously, it is the one-dimension-higher representation of a topological term of the form
\begin{equation}
\int_{M^2} \phi^*\omega,
\end{equation}
where $\omega$ is the canonical two-form of the Grassmannian. Thus, even before taking any continuum limit, one sees explicitly that the $SU(N)$ antiferromagnet carries the same type of geometric term that appeared in the $SU(2)$ spin-coherent-state action, now written in its natural general form.

 In the antiferromagnet, however, one must distinguish between the two sublattices. The local moments on sublattice $A$ and $B$ transform in conjugate representations, which results in opposite signs for the Berry phase. Thus, the full topological contribution takes the form
\begin{equation}
S_B
=
\frac{n_c}{4}
\sum_{i\in A}\int d\tau du\;
\mathrm{Tr}\!\left(
\tilde Q_i\,\partial_u \tilde Q_i\,\partial_\tau \tilde Q_i
\right)
-
\frac{n_c}{4}
\sum_{i\in B}\int d\tau du\;
\mathrm{Tr}\!\left(
\tilde Q_i\,\partial_u \tilde Q_i\,\partial_\tau \tilde Q_i
\right).
\end{equation}
Then, the parallel with the previous subsection becomes very close. In the antiferromagnetic phase, one again decomposes the lattice field into staggered and smooth components, now in terms of a slowly varying Grassmannian field $\Omega$ and a small uniform fluctuation $L$: \begin{equation}
    Q(i) \approx \Omega_i \sqrt{1-a^2 L_i^2}+\eta_i a L_i\:\:.
\end{equation}
In this line, and after performing the gradient expansion and integrating out $L$, one obtains a $(2+1)$-dimensional nonlinear sigma model together with a residual lattice Berry phase,
\begin{equation}
S
=
S_B'
+
\frac12\int_0^\beta d\tau\int d^2x\;
\frac{\rho_s}{2}\,
\mathrm{Tr}\!\left[
(\nabla\Omega)^2+\frac{1}{c^2}(\partial_\tau\Omega)^2
\right],
\end{equation}
with
\begin{equation}
S_B'
=
i n_c\sum_j \eta_j\,\omega_j,
\qquad
\omega_j=
\frac{1}{4i}\int_0^\beta d\tau\int_0^1 du\;
\mathrm{Tr}\!\left(
\tilde \Omega_j\,\partial_u\tilde\Omega_j\,\partial_\tau\tilde\Omega_j
\right).
\end{equation}
This is precisely the Grassmannian analog of the $SU(2)$ result in which the smooth long-wavelength fluctuations are governed by an NLSM, while the Berry phase survives as an additional staggered lattice contribution. Consequently, the physical implications are also the same. For configurations that remain smooth on the scale of the lattice spacing, the residual term $S_B'$ vanishes, and the theory reduces to the sigma-model dynamics alone. However, nontrivial Berry-phase effects arise only from space-time singularities of the order-parameter field --- the analogs of hedgehogs --- which interpolate between configurations belonging to different topological sectors, i.e. configurations with different values of the integer
\begin{equation}
W[\Omega]\in\mathbb{Z}.
\end{equation}
 Therefore, one can show that a tunneling event with skyrmion-number change $\Delta W$ located on one of the four dual-sublattice plaquettes contributes a phase
\begin{equation}
(\xi_j)^{\,n_c\Delta W},
\qquad
\xi_j\in\{1,-1,i,-i\},
\end{equation}
so that, once again, the Berry phase turns the gas of topological tunneling events into an interference problem. In other words, the entire mechanism discussed before for $SU(2)$ --- vanishing contribution from smooth configurations, nontrivial phases attached to singular events, and the selection of the quantum-disordered phase through interference --- survives in the $SU(N)$ antiferromagnet with the replacement
\begin{equation}
S^2 \;\longrightarrow\; \mathrm{Gr}(m,\mathbb C^N),
\qquad
(\zeta_n)^{2SQ_n}
\;\longrightarrow\;
(\xi_j)^{\,n_c\Delta W}.
\end{equation}
This is precisely the point where the $SU(N)$ formulation becomes valuable. It does not alter the Berry-phase mechanism itself; rather, it embeds it into a framework where it can be controlled. While at the level of the NLSM the structure closely parallels the $SU(2)$ case, the real advantage emerges once the theory is reformulated in terms of partons. In that description, fluctuations are organized in powers of $1/N$, and both the smooth sector and the topological tunneling events --- including their Berry phases --- can be treated on equal footing within a controlled expansion. In this way, the interference mechanism that in $SU(2)$ appears as a semiclassical insight becomes part of a systematic large-$N$ description, as we will see later.

However, before turning to the large-$N$ parton treatments, it is important to keep in mind what has, and has not, been established so far: the Grassmannian coherent-state formulation shows that in $SU(N)$ antiferromagnets, as in the $SU(2)$ problem, smooth configurations are governed by a NLSM, while singular tunneling events carry nontrivial Berry phases whose interference \emph{can} favor lattice-symmetry breaking. Nevertheless, this general mechanism does not by itself imply that every square-lattice $SU(N)$ AFM-like Hamiltonian must have a VBS groundstate. What one can say in general is more modest. The Berry phases constrain the allowed topological tunneling events and therefore strongly restrict the possible structure of a quantum-disordered phase, should such a phase be reached. But whether the system actually enters such a phase, and whether that phase is a fourfold VBS, a twofold bond-ordered state, or something less conventional, depends on 
\begin{center}
    (1) the representation, \\
    (2) the microscopic family of Hamiltonians under consideration.
\end{center}
In particular, the original large-$N$ analyses do not establish a universal outcome for all these $SU(N)$ Hamiltonians. Rather, they identify controlled limits in which the ground state can be determined explicitly. Historically, one first sees this logic already in the $SU(2)$ problem: Arovas and Auerbach \cite{ArovasAuerbach1988PRB,AuerbachArovas1988PRL} used Schwinger-boson large-$N$ methods to argue that the square-lattice antiferromagnet remains N\'eel ordered for the physically relevant large-spin cases $S=\frac{1}{2},1,\frac{3}{2},\dots\:\:$. In fact, for a general $SU(N)$ model in the large-$N$ limit, one finds
\begin{equation}
    (\text{Paramagnet}) \quad
    \left(\frac{n_c}{N}\right)
    <
    \left(\frac{n_c}{N}\right)_{\text{crit}}
    \approx 0.19
    \hspace{60pt}
    (\text{N\'eel}) \quad
    \left(\frac{n_c}{N}\right)
    >
    0.19\:\:,
\end{equation}
signaling two different regimes: a quantum paramagnet and a N\'eel-ordered phase controlled by the ratio $n_c/N$. A sharper finite-$N$ version of this statement was later obtained numerically through Quantum Monte Carlo simulations of the $SU(N)$ square-lattice nearest-neighbor model (the \emph{$J$-model}) in the $\bm{N}-\bm{\bar N}$ representation ($m=1,\; n_c=1$). While large-$N$ arguments indicate that increasing $N$ favors valence-bond order, they do not fix the precise behavior at finite $N$. This was clarified in 2003 by Kawashima \emph{et al.} \cite{HaradaKawashimaTroyer2003}, who found that the $J$-model exhibits
\begin{equation}
\begin{cases}
N \leq 4 & \text{N\'eel-ordered state},\\
N \geq 5 & \text{spin-Peierls / VBS state},
\end{cases}
\end{equation}
with no other intermediate states. Subsequently, Beach \emph{et al.} \cite{BeachAletMambriniCapponi2009} employed a continuous-$N$ projector QMC method and located the transition more precisely at
\begin{equation}
    N_c \approx 4.57,
\end{equation}
separating the N\'eel groundstate from a columnar valence-bond solid one (see $n_c=1$ of Fig.\ref{fig:groundstates_nc_mod_ph_lattice}). Thus, in the plain nearest-neighbor model, changing $N$ places different members of the family on different sides of the phase diagram; but it does not provide, at fixed integer $N$, a conventional tunable N\'eel--VBS transition.

 This observation suggests a natural next step. Instead of varying $N$, one may fix the symmetry group --- for instance $SU(2)$, $SU(3)$, or $SU(4)$ --- and modify the microscopic Hamiltonian in order to destabilize the N\'eel groundstate. In practice, there are several standard ways of doing this. In the physical $SU(2)$ case, where the local generators can be written as the spin vector $\vec S_i$, one possibility is to add frustrating further-neighbor exchanges,
\begin{equation}
H_{J_1J_2J_3}
=
J_1\sum_{\langle ij\rangle}\vec S_i\cdot \vec S_j
+
J_2\sum_{\langle\langle ij\rangle\rangle}\vec S_i\cdot \vec S_j
+
J_3\sum_{\langle\langle\langle ij\rangle\rangle\rangle}\vec S_i\cdot \vec S_j\:.
\end{equation}
Here the $J_2$ and $J_3$ terms compete with the nearest-neighbor antiferromagnetic order and may suppress the N\'eel phase. However, this does not by itself determine the nature of the resulting nonmagnetic phase: depending on the microscopic parameters, the system may develop valence-bond order, stripe or spiral order, or even a spin-liquid-like regime. Thus, frustrating exchanges should be viewed as a general mechanism for destabilizing N\'eel order, not as a direct construction of a VBS phase.

\noindent Another possibility is to add ring-exchange interactions, schematically
\begin{equation}
H_{\mathrm{ring}}
=
K\sum_{\square}
\left(P^{\mathrm{perm}}_{1234}+(P^{\mathrm{perm}}_{1234})^{-1}\right),
\end{equation}
where $P^{\mathrm{perm}}_{1234}$ cyclically permutes the local states around an elementary plaquette. Such terms favor resonances between local bond configurations and can also destabilize conventional magnetic order, although again the resulting phase depends on the lattice and couplings.

Now, translating these ideas to the quark--antiquark $SU(N)$ models discussed here, it is more convenient to phrase the same mechanism in terms of singlet projectors. On a bipartite lattice with $\mathbf N$ on sublattice $A$ and $\overline{\mathbf N}$ on sublattice $B$, the nearest-neighbor exchange is proportional, up to an additive constant, to the projector onto the unique $SU(N)$ singlet channel, $\Pi^{\mathrm{sing}}_{ij}\equiv \Pi_{\mathbf 1}^{SU(N)}$.  Equivalently, using the generator notation introduced above,
\begin{equation}
\frac{1}{N}
\sum_{\alpha,\beta}
S^\beta_{\alpha}(i)\,
\bar S^\alpha_{\beta}(j)
=
-\Pi^{\mathrm{sing}}_{ij}
+\text{constant}.
\end{equation}
Thus the nearest-neighbor quark--antiquark Hamiltonian may be written, after dropping constants, as
\begin{equation}
H_J
=
-J\sum_{\langle ij\rangle}
\Pi^{\mathrm{sing}}_{ij},
\qquad J>0,
\end{equation}
where the minus sign simply means that the energy is lowered when a bond forms an $SU(N)$ singlet. Then, for the specific purpose of studying a controlled N\'eel--VBS transition, it is more useful to introduce interactions that favor singlet formation directly; for instance, adding products of nearby singlet projectors. This is precisely the idea behind the $J$--$Q$ models:
\begin{equation}
H_{JQ}
=
-J\sum_{\langle ij\rangle}\Pi^{\mathrm{sing}}_{ij}
-
Q\sum_{\langle ijkl\rangle}
\Pi^{\mathrm{sing}}_{ij}\Pi^{\mathrm{sing}}_{kl}
+\cdots .
\end{equation}
Hence, the $Q>0$ term favors the simultaneous formation of nearby singlet bonds, and depending on the geometry chosen for the pair of bonds $(ij)$ and $(kl)$, it energetically favors columnar, plaquette, or related valence-bond patterns.

In this way, one obtains a tunable family of Hamiltonians,
\begin{equation}
H(g)
=
H_J
+
g\,H_{\mathrm{singlet}},
\end{equation}
where $g$ may stand for a frustrating exchange, a ring-exchange coupling, or a multi-singlet-projector interaction such as $Q/J$. For small $g$, the system may remain in the N\'eel phase, while for large enough $g$ the N\'eel order can be destroyed and a quantum-disordered phase may appear. To analyze the nature of such a transition, it is important to first identify the appropriate order parameters. In the $SU(2)$ case, the long-wavelength description of the antiferromagnet is formulated in terms of the staggered field $\mathbf n(x)$ appearing in the nonlinear sigma model,
\begin{equation}
|\mathbf n(x)|^2=1 \quad \Rightarrow \quad \expval{\mathbf n} \neq 0 \:\:\:(\text{N\'eel order}) 
\end{equation}
which encodes the N\'eel order: A nonzero expectation value of $\mathbf n$ signals spontaneous breaking of $SU(2)$ spin-rotation symmetry. On the other hand, a valence-bond solid is characterized by a modulation of bond energies. On the square lattice, this may be described by a $Z_4$ complex field
\begin{equation}
\psi_i = (-1)^{x_i}\,\mathbf S_i\cdot \mathbf S_{i+\hat x}
+ i(-1)^{y_i}\,\mathbf S_i\cdot \mathbf S_{i+\hat y},
\end{equation}
which breaks lattice symmetries while preserving spin rotation. A completely analogous structure emerges in the $SU(N)$ case. The N\'eel phase is described by the slowly varying Grassmannian field $\Omega(x)$ appearing in the sigma-model action derived above, while VBS order is again associated with symmetry-breaking patterns of bond operators.

With these order parameters in hand, the central question can now be formulated sharply: can the transition
\begin{equation}
\text{N\'eel}
\quad\longleftrightarrow\quad
\text{VBS}
\end{equation}
be continuous when driven by a microscopic tuning parameter $g$? From a conventional Landau--Ginzburg perspective, this appears unlikely. The N\'eel order parameter $\mathbf n(x)$ and the VBS order parameter $\psi$ transform under entirely different symmetry groups: $\mathbf n$ carries spin-rotation quantum numbers, while $\psi$ transforms nontrivially under lattice symmetries but is a spin singlet. Hence, a natural effective theory describing their competition is therefore of the form
\begin{equation}
\mathcal L(\mathbf n,\psi)
=
\mathcal L_{\mathrm{NLSM}}(\mathbf n)
+ |\partial_\mu \psi|^2
+ r_\psi |\psi|^2
+ u_\psi |\psi|^4
+ \lambda |\mathbf n|^2 |\psi|^2
+\cdots,
\end{equation}
where $\mathcal L_{\mathrm{NLSM}}(\mathbf n)$ is the usual AFM nonlinear sigma-model functional. Now the whole point of the story is that, generically, such a theory predicts either a first-order transition or an intermediate phase in which both orders coexist, and therefore,  a direct continuous transition between the N\'eel and VBS phases would require fine-tuning to a multicritical point.

However, the central insight of \textbf{Deconfined Quantum Criticality (DQCP)} \cite{DQCP2004Senthil, DQCP_Review} is that this conclusion can fail. In particular, it has been argued that for a broad class of quantum antiferromagnets --- including the models discussed above --- a direct second-order transition between N\'eel and VBS phases can occur without fine-tuning. The crucial point is that the critical theory is not naturally expressed in terms of the order parameters $\mathbf n$ or $\psi$. Instead, it is formulated in terms of \emph{fractionalized degrees of freedom} --- \textbf{spinons} --- coupled to an \emph{emergent gauge field}. But, from the perspective developed in the previous subsection, this result is not entirely unexpected. The Berry-phase analysis already shows that the relevant topological excitations are singular tunneling events (hedgehogs or monopoles), whose interference properties cannot be captured within a local order-parameter description. The parton constructions introduced earlier provide precisely the framework in which these excitations become explicit dynamical variables. In this sense,
\begin{center}
\emph{the transition to the parton description is not merely a convenient reformulation of the problem, but rather a natural step toward identifying the correct low-energy degrees of freedom governing the critical point}.
\end{center}

\noindent This general picture is realized explicitly in lattice models where a direct N\'eel--VBS transition can be accessed in a controlled way. The simplest example is the $SU(2)$ $J$--$Q$ model introduced above: here the singlet-projector structure is used to tune between antiferromagnetic and VBS phases. Numerical simulations show that this interpolation does not follow the conventional Landau scenario, but instead exhibits scaling behavior consistent with a continuous transition, together with an emergent $U(1)$ symmetry of the VBS order parameter near criticality \cite{Sandvik2007}. For larger $SU(N)$, the same idea must be implemented differently. Since the nearest-neighbor model is already VBS ordered for $N\geq5$, one instead introduces competing interactions that restore N\'eel order. A particularly useful construction is the $SU(N)$ $J_1$--$J_2$ model of Kaul and Sandvik \cite{KaulSandvik2012}, where $J_1$ corresponds to the quark--antiquark singlet-projector interaction and $J_2$ to a same-sublattice permutation term. In this case, increasing $J_2/J_1$ drives a transition from a VBS phase to a N\'eel phase, again providing a controlled setting to study their competition. Numerical results indicate that both orders become critical at the same point, with behavior approaching that of the noncompact $\mathbb{CP}^{N-1}$ theory at large $N$.

The key lesson is therefore not tied to a specific Hamiltonian, but to a structural feature: when a model allows for a direct interpolation between N\'eel and VBS phases without introducing additional competing orders, the resulting transition can fall outside the Landau--Ginzburg paradigm. In such cases, the critical theory is not governed by fluctuations of $\mathbf n$ or $\psi$, but by fractionalized spinons coupled to an emergent gauge field. This naturally motivates the parton formulation, to which we  turn in the next section, where these degrees of freedom arise explicitly and where the compact (and noncompact) $\mathbb{CP}^{N-1}$ descriptions can be derived in a systematic way.

\noindent Finally, it is worth emphasizing that the Néel--VBS transition is only one representative example of a broader class of \emph{Landau-forbidden} quantum critical phenomena. Within quantum magnets, one can also find proposed direct transitions from N\'eel order to algebraic spin-liquid phases, where the critical theory involves fermionic spinons coupled to an emergent gauge field rather than only the N\'eel order parameter \cite{GhaemiSenthil2006}. Another closely related class consists of transitions between different VBS phases: in some cases these are controlled by deconfined critical points with emergent spin-$\frac12$ excitations, while in quantum-dimer realizations near Rokhsar--Kivelson points the physics can instead involve height fields, gapless photons, and ``Cantor deconfinement'' \cite{VishwanathBalentsSenthil2004,FradkinHuseMoessnerOganesyanSondhi2004}. A more exotic example appears in topological spin-Hall systems, where skyrmions of the magnetic order carry electric charge; condensing them can produce superconductivity, allowing a direct spin-Hall-insulator--superconductor transition despite the unrelated broken symmetries \cite{GroverSenthil2008}. Thus, the common lesson is not that all these transitions are identical, but that the correct critical variables are often defects, fractionalized fields, and emergent gauge structures rather than the conventional order parameters themselves.

\noindent We now turn in the next section to the nearest-neighbor $SU(N)$ antiferromagnet itself, i.e. the $J$-model, as the simplest setting where the parton logic can be made explicit. The purpose there will not be to derive the full DQCP field theory previously described, but rather to understand how the same microscopic Hamiltonian can be analyzed in two complementary large-$N$ limits: the fermionic and bosonic formulations. 

\subsubsection*{Parton Large-N calculation of J-model}
With all the ingredients in place, we now return to the simplest nontrivial setting: the nearest-neighbor $SU(N)$ antiferromagnet, i.e. the $J$-model. The goal of this subsection is not to revisit its definition, but rather to show how the same microscopic Hamiltonian admits two complementary large-$N$ descriptions, depending on how the parton degrees of freedom are organized. The key point is that the fermionic and bosonic constructions are not merely different representations of the same algebra. Instead, they naturally reorganize the Hilbert space in ways that emphasize different physical regimes of the model.

On the fermionic side, the natural large-$N$ saddle is expressed in terms of bond Hubbard--Stratonovich fields. Schematically, the quartic interaction generated by the exchange term can be rearranged into a product of inter-site bilinears,
\begin{equation}
\sum_{\langle ij \rangle} \sum_{\alpha \beta} \sum_{a b}
c^{\dagger}_{\alpha a}(i)\,c_{\beta a}(i)\;
\frac{J}{N}\;
c^{\dagger}_{\beta b}(j)\,c_{\alpha b}(j)
=
- \sum_{\langle ij \rangle} \sum_{\alpha \beta a b}
c^{\dagger}_{\alpha a}(i)\,c_{\beta b}^\dagger(j)\;
\frac{J}{N}\;
c_{\beta a}(i)\,c_{\alpha b}(j)
\label{eq:quartic_fermions}
\end{equation}
and then decoupled by a complex matrix field $\chi_{ab}(ij)$. In the path-integral language, $\chi_{ab}(ij)$ is not an operator, but an auxiliary bond field whose saddle-point value measures the tendency of the system to form a singlet in the link $(ij)$:
\begin{equation}
\chi_{ab}(ij)
\quad\longleftrightarrow\quad
 \sum_{\alpha} c_{\alpha b}^{\dagger}(j) c_{\alpha a}(i).
\end{equation}
Thus, when the large-$N$ saddle develops a nonzero pattern of $\chi_{ab}(ij)$, the mean-field state is organized around bond amplitudes rather than local magnetic moments. This is why the fermionic formulation is naturally adapted to deeply quantum-disordered phases: the dominant variables are already the valence-bond degrees of freedom, and different saddle-point patterns of $\chi_{ij}$ directly describe different spin-Peierls or VBS states.

On the bosonic side, the starting point is formally very similar. The exchange interaction again generates quartic terms in the parton fields, which can be rearranged into singlet bilinears across a bond. Using the bosonic generators, and ignoring the traceless subtractions, since they only produce a constant shift in the Hamiltonian, one obtains schematically
\begin{equation}
H_{i j}
=
-\frac{J}{N}
\sum_{a, b}
\sum_{\alpha, \beta}
b_{\alpha a}^{\dagger}(i)\, b_{\beta a}(i)\;
\bar{b}_{\alpha b}^{\dagger}(j)\, \bar{b}_{\beta b}(j).
\end{equation}
Hence, as in the fermionic case, this interaction can be decoupled by introducing a complex Hubbard--Stratonovich field $Q_{ab}(ij)$, such that
\begin{equation}
Q_{ab}(ij)
\quad\longleftrightarrow\quad
\sum_{\alpha} b_{\alpha a}(i) \bar{b}_{\alpha b}(j),
\end{equation}
encoding the amplitude for forming a singlet between sites $i$ and $j$. In this sense, $Q_{ij}$ plays a role analogous to the fermionic bond field $\chi_{ij}$, describing short-range valence-bond correlations. However, a crucial difference with the fermionic formulation now emerges. While the fermionic theory is organized at the saddle-point level entirely in terms of bond variables, the bosonic path integral admits an additional class of saddle points. Because the bosonic fields are ordinary complex variables, the action allows for configurations in which the partons themselves acquire a coherent expectation value,
\begin{equation}
\langle b_{\alpha a}(i) \rangle \neq 0,
\qquad
\langle \bar b_{\alpha a}(j) \rangle \neq 0.
\end{equation}
Such configurations correspond to a macroscopic occupation of a single mode and are fully consistent with the local constraints. The key observation is that, in this regime, the physical $SU(N)$ generators develop a nonzero expectation value,
\begin{equation}
\langle S^\beta_{\alpha}(i)\rangle
=
\sum_a
\langle b^\dagger_{\alpha a}(i)\rangle
\langle b_{\beta a}(i)\rangle,
\end{equation}
so that the condensation of the bosonic partons directly produces a classical order parameter. In other words, \emph{long-range N\'eel order appears as a Bose condensation of spinons}.

\noindent Therefore, in the bosonic formulation the large-$N$ saddle admits two qualitatively distinct structures:
\begin{align}
\langle Q_{ij} \rangle \neq 0
&\quad\text{: singlet-pair amplitude (valence-bond correlations)},\\[4pt]
\langle b_{\alpha}(i) \rangle \neq 0
&\quad\text{: spinon condensation (N\'eel order)},
\end{align}
so that, in the large-$N$ limit, fluctuations around these saddle points are suppressed, and the physics is controlled by which of these structures dominates. The quantum-disordered regime corresponds to gapped bosons with finite $Q_{ij}$, while the N\'eel phase is characterized by spinon condensation and long-range magnetic order.

\medskip

In this way, although both fermionic and bosonic formulations introduce bond Hubbard--Stratonovich fields, their physical content is fundamentally different: the fermionic theory naturally selects saddle points organized in terms of valence-bond amplitudes, whereas the bosonic theory retains direct access to magnetic order through spinon condensation. In what follows, we will focus primarily on the fermionic formulation. This choice is motivated by the fact that its saddle-point structure directly captures the physics of quantum-disordered phases and provides a natural setting to understand how different valence-bond patterns—and their associated degeneracies—emerge. In this sense, it offers a complementary perspective to the Berry-phase analysis discussed previously, where such degeneracies arise from interference between hedgehog tunneling events. 

So, we now return to the quartic interaction in Eq.\ref{eq:quartic_fermions}. After rewriting it in terms of inter-site bilinears, the interaction can be decoupled by a Hubbard--Stratonovich transformation in the bond channel. At the same time, the local constraints on the fermionic Hilbert space must be enforced through Lagrange multiplier fields. Passing then to the path-integral formulation, the partition function is therefore expressed in terms of Grassmann fields $c_{\alpha a},\,c^\dagger_{\alpha a}$, together with auxiliary bosonic fields $\chi_{ab}(ij)$ and Lagrange multipliers $\lambda^{A,B}_{ab}$. Importantly, in this formulation the fields $\chi_{ab}(ij)$ and $\chi^\dagger_{ba}(ji)$ are treated as \emph{independent complex variables}; only at the saddle point one imposes the Hermiticity condition
\begin{equation}
\chi^\dagger_{ba}(ji) = \chi^*_{ab}(ij).
\end{equation}
With these ingredients, the effective action takes the form
\begin{equation}
\begin{aligned}
S_{\text{eff}} = \int_0^\beta d\tau \Bigg[
&\sum_{i,\alpha,a} 
c^\dagger_{\alpha a}(i)\,\partial_\tau\,c_{\alpha a}(i) \quad + \sum_{i \in A, a,b}
i \lambda^{A}_{ab}
\left(
\sum_\alpha c^\dagger_{\alpha a}(i)\,c_{\alpha b}(i)
- \delta_{ab} m
\right) \\[4pt]
&+ \sum_{i \in B, a,b}
i \lambda^{B}_{ab}
\left(
\sum_\alpha c^\dagger_{\alpha a}(i)\,c_{\alpha b}(i)
- (N-m)\delta_{ab}
\right) \\[6pt]
&+ \sum_{\langle ij \rangle, a, b}
\Bigg\{
\frac{N}{J}\,\chi_{ab}(ij)\,\chi^{\dagger}_{ba}(ji) \\
&\qquad\qquad \qquad
+ \sum_{\alpha}
\Big[
\chi_{ab}(ij)\,c^\dagger_{\alpha a}(i)\,c_{\alpha b}(j)
+
\chi^{\dagger}_{ba}(ji)\,c^\dagger_{\alpha b}(j)\,c_{\alpha a}(i)
\Big]
\Bigg\}
\Bigg].
\end{aligned}
\end{equation}
At this stage, the problem has been reduced to a theory of fermionic partons coupled to static auxiliary fields. In the large-$N$ limit, fluctuations of these fields are suppressed, and the path integral is dominated by saddle-point configurations. One therefore treats $\chi_{ab}(ij)$ and $\lambda_{ab}$ as static mean fields and determines them self-consistently by minimizing the effective action. This yields the mean-field equations
\begin{equation}
\begin{aligned}
    &\frac{\delta S_{\text{eff}}}{\delta\lambda_{ab}^i}=0\qquad \Rightarrow \qquad 
\sum_\alpha \langle c^\dagger_{\alpha a}(i)\,c_{\alpha b}(i) \rangle = \begin{cases}
    m\,\delta_{ab} \qquad \qquad \quad i\in A\\ (N-m)\,\delta_{ab} \qquad \:i\in B
\end{cases}\\[6pt]
&\frac{\delta S_{\text{eff}}}{\delta\chi_{ab}^*}=0\qquad \Rightarrow \qquad 
\chi_{ab}(ij)
=
- \frac{J}{N} \sum_\alpha
\langle c^\dagger_{\alpha b}(j)\,c_{\alpha a}(i) \rangle.
\end{aligned}
\label{eq:saddle_point_eq}
\end{equation}
Solving these equations determines the structure of the ground state. In particular, different saddle-point patterns of $\chi_{ij}$ correspond to different arrangements of bond amplitudes, allowing one to identify uniform, symmetry-broken (VBS), or in more general contexts, exotic spin-liquid-like states depending on their spatial structure and symmetry properties.

For a complementary route, it is convenient to integrate out the fermionic degrees of freedom. Since the action is quadratic in the Grassmann fields, this can be done exactly, leading to an effective action purely in terms of the auxiliary fields $\chi_{ab}(ij)$ and $\lambda_{ab}^i$:
\begin{equation}
\begin{aligned}
\frac{S_{\text{eff}}[\chi,\lambda]}{N}
=
\int_0^\beta d\tau \Bigg[
&\sum_{\langle ij\rangle,ab}
\frac{1}{J}\,|\chi_{ab}(ij)|^2
- \frac{i}{N}\sum_i \mathrm{tr}\big(\lambda_i G_i\big)
- \mathrm{tr}\,\ln\big(\partial_\tau + \lambda + \chi\big)
\Bigg],
\end{aligned}
\end{equation}
where the trace now runs over spatial, color, and imaginary-time indices, and we have introduced the matrix $G_i=\operatorname{diag}_{\text {col }}\left(\delta_{i \in A} m+\delta_{i \in B}(N-m)\right)$ that encodes the local constraints on each sublattice. Now, a particularly important simplification occurs in the particle--hole symmetric case. For even $N$, choosing the representation with $m = \frac{N}{2}$, the constraints on the two sublattices become identical. In this situation, the action is invariant under a particle--hole transformation that exchanges the two sublattices, and one finds that the saddle-point solution is given by $\lambda_{ab}^{A} = \lambda_{ab}^{B} = 0$. Physically, this reflects the absence of any local chemical potential term: the constraint is satisfied symmetrically without the need for a nontrivial Lagrange multiplier background.

\noindent Under these conditions, the effective action simplifies considerably and reduces to a functional depending only on the bond fields,
\begin{equation}
\boxed{
\frac{S_{\text{eff}}[\chi]}{N}
=
\int_0^\beta d\tau \left[
\sum_{\langle ij\rangle,ab}
\frac{1}{J}\,|\chi_{ab}(ij)|^2
-
\mathrm{tr}\,\ln\big(\partial_\tau + \chi\big)
\right].
}
\end{equation}
This form makes the large-$N$ structure particularly transparent: the competition between different phases is entirely encoded in the saddle-point configurations of the bond field $\chi_{ij}$, whose spatial pattern determines the nature of the ground state. However, the problem remains in the fact that the saddle-point equation (second in Eq.\ref{eq:saddle_point_eq}), admits several classes of solutions. A first natural possibility is given by \emph{translationally invariant} ans\"atze, in which the magnitude of the bond field is uniform on all links,
while its phases define a gauge-invariant flux through each plaquette,
\begin{equation}
|\chi_{ij}| = \chi\:,\hspace{40pt} \Phi_{\square}
=
\arg\!\left(\prod_{\square} \chi_{ij}\right).
\end{equation}
A particularly important example is the $\pi$-flux phase, characterized by $\Phi_{\square} = \pi$,
which corresponds to a uniform saddle with a gapless fermionic spectrum featuring Dirac nodes at $(\pm \pi/2,\pm \pi/2)$. This state preserves all lattice symmetries and describes a delocalized, spin-liquid-like configuration.
\begin{figure}
    \centering
    \includegraphics[width=0.93\linewidth]{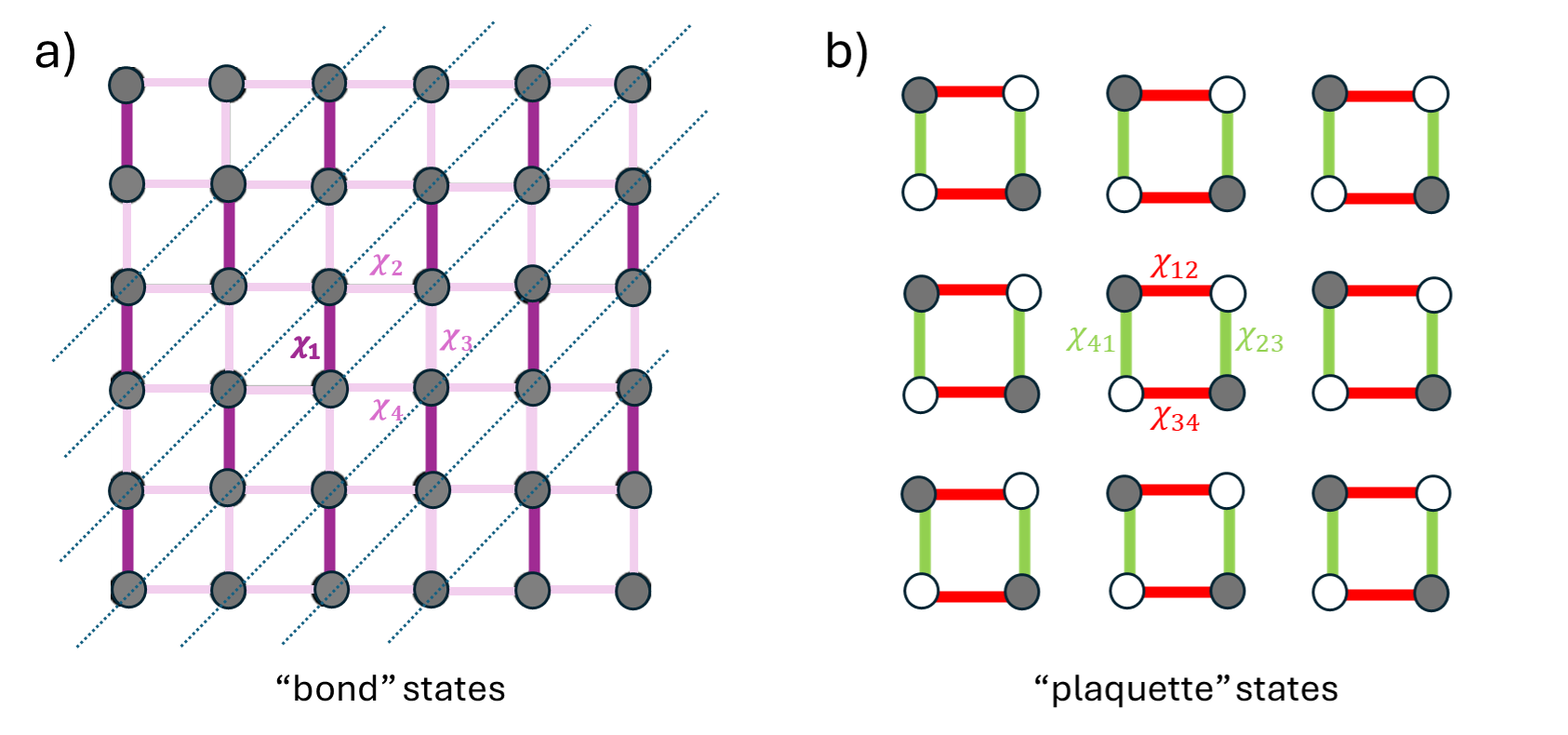}
    \caption{Groundstate configurations of a $\bm{N/2}-\bm{N/2}$ $SU(N)$ single color ($n_c=1$) Heisenberg model. (a)  Peierls phase/ Bond states, where $\chi_1\neq0,\: \chi_{2,3,4}=0$, and Flux phase, where $\chi_1=\chi_{2,3,4}$. (b) Alternative degenerate groundstate configurations giving rise to disconnected dimerized plaquettes.} 
    \label{fig:groundstate_particle_hole}
\end{figure}
But, the crucial observation made by Affleck and Marston is that such uniform solutions do not minimize the energy in the present large-$N$ limit. Instead, the system lowers its energy by developing \emph{spatially inhomogeneous} bond configurations, in which $\chi_{ij}$ becomes nonzero only on a subset of links \cite{affleck1988large}. These are the so-called \emph{bond states}, where each site participates in a fixed number of strong singlet bonds, forming a pattern that breaks lattice symmetries. A simple example is shown in Fig.\ref{fig:groundstate_particle_hole}a, where the bond field takes the form
\begin{equation}
\text{(bond states)}\hspace{25pt}\chi_{ij}
=
\begin{cases}
\chi=J/2 \neq 0, & \text{on selected dimer links},\\
0, & \text{otherwise},
\end{cases}
\end{equation}
so that each site forms a singlet with a single neighbor. The key point is that, at $N=\infty$, these inhomogeneous bond configurations are energetically degenerate at the level of the saddle point and define a large manifold of mean-field ground states. But moreover, these solutions correspond to localized singlet formation and thus naturally describe valence-bond-solid (VBS) order, providing a microscopic realization of the quantum-disordered phases discussed earlier.

Now, even though Affleck and Marston first proposed these VBS bond states, their work focused on the single-color case $n_c=1$. Therefore, one year later, in 1989, Read and Sachdev \cite{read1989some} generalized their work for an arbitrary $n_c$ number of colors and the fundamental-antifundamental $\bm{N-\overline{N}}$ lattice (not only $m=N-m=N/2$). In this vein, they showed that for an arbitrary number of colors $n_c$ (physically this will be the number of valence bonds), the $n_c\times n_c$ matrix field $[\chi_{ab}(ij)]_{ab}$ becomes diagonal in color space in the large $N$ limit. This means that, taking into account the decoupling of each color bond, the construction of groundstate configurations is equivalent to the single color calculations done by Affleck and Marston, with the difference that in the end one has to take the tensor product of all the colored lattice states. Now, one of the key contributions of this seminal work was the proposal of VBS states that break lattice symmetry in nontrivial ways. Particularly, they found that there is an additional continuous family of mean field solutions which is degenerate with the bond states in the $N \rightarrow \infty$ limit. This family consists of single color bond configurations of disjoint plaquettes, where the mean field values of $\chi_{a}(ij)$ are non-zero only along the links surrounding the plaquette (see Fig.\ref{fig:groundstate_particle_hole}b), and the  specific constraints of $\chi^{(a)}_\square$ (change in notation for simplicity) are 
\begin{equation}
    \begin{gathered}
    \left|\chi^{(a)}_{12}\right|=\left|\chi^{(a)}_{34}\right|\equiv h, \qquad\left|\chi^{(a)}_{41}\right|=\left|\chi^{(a)}_{23}\right|\equiv v, \qquad h^2+v^2=\left(\frac{J}{2}\right)^2, \\
\chi^{(a)}_{12} \chi^{(a)}_{23} \chi^{(a)}_{34}\chi^{(a)}_{41}=-h^2v^2\\[6pt] (\text{plaquette states}) .
\end{gathered}
\end{equation}
Here, $h,v\in \mathbb{R}$ denote the bond amplitudes on the horizontal and vertical links of the plaquette, respectively. Interestingly, note that the product of the bond fields around the loop in last expression defines a gauge-invariant flux, and the negative sign corresponds precisely to a phase $\Phi_\square=\pi$. However, this should not be confused with the uniform $\pi$-flux state discussed earlier. In the present case, the $\pi$-flux is realized \emph{locally} on isolated plaquettes, with $\chi_{ij}$ vanishing on all other links. 

At this stage, it is important to pause and clarify the meaning of the large-$N$ saddle-point solutions. 
As we have seen, the minimization of the effective action at $N=\infty$ does not select a unique ground state, but rather a highly degenerate manifold of configurations. These configurations include coverings of the lattice by \emph{colorful disconnected dimers}, and more general families of \emph{colorful disconnected plaquettes}, where the bond amplitudes are nonzero only along the edges of isolated loops. Here, the term ``overlapping'' refers to the fact that different colors are decoupled in the large-$N$ limit, so that each color can realize an independent bond configuration. As a result, bonds of different colors are not mutually exclusive and may occupy the same links without additional constraints.

\noindent Now, the crucial point is that, at the saddle-point level, the specific arrangement of these bonds is arbitrary: different dimer coverings or plaquette tilings all have the same energy. In other words, the large-$N$ theory identifies a vast manifold of degenerate mean-field states, rather than a single ordered configuration. This degeneracy is not accidental, but reflects the fact that at leading order the theory is sensitive only to the \emph{local formation of singlets}, and not to their global spatial arrangement. However, this also implies an important limitation: these saddle-point configurations should not be interpreted as exact ground states of the microscopic Hamiltonian. Rather, they should be understood as a set of \emph{low-lying states} that become exactly degenerate only in the strict $N\to\infty$ limit. At any finite $N$, fluctuations around the saddle point lift this degeneracy and select a subset of configurations as true ground states.

\begin{figure} \centering \includegraphics[width=0.95\linewidth]{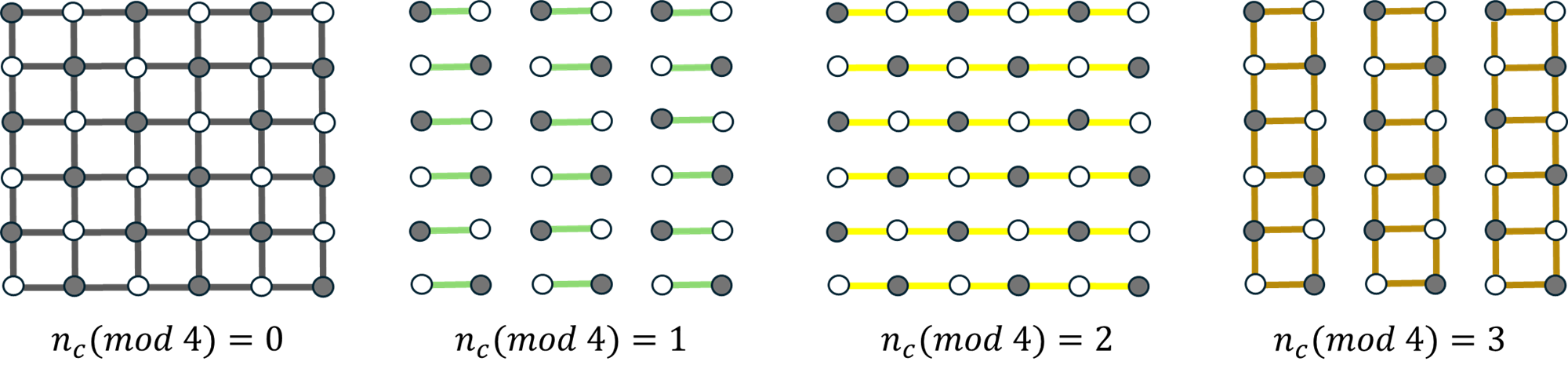} \caption{Possible metastable configurations of the $\bm{N/2-N/2}$ square bipartite lattice. Note how the degeneracy follows the lower bond rule of 1,4,2,4 for $n_c$ values of 0,1,2,3 (mod 4), respectively. } \label{fig:groundstates_nc_mod_ph_lattice} \end{figure}

\noindent Indeed, the structure of these corrections can be analyzed systematically in a $1/N$ expansion. As shown by Read and Sachdev, the leading $1/N$ corrections arise from fluctuations of the bond fields $\chi_{ij}$ and depend sensitively on the relative arrangement of neighboring bonds. In particular, in the case of the $\bm{N/2-N/2}$ square lattice, configurations with a larger number of \emph{parallel bonds} gain more energy from fluctuations than more isotropic configurations (such as plaquette states). As a consequence, the degeneracy of the saddle-point manifold is lifted in favor of specific symmetry-breaking patterns. For instance, for $n_c=1$, one can show that $1/N$ corrections lower the states' energies according to the number of parallel bonds in the configuration, selecting a specific family of \emph{columnar valence-bond-solid} states shown in Fig.\ref{fig:groundstates_nc_mod_ph_lattice}. More generally, for arbitrary $n_c$, the same interplay between bond geometry and fluctuations determines the structure of the lowest-energy configurations (see others $n_c$ in Fig.\ref{fig:groundstates_nc_mod_ph_lattice}).

Here is precisely where the interesting part comes into the game. Looking again at Fig.\ref{fig:groundstates_nc_mod_ph_lattice}, we note that the groundstate degeneracies of systems with $n_c(\text{mod 4})=0,1,2,3$ are exactly 1,4,2,4, respectively:
\begin{itemize}
    \item For $n_c\equiv_4 0$, the net cannot be altered without changing the configuration, making it the only possible solution.
    \item In the case of $n_c\equiv_4 2$, another state is possible by rotating $\pi/2$ all the bonds, causing a double degeneracy.
    \item And finally, for $n_c\equiv_4 1,3$, in addition to the $\pi/2$ rotation, there are two other possible configurations obtained by the shift of one sublattice position in all bonds. Therefore, their degeneracy: $2\times 2=4$. 
\end{itemize}

\noindent However, this is not a coincidence. The pattern of degeneracies observed, is in fact the direct imprint of the Berry-phase structure discussed previously in the $SU(2)$ and $SU(N)$ cases, now realized within the large-$N$ framework. In the semiclassical description, the proliferation of hedgehog tunneling events is strongly constrained by quantum interference: Berry-phase factors lead to destructive interference for most processes, allowing only those with specific periodicities. This imposes a selection rule on the allowed configurations, which in turn fixes the degeneracy of the low-energy states. From the present large-$N$ perspective, the same mechanism appears in a complementary form. The saddle-point analysis identifies the manifold of competing singlet configurations, while the $1/N$ corrections act as an interference mechanism that selects among them. Remarkably, the subset of configurations that survives this selection --- and their symmetry properties --- precisely reproduces the degeneracy pattern dictated by Berry-phase interference. 

\noindent In this sense, the large-$N$ construction provides a controlled and transparent realization of the same mechanism previously identified through Berry phases. This highlights the deeper role of the $SU(N)$ generalization. Rather than simply enhancing quantum fluctuations, it reorganizes the problem so that the relevant low-energy degrees of freedom --- singlet bonds, their spatial patterns, and their collective fluctuations --- become explicit. In particular, the parton construction naturally captures these features and provides a unified framework in which both ordered and quantum-disordered phases, as well as the mechanisms selecting between them, can be analyzed on equal footing. Therefore, the passage to $SU(N)$ and the introduction of parton degrees of freedom should be understood not as formal extensions, but as essential tools for revealing the geometric and topological structure underlying strongly correlated systems. 

\subsubsection{SU(4) Spin-Orbit and Spin-Pseudospin Symmetries}\label{sec:su4_kk_model}
In the previous subsection, we considered $SU(N)$ generalizations of the Heisenberg model primarily as a heuristic tool —-- a convenient, though not directly physical, framework to gain insight into the $SU(2)$ case. We now turn to systems where $SU(N)$ symmetries are not merely theoretical constructs but can emerge in actual materials, such as in $SU(4)$ spin-orbital and spin-pseudospin models relevant to transition metal compounds and graphene-based systems. Now, before proceeding, a word of caution is in order: one might be reluctant to accept $SU(N)$ symmetries with $N>2$ in condensed matter systems, given that there are no fundamental degrees of freedom beyond spin and charge that are strictly conserved. However, such symmetries should be understood as emergent or approximate, valid within an effective low-energy description where symmetry-breaking perturbations are either absent or irrelevant at the energy scales of interest. In this sense, $SU(N)$ models serve as controlled approximations that capture the essential physics while filtering out subleading effects.

\subsubsection*{The SU(4) Kugel-Khomskii model}

In this way, the first relevant example for us is the so-called Kugel-Khomskii (KK) model, in which, apart from spin and charge, the orbital degrees of freedom of the electron play a significant role. Similar to the case of a Mott insulator, this model is based on the fact that strong Coulomb interactions suppress charge fluctuations, so that spin and orbital remain as the dominant low-energy degrees of freedom. Hence, the Kugel-Khomskii model provides an effective description (indeed, in 2nd-order perturbation theory) of spin-orbital exchange interactions, typically derived in the strong-coupling limit of multi-orbital Hubbard models.

Furthermore, the KK model applies to a broad class of strongly correlated materials, particularly transition metal oxides and certain rare-earth compounds, where partially filled $d$ - or $f$-shells give rise to complex electronic behavior. A paradigmatic example is $\mathrm{LaMnO}_3$, a perovskite manganite with one $e_g$ electron per Mn site (explained in detail later), where the orbital ordering and magnetic structure are well captured by a mean-field treatment of the Kugel-Khomskii Hamiltonian. In contrast, compounds such as $\mathrm{CeB}_6$ go beyond the mean-field (MF) picture due to the strong entanglement between spin and orbital degrees of freedom, as well as enhanced quantum fluctuations arising from the localized $f$-electrons. In such systems, the interplay between spin and orbital dynamics leads to multipolar ordering and nontrivial collective excitations, which require more sophisticated theoretical approaches beyond static MF theory described in the previous pages.

\begin{figure}[ht]
    \centering
    \includegraphics[width=1\linewidth]{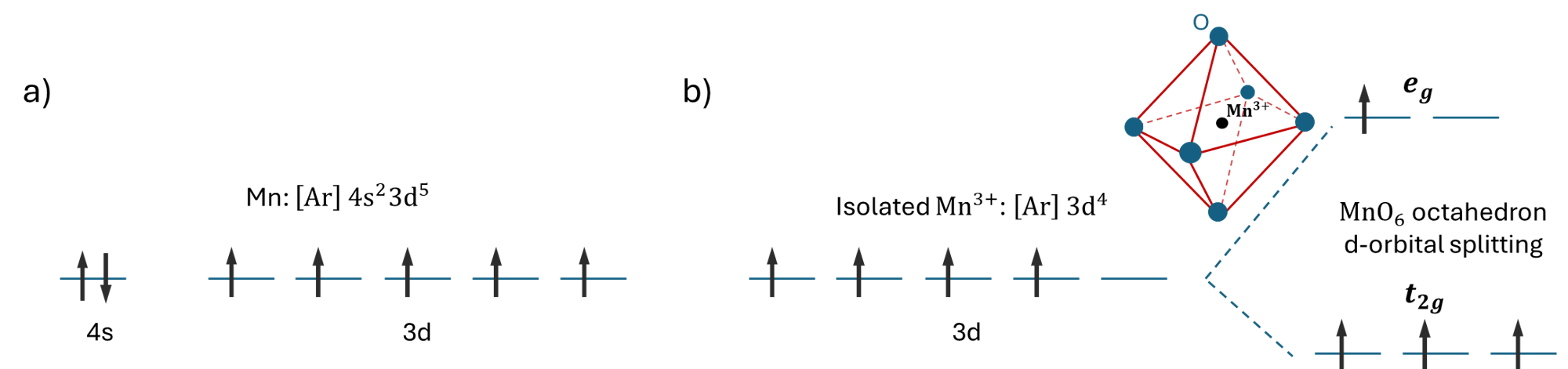}
    \caption{ Electron orbital filling in neutral degenerate (left) and ionic split (right) energy levels.}
    \label{fig:Hunds_Mn}
\end{figure}

As mentioned before, the starting point of the KK derivation is a multi-orbital Hubbard model with both on-site Coulomb repulsion and orbital degeneracy. Specifically, one considers electrons occupying degenerate orbitals with strong local interactions that suppress double occupancy. This is written in the most general form in the Hamiltonian
\begin{equation}
    \begin{split}
        H=  \underbrace{\sum_{i \alpha \sigma}\left(-t_{i, i+1}^{\alpha} c_{i \alpha \sigma}^{\dagger} c_{i+1 \alpha \sigma}+\text {h.c}\right)}_{H_t \: \equiv \: \text{intra-orbital hopping}} \underbrace{+\:\:\frac{U}{2}\sum_{\substack{i,\alpha\alpha',\sigma\sigma' \\(\alpha,\sigma) \neq (\alpha',\sigma')}}
n_{i \alpha \sigma} n_{i \alpha' \sigma'} }_{H_U \: \equiv \:\text{on-site Coulomb repulsion}} \: \underbrace{- \: \:\: J \sum_i\left(2 \vec{S}_{i 1} \cdot \vec{S}_{i 2}+\frac{1}{2}\right)}_{H_J \: \equiv \:\text{Hund's coupling}}
    \end{split}
    \label{eq:multi_orbital_Hubbard}
\end{equation}
where
\vspace{-0.38cm}
\begin{itemize}
    \item[-] $c_{i \alpha \sigma}^{\dagger}\left(c_{i \alpha \sigma}\right)$ represents an electron creation (annihilation) operator with orbital $\alpha=1,2$ and spin $\sigma=\uparrow,\downarrow$ at the $i$-th site; and as usual $n_{i \alpha \sigma}=c_{i \alpha \sigma}^{\dagger} c_{i \alpha \sigma}$ is the occupation number operator. 
    \item[-] $\vec{S}_{i \alpha}$ denotes the on-site electron spin operator within the orbital $\alpha$ at the $i$-th site.
    
    \item[-] the first term $H_t$ describes intra-orbital hopping processes, where electrons move between neighboring lattice sites while conserving both their spin $\sigma$ and orbital index $\alpha$. That is, hopping is only allowed between the same orbitals on adjacent sites.

    \item[-] the second term $H_U$ accounts for on-site $SU(4)$-symmetric Coulomb repulsion between distinct spin-orbital flavors. The prefactor $1/2$ prevents double counting of interaction pairs, and the condition $(\alpha,\sigma) \neq (\alpha',\sigma')$ is imposed to exclude self-interaction. Note that this term assumes that all possible on-site interactions occur with the same strength $U$ regardless of the specific spin and orbital configurations (see Fig.\ref{fig:onsite_term}).
\begin{figure}[ht]
    \centering
    \includegraphics[width=0.9\linewidth]{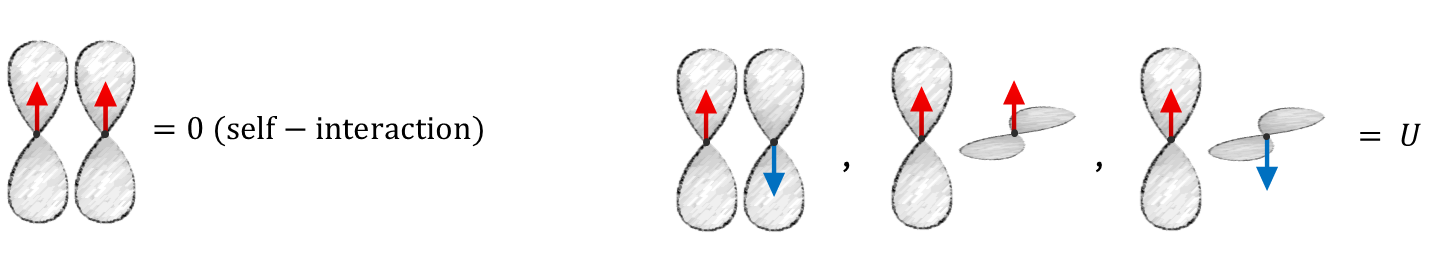}
    \caption{Uniform on-site interaction strength $U$. Here, orbitals $\alpha=1,2$ are diagrammatically represented by $p_z,p_x$ atomic levels.}
    \label{fig:onsite_term}
\end{figure}

    \item[-] the third term $H_J$ is the so-called Hund's coupling, which energetically favors the alignment of electron spins in different orbitals at the same site. This term arises from exchange interactions and reflects the tendency of electrons to maximize their total spin in accordance with Hund's first rule.
\end{itemize}
Recall that Hund’s rules provide a qualitative understanding of how electrons fill degenerate orbitals in an atom in order to minimize the total energy due to electron–electron interactions. In particular, they can be stated as:
\vspace{-0.2cm}
\begin{enumerate}
\item **First rule**: Electrons prefer to maximize their total spin $\bm{S}$ by occupying different orbitals with parallel spins. This reduces Coulomb repulsion via the Pauli exclusion principle and is directly associated with the Hund’s exchange term $H_J$ in the Hamiltonian.
\item **Second rule**: For a given spin configuration, electrons arrange themselves to maximize their total orbital angular momentum $\bm{L}$, reflecting additional degeneracy lifting due to Coulomb interactions.
\item **Third rule**: In atoms with less than half-filled shells, the total angular momentum $\bm{J = |L - S|}$ is favored, while in more than half-filled shells, $\bm{J = L + S}$ is energetically preferred. This rule becomes relevant in the presence of spin–orbit coupling, which we will not consider in this discussion.
\end{enumerate}
To put this into practice, it is imperative to study the case of, for example, Mn: $[\mathrm{Ar}] \:3 d^5 4 s^2$. In neutral atoms, electrons fill orbitals in order of increasing effective energy
\begin{equation}
    1 s<2 s<2 p<3 s<3 p<4 s<3 d<4 p \ldots
\end{equation}
so $4s$ fills before $3 d$, and hence the electronic configuration for Mn is $3 d^5 4 s^2$, with Hund's spin-orbital setting as shown in Fig.\ref{fig:Hunds_Mn}a.

In ions, however, the 3d orbitals become more stable than the 4s orbital. This happens because first, after filling, the 3d electrons are more strongly attracted to the nucleus and screen poorly; and second, the 4s orbital is more diffuse (further from the nucleus), so it’s more weakly bound once other electrons are present. As a result, when electrons are removed, they come from the highest-energy (least bound) orbitals first --- which turn to be the 4s electrons. So for the isolated ion $\mathrm{Mn}^{3+}$, we have:
\begin{equation}
    \mathrm{Mn}:[\mathrm{Ar}]\: 3 d^5 4 s^2 \:\longrightarrow \: \mathrm{Mn}^{3+}:[\mathrm{Ar}] \: 3 d^4.
\end{equation}
Now, this is not the end of the story. Due to the presence of nearby oxygen ions, the potential felt by the Mn electrons in the d-orbitals is modified: some orbitals point directly toward the oxygen atoms and experience stronger electrostatic repulsion, while others point between them and feel less. This effect is known as \textbf{crystal field splitting}, and it lifts the degeneracy of the five 3d orbitals. In an octahedral crystal field (from the $\mathrm{MnO}_6$ octahedron within the $\mathrm{LaMnO}_3$ structure - see Fig.\ref{fig:Hunds_Mn}b), the Mn-3d levels split into two sets: a lower-energy triplet called $t_{2 g}\left(d_{x y}, d_{y z}, d_{z x}\right)$ and a higher-energy doublet called $e_g\left(d_{x^2-y^2}, d_{3 z^2-r^2}\right)$. As a result, the $3 d^4$ electrons of $\mathrm{Mn}^{3+}$ occupy these levels according to Hund's rules and the strength of the crystal field, leading to the configuration: $t_{2 g}^{\uparrow \uparrow \uparrow} e_{g}^{\uparrow}$.

\noindent Hence, when referring to the orbital index $\alpha=1,2$ in the Hubbard Hamiltonian (Eq.\ref{eq:multi_orbital_Hubbard}), we are specifically labeling these two $e_g$ orbitals. That is,
$$
\alpha=1 \Rightarrow d_{3 z^2-r^2}, \quad \alpha=2 \Rightarrow d_{x^2-y^2}
$$
This identification is crucial because it restricts the active Hilbert space to a two-orbital system per site, consistent with the experimentally relevant degrees of freedom in $\mathrm{LaMnO}_3$. Particularly, while the $t_{2 g}$ orbitals, being singly occupied and lower in energy, are often treated as an inert localized core spin; the dynamics of the mobile $e_g$ electron, on the other hand, is governed by the interplay between its spin and orbital index, both of which are coupled via superexchange interactions (as we will see) arising from virtual hopping processes between neighboring sites.

Before proceeding, let's cover one last aspect of the Hubbard Hamiltonian. On a first look, one might be puzzled by the particular structure of the Hund's coupling term
$$
H_J\:=\:-J \sum_i\left(2 \vec{S}_{i 1} \cdot \vec{S}_{i 2}+\frac{1}{2}\right).
$$
Why doesn't it resemble a standard Heisenberg-type interaction like $J_H \sum_i \vec{S}_{i 1} \cdot \vec{S}_{i 2}$ ? Why this specific 2 prefactor and 1/2 additive constant?

\noindent The reason lies in a convenient redefinition of the exchange constant $J_H$, such that the term shown above energetically favors configurations with maximal total spin (i.e., parallel spins in different orbitals), treating both aligned (triplet) and anti-aligned (singlet) configurations symmetrically. To make this more transparent, recall that for two spin- $\frac{1}{2}$ operators $\vec{S}_{i 1}$ and $\vec{S}_{i 2}$ at the same site $i$, the total spin squared satisfies:
$$
\left(\vec{S}_{i 1}+\vec{S}_{i 2}\right)^2=\vec{S}_{i 1}^2+\vec{S}_{i 2}^2+2 \vec{S}_{i 1} \cdot \vec{S}_{i 2}
$$
Each individual spin-$\frac{1}{2}$ has $\vec{S}^2=\frac{3}{4}$, so we get:
\begin{equation}
    - J_H\:\vec{S}_{i 1} \cdot \vec{S}_{i 2}= -J_H\left[\frac{1}{2}\left(\vec{S}_{i 1}+\vec{S}_{i 2}\right)^2-\frac{3}{2}
    \right]= 
    \left\{
    \begin{array}{ll}
        +\frac{3}{4} \scriptstyle J_H \:\:, & \text{singlet } (S=0)\\[3pt]
        -\frac{1}{4} \scriptstyle J_H \:\:, & \text{triplet } (S=1) 
    \end{array}
    \right.
\end{equation}
However, this is not symmetric. So we can construct a term that projects onto the spin triplet state, specifically satisfying:
$$
\begin{cases}\text { Triplet }(S=1): & 2 \vec{S}_{i 1} \cdot \vec{S}_{i 2}+\frac{1}{2}=+1 \\ \text { Singlet }(S=0): & 2 \vec{S}_{i 1} \cdot \vec{S}_{i 2}+\frac{1}{2}=-1\end{cases}
$$
This naturally gives rise to the Hund's coupling term in the form:
\begin{equation}
   H_J=-J \sum_i\left(2 \vec{S}_{i 1} \cdot \vec{S}_{i 2}+\frac{1}{2}\right)=\left\{
    \begin{array}{ll}
        + J \:\:, & \text{singlet } (S=0)\\[3pt]
        -J\:\:, & \text{triplet } (S=1) 
    \end{array}
    \right. 
\end{equation}
which assigns an energy of $-J$ to triplet states and $+J$ to singlet states. The form is particularly convenient for modeling because it directly reflects Hund's first rule: states with aligned spins (maximum total spin) are energetically favored. In conclusion, the additive constant $\frac{1}{2}$ simply ensures that the energy difference between the singlet and triplet sectors is exactly $2J$, consistent with the level splitting obtained from atomic physics. This symmetric normalization simplifies then the analysis of spin-orbital models and helps clarify how the interaction energetically selects spin-aligned states.

With all the ingredients ready, we can now start with the \textbf{derivation of the SU(4) KK model}. In the following, we consider:
\begin{center}
\fbox{%
\parbox{0.8\textwidth}{%
\begin{itemize}
    \item[\checkmark] One electron per lattice site (or equivalently $\frac{1}{4}$-filling)
    \item[\checkmark] Hund's coupling $J=0$
    \item[\checkmark] No pair-hopping and spin-flip terms (i.e. hopping conserves $\alpha,\sigma$)
\end{itemize}
}}
\end{center}
such that (1) the only interaction is an on-site repulsion $U$ that penalizes double occupancy; (2) all spin-orbital states are treated symmetrically: $SU(4)$ symmetry; and (3) no terms distinguish between spin-aligned and spin-antialigned configurations in doubly occupied states.

The question now is, why do we consider this situation? The reason is that the $SU(4)$ symmetric limit provides a clear and tractable theoretical framework in which spin and orbital degrees of freedom are fully entangled and treated equally. This limit captures the essential structure of Kugel-Khomskii physics, especially when Hund’s coupling is much smaller than the Mott gap ($J\ll U$). Moreover, the $SU(4)$ model acts as a low-energy fixed point from which one can systematically study more realistic effects such as symmetry breaking, quantum fluctuations, and deviations beyond mean-field theory.

To proceed, let us briefly recall the general framework of deriving effective low-energy Hamiltonians using second-order perturbation theory. Consider a full Hamiltonian $H=H_0+V$, where $H_0$ is exactly solvable and has a well-defined low-energy subspace $\mathcal{L}$ separated by an energy gap from higher-energy states. We denote by $P$ the projector onto this low-energy subspace and by $Q=1-P$ its orthogonal complement. When the perturbation $V$ is small compared to the energy scale of the gap, the low-energy dynamics can be described by an effective Hamiltonian acting only within $\mathcal{L}$. To second order in $V$, this effective Hamiltonian can be formally written as
\begin{equation}
    H_{\mathrm{eff}}=P H_0 P\:+\:P V P\:-\:P V Q \frac{1}{H_0 -E_0} Q V P
\end{equation}
where $E_0$ is the ground state energy of $H_0$ within $\mathcal{L}$. The first two terms represent the projection of the unperturbed Hamiltonian and the perturbation onto the low-energy subspace, while the third term encodes virtual processes where the system is temporarily excited out of $\mathcal{L}$ by $V$ and then returns. The operator $\frac{1}{H_0 -E_0}$ acts as a resolvent projecting onto the high-energy sector and accounts for the energy cost of these virtual excitations (also called propagator/Green function).

Having recalled the general formalism of second-order perturbation theory for deriving effective low-energy Hamiltonians, we now apply this technique to our multi-orbital Hubbard model with spin and orbital degrees of freedom. In this context:
\begin{equation}
        \boxed{\begin{split}&H= \sum_i \Bigg\{ \:\underbrace{\frac{U}{2}\sum_{\substack{
                          (\alpha, \sigma) \neq (\alpha',  \sigma')}}
n_{i \alpha \sigma} n_{i \alpha' \sigma'} }_{\text{exactly solvable}\:H_0} \: + \: \underbrace{\sum_{\alpha \sigma j}\left(-t_{i j}\: c_{i \alpha \sigma}^{\dagger} c_{j \alpha \sigma}+\text {h.c}\right)}_{\text{small perturbation}\: V} \:\Bigg\}\\ \\
&\:\:\:\:\:\mathcal{L}_i=\Big\{|1^\uparrow\rangle_i,\:\:|1^\downarrow\rangle_i,\:\:|2^\uparrow\rangle_i,\:\:|2^\downarrow\rangle_i\Big\}_{\text{site }i} \Longrightarrow \mathcal{L}=\operatorname{span}\left( \bigotimes_i \mathcal{L}_i\right) \\ \\
&\:\:\:(\mathcal{H}\setminus \mathcal{L}_i)_j=\Big\{|1^\uparrow,1^\downarrow\rangle_j,\:\:|2^\uparrow,2^\downarrow\rangle_j,\:\:|1^\updownarrow,2^\updownarrow\rangle_j\Big\}_{\text{site }j}\otimes \:\emptyset_{\text{site }i}\\ 
&\qquad\qquad\qquad\qquad\qquad\qquad\Longrightarrow
\mathcal{H} \setminus \mathcal{L} = \operatorname{span}\left(
\bigcup_{\langle i,j \rangle} (\mathcal{H} \setminus \mathcal{L}_i)_{j}
\right)
    \end{split}}
    \label{eq:symmetric_Hubbard_manifold}
\end{equation}
\begin{itemize}
    \item $H_0$ corresponds to the on-site interaction term, which sets the high-energy scale by penalizing double occupancy.
    \item $V$ is the electron hopping term, which acts as a perturbation mixing states with different occupancies.
    \item The low-energy subspace $\mathcal{L}$, onto which the projector $P$ projects, is defined by states with exactly one electron per site.
    \item The complementary projector $Q=1-P$ projects onto states with at least one doubly occupied site, which are energetically costly due to $U$.
\end{itemize}
By substituting these elements into the effective Hamiltonian formula,
\begin{equation}
    H_{\mathrm{eff}}=-P V Q \frac{1}{H_0 -E_0} Q V P,
    \label{eq:effective_hamiltonian}
\end{equation}
we obtain an effective low-energy model that captures the interplay of spin and orbital degrees of freedom mediated by virtual hopping processes. Let's see how it works:
\begin{enumerate}[label=(\roman*)]
    \item The projectors $P$ are only there to recall that this effective Hamiltonian is only valid in the low-energy subpace $\mathcal{L}$. Hence, once we pick up one state $\ket{\psi}\in\mathcal{L}$, we can forget about its action ($P\ket{\psi}=\ket{\psi}$). In this way, let's take an arbitrary 
    \begin{equation}
|\psi\rangle=\:...\otimes\:\ket{a}_i\otimes\ket{b}_j\: \otimes \:... \in \mathcal{L} \qquad\text{ with }\quad a, b \in\{1^\uparrow, 1^\downarrow, 2^\uparrow, 2^\downarrow\}.
    \end{equation}

    \item We now act with the hopping term $V$ on $|\psi\rangle$, having in mind that there are two types of processes: (a) Electron from site $i$ hops to site $j$; and (b) electron from site $j$ hops to site $i$. This is written as 
    \begin{equation}
    \begin{split}
        (V)_{ij}\ket{\psi}&=-\sum_{\mu}\left(t_{i j}\: c_{i \mu}^{\dagger} c_{j \mu}\:+\:t_{ij}^*\:  c_{j \mu}^{\dagger}c_{i \mu}\right)\left[\:...\otimes\:\ket{a}_i\otimes\ket{b}_j\: \otimes \:...\right]\\&=-\sum_{\mu}\Big[t_{ij}\left(...\otimes\:\ket{a,b}_i\otimes\emptyset_j\: \otimes \:...\right)\delta_{\mu,b}\:+\:t_{ij}^*\left(...\otimes\:\emptyset_i\otimes\:\ket{a,b}_j \: \otimes \:...\right)\delta_{\mu,a}\Big]\\&=-\Big[t_{ij}\left(...\otimes\:\ket{a,b}_i\otimes\emptyset_j\: \otimes \:...\right)\:+\:t_{ij}^*\left(...\otimes\:\emptyset_i\otimes\:\ket{a,b}_j \: \otimes \:...\right)\Big].
    \end{split}
    \end{equation}
    Note that the resulting state now lives in $\mathcal{H}\setminus \mathcal{L}$ (since it is a superposition of two states with a doubly occupied site), and hence the projector $Q$ does nothing.

    \item Continuing with Eq.\ref{eq:effective_hamiltonian}, we have to apply $(\hat{H}_0-E_0)^{-1}$, where $E_0$ is defined by $H_0\ket{\psi}=E_0\ket{\psi}$. However, since $\ket{\psi}$ is a state with single occupied sites, there is no on-site energy and therefore $E_0=0$. On the other hand, for the action of $\hat{H}_0$, we have 
    \begin{equation}
    \begin{split}
        &\hat{H}_0\:(...\otimes\:\ket{a,b}_i\otimes\emptyset_j\: \otimes \:...)=U \:(...\otimes\:\ket{a,b}_i\otimes\emptyset_j\: \otimes \:...)\\
        &\hat{H}_0\:(...\otimes\:\emptyset_i\otimes\ket{a,b}_j \: \otimes \:... )= U \:(...\otimes\:\emptyset_i\otimes\ket{a,b}_j \: \otimes \:... ),
    \end{split}
    \end{equation}
    due to the double occupancy of sites $i$ and $j$, respectively. This means that up to this moment, we have
    \begin{equation}
    \begin{aligned}
         -\left(Q \frac{1}{H_0 -E_0} Q V P\right)_{ij} \ket{\psi}&=\frac{1}{U}\Big[t_{ij}\left(...\:\ket{a,b}_i\otimes\emptyset_j\:...\right)\:+\:t_{ij}^*\left(...\:\emptyset_i\otimes\ket{a,b}_j\:...\right)\Big].
        \end{aligned}
    \end{equation}
    \item For the final step, we now realize that for coming back to $\mathcal{L}$ (so that the last projector $P$ doesn't give 0), the only term in $V$ which doesn't give a vanishing answer is the hopping which reverses the double occupancy of our intermediate state. This is exactly 
    \begin{equation}
    \begin{split}
    &-\sum_{\mu}\left(t_{ji}\, c_{j\mu}^{\dagger}c_{i\mu}+t_{ji}^{*}\,c_{i\mu}^{\dagger}c_{j\mu}\right)\left[\frac{t_{ij}}{U}\left(\ket{a,b}_i\otimes\emptyset_j\right)+\frac{t_{ij}^{*}}{U} \left(\emptyset_i\otimes\ket{a,b}_j\right)\right]\\[4pt]
    &=-\frac{t_{ji}t_{ij}}{U}\sum_{\mu}c_{j\mu}^{\dagger}c_{i\mu}\left(\ket{a,b}_i\otimes\emptyset_j\right)-\frac{t_{ji}^{*}t_{ij}^{*}}{U}\sum_{\mu}c_{i\mu}^{\dagger}c_{j\mu}\left(\emptyset_i\otimes\ket{a,b}_j\right) \\[4pt]
    &=-\frac{t_{ji}t_{ij}}{U}\Big[\underbrace{\ket{a}_i\ket{b}_j}_{\mu=b}-\underbrace{\ket{b}_i\ket{a}_j}_{\mu=a}\Big]-\frac{t_{ji}^{*}t_{ij}^{*}}{U}\Big[\underbrace{\ket{a}_i\ket{b}_j}_{\mu=a}-\underbrace{\ket{b}_i\ket{a}_j}_{\mu=b}\Big]
    \\[4pt]
    &=
    -\frac{2|t_{ij}|^2}{U}
    \left[
    \ket{a}_i\ket{b}_j
    -
    \ket{b}_i\ket{a}_j
    \right].
    \end{split}
    \end{equation}
\end{enumerate}
Then, the action of the effective Hamiltonian on the arbitrary singly occupied state is 
\begin{equation} (H_{\text{eff}})_{ij}\big[...\ket{a}_i\ket{b}_j...\big]=-\frac{2|t_{ij}|^2}{U}\Big[(...\ket{a}_i\ket{b}_j...)-(...\ket{b}_i\ket{a}_j...)\Big]
\end{equation}
which means that the effective Hamiltonian on itself can be written as 
\begin{equation}
    (H_{\text{eff}})_{ij}=-\frac{2|t_{ij}|^2}{U}\Big(\mathbb{1}_{\text{spin-orbital}}-P_{ij}^{\text{spin-orbital}}\Big)
    \label{eq:effective_hamiltonian_entrances}
\end{equation}
where the so-called \textit{exchange operator} $P_{ij}^{\text{spin-orbital}}$ is defined as 
\begin{equation}
P_{ij}\big(...\ket{a}_i\ket{b}_j...\big)=...\ket{b}_i\ket{a}_j...\qquad .
\end{equation}
Hence, by dropping the constant term (i.e. the identity term) in Eq.\ref{eq:effective_hamiltonian_entrances}, we find that the \textbf{effective KK Hamiltonian} is no more than
\begin{equation}
    \boxed{H_{\text{eff}}^{KK}=\sum_{<ij>} \frac{2|t_{ij}|^2}{U}P_{ij}^{\text{spin-orbital}}}\:.
    \label{eq:kk_hamiltonian_exchange}
\end{equation}
At this point, various questions should pop up:
\begin{itemize}
    \item \textit{What is the exchange operator $P_{ij}$ telling us?} \\ \noindent This operator favors entanglement between the internal degrees of freedom of neighboring sites, which is the quantum analog of magnetic/orbital correlations. Therefore, it is the responsible object for generating the effective interactions, the so-called \textbf{superexchange}, between spin-orbital degrees of freedom in the Mott-insulating regime. This reflects the fact that, even though charge fluctuations are suppressed due to the large on-site repulsion $ U $, virtual hopping processes of order $t^2 / U$ allow an electron to briefly hop to a neighbor and return, effectively generating an interaction between localized spin-orbitals.
    \item \textit{Where is the SU(4) Symmetry?}\\ \noindent The key lies in the structure of the local Hilbert space. At quarter filling, each site is singly occupied by one of four possible states, given by the $\set{\ket{1^\uparrow},\ket{1^\downarrow},\ket{2^\uparrow},\ket{2^\downarrow}}\simeq \mathbb{C}^4$. Then, the exchange operator $ P_{ij} $ treats all these four states on equal footing, and is invariant under global $SU(4)$ rotations in this combined space. To be more specific, although each bond operator $P_{i j}$ acts on a two-site Hilbert space $\mathbb{C}^4 \otimes \mathbb{C}^4$, it is invariant only under the diagonal action of $SU(4)$, corresponding to the same unitary acting on both sites:
    \begin{equation}
        P_{i j}(U \otimes U)=(U \otimes U) P_{i j}, \quad \forall U \in SU(4) .
    \end{equation}
    So, the Hamiltonian built from exchange operators commutes with global $SU(4)$ transformations of the form,
    \begin{equation}
    |a\rangle_i \mapsto U|a\rangle_i \quad \text { for all } i,
    \end{equation}
    and hence, we have a global $SU(4)$ symmetry across the lattice.
    \item \textit{What would have happened if we would have included Hund's coupling?}\\ \noindent While the pure Kugel-Khomskii Hamiltonian in Eq.\ref{eq:kk_hamiltonian_exchange} emerges in the limit $ J_H = 0 $, turning on Hund’s coupling explicitly \textbf{breaks the SU(4) symmetry}. This is because it energetically differentiates spin configurations depending on their orbital content, thus treating the four internal states unequally. As a result, the $SU(4)$-invariant exchange operators $ P_{ij} $ get replaced by more intricate spin-orbital interaction terms. By defining, similar to the spin operators
    \begin{equation}
        \vec{S}_i \equiv \frac{1}{2} \sum_{\alpha \sigma \sigma^{\prime}}\left(c_{i \alpha \sigma}^{\dagger} \vec{\tau}_{\sigma \sigma^{\prime}} c_{i \alpha \sigma^{\prime}}\right),
    \end{equation}
    the \textbf{pseudospin operators} 
    \begin{equation}
        \vec{\mathcal{T}}_i \equiv \frac{1}{2} \sum_{\sigma \alpha \alpha^{\prime}}\left(c_{i \alpha \sigma}^{\dagger} \vec{\tau}_{\alpha \alpha^{\prime}} c_{i \alpha^{\prime} \sigma}\right),
    \end{equation}
    one can show that the 2nd-order Hund's KK effective Hamiltonian is given by
    \begin{equation}
        \begin{split}
        H_{\mathrm{eff}}= & \sum_{<ij>}\left\{\frac{4 t^2}{U}\left(\vec{S}_i \cdot \vec{S}_{j}-\frac{1}{4}\right)\left(2 \mathcal{T}_i^z  \mathcal{T}_{j}^z+\frac{1}{2}\right)\right. \\ & +\frac{4 t^2}{U+J}\left(\vec{S}_i \cdot \vec{S}_{j}-\frac{1}{4}\right)\left(\vec{\mathcal{T}}_i \cdot \vec{\mathcal{T}}_{j}-2 \mathcal{T}_i^z  \mathcal{T}_{j}^z+\frac{1}{4}\right) \\ & \left.+\frac{4 t^2}{U-J}\left(\vec{S}_i \cdot \vec{S}_{j}+\frac{3}{4}\right)\left(\vec{\mathcal{T}}_i \cdot \vec{\mathcal{T}}_{j}-\frac{1}{4}\right)\right\}.
        \end{split}
        \label{eq:nonsym-KK-hamiltonian}
    \end{equation}

    This leads to a reduced symmetry, often $SU(2)_{\text{spin}}\times SU(2)_{\text{orbital}}$ or even a lower one of $SU(2)_{\text {spin}} \times \mathbb{Z}_2^{\text {orbital}}$, and generates competition between various types of ordering. Physically, this correction reflects the more realistic behavior of transition metal ions, where spin and orbital moments interact differently due to intra-atomic exchange.

    To gain insight into this Hamiltonian and its reduced symmetries, one must understand the physical meaning of the pseudospin operators. Just like for real spin $\vec{S}$, we can describe the two-level orbital system $\{\ket{1},\ket{2}\}\simeq \mathbb{C}^2$ using Pauli matrices $\vec{\tau}=(\tau_x,\tau_y,\tau_z)$ --- more precisely, through the unique irreducible and faithful representation of the symmetry group SU(2), which governs all two-level quantum systems. In this picture, the pseudospin operator $\vec{T}_i$ then acts in this orbital space as:
    \begin{equation}
        \left\{ 
    \begin{aligned} 
   & \Rightarrow \: \mathcal{T}_i^z=+\frac{1}{2} \quad \text{means orbital } \ket{1}\text{ is occupied}\\  
   & \Rightarrow \: \mathcal{T}_i^z=-\frac{1}{2} \quad \text{means orbital } \ket{2}\text{ is occupied}\\
   & \Rightarrow\ \ \ \mathcal{T}_i^x, \mathcal{T}_i^y \quad \  \text{are useful for superpositions or tunneling between orbitals}.
\end{aligned} \right.
    \end{equation}
So, the vector $\vec{\mathcal{T}}_i$ tells us: which orbital is occupied (via $\mathcal{T}^z$), and whether there are coherent fluctuations between orbitals (via $\mathcal{T}^x, \mathcal{T}^y$).

In this light, the terms in the effective Hamiltonian can now be understood as coupling spin and orbital degrees of freedom in different ways, depending on the nature of the virtual processes involved. Each line in the Hamiltonian reflects a distinct physical channel governed by the strength of on-site interactions $U$ and the Hund's coupling $J$. Let us now examine each term in detail:
\begin{itemize}
    \item[-]The \textbf{1st term}, proportional to $\frac{4 t^2}{U}$, involves a spin-singlet projector $\left(\vec{S}_i \cdot \vec{S}_j-\frac{1}{4}\right)$ and an Ising-like coupling in the orbital sector, $\left(2 \mathcal{T}_i^z \mathcal{T}_j^z+\frac{1}{2}\right)$. Hence, this term favors the formation of spin singlets (spin AFM) whenever the orbitals are either both $\ket{1}$ or both $\ket{2}$ in the respective bond (orbital FM). Note how the symmetry here is anisotropic in orbital space - breaks full $SU(2)_{\text {orbital }}$ to $\mathbb{Z}_2^\text{orbital}$.
    \item[-] The \textbf{2nd term}, proportional to $\frac{4 t^2}{U+J}$, again favors spin-singlet configurations but introduces transverse orbital interactions via $\vec{\mathcal{T}}_i \cdot \vec{\mathcal{T}}_j-2 \mathcal{T}_i^z \mathcal{T}_j^z$, allowing for orbital fluctuations and mixing. Hence, we again have a formation of spin AFM order but now coupled to orbital order that penalizes $\mathcal{T}^z \mathcal{T}^z$ alignment — indicating a preference for transverse orbital fluctuations. As the 1st term, here we also break $SU(4)$. The $SU(2)_{\text {spin }}$ symmetry is still there, but in the orbital part, only a $U(1)_{\text {orbital }}$ symmetry, corresponding to rotations about the $z$-axis in orbital space, remains. 
    \item[-] Finally, the \textbf{3rd term}, proportional to $\frac{4 t^2}{U-J}$, weights the spin-triplet sector \\ \noindent $\left(\vec{S}_i \cdot \vec{S}_j+\frac{3}{4}\right)$, and couples it to full $SU(2)$ orbital interactions $\left(\vec{\mathcal{T}}_i \cdot \vec{\mathcal{T}}_j-\frac{1}{4}\right)$. This term arises due to Hund's rule favoring parallel spins, and introduces entanglement between triplet spin configurations and orbital fluctuations. Hence, in this case, we have the formation of an FM spin exchange together with AFM-orbital interaction. The full term is $SU(2)_{\text {spin }} \times SU(2)_{\text{orbital }}$ symmetric, but it still breaks $SU(4)$ (which would mix spin and orbital sectors nontrivially).
\end{itemize}
Indeed, this contrary pattern—where antiferromagnetic orbital order tends to favor ferromagnetic spin order and vice versa—was one of the first key insights drawn in the early days of spin-orbital physics. In the 1970s, Kugel and Khomskii \cite{kugel1973} and Inagaki \cite{Inagaki1975} studied orbitally degenerate Hubbard-type models within a mean-field framework to understand the magnetic structures of transition metal compounds. Their analyses revealed that the interplay between spin and orbital degrees of freedom naturally leads to such inverse ordering tendencies. 
\end{itemize}
Coming back to the $SU(4)$ symmetric case (for the homogeneous hopping situation)
\begin{equation*}
    H_{\text{SU(4)}}^{KK}=\frac{2|t|^2}{U}\sum_{<ij>} P_{ij}^{\text{spin-orbital}},
\end{equation*}
    we now want to see how this is equivalent to the $SU(4)$ Heisenberg Hamiltonian. First, let's note that our exchange operator $P_{ij}^{\text{spin-orbital}}$ defined by  $P_{i j}\ket{b}_i \ket{a}_j=\ket{a}_i\ket{b}_j$ can be rewritten in fermionic operators as 
\begin{equation}
    P_{i j}=\sum_{a, b=1}^4 c_{i a}^{\dagger} c_{i b} c_{j b}^{\dagger} c_{j a} \qquad (\text{quartic interaction}),
\end{equation}
    acting on any basis $a,b\in\set{1,2,3,4}$ of $\mathbb{C}^4_{\text{spin-orbital}}$ (not necessarily $\set{1^{\uparrow},1^{\downarrow},2^{\uparrow},2^{\downarrow}}$). Define the matrix operators 
\begin{equation}
    E_{a b}^{(i)} \equiv c_{i a}^{\dagger} c_{i b} \qquad \xrightarrow[]{ \:\: \:\:\text{so that \:\:}} \qquad \:\: P_{ij}=\sum_{a,b} E^{(i)}_{ab}E_{ba}^{(j)}.
\end{equation}
    and, as discussed before, $\hat{E}^{(i)}$ operators generate $U(4)$ rather than $SU(4)$, since \begin{equation}
        \operatorname{Tr} \hat{E}^{(i)}=\sum_a E^{(i)}_{aa}=\sum_ac_{i a}^{\dagger} c_{i a}=\hat{n}_i\neq 0 \qquad \text{(1 fermion per site)}.
    \end{equation}
    Therefore, one needs to incorporate the standard  traceless version of $SU(4)$ generators 
\begin{equation}
    T_{a b}^{(i)} \equiv c_{i a}^{\dagger} c_{i b}-\frac{n_i}{4} \delta_{a b}\:\stackrel{(n_i=1)}{=} \: E^{(i)}_{ab}-\frac{1}{4} \delta_{a b} \quad \xrightarrow[]{\:\:\text{ so that }\:\:} \:\:P_{ij}=\sum_{a,b}\left(T_{a b}^{(i)}+\frac{\delta_{ab}}{4}\right)\left(T_{ba}^{(j)}+\frac{\delta_{ab}}{4}\right),
\end{equation}
    and by expanding the product
\begin{equation}
    \begin{aligned}
    P_{ij}=\sum_{a,b}T_{a b}^{(i)}T_{ba}^{(j)}+\sum_a\left(\cancel{\frac{T_{a a}^{(i)}}{4}+\frac{T_{aa}^{(j)}}{4}}+\frac{1}{16}\right)=\sum_{a,b}T_{a b}^{(i)}T_{ba}^{(j)}+\frac{1}{4}\mathbb{1}\:,
\end{aligned}
\end{equation}
    where we have used the tracelessness of the $SU(4)$ generators. In conclusion, we obtain 
\begin{equation}
    H^{KK}_{\text{eff}}=\sum_{<ij>} \frac{2|t|^2}{U}\left(\sum_{a,b}T_{a b}^{(i)}T_{ba}^{(j)}+\frac{1}{4}\mathbb{1}\right)\:,
\end{equation}
and by dropping (as always) the constant energy term per bond, we get the familiar form of the $\bm{SU(4)}$ \textbf{Heisenberg Hamiltonian}:
\begin{equation}
\boxed{H^{KK}_{\text{eff}}=\frac{2|t|^2}{U}\sum_{<ij>}\sum_{a,b=1}^4 T_{a b}^{(i)}T_{ba}^{(j)}}\:\:.
    \label{eq:sym-KK-hamiltonian}
\end{equation}
Hence, as promised at the beginning of the section, the $SU(4)$-symmetric limit (one electron per site, 
$J_H$=0, homogeneous hopping), of the two-orbital Hubbard model efficiently reduces to the $SU(4)$ Heisenberg Hamiltonian in the fundamental representation. In other words, the superexchange mechanism in a two-orbital Mott insulator naturally generates a nearest-neighbor $SU(4)$ quantum magnet. It is worth emphasizing, however, that the $SU(4)$ model obtained here is not identical to the $SU(N)$ Heisenberg models introduced earlier in Sec.\ref{sec:sun_heisenberg_models}. In particular, the Hamiltonian in Eq.(\ref{eq:su_N_heisenberg}), originally studied by Read and Sachdev within a large-$N$ expansion, assumes that the two sublattices of a bipartite lattice carry conjugate representations of $SU(N)$, commonly interpreted as ``quarks'' and ``antiquarks''. This structure ensures that nearest-neighbor bonds can form local $SU(N)$ singlets and is crucial for the large-$N$ analytical treatment of the model.

\noindent By contrast, the $SU(4)$ Kugel–Khomskii Hamiltonian derived above corresponds to a different realization of $SU(N)$ magnetism, where all lattice sites transform in the same fundamental representation. In this case, the interaction is naturally expressed in terms of the permutation operator acting on the combined spin–orbital Hilbert space. As emphasized in Ref.~\cite{LiMa1998}, this distinction leads to qualitatively different physical behavior from the large-$N$ models of Read and Sachdev, and the two formulations should therefore be regarded as complementary realizations of $SU(N)$ quantum magnetism rather than equivalent ones.

This is how, in the first decade of the 2000s, this symmetric fixed point of spin–orbital exchange was extensively studied across different lattice geometries and spatial dimensions. Early theoretical work recognized that such an $SU(4)$-symmetric model is qualitatively "more quantum" than its $SU(2)$ analogue, with stronger fluctuations and unconventional ordering tendencies \cite{LiMa1998}. In one dimension, the $SU(4)$ chain exhibits critical behavior with multiple gapless modes consistent with $SU(4)_1$ Wess–Zumino–Witten universality \cite{AzariaGogolinNersesyan1999} (a clear departure from simple $SU(2)$ spin chains) and associated characteristic four-site correlations \cite{FrischmuthBeatMila1999}. More striking deviations from semiclassical behavior occur in two dimensions. On the square lattice, variational and tensor-network studies reveal that the ground state spontaneously dimerizes \cite{CorbozMila2011}. In this phase, the dimers do not form $SU(4)$ singlets but instead transform in a six-dimensional irreducible representation, leading to a further breaking of the global $SU(4)$ symmetry. Interestingly, numerical evidence suggests that this dimerization coexists with a gapless spectrum, indicating a fundamentally different organization of low-energy degrees of freedom than in conventional $SU(2)$ antiferromagnets. On the honeycomb lattice, the situation is even more remarkable. Tensor-network, flavor-wave, and exact diagonalization studies provide “clean evidence” that the $SU(4)$ Kugel–Khomskii model supports a spin–orbital quantum liquid with no apparent symmetry breaking \cite{CorbozMila2012}. The resulting algebraic correlations are accurately captured by a projected $\pi$-flux Dirac parton ansatz at quarter filling, suggesting that the model realizes a stable algebraic spin–orbital liquid phase.

However, treating the $SU(4)$ Kugel–Khomskii Hamiltonian as the end of the story would miss its main practical value: it serves as a symmetric starting point from which one can organize the effects of physically unavoidable perturbations. In transition-metal oxides, orbital degeneracy is not just an abstract labeling — it is tied to lattice geometry, crystal-field splittings, and directional hopping, which often introduce anisotropies and Hund-driven symmetry breaking \cite{TokuraNagaosa2000,Khaliullin2005}. A particularly important development in this direction is the observation that spin–orbit-entangled $d^1$ Mott systems on edge-sharing octahedra can admit an emergent $SU(4)$ description after an appropriate change of basis: bond-dependent microscopic hopping can be “gauged” into a form with enlarged symmetry, yielding an $SU(4)$-symmetric Hubbard model and, in the Mott limit, an $SU(4)$ Heisenberg-type exchange \cite{YamadaOshikawaJackeli2018}. This is conceptually important because it shows that $SU(4)$ can be seen as a potentially robust effective description enforced by lattice and orbital structure, and not as a fine-tuned accident. Building on this, subsequent work broadened the model-building perspective and clarified how lattice symmetries and spin–orbital entanglement can stabilize families of $SU(4)$-symmetric spin–orbital liquids on different geometries \cite{YamadaOshikawaJackeli2021}. Complementarily, once one accepts that real systems will sit only near the $SU(4)$ point, a key question becomes which phases survive under controlled $SU(4)$-breaking perturbations (Hund’s coupling, anisotropic exchanges, further-neighbor terms). In this vein, systematic studies of exchange-frustrated perturbations on the honeycomb lattice suggest that the liquid-like regime can persist over an extended neighborhood of the symmetric point, while identifying the dominant competing orders \cite{NatoriNutakkiPereiraAndrade2019}.

Finally, the last decade has significantly improved our understanding of what we mean by an “$SU(4)$ spin–orbital liquid,” especially in two dimensions. On the honeycomb lattice, the projected $\pi$-flux Dirac parton description mentioned earlier has been stress-tested from multiple angles. Variational analyses directly addressed the leading competing instability — formation of four-site singlet plaquettes (“tetramerization”) — and found that the algebraic liquid is stable in a substantial region, with plaquette order requiring sufficiently strong symmetry-allowed perturbations \cite{LajkoPenc2013}. More recent numerical work has also clarified small finite-size and geometry effects (e.g. cylinders) that can make the spectrum appear gapped, while remaining consistent with a gapless Dirac-like algebraic state in the two-dimensional thermodynamic limit \cite{JinNatoriKnolle2023}. Perhaps most importantly for connecting theory to measurable signatures, explicit calculations of dynamical structure factors for the $SU(4)$ algebraic liquid now provide qualitative clear predictions — broad continua and characteristic momentum–frequency patterns expected from fractionalized partons rather than magnons — and quantify how Gutzwiller projection reshapes spectral weight relative to mean-field parton theory \cite{VorosPenc2023}.

The triangular lattice, by contrast, appears to realize a rather different organization of low-energy degrees of freedom. Instead of a Dirac algebraic liquid, numerical and variational studies increasingly point toward critical liquid behavior tied to an emergent parton Fermi surface, accompanied by a spontaneous reduction of lattice rotational symmetry (a “stripy” or nematic organization) while preserving global $SU(4)$ \cite{JinSunTuZhou2022}. In parallel, alternative viewpoints based on dimer/effective descriptions help to organize candidate low-energy manifolds and provide complementary intuition for why conventional symmetry breaking is avoided in this highly quantum regime \cite{KeselmanSavaryBalents2020}. Very recent large-scale variational Monte Carlo work supports this triangular-lattice picture by energetically selecting specific stripy/nematic liquid ansätze and characterizing their parton Fermi-surface structure on large clusters \cite{ZhangJinZhou2024}. Overall, these results suggest that while the $SU(4)$ point is a unifying symmetry principle, the geometry of the lattice strongly influences which gapless liquid phenomenology is realized (Dirac-type on honeycomb versus Fermi-surface-type on triangular), and therefore which probes are most useful to phase identification.

\subsubsection{Multipolar Exchange Interactions}\label{sec:multipolar_su_n}
After establishing the $SU(4)$ symmetric form of the spin-orbital exchange Hamiltonian, it is natural to ask how the underlying interaction can be interpreted in terms of physically observable operators. While the exchange operator $P_{ij}$ provides a compact and symmetry-transparent description, it obscures the structure of the local degrees of freedom that participate in the interaction. In particular, the four-dimensional on-site Hilbert space can equally be viewed as a spin- $\frac{3}{2}$ manifold, whose operator algebra admits a natural decomposition in terms of multipole moments, as we will see. Instead of working with the fifteen generators of the $\mathfrak{su}(4)$ algebra, one may reorganize them according to their transformation properties under spin rotations. In this language, the operator space decomposes into irreducible spherical tensors corresponding to dipolar, quadrupolar, and octupolar moments \cite{Biedenharn1984, Balla2014, Lindgard1974tables, Jensen1991rare}. Such a classification is familiar from atomic and nuclear physics, where higher-order multipoles characterize anisotropic distributions of charge or magnetization. In the context of strongly correlated systems, however, these multipoles describe collective spin-orbital fluctuations that can serve as order parameters of unconventional phases.

Indeed, this perspective is particularly useful in systems where conventional magnetic order fails to capture the relevant correlations. In several transition-metal and rare-earth compounds, the primary ordering tendency is not associated with dipolar magnetism but rather with quadrupolar or octupolar degrees of freedom, giving rise to the so-called \textit{multipolar} or \textit{hidden-order phases} \cite{SuzukiIkedaOppeneer2018, Toth2011, Kuramoto2009}. In the following, we briefly review the construction of irreducible spherical tensor operators and show how the $SU(N)$ exchange Hamiltonian can be expressed in terms of multipolar interactions, which not only provides a transparent physical interpretation of the previously studied $SU(N)$ models, but also establishes a direct link between $SU(4)$ magnetism and the broader framework of multipolar ordering phenomena.

Let's start by recalling that any fundamental representation space $\mathbb{C}^N$ of $SU(N)$ can be regarded as the spin $S=\frac{N-1}{2}$ space:
\begin{equation}
    \mathbb{C}^N\cong \mathcal{H}_S= \mathbb{C}^{2S+1}.
\end{equation}
Then, the canonical identification of the fundamental representation space $\mathbb{C}^N$ with the basis of $\mathcal{H}_S$ is given by 
\begin{equation}
\begin{aligned}
    &\ket{\alpha}\:_{(\alpha=1,...,N)} \quad\: \xleftrightarrow[]{\quad\quad} \quad \ket{m} \in \mathcal{H}_S \:\:\: \text{ with } \:\: (m=S+1-\alpha) \\
    S=3/2 \text{: }\:\:& \\
    & \qquad \begin{pmatrix}
\ket{1} \\
\ket{2} \\
\ket{3}\\
\ket{4}
\end{pmatrix} \qquad \xleftrightarrow[]{\quad\quad} \qquad \begin{pmatrix}
\ket{3/2} \\
\ket{1/2} \\
\ket{-1/2} \\
\ket{-3/2}
\end{pmatrix}.
\end{aligned}
\end{equation}
Now, the operator spaces acting on these representation spaces are the linear endomorphisms  $\text{End}(\mathbb{C}^N)$ and $\text{End}(\mathcal{H}_S)$. Equivalently, we can express these last two as matrix spaces given by $\text{End}(\mathbb{C}^N)=\mathfrak{su}(N)\oplus\mathbf{1}$ and $\text{End}(\mathcal{H}_S)=\mathcal{H}_S\otimes \mathcal{H}_S^*\cong\mathcal{H}_S\otimes \mathcal{H}_S$ (using the self-duality of $SU(2)$-irreps). We then have the situation: 
\begin{figure}
    \centering
    \includegraphics[width=0.65\linewidth]{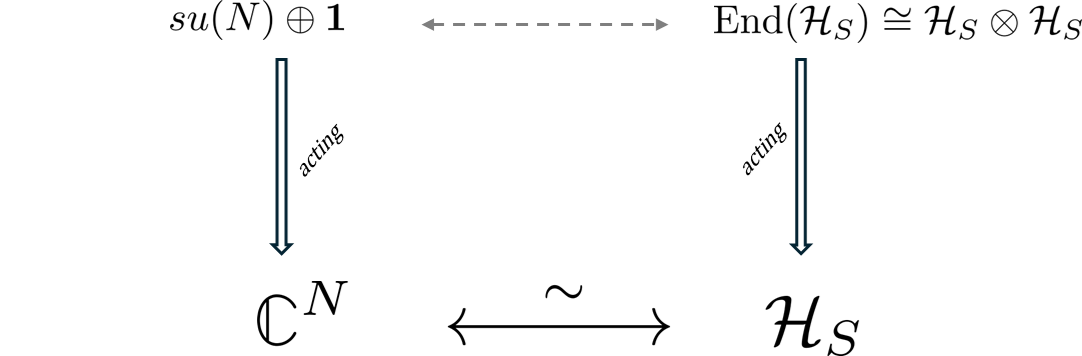}
\end{figure}
Here, we note that the tensor product $\mathcal{H}_S\otimes \mathcal{H}_S$ can be regarded as ``addition of angular momentum'' such that we have the decomposition in irreducible representations applies, and therefore:  
\begin{equation}
    \operatorname{End}(\mathcal{H}_S)\:\cong \:\mathcal{H}_S \otimes\mathcal{H}_S= \bigoplus_{k=0}^{2 S} \mathcal{H}_k \quad \xRightarrow[]{\quad} \quad \operatorname{End}(\mathcal{H}_S)=\bigoplus_{k=0}^{2 S}\mathcal{M}_k\:\:.
\end{equation}
The space of operators decomposes into irreducible $SU(2)$ sectors $\mathcal{M}_k$ labeled by an integer ``spin'' $k$. Each irreducible component is then a matrix space of dimension $\operatorname{dim}(\mathcal{M}_k)=2k+1$, and has an operator basis given by the so-called \textit{spherical tensor operators} of rank $k$. We denote such basis set as $\{T^{(k)}_q\}_{q\:=\:-k,...,k}$. Now, since we are dealing with integer spins $k$, the basis elements $T^{(k)}_q$ follow the same transformation properties of the spherical harmonics $Y^{l}_m$; this is:
\begin{equation}
    \begin{aligned}
    &Y^l_{m^{\prime}}\left(\theta^{\prime}, \phi^{\prime}\right)=\sum_{m=-l}^l D_{m^{\prime} m}^l(\alpha, \beta, \gamma) Y^l_m(\theta, \phi) \qquad [SO(3) \text{ rotation } \rightarrow SU(2) \text{ induced}]\\
    & \qquad \qquad \qquad \qquad \qquad\qquad\Downarrow \\
     &\qquad U_R \:T_q^{(k)} \:U^{\dagger}_R\:=\sum_{q^{\prime}=-k}^k D_{q^{\prime} q}^{(k)}(R)\: T_{q^{\prime}}^{(k)} \qquad  \quad [SU(2) \text{ rotation } \rightarrow SO(3) \text{ induced}]\:\:.
    \end{aligned}
\end{equation}
Or equivalently, they satisfy the commutation relations
\begin{equation}
\begin{aligned}
    &{\left[S_z, T_q^{(k)}\right]=q\: T_q^{(k)}} \\
    &{\left[S_{ \pm}, T_q^{(k)}\right]=\sqrt{(k \mp q)(k \pm q+1)} T_{q \pm 1}^{(k)}}\:\:.
\end{aligned}
\end{equation}
Hence, spherical tensor operators are no more than matrices that transform under an integer irreducible representation $k$ of $SU(2)$. However, irreducible representations of the rotational group $SO(3)$ correspond exactly to $k=0,1,2,3,...$, which are the scalar, vector, quadrupole, octupole… tensors. In this way, what we have finally found is that the operator space of $\mathcal{H}_S$ naturally decomposes into multipole tensors,
\begin{equation}
    \text{End}(\mathcal{H}_S)= \mathbf{1} \oplus  \:\underbrace{\text{span}\left(T^{(1)}_{-1,0,1}\right)}_{\text{3 dipoles}}\: \oplus\:  \underbrace{\text{span}\left(T^{(2)}_{-2,...,2}\right)}_{\text{5 quadrupoles}}\:\oplus\:\underbrace{  \text{span}\left(T^{(3)}_{-3,...,3}\right)}_{\text{7 octupoles}}\: \oplus\:...\: \oplus\: \text{span}\left(T^{(2S)}_{-2S,...,2S}\right),
\end{equation}
and by comparing with $\text{End}(\mathbb{C}^N)=\mathbf{1}\oplus \mathfrak{su}(N)$, we conclude
\begin{equation}
    \boxed{
    \mathfrak{su}(N)= \bigoplus_{k=1}^{N-1}\:\text{span}\left(T^{(k)}_{-k,...,k}\right)=\bigoplus_{k=1}^{N-1}\mathcal{M}_k}\:\:.
\end{equation}
In other words, the generators of $SU(N)$ can be chosen according to how they transform under $SU(2)$, forming sets of operators that correspond to multipoles (dipoles, quadrupoles,...). For completeness, and to avoid interrupting the main discussion, the explicit construction of these generators, together with the normalization conventions for the different multipolar sectors, is collected in Appendix~\ref{app:suN-multipoles}.

Nevertheless, this is only half of the story. So far, we have only described the operators acting on a single site of our lattice model. The $SU(N)$ Hamiltonians, however, are built from a different operator space. What ultimately matters is the matrices acting on each bond space $\mathbb{C}^N_{\text{site }i\:}\otimes \mathbb{C}^N_{\text{site }j}$, and therefore we must study 
\begin{equation}
\begin{aligned}
    \text{End}(\mathbb{C}^N_{\text{site }i\:}\otimes \mathbb{C}^N_{\text{site }j})=\text{End}(\mathbb{C}^N_{\text{site }i\:})\otimes \text{End}(\mathbb{C}^N_{\text{site }j})&=\left(\bigoplus_{k=0}^{N-1}\mathcal{M}^{(i)}_k\right)\otimes \left(\bigoplus_{k'=0}^{N-1}\mathcal{M}_{k'}^{(j)}\right)\\\\
    &=\bigoplus_{k,k'=0}^{N-1} \mathcal{M}^{(i)}_k\otimes \mathcal{M}^{(j)}_{k'}\:\:.
\end{aligned}
\end{equation}
Now, under global $SU(2)$ rotations on the bond (i.e. $U_{ij}=U\otimes U$), the correct covariant bond operators are the coupled tensors (``basis of total bond angular momentum'')
\begin{equation}
    \mathbb{T}^{(K)}_Q\equiv\left[T^{(k)}(i) \otimes T^{\left(k^{\prime}\right)}(j)\right]_Q^{(K)}=\sum_{q, q^{\prime}} C_{k q k^{\prime} q^{\prime}}^{K Q} \:\:T_q^{(k)}(i)\otimes T_{q^{\prime}}^{(k^{\prime})}(j)\:\:,
\end{equation}
where $K={|k-k'|,...,k+k'}$ is the total multipole rank of the bond operator, $Q=-K,...,K$ its projection component index, and $C_{k q k^{\prime} q^{\prime}}^{K Q}$ the corresponding Clebsch–Gordan coefficients (this is no more than the change of basis $|K Q\rangle=\sum_{q, q^{\prime}} C_{k q k^{\prime} q^{\prime}}^{K Q}|k q\rangle\left|k^{\prime} q^{\prime}\right\rangle$). This means that on the bond, we can construct new spherical tensor operators $\mathbb{T}^{(K)}_Q$ as a linear combination of pure tensor products of the single-site ones, such that the resulting $\mathbb{T}^{(K)}_Q$ transform in the $K$ multipole expansion of $SU(2)\text{ (or better }SO(3))$. Hence the total expansion for the bond operator space is of the form
\begin{equation}
\begin{aligned}
&\boxed{\operatorname{End}\left(\mathcal{H}_i \otimes \mathcal{H}_j\right) \:\:\cong\:\:\: \bigoplus_{k, k^{\prime}=0}^{N-1} \:\:\:\bigoplus_{K=\left|k-k^{\prime}\right|}^{k+k^{\prime}} \mathcal{M}_K \:\: = \ \bigoplus_{K=0}^{2(N-1)} (\mathcal{M}_{K})^{\oplus m_K}}\:\:\\
    & \text{ where }  \:\:m_K=\#\big\{(k,k')\in[0,N-1]^2 \text{ fulfilling } |k-k'|\leq K \leq k+k'\big\}\:\:.
\end{aligned}
\end{equation}

Now we turn to the actual operators that appear in $SU(N)$ lattice Hamiltonians. By construction, these Hamiltonians are globally $SU(N)$-invariant, meaning that for a global transformation $U \in SU(N)$ acting on every site (in whatever on-site representation $V_R$ we choose),
\begin{equation}
    U_{\text {glob }}=\bigotimes_{\ell} U_{\ell}^{(R)}, \quad U_{\text {glob }} H U_{\text {glob }}^{\dagger}=H \quad \Longleftrightarrow \quad\left[H, \sum_{\ell} T_{\ell}^A\right]=0 .
\end{equation}
In particular, the standard nearest-neighbor exchange interactions are built by contracting the $SU(N)$ generators (or, equivalently, by permutation/projectors), and therefore form $SU(N)$-invariant operators on each bond.

\begin{itemize}
    \item[$\blacktriangleright$] \textbf{Proof of SU(N) Symmetry} (independent of representation): \\
     Consider a Heisenberg Hamiltonian acting on the Hilbert space constrained by the local occupancy condition. After dropping the density term proportional to the identity, the Hamiltonian can be written as
    \begin{equation}
        H=\sum_{<ij>}H_{ij}=\sum_{<ij>} \left\{ J_{ij}\sum_{\alpha,\beta=1}^{N}T_{\alpha \beta}^{(i)} T_{\beta \alpha}^{(j)}\right\} \:\:.
        \label{eq:Heisenberd-any-irrep}
    \end{equation}
    The operators $\{T_{\alpha\beta}^{(i)}\}_{\alpha\beta}$ denote the generators of $su(N)$ acting on site $i$, obtained by applying the local Lie-algebra homomorphism
\begin{equation}
\begin{aligned}
    \rho_i : su(N) \to \text{End}(\mathcal{H}_i) \:\:\:;\qquad &\widetilde{T}_{\alpha \beta}\equiv E_{\alpha\beta}-\frac{\delta_{\alpha\beta}}{N}\mathbb{1} \:\:\:;\qquad E_{\alpha\beta}=\ket{e_\alpha}\bra{e_\beta}\\ 
    &\  T_{\alpha \beta}^{(i)}\equiv \rho_i\left(\widetilde{T}_{\alpha \beta}\right)\:\:
\end{aligned}
\end{equation}
Here $\{e_\alpha\}_{\alpha=1}^N$ is the canonical basis of $\mathbb C^N$. The matrices $\widetilde T_{\alpha\beta}$ form a basis of $su(N)$ in the fundamental representation, while $T_{\alpha\beta}^{(i)}$ denote their realization on the local Hilbert space $\mathcal H_i$. The map $\rho_i$ corresponds to any representation homomorphism, so the construction is completely independent of the specific choice of on-site representation spaces.

Under these definitions, we see that the abstract-defined unit matrices fulfill
\begin{equation}
    \left[E_{\alpha \beta}, E_{\gamma \delta}\right]=\delta_{\beta \gamma} E_{\alpha \delta}-\delta_{\alpha \delta} E_{\gamma \beta} \quad \xrightarrow[]{\text{traceless ones}} \quad \left[\widetilde{T}_{\alpha \beta}, \widetilde{T}_{\gamma \delta}\right]=\delta_{\beta \gamma} \widetilde{T}_{\alpha \delta}-\delta_{\alpha \delta} \widetilde{T}_{\gamma \beta}
\end{equation}
and therefore, on the on-site representations
\begin{equation}
\begin{cases}
    \left[T_{\alpha \beta}^{(i)}, T_{\gamma \delta}^{(i)}\right]=\delta_{\beta \gamma} T_{\alpha \delta}^{(i)}-\delta_{\alpha \delta} T_{\gamma \beta}^{(i)},\\
\left[T_{\alpha \beta}^{(i)}, T_{\gamma \delta}^{(j)}\right]=0 \qquad(i \neq j)\:\:.
\end{cases}
\end{equation}
Now, a global $SU(N)$ rotation is defined as 
\begin{equation}
\begin{aligned}
    U(g)=&\bigotimes_i \rho_i(g) = \bigotimes_i \rho_i\left(e^{-i\epsilon ^{a}T^{a}}\right)\:\:,\\
    &\text{but, since the identity}\\
    e^A \otimes e^B=e^{A \otimes I+I \otimes B} \quad \xrightarrow[]{ \text{ implies } } \quad & \rho_i\left(e^{-i\epsilon ^{a}T^{a}}\right)\otimes \rho_j\left(e^{-i\epsilon ^{a}T^{a}}\right)=e^{-i\epsilon^{a}\left[\rho_i(T^{a})\otimes\mathbb{1}\:+\:\mathbb{1}\otimes\rho_j(T^{a})\right]}\:\:,
\end{aligned}
\end{equation}
we only care about the generator $Q^a$ of global rotations
\begin{equation}
    U_a=e^{-i\epsilon^{a}\sum_{i}\rho_i(T^a)}=e^{-i\epsilon^{a}Q^a} \quad ; \qquad Q^{a}\equiv \sum_{i}\rho_i(T^a) \:\:.
\end{equation}
Note, however, that we can always choose the traceless unit matrix basis of $su(N)$, and therefore we are only interested in generators 
\begin{equation}
Q_{\mu\nu}=\sum_i\rho_i(\widetilde{T}_{\mu\nu})=\sum_i T_{\mu\nu}^{(i)} \quad \xrightarrow[]{\text{ such that }}\quad [Q_{\mu\nu}\:, \:H]\overset{!}{=}0\:\:.
\end{equation}
Having in mind that the Hamiltonian is invariant if the local bond-terms $H_{ij}$ are invariant, we then compute
\begin{equation}
    \begin{aligned}        \left[Q_{\mu\nu}\:,\:H_{ij}\right]&\propto\Big[\sum_k T^{(k)}_{\mu\nu}\:\:\:,\:\:  \sum_{\alpha\beta} T^{(i)}_{\alpha\beta}T^{(j)}_{\beta\alpha}\Big]=\Big[ T^{(i)}_{\mu\nu}+T^{(j)}_{\mu\nu}\:\:\:,\:\: \sum_{\alpha\beta} T^{(i)}_{\alpha\beta}T^{(j)}_{\beta\alpha}\Big]\\
    &=\sum_{\alpha, \beta}\left[T_{\mu \nu}^{(i)}, T_{\alpha \beta}^{(i)}\right] T_{\beta \alpha}^{(j)}+ \sum_{\alpha, \beta} T_{\alpha \beta}^{(i)}\left[T_{\mu \nu}^{(j)}, T_{\beta \alpha}^{(j)}\right]\\
    &=\sum_{\alpha, \beta}\left(\delta_{\nu \alpha} T_{\mu \beta}^{(i)}-\delta_{\mu \beta} T_{\alpha \nu}^{(i)}\right) T_{\beta \alpha}^{(j)} \:\: + \:\: \sum_{\alpha, \beta} T_{\alpha \beta}^{(i)}\left(\delta_{\nu \beta} T_{\mu \alpha}^{(j)}-\delta_{\mu \alpha} T_{\beta \nu}^{(j)}\right)\\
    &=\cancel{\sum_\beta T_{\mu \beta}^{(i)} T_{\beta \nu}^{(j)}}-\bcancel{\sum_\alpha T_{\alpha \nu}^{(i)} T_{\mu \alpha}^{(j)}} + \bcancel{\sum_\alpha T_{\alpha \nu}^{(i)} T_{\mu \alpha}^{(j)}} -\cancel{\sum_\beta T_{\mu \beta}^{(i)} T_{\beta \nu}^{(j)}} \quad = \: 0 \:\:,
    \end{aligned}
\end{equation}
and conclude that the family of Hamiltonians in Eq.\ref{eq:Heisenberd-any-irrep} constructed from different representations are invariant under global $SU(N)$ rotations.  
\end{itemize}

\noindent However, $\:SU(N)$ invariance does not fix how the same bond operator decomposes with respect to the embedded $SU(2)$ subgroup used above to organize multipoles. Once we rewrite the bond Hamiltonian in the coupled-tensor basis $\mathbb{T}_Q^{(K)}$, it generally contains several $SU(2)$ channels $K$, i.e. it becomes a sum of multipole-multipole couplings on the bond. Which $K$-sectors are present (and how they are weighted) depends on the choice of on-site representation. For instance, whether the two sites carry the same representation $(\bm{N},\bm{N})$ or a conjugate pair $(\bm{N},\bm{\bar{N}})$. In this sense, the Hamiltonian is always an $SU(N)$ scalar, but it can look very different when viewed through the $SU(2)$ multipole decomposition. 

\noindent To illustrate this distinction explicitly, let's work on the $SU(4)$ cases:
\begin{itemize}
    \item[$\blacktriangleright$] $\bm{(N,N)}$ \textbf{isotropic lattice}
    
    Here, we are thinking in our symmetric KK Hamiltonian derived in Eq.\ref{eq:sym-KK-hamiltonian}. To treat the general case in which one doesn't have $SU(4)$ symmetry, but only $SU(2)$-spin invariance (Eq.\ref{eq:nonsym-KK-hamiltonian}), let's first analyze what happens to our general multipole decomposition in the case of only a $SU(2)$ symmetry. 
    Having that $\mathcal{H}_i=\mathcal{H}_j=\mathbb{C}^N$, we then have 
    $$\operatorname{End}\left(\mathcal{H}_i \otimes \mathcal{H}_j\right) \:\cong\:\bigoplus_{k, k^{\prime}=0}^{N-1} \:\:\bigoplus_{K=\left|k-k^{\prime}\right|}^{k+k^{\prime}} \mathcal{M}_K \: = \: \bigoplus_{K=0}^{2(N-1)} (\mathcal{M}_{K})^{\oplus m_K}\:\:.$$
    However, since our Hamiltonian is a scalar under global $SU(2)$ rotations, it must live in the $K=0$ sector, and therefore only the term-spaces with $k=k'$ can appear. This is 
    \begin{equation}
    [\operatorname{End}\left(\mathcal{H}_i \otimes \mathcal{H}_j\right)]_{SU(2)-\text{scalar}} \:\cong\:\mathbf{1} \: \oplus\:\underbrace{\mathbf{1} \: \oplus \:...\: \oplus \mathbf{1}}_{N-1 \text{ times}}\:,
    \end{equation}
    where each term-space $\mathbf{1}$ (discarding the trivial identity contribution) is spanned by only one tensor constructed from 
    \begin{equation}
        \left[T^{(k)}(i) \otimes T^{(k)}(j)\right]_0^{(0)}=\sum_{q=-k}^k(-1)^q \:T_q^{(k)}(i) T_{-q}^{(k)}(j)\:\:.
    \end{equation}
    Here we note something remarkable: 
    \begin{center}
        \textit{when we have } $SU(2)$ \textit{ symmetry, we can only have interactions of the form dipole--dipole, quadrupole--quadrupole, octupole--octupole, etc...}
    \end{center}
    So for an $SU(2)$-invariant two-site interaction (no preferred spatial direction), the scalar bilinears are only
$$
H_{i j}^{SU(2), \mathrm{iso}}=\sum_{k=1}^{N-1} J_k\sum_{q=-k}^k(-1)^q T_q^{(k)}(i) T_{-q}^{(k)}(j)
$$
where $J_k$ is the coupling constant associated with the equal rank-$k$ multipole--multipole interaction. Mixed couplings (e.g. dipole--quadrupole) can only arise if additional structures besides 
spin rotations are present. For instance, if the bond carries a spatial direction 
$\hat{\mathbf r}_{ij}$, or if spin--orbit coupling ties spin operators to spatial tensors, 
one may form scalar combinations of the form
\[
(\text{spin tensor}) \times (\text{spatial tensor}) .
\]
In such situations, the total angular momentum of the spin operators may be 
$K=1,2,3,\ldots$, which can then be contracted with the appropriate spatial tensor to 
produce a scalar interaction. Therefore, dipole--quadrupole terms are not ``forbidden'' in general; they are simply 
absent for purely spin $SU(2)$-invariant interactions on isotropic bonds constructed 
only from on-site operators.

Let us now move to the case in which we consider an isotropic lattice with $\mathcal{H}_i=\mathcal{H}_j=\mathbb{C}^4$. Therefore, the $SU(2)$ invariant Hamiltonian is of the form
\begin{equation}
    H_{i j}=J_1 \sum_{q=-1}^1 (-1)^q T_q^{(1)}(i) T_{-q}^{(1)}(j)+J_2\sum_{q=-2}^2 (-1)^qT_q^{(2)}(i) T_{-q}^{(2)}(j)+J_3\sum_{q=-3}^3(-1)^q T_q^{(3)}(i) T_{-q}^{(3)}(j)\:\:.
    \label{eq:iso-su2-hamilto}
\end{equation}
If the Hamiltonian is $SU(4)$-invariant, the only invariant quadratic operator is
$$
H_{i j} = J_{ij}  \sum_{A=1}^{15} T_i^A T_j^A \:\:,
$$
and rewriting in the multipolar basis gives
$$
H_{i j} = J_{ij} \left[\sum_{q=-1}^1 (-1)^qT_q^{(1)}(i) T_{-q}^{(1)}(j)+\sum_{q=-2}^2 (-1)^q T_q^{(2)}(i) T_{-q}^{(2)}(j)+\sum_{q=-3}^3 (-1)^q T_q^{(3)}(i) T_{-q}^{(3)}(j) \right] .
$$
where dipole-dipole, quadrupole-quadrupole, and octupole-octupole appear with the same coefficient. That equality of couplings is precisely the $SU(4)$ symmetry condition.
In this line of ideas, we conclude with the 
\begin{center}
\fbox{
\parbox{0.85\linewidth}{
\centering
\textbf{Key observation:} 
the $SU(4)$ Heisenberg Hamiltonian on an isotropic $\mathbb{C}^4$ lattice can be viewed as a spin-$\tfrac{3}{2}$ exchange model in which the dipole--dipole, quadrupole--quadrupole, and octupole--octupole interactions all appear with the \emph{same coupling constant}:
\[
H_{ij}\propto 
(\text{dipole--dipole})+
(\text{quadrupole--quadrupole})+
(\text{octupole--octupole})
\]
}
}
\end{center}

\vspace{0.58cm}
\item[$\blacktriangleright$] $\bm{(N,\bar{N})}$ \textbf{quark-antiquark lattice}

Although $\overline{\mathbf{N}} \cong \Lambda^{N-1} \mathbf{N}$ is not equivalent to the fundamental as an $SU(N)$ representation, under the embedded $SU(2)$-spin both $\mathbf{N}$ and $\overline{\mathbf{N}}$ realize the same spin- $S=(N-1) / 2$ irrep. This is 
\begin{equation}
    \overline{\mathbf{N}} \cong \Lambda^{N-1} \mathbf{N}=  \Lambda^{N-1} \mathbb{C}^N \cong \mathbb{C}^N \cong \mathcal{H}_S \quad \left(\text{with }\:\:S=\frac{N-1}{2}\right)\:\:.
\end{equation}
Hence, the bond operator space still decomposes into the same $k=1,2,3,...$ multipole sectors, but with the only modification that on the $\overline{\mathbf{N}}$ sublattice the generators are taken in the conjugate representation:
\begin{equation}
    H_{i j}^{SU(2), \text {conj }}=\sum_{k=0}^{N-1} J_k \sum_{q=-k}^k(-1)^q T_q^{(k)}(i) \bar{T}_{-q}^{(k)}(j)\:\:.
\end{equation}
Using $\bar{T}_{-q}^{(k)}=(-1)^q T_q^{(k)}$, this may be rewritten as
\begin{equation}
    H_{i j}^{SU(2), \text {conj }}=\sum_{k=0}^{N-1} J_k \sum_{q=-k}^k T_q^{(k)}(i) T_{q}^{(k)}(j)\:\:.
\end{equation}
However, this is \emph{not} the standard $SU(2)$ scalar contraction of two rank-$k$ tensors, which would instead involve the combination
\[
\sum_{q=-k}^{k}(-1)^q\,T_q^{(k)}(i)T_{-q}^{(k)}(j).
\]
Therefore, the above rewriting should not be interpreted as a genuine multipole expansion in the usual rotational sense, but only as a \emph{pseudo-multipole} decomposition: it organizes the Hamiltonian according to the same $SU(2)$ rank sectors, while the actual bond contraction is modified by the conjugate nature of the second representation.
In particular, the $SU(N)$ quark--antiquark Heisenberg model can be formally rewritten as
\begin{equation}
    H_{i j}^{SU(N), \text {conj }}=J_{ij}\sum_{k=0}^{N-1}\sum_{q=-k}^k T_q^{(k)}(i) T_{q}^{(k)}(j) \:\:,
\end{equation}
which is multipolar only in this formal, or ``pseudo'', sense.
\end{itemize}

Now, a more useful way (especially for numerics) to express this last conjugated Heisenberg model is by using the $SU(2)$ bilinear form $x=\bm{S}_i \cdot \bm{S}_j$. In the following, we will explain how the $SU(N) \:\:(\bm{N,\overline{N}})$ Hamiltonian can be rewritten in terms of $x$. Let's start by unifying the decomposition of the ($\bm{N,N}$) and ($\bm{N,\overline{N}}$) bond spaces under two different pictures:
\begin{equation}
\begin{aligned}
     \bullet \quad\text{For }(\bm{N,N}): &\\
     & \bm{N}\otimes\bm{N}\:=\:\mathrm{Sym}^{2}(\mathbb{C}^N)\oplus \Lambda^{2}(\mathbb{C}^N) \qquad \text{(decomposition in irreps of }SU(N))\\
     &\bm{N}\otimes\bm{N}\:=\:\bigoplus_{J=0}^{N-1}\mathcal{H}_J \qquad\:\: (\text{decomposition in spin}-J \text{ irreps of } SU(2))
\end{aligned}
\end{equation}
\begin{equation}
\begin{aligned}
     \bullet \quad \text{For }(\bm{N,\overline{N}}): &\\
     & \bm{N}\otimes\bm{\overline{N}}\:=\:\mathbf{1} \oplus \text { Adj}\qquad \text{(decomposition in irreps of }SU(N))\qquad  \quad \\
     &\bm{N}\otimes\bm{\overline{N}}\:=\:\bigoplus_{J=0}^{N-1}\mathcal{H}_J \:\qquad\:\: (\text{in irreps of } SU(2))\:\:.
\end{aligned}
\end{equation}
Here, we note something extremely important:
\begin{center}
    \textit{Under the }$SU(N)$\textit{ irrep decomposition, both lattice bond spaces decompose into \textbf{only two} irreducible sectors of the form}
    $$
\boxed{\mathcal{H}_i\otimes\mathcal{H}_j=\mathrm{Irrep_A}\oplus \mathrm{Irrep}_B}
    $$
\end{center}
Therefore, any $SU(N)$-invariant two-site operator must be a linear combination of projectors $\set{\Pi_A,\Pi_B}$ in each irrep. This is
\begin{equation}
\begin{aligned}
    H_{i j}=a \Pi_{\mathrm{A}}+b &\Pi_{\mathrm{B}}\qquad \quad 
    (\text{with } \:\:\mathbb{1}=\Pi_{\mathrm{A}}+ \Pi_{\mathrm{B}})\:\:.
\end{aligned}
\end{equation}
\begin{equation}
\begin{aligned}
     \bullet \quad \underline{\text{For }(\bm{N,N})}: &\\
     \mathrm{Sym}^{2}(\mathbb{C}^N)&\oplus \Lambda^{2}(\mathbb{C}^N) \:\:\xrightarrow[]{\text{invariant bond op.}}\:\: H_{i j}=a \Pi_{\mathrm{sym}}+b \Pi_{\mathrm{asym}} \\
     &\text{redefining linear combination in terms of}\\ &\hspace{2cm}\mathbb{1}=\Pi_{\mathrm{sym}}+\Pi_{\mathrm{asym}}, \quad P_{i j}=\Pi_{\mathrm{sym}}-\Pi_{\mathrm{asym}};\\
     &\text{we get } \qquad \qquad  H_{ij}=a'\:\mathbb{1} + b'P_{ij}\:\:,\\
     &\text{and by dropping the identity term }\\
        & \hspace{6cm} \boxed{H_{ij}^{(\bm{N},\bm{N})}\propto P_{ij}\sim \sum_A T_i^A T_j^A} \hspace{2cm}
\end{aligned}
\end{equation}
\begin{equation}
\begin{aligned}
     \bullet \quad \underline{\text{For }(\bm{N,\overline{N}})}: &\\
     \mathbf{1} \oplus  \text { Adj} &\:\:\xrightarrow[]{\text{invariant bond op.}}\:\: H_{i j}=a \Pi_{\mathbf{1}}+b \Pi_{\mathrm{Adj}} \\
     &\text{redefining linear combination in terms of}\\ &\hspace{1.6cm}\mathbb{1}=\Pi_{\mathbf{1}}+\Pi_{\mathrm{Adj}} \:\: , \quad\text{ so we get } \\
     & \hspace{1.9cm} H_{ij}=a \Pi_{\mathbf{1}}+b\left( \Pi_{\mathrm{Adj}}+\Pi_{\mathbf{1}}\right)-b\Pi_{\mathbf{1}}=b\:\mathbb{1}+(a-b)\Pi_{\mathbf{1}}\:\:\\
     &\text{and by dropping the identity term }\\
        & \hspace{6cm} \boxed{H_{ij}^{(\bm{N},\bm{\overline{N}})}\propto \Pi_\mathbf{1}\sim \sum_A T_i^A \bar{T}_j^A} \hspace{2cm}
\end{aligned}
\end{equation}
This is exactly the group theoretical reason why our isotropic and quark-antiquark $SU(N)$ Heisenberg models were respectively proportional to the permutation operator $P_{ij}$ and the projector into the singlet $\Pi_\mathbf{1}$. 

At this point, we will focus on the conjugated $(\mathbf N,\overline{\mathbf N})$ lattice, since in that case the bond Hamiltonian is proportional to the projector onto the unique $SU(N)$ singlet, and this allows for a particularly simple rewriting in terms of the $SU(2)$ bilinear $x=\mathbf S_i\cdot \mathbf S_j$. The $(\mathbf N,\mathbf N)$ isotropic lattice can be treated analogously afterwards, although its invariant bond operator, the permutation $P_{ij}$, will have a dependence on $x=\mathbf S_i\cdot \mathbf S_j$ much more complicated.

The crucial observation is that the singlet subspace of $SU(N)$ inside $\mathbf N\otimes \overline{\mathbf N}$ coincides with the total-spin $J=0$ sector of the embedded $SU(2)$. Indeed, choosing dual bases $\{\ket{\alpha}\}_{\alpha=1}^N$ for $\mathbf N$ and $\{\ket{\bar\alpha}\}_{\alpha=1}^N$ for $\overline{\mathbf N}$, the normalized $SU(N)$ singlet is
\begin{equation}
    \ket{\Omega}_{ij}=\frac{1}{\sqrt N}\sum_{\alpha=1}^N \ket{\alpha}_i\otimes \ket{\bar\alpha}_j .
\end{equation}
By construction, this state is invariant under all global $SU(N)$ transformations on the bond,
\begin{equation}
    \left(U\otimes \overline U\right)\ket{\Omega}_{ij}=\ket{\Omega}_{ij}\qquad \forall\,U\in SU(N).
\end{equation}
In particular, it is invariant under the embedded $SU(2)\subset SU(N)$. Therefore $\ket{\Omega}_{ij}$ must belong to the $SU(2)$-invariant subspace of the bond. But under the $SU(2)$ point of view we have
\begin{equation}
    \mathbf N\otimes \overline{\mathbf N}\cong \mathcal H_S\otimes \mathcal H_S
    =\bigoplus_{J=0}^{2S}\mathcal H_J,
    \qquad S=\frac{N-1}{2},
\end{equation}
and this decomposition contains a unique one-dimensional invariant sector, namely $J=0$. Hence the $SU(N)$ singlet and the $SU(2)$ singlet must span the same subspace:
\begin{equation}
    \boxed{\Pi_{\mathbf 1}^{SU(N)}=\Pi_{J=0}^{SU(2)}}.
\end{equation}

This immediately implies that the quark--antiquark $SU(N)$ Heisenberg bond Hamiltonian can be written as
\begin{equation}
    H_{ij}^{(\mathbf N,\overline{\mathbf N})}\propto \Pi_{\mathbf 1}
    =\Pi_{J=0}.
\end{equation}
We now want to express this projector as a polynomial in the scalar operator $x\equiv \mathbf S_i\cdot \mathbf S_j$. Since the total bond spin is
$\mathbf J=\mathbf S_i+\mathbf S_j$, we have the standard identity
\begin{equation}
    \mathbf J^2=\mathbf S_i^2+\mathbf S_j^2+2\,\mathbf S_i\cdot \mathbf S_j
    =2S(S+1)+2x,
\end{equation}
and therefore on the total-spin $J$ sector the operator $x$ has eigenvalue
\begin{equation}
    x_J=\frac12\Big(J(J+1)-2S(S+1)\Big),\qquad J=0,1,\dots,2S=N-1.
\end{equation}
Because the decomposition
\begin{equation}
    \mathcal H_S\otimes\mathcal H_S=\bigoplus_{J=0}^{2S}\mathcal H_J
\end{equation}
is multiplicity-free, every $SU(2)$-invariant bond operator can be expanded spectrally as
\begin{equation}
    H_{ij}=\sum_{J=0}^{2S} c_J\,\Pi_J,
\end{equation}
with $\Pi_J$ the projector onto the total-spin $J$ subspace. In particular, each $\Pi_J$ can be reconstructed from the operator $x$ by Lagrange interpolation:
\begin{equation}
    \Pi_J=\prod_{J'\neq J}\frac{x-x_{J'}}{x_J-x_{J'}} \qquad \xrightarrow[]{\text{for }J=0\:} \qquad \Pi_{J=0} =\prod_{J'=1}^{2S}\frac{x-x_{J'}}{x_0-x_{J'}} 
\end{equation}
Using
\begin{equation}
    x_0=-S(S+1),
    \qquad
    x_{J'}=\frac12\Big(J'(J'+1)-2S(S+1)\Big),
\end{equation}
one finds
\begin{equation}
    \frac{x-x_{J'}}{x_0-x_{J'}}
    =
    1-\frac{x+S(S+1)}{\frac12\,J'(J'+1)}.
\end{equation}
Hence the projector onto the $SU(N)$ singlet can finally be written as
\begin{equation}
    \boxed{
    \Pi_{\mathbf 1}
    =
    \Pi_{J=0}
    =
    \prod_{J'=1}^{2S}
    \left[
    1-\frac{2\big(\mathbf S_i\cdot \mathbf S_j+S(S+1)\big)}{J'(J'+1)}
    \right]
    }.
\end{equation}
Equivalently, recalling that $2S=N-1$,
\begin{equation}
    \boxed{
    H_{ij}^{(\mathbf N,\overline{\mathbf N})}
    \propto
    \Pi_{\mathbf 1}
    =
    \prod_{J'=1}^{N-1}
    \left[
    1-\frac{2\big(\mathbf S_i\cdot \mathbf S_j+S(S+1)\big)}{J'(J'+1)}
    \right].
    }
\end{equation}
This result provides an explicit polynomial expression for the $SU(N)$ quark--antiquark Heisenberg interaction in terms of the $SU(2)$ bilinear $x=\mathbf S_i\cdot \mathbf S_j$. In other words, although the Hamiltonian is defined through $SU(N)$ symmetry, in the conjugated lattice the interaction can be written exactly as the projector onto the total-spin $J=0$ sector of two spins $S=(N-1)/2$.

To illustrate the structure of this formula, let us compute the interaction explicitly for small values of $N$.
\begin{itemize}

\item \underline{$(N=2)\Rightarrow (S=\tfrac12)$}:

Here $2S=1$, so only $J'=1$ appears in the product,
\begin{equation}
    H_{ij}\propto
1-\frac{2(\mathbf S_i\cdot\mathbf S_j+S(S+1))}{1\cdot2} \:\:\:\xrightarrow[]{\:\:\text{ with } \:S(S+1)=\tfrac34\:\:} \quad \boxed{
    H_{ij}^{(\mathbf 2,\overline{\mathbf 2})}\propto
\frac14-\mathbf S_i\cdot\mathbf S_j}
\end{equation}
which is precisely the familiar projector onto the singlet of two spin-$\tfrac12$ particles.

\item \underline{$(N=3)\Rightarrow (S=1)$}:

Now $2S=2$, and the product runs over $J'=1,2$,
\begin{equation}
    \Pi_0=
\left[1-(\mathbf S_i\cdot\mathbf S_j+2)\right]
\left[1-\frac{\mathbf S_i\cdot\mathbf S_j+2}{3}\right] \:\: \xrightarrow[]{\:\text{expand}\:}  \quad \boxed{
    H_{ij}^{(\mathbf 3,\overline{\mathbf 3})}\propto\frac{1}{3}\left[\left(\mathbf{S}_i \cdot \mathbf{S}_j\right)^2-1\right]}
\end{equation}

\item \underline{$(N=4)\Rightarrow (S=\tfrac32)$}:

Here $2S=3$, and the product runs over $J'=1,2,3$,
\begin{equation*}
    \Pi_0=\left[1-\frac{\mathbf S_i\cdot\mathbf S_j+\tfrac{15}{4}}{1}\right]
\left[1-\frac{\mathbf S_i\cdot\mathbf S_j+\tfrac{15}{4}}{3}\right]
\left[1-\frac{\mathbf S_i\cdot\mathbf S_j+\tfrac{15}{4}}{6}\right]
\end{equation*}
Expanding this gives a cubic polynomial in $\mathbf S_i\cdot\mathbf S_j$,
\begin{equation}
   \boxed{
    H_{ij}^{(\mathbf 4,\overline{\mathbf 4})}\propto\frac{33}{128} +\frac{31}{96}\,\mathbf S_i\cdot\mathbf S_j-\frac{5}{72}(\mathbf S_i\cdot\mathbf S_j)^2-\frac{1}{18}(\mathbf S_i\cdot\mathbf S_j)^3}\:\:.
\end{equation}
\end{itemize}
Let us now return to the isotropic $(\mathbf N,\mathbf N)$ lattice. In this case, the canonical $SU(N)$-invariant bond operator is not the singlet projector, since $\mathbf N\otimes \mathbf N$ does not contain the trivial $SU(N)$ representation. Instead, after dropping the identity contribution, the unique nontrivial invariant bond operator is the permutation
\begin{equation}
    H_{ij}^{(\mathbf N,\mathbf N)}\propto P_{ij}.
\end{equation}
We now want to understand how this operator looks when rewritten in terms of the embedded $SU(2)$ bilinear
$x=\mathbf S_i\cdot \mathbf S_j$. Under the $SU(2)$ point of view, we again identify $\mathbf N\cong \mathcal H_S$ with $S=(N-1)/2$, so that
\begin{equation}
    \mathbf N\otimes \mathbf N \cong \mathcal H_S\otimes \mathcal H_S
    =\bigoplus_{J=0}^{2S}\mathcal H_J.
\end{equation}
Since the permutation operator $P_{ij}$ commutes with global $SU(2)$ rotations, it must act diagonally on each total-spin sector $\mathcal H_J$. Therefore it admits the spectral decomposition
\begin{equation}
    P_{ij}=\sum_{J=0}^{2S} p_J\,\Pi_J,
\end{equation}
where $\Pi_J$ projects onto the total-spin $J$ subspace. The eigenvalue $p_J$ is fixed by exchange symmetry. Indeed, for two identical spins $S$, the permutation operator acts on the coupled state of total spin $J$ as
\begin{equation}
    P_{ij}\big|_J=(-1)^{2S-J}.
\end{equation}
Since $2S=N-1$, this becomes
\begin{equation}
    \boxed{
    P_{ij}=\sum_{J=0}^{N-1}(-1)^{N-1-J}\,\Pi_J=\sum_{J=0}^{2S}
    (-1)^{2S-J}
    \prod_{J'\neq J}\frac{x-x_{J'}}{x_J-x_{J'}}
    }.
\end{equation}
after using the interpolation formula again
\begin{equation}
    \Pi_J=\prod_{J'\neq J}\frac{x-x_{J'}}{x_J-x_{J'}},
    \qquad
    x_J=\frac12\Big(J(J+1)-2S(S+1)\Big)\:\:.
\end{equation}
Hence the isotropic $SU(N)$ Heisenberg interaction can be written as
\begin{equation}
    \boxed{
    H_{ij}^{(\mathbf N,\mathbf N)}
    \propto
    \sum_{J=0}^{2S}
    (-1)^{2S-J}
    \prod_{J'\neq J}\frac{x-x_{J'}}{x_J-x_{J'}}
    }.
\end{equation}
In contrast to the conjugated $(\mathbf N,\overline{\mathbf N})$ case, where the Hamiltonian projects only onto the $J=0$ sector, here all total-spin sectors contribute, with coefficients fixed by the exchange parity $(-1)^{2S-J}$. Therefore the $(\mathbf N,\mathbf N)$ Hamiltonian is again a polynomial in $x=\mathbf S_i\cdot \mathbf S_j$, but now involving all projectors $\Pi_J$ rather than only $\Pi_{J=0}$.

To illustrate this structure, let us again consider the first few values of $N$.

\begin{itemize}

\item \underline{$(N=2)\Rightarrow (S=\tfrac12)$}:

Here $J=0,1$, and
\begin{equation}
    P_{ij}=-\Pi_0+\Pi_1.
\end{equation}
Using $\Pi_0=\frac14-\mathbf S_i\cdot \mathbf S_j$, one finds
\begin{equation}
    \boxed{
    H_{ij}^{(\mathbf 2,\mathbf 2)}
    \propto
    P_{ij}
    =
    2\,\mathbf S_i\cdot \mathbf S_j+\frac12
    }.
\end{equation}
After dropping the constant, this is just the ordinary spin-$\tfrac12$ Heisenberg interaction
$\mathbf S_i\cdot \mathbf S_j$.

\item \underline{$(N=3)\Rightarrow (S=1)$}:

Now $J=0,1,2$, and the exchange parity gives
\begin{equation}
    P_{ij}=+\Pi_0-\Pi_1+\Pi_2.
\end{equation}
Using the eigenvalues $(x_0=-2), \:\:(x_1=-1), \:\:(x_2=1)$; the unique quadratic polynomial satisfying
\[
P(x_0)=+1,\qquad P(x_1)=-1,\qquad P(x_2)=+1
\]
is
\begin{equation}
    \boxed{
    H_{ij}^{(\mathbf 3,\mathbf 3)}
    \propto
    P_{ij}
    =
    (\mathbf S_i\cdot \mathbf S_j)^2+\mathbf S_i\cdot \mathbf S_j-1
    }.
\end{equation}

\item \underline{$(N=4)\Rightarrow (S=\tfrac32)$}:

Here $J=0,1,2,3$, and therefore
\begin{equation}
    P_{ij}=-\Pi_0+\Pi_1-\Pi_2+\Pi_3.
\end{equation}
The corresponding eigenvalues of $x=\mathbf S_i\cdot \mathbf S_j$ are
\[
x_0=-\frac{15}{4},\qquad
x_1=-\frac{11}{4},\qquad
x_2=-\frac{3}{4},\qquad
x_3=\frac{9}{4}.
\]
Interpolating the cubic polynomial with values
\[
P(x_0)=-1,\qquad P(x_1)=+1,\qquad P(x_2)=-1,\qquad P(x_3)=+1,
\]
one obtains
\begin{equation}
    \boxed{
    H_{ij}^{(\mathbf 4,\mathbf 4)}
    \propto
    P_{ij}
    =
    -\frac{67}{32}
    -\frac{9}{8}\,\mathbf S_i\cdot \mathbf S_j
    +\frac{11}{18}(\mathbf S_i\cdot \mathbf S_j)^2
    +\frac{2}{9}(\mathbf S_i\cdot \mathbf S_j)^3
    }.
\end{equation}
Up to an additive constant and an overall normalization, this is the spin-$\tfrac32$ form of the $SU(4)$-symmetric isotropic exchange discussed before.

\end{itemize}

\noindent We therefore arrive at the following parallel picture. In the conjugated $(\mathbf N,\overline{\mathbf N})$ lattice, the $SU(N)$ Heisenberg bond Hamiltonian is the projector onto the unique $J=0$ sector, and so it is represented by a single projector polynomial $\Pi_0(x)$. In the isotropic $(\mathbf N,\mathbf N)$ lattice, by contrast, the Hamiltonian is the permutation operator $P_{ij}$, which receives contributions from all total-spin sectors with alternating signs. Thus both cases admit an exact rewriting as polynomials in $x=\mathbf S_i\cdot \mathbf S_j$, but the conjugated case isolates one spin sector whereas the isotropic one combines all of them.

At this point, it is useful to clarify the relation between the multipolar basis and the more familiar polynomial expansion in
$x=\mathbf S_i\cdot \mathbf S_j$. Let's recall that in both lattices
$$
\begin{cases}
\mathbf{N}\otimes\mathbf{\overline{N}}\cong\mathbf{N}\otimes\mathbf{N}\cong\mathcal{H}_{S}\otimes\mathcal{H}_S\cong\bigoplus_{J=0}^{2S}\mathcal{H}_J \qquad \left(\text{with }\:\: S=\frac{N-1}{2}\right)\\ 
\\
\text{End}^{SU(2)}_{\text{invariant}}(\mathcal{H}_{S}\otimes\mathcal{H}_S)=\text{span}\left(\mathbb{M}_{k=0,...,2S}\right) \qquad \left[\text{with }\:\:\mathbb{M}_k\equiv\sum_{q=-k}^{k}(-1)^q\,T_q^{(k)}(i)T_{-q}^{(k)}(j)\right]\:\:.
\end{cases}
$$
Using the first line, we know that there are two ways to express the $SU(2)$-invariant bond operators, namely $(\mathrm{i})\: \text{projectors }\left\{\Pi_J\right\}_{J=0,...,2S}$ and $  (\mathrm{ii})\:\text{monomials } \left\{1,x,x^2,...,x^{2S}\right\}$.
However, by using the decomposition in spherical tensors, we have a third way
\begin{equation}
\boxed{
    (\mathrm{i})\  \text{projectors } \left\{\Pi_J\right\}_{J=0}^{2S} \: \hspace{0.8cm} (\mathrm{ii})\  \text{monomials} \left\{1,x,x^2,...,x^{2S}\right\}\: \hspace{0.8cm} (\mathrm{iii})\ \text{multipoles} \left\{\mathbb{M}_k\right\}_{k=0}^{2S}}\:\:.
\end{equation}
Now, since all $\set{\mathbb{M}_k\:, x^n}$ operators are $SU(2)$-scalars, then they must be diagonal in the total-spin-$J$ spaces. Meaning that they are linear combinations of the projectors $\Pi_J$:
\begin{equation}
    \begin{aligned}
        x=\sum_{J=0}^{2 S} x_J \Pi_J \quad \xrightarrow[]{\quad } \quad &x^n=\sum_{J=0}^{2 S} (x_J)^n \:\Pi_J \qquad \quad \left[\text{with }\:\:x_J=\frac{1}{2}(J(J+1)-2 S(S+1))\right]\\
        &\:\mathbb{M}_k=\sum_{J=0}^{2S} \mu_J^{(k)} \:\Pi_J
    \end{aligned}
\end{equation}
where the coefficients $\mu_{J}^{(k)}$ depend on the normalization of the spherical tensors $T_q^{(k)}$ and, in standard conventions, are determined by Wigner $6j$-symbols. Therefore both $\left\{x^n\right\}$ and $\left\{\mathbb{M}_k\right\}$ are different basis sets for the $SU(N)$-invariant bond operators, and thus they are related by 
$$
\mathbb{M}_k=\sum_{n=0}^{2 S} A_{kn} \:x^n\:\:\:\:\:\:\:, \qquad x^n=\sum_{k=0}^{2 S} B_{n k} \:\mathbb{M}_k,
$$
for some invertible matrices $A$ and $B$. Indeed, a further fact is that each coefficient $\mu_{J}^{(k)}$ is actually a polynomial of degree $k$ in $J(J+1)$, and therefore the scalar multipoles $\mathbb{M}_k$ are polynomials of the same degree $k$ in $x=\bm{S}_i \cdot \bm{S}_j$. This is written as 
\begin{equation}
    \mathbb{M}_k=P_k(x)\quad \xrightarrow[]{\:\text{ say ($N$=4)(2S=3) }\:} \ \ \begin{cases}
        \mathbb{M}_1=a_{10}+a_{11} x \\
        \mathbb{M}_2=a_{20}+a_{21} x+a_{22} x^2\\
        \mathbb{M}_3=a_{30}+a_{31} x+a_{32} x^2+a_{33} x^3
    \end{cases}
\end{equation}
from where we can clearly see that the change of basis is done by a triangular matrix
\begin{equation}
    \left(\begin{array}{l}
\mathbb{M}_1 \\
\mathbb{M}_2 \\
\mathbb{M}_3
\end{array}\right)=\left(\begin{array}{lll}
* & 0 & 0 \\
* & * & 0 \\
* & * & *
\end{array}\right)\left(\begin{array}{c}
x \\
x^2 \\
x^3
\end{array}\right)+\text { constants . }
\end{equation}
This observation allows us to connect the multipolar description of the interaction with the more common polynomial form used in the spin-model literature. For our prototypical example of $N=4$ (corresponding to $S=3/2$), the scalar multipole contractions $\mathbb{M}_1,\mathbb{M}_2,\mathbb{M}_3$ correspond respectively to the dipole--dipole, quadrupole--quadrupole, and octupole--octupole couplings. Since each $\mathbb{M}_k$ is a polynomial of degree $k$ in $x=\mathbf S_i\cdot\mathbf S_j$, a general $SU(2)$-invariant Hamiltonian can be expressed equivalently either in the multipolar basis
\begin{equation}
H_{ij}^{(N=4)}\propto g_1\,\mathbb{M}_1+g_2\,\mathbb{M}_2+g_3\,\mathbb{M}_3 \qquad \Big[ g_1=g_2=g_3 \:\:\text{if $SU(4)$}   \Big]
\end{equation}
or in the polynomial basis
\begin{equation}
H_{ij}^{(N=4)}\propto \:\:c_1 \mathbf S_i\cdot\mathbf S_j+c_2 (\mathbf S_i\cdot\mathbf S_j)^2+c_3 (\mathbf S_i\cdot\mathbf S_j)^3
\:\:  .
\end{equation}
In the latter form, the three nontrivial terms correspond to interactions that are respectively linear, quadratic, and cubic in the scalar product of the spins. For this reason, Hamiltonians of this type are commonly referred to as \textbf{\emph{bilinear--biquadratic--bicubic} (BLBQ-BC) spin models}. In this broader sense, our $SU(N)$ Heisenberg Hamiltonians are just very special fine-tuned points inside larger families of Hamiltonians for spins $S=(N-1)/2$ respectively. In the multipolar language, they correspond to the highly symmetric points where all multipole exchange channels become exactly degenerate, thereby enhancing quantum fluctuations and giving rise to unusual competing phases.

\subsubsection{Higher Spin models and different Multipolar Orders}
As pointed out before, the polynomial representation derived at the end of the previous section is not merely a convenient rewriting, but places our $SU(N)$ exchange models within a broader class of higher-spin Hamiltonians that have been extensively studied in the literature. In general, for a spin-$S$ system, the most general isotropic/conjugated two-site interaction can be written as a polynomial in $x=\mathbf S_i\cdot\mathbf S_j$ of degree up to $2S$:
\begin{equation}
    \boxed{H_{ij}^{(S)}\propto \:\:c_1 \mathbf S_i\cdot\mathbf S_j+c_2 (\mathbf S_i\cdot\mathbf S_j)^2+...+c_{2S} \:(\mathbf S_i\cdot\mathbf S_j)^{2S}} \:\:.
\end{equation}
The well-known spin-$1$ case already illustrates this structure: in addition to the usual bilinear Heisenberg interaction, a biquadratic term $(\mathbf S_i\cdot\mathbf S_j)^2$ naturally appears and plays an essential role in determining the phase diagram. 

A natural question, however, is whether these higher-order terms are merely allowed by symmetry, or whether they can also arise microscopically with appreciable strength. In the spin-$1$ case, this issue was addressed explicitly by Mila and Zhang, who showed that a sizable biquadratic coupling can emerge from superexchange in electron systems with quasi-degenerate orbitals \cite{Mila2000}, or more generally from situations where spins and orbital degrees of freedom are strongly entangled by SOC \cite{PiSavrasov2014}. The key point is that, once low-lying orbital excitations are present, the effective exchange is no longer exhausted by the usual bilinear Heisenberg term, and additional interactions proportional to $(\mathbf S_i\cdot \mathbf S_j)^2$ naturally appear. From this perspective, the bilinear--biquadratic (BLBQ) Hamiltonian is not just a formal extension of the Heisenberg model, but a genuine low-energy effective description of certain multiorbital Mott systems. 

More broadly, recent work has emphasized that realistic spin models derived from electronic Hamiltonians often contain sizable non-Heisenberg interactions. Using first-principles electronic-structure calculations combined with downfolding techniques, Hoffmann and Blügel show that biquadratic and other higher-order exchange terms are frequently required to reproduce experimental magnetic properties \cite{HoffmannBlugel2020}. Indeed, in systems with larger local Hilbert spaces, this trend becomes even more pronounced: multiorbital Hubbard models with three active orbitals per site naturally lead to an effective spin-$\tfrac{3}{2}$ Hamiltonian containing bilinear, biquadratic, and bicubic interactions \cite{SoniMoreo2022}. In this case, the cubic term $(\mathbf S_i\cdot\mathbf S_j)^3$ appears on equal footing with the lower-order couplings, providing a microscopic realization of the BLBQ-BC spin models introduced above.

Now, for the generic case of $S\geq 1$, the inclusion of the biquadratic interaction has important consequences for the physics of quantum spin systems. Already in the spin-$1$ context, the BLBQ Hamiltonian exhibits a remarkably rich phase diagram that goes far beyond conventional magnetic ordering. Depending on the relative strength of the BL and BQ couplings, the system can host antiferromagnetic phases, dimerized states, topological Haldane phases, and quadrupolar (aka. spin-nematic) states in which the dipolar magnetic moment does not order while non-vanishing higher-rank spin correlations show up \cite{PencLauchli2011}. A particularly important point in this phase diagram is the celebrated Affleck–Kennedy–Lieb–Tasaki (AKLT) Hamiltonian \cite{AKLT1987}, which provides an exactly solvable realization of the Haldane phase and plays a central role in modern discussions of symmetry-protected topological order \cite{Tasaki2020, PollmannOshikawa2012}. This is how, in the last two decades, the resulting rich landscape of phases, including AKLT-related ones, has been studied extensively in one dimension and higher dimensions using a variety of analytical and numerical methods \cite{LauchliTrebst2006,DeChiaraLewenstein2011,NiesenCorboz2017,KartsevSantos2020,RenWu2023,LianLi2026}. In this sense, the biquadratic interaction places dipolar and quadrupolar correlations on a comparable footing and provides a concrete realization of what we referred to as \emph{multipolar/nematic order}.

Importantly, in the BLBQ model such multipoles arise directly from the exchange interaction between high-spin degrees of freedom, without requiring additional crystal-field or spin–orbit mechanisms. This corresponds precisely to what we will later call the route of \emph{high-spin exchange multipoles}. Examples include the quadrupolar correlations observed in triangular-lattice spin-$1$ models \cite{StoudenmireBalents2009} and the nematic phases appearing in frustrated BLBQ systems \cite{PohleMotome2023}. In these cases the multipolar order parameter is constructed purely from spin operators and therefore emerges naturally from the polynomial Hamiltonians discussed above.

At the same time, the concept of multipolar order also appears in other contexts that are conceptually distinct from the high-spin exchange mechanism considered here. In systems with strong spin–orbit coupling, for instance, the relevant local degrees of freedom are spin–orbit entangled moments whose exchange interactions are naturally expressed in terms of multipolar operators \cite{IwaharaChibotaru2022}. This mechanism underlies a number of proposed multipolar phases in $d$- and $f$-electron compounds, including double perovskites and related materials \cite{ChenBalents2010,HiraiChen2020, Tagaki2019}. Similarly, in rare-earth systems, the multipolar degrees of freedom often originate from crystal-field-split $f$-electron states, leading to multipolar ordering phenomena such as those observed in compounds like CeB$_6$ or actinide dioxides \cite{Kuramoto2009, Shiina1997, Santini2009}. 

The examples discussed above illustrate that the notion of multipolar order appears in a variety of seemingly unrelated physical settings. In the literature, however, the same terminology is often used to describe phenomena that originate from rather different microscopic mechanisms. In order to avoid potential confusion, it is useful to distinguish the main routes through which multipole degrees of freedom emerge in condensed-matter systems. Broadly speaking, three mechanisms can be identified:
\begin{equation*}
    \begin{aligned}
        (\mathrm{i)} \ &\text{Crystal-field Splitting} \quad \xleftarrow[]{\qquad\qquad} (f-\textit{electrons})\\
        (\mathrm{ii)} \ &\text{Strong spin–orbit coupling} \:\:\xleftarrow[]{\qquad\:\:} (j_{\text{eff}}/d^1-\textit{systems})\\
        (\mathrm{iii)} \ &\text{Exchange  between high-spin-S} \:\:\xleftarrow[]{\quad} \:\:\:(\textit{our case }\text{- discussed above})
    \end{aligned}
\end{equation*}
\begin{itemize}
    \item \underline{Crystal-field multipoles (Stevens operators)}
    
    In rare-earth ions, because of strong atomic spin-orbit coupling, the atomic Hamiltonian first produces a multiplet with fixed atomic angular momentum $J$ $(\mathbf{J}=\mathbf{L}+\mathbf{S})$, and therefore the relevant allowed states are
$$
\Big\{|J, m\rangle\Big\}_{m=-J, \ldots, J }\quad \:\: \xrightarrow[]{\quad \text{hence}\quad} \quad \mathcal{H}_J \:=\:\text{ local Hilbert space}\:\:.
$$
Now, in the crystallographic lattice picture, the surrounding ions generate, over an arbitrary atom in which we are interested, an electrostatic potential $V(\mathbf{r})$; which can always be expanded in spherical harmonics:
$$
V(\mathbf{r})=\sum_{k, q} A_{k q} r^k Y_{k q}(\theta, \phi)
$$
However, when one projects this potential onto the $J$-multiplet, the spatial tensors $Y_{k q}$ become operators acting on the on-site Hilbert space $\mathcal{H}_J$. Indeed, using the Wigner-Eckart theorem,
$$
r^k Y_{k q} \rightarrow O_k^q(\mathbf{J})
$$
where $O_k^q$ are the so-called \emph{Stevens operators}. These form a set of irreducible spherical tensor operators acting within the $J$-multiplet and can be expressed as polynomials in the angular-momentum operators $J_x,J_y,J_z$. The explicit expressions for the lowest ranks are listed in Table~\ref{tab:stevens}.
\begin{table}[!h]
\centering
\begin{tabular}{c c l}
\hline
Rank $k$ & Component $q$ & Operator $O_k^{\,q}(\mathbf{J})$ \\
\hline

\multirow{3}{*}{$k=1$ (dipoles)}
& $0$ & $O_1^{\,0} = J_z$ \\
& $1$ & $O_1^{\,1} = J_x$ \\
& $-1$ & $O_1^{\,-1} = J_y$ \\

\hline

\multirow{5}{*}{$k=2$ (quadrupoles)}
& $0$ & $O_2^{\,0} = 3J_z^2 - X$ \\
& $1$ & $O_2^{\,1} = \tfrac12\left(J_xJ_z + J_zJ_x\right)$ \\
& $-1$ & $O_2^{\,-1} = \tfrac12\left(J_yJ_z + J_zJ_y\right)$ \\
& $2$ & $O_2^{\,2} = J_x^2 - J_y^2$ \\
& $-2$ & $O_2^{\,-2} = J_xJ_y + J_yJ_x$ \\

\hline

\multirow{7}{*}{$k=3$ (octupoles)}
& $0$ & $O_3^{\,0} = 5J_z^3 - (3X-1)J_z$ \\
& $1$ & $O_3^{\,1} = \tfrac14\!\left[(5J_z^2 - X - \tfrac12)J_x + J_x(5J_z^2 - X - \tfrac12)\right]$ \\
& $-1$ & $O_3^{\,-1} = \tfrac14\!\left[(5J_z^2 - X - \tfrac12)J_y + J_y(5J_z^2 - X - \tfrac12)\right]$ \\
& $2$ & $O_3^{\,2} = \tfrac12\left[(J_x^2 - J_y^2)J_z + J_z(J_x^2 - J_y^2)\right]$ \\
& $-2$ & $O_3^{\,-2} = \tfrac12\left[(J_xJ_y + J_yJ_x)J_z + J_z(J_xJ_y + J_yJ_x)\right]$ \\
& $3$ & $O_3^{\,3} = J_x^3 - 3J_xJ_y^2$ \\
& $-3$ & $O_3^{\,-3} = J_y^3 - 3J_yJ_x^2$ \\

\hline
\end{tabular}
\caption{Stevens operator equivalents $O_k^{\,q}$ written in terms of angular momentum operators $J_x,J_y,J_z$, with $X\equiv J(J+1)$. Adapted from \cite{rotter_mcphase_manual_node132}.}
\label{tab:stevens}
\end{table}
In this representation, the so-called \emph{crystal-field Hamiltonian}, that accounts for the electrostatic potential $V(\mathbf{r})$ due to the ionic environment, takes the form
\begin{equation}
    H_{\mathrm{CF}}=\sum_{k, q} B_{k q} O_k^q(\mathbf{J})\:\: \qquad \Big[\text{multipole expansion of }\:\text{End}(\mathcal{H}_J)\Big]\:,
\end{equation}
where the coefficients $B_{k q}$ are known as the crystal-field parameters that encode the geometry/symmetries of the ionic environment around the ion. However, this is just a single-ion Hamiltonian, and it does not yet produce multipolar order. The many-body multipolar order appears when inter-site exchange couples these multipoles:
\begin{equation}
    H=\sum_i H_{\mathrm{CF}}(i)+\sum_{<ij>} \sum_{k q} J_k^{i j} \:\: O_k^q(i) O_k^{-q}(j)\:\:.
\end{equation}
This is exactly how, for instance, a system can have 
\begin{equation}
    \langle \mathbf{J}\rangle=0 \qquad \:\:\text{but } \qquad \left\langle O_2^0\right\rangle \neq 0\quad \text{(quadrupolar order)}
\end{equation}

\item \underline{Spin-orbit-entangled multipoles}

In many $4d$ and $5d$ compounds, the local degrees of freedom are not the true atomic total angular moment $J$. Instead, the effective on-site Hilbert spaces, and hence the effective low-energy states, are obtained after several projections, resulting in highly entangled spin-orbit spaces. To illustrate such a situation, let's consider the case in which the crystal field first selects a low-energy orbital manifold, most commonly the $t_{2g}$ sector of the $d$ shell. Thus, rather than starting from the full atomic $(L,S)$ multiplet and then projecting onto a fixed $J$, one first restricts the local Hilbert space to
\begin{equation}
    \mathcal H_{\rm loc}\;\equiv\;\text{span}\big(d_{xy},d_{yz},d_{zx}\big)\otimes \mathcal H_{S},
\end{equation}
where the three $t_{2g}$ orbitals transform as an effective orbital angular momentum $L_{\rm eff}=1$. Within this reduced manifold, the on-site spin--orbit coupling takes the form
\begin{equation}
    H_{\rm SOC}=\lambda\,\mathbf L_{\rm eff}\cdot \mathbf S
\end{equation}
Diagonalizing $H_{\rm SOC}$ then produces local spin--orbit-entangled states, such as $j_{\rm eff}=1/2$ and $j_{\rm eff}=3/2$ multiplets, or related Kramers/non-Kramers pseudospin degrees of freedom depending on filling and symmetry.

The essential point is that the resulting local moments are no longer pure spins, but spin--orbital entangled objects. Consequently, the inter-site exchange interaction, once projected onto this entangled low-energy manifold, is not exhausted by a simple dipole--dipole Heisenberg term. Instead, it generally acquires anisotropic and multipolar components, which can be organized in terms of irreducible tensor operators acting on the local pseudospin-orbital manifold. Schematically, one obtains an exchange Hamiltonian of the form
\begin{equation}
    H_{\rm ex}=\sum_{\langle ij\rangle}\sum_{KQ,K'Q'} \mathcal J^{\,ij}_{KQ,K'Q'}\,
\mathcal O^{(K)}_Q(i)\,\mathcal O^{(K')}_{Q'}(j)\:\:,
\end{equation}
where the operators $\mathcal O^{(K)}_Q$ are multipoles of the spin--orbit-entangled local space. In this route, therefore, the multipolar structure does not arise because a crystal field splits a pre-existing atomic $J$ multiplet, but because spin--orbit coupling entangles spin and orbital degrees of freedom inside a crystal-field-selected manifold, and the projected exchange between these entangled moments naturally takes a multipolar form.
\end{itemize}

%% file: chapter4_5_SuN_generalization.tex
\section{0+1 SU(N)-Superspin Theory} \label{sec:su_n_superspin}
Having established the physical relevance of $SU(N)$ degrees of freedom in strongly correlated systems, we now turn to the construction of their coherent-state path integral. Our goal in this chapter is to derive the classical action governing the imaginary-time dynamics, and in particular to understand how the Wess--Zumino term generalizes beyond $SU(2)$. As we will see, the relevant geometric arena is no longer the two-sphere
$S^2\simeq \mathbb{CP}^1$, but its natural $SU(N)$ generalization, the
complex projective space $\mathbb{CP}^{N-1}$. In group-theoretic terms, this
manifold arises as the coherent-state orbit
\begin{equation}
    \mathbb{CP}^{N-1}
    \simeq
    \frac{SU(N)}{S(U(N-1)\times U(1))},
\end{equation}
where $S(U(N-1)\times U(1))$ is the stabilizer of a reference state. However,
although this stabilizer contains a non-Abelian $SU(N-1)$ factor, the
topological part of the Wess--Zumino term for fundamental coherent states, and
more generally for symmetric irreducible representations, is controlled by the
associated $U(1)$ Berry line bundle. In particular, its Berry connection will play the role of a
generalized monopole potential, while its curvature will be proportional to the
Fubini--Study Kähler form on $\mathbb{CP}^{N-1}$. 

\noindent In this section we keep the construction as basis-independent as possible. 
When an explicit set of generators is needed, the standard Gell--Mann basis is a convenient algebraic choice, while physically adapted bases are discussed separately. 
The multipolar organization of $\mathfrak{su}(N)$ is reviewed in Appendix~\ref{app:suN-multipoles}, and several useful $SU(4)$ embeddings are summarized in Appendix~\ref{app:su4-embeddings-parametrizations}.

\subsection{SU(N) Coherent States and \texorpdfstring{$\mathbb{CP}^{N-1}$}{CP{N-1}}}
\subsubsection{SU(N) Representation Theory}\label{Sec:su_n_representation}
We begin our discussion by recalling that the $SU(N)$ groups are defined (in the fundamental irrep) as 
\begin{equation}
    SU(N) = \Set{g\in M_{N\times N}(\mathbb{C}) \ | \  g^\dagger g = \mathbb{1}_{N} \ \text{ and} \ \det(g)=1},
\end{equation}
so that their role in physics is to describe the set of all unitary basis transformations with $\det=1$ of an $N$-level quantum system. Moreover, as we have already mentioned many times, $SU(N)$ is indeed a Lie group with its associated Lie algebra $\mathfrak{su}(N)$ generated by $N^2-1$ generators $\{T^1, \dots, T^{N^2-1}\}$. For instance, in the case of $SU(3)$, the famous eight Gell-Mann matrices form one such set of generators, while for $SU(4)$, a total of 15 generators are needed. 

\noindent However, there is a way to standardize the formulation of such generators without resorting to a direct generalization of the Gell-Mann construction in $N\times N$ matrices. In this approach, we take advantage of the $N^2-N$ raising and lowering operators together with the $N-1$ Cartan operators. Therefore, in the fundamental irrep (to be properly defined later), one convenient set of generators can be defined from the matrices 
\begin{equation}
    \begin{aligned}
\hat{g}_{i j}(i \neq j)\equiv \ket{i}\bra{j}, \quad \hat{H}_1 & =\frac{1}{2}\left(\hat{g}_{11}-\hat{g}_{22}\right),\: \ldots , \:\hat{H}_{N-1} =\frac{1}{2}\left(\hat{g}_{N-1 N-1}-\hat{g}_{N N}\right),
\end{aligned}
\end{equation}
where $i,j=1,\dots,N$. Using the commutation relations $\left[\hat{g}_{i j}, \hat{g}_{k l}\right]=\delta_{k j} \hat{g}_{i l}-\delta_{i l} \hat{g}_{k j}$ and the Cartan/diagonals $[\hat{H}_k,\hat{H}_l]=0$, we can easily identify the $\hat{H}$ operators as the $N-1$ Cartan generators, and the matrices $\hat{g}_{ij}$ with $i<j$ ($i>j$) as the raising (lowering) operators. 

Under this picture, we aim to generalize the notion of the maximally polarized state $\ket{\uparrow}$ that we have for $SU(2)$. Hence, we define the so-called \textbf{highest-weight state} $\ket{\uparrow} \in \mathcal{H}_N$ by the condition
\begin{equation}
    \hat{g}_{i j}\ket{\uparrow} = 0 \quad \forall\, i<j,
\end{equation}
that is, the state annihilated by all raising operators. The advantage of this definition is that, when considering other $SU(N)$ representations, we can still identify such a $\ket{\uparrow}$ in the corresponding representation spaces $\mathcal{H}_m$ (being $m$ a labelling index). Therefore, even though our definition was introduced in the fundamental irrep, it remains valid in any arbitrary irrep. 

\noindent Furthermore, noticing that $\ket{\uparrow}$ is also a common eigenvector of the Cartan generators,
\begin{equation}
    \hat{H}_1\ket{\uparrow}=\frac{1}{2} \lambda_1\ket{\uparrow}\:, \quad \ldots\quad,\: \hat{H}_{N-1}\ket{\uparrow}=\frac{1}{2} \lambda_{N-1}\ket{\uparrow},
\end{equation}
the $N-1$ eigenvalues $\left[\lambda_1, \ldots, \lambda_{N-1}\right]$, known as the \emph{Dynkin labels} (defined with respect to the standard Cartan basis), can be used to label the irreducible representations of $SU(N)$, with their dimensions computed from this tuple (via Weyl’s dimension formula). In this way, we move from labeling irreps through Casimir eigenvalues to labeling them through the Cartan eigenvalues of the highest-weight state $\ket{\uparrow}$. Once the Dynkin labels $\left[\lambda_1, \ldots, \lambda_{N-1}\right]$ of an irrep are fixed, the eigenvalues $c_2, \ldots, c_N$ of the Casimir operators are fixed as well. In other words, the Casimir eigenvalues are functions of the Dynkin labels.

\noindent For instance, for $\mathfrak{su}(3)$, the irreducible representations are labeled by $\lambda_1$ and $\lambda_2$, the quadratic Casimirs (with the standard normalization $\operatorname{Tr}\left(T^a T^b\right)=\delta^{a b}/ 2$) are computed from 
\begin{equation}
    c_2\left(\left[\lambda_1, \lambda_2\right]\right)=\frac{1}{3}\left(\lambda_1^2+\lambda_2^2+\lambda_1 \lambda_2+3 \lambda_1+3 \lambda_2\right)\:,
\end{equation}
and the dimension of the $\left[\lambda_1, \lambda_2\right]$ irrep is given by
\begin{equation}
\operatorname{dim}\left[\lambda_1, \lambda_2\right]=\frac{1}{2}\left(\lambda_1+1\right)\left(\lambda_2+1\right)\left(\lambda_1+\lambda_2+2\right).
\end{equation}
This means that for a three-level system equipped with the Cartesian basis, the generators of the Cartan subalgebra are 
\begin{equation}
    \hat{H}_1=\frac{1}{2}\begin{pmatrix}
1 & 0 & 0 \\
0 & -1 & 0 \\
0 & 0 & 0
\end{pmatrix}, \quad
    \hat{H}_2=\frac{1}{2}\begin{pmatrix}
0 & 0 & 0 \\
0 & 1 & 0 \\
0 & 0 & -1
\end{pmatrix}
\end{equation}
and the highest-weight state is
\begin{equation}
    \ket{\uparrow}=\begin{pmatrix}
1 \\
0 \\
0  
\end{pmatrix},
\end{equation}
implying that the fundamental representation has $\operatorname{dim}[1,0]=3$. 

Nevertheless, note that the fundamental representation is a special example of what is often called a \emph{degenerate representation}, in the sense that only one Dynkin label is nonzero. Previously, we said that the fundamental representation of $SU(2)$ is defined by the minimal dimension in which the group can be represented nontrivially as matrices. In that case, the fundamental irrep is unique and corresponds to spin $S=1/2$, with $\dim\mathcal H_{1/2}=2$. The situation for $SU(3)$ is slightly more subtle: there are two inequivalent three-dimensional irreps, namely $[1,0]$ and $[0,1]$. These correspond to the fundamental and antifundamental representations, respectively. Therefore, when discussing the fundamental irrep of $SU(N)$, one should specify which degenerate highest-weight representation is being used. In this work, by the fundamental representation we will mean the irrep with Dynkin labels $[1,0,\ldots,0]$, whereas the conjugate, or antifundamental, representation is labeled by $[0,\ldots,0,1]$. For the purposes of the present construction, we will always work with the first choice, $[1,0,\ldots,0]$, such that, in the standard basis of $\mathbb C^N$, the highest-weight state takes the simple form
\begin{equation} \label{eq:max_weight_state}
    \boxed{
    \ket{\uparrow}_N=
    \begin{pmatrix}
    1\\
    0\\
    \vdots\\
    0
    \end{pmatrix}}
    \ \ .
\end{equation}
Furthermore, it is important to note that this notation also makes contact with the usual symmetric and antisymmetric tensor representations. The representations $[p, 0, \ldots, 0]$ correspond to the fully symmetric powers $\operatorname{Sym}^p\left(\mathbb{C}^N\right)$, while the fully antisymmetric $p$-index representation $\wedge^p \mathbb{C}^N$ is labeled by a single 1 in the $p$-th Dynkin position. In particular, $\wedge^{N-1} \mathbb{C}^N$ has labels $[0, \ldots, 0,1]$, and is the antifundamental representation.

In the next subsection, we will exploit the expression in Eq.\ref{eq:max_weight_state}, but for now, let us present another fact that will be used repeatedly in the parallel computations. An arbitrary element $g \in SU(N)$ in the $N\times N$ matrix representation can be parameterized as 
\begin{equation}
    \boxed{g= \begin{array}{ccc}
         \left(\begin{array}{cccc}
1 & 0 & \cdots & 0 \\
0 & & & \\
\vdots & & X_{N-1} & \\
0 & &
\end{array}\right)
         & 
\left(\begin{array}{ccc}
e^{i \varphi} \cos \theta & -\sin \theta & 0 \\
\sin \theta & e^{-i \varphi} \cos \theta & 0 \\
0 & 0 & I_{N-2}
\end{array}\right)
         &
\begin{array}{ccc}
         \left(\begin{array}{cccc}
1 & 0 & \cdots & 0 \\
0 & & & \\
\vdots & & Y_{N-1} & \\
0 & &
\end{array}\right)
\\
    \end{array} 
    \end{array}}
    \label{eq:SU(N)_parametrization}
\end{equation}
where $X_{N-1}, Y_{N-1}$ are $(N-1)\times(N-1)$ matrices representing elements of $SU(N-1)$, $I_k$ is the $k \times k$ identity matrix, and the angles are defined as $\theta\in(0,\pi/2)$ and $\varphi\in(0,2\pi)$. For a detailed proof of this result, the reader is referred to Appendix B of \cite{Nemoto_2000}. Additionally, the parametrization above is used here only as a tool for constructing coherent states and then quotienting by the stabilizer of the reference ray. It should be distinguished from Euler-angle parametrizations of full group elements, where the object is a complete unitary matrix rather than a point in $\mathbb{CP}^{N-1}$ (see Appendix~\ref{app:su4-euler-parametrizations}).

Note also that in this parametrization we have explicitly excluded the boundary values $\theta=0$ and $\theta=\pi/2$. These special points correspond to degenerate cases where the middle block loses its mixing character, and part of the transformation can be reabsorbed into the $SU(N-1)$ matrices $X_{N-1}$ and $Y_{N-1}$. In this sense, we are working within a regular coordinate patch, where the parametrization is smooth and non-redundant. The useful point is that the excluded elements form a set of measure zero and do not affect the construction that follows. In fact, while this single patch does not provide a globally unique parametrization of all elements of $SU(N)$, it is sufficient to describe a dense subset of the group. Moreover, additional patches could be introduced if a fully global description were needed.

\subsubsection{SU(3) Redundancies and \texorpdfstring{$\mathbf{\mathbb{CP}^2}$}{CP2}}

Once we have the above results, we can start constructing $SU(3)$-coherent states in analogy to the $SU(2)$ case. Therefore, we begin from the expression of the ``redundant'' coherent states $\ket{\tilde{g}}=\tilde{g}\ket{\uparrow}$ and immediately realize that, when we have the first action
\begin{equation}
\tilde{g}\ket{\uparrow}_3=(...)(...)\left(\begin{array}{ccc}
1 & 0 & 0 \\
0 & e^{i \varphi_3} \cos \xi_2 & -e^{-i \varphi_4} \sin \xi_2 \\
0 & e^{i \varphi_4} \sin \xi_2 & e^{-i \varphi_3} \cos \xi_2
\end{array}\right)\left(\begin{array}{l}
1 \\
0 \\
0
\end{array}\right),
\end{equation}
the lower right $SU(2)$ matrix $Y_2(\xi_2,\varphi_3,\varphi_4)$ leaves the reference state invariant. This means that these $SU(2)$ transformations belong to the isotropy subgroup of $\ket{\uparrow}_3$ and therefore must be treated as redundancies in the path integration. In other words, continuous deformations along these directions do not generate new physical paths of coherent states.

Having in mind this redundancy, and continuing with the matrix products (now parametrizing $X_2[\xi_1,\varphi_1,\varphi_2]$ with $\xi_1\in(0,\pi/2)$ and $\varphi_{1,2}\in(0,2\pi)$), we find 
\begin{equation}
    \begin{aligned}
\left|\tilde{\mathbf{n}}_3\right\rangle \equiv \tilde{g}\left|\uparrow\right\rangle 
& = \left(\begin{array}{ccc}
1 & 0 & 0 \\
0 & e^{i \varphi_1} \cos \xi_1 & -e^{-i \varphi_2} \sin \xi_1 \\
0 & e^{i \varphi_2} \sin \xi_1 & e^{-i \varphi_1} \cos \xi_1
\end{array}\right)\left(\begin{array}{c}
e^{i\varphi} \cos\theta\\
\sin \theta \\
0
\end{array}\right) \\[7pt]
& =\left(\begin{array}{c}
e^{i \varphi} \cos \theta \\
0 \\
0
\end{array}\right)+\sin \theta\binom{0}{\left|\tilde{\mathbf{n}}_2\right\rangle}.
\end{aligned}
\end{equation}
At this point, we still have the usual projective $U(1)$ redundancy of quantum states: two vectors that differ only by an overall phase represent the same physical state,
\begin{equation}
    \ket{\psi}\sim e^{i\alpha}\ket{\psi}.
\end{equation}
Since we are working in the regular patch $\theta\in(0,\pi/2)$, the first component is nonzero, $\cos\theta\neq0$, and we can use this freedom to choose the representative whose first component is real and positive. Equivalently, we fix the phase by taking $e^{i\varphi}=1$. This leaves us with the expression for the \textbf{$SU(3)$ coherent states}
\begin{equation}\label{eq:su3_coherent}
    \boxed{\ket{\mathbf{n}_3}\equiv\ket{g(\theta,\xi, \varphi_1, \varphi_2)}=\left(\begin{array}{c}
\cos \theta \\
\sin \theta \ e^{i\varphi_1} \cos \xi \\
\sin \theta \ e^{i\varphi_2} \sin \xi
\end{array}\right)} \ \ .
\end{equation}
At this point, one might be tempted to conclude that, since $SU(2)\times U(1)$ act as redundancies in the path integral, the coherent states manifold should simply be obtained by taking the quotient with respect to this group. However, a small subtlety arises here. More precisely, the stabilizer of the ray generated by $\ket{\uparrow}_3$ inside $SU(3)$ is not written strictly as $SU(2)\times U(1)$, but rather as
\begin{equation}
    S(U(2)\times U(1))\subset SU(3),
\end{equation}
where the symbol $S$ reminds us that the total determinant must remain equal to one. In other words, the phase transformation and the $SU(2)$ rotation are not completely independent, but must combine in such a way that the resulting matrix still belongs to $SU(3)$. This is the mathematically precise way of saying that the transformations acting inside the lower two-dimensional subspace, together with a compensating phase on the first component, do not change the physical ray $[\ket{\uparrow}_3]$.
With this clarification in mind, taking the quotient by the redundant transformations, we find that the coherent-state manifold is
\begin{equation}
    \boxed{\frac{SU(3)}{S(U(2)\times U(1))} \simeq S^5/S^1 \simeq \mathbb{CP}^2}\:\:.
\end{equation}
Thus, the space of physical coherent states is the complex projective plane $\mathbb{CP}^2$, which has complex dimension $2$ and real dimension $4$. The question now is what consequences this fact brings. For $SU(2)$, we had $S^2\simeq \mathbb{CP}^1$, and now we have $\mathbb{CP}^2$. So, what is the difference? We will go deeper into these aspects once we consider the respective Wess--Zumino terms, but for now, let us first answer how the decomposition of the identity in such coherent states should look. In this way, we use again the fact that the projectors $\ket{\tilde{g}}\bra{\tilde{g}}=\ket{g}\bra{g}$ are $U(1)$-$S^1$ invariant, such that we can first focus on the geometry before taking the $U(1)$ quotient. In other words, it is enough to consider normalized states in $\mathbb{C}^3$, which form a five-dimensional sphere. This means that we can use the identification
\begin{equation}
    SU(3)/SU(2)\simeq S^5,
\end{equation}
and therefore obtain the round metric (coming from the $\mathbb{R}^6$ embedding) of $S^5$    
\begin{equation}\label{eq:S_5mesaure}
    \begin{gathered}
\left|d s_3\right|^2=d \theta^2+\cos ^2 \theta \ d \varphi^2+\sin ^2 \theta\left(d \xi^2+\cos ^2 \xi \ d \varphi_1^2+\sin ^2 \xi \ d \varphi_2^2\right), \\
d \mu_3= \sqrt{\det G_{S^5}} \ d(\text{angles})=\cos \theta \sin ^3 \theta \cos \xi \sin \xi \ d \theta d \xi d \varphi d \varphi_1 d \varphi_2,
\end{gathered}
\end{equation}
and therefore have the measure of $S^5$ as above. At this point, one might be tempted to think that the decomposition of the identity is simply obtained by normalizing with the total volume of $S^5$. However, this is not quite correct. Indeed, by symmetry, the integral over all coherent-state projectors must be proportional to the identity operator, but not necessarily equal to it. A simple trace argument shows that
\begin{equation}
    \int d\mu_3 \ \ket{\mathbf{n}_3}\bra{\mathbf{n}_3}
    = \frac{\text{vol}(S^5)}{3}\,\mathbb{1},
\end{equation}
and therefore the correct $SU(3)$ decomposition of the identity is
\begin{equation}
    \boxed{
    \mathbb{1}= \frac{3}{\text{vol}(S^5)} \int d\mu_3 \ \ket{\mathbf{n}_3}\bra{\mathbf{n}_3}
    }.
\end{equation}
Indeed, since $\text{vol}(S^5)=\pi^3$, the normalization constant can also be written as $\frac{3}{\pi^3}$. Now, having this result, we can repeat the same procedure as in the spin path integral section and arrive at the same results of Eq.\ref{eq:spin_path_integral_gener} and Eq.\ref{eq:spin_action}. But, before moving on, one last comment about the metric and the measure on $\mathbb{CP}^2$ (and in general $\mathbb{CP}^{N-1}$) should be made. For this purpose, let us illustrate the case of $SU(2)$, which is equivalent to considering the Hopf fibration
\begin{equation}
    S^1 \to S^3 \to S^2\simeq\mathbb{CP}^1.
\end{equation}
In this situation, the $S^1$ acts on the points of $S^3$, allowing us to take the quotient. As a consequence, the metric of $S^3$ induces a metric on $S^2\simeq\mathbb{CP}^1$, known as the \textbf{Fubini--Study metric}. In particular, the measure on $\mathbb{CP}^1$ can be obtained from the measure on $S^3$ (Eq.\ref{eq:S3_measure}) by integrating out the fiber direction (i.e. the $S^1$ angle):
\begin{equation}
     d \mu_2=\cos(\theta) \sin(\theta) \ d \theta d \varphi_1 d \varphi_2 
     \;\longrightarrow\; 
     d \mu_{CP^1}=\cos(\theta) \sin(\theta) \ d \theta  d \varphi_2.
\end{equation}
The important point here is that the measure on $\mathbb{CP}^1$ is inherited from that of $S^3$ after removing the $U(1)$ fiber, rather than being the standard round measure coming from the $\mathbb{R}^3$ embedding. Indeed, these two measures are not equivalent: while the Fubini--Study metric gives $\text{vol}(\mathbb{CP}^1)=\pi$, the usual round sphere has volume $4\pi$. This difference reflects a different normalization of the metric, not a different geometry. 

\noindent Having said that, for the case of $SU(3)$ and $\mathbb{CP}^2$, the measure of the coherent-state manifold (with the Fubini--Study metric) can be written as 
\begin{equation}
    d \mu_{CP^2}=\cos \theta \sin ^3 \theta \cos \xi \sin \xi \ d \theta d \xi d \varphi_1 d \varphi_2,
\end{equation}
where we have integrated out the $\varphi$-angle in Eq.\ref{eq:S_5mesaure}, corresponding to the $U(1)$ fiber.

\subsubsection{SU(4) Redundancies and \texorpdfstring{$\mathbf{\mathbb{CP}^3}$}{CP3}}
In the spirit of the previous section, we repeat all the calculations now for the $SU(4)$ group. This way, we start from
\begin{equation}
 \tilde{g}|\uparrow\rangle=\left(\begin{array}{cccc}
1 & 0 & \cdots & 0 \\
0 & & & \\
\vdots & & X_3& \\
0 & &
\end{array}\right)\left(\begin{array}{cccc}
e^{i \varphi} \cos \theta & -\sin \theta & 0  \\
\sin \theta & e^{-i \varphi} \cos \theta & 0 \\
0&0  & I_2
\end{array}\right)\left(\begin{array}{cccc}
1 & 0 & \cdots & 0 \\
0 & & & \\
\vdots & & Y_3 & \\
0 & & &
\end{array}\right)\left(\begin{array}{l}
1 \\
0 \\
0 \\
0
\end{array}\right)
\label{eq:SU(4)_parametrization}
\end{equation}
and see that all matrices $Y_3 \in SU(3)$ are redundancies. Therefore, continuing with the calculation
\begin{equation}
    \begin{aligned}
\left|\tilde{\mathbf{n}}_4\right\rangle & = 
\left(\begin{array}{cccc}
1 & 0 & \cdots & 0 \\
0 & & & \\
\vdots & & X_3& \\
0 & &
\end{array}\right) \left(\begin{array}{c}
e^{i \varphi} \cos \theta \\
\sin \theta \\
0 \\
0
\end{array}\right) 
\\
&= \left(\begin{array}{cc}
I_2 & 0 \\
0 & X_2
\end{array}\right)\left(\begin{array}{cccc}
1 & 0 & 0 & 0 \\
0 & e^{i \varphi_1} \cos \xi_1 & -\sin \xi_1 & 0 \\
0 & \sin \xi_1 & e^{-i \varphi_1} \cos \xi_1 & 0 \\
0 & 0 & 0 & 1
\end{array}\right)\left(\begin{array}{ll}
I_2 & 0 \\
0 & Y_2
\end{array}\right)\left(\begin{array}{c}
e^{i \varphi} \cos \theta \\
\sin \theta \\
0 \\
0
\end{array}\right)
\\
& =\left(\begin{array}{c}
e^{i \varphi} \cos \theta \\
0 \\
0 \\
0
\end{array}\right)+\sin \theta\binom{0}{\left|\tilde{
\mathbf{n}}_3\right\rangle} ,
\end{aligned}
\end{equation}
we find that the \textbf{SU(4) coherent states} are written as
\begin{equation}
    \boxed{\ket{\mathbf{n}_4}\equiv\ket{g(\theta,\xi_1, \xi_2, \varphi_1, \varphi_2, \varphi_3)}=\left(\begin{array}{c}
\cos \theta \\
\sin \theta \ e^{i\varphi_1} \cos \xi_1 \\
\sin \theta \ \sin\xi_1 \ e^{i\varphi_2} \cos \xi_2 \\
\sin \theta \ \sin\xi_1 \ e^{i\varphi_3} \sin \xi_2
\end{array}\right)}\ ,
\end{equation}
such that the coherent state manifold is
\begin{equation}
    \boxed{\frac{SU(4)}{S(U(3)\times U(1))} \simeq S^7/S^1 \simeq \mathbb{CP}^3} \ .
\end{equation}
In fact, the cases $SU(2)$ and $SU(4)$ are exceptional: owing to the accidental low-rank Lie-algebra isomorphisms $\mathfrak{su}(2)\simeq\mathfrak{so}(3)$ and $\mathfrak{su}(4)\simeq\mathfrak{so}(6)$, their coherent-state manifolds admit alternative descriptions in terms of orthogonal groups. More precisely, $\mathbb{CP}^{1}\simeq SO(3)/SO(2)$, while the present manifold may equivalently be written as $\mathbb{CP}^{3}\simeq SO(6)/U(3)$. The precise global group relations, their formulation in terms of spin groups, and the representation-theoretic origin of these identifications are discussed in Appendix~\ref{app:low_rank_su_so}.

\noindent Now, for this case of $SU(4)$, the relevant round metric is the one corresponding to $S^7$
\begin{equation}
    \begin{split}
    \left|d s_4\right|^2=d \theta^2+\cos ^2\theta d \varphi^2+\sin ^2\theta \Big[ d \xi_1^2 & +\cos ^2\xi_1 \:d\varphi_1^2+\sin ^2\xi_1 \\ &\left(d \xi_2^2+\cos ^2\xi_2 \:d \varphi_2^2+\sin^2 \xi_2 \:d \varphi_3^2\right)\Big]\:\:,
\end{split}
\end{equation}
which has associated measure 
\begin{equation}
d \mu_4=(\cos\theta) \:(\sin ^5\theta) \: (\cos\xi_1) \: (\sin ^3\xi_1) \: (\cos\xi_2) \:(\sin \xi_2) \:\: d\theta \: d\xi_1 \: d\xi_2 \: d\varphi \: d\varphi_1 \: d\varphi_2 \: d\varphi_3. 
\end{equation}
Again, this measure is the one we use to find the decomposition of the identity. As in the $SU(3)$ case, one should not simply divide by the total volume of $S^7$. By symmetry, the integral over all coherent-state projectors must be proportional to the identity operator, and taking the trace gives
\begin{equation}
    \int d\mu_4 \ \ket{\mathbf{n}_4}\bra{\mathbf{n}_4}
    = \frac{\text{vol}(S^7)}{4}\,\mathbb{1}.
\end{equation}
Therefore, the correct $SU(4)$ decomposition of the identity is
\begin{equation}
    \boxed{
    \mathbb{1}= \frac{4}{\text{vol}(S^7)}
    \int d\mu_4 \ \ket{\mathbf{n}_4}\bra{\mathbf{n}_4}
    }.
\end{equation}
Since for the unit seven-sphere one has $\text{vol}(S^7)=\pi^4/3$, the normalization can also be written as $\frac{12}{\pi^4}$. Finally, by integrating out the $\varphi$-angle in the measure $d\mu_4$, corresponding to the $U(1)$ fiber, we obtain the Fubini--Study measure on $\mathbb{CP}^3$:
\begin{equation}
d \mu_{\mathbb{CP}^3}
=
(\cos\theta)(\sin ^5\theta)(\cos\xi_1)(\sin ^3\xi_1)(\cos\xi_2)(\sin \xi_2)
\, d\theta \, d\xi_1 \, d\xi_2 \, d\varphi_1 \, d\varphi_2 \, d\varphi_3.
\end{equation}

\subsubsection{SU(N) Generalization and \texorpdfstring{$\mathbf{\mathbb{CP}^{N-1}}$}{CP{N-1}}}

Lastly, considering the general case of $SU(N)$, we can now see the pattern behind the previous examples. Starting from
\begin{equation}
    \tilde{g}\ket{\uparrow}_N= (...) (...)
\begin{array}{ccc}
         \left(\begin{array}{cccc}
1 & 0 & \cdots & 0 \\
0 & & & \\
\vdots & & Y_{N-1} & \\
0 & &
\end{array}\right)
\\
    \end{array} 
    \left(\begin{array}{c}
1 \\
0 \\
\vdots \\
0
\end{array}\right),
\end{equation}
we immediately notice that the matrix $Y_{N-1}\in SU(N-1)$ leaves the reference state invariant. Therefore, as before, these transformations must be interpreted as redundancies in the construction of coherent states. Having this in mind, the redundant coherent states can be obtained recursively as
\begin{equation}
   \ket{\tilde{\mathbf{n}}_N}
   =
   \left(\begin{array}{c}
e^{i \varphi} \cos \theta \\
0 \\
\vdots \\
0
\end{array}\right)
+\sin \theta
\binom{0}{\left|\tilde{\mathbf{n}}_{N-1}\right\rangle}.
\end{equation}
After removing the usual projective $U(1)$ redundancy of quantum states, we can choose the representative whose first component is real and positive. This gives the physical coherent states $\ket{\mathbf n_N}$. Hence, as in the previous cases, one should be a little careful with the quotient. Although one may first say that the redundancies are $SU(N-1)\times U(1)$, the mathematically precise stabilizer of the ray $[\ket{\uparrow}_N]$ inside $SU(N)$ is
\begin{equation}
    S(U(N-1)\times U(1)).
\end{equation}
The symbol $S$ reminds us that the full matrix must still have determinant one. Therefore, the coherent-state manifold is
\begin{equation}
    \boxed{
    \frac{SU(N)}{S(U(N-1)\times U(1))}
    \simeq S^{2N-1}/S^1
    \simeq \mathbb{CP}^{N-1}
    }\: .
\end{equation}
In this way, before taking the $U(1)$ quotient, we work with normalized vectors in $\mathbb C^N$, which form the sphere $S^{2N-1}$. The corresponding measure is
\begin{equation}
\begin{split}
    d \mu_N
    =&\,
    \cos \theta \sin ^{2N-3}\theta \,
    \cos \xi_1 \sin ^{2N-5}\xi_1 \,
    \cos \xi_2 \sin ^{2N-7}\xi_2
    \cdots
    \cos \xi_{N-2}\sin \xi_{N-2}
    \\
    &\hspace{160pt}\cdot
    d \theta \, d \xi_1 \cdots d \xi_{N-2}
    \, d \varphi \, d \varphi_1 \cdots d \varphi_{N-1}.
\end{split}
\end{equation}
Then, the decomposition of the identity is not obtained by simply dividing by the volume of $S^{2N-1}$. By symmetry, the integral over all coherent-state projectors must be proportional to the identity operator, and the trace fixes the proportionality constant:
\begin{equation}
    \int d\mu_N \ \ket{\mathbf n_N}\bra{\mathbf n_N}
    =
    \frac{\operatorname{vol}(S^{2N-1})}{N}\,\mathbb{1}_N .
\end{equation}
Therefore,
\begin{equation}
    \boxed{
    \mathbb{1}_N
    =
    \frac{N}{\operatorname{vol}(S^{2N-1})}
    \int d\mu_N \ \ket{\mathbf n_N}\bra{\mathbf n_N}
    }\:\:,
\end{equation}
where by using $\operatorname{vol}(S^{2N-1})=\frac{2\pi^N}{(N-1)!}$,
the normalization constant can also be written as $\frac{N!}{2\pi^N}$.
Finally, by integrating out the $\varphi$-angle, corresponding to the $U(1)$ fiber, we obtain the Fubini--Study measure on $\mathbb{CP}^{N-1}$:
\begin{equation}
\begin{split}
    d \mu_{\mathbb{CP}^{N-1}}
    =&\,
    \cos \theta \sin ^{2N-3}\theta \,
    \cos \xi_1 \sin ^{2N-5}\xi_1 \,
    \cos \xi_2 \sin ^{2N-7}\xi_2
    \cdots
    \cos \xi_{N-2}\sin \xi_{N-2}
    \\
    &\hspace{180pt}\cdot 
    d \theta \, d \xi_1 \cdots d \xi_{N-2}
    \, d \varphi_1 \cdots d \varphi_{N-1}.
\end{split}
\end{equation}
So, in this normalization, its total volume is $\operatorname{vol}(\mathbb{CP}^{N-1})=\frac{\pi^{N-1}}{(N-1)!}$.

\section{WZ terms, Lagrangian and Classical Equations} \label{sec:suN-wz-lagrangian-eom}
As we already mentioned, following the same procedure as in Sec.\ref{sec:construction_superspin_path_integral}, we can finally find an action that looks like Eq.\ref{eq:spin_action}. We only need to replace the coherent states and adapt the Hamiltonian (see the next section). Therefore, we write the $SU(N)$ superspin action as 
\begin{equation}\label{eq:spin_action_gene}
    S[\mathbf{n}_N(\tau)]=\int_0^\beta d \tau\left(-\braket{\partial_\tau \mathbf{n}_N  | \mathbf{n}_N}+\langle \mathbf{n}_N|\mathbf{B} \cdot \hat{\mathbf{T}}| \mathbf{n}_N\rangle\right).
\end{equation}
Here $\mathbf B\cdot\hat{\mathbf T}$ denotes a general traceless Hermitian single-site Hamiltonian expanded in a chosen basis of $\mathfrak{su}(N)$. The coefficients $B_a$ should therefore be understood as generalized source fields. 
Only after choosing a physical embedding do some of them acquire the interpretation of ordinary magnetic fields, orbital fields, multipolar fields, or other physical couplings.
For the time being, we aim to study the inherent topology coming from the Wess--Zumino term $-\braket{\partial_\tau \mathbf{n}_N  | \mathbf{n}_N}$. Later on, in the next section, we will focus on the second term in order to derive the complete local representation of the $SU(N)$ action.  

\subsection{WZ for SU(3)}

Using Eq.\ref{eq:su3_coherent}, it is easy to compute $-\braket{\partial_\tau \mathbf{n}_3  | \mathbf{n}_3}$. In this way, we find that the \textbf{local representation of the $\bm{SU(3)}$ WZ term} is given by
\begin{equation}
    \boxed{
    S_{WZ}^{(3)}[\theta, \xi, \varphi_1, \varphi_2]
    =
    i\int_{0}^\beta d\tau \ \sin^2\theta 
    \left(\cos^2 \xi \ \dot{\varphi}_1+\sin^2 \xi \ \dot{\varphi}_2\right)
    } \ .
\end{equation}
Notice that if $\xi=0$, the third component of the coherent state is switched off, and we recover the usual $SU(2)$ WZ term, after using $\sin^2\theta=\frac{1-\cos\theta^\prime}{2}$.
Now, to understand the coordinate-patch structure and have an idea of the topology of $\mathbb{CP}^2$, we introduce the homogeneous notation
\begin{equation}
    \left(\begin{array}{c}
\cos \theta \\
\sin \theta \ e^{i\varphi_1} \cos \xi \\
\sin \theta \ e^{i\varphi_2} \sin \xi
\end{array}\right) = \left(\begin{array}{c}
z_0 \\
z_1 \\
z_2
\end{array}\right),
\end{equation}
such that, the standard charts of $\mathbb{CP}^2$ are given by
\begin{equation}
\begin{split}
    U_i&=\left\{\left[z_0:z_1:z_2\right]\in \mathbb{CP}^2 \mid z_i\neq 0\right\},\\
    \Phi_i &: U_i \rightarrow \mathbb{C}^2, \qquad 
    \Phi_i\left(\left[z_0:z_1:z_2\right]\right)
    =
    \left(\frac{z_j}{z_i},\frac{z_k}{z_i}\right),
    \qquad j,k\neq i,\ j\neq k .
\end{split}
\end{equation}
Thus, $\mathbb{CP}^2$ needs three standard coordinate charts to describe its differential structure. This is the natural generalization of the $SU(2)$ case, where $\mathbb{CP}^1\simeq S^2$ cannot be covered by a single chart. However, in $\mathbb{CP}^2$ it is better not to speak literally of ``three poles''. What we have instead are three coordinate patches, each one adapted to the region where one homogeneous coordinate is nonzero. In general, $\mathbb{CP}^{N-1}$ is covered by $N$ such standard charts.

Now, in the present parametrization, we fixed the projective phase by choosing the representative with first component real and positive. Therefore, the expression above corresponds to the chart $U_0$, where
\begin{equation}
    z_0=\cos\theta\neq 0.
\end{equation}
In other words, this local representation is valid away from the region where $\theta=\pi/2$. On the other hand, the point $\theta=0$ does not create a singularity in the WZ term itself: although the angular variables $\varphi_1$ and $\varphi_2$ lose their meaning there, their coefficients vanish, and the expression remains regular. Indeed, rewriting the $SU(3)$ WZ action as 
\begin{equation}
    S_{WZ}^{(3)}[\theta, \xi, \varphi_1, \varphi_2]
    =
    \frac{i}{4}\int_{0}^\beta d\tau 
    \Big[ 
    2\cos^2\xi (1-\cos2\theta) \dot{\varphi}_1 
    + (1-\cos2\theta)(1-\cos2\xi)\dot{\varphi}_2
    \Big],
\end{equation}
we see explicitly that the potentially problematic phase derivatives are harmless at $\theta=0$. This is the analog of what happened in the $SU(2)$ case: the local WZ potential is regular in one patch, while a different local expression is needed in the other patches. Therefore, the expressions in the charts $U_1$ and $U_2$ should be obtainable by choosing different projective gauges and, equivalently, by adding suitable total derivative terms to the local Lagrangian.

Moving on, we would like to specify the topological structure behind this Wess--Zumino term. At first sight, one might be tempted to associate the construction with the full stabilizer group $S(U(2)\times U(1))$, since this is the group that appears in the coset representation of $\mathbb{CP}^2$. However, the WZ term of a coherent state is more specific: it is the Berry phase of a single quantum ray. Therefore, the relevant bundle is not the full non-Abelian stabilizer bundle, but the canonical $U(1)$ line bundle over $\mathbb{CP}^2$. In this language, the local Berry connection is
\begin{equation}
    \mathcal A = -\braket{d\mathbf n_3|\mathbf n_3} =\braket{\mathbf n_3|d\mathbf n_3} \quad \:\:\xRightarrow[]{\:\:\text{WZ term}\:\:}\quad \:\:S^{(3)}_{WZ}=\int_{S^1}\mathcal A \:\:.
\end{equation}
Since the connection is only locally defined, its curvature $\mathcal F=d\mathcal A$ is the globally defined object. This curvature is proportional to the Fubini--Study Kähler form on $\mathbb{CP}^2$, and its first Chern number satisfies
\begin{equation}
    W=\frac{1}{2\pi i}\int_{\Sigma}\mathcal F \:\:\:\in \mathbb Z\:\:,
    \qquad \:\:\Sigma\simeq S^2.
\end{equation}
Thus, the topological classification relevant for the WZ term is $\pi_2(\mathbb{CP}^2)=\mathbb Z$, or equivalently the first Chern class of the canonical $U(1)$ Berry bundle.

\noindent To clarify this part, let us examine in detail each statement that was made before:
\begin{itemize}
    \item \underline{\textbf{U(1) Berry Phase}}\\
    The question now is why the WZ term does not contain a non-Abelian $SU(2)$ connection, even though the coset description of $\mathbb{CP}^2$ involves the stabilizer group $S(U(2)\times U(1))$. The concrete reason is that, in the fundamental representation, the coherent state $g\ket{\uparrow}_3$ only remembers the first column of the matrix $g$. Indeed, writing schematically Eq.\ref{eq:SU(N)_parametrization} for $SU(3)$
    \begin{equation}
    g=\begin{pmatrix}
    1&0\\
    0&X_2
    \end{pmatrix}
    M(\theta,\varphi)
    \begin{pmatrix}
    1&0\\
    0&Y_2
    \end{pmatrix},
    \end{equation}
    we immediately have
    \begin{equation}
    \begin{pmatrix}
    1&0\\
    0&Y_2
    \end{pmatrix}
    \ket{\uparrow}_3
    =
    \ket{\uparrow}_3.
    \end{equation}
    Therefore, the matrix $Y_2$ disappears before taking any derivative in $-\braket{\partial_\tau \mathbf{n}_3  | \mathbf{n}_3}$. In particular, no term of the form $Y_2^{-1}dY_2$ can appear in the Berry connection/WZ term. On the contrary, the remaining $SU(2)$ matrix $X_2$ does act nontrivially, but only through the normalized vector obtained from its first column,
    \begin{equation}
    X_2\begin{pmatrix}1\\0\end{pmatrix}
    =
    \ket{\mathbf{\tilde{n}}_2}.
    \end{equation}
    Thus, the full coherent state has the recursive form
    \begin{equation}\label{eq:select_direction_in_C^3}
    \ket{\mathbf n_3}
    =
    \begin{pmatrix}
    \cos\theta\\
    \sin\theta\,\ket{\mathbf{\tilde{n}}_2}
    \end{pmatrix}\:\:\in\:\mathbb{C}^3.
    \end{equation}
    Consequently,
    \begin{equation}
        \braket{\mathbf n_3|d\mathbf n_3}=\sin^2\theta\,\braket{\mathbf{\tilde{n}}_2|d|\mathbf{\tilde{n}}_2},
    \end{equation}
    where the remaining terms cancel because $\braket{\mathbf n_3|\mathbf n_3}=1$. This object is a scalar one-form, not a matrix-valued one. Hence, the Berry connection entering the WZ term is a $U(1)$ connection.

    So, the deepest reason behind the absence of a non-Abelian contribution is that in the fundamental (and more generally, totally symmetric) representation, the coherent state $g\ket{\uparrow}$ selects a single direction in $\mathbb{C}^3$, i.e. a one-dimensional subspace (Eq.\ref{eq:select_direction_in_C^3}). Therefore, all the information of the state is contained in a single normalized vector, and any transformation acting on the orthogonal complement (such as the $SU(2)$ matrices above -- $\ket{\mathbf{\tilde{n}}_2}$ in Eq.\ref{eq:select_direction_in_C^3}) only serves to complete this vector into a full basis of $\mathbb{C}^3$. Since the WZ term depends only on the evolution of the state itself, and not on how this state is embedded into a larger frame, these extra degrees of freedom necessarily drop out. As a consequence, the Berry connection can only be a scalar one-form, i.e. a $U(1)$ connection.
    
    However, this situation changes dramatically for other representations. In particular, for a rank-$m$ antisymmetric irrep of $SU(N)$, the highest-weight state no longer selects a single direction, but rather an $m$-dimensional subspace of $\mathbb{C}^N$. In that case, the coherent-state construction naturally involves a local $m$-frame, and changes of this frame are redundant. This is precisely what gives rise to a $U(m)$ gauge structure and, consequently, to a \emph{non-Abelian Berry connection}\cite{SachdevRead_Bosonic_1990}. Thus, the absence of the $SU(2)$ contribution in the present case is not accidental, but a direct consequence of the fact that we are describing a one-dimensional object (a line) rather than a higher-dimensional subspace.

   \item \underline{\textbf{Fubini--Study Kähler form}}\\
    Let us now understand the geometric origin of the curvature appearing in the WZ term. For this purpose, we momentarily forget about our specific parametrization and recall a general fact about the geometry of complex projective spaces. On $\mathbb{CP}^2$, one can introduce local coordinates by choosing the chart  
    $U_0=\{z_0\neq 0\}$ and defining the affine coordinates
    \begin{equation}
        w_1=\frac{z_1}{z_0}, \qquad
        w_2=\frac{z_2}{z_0}.
    \end{equation}
    In this patch, every point of $\mathbb{CP}^2$ can be represented by a normalized vector of the form
    \begin{equation}
        \ket{n(w)}=
        \frac{1}{\sqrt{1+|w|^2}}
        \begin{pmatrix}
        1\\
        w_1\\
        w_2
        \end{pmatrix},
        \qquad
        |w|^2=|w_1|^2+|w_2|^2.
    \end{equation}
    This is a completely general parametrization of rays in $\mathbb{C}^3$, independent of our coherent-state construction. From this representation, one defines the \textbf{Fubini--Study Kähler potential}
    \begin{equation}
        K_{FS}=\log(1+|w|^2),
    \end{equation}
    which encodes the intrinsic geometry of $\mathbb{CP}^2$, and whose associated Kähler form is
    \begin{equation}\label{eq:FSK_form_CP2}
        \omega_{FS}
        =
        i\,\partial\bar\partial K_{FS}
        =
        i\,
        \frac{
        (1+|w|^2)\delta_{ab}-\bar w_a w_b
        }{
        (1+|w|^2)^2
        }
        dw_a\wedge d\bar w_b
        \qquad (a,b=1,2).
    \end{equation}
    This object is globally well-defined and captures the curvature of the canonical line bundle over $\mathbb{CP}^2$ (fiber is $\mathbb{C}$ with structure group $U(1)$).
    
    \medskip
    
    Now, we return to our coherent states. The key observation is that our states in Eq.\ref{eq:su3_coherent} are already in this standard form, once we identify
    \begin{equation}
    \begin{aligned}
        w_1=\tan\theta\,\cos\xi\,e^{i\varphi_1}
        \\
        w_2=\tan\theta\,\sin\xi\,e^{i\varphi_2}
    \end{aligned}\quad\: \xRightarrow[]{\qquad}\quad \left(\begin{array}{c}
\cos \theta \\
\sin \theta e^{i \varphi_1} \cos \xi \\
\sin \theta e^{i \varphi_2} \sin \xi
\end{array}\right)=\frac{1}{\sqrt{1+|w|^2}}\left(\begin{array}{c}
1 \\
w_1 \\
w_2
\end{array}\right)
    \end{equation}
    which coincides exactly with our coherent state $\ket{\mathbf n_3}$. Hence, we can compute the Berry connection in full generality, and obtain
    \begin{equation}
        \mathcal A
        =
        \braket{n(w)|d n(w)}
        =
        \frac{1}{2}
        \frac{
        \bar w_a\,dw_a - w_a\,d\bar w_a
        }{
        1+|w|^2
        }.
    \end{equation}
    Then, taking the exterior derivative gives
    \begin{equation}
        \mathcal F = d\mathcal A
        =
        -
        \frac{
        (1+|w|^2)\delta_{ab}-\bar w_a w_b
        }{
        (1+|w|^2)^2
        }
        dw_a\wedge d\bar w_b\:\:,
    \end{equation}
    and comparing with the Fubini--Study Kähler form (Eq.\ref{eq:FSK_form_CP2}), we find
    \begin{equation}
        \boxed{
        \mathcal F = i\,\omega_{FS}
        }
        \qquad\Longleftrightarrow\qquad
        \boxed{
        \omega_{FS}=-i\,\mathcal F
        }\:\:,
    \end{equation}
    which shows that the Berry curvature is the pullback of the Fubini--Study Kähler form on $\mathbb{CP}^2$, while the WZ term is obtained from its associated connection.
\end{itemize}

\subsection{WZ for SU(4)}
Repeating the same calculation as above, we find that the \textbf{local representation of the $\bm{SU(4)}$ WZ term} is given by
\begin{equation}
    \boxed{
    S_{WZ}^{(4)}[\theta,\xi_1,\xi_2,\varphi_1,\varphi_2,\varphi_3]
    =
    i\int_0^\beta d\tau\,
    \sin^2\theta
    \left[
    \cos^2\xi_1\,\dot\varphi_1
    +
    \sin^2\xi_1
    \left(
    \cos^2\xi_2\,\dot\varphi_2
    +
    \sin^2\xi_2\,\dot\varphi_3
    \right)
    \right]
    } .
\end{equation}
This expression is the natural $SU(4)$ analogue of the $SU(3)$ result. It is written in the local chart $U_0\subset\mathbb{CP}^3$, where the first homogeneous coordinate is nonzero, i.e. $z_0=\cos\theta\neq0$. Therefore, the expression is valid away from the region $\theta=\pi/2$. On the other hand, the point $\theta=0$ does not produce a singularity in the WZ term itself, since the angular variables lose their meaning there but their coefficients vanish.

\noindent Recalling that in this case the coherent-state manifold is $\mathbb{CP}^3$, which has complex dimension $3$ and real dimension $6$, we have now four standard coordinate charts,
\begin{equation}
    U_i=\{[z_0:z_1:z_2:z_3]\in\mathbb{CP}^3\mid z_i\neq0\},
    \qquad i=0,1,2,3.
\end{equation}
Therefore, to pass to another chart $U_i$, one fixes the phase so that $z_i$ is real and positive. This corresponds to a local $U(1)$ gauge transformation, under which the WZ term changes by a total derivative.

Moreover, as in the $SU(3)$ case, the WZ term is not associated with the full stabilizer group $S(U(3)\times U(1))$. Although this group appears in the coset description of $\mathbb{CP}^3$, the coherent state in the fundamental representation only defines a single complex ray in $\mathbb C^4$. Therefore, the relevant Berry connection is again the Abelian connection of the canonical $U(1)$ line bundle over $\mathbb{CP}^3$:
\begin{equation}
    \mathcal A
    =
    \braket{\mathbf n_4|d\mathbf n_4},
    \qquad
    S_{WZ}^{(4)}=\int_{S^1}\mathcal A \:\:,
\end{equation}
whose associated Berry curvature is a 2-form proportional to the Fubini--Study Kähler form on $\mathbb{CP}^3$
\begin{equation}
    \mathcal F=d\mathcal A=i\,\omega_{FS}\:.
\end{equation}
Consequently, the relevant topological invariant is still the first Chern class of the canonical Berry line bundle, not a non-Abelian second or third Chern class. In particular, for any embedded two-cycle $\Sigma\simeq\mathbb{CP}^1\subset\mathbb{CP}^3$, one has
\begin{equation}
    c_1
    =
    \frac{1}{2\pi i}\int_{\Sigma}\mathcal F
    =
    \frac{1}{2\pi}\int_{\Sigma}\omega_{FS}
    \in\mathbb Z .
\end{equation}

\subsection{SU(3) Total Lagrangian and Classical Equations}
We complete our discussion by computing the second term on Eq.\ref{eq:spin_action_gene} for the cases of $SU(3)$ and $SU(4)$. Here, we aim to have an intuition about the classical analog of superspin dynamics once we compute the expectation values of the Hamiltonian and the Euler-Lagrange equations derived from the total Lagrangian (WZ+dynamical).  

We start by clarifying that $\mathbf B$ is now an eight-component generalized source field that couples to the eight Gell--Mann matrices $\{\lambda_i\}$ in the Hamiltonian. Therefore, the matrix representation of the Hamiltonian is written as
\begin{equation}
    H=\mathbf{B} \cdot \hat{\bm{\lambda}}= \left(\begin{array}{ccc}
B_3+\frac{B_8}{\sqrt{3}} & B_1-iB_2 & B_4-i B_5 \\
B_1+iB_2 & -B_3+\frac{B_8}{\sqrt{3}} & B_6-iB_7 \\
B_4+iB_5 & B_6+i B_7 & -\frac{2 B_8}{\sqrt{3}}
\end{array}\right) .
\end{equation}
At this point, it is important to stress that this construction is basis-dependent only at the level of notation. The Hamiltonian $H$ is a fixed operator in $\mathrm{End}(\mathbb{C}^3)$, while the decomposition $H=\sum_{a} B_a T_a$ depends on the chosen basis $\{T_a\}$ of $\mathfrak{su}(3)$. If we change to another basis $T'_a = R_{ab} T_b$, the coefficients transform as $B'_a = (R^{-1})_{ab} B_b$, so that the operator $H$ itself remains unchanged. Since the expectation value $\langle \mathbf n_3|H|\mathbf n_3\rangle$ depends only on the operator $H$ and the state $\ket{\mathbf n_3}$, it is invariant under such changes of basis. Therefore, choosing the Gell--Mann matrices is merely a convenient parametrization of a general traceless Hermitian Hamiltonian, and one could equivalently work with the matrix units $\hat g_{ij}$ and Cartan generators introduced above, dipole/quadrupole basis, or any other basis of $\mathfrak{su}(3)$. If $\mathbb C^3$ is interpreted as a spin-$1$ Hilbert space, the same algebra can instead be reorganized into dipolar and quadrupolar sectors, as reviewed in Appendix~\ref{app:su3-multipoles}.

\noindent In this way, we can safely compute the expectation values $\braket{\mathbf{n}_3|\mathbf{B} \cdot \hat{\bm{\lambda}}|\mathbf{n}_3}$ resulting in 
\begin{equation}
    \begin{split}
        \braket{\mathbf{n}_3|\mathbf{B} \cdot \hat{\bm{\lambda}}|\mathbf{n}_3}= \frac{1}{6} \Bigg\{ &
        2\left(3B_3 +\sqrt{3}B_8\right) \cos^2\theta
        + \sin^2\theta \Big[-6B_3 \cos^2\xi +\sqrt{3} B_8 \left(3\cos(2\xi)-1\right) \\ & + 6\sin(2\xi)\big(B_6\cos(\varphi_1-\varphi_2) -B_7\sin(\varphi_1-\varphi_2)\big)
        \Big]
        +6\sin(2\theta)\Big[\cos(\xi)\\ & \ \ \ \ \  \left(B_1\cos\varphi_1 + B_2\sin\varphi_1\right)  + \sin(\xi)\left(B_4\cos\varphi_2 + B_5\sin\varphi_2\right)
        \Big]
        \Bigg\}.
    \end{split}
\end{equation}
The expression also makes clear that the components $B_4,B_5$ play a role analogous to $B_1,B_2$: both pairs mix the first component of the coherent state with the remaining two components. 
In the case of $SU(2)$, the transverse components can be set to zero by a suitable choice of quantization axis, reflecting the fact that the magnetic field transforms as a vector under physical rotations. For the full $SU(3)$ problem, however, the eight Gell--Mann coefficients are not the components of a single physical vector. For example, in a spin-$1$ embedding they split into dipolar and quadrupolar sectors, and therefore one cannot eliminate arbitrary components by a simple spatial rotation.. Instead, we consider a restricted class of Hamiltonians in which only a subset of the couplings is present. In particular, for simplicity, we focus on the case $B_1=B_2=B_4=B_5=0$, which selects a reduced sector of the full $SU(3)$ parameter space while still capturing nontrivial dynamics through the remaining components. In such a case, the dynamical $SU(3)$ action $S_0^{(3)}$ reduces to
\begin{equation}
\boxed{
\begin{aligned}
    S_0^{(3)}[\theta,\xi,\varphi_{1,2}]=\frac{1}{6} \int_0^\beta d\tau \ \Big\{ &
        2\left(3B_3 +\sqrt{3}B_8\right) \cos^2\theta
        + \sin^2\theta \Big[-6B_3 \cos^2\xi \\ & +\sqrt{3} B_8 \left(3\cos(2\xi)-1\right) + 6\sin(2\xi)\\ & \ \ \ \ \ \ \ \ \ \ \ \ \ \big(B_6\cos(\varphi_1-\varphi_2) -B_7\sin(\varphi_1-\varphi_2)\big)
        \Big]
        \Big\}
    \end{aligned}},
\end{equation}
where we see how the components $B_6,B_7$ couple the angles $\varphi_1,\varphi_2$. Thus, having the complete action $S_0^{(3)} + S_{WZ}^{(3)}$, we can impose $\delta S=0$ which results in the following four equations of motion
\begin{equation}
\boxed{
    \left\{ 
    \begin{aligned} 
    &i \big( \dot{\varphi}_1 \cos^2\xi  + \dot{\varphi}_2 \sin^2\xi \big)    = \Big[B_3 (1+\cos^2\xi)+\sqrt{3}B_8\sin^2\xi\Big] + \sin(2\xi) \Big[B_7\sin(\varphi_1-\varphi_2)\\ & \ \ \ \ \ \ \ \ \ \ \ \ \ \ \ \ \ \ \ \ \ \ \ \ \  \ \ \ \ \ \ \ \ \ \ \ \ \ \ \ \ \ \ \ \ \ \  \ \ \ \ \ \ \ \ \  \ \ \ \ \ \ \ \ \ \ \ \ \ \ \ \ \ \ \ \ \ \  -B_6\cos(\varphi_1-\varphi_2)\Big]                     \\
    &i \big(\dot{\varphi}_1 - \dot{\varphi}_2 \big)= \Big[B_3-\sqrt{3}B_8\Big] + 2\cot(2\xi) \Big[B_6\cos(\varphi_1-\varphi_2)-B_7\sin(\varphi_1-\varphi_2)\Big]
    \\
    &i\big(\dot{\theta}\cos\theta \cos\xi  - \dot{\xi} \sin\theta \sin\xi \big)= - \sin\theta \sin\xi \Big[B_7\cos(\varphi_1-\varphi_2)+B_6\sin(\varphi_1-\varphi_2)\Big]
    \\
    &i\big(\dot{\theta}\cos\theta \sin\xi  + \dot{\xi} \sin\theta \cos\xi \big)=\sin\theta \cos\xi \Big[B_7\cos(\varphi_1-\varphi_2)+B_6\sin(\varphi_1-\varphi_2)\Big]
\end{aligned} \right. }
\end{equation}
These equations should be understood as the classical superspin equations of motion on
$\mathbb{CP}^2$. Their nonlinearity is a direct consequence of the curvilinear
coordinates used to parametrize the coherent-state manifold, together with the
coupling between the two relative phases induced by $B_6$ and $B_7$. In this sense, they provide the natural $SU(3)$ generalization of the usual spin-precession equations.

\subsection{SU(4) Total Lagrangian}
Moving on to an abstract $SU(4)$ superspin system, one may proceed exactly as in the $SU(3)$ case and expand the Hamiltonian in any convenient basis of $\mathfrak{su}(4)$. 
The most transparent choice, however, depends on the physical interpretation of the four-dimensional local Hilbert space. 
For a spin-$3/2$ realization, the multipolar basis $15=3+5+7$ is natural; for a spin-orbital or spin-valley realization, the tensor-product basis is natural; and for singlet-triplet physics, an $SO(4)$-adapted organization is often more useful. 
These different possibilities are reviewed in Appendices~\ref{app:su4-multipoles} and~\ref{app:su4-embeddings-parametrizations}.

Consequently, the generalized Gell--Mann expression below should be understood as an abstract $SU(4)$ parametrization of the same Hamiltonian. 
It is algebraically equivalent to the physically adapted bases discussed above, but it does not by itself distinguish spin, orbital, multipolar, singlet-triplet, or channel degrees of freedom. 
The explicit relation between the generalized Gell--Mann and spin-orbital tensor-product bases is given in Appendix~\ref{app:su4-spin-orbital-basis}. 
Having this in mind, in the Gell--Mann basis we have
\begin{equation}
     H=\mathbf{B} \cdot \hat{\bm{\lambda}}= \left(\begin{array}{cccc}
\frac{B_{15}}{\sqrt{6}}+B_3+\frac{B_8}{\sqrt{3}} & B_1-i B_2 & B_4-i B_5 & B_9-iB_{10} \\
B_1+iB_2 & \frac{B_{15}}{\sqrt{6}}-B_3+\frac{B_8}{\sqrt{3}} & B_6-iB_7 & B_{11}-iB_{12} \\
B_4+iB_5 & B_6+iB_7 & \frac{B_{15}}{\sqrt{6}}-\frac{2 B_8}{\sqrt{3}} & B_{13}-i B_{14} \\
B_9+iB_{10} & B_{11}+iB_{12} & B_{13}+iB_{14} & -\sqrt{\frac{3}{2}} B_{15}
\end{array}\right).
\end{equation}
Therefore, after setting $B_1=B_2=B_4=B_5=B_9=B_{10}=0$, we find that the dynamical expectation value is
\begin{equation}
    \begin{split}
        \braket{\mathbf{n}_4|\mathbf{B} \cdot \hat{\bm{\lambda}}|\mathbf{n}_4}=
        \frac{1}{6}\Bigg\{ &
            \Big[6B_3+2\sqrt{3}B_8+\sqrt{6}B_{15}\Big]\cos^2\theta + \sin^2\theta \Big[
            \big(-6B_3+2\sqrt{3}B_8\big) \cos^2\xi_1 \\ & + \sqrt{6}B_{15} \big(\cos(2\xi_1)+2\sin^2\xi_1\cos(2\xi_2)\big) + 4\sin^2\xi_1\cos\xi_2 \\ & \Big\{-\sqrt{3}B_8\cos\xi_2+3\sin\xi_2\big(B_{13}\cos(\varphi_3-\varphi_2)+B_{14}\sin(\varphi_3-\varphi_2)\big)\Big\} \\ & + 12\cos\xi_1 \sin\xi_1 \Big\{ \sin\xi_2 \big( B_{11}\cos(\varphi_3-\varphi_1) + B_{12}\sin(\varphi_3-\varphi_1) \big) \\ & +\cos\xi_2 \big( B_{6}\cos(\varphi_2-\varphi_1) + B_{7}\sin(\varphi_2-\varphi_1)
       \big)  \Big\}
            \Big]
        \Bigg\},
    \end{split}
\end{equation}
where we now highlight the role of $B_6,B_7$ coupling $(\varphi_1,\varphi_2)$, $B_{11},B_{12}$ coupling $(\varphi_1,\varphi_3)$, and $B_{13},B_{14}$ coupling $(\varphi_2,\varphi_3)$. The corresponding Euler--Lagrange equations can be obtained in the same way as for $SU(3)$. They describe classical motion on $\mathbb{CP}^3$ and are more cumbersome because the three independent relative phases are coupled by the off-diagonal components of the Hamiltonian. For this reason, we do not display them explicitly here.

%% file: appendix1_multipolar_generators.tex
\appendix

\section{Multipolar generator basis for \texorpdfstring{$\bm{SU(N)}$}{SU(N)}}
\label{app:suN-multipoles}
In Sec.~\ref{sec:multipolar_su_n}, we used the fact that the fundamental representation space of $SU(N)$ can be identified with the Hilbert space of a spin $S=(N-1)/2$, namely $\mathbb C^N\cong \mathcal H_S$. This identification allows one to organize the operator algebra acting on $\mathbb C^N$ according to its transformation properties under the embedded spin-rotation subgroup $SU(2)\subset SU(N)$.

\noindent More precisely, the space of operators acting on $\mathcal H_S$ decomposes into irreducible $SU(2)$ tensor sectors $\mathcal M_k$, with $k=0,\ldots,2S$. The $k=0$ sector corresponds to the identity operator, while the traceless part gives the Lie algebra of $SU(N)$. Hence,
\begin{equation}\label{eq:decomposition_suN}
\mathfrak{su}(N)=\bigoplus_{k=1}^{N-1}\mathcal M_k .
\end{equation}
In this way, the generators of $SU(N)$ may be grouped into dipolar, quadrupolar, octupolar, and higher multipolar sectors. Since $\dim\mathcal M_k=2k+1$, this decomposition accounts for all $N^2-1$ generators through $N^2-1=3+5+7+\cdots+(2N-1)$.

Strictly speaking, the construction begins with an embedded Lie algebra $\mathfrak{su}(2)\subset\mathfrak{su}(N)$, so that one could tentatively think that the global image of the corresponding embedding would depend on the spin-$S$ representation: isomorphic to $SU(2)$ for half-integer $S$ and to $SO(3)$ for integer $S$. Nevertheless, the induced conjugation action on $\operatorname{End}(\mathcal H_S)$ always factors through $SO(3)$, since the  elements $g$ and $-g$ act identically on every operator $A\in\operatorname{End}(\mathcal H_S)$,
\begin{equation}
    U_S(-g)A\,U_S(-g)^\dagger
    =
    U_S(g)A\,U_S(g)^\dagger.
\end{equation}
Consequently, all operator sectors $\mathcal M_k$ carry genuine integer-rank representations of $SO(3)$, independently of whether the states in $\mathcal H_S$ have integer or half-integer spin.

Now, the purpose of this appendix is to make the decomposition in Eq.\ref{eq:decomposition_suN} explicit. In the main text, we used the abstract spherical-tensor notation $T_q^{(k)}$, which is the most convenient language for discussing transformation properties under rotations. Here, instead, we present a real Cartesian multipolar basis, denoted by $\mu_a$. This basis is not identical to the spherical basis $T_q^{(k)}$, but it spans the same irreducible subspaces $\mathcal M_k$. Equivalently, the two descriptions are related by a change of basis inside each multipolar sector.

\noindent Throughout this appendix, $S_x,S_y,S_z$ denote the generators of the embedded spin-$S$ representation of $SU(2)$, with $S=(N-1)/2$. The multipolar generators $\mu_a$ are chosen with the normalization
\begin{equation}
\operatorname{Tr}(\mu_a\mu_b)=2\delta_{ab}.
\end{equation}
This is the same normalization as the standard Gell-Mann matrices. If one instead wants to use generators normalized as $\operatorname{Tr}(T^aT^b)=\delta^{ab}/2$, one simply defines $T^a=\mu_a/2$. We now list the first few cases explicitly, beginning with $SU(2)$, $SU(3)$, and $SU(4)$, and then summarize the general $SU(N)$ normalization structure.

The construction works as follows. Once the fundamental space $\mathbb C^N$ is identified with the spin-$S$ Hilbert space $\mathcal H_S$, with $S=(N-1)/2$, we can use the usual spin matrices $S_x,S_y,S_z$ as building blocks for the whole algebra $\mathfrak{su}(N)$. The rank-$1$ sector is generated by operators linear in $S_x,S_y,S_z$, and corresponds to the usual dipolar degrees of freedom. The rank-$2$ sector is generated by traceless quadratic combinations of the spin operators, and corresponds to quadrupoles. Similarly, the rank-$3$ sector is generated by traceless cubic combinations and corresponds to octupoles, and so on. In this way, the multipolar basis is constructed from symmetrized polynomials in the spin operators. The word ``symmetrized'' is important because the spin components do not commute. For instance, products such as $S_xS_y$ must be replaced by the Hermitian combination $S_xS_y+S_yS_x$. This guarantees that the resulting generators are Hermitian, as required for a basis of $\mathfrak{su}(N)$.

\subsection{\texorpdfstring{$\bm{SU(2)}$}{SU(2)}: dipolar sector}
\label{app:su2-multipoles}

The first example is $SU(2)$. In this case $N=2$, so $S=1/2$, and the algebra contains only the rank-$1$ sector. Therefore, there are only three generators, corresponding to the usual spin dipoles. With the normalization $\operatorname{Tr}(\mu_a\mu_b)=2\delta_{ab}$, one may choose
\begin{equation}
\boxed{\mu_1=2S_x,\qquad \mu_2=2S_y,\qquad \mu_3=2S_z}\:.
\end{equation}
For $S=1/2$, these are precisely the Pauli matrices. Thus, in the simplest case, the multipolar construction reduces to the ordinary $SU(2)$ spin algebra.

\subsection{\texorpdfstring{$\bm{SU(3)}$}{SU(3)}: dipolar and quadrupolar sectors}
\label{app:su3-multipoles}

The first genuinely multipolar case is $SU(3)$ \cite{Sutherland1975}. Here $N=3$, so $S=1$, and the eight generators split as $8=3+5$. The first three generators form the dipolar sector $\mathcal M_1$, while the remaining five form the quadrupolar sector $\mathcal M_2$. For the dipolar sector, the construction is immediate: the three spin matrices themselves transform as a vector under the embedded $SU(2)$. Thus, in the spin-1 representation relevant for $SU(3)$, one may choose
\begin{equation}
\boxed{\mu_1=S_x, \quad \mu_2=S_y, \quad \mu_3=S_z }\:.
\end{equation}
The absence of an additional numerical factor, compared with the $\operatorname{SU}(2)$ case where $\mu_a=2 S_a$, is simply a consequence of the normalization convention. In general, the spin- $S$ matrices satisfy $\operatorname{Tr}\left(S_a S_b\right)=\frac{1}{3} S(S+1)(2 S+1) \delta_{a b}$. Therefore, if the dipolar generators are written as $\mu_a=\alpha_1(N) S_a$, the condition $\operatorname{Tr}\left(\mu_a \mu_b\right)=2 \delta_{a b}$ fixes
\begin{equation}\label{eq:dipole_normalization}
\alpha_1(N)=\sqrt{\frac{24}{N\left(N^2-1\right)}} .
\end{equation}
For $SU(2)$, this gives $\alpha_1(2)=2$, while for $SU(3)$ it gives $\alpha_1(3)=1$. Hence the normalized dipolar generators in the $SU(3)$ case are exactly the spin-1 matrices themselves.

Now, to construct the quadrupolar sector explicitly, we start from the products of two spin operators. However, an arbitrary product $S_a S_b$, with $a, b=x, y, z$, is not yet a pure quadrupole. Since the spin components do not commute, such a product naturally contains three different pieces: a scalar part, an antisymmetric part, and a symmetric traceless part. More explicitly,
\begin{equation}
S_a S_b=\frac{1}{3} \delta_{a b} \mathbf{S}^2+\left[\frac{1}{2}\left\{S_a, S_b\right\}-\frac{1}{3} \delta_{a b} \mathbf{S}^2\right]+\frac{1}{2}\left[S_a, S_b\right] .
\end{equation}
The first term is proportional to $\mathbf{S}^2$, and therefore belongs to the scalar sector. The last term is antisymmetric in the indices $a, b$. Using the spin commutation relations $\left[S_a, S_b\right]=i \epsilon_{a b c} S_c$, we see that this part is again proportional to a spin operator, and hence belongs to the dipolar sector. Therefore, the genuinely new rank-2 object is the symmetric traceless combination:
\begin{equation}\label{eq:quadropole_expression}
Q_{a b}=\frac{1}{2}\left\{S_a, S_b\right\}-\frac{1}{3} \delta_{a b} \mathbf{S}^2 .
\end{equation}
This object is symmetric, $Q_{a b}=Q_{b a}$, and traceless. Therefore, it can be represented as the operator-valued matrix
$$
Q=\left(\begin{array}{lll}
Q_{x x} & Q_{x y} & Q_{x z} \\
Q_{x y} & Q_{y y} & Q_{y z} \\
Q_{x z} & Q_{y z} & Q_{z z}
\end{array}\right), \quad Q_{x x}+Q_{y y}+Q_{z z}=0 .
$$
The symmetry condition leaves six independent entries, and the traceless condition removes one of them. Hence, only five independent components remain. These five components are precisely the quadrupolar generators. It is important to clarify the meaning of the indices $a, b=x, y, z$. They are not labels of the $SU(3)$ generators themselves. Rather, they are vector indices associated with the embedded spin-rotation subgroup $SU(2) \subset SU(3)$. Each entry $Q_{a b}$ is still a $3 \times 3$ operator acting on the spin-1 Hilbert space $\mathcal{H}_1$, but the pair of indices $(a, b)$ tells us how this operator transforms under spin rotations. More explicitly, if $U_R\in SU(2)$ is the spin-1 representation of a rotation $R \in SO(3)$, then
\begin{equation}
\left[Q \longrightarrow R Q R^T\right]\qquad U_R Q_{a b} U_R^{\dagger}=\sum_{c, d} R_{a c} R_{b d} Q_{c d}\:.
\end{equation}
This transformation preserves both properties: if $Q$ is symmetric and traceless before the rotation, it remains symmetric and traceless after the rotation. Thus, the five-dimensional space of symmetric traceless tensors is closed under rotations, and this is precisely the rank-2 irreducible sector $\mathcal{M}_2$.

\noindent A convenient real Cartesian basis for this five-dimensional space is then
\begin{equation}
\boxed{\:\begin{aligned}
\mu_4&=S_x^2-S_y^2,\\
\mu_5&=S_xS_y+S_yS_x,\\
\mu_6&=S_xS_z+S_zS_x,\\
\mu_7&=S_yS_z+S_zS_y,\\
\mu_8&=\frac{1}{\sqrt{3}}\left(3S_z^2-\mathbf S^2\right).\:
\end{aligned}}
\end{equation}
In terms of $Q_{a b}$, these operators correspond to the off-diagonal components $\mu_5=2 Q_{x y}$, $\mu_6=2 Q_{x z}$, $\mu_7=2 Q_{y z}$, and the two diagonal traceless combinations may be chosen as $\mu_4=Q_{x x}-Q_{y y}$, $\mu_8=\sqrt{3} Q_{z z}$. Equivalently, the full tensor $Q$ can be reconstructed from the five generators as
$$
Q=\left(\begin{array}{ccc}
\frac{1}{2}\left(\mu_4-\frac{\mu_8}{\sqrt{3}}\right) & \frac{\mu_5}{2} & \frac{\mu_6}{2} \\
\frac{\mu_5}{2} & -\frac{1}{2}\left(\mu_4+\frac{\mu_8}{\sqrt{3}}\right) & \frac{\mu_7}{2} \\
\frac{\mu_6}{2} & \frac{\mu_7}{2} & \frac{\mu_8}{\sqrt{3}}
\end{array}\right) .
$$

\noindent This also explains the relation with the spherical tensor notation used in the main text. The operators $\mu_4, \ldots, \mu_8$ form a real Cartesian basis of the quadrupolar sector $\mathcal{M}_2$, whereas the operators $T_q^{(2)}$, with $q=-2, \ldots, 2$, form a spherical basis of the same space. The spherical basis is obtained by taking complex linear combinations of the Cartesian one. Schematically,
\begin{equation}
T_{ \pm 2}^{(2)} \sim \mu_4 \pm i \mu_5, \quad T_{ \pm 1}^{(2)} \sim \mu_6 \pm i \mu_7, \quad T_0^{(2)} \sim \mu_8,
\end{equation}
up to conventional signs and normalization factors. Equivalently, one may generate the same operators through the highest-weight construction. The highest component of the rank-2 tensor is proportional to $S_{+}^2$. Since
\begin{equation}
S_{+}^2=\left(S_x+i S_y\right)^2=S_x^2-S_y^2+i\left(S_x S_y+S_y S_x\right)
\end{equation}
we see explicitly that the $q=2$ spherical quadrupole is the complex combination of $\mu_4$ and $\mu_5$. Repeated commutators with $S_{-}$ then generate the remaining components: the $q= \pm 1$ components are related to $\mu_6,\: \mu_7$, while the $q=0$ component is related to $\mu_8$. Thus, the five Cartesian quadrupoles above are the real form of the rank-2 spherical tensor multiplet introduced in the main text.

As a final consistency check, let us record the quadratic identities satisfied by the two multipolar blocks. Since the first three generators are simply the spin-1 matrices, their squared sum is fixed by the ordinary $SU(2)$ Casimir:
\begin{equation}
\sum_{a=1}^3 \mu_a^2=S_x^2+S_y^2+S_z^2=\mathbf{S}^2=S(S+1) \hat{\mathbb{1}}=2\: \hat{\mathbb{1}} .
\end{equation}
On the other hand, the full set $\left\{\mu_a\right\}_{a=1}^8$ is just a basis of $\mathfrak{su}(3)$ in the fundamental representation, with the Gell-Mann normalization $\operatorname{Tr}\left(\mu_a \mu_b\right)=2 \delta_{a b}$. Therefore, its quadratic Casimir is fixed:
\begin{equation}
\sum_{a=1}^8 \mu_a^2=\frac{16}{3} \hat{\mathbb{1}}=
4c_2([1,0])\:\hat{\mathbb{1}}\:.
\end{equation}
Note that this value is the same quadratic Casimir that follows from the Dynkin-label classification from Sec.\ref{Sec:su_n_representation}. In the standard convention $T^a=\mu_a/2$, one has $\operatorname{Tr}(T^aT^b)=\delta^{ab}/2$, and the fundamental irrep of $SU(3)$ has Dynkin label $[1,0]$. Its quadratic Casimir is $c_2([1,0])=\frac{4}{3}$. Now, since the present basis uses $\mu_a=2T^a$, the Casimir is multiplied by a factor of four. 

\noindent In this way, the quadrupolar contribution is then obtained by subtracting the dipolar block from the full $SU(3)$ Casimir:
\begin{equation}
    \sum_{a=4}^8 \mu_a^2=\frac{16}{3} \hat{\mathbb{1}}-2 \hat{\mathbb{1}}=\frac{10}{3} \hat{\mathbb{1}}\: .
\end{equation}
Thus, in the multipolar basis, the fixed quadratic Casimir of the fundamental $SU(3)$ representation decomposes as
\begin{equation}
    \underbrace{\sum_{a=1}^8 \mu_a^2}_{SU(3) \text { Casimir }}=\underbrace{\sum_{a=1}^3 \mu_a^2}_{\text {dipolar block }}+\underbrace{\sum_{a=4}^8 \mu_a^2}_{\text {quadrupolar block }}=2 \hat{\mathbb{1}}+\frac{10}{3} \hat{\mathbb{1}} \:.
\end{equation}
These relations should not be interpreted as additional labels of the representation, since the representation has already been fixed to be the fundamental. Rather, they are operator identities inside this fixed representation. They express how the total quadratic Casimir is distributed between the dipolar sector $\mathcal{M}_1$ and the quadrupolar sector $\mathcal{M}_2$. For this reason, we will refer to them as \emph{block-Casimir identities}.

\subsection{\texorpdfstring{$\bm{SU(4)}$}{SU(4)}: dipolar, quadrupolar and octupolar sectors}
\label{app:su4-multipoles}
Here, the fundamental representation space $\mathbb{C}^4$ can be identified with the spin $S=3/2$ Hilbert space $\mathcal{H}_{3/2}$. The traceless algebra now decomposes as $\mathfrak{su}(4)=\mathcal{M}_1 \oplus \mathcal{M}_2 \oplus \mathcal{M}_3$, and the fifteen generators split into $15=3+5+7$ corresponding respectively to dipolar, quadrupolar, and octupolar sectors. As in the previous case, the dipolar sector is generated by the spin matrices themselves, with normalization following Eq.\ref{eq:dipole_normalization},
\begin{equation}
    \boxed{\:\mu_1=\sqrt{\frac{2}{5}} S_x, \quad \mu_2=\sqrt{\frac{2}{5}} S_y, \quad \mu_3=\sqrt{\frac{2}{5}} S_z\: }\:\:.
\end{equation}

Similarly, the quadrupolar sector is obtained exactly as in the $SU(3)$ case. Namely, one uses the same symmetric traceless tensor $Q$ (Eq.\ref{eq:quadropole_expression}).
The tensor structure is therefore unchanged: $Q_{ab}$ is still the rank-$2$ irreducible tensor of the embedded $SU(2)$. What changes is the representation in which the spin matrices are evaluated. In the present case, $S_a$ are spin-$3/2$ matrices acting on $\mathcal H_{3/2}$, and therefore the trace norm of the quadrupolar operators differs from the spin-$1$ case.

\noindent For general $N$, the normalization factor for the rank-$2$ sector is fixed by imposing again the convention $\operatorname{Tr}(\mu_a\mu_b)=2\delta_{ab}$. This gives
\begin{equation}\label{eq:quadrupole_normalization}
\alpha_2(N)=\sqrt{\frac{5!}{N(N^2-1)(N^2-4)}}\:.
\end{equation}
This is the quadrupolar analogue of the dipolar normalization factor $\alpha_1(N)$. For $N=3$, one finds $\alpha_2(3)=1$, which is why no additional factor appeared in the $SU(3)$ quadrupoles. For $N=4$, instead, $\alpha_2(4)=\frac{1}{\sqrt 6}$. Therefore, using the same Cartesian basis as before, the normalized quadrupolar generators in the $SU(4)$ case are
\begin{equation}
\boxed{\:\begin{aligned}
\mu_4&=\frac{1}{\sqrt6}\left(S_x^2-S_y^2\right),\\
\mu_5&=\frac{1}{\sqrt6}\left(S_xS_y+S_yS_x\right),\\
\mu_6&=\frac{1}{\sqrt6}\left(S_xS_z+S_zS_x\right),\\
\mu_7&=\frac{1}{\sqrt6}\left(S_yS_z+S_zS_y\right),\\
\mu_8&=\frac{1}{\sqrt6}\frac{1}{\sqrt3}\left(3S_z^2-\mathbf S^2\right).
\end{aligned}\:}
\end{equation}
Thus, the five quadrupolar generators of $SU(4)$ have exactly the same form as the $SU(3)$ quadrupoles, but multiplied by the new normalization factor.

The genuinely new sector in $SU(4)$ is the octupolar sector $\mathcal M_3$. By analogy with the quadrupoles, which were obtained from quadratic products of spin matrices, the octupoles are obtained from their cubic products. However, just as an arbitrary product $S_aS_b$ is not a pure quadrupole, an arbitrary cubic product $S_aS_bS_c$ is not automatically a pure octupole. It also contains lower-rank pieces. Therefore, the rank-$3$ sector is obtained by taking the fully symmetric traceless part of such products. 

\noindent More explicitly, one introduces an operator-valued rank-$3$ tensor
\begin{equation}
O_{abc}
=
S_{(a}S_bS_{c)}
-
\frac{3\mathbf S^2-\hat{\mathbb 1}}{15}
\left(
\delta_{ab}S_c+\delta_{ac}S_b+\delta_{bc}S_a
\right),
\end{equation}
where $S_{(a}S_bS_{c)}$ denotes the fully symmetrized product,
\begin{equation}
S_{(a}S_bS_{c)}
=
\frac{1}{6}
\sum_{\pi\in S_3}
S_{\pi(a)}S_{\pi(b)}S_{\pi(c)} .
\end{equation}
The subtraction term removes the trace, and therefore removes the lower-rank dipolar component contained in the cubic product. Hence $O_{abc}$ is symmetric in all its indices and traceless in the sense that
\begin{equation}
O_{abc}=O_{(abc)},
\qquad
\sum_a O_{aab}=0 .
\end{equation}
This is the direct analog of the quadrupolar tensor $Q_{ab}$, but now for rank $3$. The counting also works in the expected way. A fully symmetric rank-$3$ tensor in three dimensions has $\binom{3+3-1}{3}=10$
independent components, but the tracelessness condition removes three of them, leaving $7$ independent components. These seven components form precisely the octupolar sector $\mathcal M_3=\operatorname{span}\{\mu_9,\ldots,\mu_{15}\}$. Furthermore, under a spin rotation, the tensor transforms as
\begin{equation}
U_R O_{abc}U_R^\dagger
=
\sum_{d,e,f}
R_{ad}R_{be}R_{cf}O_{def}\:,
\end{equation}
and, therefore, the seven-dimensional space of symmetric traceless rank-$3$ tensors is closed under rotations and realizes the rank-$3$ irreducible representation of the embedded $SU(2)$.

\noindent Finally, the normalization follows the same logic as before. For general $N$, the rank-$3$ normalization factor is fixed by imposing the convention $\operatorname{Tr}(\mu_a\mu_b)=2\delta_{ab}$, giving
\begin{equation}\label{eq:octupole_normalization}
\alpha_3(N)
=
\sqrt{
\frac{6!}
{N(N^2-1)(N^2-4)(N^2-9)}
}\:.
\end{equation}
Hence, for $N=4$, $\alpha_3(4)=\frac{1}{\sqrt 7}$, and a convenient choice for the seven generators $\mu_9,\ldots,\mu_{15}$ is given by
\begin{equation}
\boxed{
\begin{aligned}
\mu_9&=
\frac{1}{6}
\left[
2\left\{S_x,S_x^2-S_y^2\right\}
-\left\{S_z,\left\{S_x,S_z\right\}\right\}
+\left\{S_x,3S_z^2-\mathbf S^2\right\}
\right],
\\[3pt]
\mu_{10}
&=
\frac{1}{6}
\left[
2\left\{S_y,S_x^2-S_y^2\right\}
+\left\{S_z,\left\{S_y,S_z\right\}\right\}
-\left\{S_y,3S_z^2-\mathbf S^2\right\}
\right],
\\[3pt]
\mu_{11}
&=
\frac{1}{\sqrt6}
\left\{S_z,S_x^2-S_y^2\right\},
\\[3pt]
\mu_{12}
&=
\frac{1}{\sqrt6}
\left\{S_z,\left\{S_x,S_y\right\}\right\},
\\[3pt]
\mu_{13}
&=
\frac{1}{2\sqrt{15}}
\left[
\left\{S_z,\left\{S_x,S_z\right\}\right\}
+
\left\{S_x,3S_z^2-\mathbf S^2\right\}
\right],
\\[3pt]
\mu_{14}
&=
\frac{1}{2\sqrt{15}}
\left[
\left\{S_z,\left\{S_y,S_z\right\}\right\}
+
\left\{S_y,3S_z^2-\mathbf S^2\right\}
\right],
\\[3pt]
\mu_{15}
&=
\frac{\sqrt{10}}{3}
\left[
S_z^3
-
\frac{1}{5}
\left(
3\mathbf S^2-\hat{\mathbb 1}
\right)S_z
\right]\:,
\end{aligned}
}
\end{equation}
where $\{A,B\}=AB+BA$, and all spin operators are evaluated in the spin-$3/2$ representation. 

As a last consistency check, let us record the block-Casimir identities for the $SU(4)$ basis. In the standard normalization $T^a=\mu_a/2$, the fundamental irrep $[1,0,0]$ has $c_2([1,0,0])=15/8$. Therefore, in the present $\mu_a$-normalization,
\begin{equation}
\sum_{a=1}^{15}\mu_a^2
=
4c_2([1,0,0])\hat{\mathbb 1}
=
\frac{15}{2}\hat{\mathbb 1}.
\end{equation}
This total Casimir decomposes into the dipolar, quadrupolar, and octupolar blocks as
\begin{equation}
\sum_{a=1}^{3}\mu_a^2=\frac{3}{2}\hat{\mathbb 1},
\qquad
\sum_{a=4}^{8}\mu_a^2=\frac{5}{2}\hat{\mathbb 1},
\qquad
\sum_{a=9}^{15}\mu_a^2=\frac{7}{2}\hat{\mathbb 1}.
\end{equation}
Thus, $\frac{15}{2}=\frac{3}{2}+\frac{5}{2}+\frac{7}{2}$, showing how the fixed quadratic Casimir of the fundamental $SU(4)$ representation is distributed among the three multipolar sectors.

\subsection{\texorpdfstring{$\bm{SU(N)}$}{SU(N)}: general multipolar sectors}
\label{app:sun_general-multipoles}
We now extract the general pattern behind the explicit $SU(2), SU(3)$, and $SU(4)$ constructions. Having the decomposition of $\mathfrak{su}(N)$ into irreducible multipolar sectors $\mathfrak{su}(N)=\bigoplus_{k=1}^{N-1} \mathcal{M}_k$,
where $\mathcal{M}_k$ is the rank-$k$ multipolar sector of dimension $2 k+1$, in the Cartesian basis used in this appendix, the generators are ordered block by block as
\begin{equation}\label{eq:multipolar_basis_by_sectors}
\mathcal{M}_k=\operatorname{span}\left\{\mu_{k^2}, \mu_{k^2+1}, \ldots, \mu_{(k+1)^2-1}\right\} \qquad [k=1, \ldots, N-1]\: .
\end{equation}
Thus, $\mathcal{M}_1=\operatorname{span}\left\{\mu_1, \mu_2, \mu_3\right\}$, $ \mathcal{M}_2=\operatorname{span}\left\{\mu_4, \ldots, \mu_8\right\}$, $\mathcal{M}_3=\operatorname{span}\left\{\mu_9, \ldots, \mu_{15}\right\}$, and so on. At this point, an important remark should be kept in mind. As it was introduced in Sec.\ref{Sec:su_n_representation}, a natural basis for the $\mathfrak{h}_{\mathfrak{su}(N)}$ Cartan algebra of $\mathfrak{su}(N)$ is given in terms of the diagonal matrices
$H_i=\frac{1}{2}\left(g_{i i}-g_{i+1, i+1}\right)$, making it well adapted to the computation of roots, weights, and Dynkin labels. Now, since in the present multipolar basis (Eq.\ref{eq:multipolar_basis_by_sectors}), the last element of each multipolar block, $\mu_{(k+1)^2-1}$, is the $q=0$ component of the rank-$k$ tensor, and therefore diagonal in the $S_z$ basis, the collection $\set{\mu_3, \mu_8, \mu_{15}, \mu_{24}, \ldots, \mu_{N^2-1}}$ of all these block-last elements form a new basis for the Cartan algebra but adapted to the multipolar decomposition. In other words, such a set spans the same Cartan subalgebra as the standard generators $H_i$, but it corresponds to a different basis:
\begin{equation}
    \mathfrak{h}_{\mathfrak{su}(N)}\:=\:\operatorname{span}\left\{H_1, \ldots, H_{N-1}\right\}=\operatorname{span}\left\{\mu_3, \mu_8, \mu_{15}, \ldots, \mu_{N^2-1}\right\}
\end{equation}
So, the two bases organize the algebra differently, but they span the same diagonal subalgebra. Importantly, the diagonal (q=0) generator in each multipolar block can be viewed as a traceless polynomial in $S_z$, such that, imposing $\operatorname{Tr}(\mu_a\mu_b)=2\delta_{ab}$ on these diagonal representatives gives the same normalization coefficients $\alpha_k(N)$ quoted below. In the following, we collect the main general properties of such a multipolar basis.
\begin{itemize}
    \item \underline{Block-Casimir identities}\\
    For each multipolar block $\mathcal M_k$, define the quadratic operator
    \begin{equation}
    C_k^{\mathrm{block}}
    =
    \sum_{a=k^2}^{(k+1)^2-1}\mu_a^2 .
    \end{equation}
Since $\mathcal M_k$ is an irreducible rank-$k$ tensor sector under the embedded $SU(2)$, the sum $C_k^{\mathrm{block}}$ is an $SU(2)$-scalar. Therefore, in the irreducible spin-$S=(N-1)/2$ Hilbert space, Schur's lemma implies $C_k^{\mathrm{block}}=c_k(N)\hat{\mathbb 1}$.
The constant is fixed immediately by taking the trace. Using $\operatorname{Tr}(\mu_a\mu_b)=2\delta_{ab}$ and $\dim\mathcal M_k=2k+1$, one obtains
    \begin{equation}
    Nc_k(N)
    =
    \operatorname{Tr}C_k^{\mathrm{block}}
    =
    2(2k+1).
    \end{equation}
Hence,
\begin{equation}
\boxed{
\sum_{a=k^2}^{(k+1)^2-1}\mu_a^2
=
\frac{2(2k+1)}{N}\hat{\mathbb 1},
\qquad
k=1,\ldots,N-1.
}
\end{equation}
This is what we call the block-Casimir identity. It should not be confused with an independent Casimir of the full $SU(N)$ algebra: the block $\mathcal M_k$ is not itself a Lie subalgebra. Rather, the identity tells us how the quadratic Casimir of the fundamental representation is distributed among the irreducible multipolar sectors. Indeed, summing over all blocks gives
\begin{equation}
\sum_{a=1}^{N^2-1}\mu_a^2
=
\frac{2}{N}
\sum_{k=1}^{N-1}(2k+1)\hat{\mathbb 1}
=
\frac{2(N^2-1)}{N}\hat{\mathbb 1}.
\end{equation}

\item \underline{Normalization coefficients}\\
The next general ingredient is the normalization coefficient multiplying each rank-$k$ multipolar block. As in the explicit $SU(3)$ and $SU(4)$ examples, the rank-$k$ generators are obtained from degree-$k$ symmetrized products of spin matrices, with all lower-rank traces subtracted. Schematically,
\begin{equation}
    \mathcal M_k
    \sim
    \operatorname{SymTraceless}
    \left[
    S_{a_1}S_{a_2}\cdots S_{a_k}
    \right].
\end{equation}
However, the trace norm of such a polynomial depends on the spin representation in which the matrices are evaluated. Since $S=\frac{N-1}{2}$, the same formal spin polynomial has different normalization for different values of $N$. Therefore, for every rank $k$, one needs an $N$-dependent coefficient $\alpha_k(N)$ such that the final generators satisfy $\operatorname{Tr}(\mu_a\mu_b)=2\delta_{ab}$. The general result may be written in the compact factorial form
\begin{equation}
    \alpha_k(N)
    =
    \sqrt{
    \frac{(k+3)!(N-k-1)!}{(N+k)!}
    },
    \qquad
    1\leq k\leq N-1.
\end{equation}
Equivalently, and more transparently for the first few multipolar sectors,
\begin{equation}\label{eq:general_alpha_k}
\boxed{
    \alpha_k(N)=\sqrt{\frac{(k+3)!}{ N(N^2-1)(N^2-4)\cdots(N^2-k^2)}
    },
    \qquad
    1\leq k\leq N-1.
}
\end{equation}
For $k=1,2,3$, this formula reproduces the normalization factors used above for the dipolar, quadrupolar, and octupolar sectors, respectively (see Eqs.~\ref{eq:dipole_normalization}, \ref{eq:quadrupole_normalization}, \ref{eq:octupole_normalization}).

\item \underline{Large-N scaling}\\
From Eq.\ref{eq:general_alpha_k}, for fixed multipole $k\ll N$, the denominator scales as $N\:\prod_{j=1}^k(N^2-j^2)$ $\sim N^{2k+1}$, and therefore  
\begin{equation}\label{eq:alpha_large_N_scaling}
\boxed{
    \alpha_k(N)
    \sim
    \sqrt{(k+3)!}\,N^{-k-\frac{1}{2}},
    \qquad
    N\gg k .
}
\end{equation}
In this line, higher-rank multipoles require increasingly strong normalization factors as $N$ grows. However, this scaling should be understood as a normalization statement. A rank-$k$ multipole is constructed from a degree-$k$ spin polynomial, and since $S=\tfrac{(N-1)}{2}$, such a polynomial grows roughly as $S^k\sim N^k$. Moreover, the trace norm involves a sum over $N$ states, so roughly $\sqrt{\operatorname{Tr}(S^{2k})}\sim N^{k+1/2}$. Hence, the coefficient $\alpha_k(N)\sim N^{-k-1/2}$ precisely compensates this growth, ensuring that the final $SU(N)$ generators satisfy the fixed normalization $\operatorname{Tr}(\mu_a\mu_b)=2\delta_{ab}$. The scaling of the ratios $\alpha_k(N)/\alpha_1(N)$ is shown in Fig.\ref{fig:alpha_scaling}, illustrating the rapid decrease of higher-rank normalization factors relative to the dipolar one. 
\begin{figure}[h!]
    \centering
    \includegraphics[width=0.6\linewidth]{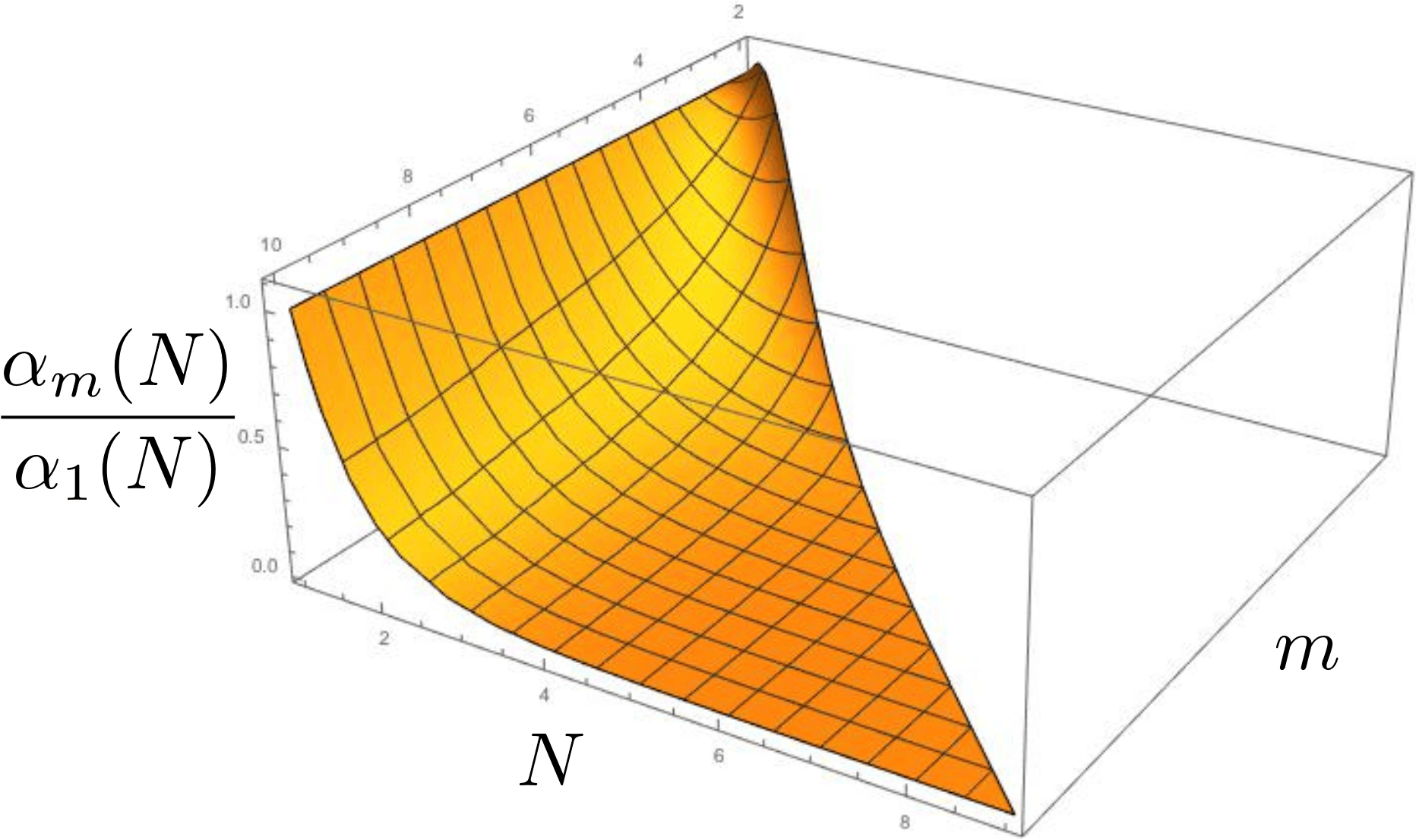}
    \caption{
    Scaling of the normalization coefficients $\alpha_k(N)$ relative to the dipolar coefficient $\alpha_1(N)$. Higher-rank multipolar sectors are increasingly suppressed by the trace normalization as $N$ grows.
    }
    \label{fig:alpha_scaling}
\end{figure}

Here, it is important to note that the role of $\alpha_k(N)$ is to remove this representation-dependent growth. It should therefore not be interpreted by itself as a statement about the dynamical or RG relevance of a multipolar interaction; that depends on the low-energy theory and on which fields become critical.

Additionally, a useful recursion relation follows directly from Eq.\ref{eq:general_alpha_k}:
\begin{equation}\label{eq:alpha_recursion_relation}
    \frac{\alpha_k(N)}{\alpha_k(N-1)}=\sqrt{\frac{N-1-k}{N+k}} .
\end{equation}
This relation shows explicitly that, for fixed $k$, the normalization decreases as the Hilbert-space dimension is increased. At the upper end of the multipolar tower, the highest available rank is $k=N-1$, for which
\begin{equation}
    \alpha_{N-1}(N)=\sqrt{\frac{(N+2)!}{(2N-1)!}}\:\:
    \xrightarrow[]{\text{ Stirling's approx. }}\:\: \propto
    \exp\left[-\frac{N}{2}\log N\right]\:,
\end{equation}
after using Stirling's approximation up to subleading prefactors. Thus, the highest-rank multipoles are the most strongly rescaled in the large-$N$ limit.
\end{itemize}

%% file: appendix2_SU4_embeddings.tex
\section{Embeddings and parameterizations of SU(4)} \label{app:su4-embeddings-parametrizations}
In the main text and in Appendix~\ref{app:su4-multipoles}, we have mostly used one particular physical realization of the fundamental representation of $SU(4)$: the identification of the four-dimensional local Hilbert space with a single spin-$\frac32$ multiplet,  $\mathbb C^4 \simeq \mathcal H_{S=3/2}$. Therefore, with respect to the physical $SU(2)$ subgroup generated by the spin-$\frac32$ operators, the algebra then decomposes into irreducible multipolar sectors, $\mathcal M_1\oplus\mathcal M_2\oplus\mathcal M_3$,  corresponding respectively to dipolar, quadrupolar, and octupolar operators. This is the natural basis when the four local states of $\mathbb{C}^4$ should be interpreted as a single higher-spin object.

However, this is not the only physically useful way of organizing the same algebra. The same fundamental representation space $\mathbb C^4$ may arise from rather different physical constructions. For instance, it may describe a spin-orbital Hilbert space $(\mathbb C^2_{\rm spin}\otimes\mathbb C^2_{\rm orb})$, a two-qubit system $(\mathbb C^2_{\rm qbt-1}\otimes\mathbb C^2_{\rm qbt-2})$, a singlet-triplet manifold of such two-qubit system $(\mathcal H_1^{\rm sym}\oplus\mathcal H_0^{\rm anti})$, or the internal space of a four-terminal scattering matrix. As vector spaces, one may write schematically,
\begin{equation} 
\mathbb C^4 \:\:\simeq\:\: \mathcal H_{S=3/2}  \:\:\simeq\:\:  (\mathbb C^2_{\rm spin}\otimes\mathbb C^2_{\rm orb})  \:\:\simeq\:\:  (\mathbb C^2_{\rm qbt-1}\otimes\mathbb C^2_{\rm qbt-2}) \: \:\simeq\:\:  (\mathcal H_1^{\rm sym}\oplus\mathcal H_0^{\rm anti}), 
\end{equation}
but these identifications are not physically equivalent. Each one singles out a different natural embedded subgroup of $SU(4)$, meaning the subgroup whose action preserves the physical interpretation assigned to the four local states. And moreover, once this subgroup is fixed, the fifteen $\mathfrak{su}(4)$ generators are naturally organized according to how they transform under it.

\noindent For example, 
\begin{itemize}
    \item the spin-$\frac32$ interpretation selects the embedded spin-rotation group $SU(2)\subset SU(4)$, leading to the multipolar decomposition $15=3+5+7$.
    \item The spin-orbital interpretation $\mathbb C^2_{\rm spin}\otimes\mathbb C^2_{\rm orb}$ instead selects $SU(2)_{\rm spin}\times SU(2)_{\rm orb}\subset SU(4)$, for which the natural generators are $\sigma_a\otimes\mathbb 1$, $\mathbb 1\otimes\tau_b$, and $\sigma_a\otimes\tau_b$, giving $15=3+3+9$.
    \item Finally, the Clebsch--Gordan organization
$\mathbb C^2\otimes\mathbb C^2\simeq
\mathcal H_1^{\rm sym}\oplus\mathcal H_0^{\rm anti}$
is adapted to singlet-triplet physics, or more precisely to the
symmetric and antisymmetric sectors of two qubits, and naturally
leads to an $\mathfrak{so}(4)\simeq
\mathfrak{su}(2)\oplus\mathfrak{su}(2)$ structure.
\end{itemize}   
Thus, the preferred basis of $\mathfrak{su}(4)$ is not selected by the abstract algebra alone, but by the physical meaning assigned to the four basis states. This distinction is important because $SU(4)$ appears in several condensed-matter and quantum-information settings. As discussed in Sec.~\ref{sec:su4_kk_model}, in spin-orbital systems one has $\mathbb C^2_{\rm spin}\otimes\mathbb C^2_{\rm orbital}$, and $SU(4)$ arises as the symmetric limit of the Kugel--Khomskii superexchange in two-orbital Hubbard models at one particle per site. A closely related structure appears in graphene and moir\'e systems, where the orbital degree of freedom is replaced by a valley pseudospin label, giving $\mathbb C^2_{\rm spin}\otimes\mathbb C^2_{\rm valley}$ and leading to approximate spin-valley $SU(4)$ descriptions~\cite{Yang2006,Young2012,ZhangMao2020}.

\noindent Alternatively, in quantum information, $SU(4)$ is the natural group of unitary two-qubit operations acting on $\mathbb C^2_{\rm qbt\text{-}1}\otimes\mathbb C^2_{\rm qbt\text{-}2}$, up to an overall phase. In this setting, it is often useful to reorganize the two-qubit Hilbert space according to its symmetry under the exchange of the two qubits,
\begin{equation}
    \mathbb C^2_{\rm qbt\text{-}1}\otimes\mathbb C^2_{\rm qbt\text{-}2}
    =
    \operatorname{Sym}^2(\mathbb C^2)\oplus \wedge^2(\mathbb C^2),
\end{equation}
where $\operatorname{Sym}^2(\mathbb C^2)$ is the three-dimensional symmetric subspace (the triplet) and $\wedge^2(\mathbb C^2)$ is the one-dimensional antisymmetric subspace (the singlet). This decomposition is useful because it provides a symmetry-adapted way of classifying two-qubit gates: operations that respect exchange symmetry act within these sectors, while the remaining operations mix them. In this way, one can distinguish sector-preserving gates from gates that transfer amplitude between the symmetric and antisymmetric subspaces. 

\noindent Finally, in mesoscopic transport, $SU(4)$ appears in a somewhat different way: not necessarily as an internal symmetry, but as a parametrization of scattering between four coherent transport channels. If $\psi^{\rm in}_b$ and $\psi^{\rm out}_a$ denote the incoming and outgoing wave amplitudes, a four-channel scatterer is described by a $4\times4$ matrix $\mathbb S$ satisfying
\begin{equation}
    \psi^{\rm out}_a=\sum_{b=1}^{4}\mathbb S_{ab}\psi^{\rm in}_b .
\end{equation}
Conservation of probability current makes $\mathbb S$ unitary, $\mathbb S\in U(4)$, and after removing an overall phase the nontrivial part belongs to $SU(4)$. In this setting, an $SU(4)$ parametrization gives a systematic way to describe reflection and transmission amplitudes among the four channels, for example in junctions of quantum wires or related mesoscopic devices \cite{TilmaByrdSudarshan2002,TilmaSudarshan2002, AristovNiyazov2015}.

In this line, the different appearances of $SU(4)$ can be summarized as different coordinate systems on the same algebra:
\begin{center}
\renewcommand{\arraystretch}{1.25}
\begin{tabular}{c|c|c|c}
\text{Interpretation} $\mathbb C^4$ & \text{Structure} & \text{Adapted basis} & \text{Organization} $\mathfrak{su}(4)$\\
\hline
$\mathcal H_{S=3/2}$ & $SU(2)$ & \text{multipoles} & $15=3+5+7$ \\
$\mathbb C^2_{\rm spin}\otimes\mathbb C^2_{\rm orb/valley}$ & $SU(2)\times SU(2)$ & \text{tensor products} & $15=3+3+9$ \\
$\operatorname{Sym}^2(\mathbb C^2)\oplus \wedge^2(\mathbb C^2)$ &  $\mathfrak{so}(4)$ & \text{symm./anti. blocks} & $15=8+1+6$ \\
$\mathbb C^4_{\rm channels}$ & \text{channel and rephasing} & \text{Euler parameters} & $6_{\text{reph}}+9_{\text{scatt}} \text{ angles}$
\end{tabular}
\end{center}
The purpose of this appendix is therefore not to introduce different algebras, but to compare physically adapted bases for the same algebra $\mathfrak{su}(4)$. The generalized Gell--Mann matrices provide a convenient algebraic reference basis, closely tied to the usual Cartan and ladder construction of $\mathfrak{su}(N)$. However, in applications one often reorganizes the same fifteen generators according to the physical subgroup, symmetry, or observable structure that is most relevant.

\subsection{Generalized Gell-Mann basis} \label{app:su4-gellmann-basis}

After emphasizing that different physical realizations of $\mathbb C^4$ lead to different natural bases, it is useful to start from the most standard algebraic one: the generalized Gell--Mann basis. This basis is not adapted to spin, orbital, or singlet-triplet physics. Rather, it is adapted to the canonical matrix-unit construction of $\mathfrak{su}(N)$, or equivalently to the usual ladder and Cartan decomposition.

\noindent The familiar reference point is $SU(3)$. In the canonical basis $\{\ket{1},\ket{2},\ket{3}\}$, let
\begin{equation}
    E_{pq}\equiv \ket{p}\bra{q}.
\end{equation}
The usual Gell--Mann matrices are organized according to the three pairs of levels $(1,2)$, $(1,3)$, and $(2,3)$, together with two diagonal Cartan generators:
\begin{equation}
\begin{aligned}
    \lambda_1 &= E_{12}+E_{21},
    &
    \lambda_2 &= -iE_{12}+iE_{21},
    &
    \lambda_3 &= E_{11}-E_{22},
    \\[3pt]
    \lambda_4 &= E_{13}+E_{31},
    &
    \lambda_5 &= -iE_{13}+iE_{31},
    \\[3pt]
    \lambda_6 &= E_{23}+E_{32},
    &
    \lambda_7 &= -iE_{23}+iE_{32},
    &
    \lambda_8 &= \frac{1}{\sqrt3}\left(E_{11}+E_{22}-2E_{33}\right).
\end{aligned}
\end{equation}
Equivalently, $\lambda_1,\lambda_4,\lambda_6$ are real symmetric off-diagonal matrices, $\lambda_2,\lambda_5,\lambda_7$ are imaginary antisymmetric off-diagonal matrices, and $\lambda_3,\lambda_8$ span the Cartan subalgebra. With the standard normalization,
\begin{equation}
    \operatorname{Tr}(\lambda_a\lambda_b)=2\delta_{ab}.
\end{equation}

The passage from $SU(3)$ to $SU(4)$ is completely systematic. First, one embeds the eight $SU(3)$ Gell--Mann matrices into the upper-left $3\times3$ block of a $4\times4$ matrix,
\begin{equation}
    \lambda_a^{(4)}
    =
    \begin{pmatrix}
        \lambda_a^{(3)} & 0\\
        0 & 0
    \end{pmatrix},
    \qquad a=1,\ldots,8.
\end{equation}
The remaining generators are those that connect the fourth basis state $\ket{4}$ to the first three states. Thus, for the pairs $(1,4)$, $(2,4)$, and $(3,4)$, one adds again one real symmetric and one imaginary antisymmetric generator:
\begin{equation}
\begin{aligned}
    \lambda_9 &= E_{14}+E_{41},
    &
    \lambda_{10} &= -iE_{14}+iE_{41},
    \\[3pt]
    \lambda_{11} &= E_{24}+E_{42},
    &
    \lambda_{12} &= -iE_{24}+iE_{42},
    \\[3pt]
    \lambda_{13} &= E_{34}+E_{43},
    &
    \lambda_{14} &= -iE_{34}+iE_{43}.
\end{aligned}
\end{equation}
Finally, since the rank of $SU(4)$ is three, one needs one more diagonal Cartan generator beyond $\lambda_3$ and $\lambda_8$. A standard normalized choice is
\begin{equation}
    \lambda_{15}
    =
    \frac{1}{\sqrt6}
    \left(
    E_{11}+E_{22}+E_{33}-3E_{44}
    \right)
    =
    \frac{1}{\sqrt6}
    \operatorname{diag}(1,1,1,-3).
\end{equation}
This keeps exactly the same pattern as in $SU(3)$: for every pair $p<q$, there is a real symmetric generator $E_{pq}+E_{qp}$ and an imaginary antisymmetric generator $-iE_{pq}+iE_{qp}$, while the remaining generators are diagonal Cartan matrices. In the $SU(4)$ case, the counting is therefore
\begin{equation}
    6\ \text{real symmetric}
    \quad+\quad
    6\ \text{imaginary antisymmetric}
    \quad+\quad
    3\ \text{Cartan}
    \quad=\quad
    15.
\end{equation}
More generally, for $SU(N)$ one starts from the canonical matrix units $E_{pq}=\ket{p}\bra{q}$, with $p,q=1,\ldots,N$. For each pair $1\leq p<q\leq N$, define
\begin{equation}
    \lambda^{(S)}_{pq}=E_{pq}+E_{qp},
    \qquad
    \lambda^{(A)}_{pq}=-iE_{pq}+iE_{qp}.
\end{equation}
These give $\frac{N(N-1)}{2}$ real symmetric and $\frac{N(N-1)}{2}$ imaginary antisymmetric generators. The remaining $N-1$ diagonal Cartan generators may be chosen as
\begin{equation}
    \lambda^{(D)}_{\ell}
    =
    \sqrt{\frac{2}{\ell(\ell+1)}}
    \left(
    \sum_{r=1}^{\ell}E_{rr}
    -
    \ell E_{\ell+1,\ell+1}
    \right),
    \qquad
    \ell=1,\ldots,N-1.
\end{equation}
Thus the total number of generators is
\begin{equation}
    \frac{N(N-1)}{2}
    +
    \frac{N(N-1)}{2}
    +
    (N-1)
    =
    N^2-1,
\end{equation}
as required for $\mathfrak{su}(N)$. With the normalization above, the generalized Gell--Mann matrices satisfy $\operatorname{Tr}(\lambda_a\lambda_b)=2\delta_{ab}$. The important point is that this basis is canonical and algebraically convenient, but it does not yet encode the physical meaning of the labels $1,\ldots,N$. In the $SU(4)$ case, the four labels may denote spin-$\frac32$ states, spin-orbital states, two qubits, singlet-triplet states, or scattering channels. The generalized Gell--Mann basis treats them simply as four abstract components of $\mathbb C^4$. Therefore, a Hamiltonian written as
\begin{equation}
    H=\sum_{a=1}^{15}B_a\lambda_a
\end{equation}
is a perfectly general traceless Hermitian $4\times4$ Hamiltonian, but the coefficients $B_a$ should be interpreted as algebraic coordinates or generalized source fields. They become physical magnetic fields, orbital fields, singlet-triplet couplings, or scattering parameters only after one specifies which physical embedding of $\mathbb C^4$ is being used. In this sense, the generalized Gell--Mann basis provides a neutral reference point, while the following subsections reorganize the same fifteen generators in bases adapted to more specific physical interpretations.

\subsection{Spin-orbital tensor-product basis}\label{app:su4-spin-orbital-basis}
We now turn to the first physically adapted basis of $\mathfrak{su}(4)$. Unlike the generalized Gell--Mann basis, which treats the four states as abstract components of $\mathbb C^4$, the spin-orbital basis uses the factorization already encountered in the Kugel--Khomskii discussion,
\begin{equation}
    \mathcal H_{\rm loc}
    =
    \operatorname{span}
    \left\{
    |1^\uparrow\rangle,\,
    |1^\downarrow\rangle,\,
    |2^\uparrow\rangle,\,
    |2^\downarrow\rangle
    \right\}
    \simeq
    \mathbb C^2_{\rm orb}\otimes\mathbb C^2_{\rm spin}.
\end{equation}
Here the first label denotes the orbital degree of freedom and the second label the physical spin. In graphene or moir\'e systems, the same algebraic structure appears with the replacement $\mathbb C^2_{\rm orb}\to\mathbb C^2_{\rm valley}$. For explicit matrices we use the ordered basis $(|1^\uparrow\rangle,|1^\downarrow\rangle,|2^\uparrow\rangle,|2^\downarrow\rangle)$, i.e. the convention in which the orbital factor comes first and the spin factor second.

Let $\sigma_a$, with $a=x,y,z$, act on the spin factor, and let $\tau_b$, with $b=x,y,z$, act on the orbital factor. Then the physical spin and orbital pseudospin operators are
\begin{equation}
    S_a=\frac12\,\mathbb 1_{\rm orb}\otimes\sigma_a,
    \qquad
    \mathcal T_b=\frac12\,\tau_b\otimes\mathbb 1_{\rm spin}.
\end{equation}
The corresponding tensor-product basis of $\mathfrak{su}(4)$ is obtained from all nontrivial tensor products of the two Pauli algebras,
\begin{equation}
    \mathbb 1\otimes\sigma_a,
    \qquad
    \tau_b\otimes\mathbb 1,
    \qquad
    \tau_b\otimes\sigma_a
    \qquad
    [a,b=x,y,z]\:\:.
\end{equation}
Thus the fifteen generators are organized as $15=3_{\rm spin}+3_{\rm orb}+9_{\rm spin\text{-}orb}$. The first three generators rotate the physical spin without changing the orbital label. The next three rotate the orbital pseudospin without changing the spin. The remaining nine are mixed spin-orbital operators, which measure or couple the two sectors simultaneously. This is precisely why this basis is natural for spin-orbital and spin-valley problems: it separates spin, orbital/valley, and genuinely mixed spin-orbital correlations from the beginning.

In block form, where each block acts on the spin sector and the $2\times2$ block structure refers to the orbital sector, the spin generators are
\begin{equation}
    \mathbb 1\otimes\sigma_a
    =
    \begin{pmatrix}
        \sigma_a & 0\\
        0 & \sigma_a
    \end{pmatrix}.
\end{equation}
The orbital generators are
\begin{equation}
    \tau_x\otimes\mathbb 1
    =
    \begin{pmatrix}
        0 & \mathbb 1\\
        \mathbb 1 & 0
    \end{pmatrix},
    \qquad
    \tau_y\otimes\mathbb 1
    =
    \begin{pmatrix}
        0 & -i\mathbb 1\\
        i\mathbb 1 & 0
    \end{pmatrix},
    \qquad
    \tau_z\otimes\mathbb 1
    =
    \begin{pmatrix}
        \mathbb 1 & 0\\
        0 & -\mathbb 1
    \end{pmatrix}.
\end{equation}
Finally, the mixed spin-orbital generators are
\begin{equation}
    \tau_x\otimes\sigma_a
    =
    \begin{pmatrix}
        0 & \sigma_a\\
        \sigma_a & 0
    \end{pmatrix},
    \qquad
    \tau_y\otimes\sigma_a
    =
    \begin{pmatrix}
        0 & -i\sigma_a\\
        i\sigma_a & 0
    \end{pmatrix},
    \qquad
    \tau_z\otimes\sigma_a
    =
    \begin{pmatrix}
        \sigma_a & 0\\
        0 & -\sigma_a
    \end{pmatrix}.
\end{equation}
These fifteen matrices are all traceless and Hermitian, and therefore form a basis of $\mathfrak{su}(4)$. The important point is not the normalization convention, but the physical organization: the basis directly separates spin rotations, orbital rotations, and mixed spin-orbital operators.

The advantage of this basis is that physically meaningful one-site observables are immediately visible. A general one-site density matrix can be expanded as
\begin{equation}
    \rho
    =
    \frac14
    \left[
    \mathbb 1\otimes\mathbb 1
    +
    \sum_a m_a^s\,\mathbb 1\otimes\sigma_a
    +
    \sum_b m_b^o\,\tau_b\otimes\mathbb 1
    +
    \sum_{b,a} C_{ba}\,\tau_b\otimes\sigma_a
    \right].
\end{equation}
The coefficients $m_a^s$ describe spin polarization, $m_b^o$ describes orbital or valley polarization, and $C_{ba}$ describes local spin-orbital correlations. More explicitly,
\begin{equation}
    \langle S_a\rangle=\frac12 m_a^s,
    \qquad
    \langle \mathcal T_b\rangle=\frac12 m_b^o,
    \qquad
    \langle \mathcal T_b S_a\rangle=\frac14 C_{ba}.
\end{equation}
Thus, in this basis, one can directly diagnose magnetic order, orbital or valley order, and spin-orbital entanglement. In a lattice problem, the corresponding two-site observables include $\langle \mathbf S_i\cdot\mathbf S_j\rangle$, $\langle\boldsymbol{\mathcal T}_i\cdot\boldsymbol{\mathcal T}_j\rangle$, and mixed quantities such as $\langle S_i^a\mathcal T_i^b S_j^c\mathcal T_j^d\rangle$. These are precisely the natural observables for distinguishing spin order, orbital order, and spin-orbital liquid behavior.

\noindent A general traceless local Hamiltonian also takes a transparent form:
\begin{equation}
    H_{\rm loc}
    =
    \sum_a h_a^s\,\mathbb 1\otimes\sigma_a
    +
    \sum_b h_b^o\,\tau_b\otimes\mathbb 1
    +
    \sum_{b,a} h_{ba}^{so}\,\tau_b\otimes\sigma_a.
\end{equation}
The first term is a spin field, such as a Zeeman coupling. The second term is an orbital or valley field, describing for example an orbital splitting, a valley polarization, or tunneling between orbital states. And the third term couples spin and orbital sectors and may describe spin-orbit coupling, spin-valley coupling, or more general local anisotropies. This is one of the main differences with the Gell--Mann form $H=\sum_a B_a\lambda_a$: in the tensor-product basis, the coefficients have an immediate physical interpretation once the factorization of $\mathbb C^4$ is fixed.

The same simplification appears at the level of interactions. In the $SU(4)$-symmetric Kugel--Khomskii limit derived in Sec.~\ref{sec:su4_kk_model}, the nearest-neighbor interaction is proportional to the permutation operator $P_{ij}$ acting on the full spin-orbital flavor. In the tensor-product basis, this permutation factorizes into spin and orbital permutations. In particular, one obtains
\begin{equation}
\begin{aligned}
    P_{ij}&=\left(\frac12+2\,\mathbf S_i\cdot\mathbf S_j\right)\left(\frac12+2\,\boldsymbol{\mathcal T}_i\cdot\boldsymbol{\mathcal T}_j\right)\\[4pt]
    &=
    \frac14 \:+\:\mathbf S_i\cdot\mathbf S_j\:+\:\boldsymbol{\mathcal T}_i\cdot\boldsymbol{\mathcal T}_j\:+\:4(\mathbf S_i\cdot\mathbf S_j)(\boldsymbol{\mathcal T}_i\cdot\boldsymbol{\mathcal T}_j).
\end{aligned}
\end{equation}
This expression shows explicitly how the $SU(4)$-symmetric exchange entangles spin and orbital correlations. The interaction is not simply a spin exchange plus an orbital exchange; it also contains the mixed product $(\mathbf S_i\cdot\mathbf S_j)(\boldsymbol{\mathcal T}_i\cdot\boldsymbol{\mathcal T}_j)$, which is essential for treating the four spin-orbital flavors on equal footing.

This basis becomes even more useful away from the $SU(4)$-symmetric point. For example, when Hund's coupling and orbital anisotropies are included, the effective Kugel--Khomskii Hamiltonian contains terms of the form
\begin{equation}
\begin{aligned}
    \left(\mathbf S_i\cdot\mathbf S_j-\frac14\right)
    \left(2\mathcal T_i^z\mathcal T_j^z+\frac12\right),\qquad\\[4pt]
    \quad\left(\mathbf S_i\cdot\mathbf S_j-\frac14\right)
    \left(
    \boldsymbol{\mathcal T}_i\cdot\boldsymbol{\mathcal T}_j
    -
    2\mathcal T_i^z\mathcal T_j^z
    +
    \frac14
    \right),\\[4pt]
    \left(\mathbf S_i\cdot\mathbf S_j+\frac34\right)
    \left(
    \boldsymbol{\mathcal T}_i\cdot\boldsymbol{\mathcal T}_j
    -
    \frac14
    \right).\qquad
\end{aligned}
\end{equation}
In a generic Gell--Mann basis, these terms would appear as a complicated collection of couplings among the fifteen generators. In the tensor-product basis, their meaning is immediate: they distinguish spin singlet and spin triplet channels, and they distinguish longitudinal and transverse orbital correlations. Thus, the spin-orbital basis is not only useful at the $SU(4)$-symmetric point; it is also the natural language for controlled $SU(4)$-breaking perturbations such as Hund coupling, orbital anisotropy, valley splitting, or spin-orbit coupling.

Let us now relate this basis to the generalized Gell--Mann basis, and express the tensor-product generators as fixed linear combinations of the fifteen $\lambda$-matrices. For the spin sector, one finds
\begin{equation}
\begin{aligned}
    \mathbb 1\otimes\sigma_x&=\lambda_1+\lambda_{13},
    \qquad
    \mathbb 1\otimes\sigma_y=\lambda_2+\lambda_{14},\\[3pt]
    &\mathbb 1\otimes\sigma_z=\lambda_3-\frac{1}{\sqrt3}\lambda_8+\frac{\sqrt6}{3}\lambda_{15}.
\end{aligned}
\end{equation}
For the orbital sector,
\begin{equation}
\begin{aligned}
    \tau_x\otimes\mathbb 1=&\lambda_4+\lambda_{11},
    \qquad
    \tau_y\otimes\mathbb 1=\lambda_5+\lambda_{12},\\[3pt]
    &\tau_z\otimes\mathbb 1
    =
    \frac{2}{\sqrt3}\lambda_8
    +
    \frac{\sqrt6}{3}\lambda_{15}.
\end{aligned}
\end{equation}
The mixed generators are similarly obtained as
\begin{equation}
\begin{array}{lll}
    \tau_x\otimes\sigma_x=\lambda_6+\lambda_9,
    &
    \tau_x\otimes\sigma_y=\lambda_{10}-\lambda_7,
    &
    \tau_x\otimes\sigma_z=\lambda_4-\lambda_{11},
    \\[4pt]
    \tau_y\otimes\sigma_x=\lambda_7+\lambda_{10},
    &
    \tau_y\otimes\sigma_y=\lambda_6-\lambda_9,
    &
    \tau_y\otimes\sigma_z=\lambda_5-\lambda_{12},
    \\[4pt]
    \tau_z\otimes\sigma_x=\lambda_1-\lambda_{13},
    &
    \tau_z\otimes\sigma_y=\lambda_2-\lambda_{14},
    &
    \tau_z\otimes\sigma_z=
    \lambda_3+\frac{1}{\sqrt3}\lambda_8-\frac{\sqrt6}{3}\lambda_{15}.
\end{array}
\end{equation}
These relations make explicit that the Gell--Mann and tensor-product bases are not different algebras. They are related by a fixed real change of basis in the fifteen-dimensional vector space $\mathfrak{su}(4)$. The difference is interpretational: the Gell--Mann basis is organized by transitions between abstract basis states, whereas the tensor-product basis is organized by spin rotations, orbital rotations, and mixed spin-orbital correlations.

Finally, let us compare this tensor-product basis with the multipolar basis used in Appendix~\ref{app:su4-multipoles}. While the multipolar construction assumes the identification $\mathbb C^4\simeq\mathcal H_{S=3/2}$, the tensor-product basis assumes $\mathbb C^4\simeq\mathbb C^2_{\rm orb}\otimes\mathbb C^2_{\rm spin}$. In this line, the multipole approach organizes the $\mathfrak{su}(4)$ algebra according to which operators behave as dipoles, quadrupoles, and octupoles of an effective spin-$\frac32$, while the tensor-product basis asks which operators act on spin, which act on orbital/valley, and which couple the two sectors. Hence, both decompositions of the same fifteen generators are correct, but address different physical situations. Nevertheless, what one could try to do is to artificially identify the ordered spin-orbital basis with spin-$\frac32$ states:
\begin{equation}
    |1^\uparrow\rangle\leftrightarrow |3/2\rangle,
    \qquad
    |1^\downarrow\rangle\leftrightarrow |1/2\rangle,
    \qquad
    |2^\uparrow\rangle\leftrightarrow |-1/2\rangle,
    \qquad
    |2^\downarrow\rangle\leftrightarrow |-3/2\rangle,
\end{equation}
and then relate the generators. Under this picture, the spin-$\frac32$ dipole operator $J_z$ is represented by
\begin{equation}
    J_z
    \:\:\:=\:\:\:
    \operatorname{diag}
    \left(
    \frac32,\frac12,-\frac12,-\frac32
    \right)\:\:\:=\:\:\:\tau_z\otimes\mathbb 1
    +
    \frac12\,\mathbb 1\otimes\sigma_z\:\:\:=\:\:\:2\mathcal T_z+S_z\:\:,
\end{equation}
in the spin-orbital tensor-product basis. Therefore, the spin-$\frac32$ dipole $J_z$ is not the same object as the physical spin $S_z$ of the spin-orbital system. It is a particular linear combination of orbital and spin operators. This illustrates the main message of this appendix: the same $4\times4$ matrices can acquire very different physical meanings depending on the chosen embedding of $\mathbb C^4$.

\noindent The tensor-product basis is therefore the natural language for spin-orbital, spin-valley, and two-qubit realizations of $SU(4)$. It makes the physically measurable quantities $\mathbf S$, $\boldsymbol{\mathcal T}$, and $\langle S_a\mathcal T_b\rangle$ explicit, and it is the most convenient basis for writing $SU(4)$-breaking perturbations. The generalized Gell--Mann basis remains the neutral algebraic reference, while the multipolar basis becomes natural only after choosing the spin-$\frac32$ interpretation of the same four-dimensional Hilbert space.

\subsection{SO(4) singlet-triplet basis}\label{app:su4-so4-singlet-triplet-basis}
We now turn to another useful organization of the same four-dimensional space. In the previous subsection, we kept the tensor-product structure explicit and used the basis adapted to spin-orbital or spin-valley physics. Here we instead start from a more abstract two-qubit Hilbert space, $\mathbb C^2_{\rm qbt\text{-}1}\otimes\mathbb C^2_{\rm qbt\text{-}2}$, where each qubit has two states, denoted by $\ket{e}$ and $\ket{g}$. The product basis is $\ket{ee}$, $\ket{eg}$, $\ket{ge}$, and $\ket{gg}$. However, for the present purpose it is more useful to reorganize this space according to the exchange symmetry of the two qubits,
\begin{equation}
    \mathbb C^2_{\rm qbt\text{-}1}\otimes\mathbb C^2_{\rm qbt\text{-}2}
    =
    \operatorname{Sym}^2(\mathbb C^2)\oplus\wedge^2(\mathbb C^2).
\end{equation}
The three-dimensional symmetric sector and the one-dimensional antisymmetric sector are spanned, respectively, by
\begin{equation}
    \left\{\ket{S_+}=\ket{ee},
    \quad
    \ket{S_0}=\frac{\ket{eg}+\ket{ge}}{\sqrt2},
    \quad
    \ket{S_-}=\ket{gg}\right\}\qquad \left\{
    \ket{A}=\frac{\ket{eg}-\ket{ge}}{\sqrt2}\right\}.
\end{equation}
In what follows, we use the ordered basis $(\ket{S_+},\ket{S_0},\ket{S_-},\ket{A})$. The notation $S$ and $A$ refers only to exchange symmetry. Nevertheless, this decomposition is mathematically identical to the familiar Clebsch--Gordan decomposition of two spin-$\frac12$ degrees of freedom into a triplet and a singlet. Therefore, one can think of $\operatorname{Sym}^2(\mathbb C^2)$ as an effective spin-$1$ space, while $\wedge^2(\mathbb C^2)$ plays the role of a singlet. This is only a representation-theoretic analogy, but it is extremely useful because it allows us to organize operators in terms of dipoles, quadrupoles, and singlet-symmetric transitions.

\noindent To make this precise, introduce the qubit pseudospin operators, $\mathbf{s}_1=\frac12\,\boldsymbol{\sigma}\otimes\mathbb 1$ and $\mathbf{s}_2=\frac12\,\mathbb 1\otimes\boldsymbol{\sigma}$, where the Pauli matrices act on the first and second qubit factors, respectively. We then define the symmetric and antisymmetric combinations
\begin{equation}
    \mathbf J=\mathbf{s}_1+\mathbf{s}_2,
    \qquad
    \mathbf R=\mathbf{s}_2-\mathbf{s}_1.
\end{equation}
The vector $\mathbf J$ is the total qubit pseudospin. It is even under the exchange of the two qubits and therefore preserves the symmetric sector. In the basis above, $\mathbf J$ acts as the usual spin-$1$ generator on $\operatorname{Sym}^2(\mathbb C^2)$ and annihilates the antisymmetric state,
\begin{equation}
    \mathbf J:\operatorname{Sym}^2(\mathbb C^2)\to\operatorname{Sym}^2(\mathbb C^2),
    \qquad
    \mathbf J\ket{A}=0.
\end{equation}
By contrast, $\mathbf R$ is odd under exchange of the two qubits. Therefore, it maps symmetric states into the antisymmetric sector and conversely,
\begin{equation}
    \mathbf R:\operatorname{Sym}^2(\mathbb C^2)\leftrightarrow\wedge^2(\mathbb C^2).
\end{equation}
This is the precise sense in which $\mathbf R$ describes transitions between the symmetric and antisymmetric sectors. For example,
\begin{equation}
\begin{aligned}
    R_z\ket{A}=-\ket{S_0}&,
    \qquad
    R_z\ket{S_0}=-\ket{A},\\[3pt]
    R_x\ket{A}=\frac{\ket{S_+}-\ket{S_-}}{\sqrt2},
    &\qquad
    R_y\ket{A}=-\frac{i}{\sqrt2}\left(\ket{S_+}+\ket{S_-}\right).
\end{aligned}
\end{equation}
Furthermore, in the previous ordered basis $(\ket{S_+},\ket{S_0},\ket{S_-},\ket{A})$, the three $\mathbf J$ matrices are
\begin{equation}
    J_x=
    \frac{1}{\sqrt2}
    \begin{pmatrix}
        0&1&0&0\\
        1&0&1&0\\
        0&1&0&0\\
        0&0&0&0
    \end{pmatrix},
    \qquad
    J_y=
    \frac{1}{\sqrt2}
    \begin{pmatrix}
        0&-i&0&0\\
        i&0&-i&0\\
        0&i&0&0\\
        0&0&0&0
    \end{pmatrix},
    \qquad
    J_z=
    \begin{pmatrix}
        1&0&0&0\\
        0&0&0&0\\
        0&0&-1&0\\
        0&0&0&0
    \end{pmatrix},
\end{equation}
whereas the three $\mathbf R$ matrices are
\begin{equation}
    R_x=
    \frac{1}{\sqrt2}
    \begin{pmatrix}
        0&0&0&1\\
        0&0&0&0\\
        0&0&0&-1\\
        1&0&-1&0
    \end{pmatrix},
    \qquad
    R_y=
    \frac{1}{\sqrt2}
    \begin{pmatrix}
        0&0&0&-i\\
        0&0&0&0\\
        0&0&0&-i\\
        i&0&i&0
    \end{pmatrix},
    \qquad
    R_z=
    \begin{pmatrix}
        0&0&0&0\\
        0&0&0&-1\\
        0&0&0&0\\
        0&-1&0&0
    \end{pmatrix}.
\end{equation}
Hence, in the Hermitian convention used here, their commutation relations follow
\begin{equation}
    [J_\alpha,J_\beta]=i\epsilon_{\alpha\beta\gamma}J_\gamma,
    \qquad
    [J_\alpha,R_\beta]=i\epsilon_{\alpha\beta\gamma}R_\gamma,
    \qquad
    [R_\alpha,R_\beta]=i\epsilon_{\alpha\beta\gamma}J_\gamma\:,
\end{equation}
so that the the six matrices $\Set{J_\alpha,R_\alpha}_{\alpha=x,y,z}$ generate an embedded copy of $\mathfrak{so}(4)\subset\mathfrak{su}(4)$. Note how the inclusion of the $\mathbf{R}$ generators extends the $\mathfrak{su}(2)_{J}$ Lie algebra generated by $\mathbf{J}$, to the full $\mathfrak{so}(4)$ algebra. In fact, coming back to the original qubit pseudospins $\mathbf s_1$ and $\mathbf s_2$, we can see that what we really have is two copies of mutually commuting $\mathfrak{su}(2)$ algebras. Since they act on different tensor factors, their commutation relations are
\begin{equation}
    [s_{1,\alpha},s_{1,\beta}]
    =
    i\epsilon_{\alpha\beta\gamma}s_{1,\gamma},
    \qquad
    [s_{2,\alpha},s_{2,\beta}]
    =
    i\epsilon_{\alpha\beta\gamma}s_{2,\gamma},
    \qquad
    [s_{1,\alpha},s_{2,\beta}]=0.
\end{equation}
This is the concrete algebraic meaning of the local isomorphism \begin{equation}\label{eq:so4_local_isomorphism}
    \mathfrak{so}(4)\simeq\mathfrak{su}(2)_1\oplus\mathfrak{su}(2)_2\:.
\end{equation}
Therefore, the sets $\{s_{1,\alpha},s_{2,\alpha}\}$ and $\{J_\alpha,R_\alpha\}$ should not be regarded as generators of two different subalgebras, but as two bases of the same embedded $\mathfrak{so}(4)\subset\mathfrak{su}(4)$. The former makes the decomposition into two commuting $\mathfrak{su}(2)$ factors explicit, whereas the latter is adapted to the singlet--triplet organization, with $\mathbf J$ preserving the triplet sector and $\mathbf R$ connecting it to the singlet. This choice is also geometrically natural: the six generators $M_{ab}$ of $\mathfrak{so}(4)$, associated with rotations in the six coordinate planes of a four-dimensional space, can be reorganized into the two three-component vectors $J_i=\frac12\epsilon_{ijk}M_{jk}$ and $R_i=M_{i4}$. The former generate rotations among the first three directions, whereas the latter mix them with the fourth one, precisely mirroring the action of $\mathbf J$ and $\mathbf R$ on the triplet and singlet sectors.

Note, however, that Eq.\ref{eq:so4_local_isomorphism} does not imply an isomorphism between the corresponding Lie groups. At such level, one obtains
\begin{equation}\label{eq:spin4_so4_double_cover}
    \frac{SU(2)_1\times SU(2)_2}{\mathbb Z_2}
    \simeq
    SO(4)\:,
\end{equation}
meaning that $SU(2)_1\times SU(2)_2$ is the double cover of $SO(4)$. As we will formalize later (see Appendix~\ref{app:low_rank_su_so}), double covers of $SO(n)$ groups are the so-called $\operatorname{Spin}(n)$ groups, having the special feature that not always a representation of $\operatorname{Spin}(n)$, induces (or it is equivalent to) a representation of $SO(n)$. Whenever a spin representation is also an $SO(n)$ representation, we call it a \emph{tensor representation}, but when the representation fails to do that, we call it a \emph{spinor representation}. Now, for the case that we are considering here, we will have that the previous $SO(4)$ representation with representation space $\mathbb C^2_{\rm qbt\text{-}1}\otimes\mathbb C^2_{\rm qbt\text{-}2}$, standing for the fundamental $\bm 4$ irrep of $SU(4)$, is indeed a vector representation of $\operatorname{Spin}(4)$, so that it can be fully characterized from $SO(4)$ objects. Those are the two independent quadratic $\mathfrak{so}(4)$ Casimir operators constructed from (and fulfilling):
\begin{equation}\label{eq:so4_casimirs_singlet_triplet}
    \mathbf J^2+\mathbf R^2
    =
    2\left(\mathbf s_1^2+\mathbf s_2^2\right)
    =
    3,
    \qquad
    \mathbf J\cdot\mathbf R
    =
    \mathbf s_2^2-\mathbf s_1^2
    =
    0\:,
\end{equation}
Hence, these identities specify the quadratic Casimir values of the four-dimensional $SO(4)$ representation realized on the singlet--triplet space.

At this point, the six generators $\mathbf J$ and $\mathbf R$ form only an $\mathfrak{so}(4)$ subalgebra of $\mathfrak{su}(4)$, but to obtain a full basis of the fifteen traceless Hermitian $4\times4$ matrices, one must add nine further operators. The $SO(4)$-adapted basis therefore can be organized as
\begin{equation}
15\quad=\quad\underbrace{3_J+5_Q}_{8_{\rm{sym}}}\quad+\quad\underbrace{3_R+3_A}_{6_{\rm{mix}}}\quad+\quad1_{\rm{rel}},
\end{equation}
where 
\begin{itemize}
    \item the first eight generators, $\nu_{1,2,3}$ and $\nu_{4,\ldots,8}$, correspond to the dipole components of $\mathbf J$ and the five quadrupolar operators characteristic of the familiar spin-$1$ decomposition:
\begin{equation}
\begin{gathered}
    \nu_1=J_x,
    \qquad
    \nu_2=J_y,
    \qquad
    \nu_3=J_z,
    \\[0.6em]
    \nu_4=J_x^2-J_y^2,
    \qquad
    \nu_5=J_xJ_y+J_yJ_x,
    \\
    \nu_6=-(J_xJ_z+J_zJ_x),
    \qquad
    \nu_7=-(J_yJ_z+J_zJ_y),
    \\
    \nu_8=
    \frac{1}{\sqrt3}\left(\mathbf J^2-3J_z^2\right).
\end{gathered}
\end{equation}
\item the next three generators are the components of the relative pseudospin,
\begin{equation}
    \nu_9=R_x,
    \qquad
    \nu_{10}=R_y,
    \qquad
    \nu_{11}=R_z,
\end{equation}
corresponding to the first set of symmetric-antisymmetric transition operators.
\item a second set of transition operators is obtained by combining $\mathbf J$ and $\mathbf R$ into the vector $i\mathbf A=\frac12\left(\mathbf J\times\mathbf R-\mathbf R\times\mathbf J\right)$, whose components define
\begin{equation}
\begin{gathered}
    \nu_{12}=iA_x=J_yR_z+R_zJ_y=-(J_zR_y+R_yJ_z),
    \\
    \nu_{13}=iA_y=J_zR_x+R_xJ_z=-(J_xR_z+R_zJ_x),
    \\
    \nu_{14}=iA_z=J_xR_y+R_yJ_x=-(J_yR_x+R_xJ_y).
\end{gathered}
\end{equation}
Hence, the three $R_\alpha$ and the three $iA_\alpha$ account for the six real off-diagonal generators mixing $\operatorname{Sym}^2(\mathbb C^2)$ and $\wedge^2(\mathbb C^2)$.
\item finally, the fifteenth generator is
\begin{equation}
    \nu_{15}
    =
    \frac{1}{\sqrt6}\left(\mathbf J^2-\mathbf R^2\right)
    =
    \frac{1}{\sqrt6}\operatorname{diag}(1,1,1,-3),
\end{equation}
being a scalar under the $J$-rotations of the symmetric sector, but not proportional to the identity. It is traceless and simply distinguishes the symmetric block from the antisymmetric state.
\end{itemize}
This organization makes transparent the separation between the generators that preserve the exchange-symmetry decomposition and those that mix the two sectors. The operators $\mathbf J$, $\nu_4,\ldots,\nu_8$, and $\nu_{15}$ are block-preserving: $\mathbf J$ and $\nu_4,\ldots,\nu_8$ describe dipolar and quadrupolar structure inside the effective spin-$1$ symmetric manifold, while $\nu_{15}$ measures the relative symmetric/antisymmetric splitting. In contrast, $\mathbf R$ and $i\mathbf A$ are off-diagonal generators and therefore describe coherent transitions between $\operatorname{Sym}^2(\mathbb C^2)$ and $\wedge^2(\mathbb C^2)$.

Following this $\mathfrak{su}(4)$-decomposition, a general local Hamiltonian written in the $SO(4)$-adapted basis has a direct interpretation. Schematically, one may write
\begin{equation}
    H_{\rm loc}
    =
    \mathbf h_J\cdot\mathbf J
    +
    \sum_{\mu=4}^{8}q_\mu\,\nu_\mu
    +
    \mathbf h_R\cdot\mathbf R
    +
    \mathbf h_A\cdot(i\mathbf A)
    +
    \Delta\,\nu_{15}.
\end{equation}
The first term acts as a dipolar field inside the symmetric three-level manifold. The second term contains quadrupolar anisotropies, which split or mix the symmetric states without coupling them to the antisymmetric sector. The diagonal term proportional to $\nu_{15}$ controls the relative detuning between the symmetric and antisymmetric blocks. In contrast, the fields $\mathbf h_R$ and $\mathbf h_A$ drive transitions between the two sectors; they are the natural coordinates for perturbations that break the exchange-symmetry block structure or couple the antisymmetric state to the symmetric manifold.

\noindent This basis is therefore useful when the relevant perturbations are organized by a symmetric--antisymmetric splitting, especially when the low-energy Hilbert space is described in terms of singlet and triplet sectors rather than four unrelated levels. For example, the spin-rotator description of double quantum dots uses the hidden $SO(4)$ structure of the low-energy singlet--triplet manifold \cite{KikoinAvishai2002}, while related Kondo setups exploit singlet--triplet excitations and transitions in double quantum dots, where the effective low-energy dynamics can also be described in terms of an $SO(4)$ symmetry \cite{KiselevKikoinMolenkamp2003}. Furthermore, in the quantum-information setting, semiconductor singlet--triplet qubits use the same sector organization as their computational structure, from the coherent control of two-electron spin states in double quantum dots to recent multi-qubit quantum-dot arrays \cite{Petta2005,Zhang2025}.

So, by this point, the conceptual role of $SO(4)$ must be clear. The six generators $\mathbf J$ and $\mathbf R$ form the embedded $\mathfrak{so}(4)$ algebra: $\mathbf J$ generates rotations inside the symmetric sector, while $\mathbf R$ transforms as a vector under these rotations and connects the symmetric and antisymmetric sectors. The remaining nine generators complete this $SO(4)$ structure to the full $\mathfrak{su}(4)$ algebra. Hence, the present basis is not a new algebra, but a symmetry-adapted coordinate system on the same fifteen-dimensional space of traceless Hermitian $4\times4$ matrices. The Gell--Mann basis is adapted to abstract transitions between four labels, the tensor-product basis keeps the two qubit factors explicit, and the $SO(4)$ singlet-triplet basis is adapted to the exchange-symmetry decomposition into symmetric and antisymmetric sectors.

\subsection{Euler-angle parametrizations}\label{app:su4-euler-parametrizations}
The previous subsections were concerned with different bases of the algebra $\mathfrak{su}(4)$. In each case, the main question was: given fifteen traceless Hermitian generators, which linear combinations have a transparent physical meaning? Now, this subsection serves a different purpose. We are no longer trying to find a physically adapted basis of $\mathfrak{su}(4)$, but instead, we want to parametrize full elements of the group $SU(4)$. In other words, the object of interest is not an infinitesimal generator or a local Hamiltonian, 
but a finite unitary matrix $U \in SU(4)$, written as a product of elementary transformations.

\noindent In the case of $SU(2)$, the familiar Euler-angle parametrization of any $U\in SU(2)$,
\begin{equation}
    U_{SU(2)}(\alpha,\beta,\gamma)
    =
    e^{i\alpha \sigma_z}
    e^{i\beta \sigma_y}
    e^{i\gamma \sigma_z},
    \qquad
    \alpha\in(0,\pi),\quad \beta\in(0,\pi/2), \quad \gamma\in(0,2\pi).
\end{equation}
accomplishes precisely this goal: it writes an arbitrary group element as a product of elementary transformations. Here the ranges are written in the Pauli/Gell--Mann normalization, where the generators are $\sigma_a$ rather than $S_a$. Thus, compared with Eq.\ref{eq:euler-rep}, one has $(\alpha,\beta,\gamma)$ as $(\phi/2,\theta'/2,\psi/2)$; note that the extended range $\gamma\in(0,2\pi)$ is the full $SU(2)$ covering range, accounting for the nontrivial center $\mathbb Z_2$ of $SU(2)$ (obtaining $SU(2)$ volume instead of $SU(2)/\mathbb Z_2$) \cite{TilmaByrdSudarshan2002}. 

\noindent For higher unitary groups, the same idea can be implemented recursively by choosing a subgroup $K$ together with complementary directions $P$, so that a generic element $U\in SU(N)$ can be decomposed schematically as $U=K\cdot P$. After fixing the redundancies associated with the subgroup action, this construction leads to an Euler-angle parametrization in terms of $N^2-1$ group parameters.

\noindent The simplest nontrivial example is $SU(3)$. In the Gell--Mann basis, one uses the Cartan-type splitting
\begin{equation}
\begin{aligned}
    \mathcal L(K)
    &=
    \operatorname{span}\{\lambda_1,\lambda_2,\lambda_3,\lambda_8\},
    \\
    \mathcal L(P)
    &=
    \operatorname{span}\{\lambda_4,\lambda_5,\lambda_6,\lambda_7\},
\end{aligned}
\end{equation}
which satisfies
\begin{equation}
    [\mathcal L(K),\mathcal L(K)]\subset \mathcal L(K),
    \qquad
    [\mathcal L(K),\mathcal L(P)]\subset \mathcal L(P),
    \qquad
    [\mathcal L(P),\mathcal L(P)]\subset \mathcal L(K).
\end{equation}
The subalgebra $\mathcal L(K)$ exponentiates to a $U(2)$ subgroup of $SU(3)$, acting on the two-dimensional block spanned by $\ket{1}$ and $\ket{2}$. Since $\lambda_8$ commutes with the $SU(2)$ generated by $\lambda_1,\lambda_2,\lambda_3$, a general element of this subgroup can be written as
\begin{equation}
    K{\scriptstyle(\alpha,\beta,\gamma,\phi)}= U_{{SU(2)}} \,e^{i\lambda_8\phi}
    \qquad
    \Big[U_{SU(2)}:=U_{SU(2)}^{\scriptstyle(\alpha,\beta,\gamma)}
    =
    e^{i\lambda_3\alpha}
    e^{i\lambda_2\beta}
    e^{i\lambda_3\gamma}\Big].
\end{equation}
The complementary sector $\mathcal L(P)$ contains the generators that connect the $(1,2)$ block to the third state $\ket{3}$. Instead of exponentiating all four generators independently, one chooses a representative direction, conventionally $\lambda_5$, and generates the remaining directions by conjugating with elements of $K$. Schematically, this gives
\begin{equation}
    P=K'\,e^{i\lambda_5\theta}\,K'' \:\:\: \xRightarrow[\text{ element }]{\text{ general group }} \:\:\:U=K(K'\,e^{i\lambda_5\theta}\,K'').
\end{equation}
Because $K$ is a subgroup, the product $K K'$ is again an element of $K$. Therefore, one may absorb it into a single factor, $K K'=\widetilde K$, and write
\begin{equation}
    U=\widetilde K\,e^{i\lambda_5\theta}\,K''.
\end{equation}
At this stage, however, the expression still contains one redundant parameter. A full $K\simeq U(2)$ factor has four parameters, so $\widetilde K e^{i\lambda_5\theta}K''$ would contain $4+1+4=9$ parameters, whereas $\dim SU(3)=8$. The extra parameter is a \emph{stabilizer redundancy}: different choices of the left $\widetilde K$ related by this stabilizer can be compensated by a corresponding change of the right $K''$ factor and therefore do not produce a new group element. Equivalently, the left part should be viewed not as a full $U(2)$ element, but as a representative of the coset $SU(3)/U(2)$.

\noindent In practice, this redundancy is fixed by removing the $e^{i\lambda_8\phi}$ factor from the left $\tilde{K}$ factor, while keeping the full $U(2)$ subgroup on the right $K''$. Thus one chooses
\begin{equation}
    \widetilde K
    \longrightarrow
    U_{SU(2)}^{(\alpha,\beta,\gamma)},
    \qquad
    K''
    =
    U_{SU(2)}^{(a,b,c)}\,e^{i\lambda_8\phi}.
\end{equation}
This leads to the non-redundant Euler parametrization
\begin{equation}\label{eq:SU3_euler_param}
    U_{SU(3)}^{(\alpha,\beta,\gamma,\theta,a,b,c,\phi)}
    =
    \underbrace{\:e^{i\lambda_3\alpha}e^{i\lambda_2\beta}e^{i\lambda_3\gamma} e^{i\lambda_5\theta\:}
    }_{\text{representative of }SU(3)/U(2)}
    \;
    \underbrace{
    e^{i\lambda_3 a}
    e^{i\lambda_2 b}
    e^{i\lambda_3 c}e^{i\lambda_8\phi}
    }_{U(2)\text{ subgroup}} .
\end{equation}
The first $SU(2)$ Euler block chooses a direction inside the $(1,2)$ subspace; the factor $e^{i\lambda_5\theta}$ mixes this chosen direction with $\ket{3}$; and the final $U_{SU(2)}e^{i\lambda_8\phi}$ factor keeps the full right $U(2)$ subgroup freedom~\cite{Byrd1998}. In this line, the order of generators is not arbitrary, and the parameter counting is transparent $4+4=8=\dim SU(3)$. However, one further global point is still independent of such ordering and counting. 

\noindent The product above fixes the local Euler coordinates, but the ranges of the angles must be chosen so that the corresponding Haar volume covers the group correctly. In the present normalization, a convenient covering choice is
\begin{equation}\label{eq:SU3_euler_angles_range}
    0\leq \alpha,a<\pi,
    \qquad
    0\leq \beta,\theta,b\leq \frac{\pi}{2},
    \qquad
    0\leq \gamma,c<2\pi,
    \qquad
    0\leq \phi<\sqrt{3}\pi .
\end{equation}
These ranges are obtained by computing the invariant Haar volume form associated with the Euler coordinates and matching the known $SU(3)$ group volume, with the necessary modifications coming from the embedded $SU(2)$ factors and the center $\mathbb Z_3$ (see Appendix B-C in \cite{TilmaByrdSudarshan2002}). In particular, for this parametrization, the Haar form is proportional to
\begin{equation}
    dV_{SU(3)}
    \propto
    \sin(2\beta)\,
    \sin(2b)\,
    \sin(2\theta)\,
    \sin^2\theta\,
    d\alpha\,d\beta\,d\gamma\,d\theta\,da\,db\,dc\,d\phi .
\end{equation}
Hence, with Eq.\ref{eq:SU3_euler_param} and Eq.\ref{eq:SU3_euler_angles_range}, one completes the construction of the $SU(3)$ Euler parametrization. The point to carry forward is that such construction has two crucial ingredients: (i) locally, one chooses $K$, $P$, and a stabilizer-fixed coset representative; (ii) globally, one fixes the angular domain by the Haar measure and the center of the group. With this in mind, we can now pass to $SU(4)$.

Indeed, for $SU(4)$, this same strategy becomes richer because $\dim SU(4)=15$. A complete Euler parametrization must therefore contain fifteen independent group parameters. However, there is no unique physically preferred way of arranging them. The choice of generators and the choice of subgroup $K$ determine which features of the resulting matrix are made transparent. In what follows, we discuss two useful examples. The first is the Tilma-type parametrization, which uses the canonical generalized Gell--Mann basis and follows the natural subgroup chain $SU(2)\subset SU(3)\subset SU(4)$. This construction is best viewed as a general coordinate system on $SU(4)$, and is especially useful when one wants to parametrize arbitrary two-qubit unitary transformations or density matrices.

\noindent \noindent The second is the Aristov-type parametrization (\textit{D.~N.~Aristov, private communication}), which is adapted to mesoscopic transport. In that case, the $SU(4)$ matrix is interpreted as a four-channel scattering matrix, and the parametrization is organized so that channel rephasings are separated from the physically relevant reflection and transmission data. Thus, while the Tilma construction is algebraically natural, the Aristov construction is physically adapted to scattering problems.

\subsubsection{Tilma parametrization: density-matrix adapted}

The Tilma-type parametrization is the direct $SU(4)$ extension of the $SU(3)$ construction discussed above. We again use the canonical generalized Gell--Mann basis $\{\lambda_a\}_{a=1}^{15}$, normalized as $\operatorname{Tr}(\lambda_a\lambda_b)=2\delta_{ab}$, and split the algebra into a subalgebra $\mathcal L(K)$ and a complementary sector $\mathcal L(P)$. The natural choice is
\begin{equation}
    \mathcal L(K)
    =
    \operatorname{span}\{\lambda_1,\ldots,\lambda_8,\lambda_{15}\},
    \qquad
    \mathcal L(P)
    =
    \operatorname{span}\{\lambda_9,\ldots,\lambda_{14}\}.
\end{equation}
Here $\lambda_1,\ldots,\lambda_8$ form an $SU(3)$ subalgebra acting on the first three basis states $\ket{1},\ket{2},\ket{3}$. The extra diagonal generator $\lambda_{15}$ is proportional to the identity inside this upper-left $3\times3$ block, and therefore commutes with the $SU(3)$ subalgebra. Hence
\begin{equation}
    \mathcal L(K)\simeq \mathfrak{su}(3)\oplus\mathfrak u(1),
\end{equation}
and exponentiating $\mathcal L(K)$ gives a $U(3)$-type subgroup embedded in $SU(4)$. The complementary generators $\lambda_9,\ldots,\lambda_{14}$ connect the fourth state $\ket{4}$ to the first three states, and therefore describe directions outside this $U(3)$ block. Moreover, note that the commutation relations have the same Cartan-decomposition structure as before,
\begin{equation}
    [\mathcal L(K),\mathcal L(K)]\subset \mathcal L(K),
    \qquad
    [\mathcal L(K),\mathcal L(P)]\subset \mathcal L(P),
    \qquad
    [\mathcal L(P),\mathcal L(P)]\subset \mathcal L(K).
\end{equation}
At the group level, one again starts schematically from $U=K\cdot P$. Since $K$ is now a $U(3)$-type subgroup, we can use the $SU(3)$ Euler parametrization from Eq.~\eqref{eq:SU3_euler_param} and append the extra $U(1)$ factor generated by $\lambda_{15}$. Thus a general $K$-factor can be written as
\begin{equation}
\begin{aligned}
    K&=U_{SU(3)}^{(\alpha,\beta,\gamma,\theta,a,b,c,\chi)}
     e^{i\lambda_{15}\phi}\\[4pt]
    &=
    e^{i\lambda_3\alpha}
    e^{i\lambda_2\beta}
    e^{i\lambda_3\gamma}
    e^{i\lambda_5\theta}
    e^{i\lambda_3 a}
    e^{i\lambda_2 b}
    e^{i\lambda_3 c}
    e^{i\lambda_8\chi}
    e^{i\lambda_{15}\phi}.
\end{aligned}
\end{equation}
Therefore $K$ contains $8+1=9$ parameters, as expected for a $U(3)$-type subgroup. To generate the complementary directions, one chooses a representative generator in $\mathcal L(P)$, conventionally $\lambda_{10}$, which mixes the fourth state with one chosen direction inside the first three-dimensional block. As in the $SU(3)$ case, the remaining complementary directions are generated by conjugating this representative with elements of $K$. Schematically,
\begin{equation}
    P=K'\,e^{i\lambda_{10}\psi}\,K'' \:\:\: \xRightarrow[\text{ element }]{\text{ general group }} \:\:\:U=
    K (K'\,e^{i\lambda_{10}\psi}\,K'')\:.
\end{equation}
Since $K$ is a subgroup, $K K'$ can be absorbed into a single factor, $K K'=\widetilde K$, giving
\begin{equation}
    U=\widetilde K\,e^{i\lambda_{10}\psi}\,K''.
\end{equation}
At this point, one must again fix the stabilizer freedom. A full $K\simeq U(3)$ factor has nine parameters, so the expression above would contain $9+1+9=19$ parameters, while $\dim SU(4)=15$. The four extra parameters come from the stabilizer of the chosen complementary direction inside $K$. Equivalently, the left part should not be a full $U(3)$ element; it should be a representative of the coset $SU(4)/U(3)$, whose dimension is
\begin{equation}
    \dim SU(4)-\dim U(3)=15-9=6.
\end{equation}
Thus the stabilizer quotient reduces the naive left factor $\widetilde K e^{i\lambda_{10}\psi}$ from $9+1=10$ parameters to a six-parameter representative of $SU(4)/U(3)$. This is the precise $SU(4)$ analogue of what happened in $SU(3)$, where the left $e^{i\lambda_8\phi}$ factor was omitted in order to obtain a representative of $SU(3)/U(2)$. In the Tilma parametrization, a convenient non-redundant choice of coset representative is
\begin{equation}
    C_{4/3}
    =
    e^{i\lambda_3\alpha_1}
    e^{i\lambda_2\alpha_2}
    e^{i\lambda_3\alpha_3}
    e^{i\lambda_5\alpha_4}
    e^{i\lambda_3\alpha_5}
    e^{i\lambda_{10}\alpha_6}.
\end{equation}
The first factors prepare a generic direction inside the three-dimensional block spanned by $\ket{1},\ket{2},\ket{3}$, while $e^{i\lambda_{10}\alpha_6}$ mixes that prepared direction with $\ket{4}$. The remaining factor is the full right $U(3)$-type subgroup,
\begin{equation}
    K_R
    =
    e^{i\lambda_3\alpha_7}
    e^{i\lambda_2\alpha_8}
    e^{i\lambda_3\alpha_9}
    e^{i\lambda_5\alpha_{10}}
    e^{i\lambda_3\alpha_{11}}
    e^{i\lambda_2\alpha_{12}}
    e^{i\lambda_3\alpha_{13}}
    e^{i\lambda_8\alpha_{14}}
    e^{i\lambda_{15}\alpha_{15}}.
\end{equation}
Therefore, a general element of $SU(4)$ can be written as $U=C_{4/3}K_R$, or explicitly,
\begin{equation}\label{eq:SU4_euler_param}
\begin{aligned}
U_{SU(4)}^{(\alpha_1,...,\alpha_{15})}
\:=\:\:\:&
\underbrace{
e^{i\lambda_3\alpha_1}
e^{i\lambda_2\alpha_2}
e^{i\lambda_3\alpha_3}
e^{i\lambda_5\alpha_4}
e^{i\lambda_3\alpha_5}
e^{i\lambda_{10}\alpha_6}
}_{\text{representative of }SU(4)/U(3)}
\\[4pt]
&\qquad\qquad\times
\underbrace{
e^{i\lambda_3\alpha_7}
e^{i\lambda_2\alpha_8}
e^{i\lambda_3\alpha_9}
e^{i\lambda_5\alpha_{10}}
e^{i\lambda_3\alpha_{11}}
e^{i\lambda_2\alpha_{12}}
e^{i\lambda_3\alpha_{13}}
e^{i\lambda_8\alpha_{14}}
e^{i\lambda_{15}\alpha_{15}}
}_{U(3)\text{-type subgroup}} \:,
\end{aligned}
\end{equation}
fulfilling the parameter counting $15=\dim SU(4)$. However, as in the $SU(3)$ case, local parameter counting does not yet determine the global angular ranges. These are fixed by the Haar measure. Since the generators $\lambda_j$ are Hermitian, while $U^{-1}dU$ is anti-Hermitian, we write the left-invariant Maurer--Cartan form as
\begin{equation}
    U^{-1}dU
    =
    i\sum_{j=1}^{15}\omega_j\lambda_j,
    \qquad
    \omega_j
    =
    \sum_{k=1}^{15}c_{jk}(\alpha)\,d\alpha_k,
    \qquad
    c_{jk}
    =
    \frac{1}{2i}
    \operatorname{Tr}
    \left[
        \lambda_j
        U^{-1}
        \frac{\partial U}{\partial\alpha_k}
    \right].
\end{equation}
The $\omega_j$'s are the invariant one-forms expressed in the Euler coordinates $(\alpha_1,\ldots,\alpha_{15})$. Hence the invariant volume form is their wedge product,
\begin{equation}
    \omega_1\wedge\cdots\wedge\omega_{15}
    =
    \det C(\alpha)\,
    d\alpha_1\wedge\cdots\wedge d\alpha_{15},
    \qquad
    C=(c_{jk}) .
\end{equation}
Thus $\det C$ is the Jacobian relating the invariant one-form basis to the coordinate basis. For the Tilma parametrization, this gives, up to normalization,
\begin{equation}
\begin{aligned}
    dV_{SU(4)}
    \propto{}&
    \cos^3(\alpha_4)
    \cos(\alpha_6)
    \cos(\alpha_{10})
    \sin(2\alpha_2)
    \sin(\alpha_4)
    \sin^5(\alpha_6)
    \\
    &\times
    \sin(2\alpha_8)
    \sin^3(\alpha_{10})
    \sin(2\alpha_{12})
    \,
    d\alpha_{15}\cdots d\alpha_1 \: .
\end{aligned}
\end{equation} 
The angular domain is then chosen so that the integral of this Haar form reproduces the known group volume $\operatorname{Vol}(SU(4))=\sqrt{2}\pi^9/3$. One convenient covering choice is
\begin{equation}\label{eq:SU4_euler_angles}
\begin{gathered}
    0\leq \alpha_1,\alpha_7,\alpha_{11}<\pi,
    \qquad
    0\leq
    \alpha_2,\alpha_4,\alpha_6,\alpha_8,\alpha_{10},\alpha_{12}
    \leq \frac{\pi}{2},
    \\[4pt]
    0\leq \alpha_3,\alpha_5,\alpha_9,\alpha_{13}<2\pi,
    \qquad
    0\leq \alpha_{14}<\sqrt{3}\pi,
    \qquad
    0\leq \alpha_{15}<2\sqrt{\frac{2}{3}}\,\pi \:.
\end{gathered}
\end{equation}
The triples $(\alpha_1,\alpha_2,\alpha_3)$, $(\alpha_7,\alpha_8,\alpha_9)$, and $(\alpha_{11},\alpha_{12},\alpha_{13})$ carry the usual $SU(2)$ Euler pattern $(\alpha,\beta,\gamma)\sim([0,\pi),[0,\pi/2],[0,2\pi))$, so that these embedded $SU(2)$ factors are fully covered, rather than only their $SU(2)/\mathbb Z_2$ quotients. The angle $\alpha_5$ is also extended to $2\pi$ since it belongs to the additional $SU(2)$-type factor associated with the $\lambda_{10}$ mixing. Furthermore, the remaining enlarged Cartan ranges account for the centers of the higher groups: $\alpha_{14}$ restores the $\mathbb Z_3$ center of the embedded $SU(3)$ block, while $\alpha_{15}$ restores the global center of $SU(4)$, namely $\mathbb Z_4=\{\mathbb 1,i\mathbb 1,-\mathbb 1,-i\mathbb 1\}$. These center corrections should not be confused with the $\mathbb Z_2$ centers of the embedded $SU(2)$ factors.

\noindent In this line, Eq.\ref{eq:SU4_euler_param}, together with the angular domain in Eq.\ref{eq:SU4_euler_angles}, gives an explicit Euler-angle parametrization of $SU(4)$. Now, the reason this $SU(4)$ parametrization is especially useful for quantum information is that a general two-qubit density matrix can be diagonalized as
\begin{equation}
    \rho
    =
    U\rho_d U^\dagger ,
\end{equation}
where $U\in SU(4)$ and $\rho_d=\operatorname{diag}(p_1,p_2,p_3,p_4)$, with $\sum_i p_i=1$. Thus, the eigenvalues contribute three independent real parameters. On the other hand, although $U$ contains fifteen Euler angles, the final Cartan factors generated by $\lambda_3$, $\lambda_8$, and $\lambda_{15}$ commute with $\rho_d$, and therefore do not change the density matrix. Hence, a generic two-qubit state is described by
\begin{equation}
    12\ \text{Euler angles}
    \;+\;
    3\ \text{eigenvalue parameters}
    =
    15
\end{equation}
real parameters, exactly matching the dimension of the space of $4\times4$ Hermitian trace-one matrices. In this way, the Euler parametrization gives a concrete coordinate system for the full two-qubit state space, not only for the group $SU(4)$ itself. 

This last fact is exactly the reason why the parametrization has been useful for exploring geometric questions about quantum states. The spectral decomposition $\rho=U\rho_dU^\dagger$ separates the unitary-orbit variables from the eigenvalue variables, making it useful for constructing explicit measures on spaces of mixed states, including Haar/Hurwitz and Bures/statistical-distinguishability measures~\cite{TilmaByrdSudarshan2002,Slater2002}. For two-qubit systems, this framework has been used especially in separability and entanglement geometry: the Euler/eigenvalue coordinates allow one to study separability volumes and probabilities, while criteria such as the Peres--Horodecki partial-transpose condition can be formulated directly in state-space coordinates~\cite{TilmaByrdSudarshan2002,Slater2006}. The same general problem also motivated broader developments. The recursive Euler construction was extended from $SU(N)$ to $U(N)$ and related coset spaces, giving volume formulas for unitary groups, projective spaces, and product measures for pure and mixed density matrices~\cite{TilmaSudarshan2002,TilmaSudarshan2004}. In parallel, alternative parametrizations and coset-based constructions were developed for unitary groups, density matrices, and subspaces, with the common goal of controlling redundancies while preserving positivity, invariant measures, and usefulness for optimization problems~\cite{SpenglerHuberHiesmayr2010}.

\subsubsection{Aristov parametrization: transport adapted}
The parametrization presented in this subsection follows an unpublished construction by \textit{D.N. Aristov (private communication)}. 

\noindent The Aristov-type construction starts from a different choice of subgroup $K$. In the Tilma parametrization, the distinguished decomposition was
$\mathbb C^4=\operatorname{span}\{\ket{1},\ket{2},\ket{3}\}\oplus\operatorname{span}\{\ket{4}\}$, leading to a $U(3)$-type subgroup. Here we instead use the block structure
\begin{equation}
    \mathbb C^4=\mathbb C^2_{12}\oplus\mathbb C^2_{34},
\end{equation}
which naturally singles out an $SU(2)$ acting on the $(1,2)$ block, another $SU(2)$ acting on the $(3,4)$ block, and one diagonal generator measuring the relative phase between both blocks.

\noindent To keep the notation connected with the canonical generalized Gell--Mann basis $\{\lambda_a\}_{a=1}^{15}$, we define an adapted basis $\{\tilde\lambda_a\}_{a=1}^{15}$. This is only a reordering of the off-diagonal generators, together with an orthogonal rotation in the Cartan subspace spanned by $\lambda_8$ and $\lambda_{15}$:
\begin{equation}
\renewcommand{\arraystretch}{1.3}
\begin{array}{c@{\qquad}c@{\qquad}cccc}
\tilde\lambda_1=\lambda_1
&
\tilde\lambda_4=\lambda_{13}
&
\tilde\lambda_7=\lambda_4
&
\tilde\lambda_8=\lambda_5
&
\tilde\lambda_9=\lambda_6
&
\tilde\lambda_{10}=\lambda_7
\\
\tilde\lambda_2=\lambda_2
&
\tilde\lambda_5=\lambda_{14}
&
\tilde\lambda_{11}=\lambda_9
&
\tilde\lambda_{12}=\lambda_{10}
&
\tilde\lambda_{13}=\lambda_{11}
&
\tilde\lambda_{14}=\lambda_{12}
\\
\tilde\lambda_3=\lambda_3
&
\tilde\lambda_6=-\frac{1}{\sqrt3}\lambda_8+\frac{\sqrt6}{3}\lambda_{15}
&
\multicolumn{4}{c}{
\tilde\lambda_{15}=\sqrt{\frac{2}{3}}\lambda_8+\frac{1}{\sqrt3}\lambda_{15}
}
\end{array}
\end{equation}
In this ordering, $\tilde\lambda_{1,2,3}$ generate $\mathfrak{su}(2)_{12}$, while $\tilde\lambda_{4,5,6}$ generate $\mathfrak{su}(2)_{34}$. The last Cartan generator is
$\tilde\lambda_{15}=\frac{1}{\sqrt2}\operatorname{diag}(1,1,-1,-1)$, and gives the relative $U(1)$ phase between the two blocks. Thus the adapted decomposition is
\begin{equation}
\begin{gathered}
    \mathcal L(K)
    =
    \operatorname{span}\{\tilde\lambda_1,\ldots,\tilde\lambda_6,\tilde\lambda_{15}\}
    \simeq
    \mathfrak{su}(2)_{12}\oplus\mathfrak{su}(2)_{34}\oplus\mathfrak u(1)_{\rm rel},
    \\
    \mathcal L(P)
    =
    \operatorname{span}\{\tilde\lambda_7,\ldots,\tilde\lambda_{14}\},
    \\[5pt]
    [\mathcal L(K),\mathcal L(K)]\subset\mathcal L(K), \qquad 
    [\mathcal L(K),\mathcal L(P)]\subset\mathcal L(P),
    \qquad
    [\mathcal L(P),\mathcal L(P)]\subset\mathcal L(K).
\end{gathered}
\end{equation}
Here $\mathcal L(P)$ contains the generators connecting the two blocks, $1,2\leftrightarrow3,4$. A general element of the $K$-subgroup can therefore be written as two $SU(2)$ Euler blocks and one relative $U(1)$ phase,
\begin{equation}
    K
    =
    \left(
    e^{i\alpha_1\tilde\lambda_3}
    e^{i\alpha_2\tilde\lambda_1}
    e^{i\alpha_3\tilde\lambda_3}
    \right)
    \left(
    e^{i\alpha_4\tilde\lambda_6}
    e^{i\alpha_5\tilde\lambda_4}
    e^{i\alpha_6\tilde\lambda_6}
    \right)
    e^{i\alpha_7\tilde\lambda_{15}} .
\end{equation}

\noindent At this point the construction resembles the Tilma logic: one may choose a representative $p\in\mathcal L(P)$ and try to generate the complementary directions by conjugating $e^{i\theta p}$ with $K$-factors. Since $K$ has seven parameters, the naive expression $U=K e^{i\theta p}K'$ has $7+1+7=15=\dim SU(4)$ parameters. However, for this $SU(2)\oplus SU(2)\oplus U(1)$-adapted subgroup, the parameter count is misleading. To see this, one should apply the same Haar-measure test used above. For a given parametrization $U(\alpha_1,\ldots,\alpha_{15})$, expand the left-invariant Maurer--Cartan form in the adapted basis,
\begin{equation}
    U^{-1}dU
    =
    i\sum_{j=1}^{15}\omega_j\tilde\lambda_j,
    \qquad
    \omega_j
    =
    \sum_{k=1}^{15}c_{jk}(\alpha)\,d\alpha_k,
    \qquad
    c_{jk}
    =
    \frac{1}{2i}\operatorname{Tr}
    \left[
    \tilde\lambda_j
    U^{-1}
    \frac{\partial U}{\partial\alpha_k}
    \right].
\end{equation}
The invariant volume form is again controlled by the determinant of the coefficient matrix $C=(c_{jk})$,
\begin{equation}
    \omega_1\wedge\cdots\wedge\omega_{15}
    =
    \det C(\alpha)\,
    d\alpha_1\wedge\cdots\wedge d\alpha_{15}.
\end{equation}
For the naive $K e^{i\theta p}K'$ ansatz, this matrix has only $\operatorname{rank}C=13$, rather than $15$. Equivalently, the Haar determinant vanishes, $\det C=0$, so the fifteen apparent parameters span only a thirteen-dimensional subset of $SU(4)$. Thus, one must use a more flexible product of $K$- and $P$-type factors, rather than a single $U=K\cdot P$ decomposition. The question is then how to choose the extra factors without losing the usefulness of the block-adapted split. A natural guide is provided by the Cartan part of $\mathcal L(K)$ itself: the three diagonal generators $\tilde\lambda_3,\tilde\lambda_6,\tilde\lambda_{15}$ act only by changing phases of the basis rays. It is therefore useful to separate these diagonal transformations from the genuinely mixing part of the group element.

\noindent Indeed, since $\ket{j}$ and $e^{i\theta_j}\ket{j}$ represent the same ray, diagonal phase transformations can be treated as rephasings of the ray representatives. After removing the overall phase, a general diagonal element of $SU(4)$ can be written as $\operatorname{diag}(e^{i\theta_j})_{j=1}^4$, with $\sum_j\theta_j=0$. In the $(12)+(34)$-adapted basis, these phases are conveniently parametrized as
\begin{equation}
\begin{gathered}
    \theta_1=\varphi_{12}+\varphi_{\rm rel},
    \qquad
    \theta_2=-\varphi_{12}+\varphi_{\rm rel},
    \qquad
    \theta_3=\varphi_{34}-\varphi_{\rm rel},
    \qquad
    \theta_4=-\varphi_{34}-\varphi_{\rm rel},
\end{gathered}
\end{equation}
so that $\varphi_{12}$ rephases the two rays inside the $(1,2)$ block oppositely, $\varphi_{34}$ does the same inside the $(3,4)$ block, and $\varphi_{\rm rel}$ measures the relative phase between both blocks. Equivalently,
\begin{equation}
\begin{aligned}
    R(\varphi_{12},\varphi_{34},\varphi_{\rm rel})
    &=
    \operatorname{diag}
    \left(
    e^{i\varphi_{12}+i\varphi_{\rm rel}},
    e^{-i\varphi_{12}+i\varphi_{\rm rel}},
    e^{i\varphi_{34}-i\varphi_{\rm rel}},
    e^{-i\varphi_{34}-i\varphi_{\rm rel}}
    \right)
    \\
    &=
    e^{\:i\:\left(\varphi_{12}\tilde\lambda_3
    +\varphi_{34}\tilde\lambda_6
    +\sqrt2\,\varphi_{\rm rel}\tilde\lambda_{15}\right)
    }\:.
\end{aligned}
\end{equation}
For a general element $U_{SU(4)}\in SU(4)$, one may act with such diagonal phase matrices from the left and from the right, $U_{SU(4)}\rightarrow R_L\,U_{SU(4)}\,R_R $. It is therefore natural to look for coordinates in which these two Cartan rephasings are isolated,
\begin{equation}
    U_{SU(4)}
    =
    R_L\,\overline U\,R_R .
\end{equation}
Each $R$ contains three Cartan angles, so the two end factors account for six parameters. The remaining core $\overline U$ must contain $15-2(4-1)=9$ parameters, giving $6_{\rm reph}+9_{\rm core}=15$. \noindent With this counting in mind, the remaining task is to choose a nine-parameter core $\overline U$ whose product with the two diagonal rephasings gives a full local parametrization of $SU(4)$. The simplest successful choice in the Aristov construction has the symbolic structure $\overline U=PKP$, rather than the naive $K e^{i\theta p}K'$. The $P$-type factor is generated not by one of the basis elements $\tilde\lambda_7,\ldots,\tilde\lambda_{14}$ alone, but by the symmetric inter-block direction
\begin{equation}
    \tilde\lambda_{16}
    :=
    \tilde\lambda_7+\tilde\lambda_9+\tilde\lambda_{11}+\tilde\lambda_{13}=\textstyle{\left(\begin{array}{llll}
0 & 0 & 1 & 1 \\
0 & 0 & 1 & 1 \\
1 & 1 & 0 & 0 \\
1 & 1 & 0 & 0
\end{array}\right)}
    \quad\in\mathcal L(P).
\end{equation}
The notation $\tilde\lambda_{16}$ does not enlarge $\mathfrak{su}(4)$; it only labels a particular direction inside the complementary sector. In terms of the decomposition $\mathbb C^4=\mathbb C^2_{12}\oplus\mathbb C^2_{34}$, it is the symmetric real generator connecting the two blocks.

\noindent Denoting the nine core parameters by $\eta_1,\ldots,\eta_9$, one may then write
\begin{equation}
    U_{SU(4)}
    =
    R_L\,
    \overline U(\eta_1,\ldots,\eta_9)\,
    R_R,
\end{equation}
with
\begin{equation}\label{eq:aristov_core_pkp}
\begin{aligned}
    \overline U(\eta_1,\ldots,\eta_9)
    =
    \underbrace{e^{i\eta_1\tilde\lambda_{16}}}_{P}
    \,
    \underbrace{
    e^{i\eta_2\tilde\lambda_3}
    e^{i\eta_3\tilde\lambda_1}
    e^{i\eta_4\tilde\lambda_3}
    e^{i\eta_5\tilde\lambda_6}
    e^{i\eta_6\tilde\lambda_4}
    e^{i\eta_7\tilde\lambda_6}
    e^{i\sqrt2\,\eta_8\tilde\lambda_{15}}
    }_{K}
    \,
    \underbrace{e^{i\eta_9\tilde\lambda_{16}}}_{P}.
\end{aligned}
\end{equation}
This realizes the desired parameter count directly: $3_L+9_{\rm core}+3_R=15$. \noindent However, the point is not only the number of parameters. In contrast with the naive $K e^{i\theta p}K'$ expression, the $R_L(PKP)R_R$ product has a nonvanishing Haar determinant on a generic region of parameter space. Therefore the two missing tangent directions of the naive ansatz are restored:
\begin{equation}
    K e^{i\theta p}K'
    \Rightarrow
    \operatorname{rank}C=13,
    \qquad
    R_L(PKP)R_R
    \Rightarrow
    \operatorname{rank}C=15
    \quad
    \text{generically}.
\end{equation}
For this parametrization, the Haar density factorizes as
\begin{equation}
\begin{gathered}
    dV
    =
    r_1(\boldsymbol\eta)\,r_2(\boldsymbol\eta)\,
    d\varphi_{\rm rel}^R\,d\varphi_{34}^R\,d\varphi_{12}^R\,
    d\eta_9\cdots d\eta_1\,
    d\varphi_{\rm rel}^L\,d\varphi_{34}^L\,d\varphi_{12}^L,
    \\[7pt]
    r_1=8\sqrt2\,\sin^3(2\eta_1)\cos(2\eta_1)
    \sin(2\eta_3)\sin(2\eta_6)\sin^3(2\eta_9) \cos(2\eta_9),
    \\
    \begin{aligned}
    r_2
    ={}&
    2\cos^2\eta_3\,
    \sin^2\eta^{(+)}_{24}
    \Big[
    \sin^2\eta^{(+)}_{57}
    \cos^2\eta_6
    \sin^2(2\eta_8)
    -
    \sin^2\eta^{(-)}_{57}
    \sin^2\eta_6
    \cos^2(2\eta_8)
    \Big]
    \\
    &\qquad +
    2\sin^2\eta_3\,
    \sin^2\eta^{(-)}_{24}
    \Big[
    \sin^2\eta^{(-)}_{57}
    \sin^2\eta_6
    \sin^2(2\eta_8)
    -
    \sin^2\eta^{(+)}_{57}
    \cos^2\eta_6
    \cos^2(2\eta_8)
    \Big]
    \\
    &\qquad\qquad-
    \sin(2\eta_3)
    \sin\eta^{(-)}_{24}
    \sin\eta^{(+)}_{24}
    \sin(2\eta_6)
    \sin\eta^{(-)}_{57}
    \sin\eta^{(+)}_{57}\:\:,
\end{aligned}
\end{gathered}
\end{equation}
with the notation $\eta^{(\pm)}_{ij}\equiv \eta_i\pm\eta_j$ for compactness.  Two features are worth keeping. First, the Haar density does not depend on the six rephasing angles contained in the two end factors $R_L$ and $R_R$, and the nontrivial Jacobian is controlled by the nine core angles $\eta_1,\ldots,\eta_9$. Second, although the Haar determinant is nonzero generically, the global angular domain is much less transparent than in the Tilma parametrization. In particular, the Haar density is periodic under shifts of the angles, but the naive integration domain $(0,\pi)^{\times15}$ is not a minimal covering. Integrating $|dV|$ over that domain overcounts the known 
$\operatorname{Vol}(SU(4))=\sqrt{2}\pi^9/3$, by a factor of $48$. A possible reduced integration domain is obtained by keeping the generic angles in $(0,\pi)$, while restricting
\begin{equation}
    \varphi_{\rm rel}^L,\eta_8,\varphi_{\rm rel}^R
    \in\left(0,\frac{\pi}{2}\right),
    \qquad
    \eta_1,\eta_9
    \in\left(0,\frac{\pi}{4}\right).
\end{equation}
However, because the zero structure of $r_2(\boldsymbol\eta)$ is complicated, this should be understood as a convenient reduced integration domain rather than a clean minimal global Euler-angle range of the type obtained in the Tilma construction. For our purposes, the important conclusion is therefore more modest: the $PKP$ construction gives a valid local Euler-angle parametrization, while the problem of choosing a nonredundant global angular domain is subtler than in the $U(3)$-adapted Tilma case.

\begin{figure}[h!]
    \centering
    \includegraphics[width=0.42\textwidth]{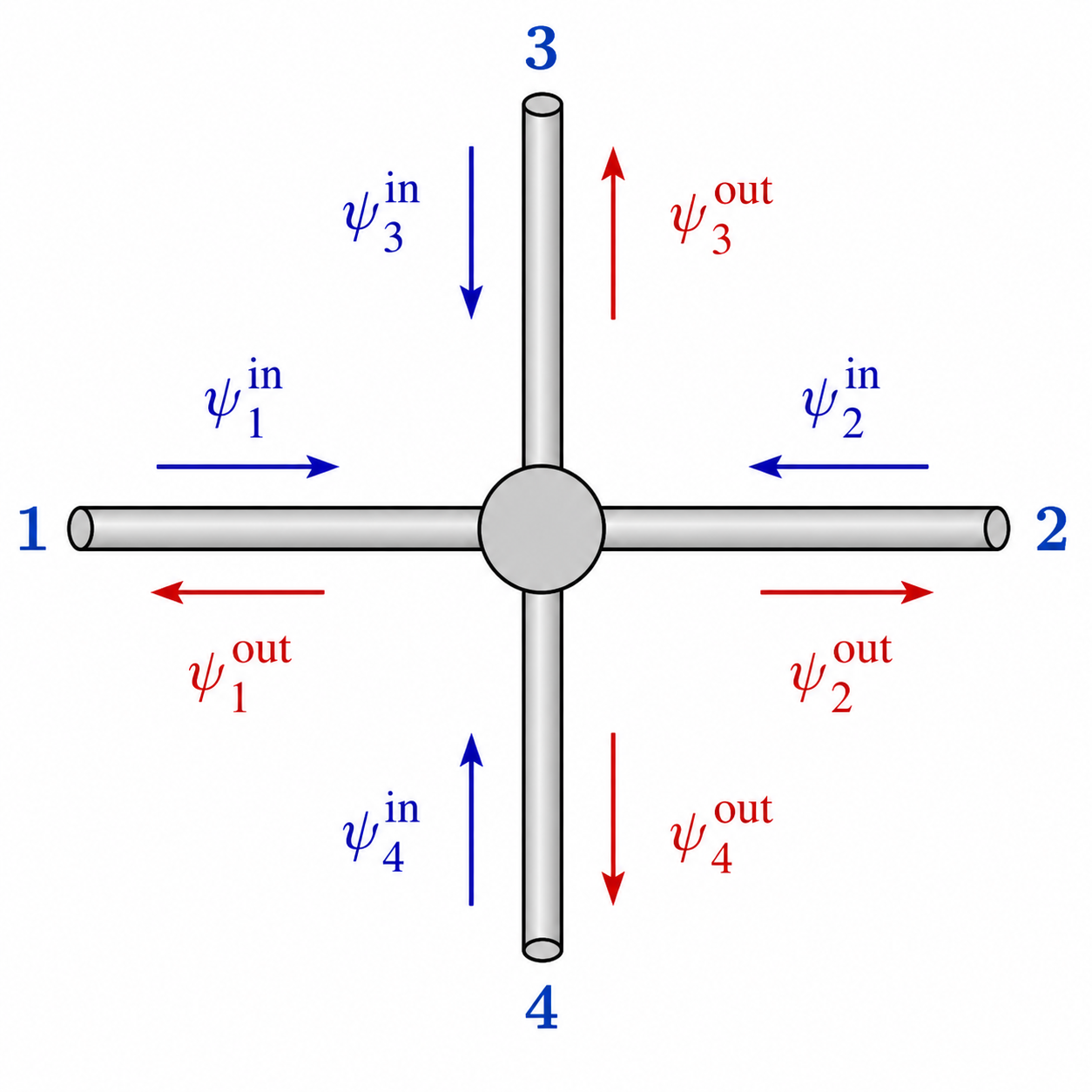}
    \caption{Four-terminal $X$-junction with channels $(1,2)$ and $(3,4)$ forming two wires.}
    \label{fig:x_junction_schematic}
\end{figure}

Now, the usefulness of this organization becomes more concrete once the four basis states are interpreted as asymptotic channels of a quantum junction. One can think of two one-dimensional wires crossing at a localized scattering region, with channel labels $1,2$ associated with one wire and $3,4$ with the other (Fig.\ref{fig:x_junction_schematic}). Far from the junction, each channel has an incoming and an outgoing wave amplitude, and at fixed energy the scattering region is described by
\begin{equation}
    \psi_i^{\rm out}
    =
    \sum_{j=1}^{4}S_{ij}\,\psi_j^{\rm in}.
\end{equation}
Thus $S_{ij}$ is the quantum amplitude for a particle entering through channel $j$ to leave through channel $i$. Current conservation implies that the full scattering matrix is unitary, $S\in U(4)$. Since its overall phase does not affect any scattering probability, one may remove it and work with the nontrivial part as an element of $SU(4)$. In this convention, the abstract group element $U_{SU(4)}$ discussed above is now interpreted as the scattering matrix $S$.

\noindent The left and right rephasings then acquire a simple meaning. They are just changes of phase convention for outgoing and incoming channel amplitudes. If $S\longrightarrow R_L S R_R$,
then each matrix element transforms as
\begin{equation}
    S_{ij}
    \longrightarrow
    e^{i\theta_i^L}\,S_{ij}\,e^{i\theta_j^R},
    \qquad
    |S_{ij}|^2
    \longrightarrow
    |S_{ij}|^2 .
\end{equation}
Therefore, any quantity depending only on scattering probabilities is invariant under the six rephasing angles. In particular, in the Landauer--Büttiker language, conductance coefficients are built from reflection and transmission probabilities; hence, up to conventional overall factors such as $e^2/h$, this dependence may be written schematically as
\begin{equation}
    G_{ij}
    =
    \delta_{ij}
    -
    |S_{ij}|^2 \:\:,
\end{equation}
 making clear that for the conductance only $|S_{ij}|^2$ are relevant, not the separate phases of $S_{ij}$. Therefore, the six end-factor parameters in $R_L$ and $R_R$ are invisible in conductance-type data. In this application, the abstract group-theoretic split $6_{\rm reph}+9_{\rm core}=15$ becomes the transport split $6_{\rm reph}+9_{\rm scatt}$, and the nine core parameters contain the rephasing-invariant scattering data. Moreover, the physical meaning of such nine parameters is especially transparent in the weak-hopping limit, where the four channels are almost detached. After an appropriate rephasing, one may write
\begin{equation}\label{eq:aristov_weak_hopping}
    S
    \simeq
    \mathbb 1+iT,
    \qquad
    T=T^\dagger,
    \qquad
    |T_{ij}|\ll1 .
\end{equation}
Here $T_{ij}$ is a small hopping amplitude between channels $j$ and $i$. Since $T=T^\dagger$, the independent off-diagonal pairs give six independent hopping magnitudes,
\begin{equation}
    |T_{12}|,\quad |T_{13}|,\quad |T_{14}|,\quad
    |T_{23}|,\quad |T_{24}|,\quad |T_{34}|.
\end{equation}
The remaining three parameters can be chosen as phases accumulated around closed loops. These phases are invariant because, under a channel rephasing, one has
\begin{equation}
    T_{ij}
    \longrightarrow
    e^{i\theta_i}T_{ij}e^{-i\theta_j},
\end{equation}
so all phase factors cancel in a closed product. A convenient choice of three independent loop phases is
\begin{equation}\label{eq:aristov_fluxes}
\begin{aligned}
    \Phi_0
    &=
    \arg\left(T_{13}T_{32}T_{24}T_{41}\right)
    =
    \arg\left(T_{13}T_{23}^*T_{24}T_{14}^*\right),
    \\
    \Phi_{132}
    &=
    \arg\left(T_{13}T_{32}T_{21}\right)
    =
    \arg\left(T_{13}T_{23}^*T_{12}^*\right),
    \\
    \Phi_{134}
    &=
    \arg\left(T_{13}T_{34}T_{41}\right)
    =
    \arg\left(T_{13}T_{34}T_{14}^*\right).
\end{aligned}
\end{equation}
These are the analogues of magnetic fluxes through independent loops of the channel graph: individual hopping phases can be changed by redefining channel phases, but the phase accumulated around a closed loop cannot. Thus, in the weak-hopping limit, the nine-dimensional core has the interpretation
\begin{equation}
    9_{\rm scatt}
    =
    6_{\rm magnitudes}
    +
    3_{\rm fluxes}.
\end{equation}
\noindent To connect this last part with the explicit core expression in Eq.\ref{eq:aristov_core_pkp}, let us expand the derived parametrization near the disconnected-channel limit. The outer rephasings can be set to the identity for this purpose, since they only change the phase convention of the channel amplitudes. Inside $\overline S$, the Cartan angles $\eta_2,\eta_4,\eta_5,\eta_7,\eta_8$ are kept finite, while the four mixing angles
$\eta_1,\eta_3,\eta_6,\eta_9$ are taken small. These are precisely the angles multiplying $\tilde\lambda_{16}$, $\tilde\lambda_1$, $\tilde\lambda_4$, and $\tilde\lambda_{16}$, and therefore they control the leading hopping processes. After a final diagonal rephasing that removes the zeroth-order diagonal phases, one obtains the weak-hopping form of Eq.\ref{eq:aristov_weak_hopping}, with independent upper-triangular entries
\begin{equation}\label{eq:aristov_T_entries}
\begin{aligned}
    T_{12}
    &=
    \eta_3,
    &
    T_{34}
    &=
    \eta_6,
    \\
    T_{13}
    &=
    e^{-i(\eta_2-\eta_5+2\eta_8)}\eta_1
    +
    e^{i(\eta_4-\eta_7)}\eta_9,
    &
    T_{14}
    &=
    e^{-i(\eta_2+\eta_5+2\eta_8)}\eta_1
    +
    e^{i(\eta_4+\eta_7)}\eta_9,
    \\
    T_{23}
    &=
    e^{i(\eta_2+\eta_5-2\eta_8)}\eta_1
    +
    e^{-i(\eta_4+\eta_7)}\eta_9,
    &
    T_{24}
    &=
    e^{i(\eta_2-\eta_5-2\eta_8)}\eta_1
    +
    e^{-i(\eta_4-\eta_7)}\eta_9 .
\end{aligned}
\end{equation}
This expression makes the $PKP$ structure visible. The angles $\eta_3$ and $\eta_6$ describe hopping inside the two blocks, $1\leftrightarrow2$ and $3\leftrightarrow4$, while $\eta_1$ and $\eta_9$ come from the two $P$-factors and generate inter-block hopping. The remaining core angles dress these hopping amplitudes with phases. In a generic situation, the resulting fluxes in Eq.\ref{eq:aristov_fluxes} are nonzero and the relation between Euler angles and microscopic hopping parameters is not especially simple.

\noindent Nevertheless, the expression simplifies considerably for a junction with identical wires and no flux. Identical wires (12) and (34) suggest equal intra-wire hopping, $\eta_3=\eta_6\equiv \alpha$.
Furthermore, a convenient fluxless choice is
\begin{equation}\label{eq:aristov_fluxless_choice}
    \eta_2=\eta_5=\eta_8=0,
    \qquad
    \eta_4=\eta_7=\frac{\pi}{2},
    \qquad
    \eta_1=\frac{\beta_1}{2},
    \qquad
    \eta_9=\frac{\beta_2}{2}\:,
\end{equation}
in which, the scattering matrix takes the symmetric form
\begin{equation}\label{eq:aristov_fluxless_smatrix}
\begin{gathered}
    S=
    \begin{pmatrix}
        r & t_1 & t_2 & t_3\\
        t_1 & r & t_3 & t_2\\
        t_2 & t_3 & r & t_1\\
        t_3 & t_2 & t_1 & r
    \end{pmatrix},\qquad \text{where}\\[11pt]
\begin{aligned}
    r
    &=
    \frac12
    \left(
    e^{i\alpha}\cos\beta_1
    +
    e^{-i\alpha}\cos\beta_2
    \right),
    &
    t_1
    &=
    \frac12
    \left(
    e^{i\alpha}\cos\beta_1
    -
    e^{-i\alpha}\cos\beta_2
    \right),
    \\
    t_2
    &=
    \frac{i}{2}
    \left(
    e^{i\alpha}\sin\beta_1
    +
    e^{-i\alpha}\sin\beta_2
    \right),
    &
    t_3
    &=
    \frac{i}{2}
    \left(
    e^{i\alpha}\sin\beta_1
    -
    e^{-i\alpha}\sin\beta_2
    \right).
\end{aligned}
\end{gathered}
\end{equation}
The amplitudes now have a direct junction interpretation. The coefficient $r$ is the reflection amplitude back into the same channel, $t_1$ is the transmission amplitude through the same wire, while $t_2$ and $t_3$ describe the two inequivalent transmissions into the other wire. The repeated pattern of entries reflects the assumed equivalence of the two wires, while the absence of a chiral phase makes the matrix symmetric under the exchange of incoming and outgoing directions in the corresponding fluxless geometry.

So, this is the strategy used in the analysis of the $X$-junction of interacting quantum wires, where the four basis states label the four asymptotic scattering channels of a four-lead junction \cite{AristovNiyazov2015}. In that problem, the $S$-matrix is the microscopic object from which the conductance matrix is constructed, but the conductances do not retain all rephasing-invariant information of the scattering problem. The split $S=R_L\,\overline S\,R_R$
therefore separates the six arbitrary channel phase conventions from the nine-parameter core $\overline S$, which contains the data relevant for the renormalization-group analysis. In particular, the ambiguity that appears when trying to formulate the flow only in terms of conductances is resolved by keeping the underlying $S$-matrix parametrization. Now, this should be viewed as part of the broader scattering-state approach to junctions of Luttinger-liquid wires. Already in three-lead $Y$-junctions, magnetic fluxes, chiral fixed points, and multi-terminal conductance matrices require a parametrization that keeps track of phase-sensitive scattering information \cite{ChamonOshikawaAffleck2003,AristovWoelfle2011}. The four-lead $X$-junction is more constrained: after removing incoming and outgoing rephasings, the remaining $SU(4)$ core has nine parameters, naturally interpreted in the weak-hopping limit as six hopping magnitudes plus three loop fluxes. 

Finally, it is worth mentioning that the parametrization studied above is not a generic two-qubit parametrization of $SU(4)$, but a transport-adapted one, organized around the block structure $(12)+(34)$ and the rephasing-invariant data of a four-channel junction. This contrasts with the more common $SU(4)$ decompositions used in two-qubit quantum information, where the physical redundancy is instead local basis rotation. There, one typically separates local $SU(2)\otimes SU(2)$ operations from a three-parameter nonlocal entangling core, as in Cartan or $KAK$ decompositions \cite{KhanejaGlaser2001,ZhangValaWhaley2003}. The same group $SU(4)$ is therefore being organized according to different physical questions: local-versus-entangling structure for two-qubit gates, and rephasing-versus-scattering structure for the $X$-junction.

%% file: appendix3_spin_covers_su-so.tex
\section{Spin covering groups and low-rank SU-SO relations}\label{app:low_rank_su_so}
Throughout the preceding sections, several exceptional relations between low-dimensional unitary and orthogonal groups have appeared in different contexts. The  $SU(2)\to SO(3)$ relation explains why quantum rotations admit both integer- and half-integer-spin representations; the decomposition $\mathfrak{so}(4)\simeq\mathfrak{su}(2)\oplus\mathfrak{su}(2)$ organizes the singlet--triplet basis of a four-dimensional Hilbert space; and the isomorphism $\mathfrak{su}(4)\simeq\mathfrak{so}(6)$ leads to the alternative coherent-state description $\mathbb{CP}^{3}\simeq SO(6)/U(3)$. Although these statements arise in different physical applications, they are manifestations of the same low-rank structure. The purpose of this appendix is to place them within a common framework by distinguishing Lie-algebra isomorphisms from global group relations, introducing the spin groups and their covering maps, and explaining how these coverings act on the relevant representations and the coset spaces representing the $\mathbb{CP}^N$ manifolds.

\subsection{Lie algebras, covering groups, and spin groups}

A Lie algebra describes the infinitesimal structure of a Lie group near the identity, but it does not completely determine its global topology. Therefore, two groups may possess isomorphic Lie algebras while still being globally different. The simplest example is $\mathfrak{su}(2)\simeq\mathfrak{so}(3)$, although $SU(2)\not\simeq SO(3)$ as Lie groups. They describe the same infinitesimal rotations, but differ globally because $SU(2)$ is simply connected, whereas $SO(3)$ is not. More generally, a smooth surjective group-manifold homomorphism 
\begin{equation}
    p:\widetilde G\to G
\end{equation}
is called a \emph{covering homomorphism} when it is locally one-to-one. Thus, sufficiently close to the identity, $\widetilde G$ and $G$ have the same group structure (an isomorphism of tangent spaces), even though globally several elements of $\widetilde G$ may correspond to the same element of $G$. The covered group can then be written as 
\begin{equation}
    G\simeq\widetilde G/\ker p\:.
\end{equation}
Now, a covering is called a \emph{double cover} when each element of $G$ has exactly two preimages. Equivalently, $\ker p\simeq\mathbb Z_2$. In the cases considered below, these two lifts will be denoted by $g$ and $-g$, even though they represent the same transformation in the covered group.

Among all possible covering groups, a distinguished role is played by the \emph{universal covering group}. For a connected Lie group $G$, this is the simply connected Lie group $\widetilde G$ that has the same Lie algebra as $G$ and covers it globally. A universal covering need not be a double covering in general; the number of lifts is determined by the topology of the original group. However, for the special orthogonal groups, one has $\pi_1(SO(n))\simeq\mathbb Z_2$ for $n\geq3$. Consequently, their universal covering groups are double covers. These groups are called the \textbf{spin groups} \cite{LawsonMichelsohn1989} and are denoted by $\operatorname{Spin}(n)$. They have the same Lie algebra as $SO(n)$, namely $\operatorname{Lie}(\operatorname{Spin}(n))\simeq\mathfrak{so}(n)$, while the covering homomorphism $\pi:\operatorname{Spin}(n)\to SO(n)$ has kernel $\{1,-1\}\simeq\mathbb Z_2$. This relation is summarized by the \emph{exact sequence}
\begin{equation}
    1
    \longrightarrow
    \mathbb Z_2
    \longrightarrow
    \operatorname{Spin}(n)
    \overset{\pi}{\longrightarrow}
    SO(n)
    \longrightarrow
    1,
\end{equation}
or, equivalently,
\begin{equation}
    SO(n)
    \simeq
    \frac{\operatorname{Spin}(n)}{\mathbb Z_2}.
\end{equation}
Thus, every rotation $R\in SO(n)$ has two lifts $g$ and $-g$ in $\operatorname{Spin}(n)$, and crucially, the physical distinction between tensor and spinor representations is totally determined by how these two lifts act. To make this precise, consider an irreducible unitary representation $\rho:\operatorname{Spin}(n)\to U(V)$. If the nontrivial element of the covering kernel acts trivially,
\begin{equation}
    \rho(-1)=\mathbb 1_V,
\end{equation}
then $\rho(-g)=\rho(g)$, and the transformation depends only on the corresponding rotation $R$. In this case, the representation also defines an ordinary representation of $SO(n)$, and we say that $\rho$ \emph {descends} to $SO(n)$. By contrast, when
\begin{equation}
    \rho(-1)=-\mathbb 1_V,
\end{equation}
the two lifts act as $\rho(-g)=-\rho(g)$, and the representation is then well defined on $\operatorname{Spin}(n)$, but not as an ordinary representation of $SO(n)$. Nevertheless, the two operators differ only by an overall phase and therefore define the same transformation on quantum-mechanical rays. Such representations are called \emph{spinorial representations}; equivalently, they define projective representations of $SO(n)$. For $\operatorname{Spin}(3)\simeq SU(2)$, this reduces to the familiar distinction between integer and half-integer spin.

The exceptional relations relevant to the present work arise because, in a few low dimensions, the spin groups are isomorphic to familiar unitary groups,
\begin{equation}
    \operatorname{Spin}(3)\simeq SU(2),
    \qquad
    \operatorname{Spin}(4)\simeq SU(2)\times SU(2),
    \qquad
    \operatorname{Spin}(6)\simeq SU(4).
\end{equation}
We now examine these three cases and their consequences for the representations and coherent-state manifolds encountered in the main text.

\subsection{\texorpdfstring{From $\operatorname{Spin}(3)$ projective spinors to the $\mathbb{CP}^{1}$ formulation of the $O(3)$ NLSM}{From Spin(3) spinors to the CP1 formulation of the O(3) NLSM}}

The simplest of the exceptional low-rank relations is
\begin{equation}
    \operatorname{Spin}(3)\simeq SU(2),
    \qquad
    SO(3)\simeq\frac{SU(2)}{\mathbb Z_2}.
\end{equation}
The corresponding double covering was already discussed in the context of quantum rotations. Here, we revisit the same relation from the viewpoint of coherent states and field theory. The central idea is that a $\operatorname{Spin}(3)$ spinor and an $SO(3)$ vector provide two complementary descriptions of the same physical spin orientation, leading to the equivalent realizations
\begin{equation}
    \mathbb{CP}^{1}
    \simeq
    \frac{SU(2)}{U(1)}
    \simeq
    \frac{SO(3)}{SO(2)}
    \simeq
    S^2.
\end{equation}
Earlier in the main text, the topology of the $SU(2)$ coherent-state manifold was already transparent from the identification $\mathbb{CP}^{1}\simeq S^2$, so homogeneous coordinates were not required. Here, however, they are useful because they make the projective and spinorial structures explicit. To expose this structure, let us return to the coherent states introduced in Eq.\ref{eq:spin_cohe_exp}. Making the representation dependence explicit, we write $\rho_S(\hat S_a)$ for the realization of the abstract $\mathfrak{su}(2)$ generator $\hat S_a$ in the spin-$S$ irrep. Thus,
\begin{equation}\label{eq:spinS_coherent_state_explicit_rep}
    \ket{g(\phi,\theta^\prime)}_S
    :=
    e^{-i\phi\rho_S(\hat S_3)}
    e^{-i\theta^\prime\rho_S(\hat S_2)}
    \ket{S,S}
    \in\mathcal H_S.
\end{equation}
The subscript $S$, which was suppressed in the main text, is included here only to indicate the representation in which the group action is evaluated. In the fundamental case, the corresponding coherent state is therefore
\begin{equation}\label{eq:fundamental_coherent_state_spinor}
    \begin{aligned}
        \ket{g(\phi,\theta^\prime)}_{\frac12}
        &=
        e^{-i\phi/2}
        \begin{pmatrix}
            \cos(\theta^\prime/2)\\
            e^{i\phi}\sin(\theta^\prime/2)
        \end{pmatrix}
        =
        e^{-i\phi/2}\bm z(\phi,\theta^\prime),
    \end{aligned}
\end{equation}
where we define (see Eq.\ref{eq:irrep_spin_cohere} with $\theta=\theta^\prime/2$ and $\varphi_2=\phi$)
\begin{equation}
    \begin{aligned}
        \bm z(\phi,\theta^\prime)
        &:=
        \begin{pmatrix}
            z_0\\
            z_1
        \end{pmatrix}
        =
        \begin{pmatrix}
            \cos(\theta^\prime/2)\\
            e^{i\phi}\sin(\theta^\prime/2)
        \end{pmatrix},
        \qquad
        \bm z^\dagger\bm z=1.
    \end{aligned}
\end{equation}
The factor $e^{-i\phi/2}$ is an overall phase and therefore does not affect the physical ray. Thus, $\bm z$ is not an additional unrelated object; it is what we previously called $\ket{\mathbf{n}_2}$. 

\noindent More generally, a point of $\mathbb{CP}^{1}$ is represented by a pair of nonzero complex numbers $[z_0:z_1]$, called \emph{homogeneous coordinates}. Two pairs $(z_0,z_1)\sim\lambda(z_0,z_1)$ represent the same projective point whenever they differ by an overall nonzero complex factor $\lambda\in\mathbb C\smallsetminus\{0\}$.
Choosing a normalized representative gives the two-component complex vector
\begin{equation}
    \bm z
    =
    \begin{pmatrix}
        z_0\\
        z_1
    \end{pmatrix},
    \qquad
    \bm z^\dagger\bm z
    =
    |z_0|^2+|z_1|^2
    =
    1.
\end{equation}
Normalization then fixes the absolute value of $\lambda$, but leaves its phase undetermined. Therefore, the normalized representative remains subject to the projective identification $\bm z\sim e^{i\alpha}\bm z$. Once said that, it is clear that the vector $\bm z(\phi,\theta^\prime)$ introduced above is precisely a particular normalized representative of the homogeneous coordinates $[z_0:z_1]$. The residual identification $\bm z\sim e^{i\alpha}\bm z$ simultaneously expresses the projective equivalence of $\mathbb{CP}^{1}$ and the irrelevance of the overall phase of the fundamental coherent states.

Now, the key aspect is that the fundamental coordinate $\bm z$ can be used to construct the coherent states of every spin-$S$ irrep. Representation-theoretically, the corresponding Hilbert space is the completely symmetric product of $2S$ fundamental spaces, $\mathcal H_S\simeq\operatorname{Sym}^{2S}\!\left(\mathbb C^2\right)$; hence, the maximum-weight state of the spin-$S$ irrep is $\ket{S,S}\leftrightarrow\ket{\uparrow}_{1/2}^{\otimes 2S}$. Since the same group element acts on each fundamental factor, it is convenient to define the corresponding spin-$S$ coherent-state representative as
\begin{equation}\label{eq:spinS_coherent_representative_z}
    \ket{\bm z,S}
    :=
    \bm z^{\otimes 2S}
    \in
    \operatorname{Sym}^{2S}\!\left(\mathbb C^2\right).
\end{equation}
In this way, $\bm z$ may be regarded as a representative of the coherent-state orientation, while $\ket{\bm z,S}$ is the corresponding representative in the spin-$S$ Hilbert space. In terms of the angular parametrization introduced above, the state constructed directly in the spin-$S$ irrep and the one constructed from $\bm z$ represent the same physical ray:
\begin{equation}\label{eq:spinS_coherent_state_from_z}
    \boxed{
        \left[\ket{g(\phi,\theta^\prime)}_S\right]
        =
        \left[\ket{\bm z(\phi,\theta^\prime),S}\right]
        =
        \left[
            \bm z(\phi,\theta^\prime)^{\otimes 2S}
        \right]
    }.
\end{equation}
The equality is stated at the level of rays because the particular vector representatives may differ by an overall phase. In fact, under the projective transformation $\bm z\to e^{i\alpha}\bm z$, one has
\begin{equation}
    \ket{\bm z,S}
    \longrightarrow
    \ket{e^{i\alpha}\bm z,S}
    =
    e^{i2S\alpha}\ket{\bm z,S},
\end{equation}
which changes the spin-$S$ coherent state only by an overall phase. Therefore, the same projective coordinate $[\bm z]$ labels the coherent states for every value of $S$, and their coherent-state manifold remains
\begin{equation}\label{eq:spinS_coherent_manifold_CP1}
    \boxed{
        \mathcal M_{\mathrm{coherent}}^{(S)}
        \simeq
        \mathbb{CP}^{1},
        \qquad
        S>0
    }.
\end{equation}
For $S=\frac12$, the full projective Hilbert space is already $\mathbb P(\mathcal H_{1/2})\simeq\mathbb{CP}^{1}$, so every state ray is coherent. For $S>\frac12$, however, the full space of rays is larger, and the coherent states form the distinguished submanifold $\mathcal M_{\mathrm{coherent}}^{(S)}\simeq\mathbb{CP}^{1}\subset\mathbb P(\mathcal H_S)\simeq\mathbb{CP}^{2S}$. Thus, the coherent-state manifold is the same for every spin, although its realization inside the corresponding projective Hilbert space depends on $S$. We can now distinguish carefully the transformation properties of the orientation representative $\bm z$, the coherent-state vector $\ket{\bm z,S}$, and the corresponding physical ray. 

Let $\pi:\operatorname{Spin}(3)\to SO(3)$ denote the double-covering map, and let $\widetilde r\in\operatorname{Spin}(3)$ be one of the two elements associated with the physical rotation $R_{\widetilde r}=\pi(\widetilde r)$. In the fundamental representation, this element is represented by the matrix
\begin{equation}
    u_{\widetilde r}
    :=
    \rho_{\frac12}(\widetilde r)
    \in SU(2),
\end{equation}
and therefore acts on the homogeneous coordinate according to
\begin{equation}\label{eq:spinor_orientation_transformation}
    \bm z
    \longrightarrow
    u_{\widetilde r}\bm z.
\end{equation}
This is the precise sense in which $\bm z$ is a \emph{spinor representative} of the coherent-state orientation: it is a two-component object transforming in the fundamental representation of $\operatorname{Spin}(3)\simeq SU(2)$. On the other hand, the corresponding coherent-state vector $\ket{\bm z,S}$, transforms in the spin-$S$ irrep. Since this representation is obtained as the $2S$-fold symmetric power of the fundamental one, $\rho_S(\widetilde r)\simeq\operatorname{Sym}^{2S}\!\left(u_{\widetilde r}\right)$, and consequently
\begin{equation}\label{eq:spinS_coherent_representative_transformation}
    \begin{aligned}
        \rho_S(\widetilde r)\ket{\bm z,S}
        &=
        \left(u_{\widetilde r}\bm z\right)^{\otimes 2S}
        \\
        &=
        \ket{u_{\widetilde r}\bm z,S}.
    \end{aligned}
\end{equation}
Therefore, the fact that $\bm z$ is a spinor does not imply that the coherent-state vector always transforms in the fundamental spinorial representation. Instead, $\ket{\bm z,S}$ transforms tensorially for integer $S$ and spinorially for half-integer $S$. This distinction becomes especially clear from the nontrivial element $-1\in\ker\pi$. Its action on the two representatives is
\begin{equation}\label{eq:Spin3_center_action_on_representatives}
    \bm z
    \longrightarrow
    -\bm z,
    \qquad
    \rho_S(-1)\ket{\bm z,S}
    =
    (-1)^{2S}\ket{\bm z,S}.
\end{equation}
Thus, integer-spin state vectors are unchanged under the nontrivial element of the covering and their representations descend to ordinary representations of $SO(3)$. Half-integer-spin state vectors, instead, acquire a minus sign and therefore carry genuine spinorial representations of $\operatorname{Spin}(3)$.

Nevertheless, coherent states describe physical rays rather than particular Hilbert-space vectors. Since vectors differing by an overall phase represent the same physical state,
\begin{equation}\label{eq:projective_center_identification}
    [\bm z]
    =
    [-\bm z],
    \qquad
    \left[
        \rho_S(-1)\ket{\bm z,S}
    \right]
    =
    \left[
        \ket{\bm z,S}
    \right].
\end{equation}
Consequently, after passing to the coherent-state rays, it no longer matters whether the chosen vector representative transforms tensorially or spinorially. The two elements $\widetilde r$ and $-\widetilde r$, which correspond to the same ordinary rotation, induce the same transformation of the projective point. Therefore, the action on the coherent-state manifold $\mathcal M_{\mathrm{coherent}}^{(S)}\simeq\mathbb{CP}^{1}$ factors through $SO(3)$ for every value of $S$. The question here must be then: Which phase-independent quantity constructed from $\bm z$ realizes this $SO(3)$ action? 

The natural starting point is not an arbitrary bilinear, but the rank-one projector associated with the normalized coordinate vector $\bm z$,
\begin{equation}\label{eq:CP1_rank_one_projector}
    P_{[\bm z]}
    :=
    \bm z\otimes\bm z^\dagger
    \in
    \mathbb C^2\otimes(\mathbb C^2)^*
    \simeq
    \operatorname{End}(\mathbb C^2).
\end{equation}
Note that this projector is insensitive to the projective phase, since
\begin{equation}
    P_{[e^{i\alpha}\bm z]}
    =
    \left(
        e^{i\alpha}\bm z
    \right)
    \otimes
    \left(
        e^{i\alpha}\bm z
    \right)^\dagger
    =
    \bm z\otimes\bm z^\dagger
    =
    P_{[\bm z]},
\end{equation}
and therefore, $P_{[\bm z]}$ depends only on the projective point $[\bm z]\in\mathbb{CP}^{1}$. Moreover, the normalization $\bm z^\dagger\bm z=1$ implies $P_{[\bm z]}^\dagger=P_{[\bm z]}$, $P_{[\bm z]}^2=P_{[\bm z]}$, $\operatorname{Tr}P_{[\bm z]}=1$, making it a Hermitian rank-one projector with unit trace. Hence, since every Hermitian $2\times2$ operator can be expanded uniquely in the basis formed by the identity and the three Pauli matrices, it admits the decomposition
\begin{equation}\label{eq:CP1_projector_spinor_vector}
        P_{[\bm z]}
        =
        \bm z\otimes\bm z^\dagger
        =
        \frac12
        \left(
            \mathbb 1_2
            +
            \vec{\mathbf n}\cdot\vec{\bm\sigma}
        \right),
\end{equation}
where the coefficient of $\mathbb 1_2$ is fixed by $\operatorname{Tr}P_{[\bm z]}=1$, while the three real coefficients multiplying the Pauli matrices define the vector
\begin{equation}\label{eq:CP1_spinor_vector_map}
    \boxed{\mathrm{n}_a
    :=
    \operatorname{Tr}
    \left(
        P_{[\bm z]}\sigma_a
    \right)
    =
    \bm z^\dagger\sigma_a\bm z,
    \qquad
    \vec{\mathbf n}
    =
    \bm z^\dagger\vec{\bm\sigma}\bm z\:}\:.
\end{equation}
Writing $\bm z=(z_0,z_1)^{\mathsf T}$ and using $\bm z^\dagger\bm z=1$, one obtains $\mathbf n^2=\left(|z_0|^2+|z_1|^2\right)^2=1$, making the fact $\mathbf n\in S^2$ explicit. In this way, the phase-independent projector associated with the ray $[\bm z]$ naturally determines a unit vector on the two-sphere. To make this correspondence explicit, let us consider the angular chart used in Eqs.\ref{eq:irrep_spin_cohere},\ref{eq:exp_value}. In this parametrization, the map $\vec{\mathbf n}=\bm z^\dagger\vec{\bm\sigma}\bm z$ becomes
\begin{equation}\label{eq:CP1_angular_spinor_vector_map}
    \bm z(\phi,\theta^\prime)
    =
    \begin{pmatrix}
        \cos(\theta^\prime/2)\\
        e^{i\phi}\sin(\theta^\prime/2)
    \end{pmatrix}
    \quad\longmapsto\quad
    \mathbf n(\phi,\theta^\prime)
    =
    \begin{pmatrix}
        \sin\theta^\prime\cos\phi\\
        \sin\theta^\prime\sin\phi\\
        \cos\theta^\prime
    \end{pmatrix},
\end{equation}
such that the resulting vector is precisely the one previously obtained from the coherent-state expectation values $\bra{\bm z,S}\rho_S(\hat{\mathbf S})\ket{\bm z,S}=S\mathbf n$. Hence, the projective spinor $[\bm z]$ and the unit vector $\mathbf n$ encode the same coherent-state orientation, while the spin quantum number $S$ fixes the magnitude of the corresponding spin expectation value.

We can now verify that this vector realizes precisely the ordinary $SO(3)$ action anticipated above. Using the same notation as before, we already know that the relation between the fundamental and vector representations is encoded in the conjugation of the Pauli matrices: $u_{\widetilde r}^\dagger \sigma_a u_{\widetilde r}=(R_{\widetilde r})_{ab}\sigma_b$. Consequently, under the spinorial transformation $\bm z\to u_{\widetilde r}\bm z$, one finds
\begin{equation}\label{eq:Spin3_induced_SO3_vector_action}
    \begin{aligned}
        \mathrm n_a
        \longrightarrow
        \mathrm n_a'
        &=
        \bm z^\dagger
        u_{\widetilde r}^\dagger
        \sigma_a
        u_{\widetilde r}
        \bm z
        \\[1mm]
        &=
        (R_{\widetilde r})_{ab}
        \bm z^\dagger\sigma_b\bm z
        =
        (R_{\widetilde r})_{ab}\mathrm n_b.
    \end{aligned}
\end{equation}
Thus, the fundamental action of $\operatorname{Spin}(3)$ on the spinor representative $\bm z$ induces the ordinary vector action of $SO(3)$ on $\mathbf  n$. In particular, the two lifts $u_{\widetilde r}$ and $-u_{\widetilde r}$ produce the same transformation, since their relative sign cancels inside the projector: $\left(-u_{\widetilde r}\bm z\right)\otimes\left(-u_{\widetilde r}\bm z\right)^\dagger=\left(u_{\widetilde r}\bm z\right)\otimes\left(u_{\widetilde r}\bm z\right)^\dagger$.

Taken together, the construction above is precisely the Hopf map $\bm z\mapsto\mathbf n$ from $S^3$ to $S^2$, whose fibers are the phase orbits $\bm z\sim e^{i\alpha}\bm z$ already identified in the projective description. The new point is that this same redundancy has a direct counterpart in the vector formulation. Indeed, $SO(3)$ acts transitively on $S^2$, while the transformations that leave a given unit vector fixed are rotations around its axis and therefore form an $SO(2)$ subgroup. Hence, the sphere can equivalently be realized as $S^2\simeq SO(3)/SO(2)$. So, combining the projective spinor and vector descriptions, one finally recovers
\begin{equation}\label{eq:CP1_SU2_SO3_cosets}
        \mathbb{CP}^{1}
        \simeq
        \frac{\operatorname{Spin}(3)}{U(1)}
        \simeq
        \frac{SU(2)}{U(1)}
        \simeq
        \frac{SO(3)}{SO(2)}
        \simeq
        S^2\:.
\end{equation}
This equivalence follows directly from the covering map $\operatorname{Spin}(3)\simeq SU(2)\to SO(3)$: the $U(1)$ subgroup preserving the spinor ray $[\bm z]$ maps onto the $SO(2)$ subgroup preserving the associated vector $\mathbf n$. Since the kernel $\{\pm\mathbb 1_2\}$ is already contained in the $U(1)$ stabilizer, it introduces no additional identification after taking the quotient, and all the isomorphisms make sense.

\subsubsection*{\texorpdfstring{Relation with bosonic partons and the $C-\mathbb{CP}^1$ model}{Relation with bosonic partons and the CCP1 model}}

Up to this point, the discussion has been purely kinematical: we have shown that the orientation of a single spin coherent state can be described equivalently by the projective spinor $[\bm z]\in\mathbb{CP}^{1}$ or by the unit vector $\mathbf n\in S^2$. To connect this geometric correspondence with the field-theoretical and parton descriptions used in Sec.\ref{sec:sun_heisenberg_models}, the next step is to allow this orientation to vary slowly in space and imaginary time. In the continuum description of a bipartite antiferromagnet, the vector $\mathbf n$ is then interpreted as the local N\'eel field, while $\bm z$ becomes its spacetime-dependent $\mathbb{CP}^{1}$ representative. Accordingly, the spinor--vector map is promoted pointwise to
\begin{equation}\label{eq:CP1_spacetime_spinor_vector_map}
    \vec{\mathbf n}(\bm x,\tau)
    =
    \bm z^\dagger(\bm x,\tau)
    \vec{\bm\sigma}
    \bm z(\bm x,\tau),
    \qquad
    \bm z^\dagger(\bm x,\tau)
    \bm z(\bm x,\tau)
    =
    1.
\end{equation}
While the physical interpretation of $\mathbf n$ is direct, due to
\begin{equation}\label{eq:Neel_field_continuum_identification}
    \left\langle
        \hat{\mathbf S}_i(\tau)
    \right\rangle
    \simeq
    \eta_i S\,
    \mathbf n(\bm x_i,\tau),
    \qquad
    \eta_i=\pm1,
\end{equation}
where $\eta_i$ distinguishes the two sublattices; the physical meaning of the projective field $\bm z(\bm x,\tau)$ could become obscure for the time being. Furthermore, once $\bm z$ is promoted to a spacetime-dependent field, it becomes important to specify again in which sense it is called a \emph{spinor field}. The answer is that the word \emph{spinor} continues to refer exclusively to the transformation of $\bm z$ under the internal spin-rotation group. More precisely, for $u_{\tilde r}\in SU(2)_{\mathrm{spin}}\simeq\operatorname{Spin}(3)$,
\begin{equation}
    \bm z(\bm x,\tau)
    \longrightarrow
    u_{\tilde r}\,\bm z(\bm x,\tau),
\end{equation}
so its two components form a doublet of the internal spin symmetry. This action must be distinguished from rotations of Euclidean spacetime. Under a  $(\bm x,\tau)\mapsto(\bm x',\tau')$ spacetime transformation, the field carries no spacetime spinor index and transforms instead as a scalar,
\begin{equation}
    \bm z'(\bm x',\tau')
    =
    \bm z(\bm x,\tau).
\end{equation}
Although in $2+1$ Euclidean dimensions the double cover of the rotation group is also abstractly isomorphic to $SU(2)$, namely $\operatorname{Spin}(3)_{\mathrm E}\simeq SU(2)_{\mathrm E}$, this is a different copy of the group: $SU(2)_{\mathrm{spin}}$ acts on the internal index of $z_\alpha$, whereas $SU(2)_{\mathrm E}$ acts on spacetime.

This distinction also removes any apparent tension between calling $\bm z$ a spinor and treating it as a bosonic field. The spin--statistics relation concerns the representation carried under spacetime rotations, or under the Lorentz group before Wick rotation, rather than quantum numbers associated with an internal symmetry. The components $z_\alpha(\bm x,\tau)$ are therefore ordinary commuting complex fields,
\begin{equation}
    z_\alpha z_\beta
    =
    z_\beta z_\alpha,
\end{equation}
and are bosonic despite transforming as a spin-$\frac12$ doublet under $SU(2)_{\mathrm{spin}}$. Thus, the terms \emph{spinorial} and \emph{bosonic} describe two independent properties of $\bm z$: the former specifies its internal transformation law, while the latter specifies its field statistics.

Once this point is clear, the appearance of such $\bm z (\bm x, \tau )$ can now be identified with the continuum limit of the bosonic parton (Schwinger-boson) representation of spin operators. As reviewed in Sec.\ref{sec:sun_heisenberg_models}, bosonic partons realize the $SU(N)$ generators through bilinears of operators carrying a flavor index and, in general, an auxiliary color index. The present $SU(2)$ case is particularly simple: flavor index only takes two values $\alpha=1,2$ and, since every spin-$S$ irrep is completely symmetric, only one color is required. Then, specializing Eqs.\ref{eq:bosonic_generators_colored},\ref{eq:bosonic_constraints_AB} to $N=2$, $m=1$, and $n_c=2S$, the generators on a given site become
\begin{equation}\label{eq:SU2_bosonic_parton_generators}
    \hat S^\beta_\alpha
    =
    b_\alpha^\dagger b_\beta
    -
    S\,\delta^\beta_\alpha,
    \qquad
    \sum_{\alpha=1}^{2}
    b_\alpha^\dagger b_\alpha
    =
    2S.
\end{equation}
At first sight, this operator-valued expression may appear unrelated to the normalized field $\bm z$. However, the connection becomes transparent after introducing the normalized bosonic operators $\hat\zeta_\alpha=\frac{b_\alpha}{\sqrt{2S}}$, 
\begin{equation}\label{eq:normalized_Schwinger_bosons}
    \frac{\hat S^\beta_\alpha}{2S}\,
    =
    \hat\zeta_\alpha^\dagger
    \hat\zeta_\beta
    -
    \frac12\delta^\beta_\alpha,
    \qquad
    \sum_{\alpha=1}^{2}
    \hat\zeta_\alpha^\dagger
    \hat\zeta_\alpha
    =
    1\:,
\end{equation}
and contracting the two-index generators with the Pauli matrices,
\begin{equation}
    \hat S^a
    =
    \frac12\:
    (\sigma^a)^\alpha_\beta\:
    \hat S^\beta_\alpha= \frac12
        b^\dagger_\alpha (\sigma^a)^{\alpha \beta}  b_\beta \quad \longrightarrow \quad \hat S^a= S  \:\bm{\hat\zeta}^\dagger \sigma^a
    \bm{\hat\zeta}
\end{equation}
The equation above makes the relation immediate. In the spin coherent-state path integral, the normalized spin operator is represented by the unit vector $\mathbf n$, whereas in the Schwinger-boson coherent-state path integral the normalized operator spinor $\bm{\hat\zeta}$ is represented by the commuting complex field $\bm z$. Then, after allowing these amplitudes to vary slowly in spacetime, the resulting dictionary is
\begin{equation}\label{eq:Schwinger_boson_CP1_dictionary}
    \boxed{
        \bm{\hat\zeta}_i
        \longrightarrow
        \bm z(\bm x,\tau),
        \qquad
        \frac{\hat{\mathbf S}_i}{S}
        \longrightarrow
        \vec{\mathbf n}(\bm x,\tau)
        =
        \bm z^\dagger(\bm x,\tau)
        \vec{\bm\sigma}
        \bm z(\bm x,\tau),
        \qquad
        \bm z^\dagger\bm z
        =
        1
    }.
\end{equation}
Thus, $\bm z$ is the normalized coherent-state amplitude of the Schwinger bosons, while the relation $b_\alpha=\sqrt{2S}\,\hat\zeta_\alpha$ restores their fixed total occupation $2S$. As a remark, note that this simple identification cannot be made with fermionic partons: one would need that the fermionic operators $c_\alpha$ are represented in the path integral by Grassmann fields $\psi_\alpha$, whereas the components of $\bm z$ are commuting complex fields.

The same dictionary also identifies the projective phase of $\bm z$ with the local phase freedom already present in the Schwinger-boson representation. Since the spin operators are bilinear in $b_{i\alpha}$, one has that the projective identification $\bm z\sim e^{i\alpha}\bm z$ becomes, in the continuum description,
\begin{equation}\label{eq:CP1_local_U1_redundancy}
    \bm z(\bm x,\tau)
    \sim
    e^{i\alpha(\bm x,\tau)}
    \bm z(\bm x,\tau),
    \qquad
    \mathbf n
    \longrightarrow
    \mathbf n.
\end{equation}
Thus, this local $U(1)$ transformation is not an additional physical symmetry, but the projective redundancy involved in representing the N\'eel vector through $\bm z$. In this sense, $z_\alpha$ is a \emph{bosonic spinon field}: it is a commuting spin-$\frac12$ doublet under $SU(2)_{\mathrm{spin}}$ and carries unit charge under the local $U(1)$ redundancy. Whether it describes a deconfined excitation remains a separate dynamical question.

Now, in the context of the Heisenberg Hamiltonian and the $O(3)$ NLSM, this local phase freedom becomes nontrivial once derivatives are introduced. To see how, recall that the long-wavelength action derived in the main text contains both a smooth dynamical contribution and the lattice Berry phase,
\begin{equation}\label{eq:O3_full_action_CP1_appendix}
\begin{gathered}
    S[\mathbf n]
    =
    S_{\mathrm{dyn}}[\mathbf n]
    +
    S_{\mathrm{top}}^{\prime}[\mathbf n],
    \\[0.5em]
    \text{with}
    \qquad
    \left\{
    \begin{aligned}
        S_{\mathrm{dyn}}[\mathbf n]
        &=
        \frac{1}{2g}
        \int_{0}^{\beta}d\tau
        \int d^{2}x\,
        \left[
            (\partial_{\tau}\mathbf n)^{2}
            +
            c^{2}(\nabla\mathbf n)^{2}
        \right],
        \\[0.4em]
        S_{\mathrm{top}}^{\prime}[\mathbf n]
        &=
        i4\pi S
        \sum_j
        (-1)^{j_x+j_y}
        W_0[
            \widetilde{\mathbf n}(\bm x_j)
        ] ,
    \end{aligned}
    \right.
    \\[0.6em]
    \text{where}
    \qquad
    W_0[
        \widetilde{\mathbf n}(\bm x_j)
    ]
    =
    \frac{1}{4\pi}
    \int_0^\beta d\tau
    \int_0^1du\,
    \widetilde{\mathbf n}
    \cdot
    \left(
        \partial_u\widetilde{\mathbf n}
        \times
        \partial_\tau\widetilde{\mathbf n}
    \right).
\end{gathered}
\end{equation}
As discussed in the main text, the first term governs the smooth fluctuations of the N\'eel field, whereas the second retains the microscopic staggered Berry phases of the lattice spins. We can now translate both contributions using $\vec{\mathbf n}=\bm z^\dagger\vec{\bm\sigma}\bm z$:

\begin{itemize}

\item For the smooth term, direct differentiation of the spinor--vector map gives
\begin{equation}\label{eq:O3_CP1_direct_kinetic_identity}
    \left(
        \partial_\mu\mathbf n
    \right)^2
    =
    4
    \left[
        (\partial_\mu\bm z)^\dagger
        (\partial_\mu\bm z)
        -
        \left|
            \bm z^\dagger\partial_\mu\bm z
        \right|^2
    \right],
    \qquad
    \mu=\tau,x,y.
\end{equation}
Note that, since normalization places $\bm z$ on $S^3$ but physical configurations are obtained after phase identification, the direction proportional to $i\bm z$ lies along the $U(1)$ fiber and does not modify the physical point $[\bm z]\in\mathbb{CP}^1$. Correspondingly, $\partial_\mu\bm z$ contains both a physical variation of the projective state and a component along this phase direction. Therefore, the quantity $\bm z^\dagger\partial_\mu\bm z$ measures precisely this phase component, and moreover, it is purely imaginary:
$\left(\bm z^\dagger\partial_\mu\bm z\right)^*
=-\bm z^\dagger\partial_\mu\bm z$.
This motivates the following definitions and transformation laws:
\begin{equation}\label{eq:CP1_connection_covariant_derivative}
\begin{aligned}
    (\text{connection})\quad
    a_\mu
    &:=
    -i\bm z^\dagger\partial_\mu\bm z,
    \hspace{1cm}
    &
    (\text{covariant derivative})\quad
    D_\mu
    &:=
    \partial_\mu-ia_\mu,
    \\[0.4em]
    (\text{gauge transf.})\quad
    a_\mu
    &\longrightarrow
    a_\mu+\partial_\mu\alpha,
    &
    (\text{covariant transf.})\quad
    D_\mu\bm z
    &\longrightarrow
    e^{i\alpha}D_\mu\bm z,
    \\[0.4em]
    (\text{horizontal proj.})\quad
    D_\mu\bm z
    &=
    \partial_\mu\bm z
    -
    \bm z
    \left(
        \bm z^\dagger\partial_\mu\bm z
    \right),
    &
    (\text{horizontality})\qquad
    \bm z^\dagger & D_\mu\bm z
    =0.
\end{aligned}
\end{equation}
Thus, $a_\mu$ records how the phase convention chosen for $\bm z$ changes from one spacetime point to another, while $D_\mu\bm z$ removes the component of $\partial_\mu\bm z$ along the phase fiber. With these definitions, Eq.\ref{eq:O3_CP1_direct_kinetic_identity} can be written as $(\partial_\mu\mathbf n)^2 =4(D_\mu\bm z)^\dagger(D_\mu\bm z)$, and the smooth part of the $O(3)$ action consequently becomes
\begin{equation}\label{eq:CP1_smooth_dynamical_action}
    S_{\mathrm{smooth}}[\bm z,a]
    =
    \frac{2}{g}
    \int_{0}^{\beta}d\tau
    \int d^{2}x\,
    \left[
        (D_\tau\bm z)^\dagger
        (D_\tau\bm z)
        +
        c^{2}
        \sum_{i=x,y}
        (D_i\bm z)^\dagger
        (D_i\bm z)
    \right].
\end{equation}

Geometrically, what is really happening is that these constructions are tied to the local spacetime realization of the principal Hopf bundle  $P\left(\mathbb{CP}^{1},U(1)\right)$. The normalized spinors $\{\bm z\}$ form the total space $S^3$, while points of the base manifold $\mathbb{CP}^{1}$ are of the form $\{[\bm z]\}$. In this line the true physical field is not a particular normalized spinor, but its projective class. We express that as $\phi:\mathcal M_{\mathrm E}\longrightarrow\mathbb{CP}^{1}$, with $\phi(\bm x,\tau)= [\bm z(\bm x,\tau)]$. In this language, since the fiber $\pi^{-1}([\bm z])$ consists of all normalized spinors $e^{i\alpha}\bm z$, a local section
$\sigma_i:U_i\subset\mathbb{CP}^{1}\to S^{3}$
selects one such representative over each point of a patch $U_i$. Consequently, wherever $\phi(\bm x,\tau)\in U_i$, the spacetime spinor field is the local lift $\bm z_\sigma(\bm x,\tau)=(\sigma\circ\phi)(\bm x,\tau)$, fulfilling $\pi\!\left(\bm z_\sigma(\bm x,\tau)\right) =\phi(\bm x,\tau)$.
Thus, the local $U(1)$ transformation introduced above is precisely the freedom to change the section used to represent the same projective field, since a different section $\sigma'$ selects another point in the same fiber and therefore in the intersection of patches we have $\bm z_{\sigma'}=e^{i\alpha(\bm x,\tau)}\bm z_\sigma$. Finally, if we denote a generic point in the total space as $\bm\xi\in S^{3}$, the canonical connection on the Hopf bundle is written as $\omega_P=-i\bm\xi^\dagger d\bm\xi$, and therefore, once a section $\sigma$ is chosen,  its local pullback to spacetime gives
\begin{equation}\label{eq:CP1_Hopf_connection_pullback}
    a_\sigma
    =
    (\sigma\circ\phi)^*\omega_P
    =
    -i\bm z_\sigma^\dagger d\bm z_\sigma
    =
    a_\mu dx^\mu.
\end{equation}
Thus, $a_\mu=-i\bm z^\dagger\partial_\mu\bm z$ is the local spacetime representative of the canonical connection, such that changing the section gives
$a_{\sigma'}=a_\sigma+d\alpha$, or equivalently
$a_\mu\to a_\mu+\partial_\mu\alpha$, exactly reproducing the transformation law obtained above.

\item For the lattice Berry-phase term, the previous discussion has already replaced the local phase freedom of $\bm z$ by the connection $a_\mu$. However, the Berry-phase contribution is sensitive to the global topological sector of the configuration rather than to the particular local section used to describe it. The appropriate quantity is therefore the curvature of this connection,
\begin{equation}\label{eq:CP1_curvature_definition}
    \mathcal F
    :=
    da,
    \qquad
    \mathcal F_{\mu\nu}
    =
    \partial_\mu a_\nu
    -
    \partial_\nu a_\mu,
\end{equation}
which remains gauge invariant. The connection with the original $O(3)$ variables follows directly from the spinor--vector map. On a fixed spatial slice, substituting
$a_i=-i\bm z^\dagger\partial_i\bm z$
and
$\vec{\mathbf n}=\bm z^\dagger\vec{\bm\sigma}\bm z$
gives $\mathcal F_{xy}
    =
    \frac12\,
    \mathbf n\cdot
    \left(
        \partial_x\mathbf n
        \times
        \partial_y\mathbf n
    \right)$. The curvature density is therefore exactly one half of the skyrmion-density integrand already encountered in the $O(3)$ description. Consequently,
\begin{equation}\label{eq:CP1_skyrmion_curvature_flux}
    Q_{\mathrm{sk}}(\tau)
    =
    \frac{1}{4\pi}
    \int d^2x\,
    \mathbf n\cdot
    \left(
        \partial_x\mathbf n
        \times
        \partial_y\mathbf n
    \right)
    =
    \frac{1}{2\pi}
    \int
    \mathcal F
    \in
    \mathbb Z.
\end{equation}
Thus, the integer skyrmion sectors of the N\'eel field are re-expressed as quantized curvature-flux sectors of the $U(1)$ connection. The monopole interpretation now follows directly from this relation. As reviewed in the main text, $Q_{\mathrm{sk}}(\tau)$ cannot change during a smooth evolution in imaginary time, so one is forced to consider instead isolated singular events at a set of spacetime points $X_n=(\bm x_n,\tau_n)$, and surround each one by a small two-sphere $S_n^2$. Finally, since their respective charges are written as
\begin{equation}\label{eq:CP1_monopole_charge}
        Q_n
        :=
        \frac{1}{2\pi}
        \int_{S_n^2}
        \mathcal F
        =
        Q_{\mathrm{sk}}(\tau_n^+)
        -
        Q_{\mathrm{sk}}(\tau_n^-)
        \in
        \mathbb Z,
\end{equation}
we can clearly recognize that what was previously introduced as instantonic/hedgehog charge in the $O(3)$ formulation is now expressed as the quantized curvature flux emerging from an isolated spacetime point. In the $\mathbb{CP}^{1}$ language, the same singularity is called a \emph{monopole event}: a charge-$Q_n$ event inserts or removes a flux $2\pi Q_n$. Locally, this can be summarized as
\begin{equation}\label{eq:CP1_monopole_Bianchi_identity}
    d\mathcal F
    =
    2\pi
    \sum_n
    Q_n\,
    \delta^{(3)}(X-X_n)\,
    d^3X,
\end{equation}
while $d\mathcal F=0$ away from the singular points. This also specifies the meaning of \emph{compact} in the $C\text{-}\mathbb{CP}^{1}$ model. The path integral includes nontrivial $U(1)$-bundle sectors around the singular points, labeled by the integer monopole charges $Q_n$. In a noncompact formulation, by contrast, $a_\mu$ is restricted to a globally defined $\mathbb R$-valued field and such flux-changing sectors are excluded.

In this line, the lattice Berry phase then retains exactly the behavior derived in the main text: it vanishes for smooth configurations and produces a position-dependent interference pattern among the singular events. The only new interpretation is that the hedgehogs carrying those Berry phases are now represented as quantized monopole-flux insertions of the compact $U(1)$ connection.

\end{itemize}

\noindent Collecting therefore both contributions, the complete correspondence takes the form
\begin{equation}\label{eq:O3_compact_CP1_full_correspondence}
    \boxed{
        S_{\mathrm{dyn}}[\mathbf n]
        +
        S_{\mathrm{top}}^{\prime}[\text{hedgehogs}]
        \quad
        \longleftrightarrow
        \quad
        S_{C\text{-}\mathbb{CP}^{1}}
        =
        S_{\mathrm{smooth}}[\bm z,a]
        +
        S_{\mathrm{top}}^{\prime}[\text{monopoles}]
    }\:,
\end{equation}
from which, we can clearly see how the low-rank relation $\operatorname{Spin}(3)\simeq SU(2)$ organizes the complete kinematical dictionary between the two formulations. The spinor bilinear reconstructs the N\'eel vector, the projective phase produces the local $U(1)$ redundancy, the skyrmion number becomes gauge flux, and hedgehog tunnelings become monopole events. Nevertheless, the microscopic coefficient of the Berry phase, its staggered lattice pattern, and the resulting interference between monopoles are not fixed by the group isomorphism alone; they additionally depend on the spin representation and on the microscopic lattice structure.

\subsection{\texorpdfstring{$\operatorname{Spin}(4)$ vectors and spinors}{Spin(4) vectors and spinors}}

In Appendix~\ref{app:su4-so4-singlet-triplet-basis}, the singlet--triplet basis of the four-dimensional $SU(4)$ space led naturally to the six generators $\{\mathbf s_1,\mathbf s_2\}$, or equivalently $\{\mathbf J,\mathbf R\}$. Their commutation relations realize the Lie-algebra identification $\mathfrak{so}(4)\simeq\mathfrak{su}(2)_1\oplus\mathfrak{su}(2)_2$. We also anticipated there that the corresponding relation at the group level is
\begin{equation}\label{eq:spin4_group_covering_relation}
    \operatorname{Spin}(4)
    \simeq
    SU(2)_1\times SU(2)_2,
    \qquad
    SO(4)
    \simeq
    \frac{SU(2)_1\times SU(2)_2}
    {\mathbb Z_2^{\mathrm{diag}}}\:,
\end{equation}
where the covering kernel is the subgroup
$\mathbb Z_2^{\mathrm{diag}}=\{(\mathbb 1_2,\mathbb 1_2),
(-\mathbb 1_2,-\mathbb 1_2)\}$. In that discussion, we used the fact that the two-qubit space
$\mathbb C^2_{\mathrm{qbt}\text{-}1}\otimes
\mathbb C^2_{\mathrm{qbt}\text{-}2}$
carries the four-dimensional tensor (vector) representation of $SO(4)$, but postponed the explanation of why the corresponding $\operatorname{Spin}(4)$ representation descends through the quotient by $\mathbb Z_2^{\mathrm{diag}}$. Here, we will make this point explicit, to then later contrast this tensor-product representation with the genuinely spinorial representations of $\operatorname{Spin}(4)$.

The most direct way to understand this $\operatorname{Spin}(4)\to SO(4)$ representation descent is to exploit the product structure in Eq.\ref{eq:spin4_group_covering_relation}. Since an irreducible representation of $SU(2)_1\times SU(2)_2$ is obtained by choosing independently an irrep of each factor, the irreducible representations of $\operatorname{Spin}(4)$ can be labeled by two spins $(j_1,j_2)$, such that their representation spaces are $\mathcal V_{(j_1,j_2)}=\mathcal V_{j_1}\otimes\mathcal V_{j_2}$, with dimension $(2j_1+1)(2j_2+1)$.

Now, the relevant point for distinguishing tensorial and spinorial representations is the action of the nontrivial element of $\mathbb Z_2^{\mathrm{diag}}$. In an $SU(2)$ spin-$j$ irrep, the central element $-\mathbb 1_2$ acts as $(-1)^{2j}$. Hence, in the general $(j_1,j_2)$ representation \cite{FultonHarris1991}
\begin{equation}\label{eq:spin4_center_action_general_irrep}
    \rho_{(j_1,j_2)}
    \left(
        -\mathbb 1_2,
        -\mathbb 1_2
    \right)
    =
    (-1)^{2(j_1+j_2)}
    \mathbb 1_{\mathcal V_{(j_1,j_2)}}.
\end{equation}
Therefore, whenever $j_1+j_2\in\mathbb Z$, the diagonal element of the covering kernel acts trivially and the representation descends to $SO(4)$. By contrast, when $j_1+j_2\in\mathbb Z+\frac12$, the two lifts differ by an overall minus sign and the representation is genuinely spinorial. Hence, when in Appendix~\ref{app:su4-so4-singlet-triplet-basis}, we have the two-qubit representation, each qubit carries the fundamental $\mathbf 2$ of one $SU(2)$, so that $j_1+j_2=1$, making that $\mathbb Z_2$ acts trivially and the representation descends to $SO(4)$. This correspondence can be summarized as
\begin{equation}\label{eq:SU4_fundamental_spin4_vector_correspondence}
    \boxed{
        \left.
        \mathbf 4_{SU(4)}
        \right|_{\operatorname{SO}(4)}
        \;\longleftrightarrow\;
        \mathbb C^2_{\mathrm{qbt}\text{-}1}
        \otimes
        \mathbb C^2_{\mathrm{qbt}\text{-}2}
        \;\simeq\;
        (\mathbf 2,\mathbf 2)_{SU(2)}
        \;\xleftarrow{\;\text{tensorial}\;}
        \mathbf 4^{\mathrm{vec}}_{\operatorname{Spin}(4)}
    }.
\end{equation}
The superscript "$\mathrm{vec}$" emphasizes that the descended tensorial representation is not an arbitrary $SO(4)$ tensor representation, but specifically its defining vector irrep. By this we simply mean the natural action of $SO(4)$ on a four-component real vector $\vec{x}=(x_i)_{i=1,...,4}\in\mathbb R^4$, namely $x_a\to R_{ab}x_b$, with $R\in SO(4)$. The relation with the $(\mathbf 2,\mathbf 2)_{SU(2)}$ representation becomes transparent by arranging the real vector $x\in\mathbb R^4$ into the $2\times2$ matrix $X(x)=x_4\mathbb 1_2+i x_k\sigma_k$. The two factors of $\operatorname{Spin}(4)\simeq SU(2)_1\times SU(2)_2$ then act as
\begin{equation}\label{eq:spin4_vector_matrix_action}
    \begin{aligned}
        &X\:
    \longrightarrow
    \:X'=U_1 X U_2^\dagger\\[-0.2cm]
     &\:\:x\:
    \longrightarrow\:
    x'
    \end{aligned}\:,
    \qquad
    \det X
    =|x|^2=|x'|^2= \det X'.
\end{equation}
Since this transformation preserves $\det X$, it preserves the norm of the corresponding four-vector $x\in \mathbb R^4$ and therefore induces an ordinary $SO(4)$ rotation. 
\vspace{0.6cm}

Now, regarding the corresponding possibilities for spinorial representations, the same product structure
$\mathcal V_{(j_1,j_2)}=\mathcal V_{j_1}\otimes\mathcal V_{j_2}$ is appealing. As mentioned above, a representation is spinorial whenever $j_1+j_2\in\mathbb Z+\frac12$. Hence, the smallest such representations are obtained by taking one of the two spins equal to $\frac12$ and the other equal to $0$. In other words, one $SU(2)$ factor acts through its fundamental irrep $\bm 2$ doublet, while the other acts trivially (as the identity). This gives two inequivalent, but equally two-dimensional, irreducible representations. At this point, it is convenient to relabel the two factors as $SU(2)_L$ and $SU(2)_R$, according to which one acts nontrivially:
\begin{equation}\label{eq:Spin4_chiral_spinors}
    \left\{\begin{aligned}
        S_L
        &\simeq
        \mathcal V_{\frac12}\otimes\mathcal V_0
        \simeq
        \mathbb C_L^2\otimes\mathbb C
        \simeq
        \mathbb C_L^2\equiv
        (\mathbf 2,\mathbf 1)
        \\[1mm]
        S_R
        &\simeq
        \mathcal V_0\otimes\mathcal V_{\frac12}
        \simeq
        \mathbb C\otimes\mathbb C_R^2
        \simeq
        \mathbb C_R^2\equiv
        (\mathbf 1,\mathbf 2)
    \end{aligned}\right.
\end{equation}
These are the two chiral spinor irreps of $\operatorname{Spin}(4)$. Thus, for
$\psi_L\in\mathbb C_L^2$ and $\psi_R\in\mathbb C_R^2$, a group element
$(U_L,U_R)\in SU(2)_L\times SU(2)_R$ acts as
\begin{equation}
\begin{cases}
    [\text{in } (\mathbf 2,\mathbf 1)] \qquad \psi_L\to(U_L\otimes 1)\psi_L=U_L\psi_L\\[1mm]
    [\text{in } (\mathbf 1,\mathbf 2)] \qquad \psi_R\to(1\otimes U_R)\psi_R=U_R\psi_R\:\:.
\end{cases}
\end{equation}
Note that, in this case, the nontrivial element $(-\mathbb 1_2,-\mathbb 1_2)$ of the covering kernel, the spin-$\frac12$ factor contributes $-1$, while the spin-$0$ factor contributes $+1$. Hence both chiral spinors acquire an overall minus sign and therefore do not descend to representations of $SO(4)$.

Furthermore, since the two chiral spinors have the same dimension but transform under different $SU(2)$ factors, it is often useful to combine them into a single four-component object rather than selecting one chirality. Their direct sum defines the Dirac spinor representation,
\begin{equation}\label{eq:Spin4_Dirac_spinor}
    \begin{pmatrix}
        \psi_L\\
        \psi_R
    \end{pmatrix}\in \mathbb C^4 \:,
    \qquad
    S_D
    =
    S_L\oplus S_R
    =
    (\mathbf 2,\mathbf 1)
    \oplus
    (\mathbf 1,\mathbf 2)\:,\qquad \rho_D(U_L,U_R)=\begin{pmatrix}
        U_L&0\\
        0&U_R
    \end{pmatrix}.
\end{equation}
which, in essence, is a four-dimensional representation, but a reducible $\operatorname{Spin}(4)$ one: its two invariant two-dimensional sectors are precisely the left- and right-chiral spinors. Here, it is important to highlight that, although $S_D$ is again four-dimensional over $\mathbb C$, it must not be confused with the vector representation discussed above. The distinction is structural:
\begin{equation}\label{eq:Spin4_vector_Dirac_comparison}
    \boxed{
    \begin{aligned}
        \mathbf 4^{\mathrm{vec}}_{\operatorname{Spin(4)}}
        &\longleftrightarrow
        (\mathbf 2,\mathbf 2)
        =
        S_L\otimes S_R,
        &
        (-1)
        &\mapsto
        +\mathbb 1_4,
        \\[1mm]
        \mathbf 4^{\mathrm{Dirac}}_{\operatorname{Spin(4)}}
        &\longleftrightarrow
        (\mathbf 2,\mathbf 1)
        \oplus
        (\mathbf 1,\mathbf 2)
        =
        S_L\oplus S_R,
        &
        (-1)
        &\mapsto
        -\mathbb 1_4.
    \end{aligned}
    }
\end{equation}
Thus, in the vector representation, the two $SU(2)$ doublets appear through a tensor product, whereas in the Dirac spinor they appear as two independent direct-sum sectors. Accordingly, our main object which is the $SU(4)$ fundamental irrep $\bm 4_{SU(4)}$ admits two inequivalent $\operatorname{Spin}(4)$ realizations: the vector one factors through $\operatorname{Spin}(4)\longrightarrow SO(4)\hookrightarrow SU(4)$, whereas the Dirac realization gives directly $\operatorname{Spin}(4)\hookrightarrow SU(4)$, without descending to $SO(4)$. 

As a last remark, it is worth emphasizing that, in the four-dimensional $SU(4)$ coherent-state problem considered here, $\operatorname{Spin}(4)$ does not provide the alternative homogeneous realization of $\mathbb{CP}^3$ that arises indeed from $\operatorname{Spin}(6)$. ts role here is instead representation-theoretic, through the different ways in which it can act on the four-dimensional $SU(4)$ space. In this line, the spinorial realizations discussed above nevertheless provide a useful contrast with the four-dimensional chiral spinor representation of $\operatorname{Spin}(6)$ appearing in the next section. 

\subsection{\texorpdfstring{$\operatorname{Spin}(6)$ and the $\mathbb{CP}^{3}$ coherent-state manifold}{Spin(6) and the CP3 coherent-state manifold}}

In the $\operatorname{Spin}(3)\simeq SU(2)$ case, the essential observation was that the normalized homogeneous-coordinate representative $\bm z\in\mathbb C^2$ of a point $[\bm z]\in\mathbb{CP}^1$ transforms in the fundamental $\mathbf 2$ of $SU(2)$, and therefore as a spinor of $\operatorname{Spin}(3)$. The coherent-state manifold, however, consists of rays rather than particular representatives, so the projective point $[\bm z]$ is insensitive to the transformation $\bm z\to-\bm z$ generated by the nontrivial element of the covering kernel. Consequently, although the representative transforms under $\operatorname{Spin}(3)$, the induced action on the coherent-state manifold descends to an ordinary $SO(3)$ action.

The $SU(4)$ fundamental representation provides a closely analogous starting point. In the main text, the coherent states were obtained explicitly as
\begin{equation}
    \ket{\mathbf{n}_4}
    =
    \begin{pmatrix}
        \cos \theta \\
        \sin \theta \,e^{i\varphi_1}\cos \xi_1 \\
        \sin \theta \,\sin\xi_1\,e^{i\varphi_2}\cos \xi_2 \\
        \sin \theta \,\sin\xi_1\,e^{i\varphi_3}\sin \xi_2
    \end{pmatrix}
    \in\mathbb C^4.
\end{equation}
Since the local Hilbert space is here the fundamental irrep $\mathbf 4_{SU(4)}$, we may simply regard this normalized coherent-state vector as a representative
\begin{equation}
    \bm z
    =
    \begin{pmatrix}
        z_0\\
        z_1\\
        z_2\\
        z_3
    \end{pmatrix}
    =
    \begin{pmatrix}
        \cos \theta \\
        \sin \theta \,e^{i\varphi_1}\cos \xi_1 \\
        \sin \theta \,\sin\xi_1\,e^{i\varphi_2}\cos \xi_2 \\
        \sin \theta \,\sin\xi_1\,e^{i\varphi_3}\sin \xi_2
    \end{pmatrix},
    \qquad
    \bm z^\dagger\bm z=1,
\end{equation}
of the homogeneous coordinates
\begin{equation}
    [\bm z]
    =
    [z_0:z_1:z_2:z_3]
    \in\mathbb{CP}^{3},
    \qquad
    \bm z\sim e^{i\alpha}\bm z.
\end{equation}
Thus, exactly as in the fundamental $SU(2)$ case, the normalized vector $\bm z$ is only a representative, whereas the actual coherent-state point is its projective class $[\bm z]$.

The exceptional relation $\operatorname{Spin}(6)\simeq SU(4)$
now gives this familiar $SU(4)$ object an additional interpretation. The fundamental $\mathbf 4_{SU(4)}$ of $SU(4)$ is one of the four-dimensional chiral spinor representations of $\operatorname{Spin}(6)$. Consequently, the same normalized vector $\bm z$ that appears above as the fundamental $SU(4)$ coherent-state representative can equivalently be regarded as a chiral $\operatorname{Spin}(6)$ spinor. The coherent-state point $[\bm z]$, on the other hand, is not itself a spinor vector, but rather the corresponding projective spinor ray. This distinction becomes important once the covering map $\pi:\operatorname{Spin}(6)\to SO(6)$ is taken into account. If $\rho_+$ denotes the chiral $\operatorname{Spin(6)}$ spinor irrep associated with our $\bm 4_{SU(4)}$, and $\{g,-g\}$ are the two lifts of the same $R\in SO(6)$, their actions on the chiral spinor representative differ by a sign,
\begin{equation}
\begin{gathered}
        \bm z
    \longrightarrow
    \rho_{+}(g)\bm z,
    \qquad
    \bm z
    \longrightarrow
    -\rho_{+}(g)\bm z\:,\\[3mm]
       [\rho_{+}(g)\bm z]
    =
    [-\rho_{+}(g)\bm z].
\end{gathered}
\end{equation}
However, after projectivization these two transformations are identical, and, although the representative $\bm z$ transforms genuinely under $\operatorname{Spin}(6)$, the induced transformation of the coherent-state point $[\bm z]\in\mathbb{CP}^3$ depends only on $R=\pi(g)$. In this sense, the action on the coherent-state manifold descends from $\operatorname{Spin}(6)$ to $SO(6)$. The natural question is then: Which phase-independent object constructed from the chiral spinor $\bm z$ realizes this descended $SO(6)$ action explicitly?

To answer this question, it is useful to return to the same construction used in the $\operatorname{Spin}(3)$ case and formulate it first in a way that applies equally to $SU(N)\simeq \operatorname{Spin}(n)$, with $N=2,4$ and $n=3,6$ respectively. Given a normalized homogeneous-coordinate representative $\bm z\in\mathbb C^N$, the most natural object that depends only on the projective point $[\bm z]$ is its rank-one projector \cite{BengtssonZyczkowski2017}
\begin{equation}\label{eq:CPN_rank_one_projector}
    P_{[\bm z]}
    :=
    \bm z\otimes\bm z^\dagger
    \in \mathbb C^N\otimes(\mathbb C^N)^*\simeq
    \operatorname{End}(\mathbb C^N).
\end{equation}
Accordingly, under $U\in SU(N)$, the projector transforms by conjugation, $P_{[\bm z]}\to U P_{[\bm z]}U^\dagger$. At the level of complex $SU(N)$ representations, the corresponding operator space decomposes as \cite{FultonHarris1991}
\begin{equation}\label{eq:fundamental_antifundamental_decomposition}
    \mathbf N\otimes\overline{\mathbf N}
    =
    \mathbf 1
    \oplus
    \mathbf{Adj},
\end{equation}
where the singlet is associated with the identity operator and the adjoint has dimension $N^2-1$. There is, however, an important point concerning the nature of the representation space relevant to the projector. Although Eq.\ref{eq:fundamental_antifundamental_decomposition} is conventionally written as a decomposition of complex representations, $P_{[\bm z]}$ is Hermitian. Choosing a basis of traceless Hermitian generators $T_A$, with $A=1,\ldots,N^2-1$, the projector can therefore be expanded as
\begin{equation}\label{eq:CPN_projector_adjoint_expansion}
    P_{[\bm z]}
    =
    \frac{1}{N}\mathbb 1_N
    +
    \sum_{A=1}^{N^2-1}
    q_A T_A,
    \qquad
    q_A\in\mathbb R.
\end{equation}
Thus, once the fixed trace has been separated, the phase-independent information carried by the projector is encoded in $N^2-1$ real coefficients,
$\bm q=(q_1,\ldots,q_{N^2-1})\in\mathbb R^{N^2-1}$, which mix among themselves under conjugation by $SU(N)$. This mixing is because the transformed generators can always be re-expanded in the same fixed basis, inducing a linear transformation among the $q_A$ coefficients [Explicitly, $UT_AU^\dagger=D_{BA}(U)T_B$, so that $q'_B=D_{BA}(U)q_A$]. In this way, \emph{the projector realizes the adjoint action on a real $(N^2-1)$-dimensional space}.

This mechanism was already present, although particularly simple, in the $\mathbb{CP}^1$ construction. For $N=2$, one has $\mathbf 2\otimes\overline{\mathbf 2}=\mathbf 1\oplus\mathbf 3$, and the three real coefficients are precisely the components $\mathrm n_a=\bm z^\dagger\sigma_a\bm z$ introduced above. Because $\operatorname{Spin}(3)\simeq SU(2)$, the real adjoint $\mathbf 3$ of $SU(2)$ coincides with the defining vector representation $\mathbb R^3$ of $SO(3)$. This special coincidence is what allowed the phase-independent projector to be expressed directly in terms of the ordinary vector $\mathbf n$.

\noindent For $N=4$, the same reasoning instead gives $\mathbf 4\otimes\overline{\mathbf 4}=\mathbf 1\oplus\mathbf{15}$. Hence, the nontrivial information contained in $P_{[\bm z]}$ is described by fifteen real coefficients $\bm q =(q_1,...,q_{15})\in \mathbb R^{15}$ forming the real adjoint representation of $SU(4)\simeq\operatorname{Spin}(6)$. Now, because the $\mathbb Z_2$-element $-1$ acts trivially in the adjoint representation $(-1)P_{[z]}(-1)=P_{[z]}$, this same $\mathbb R ^{15}$ representation of $SU(4)$ must descend to $SO(6)$. However, since the defining representation of $SO(6)$ is $\mathbb R^6$, the natural question is: What is the concrete fifteen-dimensional real representation of $SO(6)$ carried by the coefficients $(q_1,...,q_{15})$?

The answer follows naturally from the structure of the Lie algebra of $SO(6)$. Indeed, $\mathfrak{so}(6)$ is the real vector space of antisymmetric $6\times6$ matrices,
\begin{equation}
    \mathfrak{so}(6)
    =
    \left\{
        A\in M_6(\mathbb R)
        \;\middle|\;
        A^{\mathsf T}=-A
    \right\}.
\end{equation}
Such matrices are completely determined by their entries above the diagonal, and therefore contain $6(6-1)/2=15$ independent real components. Equivalently, these components may be regarded as those of an antisymmetric rank-two tensor $A_{ab}=-A_{ba}$, with $a,b=1,\ldots,6$. Hence, $\mathfrak{so}(6)\simeq \bigwedge\nolimits^2\mathbb R^6\simeq\mathbb R^{15}$. This gives precisely the real fifteen-dimensional space that we were looking for. To see how it carries a representation of $SO(6)$, let $R\in SO(6)$. Its natural action on $A\in\mathfrak{so}(6)$ is by conjugation, $A\to A'=RAR^{-1}=RAR^{\mathsf T}$. Since $A^{\mathsf T}=-A$, the transformed matrix remains antisymmetric, $(A')^{\mathsf T}=-A'$, and therefore remains inside $\bigwedge\nolimits^2\mathbb R^6\simeq \mathfrak{so}(6)$. This action of a group on its own Lie algebra is precisely its adjoint representation. In terms of the antisymmetric components, it is equivalently written as
\begin{equation}
        A_{ab}
        \longrightarrow
        A'_{ab}
        =
        R_{ac}R_{bd}A_{cd}\:,
\end{equation}
which makes clear why this representation is different from the defining vector one: an $SO(6)$ vector carries a single index and transforms as $x_a\to R_{ab}x_b$, whereas the adjoint object carries two antisymmetric vector indices and belongs to $\bigwedge^2\mathbb R^6$.

We can now return to the projector. Its traceless part, $Q_{[\bm z]}:=P_{[\bm z]}-\frac14\mathbb 1_4=\sum_{A=1}^{15}q_AT_A$, already carries the real adjoint representation of $SU(4)$. The isomorphism $\mathfrak{su}(4)\simeq\mathfrak{so}(6)$ therefore allows the same fifteen real coefficients to be reorganized into the fifteen independent components of an antisymmetric $SO(6)$ tensor,
\begin{equation}
        (q_1,\ldots,q_{15}) \in \mathbb R^{15}_{SU(4)}
        \quad\longleftrightarrow\quad
        A_{ab}=-A_{ba}\in\bigwedge\nolimits^{2}\mathbb R^6_{SO(6)}.
\end{equation}
No additional degree of freedom has been introduced in doing so: $Q_{[\bm z]}$ and $A_{ab}$ are simply two equivalent ways of representing the same phase-independent information associated with the original coherent-state ray $[\bm z]$. However, at this point, the correspondence between the two adjoint descriptions is still abstract: we know that the fifteen coefficients $q_A$ may be reorganized into an antisymmetric tensor, but we have not yet written this tensor directly in terms of the spinor representative $\bm z$. To do so, we can choose the basis of the fifteen traceless Hermitian $SU(4)$ generators so that it is adapted to the identification $\mathfrak{su}(4)\simeq\mathfrak{so}(6)$. Instead of the single adjoint label $A=1,\ldots,15$, we then label these generators by antisymmetric pairs $[ab]$, with $a,b=1,\ldots,6$, and denote them by $T_{ab}=-T_{ba}$. They can be chosen such that, for $g\in\operatorname{Spin}(6)$ and $R=\pi(g)$,
\begin{equation}\label{eq:Spin6_adjoint_generators_transformation}
    \rho_+(g)^\dagger
    T_{ab}
    \rho_+(g)
    =
    R_{ac}R_{bd}T_{cd}.
\end{equation}
Up to the fixed normalization of the generators, the same fifteen real coefficients can therefore be encoded directly in the spinor bilinears
\begin{equation}\label{eq:Spin6_spinor_tensor_map}
    \boxed{
        J_{ab}([\bm z])
        :=
        \bm z^\dagger T_{ab}\bm z,
        \qquad
        J_{ab}=-J_{ba}
    }.
\end{equation}
The projective phase cancels immediately, so $J_{ab}$ depends only on $[\bm z]$, while under $\bm z\to\rho_+(g)\bm z$ one obtains $J_{ab}\longrightarrow J'_{ab}=R_{ac}R_{bd}J_{cd}$. Thus, $J_{ab}$ is precisely the phase-independent antisymmetric $SO(6)$ tensor whose existence was inferred above. This is the direct analogue of $\mathrm n_a=\bm z^\dagger\sigma_a\bm z$ in the $\operatorname{Spin}(3)$ construction, except that the corresponding $SO(6)$ object now carries two antisymmetric vector indices rather than one.

 However, just as the vector $\mathbf n$ obtained in the $\mathbb{CP}^1$ construction did not fill the whole space $\mathbb R^3$, the tensor $J_{ab}$ obtained here cannot be an arbitrary element of $\bigwedge^2\mathbb R^6\simeq\mathbb R^{15}$. In the former case, the normalization of the spinor implied $\mathbf n^2=1$, so that the three real components of $\mathbf n$ were restricted to the two-dimensional sphere $S^2$, consistently with $\dim_{\mathbb R}\mathbb{CP}^1=2$. Here, the same dimensional argument tells us that the fifteen components of $J_{ab}$ must also satisfy constraints, since
$\dim_{\mathbb R}\mathbb{CP}^3=6$ and $\dim_{\mathbb R}\bigwedge\nolimits^2\mathbb R^6=15$. Thus, the problem is now to identify the six-dimensional subset of antisymmetric tensors that actually arises from the projective spinors $[\bm z]$.
Rather than determining these constraints independently for all fifteen components, it is simpler to use the $SO(6)$ symmetry. We choose one reference coherent-state ray $[\bm z_0]$, and denote its associated antisymmetric tensor by $J_0$. With a convenient choice of basis and normalization, $J_0$ can be brought to the form
\begin{equation}\label{eq:Spin6_reference_complex_structure}
    J_0
    =
    \begin{pmatrix}
        0&1&&&&\\
        -1&0&&&&\\
        &&0&1&&\\
        &&-1&0&&\\
        &&&&0&1\\
        &&&&-1&0
    \end{pmatrix},
    \qquad
    J_0^{\mathsf T}=-J_0,
    \qquad
    J_0^2=-\mathbb 1_6.
\end{equation}
The meaning of this matrix is particularly simple: it acts as a $90^\circ$ rotation independently in each of the three planes $(x_1,x_2)$, $(x_3,x_4)$, and $(x_5,x_6)$. Applying it twice therefore reverses every vector, which explains the relation $J_0^2=-\mathbb 1_6$. A real linear map with this property is called a \emph{orthogonal complex structure}  \cite{Salamon1989}, since it preserves the inner product $J_0^\mathsf TJ_0=\mathbb 1_6$ and plays on $\mathbb R^6$ the same algebraic role as multiplication by $i$, for which $i^2=-1$ and $i^{-1}=-i$. Indeed, one may combine the six real coordinates into
\begin{equation}
    w_1=x_1+i x_2,
    \qquad
    w_2=x_3+i x_4,
    \qquad
    w_3=x_5+i x_6,
\end{equation}
thereby regarding $\mathbb R^6$ as $\mathbb C^3$. Now, the usefulness of introducing this reference structure $J_0$ is that we do not need to construct the tensor $J$ independently for every coherent-state ray. Since $SU(4)\simeq\operatorname{Spin}(6)$ acts transitively on $\mathbb{CP}^3$, once a reference ray $[\bm z_0]$ has been chosen, any other point of the manifold can be written as
\begin{equation}
    [\bm z]
    =
    g\cdot[\bm z_0],
    \qquad
    g\in\operatorname{Spin}(6).
\end{equation}
At the same time, each ray determines its antisymmetric tensor through $J_{ab}([\bm z])=\bm z^\dagger T_{ab}\bm z$. Therefore, if $J_0$ is the tensor associated with $[\bm z_0]$, the tensor associated with the transformed ray $[\bm z]$ is obtained by the corresponding $SO(6)$ rotation $R=\pi(g)$. The two descriptions of the same transformation can be summarized as
\begin{equation}
\begin{array}{ccc}
[\bm z_0]
& \xrightarrow{\qquad g\in\operatorname{Spin}(6)\qquad} &
[\bm z]
\\[3mm]
\Big\downarrow
&&
\Big\downarrow
\\[-1mm]
J_0
& \xrightarrow{\qquad R=\pi(g)\in SO(6)\qquad} &
J=RJ_0R^{\mathsf T}.
\end{array}
\end{equation}
Thus, as the reference ray $[\bm z_0]$ is moved over the whole $\mathbb{CP}^3$ manifold, the corresponding $J_0$ is simultaneously moved through its $SO(6)$ orbit. This is precisely why the simple properties of the reference matrix become useful: every tensor obtained in this way inherits them. Indeed, using $R^{\mathsf T}R=\mathbb 1_6$,
\begin{equation}
    J^{\mathsf T}
    =
    RJ_0^{\mathsf T}R^{\mathsf T}
    =
    -J,
    \qquad
    J^2
    =
    RJ_0^2R^{\mathsf T}
    =
    -\mathbb 1_6.
\end{equation}
Consequently, the tensors associated with the coherent-state rays do not explore the whole fifteen-dimensional space $\bigwedge^2\mathbb R^6$. Rather, they lie on the particular $SO(6)$ orbit generated from $J_0$,
\begin{equation}
    \boxed{
    \begin{gathered}
        \qquad \mathbb{CP}^3\simeq\mathcal O_{J_0}
        =
        \left\{J=
            RJ_0R^{\mathsf T}
            \;\middle|\;
            R\in SO(6)
        \right\}
        \subset
        \bigwedge\nolimits^{2}\mathbb R^{6},\qquad
        \\[1mm]
        \text{whose elements satisfy }\hspace{2mm}J^{\mathsf T}
        =
        -J,
        \quad
        J^{2}
        =
        -\mathbb 1_{6}.
    \end{gathered}
    }
\end{equation}

It remains only to identify the space formed by all such $J$'s. Although every $R\in SO(6)$ produces a tensor $J=RJ_0R^{\mathsf T}$, different rotations need not produce different tensors. In particular, any rotation satisfying $RJ_0R^{\mathsf T}=J_0$ leaves the reference point unchanged. Since $R^{\mathsf T}=R^{-1}$, this condition is equivalent to $RJ_0=J_0R$. Thus, the transformations that leave $J_0$ fixed are precisely the rotations that commute with its complex structure. Once $\mathbb R^6$ is viewed as $\mathbb C^3$ through the coordinates $(w_1,w_2,w_3)$, commuting with $J_0$ means that the transformation is complex linear. At the same time, because $R\in SO(6)$, it preserves the Euclidean norm; in the complex description, this becomes preservation of the Hermitian norm, and therefore, such transformations form $U(3)$. Hence,
\begin{equation}
    \operatorname{Stab}_{SO(6)}(J_0)
    =
    U(3).
\end{equation}
and consequently, the different tensors $J$ are obtained from all of $SO(6)$, modulo those $U(3)$ transformations that leave $J_0$ unchanged \cite{BerkovitsCherkis2004},
\begin{equation}\label{eq:SO6_U3_complex_structure_orbit}
    \boxed{
        \mathbb{CP}^3\simeq\left\{
            J=RJ_0R^{\mathsf T}
        \right\}
        \simeq
        \frac{SO(6)}{U(3)}
    }.
\end{equation}
The dimension makes the identification with the original coherent-state manifold transparent, $\dim SO(6)-\dim U(3)=15-9=6 =\dim_{\mathbb R}\mathbb{CP}^3$. Thus, in direct analogy with the $\mathbb{CP}^1$ construction, the projective spinor ray may be represented by a constrained real $SO(6)$ object: for $\mathbb{CP}^1$ this was the unit vector $\mathbf n\in S^2$, while for $\mathbb{CP}^3$ it is the orthogonal complex structure $J\in\bigwedge^2\mathbb R^6$.

This geometric identification also clarifies why the $SO(6)$ description is useful beyond the group-theoretical correspondence itself. In the $\mathbb{CP}^1$ case, the same local degree of freedom could be described either by the projective spinor $[\bm z]$, with $\bm z\sim e^{i\alpha}\bm z$, or by the gauge-invariant unit vector $\mathbf n=\bm z^\dagger\boldsymbol{\sigma}\bm z$. This is precisely the relation underlying the two familiar formulations
\begin{equation}
    \mathbb{CP}^1
    \qquad\longleftrightarrow\qquad
    S^2\simeq \frac{SO(3)}{SO(2)},
\end{equation}
and, at the level of continuum fields, $\bm z(x)
    \longleftrightarrow\mathbf n(x)\:\:[
    \mathbf n^2(x)=1]$, which allow the same sigma model to be expressed in the $\mathbb{CP}^1$ or $O(3)$ language.

The same construction applies for the $SU(4)$ analogue. A slowly varying configuration of coherent-state rays may be described either by $(\mathrm{I})$ a normalized four-component spinor,
\begin{equation}
\begin{cases}
    (\mathrm{I}) \hspace{5mm}\bm z(x)\in\mathbb C^4,
    \quad
    \bm z^\dagger\bm z=1,
    \quad
    \bm z(x)\sim e^{i\alpha(x)}\bm z(x),\\[2mm]
    (\mathrm{II}) \hspace{4mm}J_{ab}(x)
    =
    \bm z^\dagger(x)T_{ab}\bm z(x),
    \quad
    J^{\mathsf T}(x)=-J(x),
    \quad
    J^2(x)=-\mathbb 1_6,
\end{cases}
\end{equation}
or, equivalently, by $(\mathrm{II})$ the real antisymmetric field $ J(x)$ whose values lie on $SO(6)/U(3)$. In particular, the $J$-description removes the arbitrary local phase of the spinor from the outset and makes the descended $SO(6)$ transformation law explicit. This can also be seen directly in the sigma-model kinetic term. In the projective description one may write
\begin{equation}
    \mathcal L_{\mathbb{CP}^3}
    \propto
    (D_\mu\bm z)^\dagger D_\mu\bm z,
    \qquad
    D_\mu\bm z
    =
    \left(
        \partial_\mu-\bm z^\dagger\partial_\mu\bm z
    \right)\bm z,
\end{equation}
or, equivalently, first in terms of the phase-independent projector and then in terms of its $SO(6)$ representative,
\begin{equation}
    \mathcal L_{\mathbb{CP}^3}
    \propto
    \operatorname{Tr}
    \left(
        \partial_\mu P\,\partial_\mu P
    \right)
    \propto
    \operatorname{Tr}
    \left[
        (\partial_\mu J)^{\mathsf T}
        (\partial_\mu J)
    \right],
\end{equation}
where the proportionality constants depend only on the normalization chosen for the generators $T_{ab}$. The dynamical field can therefore be viewed either as a projective chiral spinor or as a constrained real $SO(6)$ tensor. In this sense, the exceptional relations $\operatorname{Spin}(3)\simeq SU(2)$ and $\operatorname{Spin}(6)\simeq SU(4)$ lead to two closely parallel descriptions of the corresponding coherent-state manifolds,
\begin{equation}
\begin{aligned}
SU(2):\qquad
&\mathbb{CP}^1
\simeq
\frac{SO(3)}{SO(2)},
&&
[\bm z]\longleftrightarrow\mathbf n,
\\[2mm]
SU(4):\qquad
&\mathbb{CP}^3
\simeq
\frac{SO(6)}{U(3)},
&&
[\bm z]\longleftrightarrow J.
\end{aligned}
\end{equation}
The second formulation is therefore not merely an alternative parametrization of $\mathbb{CP}^3$: it provides a real, phase-independent language in which the $SO(6)$ symmetry and the geometry of the target space are manifest, and can consequently be useful when formulating continuum actions, symmetry constraints, order parameters, and possible geometric or topological terms.

%% file: final_remarks.tex
\section*{Final Remarks: Discussion and Perspectives}
\phantomsection
\addcontentsline{toc}{section}{Final Remarks: Discussion and Perspectives}

The discussion throughout these notes has been guided by a rather simple question: how much of the familiar geometric description of an $SU(2)$ spin survives once the local quantum degree of freedom becomes more general? Starting from the Bloch sphere, its Berry phase, and the corresponding Wess--Zumino term, we have followed this question through several complementary descriptions of $SU(N)$ systems, involving coherent-state manifolds, representation theory, parton constructions, multipolar and spin--orbital bases, as well as the special low-rank relations with orthogonal and Spin groups. Rather than representing independent constructions, these viewpoints emphasize different aspects of the same finite-dimensional quantum degrees of freedom---their symmetry, geometry, topology, or physical observables---and one of the main aims of these notes has therefore been to make the relations between them explicit. In this sense, the different approaches developed here may be regarded as a dictionary between languages that often appear separately in the literature, but which become especially useful when one tries to move between microscopic models, effective field theories, and geometrical descriptions of quantum states.

One place where such a dictionary becomes particularly useful is in systems whose local Hilbert space contains more structure than can be captured by an ordinary three-component magnetic moment. Higher-spin, spin--orbital, valley, and locally entangled degrees of freedom naturally bring in quadrupolar, octupolar, and more general multipolar observables, for which $SU(N)$ coherent states provide a systematic classical or semiclassical description. In this direction, Landau--Lifshitz dynamics has already been generalized from $SU(2)$ to $SU(N)$ \cite{classical_spin_Zhang}, while enlarged spin--orbital Hilbert spaces have also provided a natural setting for fractionalized phases and unconventional quantum criticality \cite{SeifertEtAl2020}. More recent examples include exchange-frustrated quadrupoles described through $SU(3)$ flavor variables \cite{SarkerMaSeifert2026}, $\mathbb{CP}^{2}$ skyrmions carrying intertwined dipolar and quadrupolar structure \cite{ZhangEtAl2023CP2}, and entanglement-preserving $SU(4)$ coherent states supporting $\mathbb{CP}^{3}$ skyrmion textures in frustrated dimer systems \cite{WilliamsEtAl2025}. Likewise, $SU(4)$ descriptions of $j=3/2$ Mott systems naturally organize dipolar, quadrupolar, and octupolar degrees of freedom \cite{HartSutcliffeParamekanti2026}, while the dynamics of generalized $\mathbb{CP}^{N-1}$ textures is itself now being developed for arbitrary $N$ \cite{LeeJeonHan2025}. From this perspective, the passage from the Bloch sphere to $\mathbb{CP}^{N-1}$ is not merely a formal enlargement, but provides a natural phase-space language for increasingly relevant forms of generalized quantum matter.

The same projective viewpoint also reaches considerably beyond generalized magnetism. Passing from a state vector to its projective class, or equivalently to the phase-independent projector $P=\ket{\psi}\bra{\psi}$, is precisely the starting point from which Berry curvature and quantum metric can be formulated in general multiband systems \cite{GrafPiechon2021,MitscherlingAvdoshkinMoore2025}. For a single state, this geometry is encoded by $\mathbb{CP}^{N-1}=\operatorname{Gr}(1,N)$, whereas isolated collections of states lead naturally to higher-rank projectors and Grassmannian manifolds \cite{MeraMitscherling2022}. Thus, many of the structures encountered here in the coherent-state context---projective equivalence, Berry connections, Fubini--Study geometry, Chern classes, and projector representations---also form part of the modern language of quantum geometry. From a complementary field-theoretical point of view, coherent-state homogeneous spaces, line bundles, topological terms, and the global distinction between symmetry groups and their covering groups continue to organize the relation between quantum spin systems and nonlinear sigma models \cite{ReadSaleur2026}. The projective, bundle, and $\operatorname{Spin}(n)$ constructions discussed in these notes may therefore be viewed as parts of a broader geometric framework linking coherent-state dynamics, quantum geometry, and topological field theories.

Finally, these connections point toward a more forward-looking possibility: the geometry of quantum-state space need not remain only a language for describing quantum systems, but may itself become something to manipulate and exploit. Recent proposals such as the \emph{Entanglemons} construction use generalized coherent states on $\mathbb{CP}^{3}$ to encode protected qubits in entangled internal degrees of freedom \cite{entanglemons}, while experiments on generalized Pancharatnam--Berry phases have demonstrated that geometric phases, and even transitions in their topological character, can be controlled through measurement protocols \cite{FerrerGarciaEtAl2023}. In parallel, the passage from the Abelian Berry phase associated with a single ray to non-Abelian holonomies of degenerate subspaces provides a natural geometric route toward holonomic quantum control \cite{ZhangEtAl2023Holonomic}. In this sense, the projective manifolds, Wess--Zumino terms, multipolar structures, and gauge redundancies encountered throughout these notes may be useful not only for describing generalized quantum matter, but also for understanding how geometry and topology can become active ingredients in future quantum-information settings.

%% file: acknoledgements.tex
\section*{Acknowledgments}
\phantomsection
\addcontentsline{toc}{section}{Acknowledgments}
The work of M.K. was carried out in part at the Aspen Center for Physics, which is supported by the Simons Foundation (Grant No. 1161654, Troyer). We are grateful to D. N. Aristov, R. Moessner, F. von Oppen, Y. Gefen, and M. Scheurer for fruitful discussions, valuable comments, and helpful insights on different aspects of the topics addressed in these notes. Their perspectives contributed to clarifying several conceptual points and to broadening the range of connections and applications discussed throughout the manuscript.